# π

# Multimodal Planetary Defense

Summary version submitted to ASR 1/22, accepted 9/22, published 2/23
https://www.sciencedirect.com/science/article/pii/S0273117722009395
Full (this) version linked in ASR: https://arxiv.org/abs/2110.07559 - First version 10/14/21 Current Update: Ver 42c42c – 10-7-26

**Philip Lubin - lubin@ucsb.edu**
**Physics Department, UC Santa Barbara, Santa Barbara, CA 93106**

*Abstract*— We present a practical and effective method of planetary defense that allows for extremely short mitigation time scales if required as well as long time scale mitigation. The proposed system allows a planetary defense solution using existing technologies. This one system allows for virtually any required defense mode. In general, it uses much less launch mass with vastly shorter required response time than classical deflection techniques. The method involves an array of small hypervelocity kinetic penetrators that pulverize and disassemble an asteroid or small comet. The penetrators can be passive for most cases as well as active explosive penetrators in extreme cases. In a short warning terminal mode, the threat is mitigated using the Earth's atmosphere to dissipate the energy in the fragment cloud. In this terminal mode the asteroid fragments of maximum ~<10-meter diameter allow the Earth's atmosphere to act as a "beam dump" where the fragments either burn up in the atmosphere and/or air burst, with the primary channel of energy going into spatially and temporally de-correlated shock waves. It is the de-correlated shock waves that are the key to why PI works so well in terminal situations. In longer warning time intercept scenarios, the disassembled asteroid fragments largely miss the Earth. Compared to other threat reduction scenarios, such as deflection, this approach requires much less launch mass and allows an extremely rapid response as well as using one system for virtually all cases. PI is a testable, and deployable approach with a logical roadmap of development. PI allows for effective defense against bolides up to ~1km diameter class and could virtually eliminate the threat of mass destruction caused by these threats. Terminal intercepts, closer than the Moon with intercept times of less than a few hours prior to impact are viable depending on the target size. As an example, though NOT recommended, we show that with only ~1m/s internal disruption, a 5 hour prior to impact intercept of a 50m diameter asteroid (~10Mt yield, similar to Tunguska), a few day prior to impact intercept of 100m diameter asteroid (~100Mt yield), or a 10 -30 day prior to impact intercept of Apophis (~350m diameter, ~ 2 Gt ave yield – orbit dependent) would mitigate these threats. Mitigation of a 1km diameter threat with a 60-day intercept is also viable. We show that a 20m diameter asteroid (~0.5Mt, similar to Chelyabinsk) can be mitigated with a 100 second prior to impact intercept with a 10m/s disruption and 1000 second prior to impact with a 1m/s disruption. Zero-time intercept of 20m class objects are possible due to atmospheric dispersion effects. The product of the mitigation time and the disruption speed is the key metric. Larger disruption speeds allow for shorter mitigation times. Using the Moon as a planetary defense outpost with both detection as well as mitigation (launch) capability is an option to be considered for the future as are LEO, MEO and GEO deployment. The Moon is nearly ideal, given the lack of atmosphere, allowing for long range optical/NIR LIDAR detection, and the low escape speed allows for rapid launch and interception capability using existing solid fuel boosters. For an Earth launch-based system, we show that a single heavy lift launcher such as a Falcon Heavy, Starship, SLS etc. can conservatively mitigate a multi-hundred-meter diameter asteroid such as Apophis or Bennu with passive penetrators alone. Even a smaller launcher with positive $C_3$ such as an expendable Falcon 9, can mitigate 100m class threats with passive penetrators and 1km class with passive penetrators + NED's. The system is essentially identical in both the passive only and passive + active (including NED) cases. This allows for a pre-deployed (at the ready) single launcher solution to planetary defense where only the penetrators change from passive to active for extreme threats. Pro-active mitigation of recurring threats is also an option. Having such a capability would allow humanity for the first time to take control over its destiny relative to asteroid and comet impacts. Pre-deployment of the system is key to a robust planetary defense system.

# Contents

# 1. Introduction

### *Asteroid and Comet Impact Threat*

Asteroid and comet impacts pose a continual threat to humanity. For example, on February 15, 2013, an asteroid penetrated the atmosphere over Chelyabinsk, Russia, entering at a horizon angle of approximately 18°, and releasing energy equivalent to 570 ± 150kt TNT [1]. For comparison, the nuclear weapon that was detonated approximately 0.5km above the ground in Hiroshima, Japan yielded approximately 12.5kt TNT [2]. The main airburst over Chelyabinsk occurred at an approximate altitude of 30km and created a shock wave strong enough to shatter windows out to a distance of 120km from the asteroid track, injuring over 1,200 people in Chelyabinsk city and hundreds more in nearby towns and rural areas [1]. Had the asteroid approached from a higher angle, more serious damage would have occurred from higher concentration of the impact blast energy on the ground.

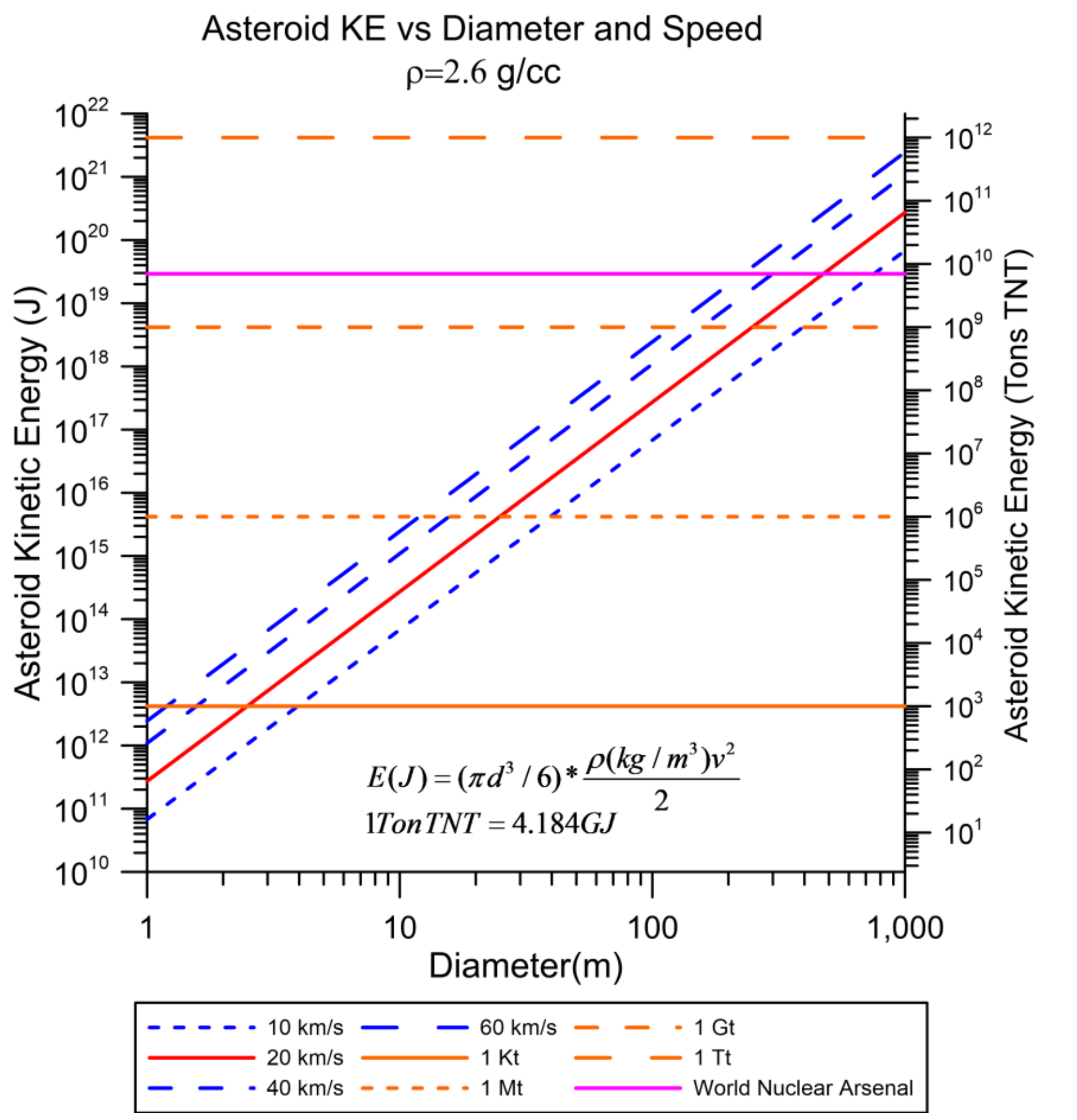

**Figure 1 –** Exo-atmospheric kinetic energy (KE) vs diameter and atmospheric entry speed. Nuclear weapon yields and total world nuclear arsenal shown for comparison.

Sixteen hours after the meteorite struck near Chelyabinsk, the 45m diameter asteroid 2012 DA14 approached to within 27,743km of Earth's surface—inside the orbit of geosynchronous satellites. If DA14 were to strike Earth, it would deliver approximately 7.2Mt TNT [3]. Although the Chelyabinsk meteorite and DA14 arrived at or near Earth on the same day, the two objects were not linked to each other, coming from completely unrelated orbits. That two such seemingly improbable events could occur within hours of each other serves as a stark reminder that humanity is continually at risk of asteroid impact.

Asteroids at least the size of DA14 (~50m diam.) are expected to strike Earth approximately every 650 years, while objects at least the size of the Chelyabinsk asteroid (~20m diam.) are expected to strike Earth approximately every 50-100 years [3]. For example, the Tunguska 1908 event could have caused large scale loss of life, but did not due to the remote area in Russia it airburst over. This event is estimated to have been a roughly 65m diameter asteroid (or possibly an atmospheric grazing comet) with an air blast energy yield of between 3 and 30Mt. This event flattened more than 2000km$^2$ of forest as seen from the 1927 Soviet Academy of Sciences expedition. Larger objects also pose a severe threat, as the total kinetic energy assuming a 20km/s impact speed associated with an impact of a 100m asteroid is equivalent to approximately 100Mt TNT, and that of the well-known 350m threat, like Apophis, would have an impact yield of approximately 1-4Gt (depending on orbital phase) TNT [3], or about ½ of the Earth's total nuclear arsenal, while a 500m threat, like Bennu, could have a yield greater than the entire Earth's nuclear arsenal. For reference, Apophis will next visit Earth on Friday April 13, 2029 and come within the geosync belt. Effective mitigation strategies are imperative to ensure humanity's continuity and future advancement.

Every day, approximately 100 tons of small debris impact the Earth and the effect is virtually undetectable. If this consisted instead of a single 100-ton asteroid with an effective diameter of about 4 meters, the effect would be quite noticeable, but not a serious threat. This is the basic idea behind our approach: reduce the threat to a large number of small objects that primarily burn up or airburst in the atmosphere, and even if small remnants impact, as in a short response threat scenario for high density asteroids, they will produce very little damage compared to the same mass in the parent asteroid. Extremely large asteroids (>1km diameter) and comets pose unique threats that we briefly discuss as a part of this program. Existential threats to humanity are very low recurrence, but are known to have happened multiple times in the past including the last mass extinction approximately 65Myr ago when an estimated 10km diameter asteroid triggered large scale life extinction, including the destruction of dinosaurs. The equivalent yield of this event is estimated to have been approximately 100 Tera-tons TNT, or about 15,000 times larger than the current nuclear arsenal of the world. **While we focus initially on asteroids, the same basic technique works with comets, though the size and speed of comets pose unique challenges. We discuss comet threats later.**

***Airburst and Energy Deposition*** **–** Stony asteroids with diameters smaller than about 85m will generally airburst in the atmosphere depending on their composition, speed, and angle of impact [4]. While this may seem to alleviate the problem as the bulk of the asteroid never reaches the Earth's surface, the problem is that the airburst altitude decreases with increased size and the damage from the airburst shock waves can be extremely dangerous, as was the case in the 1908 Tunguska (diameter estimates from 50 to 190m and 3-30Mt yield, 5-10km airburst altitude) and 2013 Chelyabinsk (~20m diameter, 19km/s, 500kt yield, 30km airburst altitude) events.

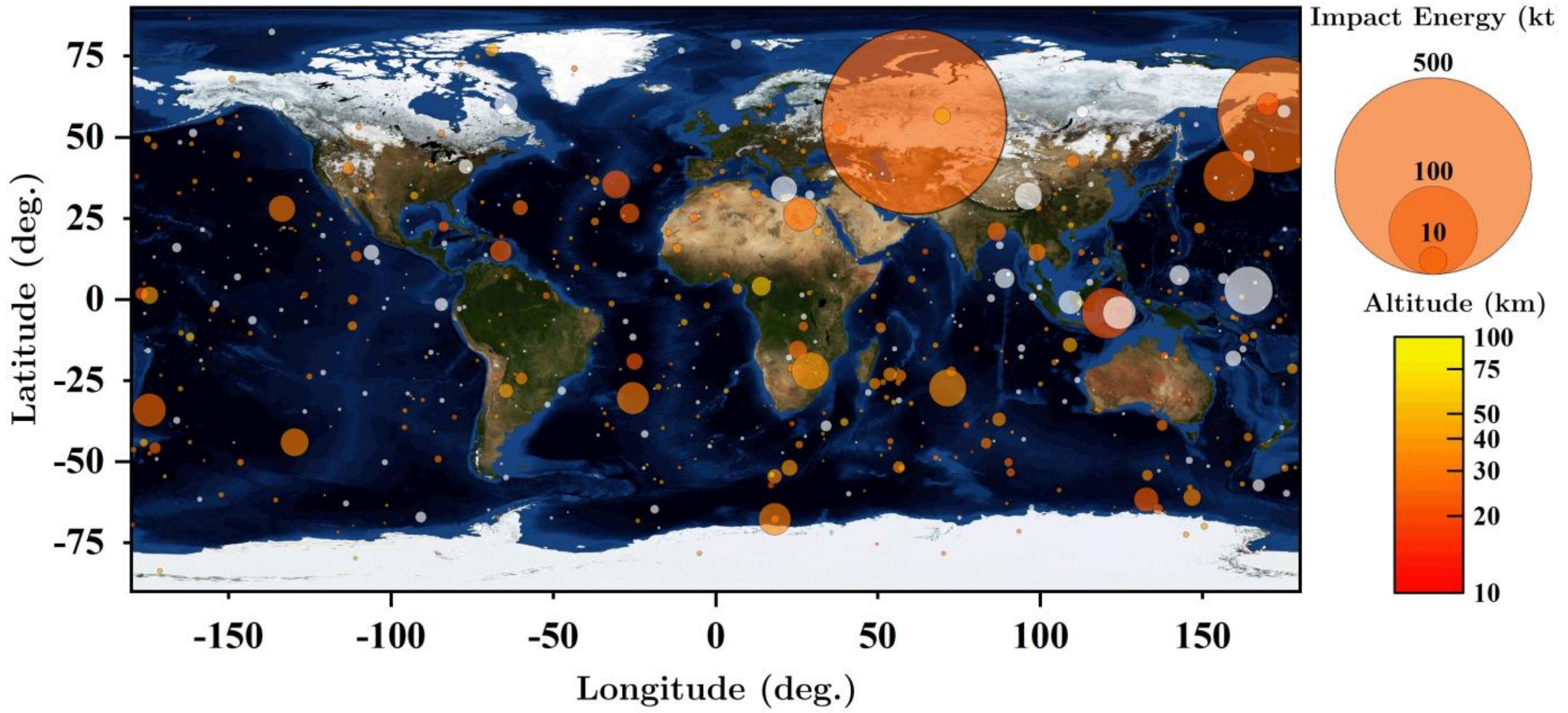

**Figure 2 –** Map of recent 873 events greater than 0.073 Kt from April 15, 1988 to Sept 29, 2021 air burst impacts from atmospheric infrasonic sensors. The Feb 15, 2013 Chelyabinsk 500 Kt event is clearly seen over Russia. It is important to note that the energy ranges of many of these events of asteroid strikes are in the ranges of tactical to strategic nuclear weapons. Data is through Sept 29, 2021. White colored point lack altitude data. Data from Alan B. Chamberlin (JPL/Caltech) **https://cneos.jpl.nasa.gov/fireballs/.**

### *Mitigation Methods*

Several concepts for asteroid deflection have been described, which can be broadly generalized into six distinct strategies:

(1) Kinetic penetrators, with or without explosive charges: an expendable spacecraft is sent to intercept the threatening object. Direct impact would modify the object's orbit through momentum transfer. Enhanced momentum transfer can be accomplished using an explosive charge, such as a nuclear weapon [5]–[8].

(2) Gradual orbit deflection by surface albedo alteration: the albedo of an object could be changed using paint [9], mirrors [10], sails [11], *etc*. As the albedo is altered, a change in the object's Yarkovsky thermal drag would gradually shift the object's orbit.
(3) Direct motive force, such as by mounting a thruster directly or indirectly to the object: thrusters could include chemical propellants, solar or nuclear powered thermal or electric drives, or ion engines [12].
(4) Indirect orbit alteration, such as gravity tractors: a spacecraft with sufficient mass would be positioned near the object, and maintain a fixed station with respect to the object using onboard propulsion. Gravitational attraction would tug the object toward the spacecraft, and modify the object's orbit [13], [14].
(5) Expulsion of surface material, *e.g.*, by robotic mining: a robot on the surface of an asteroid would repeatedly eject material from the asteroid. The reaction force from ejected material affects the object's trajectory [15].
(6) Vaporization of surface material: similar to robotic mining, vaporization on the surface of an object continually ejects the vaporized material, creating a reactionary force that pushes the object into a new path. Vaporization can be accomplished by solar concentrators [16] or by lasers [17] deployed on spacecraft stationed near the asteroid, the latter of which is proposed for the DE-STARLITE mission [18] or done using long range stand-off laser ablation (DE-STAR) [19]. During laser ablation, the asteroid itself becomes the "propellant"; thus a very modest spacecraft can deflect an asteroid much larger than would be possible with a system of similar mission mass using alternative techniques.

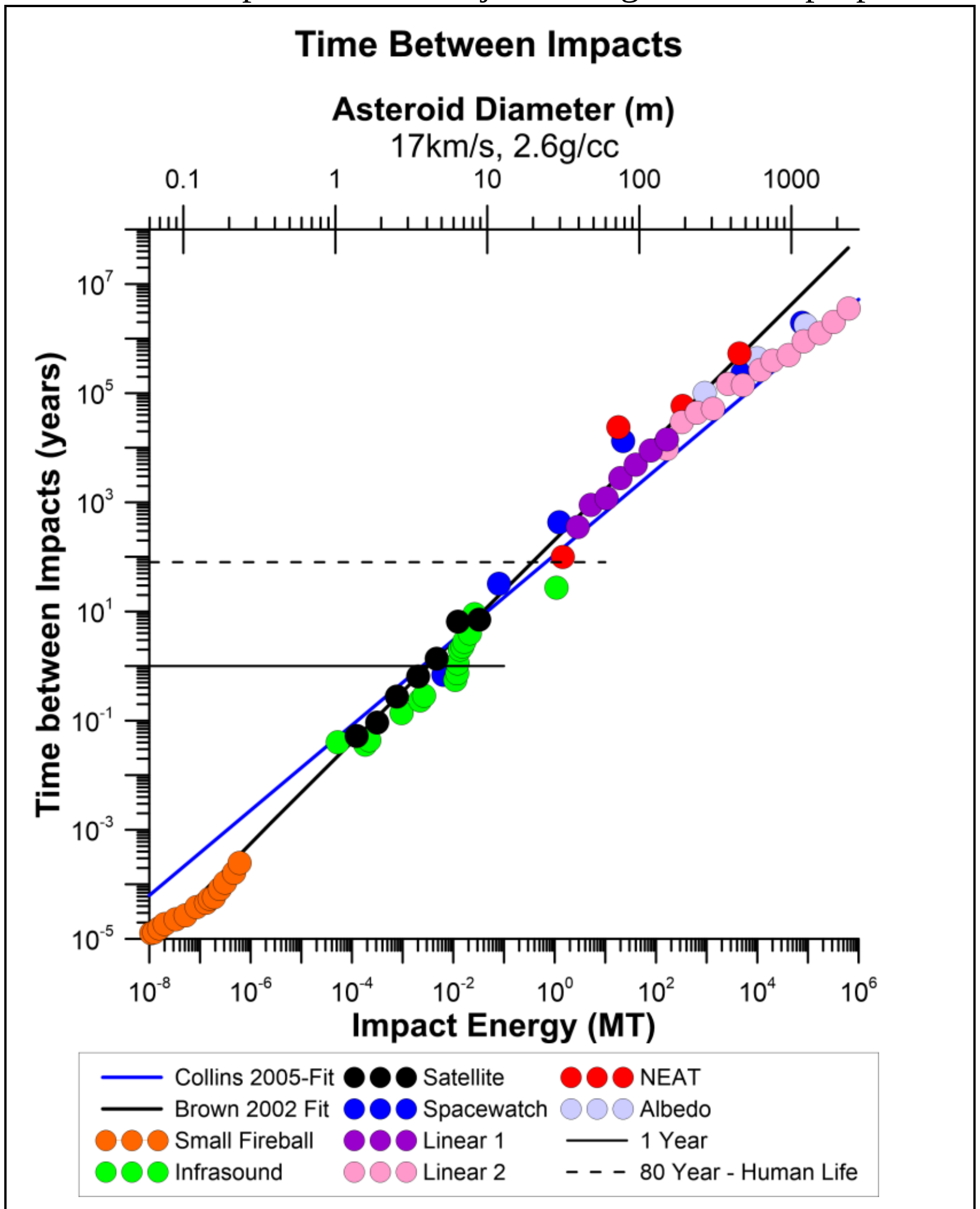

**Figure 3** – Approximate power law relationship between impact recurrence times and exo-atmospheric impact KE. Based on fitting known and extrapolated impacts from [23],[24].

## 2. Gravitational disassembly

***Gravitation Escape Speed and Disassembly Energy*** – The key to the effective use of the proposed program is the fracturing and disassembly of the parent asteroid. For this to be effective, we must overcome the self-gravitational binding energy. As will be seen, the energy required to do so is relatively small. The escape speed for a body of uniform density ρ (kg/m$^3$) with radius R (m) is:

$$v_{esc} = \sqrt{2GM/R} = \sqrt{\frac{8\pi}{3}G\rho}\,R \text{ or about } v_{esc} \simeq 2.36\times10^{-5}\sqrt{\rho}R$$

As an example, for R=100 (200m diameter) and ρ=2000 kg/m$^3$ we get $v_{esc}$ ~ 0.11 m/s.

The total gravitational binding energy of the asteroid is: $E_{BE} = 3/5\frac{GM^2}{R}$, or assuming constant density:

$$E_{BE} = \frac{16}{15} G\pi^2 \rho^2 R^5 \sim 7.02x10^{-10} \rho^2 R^5 \sim 2.19x10^{-11} \rho^2 d^5 \propto d^5$$

Note that if we calculate the KE of the entire asteroid mass at speed $v_{esc}$ :

$$KE(v = v_{esc}) = \frac{1}{2} M\, v_{esc}^{\;2} = \frac{1}{2} M \frac{8\pi}{3} G\rho R^2 = \frac{16}{9} G\pi^2 \rho^2 R^5 = \frac{5}{3} E_{BE} \text{ or } E_{BE} = \frac{3}{5} KE(v = v_{esc})$$

Note the $5^{th}$ power scaling of gravitational binding energy with diameter *d*. This is critical in understanding the scales at which various mitigation methods are feasible for a given diameter and disassembly energy.

As an example, for R=100 (200m diameter) and ρ=2000 kg/m$^3$, we get: $G_{BE} \sim 2.81x10^7 \sim 28MJ$. For comparison, one metric ton of TNT has an energy release defined to be 4.184GJ (4.2MJ/kg) or an equivalent speed of $(2E/m)^{0.5}$ =2.9km/s. Thus, a 200m diameter asteroid has a gravitational binding energy of about 6.5 kg (14 lbs) of TNT. If we assume an Apophis like asteroid has a diameter of 370m and a density of 2.6g/cc, we get a gravitational binding energy of 1.03GJ, or 240kg (540lbs) of TNT and a gravitational escape speed of 0.22m/s. These are very modest energy levels required for gravitational binding.

Another way of thinking about the scale here is to imagine a 20km/s (relative) impact speed with a 10 metric ton ($10^4$ kg) impacting payload. This has a kinetic energy (relative to the asteroid) of about $2x10^{12}$J, or about 480 tons TNT equivalent. **A one kg penetrator at a speed of 20 km/s is equivalent to about 48 kg of TNT in energy or 48x times its mass in equivalent explosive (TNT) energy.** Thu using conventional explosives only adds marginally to the kinetic energy at high closing speed (>10km/s) since the **detonation speed (which is different from and higher than the equivalent speed of $(2E/m)^{0.5}$) of chemical explosives is limited to about 10km/s (DDF, ONC) with TNT being about 6.94 km/s.** However, explosive penetrators could be extremely useful in distributing the energy laterally and in deep penetration modes IF the explosive can survive the impact in a functional state. **In low speed intercepts an explosive filled penetrator could also add significantly to the total energy.** Note that the disassembly energies are extremely low even for km-class asteroids.

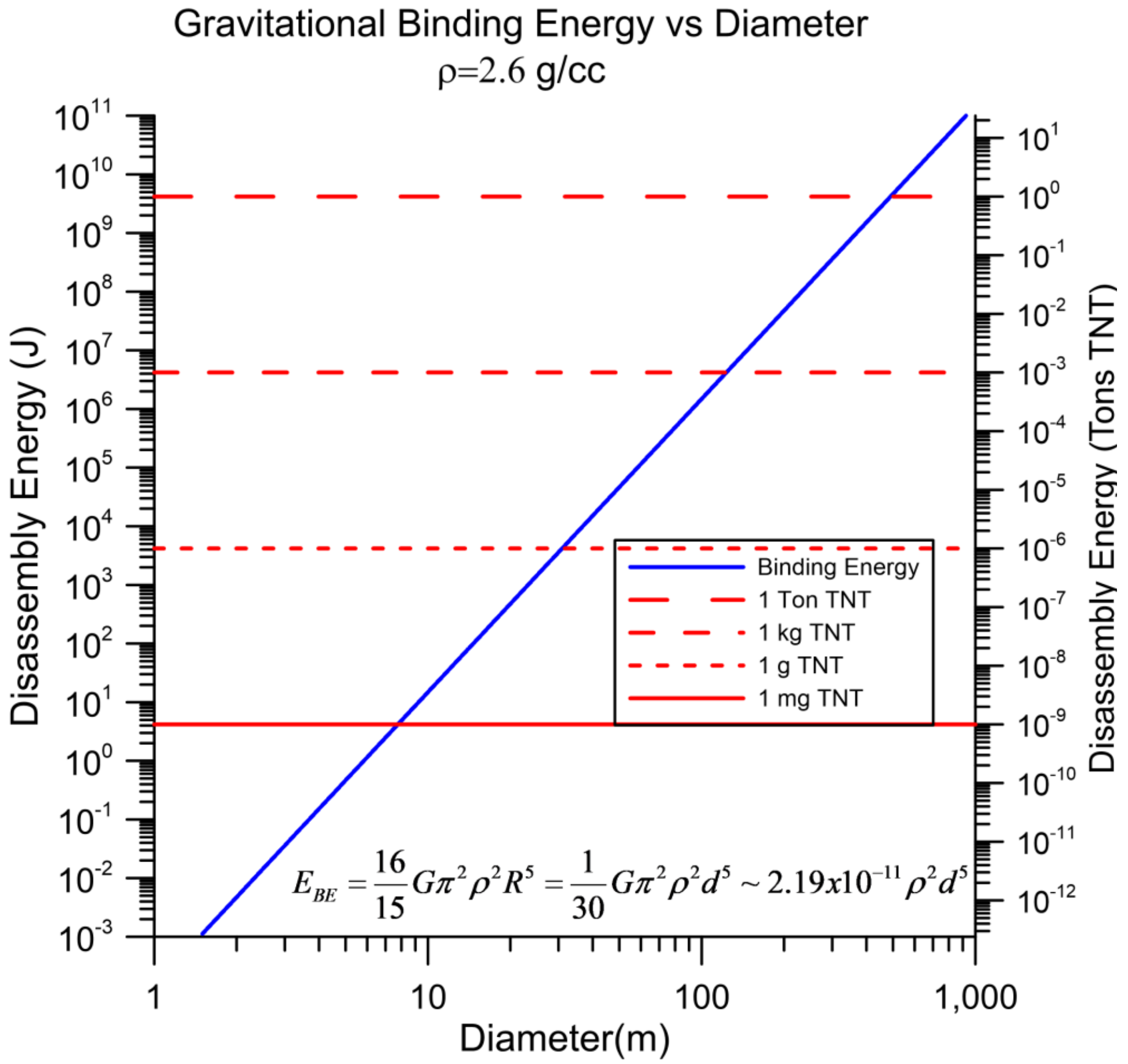

**Figure 4** - Gravitational binding energy vs diameter for density of 2.6g/cc. Note that the binding energy scales as the $5^{th}$ power of the diameter. This scaling is critical to the success of our method. Even for asteroids of 1km diameter the binding energy is only the equivalent of roughly 20 tons of TNT.

***Energy vs Momentum Transfer*** – The energy for the gravitational disassembly and for gravitational escape are relatively small, but the effective transfer of the delivered energy to the bulk macroscopic kinetic energy of the asteroid fragments is critical. The way in which the KE of the impact is converted into bulk fragment KE rather than into heat is the critical physics that must be considered and designed into an effective system. The conversion of the energetic high-speed impacts or explosives into the low-speed disassembly is one of the fundamental design constraints for a real system to work well (i.e. energy delivery is NOT the problem). The problem is the effective conversion of the delivered energy into bulk low speed kinetic energy of the resultant fragments. Common examples of this problem are airbags in cars and explosives (both conventional and nuclear) and chemical explosives used in mining and road construction in mountains. The example of an air bag and a nuclear explosion are extremely relevant. Both couple a rapid (near instantaneous) energy delivery with extremely high speed in the initial explosive. The conversion into "useful," much lower speed motion is done using gas as the mediator ($N_2$ in the case of an airbag and $N_2$, $O_2$ in the case of a chemical or nuclear weapon). This point is critical. Initial extremely high speed (high Mach #) dynamics must be efficiently converted into low-speed motion to be efficient. Our program therefore splits into two basic parts:

- The dynamics of fracturing and spreading the fragments of the parent asteroid;
- The interaction of the fragments with the Earth's atmosphere.

The first part is analogous to "strip mining" and road building, both of which use high Mach # chemical explosives that must be coupled to very slow Earth (rock etc.) motion. The second part is the hypersonic (high Mach #) shock wave of the asteroid fragments in the Earth's atmosphere. This second part of the program is very analogous to that of atmospheric nuclear weapons detonations during the testing programs of the 1940-1960's, both in energy yields and in the coupling of extremely high Mach # physics into low Mach # dynamics. We will use the data from these test programs extensively in our analysis. One point to note is that a modest altitude atmospheric (air burst) nuclear weapon delivers about 50% of its yield into the subsequent atmospheric bulk motion (wind) and shock wave at least initially. This will turn out to be analogous to the conversion of the asteroid KE into the subsequent atmospheric shock wave that delivers most of the destruction from an asteroid. Typical asteroids arrive with the equivalent of Mach 30 for a 10km/s bolide (lower limit to asteroid speed) to Mach 60 for a 20km/s one (similar to the Chelyabinsk event) to Mach 100 for a 30km/s one and Mach 200 for comets. As we will see, the lower Mach numbers are comparable to that of a 1Mt airburst at a distance of about 100m from the detonation. In our case, any arbitrary observer will be vastly further from the asteroid airburst "detonation site" and we will see that the subsequent shock speed for relevance in our case will be essentially the speed of sound (Mach 1) with the fraction of initial energy in the observed shock wave being much less than initially due to other atmospheric energy channels, especially the bulk motion of air behind the shock wave (the wind). The relative energy in the shock wave and in the wind is key to understanding energy dissipation.

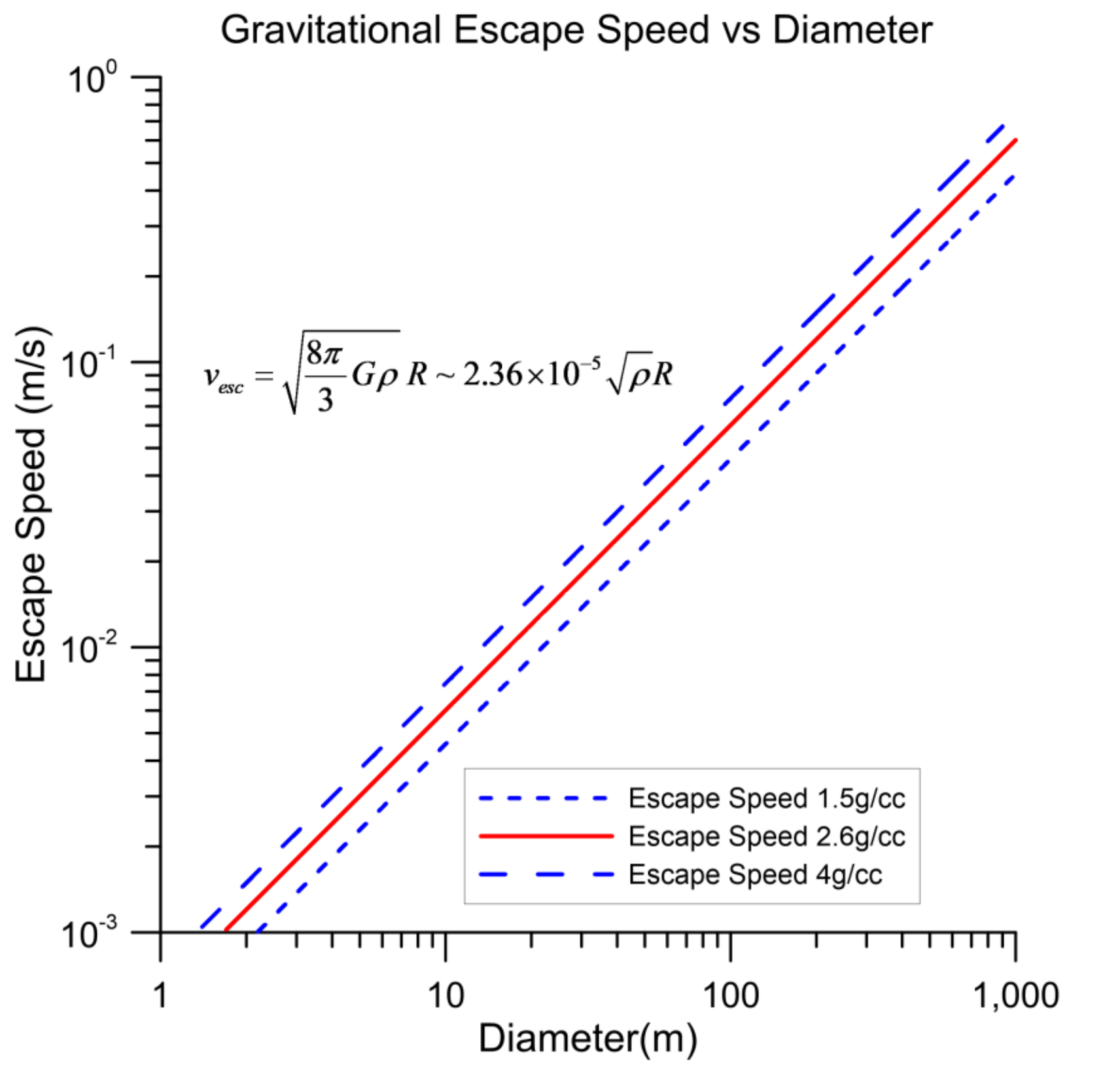

**Figure 5** – Gravitational escape speed vs asteroid diameter for density 1.5, 2.6 and 4g/cc. Note that the escape speed scale linearly with the asteroid size.

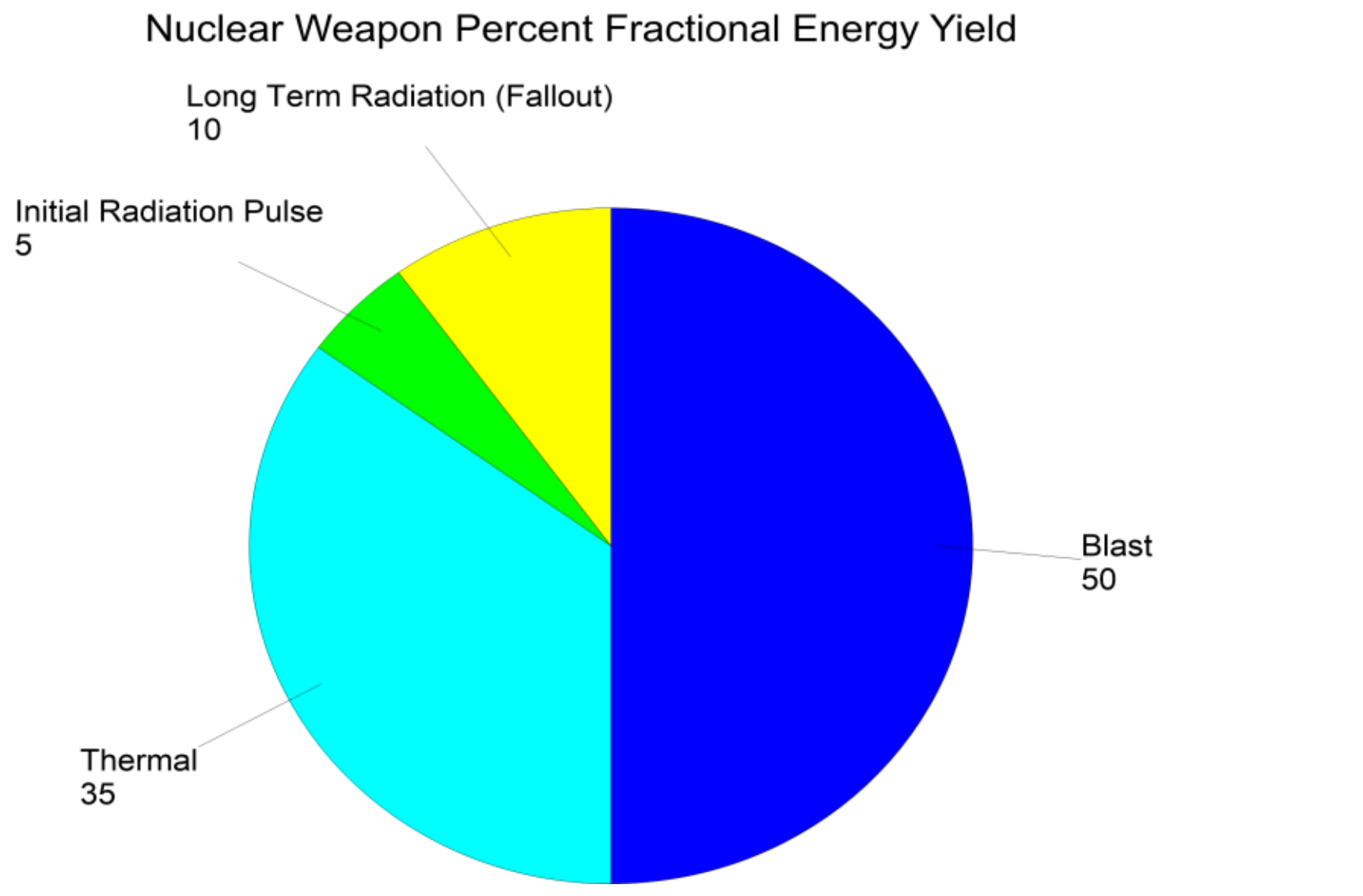

**Figure 6 –** Typical air burst nuclear weapon energy distribution. Note ~50% of the total energy yield goes into the hypersonic blast wave (incl wind) initially. The total shock wave energy decreases with distance as the shock speed decays to Mach 1.

***PI – Interceptor Penetrators*** - The proposed system uses an array of penetrating disassembly rods (PDR) to pulverize and gravitational disassembly of the asteroid. PI stands for "Pulverize It" or "Penetrator Interceptor". Since the asteroid is moving faster than achievable by chemical propulsion, there is only modest gain to be added from the rocket speed relative to Earth compared to the asteroid speed relative to Earth. In a sense, we "just get in front of the asteroid" and wait for it to hit the penetrators rather than trying to "hit the asteroid" at high speed. The spatial timing sequence of the PDR's can be handled either in waves with the outer ones hitting first or possibly by timed chemical detonation. The latter is complicated by the fact that the PDR's will largely be vaporized in the impact, though it is conceivable that a sequential or staged PDR design could survive. This is somewhat analogous to "Earth-penetrating bunker buster" weapons, BUT with a critical difference that the closing speed in our case is vastly higher than for a penetrating "bunker buster" and the impact speed is much higher than the penetrator material phonon speed. No known material can withstand the extreme frontal pressure and remain intact without a "staging" approach. High aspect ratio PDR units with various vaporization stages may be possible to allow impact survival and thus subsequent detonation. This is part of a detailed design phase that we will not discuss in this paper, but will form the basis for future research.

To visualize the general PI process, imagine pealing an onion layer by layer from the outside working inward. Optimizing the penetrator "disassembly procedure" will depend on the asteroid and the vehicle capability to deliver the PDR. Asteroids are not homogeneous objects and will generally have unknown or poorly known interior structures. Given the consequences of a failed intercept, any realistic system design will likely involve multiple interceptors and such systems will not be minimalist, but rather "over designed." A critical issue will be whether molecular binding will be dominant or not (i.e., solid or loose rubble pile) and the scale of disassembly (size distribution of fragments). If intercepted early enough, which is preferred, the broken asteroid fragments will miss the Earth altogether. If it is a terminal intercept, then the Earth's atmosphere becomes a "beam dump" or "bullet proof vest." In essentially all cases for this system, the requirement is to prevent a "ground impact" and to spread the fragments out spatially and temporally as much as possible to minimize the ground damage from the multiple fragment air bursts in the terminal interception case. We will explore this in much greater detail below.

It is important to understand the largest solid fragment from pulverization that we are willing to live with. Ideally, we would pulverize to a size scale of a meter or so, allowing the small fragments to burn up in the atmosphere with minimal shock waves. While desirable, we show that this is generally not necessary. To determine the maximum fragment size scale requires that we set a damage threshold we are willing to accept and to have some minimal understanding of the asteroid in the sense of its "material yield strength." The latter is effectively a "bulk molecular binding" energy issue. In practical terms this will mean trying to understand if we are dealing with a loose rubble pile or a solid nickel-iron asteroid. Fortunately, there are vastly more, lower density (stony/loosely bound) asteroids than there are solid nickel-iron ones. There are no know large solid nickel-iron asteroids, though we should later that we could also mitigate such a threat.

One example of this system would be a 10x10 array of large aspect ratio (meter length scale)

hardened penetrators, each of which is 100kg for a total of 10 metric tons. Delivering 10 metric tons is easily within the capability of modern launchers such as the Falcon Heavy, the SLS and with the upcoming SpaceX Starship with refueling capability, it is quite conceivable to get in excess of 100 tons to the threat. In the case of 100 tons, we could have a wide variety of options to take on asteroids well in excess of 100m diameter with the possibility of mitigating km class asteroids. Another example would be a 50x50 array of penetrators, each of which is approximately 40kg for a 100 ton total penetrator delivered. Depending on additional optimization studies, it may be more effective to have several waves of penetrators arrays on the same spacecraft that deploy sequentially to more effective pulverize the asteroid. The array distribution could be dynamically set in the spacecraft prior to impact to match the geometry of the target. Explosive filled penetrators are another option with detonation started just prior to or upon impact, or the use of a "cluster" explosive system or possible "solid balls" to macroscopically "erode" the asteroid. For large threats, waves of interceptor vehicles is yet another option. We show below that the actual total mass of the penetrators required to gravitationally disassemble and spread the fragment cloud is extremely small IF the coupling efficiency of the KE of the penetrators to the KE of the fragments is high. For unity efficiency the mass ratio of penetrators (m) to target (M) in the regime where total target energy required is dominated by the fragment KE and not gravitational binding which is true for asteroids with diameter less than ~1 km. In this case the mass ratio is $m/M = (v_{disp}/v_{\text{int}})^2$ or about $2.5 \times 10^{-9}$ for $v_{int}$= 20km/s for the asteroid closing speed and $v_{disp}$ = 1 m/s for the fragment dispersal speed relative to the parent asteroid center of mass. As a quick example, consider a 100m diameter asteroid with density 2.6 g/cc. The mass of the asteroid would be about $1.4 \times 10^9$ kg. For $v_{int}$= 20km/s (asteroid closing speed) and $v_{disp}$ = 1 m/s giving $m/M = (v_{disp}/v_{\text{int}})^2 = 2.5x10^{-9} \rightarrow m = 3.4kg$ . **This is an extremely small penetrator mass required assuming unity coupling. Of course, the coupling will NOT be unity and this is where our hypervelocity modeling simulations become critical. We are running a number of target sizes, masses and strength models along with a number intercept scenarios.**

***Nuclear Penetrator Options*** – A nuclear option is to use an array of small tactical nuclear penetrators instead of conventional penetrators. The trade space between lesser number of high yield vs large number of low yield nuclear devices needs to be simulated for various targets. The effect is basically the same. In a terminal defense situation, we focus on spreading out the impact energy into small enough pieces that they burn up in the atmosphere and do not hit the ground. The issue of radioactive contamination in the atmosphere would have to be factored in with a nuclear scenario.

***Energy Required to Spread Out Debris Cloud*** – Spreading out the fragments and imparting a "radial" dispersion pattern would greatly mitigate the threat, as we can then spatially disperse the fragments prior to hitting Earth. The minimum energy *E* (additional to the gravitational disassembly energy) required to disperse the debris at speed $v_{dis}$ for a uniform spherical asteroid of radius *R* is:

Debris cloud dispersal energy after gravitational disassembly

$$E = \frac{m v_{dis}^2}{2} = \frac{4}{3}\pi R^3 \rho \frac{v_{dis}^2}{2} = \frac{1}{6}\pi d^3 \rho \frac{v_{dis}^2}{2} \propto d^3$$

$$v_{dis} = \left[\frac{3E}{2\pi R^3 \rho}\right]^{1/2} = 1.41 km/s * \sqrt{E_{mt}/\rho_{rel} R(m)^3}$$

$$E_{mt} = dispersal\, kinetic\, energy\ E\ in\ metric\ tons\, (mt)\ TNT$$

$$1 mt\ TNT = 4.184 x 10^9\ J$$

$$\rho_{rel} = density\ relative\ to\ water$$

$$Ex: E = 10\, mt\, (10^4 kg\, TNT), \rho_{rel} = 2.6, R = 50m\, (100m\ diam) \rightarrow v_{dis} = 7.84 m/s$$

Note that the energy calculated above is the additional energy required after gravitational disassembly. For asteroids (typ. < 1km diameter), the gravitational disassembly energy is small compared to the energy required to spread the debris cloud for the typical debris cloud speeds we desire, namely for disassembly speeds ~1m/s. For large asteroids that also happen to be existential threats, with diameters >10 km, the gravitational binding energy is generally much larger than the debris spread energy (for $v$ ~1m/s). This is due to the 5th power scaling of gravitational binding energy with diameter, namely $E_{BE} \sim 2.19x10^{-11}\rho^2 d^5$, while the dispersal energy only scales as the 3rd power of the diameter. We will come back to this later. Most of this paper concerns threats from asteroids that have diameters less than 1km, and for these the gravitational binding energy is sub-dominant to the dispersal energy for $v$ ~1m/s. Another way of stating this is that the escape speed is much less than the desired dispersal speed for small asteroids (<1km diameter) while the opposite is true for large asteroids and comets (>10km diameter).

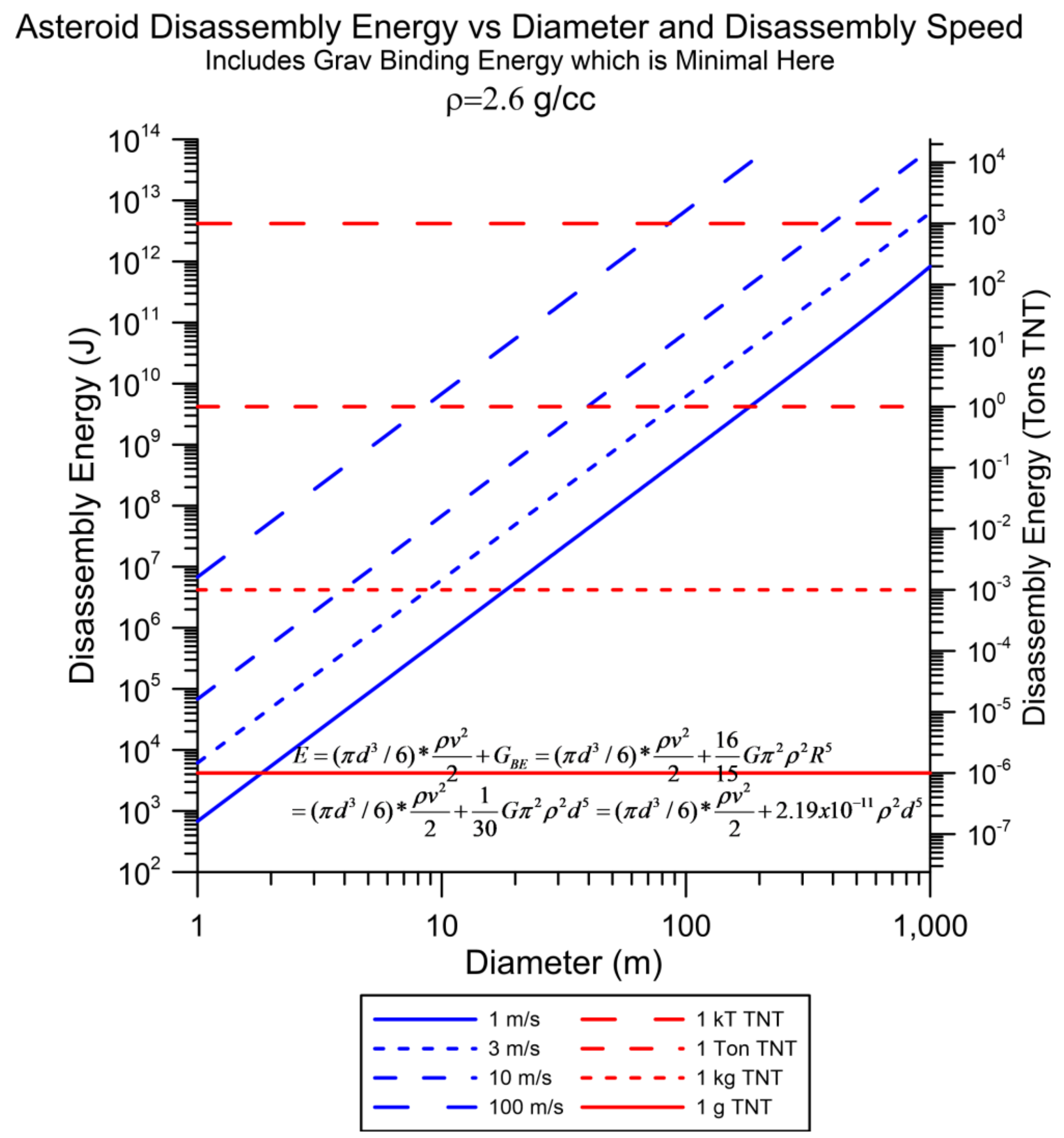

**Figure 7 –** Energy required to disassemble an asteroid at a given fragment speed relative to the center of mass for fragment speeds from 1 to 100m/s and parent asteroid diameters from 1 to 1000m. For asteroids below 10m diameter that are stony (<3g/cc), we would generally not take any action as the blast wave at the Earth's surface would be small.

If we consider a terminal defense situation with very little warning time, we can compute the debris cloud diameter as it hits the Earth's atmosphere assuming the asteroid travel time between intercept and Earth impact is $\tau_{impact}$ as simply $D_{cloud}=2v_{dis}\tau_{impact}$. For example, assuming an extremely short mitigation at **one lunar distance,** $L_{moon}=3.8x10^5$ km, and an asteroid closing speed of $v_{ast}$ =10km/s, we would have $\tau_{impact}=3.8x10^4$ s (~ 10 hr) and $D_{cloud}=2v_{dis}\tau_{impact}$=596km with the assumptions above for a 100m diameter ($R$=50m) asteroid, density 2.6g/cc, dispersal energy of $E$=10mt, and $v_{dis}$ =7.84m/s. For the same analysis, but with $v_{ast}$ = 20km/s asteroid (similar to Chelyabinsk), we have $\tau_{impact}=1.9x10^4$ s (~ 5 hr) and $D_{cloud}=2v_{dis}\tau_{impact}$=298 km.

Even for this extremely short time between intercept and hitting Earth, there is sufficient spread in the atmosphere to allow essentially independent atmospheric detonation (shock wave de-correlation) for each fragment. Even reducing the energy injected into debris spreading by a factor of 10 (1mt TNT equivalent) and the same assumptions otherwise with $v_{ast}$ = 20km/s gives $v_{dis}$ =2.5m/s and $D_{cloud}=2v_{dis}\tau_{impact}$=95km which is still more than sufficient for shock wave mitigation.

***Disruption Energy*** **–** The terminology of "disruption" has to be quantified. In this paper we use the term in a specific manner, namely the fragmentation of the target to a level such that the fragments are less than about 15m in diameter. The term "disruption" is sometimes used to refer to the general disassembly of an object (for example tidal disruption of a comet or other body) where the object undergoes a process resulting in some macroscopic pieces being separated. In the 2009 work of Jutzi et al, where the question of asteroid breakup (disruption) from impact of other smaller asteroids is studied, there are simulations of "asteroid on asteroid" impacts in which the criteria of disruption is that at least 50% of the original target mass is separated and escapes. Jutzi et al study this over a very wide range of asteroids with diameters from $6x10^{-5}$ to 200km with impacts of 3km/s and the impacting asteroids from $4x10^{-6}$ to 82km for both porous and non-porous materials. This is qualitatively very different from our case in that the Jutzi SPH simulations essentially use a "one sphere on another sphere" approach with comparable material strengths in both the impacting "projectile" and the "target", whereas we are looking at specifically designed extremely high strength high aspect ratio penetrators optimally designed to deliver energy deep inside the low strength target to break it apart. If we were to use the single equal low strength "sphere on sphere" approach studied in Jutzi et al 2009 we would generally not have an efficient use of penetrator mass as the vast majority of a "surface impact" is dissipated in local surface thermal energy rather than macroscopic scale low speed kinetic energy. The fitted simulations data suggest a disruption energy per unit target mass of about 100 J/kg for their cases. While the situation for us is very different, the comparison is still extremely instructive. The number often quoted for disruption is about 100 J/kg which corresponds to an equivalent bulk disruption speed of 14.4 m/s. This is shown in the accompanying figure for rocky asteroids up to 1km in diameter and is compared to both the (negligible) gravitational binding energy and the bulk disruption speed from 1 to 20 m/s. In this paper we will often use a desired disruption speed of 1 m/s and compared to the Jutzi et al 100 J/kg or 14.4 m/s there is a factor of 200x larger energy for disruption from the Jutzi et al study than for a theoretically optimal case that might require 1 m/s disruption as one example. Such a large difference is not unexpected given the very different impact physics but we will come back to this later in this paper when studying launch vehicle capability for various threats. We consider the Jutzi et al disruption energy of 100 J/kg to be an upper limit on non-optimized disruption systems.

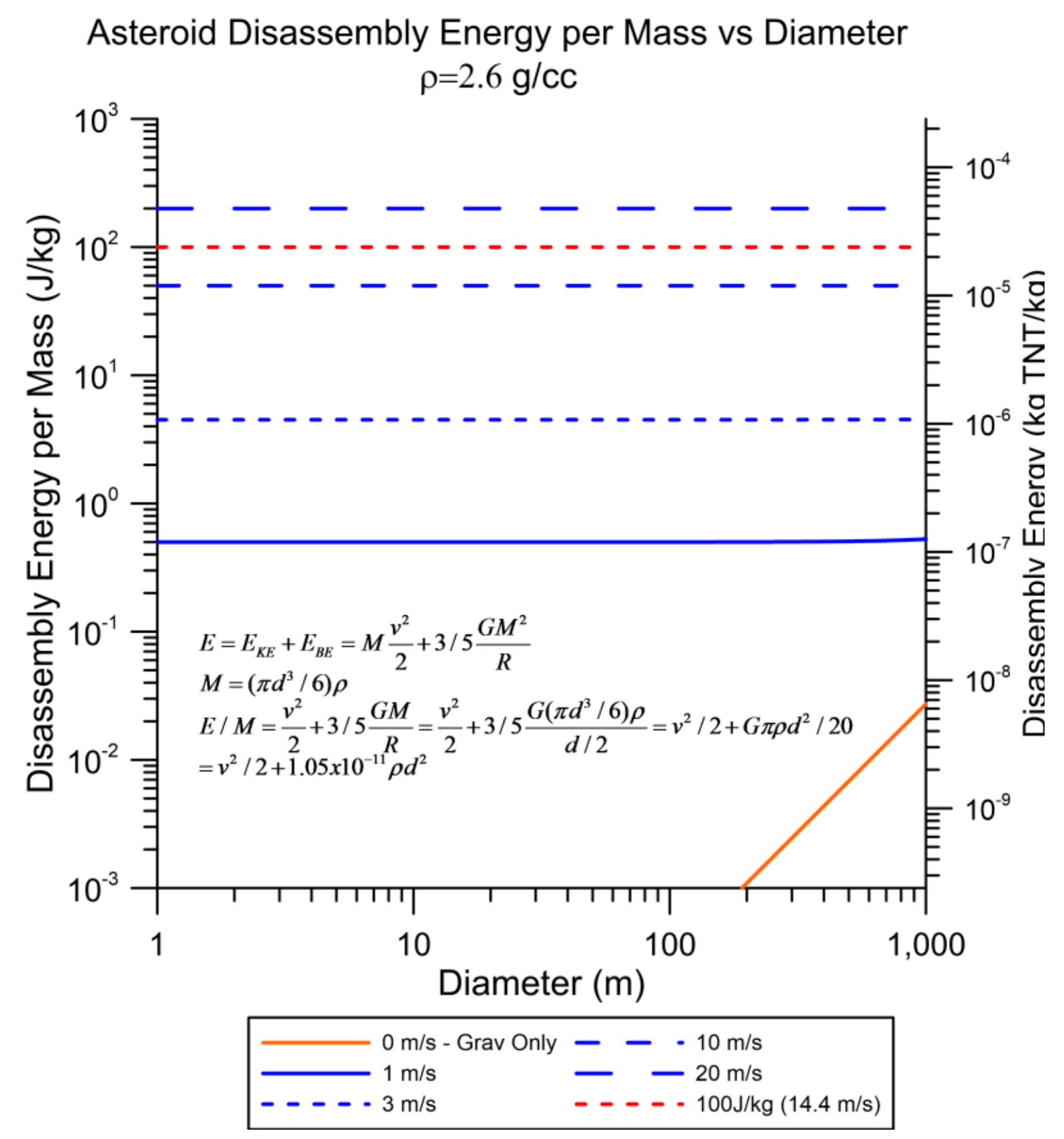

**Figure 8** – Disruption energy per unit mass of the parent object vs diameter and vs disruption speed from 0 to 20m/s. The zero-speed case is the gravitational binding energy only case. The often quoted 100 J/kg "disruption energy per mass" corresponds to 14.4 m/s for comparison.

***Bolide Heat Capacity*** – It is useful to compare the heat capacity of the bolide to the disruption energy. The specific heat of rock is ~ 1kJ/kg-C (~1/4 that of water). This is critical as whatever disruption process is used, we do NOT want to heat up the bolide by much (ie best to keep the bulk heating to <<1C). While this sounds difficult, typical hypervelocity interactions such as local vaporization form an "adiabatic gas engine" which cools as it transfers mechanical energy. An airburst nuclear weapon is an extreme example where ultimate ½ of the energy starting at millions of K, is ultimately transferred to ambient temperature air.

***Fragment Distribution Functions*** – A complex fragmentation process of the type we are considering will result in a range of fragment masses and speeds that will need to be computed based on realistic models of both the target and the multiple penetrator impact physics This will come out as part of the hypervelocity simulations and ultimately comparison to testing. One of the issues to be considered is that the energy partition amongst the fragments will depend quadratically on the speed of each fragment and the mass distribution of the fragments. In some explosive processes such as large-scale motion of material in mining and mountain "clearing" for roads, the material largely moves together while in surface detonations the shock waves will accelerate small masses to significantly higher speeds than larger masses. Similarly, in surface impacts and surface ablation technique such as laser and NED heating, there is a large tail in the speed distribution of fragments ejected. Given the nature of our strategy of deep penetration we might expect to see bulk motions of a large fraction of the material (low δ below) but this remains to be verified in numerical simulations. Since the energy in a fragment is proportion to the square of its speed relative to the center of mass, IF we have a large speed distribution among fragments of roughly equal masses, then the energy distribution will be skewed towards those with larger speeds while if we have a uniform bulk motion then the energy distribution per mass will be more uniform. This will be the focus on a number of 3D supercomputer simulations we are currently working on. These issues become particularly important for the mitigation of very large targets where the launch mass capability of boosters to deliver the penetrators is important. This issue is much less important for smaller targets (<200m diameter). We come back to this issue later in this paper.

Below we consider the case of both arbitrary distribution functions for the speed of the fragments and the particular case of a nearly Gaussian speed distribution. In the accompanying plot note the large tail on the energy PDF (probability density function) for large speed dispersion cases (low δ).

$$m = \text{fragment mass}, v = \text{speed}, \mu = \text{mean speed}, \varepsilon = v/\mu, E = KE = mv^2/2,\ v = \sqrt{2E/m}$$

$$\delta = \mu/\sigma$$

$$v, E, \varepsilon \geq 0$$

$$E(v=\mu) = m\mu^2/2$$

$$\kappa = E/E(v=\mu) = 2E/m\mu^2 = (v/\mu)^2 = \text{normalized energy}$$

$$P(v) = P(v,\mu,\sigma) = \text{speed distribution function}$$

$$P(E) = P(E,m,v) = \text{energy distribution function}$$

$$P(\varepsilon) = P(\varepsilon,\mu,\sigma) = \text{fractional speed distribution function}$$

$$P(\kappa) = P(E/E(v=\mu)) = P(E,m,v) = \text{normalized energy distribution function}$$

$$P(v)dv = P(E)dE = P(\varepsilon)d\varepsilon = P(\kappa)d\kappa$$

$$\text{Assume all unity normalized distribution functions} \int_0^\infty P(v)dv = \int_0^\infty P(E)dE = \int_0^\infty P(\varepsilon)d\varepsilon = \int_0^\infty P(\kappa)d\kappa = 1$$

$$P(E) = P(v)dv/dE = P(v)/mv = P(v)/\sqrt{2Em}$$

$$P(\varepsilon) = P(v)dv/d\varepsilon = \mu P(v) = P(v)v/\varepsilon$$

$$P(\kappa) = P(E)dE/d\kappa = P(E)E(v=\mu)$$

$$\langle v\rangle = \int_0^\infty vP(v)dv,\ \langle E\rangle = \int_0^\infty EP(E)dE,\ \langle \varepsilon\rangle = \int_0^\infty \varepsilon P(\varepsilon)d\varepsilon$$

$$\langle E\rangle = \int_0^\infty EP(E)dE = \int_0^\infty \frac{mv^2}{2}\frac{P(v)}{mv}dE = \int_0^\infty \frac{mv^2}{2}\frac{P(v)}{mv}mvdv = \frac{m}{2}\int_0^\infty v^2P(v)dv = \frac{m\mu^2}{2}\int_0^\infty \varepsilon^2 P(\varepsilon)d\varepsilon$$

$$\langle E\rangle/E(v=\mu) = \frac{m\mu^2}{2}\int_0^\infty \varepsilon^2P(\varepsilon)d\varepsilon \Big/ \frac{m\mu^2}{2} = \int_0^\infty \varepsilon^2 P(\varepsilon)d\varepsilon$$

$$(\text{Note the quadric dependence on } v)$$

$1)\ Assume\ P(v) = truncated\ (at\ zero\ speed)\ Gaussian$

$$P(v) = \frac{1}{\sqrt{2\pi}\sigma} e^{-(v-\mu)^2/2\sigma^2} \quad v \ge 0$$

$$P(v) = 0 \quad v < 0$$

$$P(E) = P(v)/mv = \frac{1}{m\sqrt{2E/m}\sqrt{2\pi}\sigma} e^{-(\sqrt{2E/m}-\mu)^2/2\sigma^2} = \frac{1}{2\sqrt{\pi m E}\sigma} e^{-(\sqrt{2E/m}-\mu)^2/2\sigma^2}$$

$Note\ that\ P(E)\ diverges\ as\ E \to 0\ while\ P(E)dE = P(v)dv\ and\ \langle E \rangle = \int_0^\infty EP(E)dE\ do\ not$

$$P(\varepsilon) = \mu P(v) = \frac{1}{\sqrt{2\pi}} \frac{\mu}{\sigma} e^{-\mu^2(\varepsilon-1)^2/2\sigma^2} = \frac{1}{\sqrt{2\pi}} \delta e^{-\delta^2(\varepsilon-1)^2/2} \quad where\ \delta = \mu/\sigma$$

$Note\ the\ linear\ dependence\ on\ \mu$

$$P(\kappa) = P(E)E(v=\mu)\frac{m\mu^2/2}{m\sqrt{2E/m}\sqrt{2\pi}\sigma} e^{-(\sqrt{2E/m}-\mu)^2/2\sigma^2} = \frac{\mu}{\sigma}\frac{\mu/2}{\sqrt{2E/m}\sqrt{2\pi}} e^{-(\sqrt{2E/m}-\mu)^2/2\sigma^2}$$

$$= \delta \frac{1/2}{\sqrt{E/m\mu^2/2}\sqrt{2\pi}} e^{-\mu^2(\sqrt{E/m\mu^2/2}-1)^2/2\sigma^2} = \delta \frac{1/2}{\sqrt{\kappa}\sqrt{2\pi}} e^{-\delta^2(\sqrt{\kappa}-1)^2/2}$$

$Note\ that\ we\ assume\ v \ge 0. This\ brings\ up\ an\ issue\ with\ normalization\ for\ a\ Gaussian\ distribution$

$$\int_{-\infty}^{\infty} \frac{1}{\sqrt{2\pi}\sigma} e^{-(v-\mu)^2/2\sigma^2} dv = 1 = \int_{-\infty}^{\mu} \frac{1}{\sqrt{2\pi}\sigma} e^{-(v-\mu)^2/2\sigma^2} dv + \int_{\mu}^{\infty} \frac{1}{\sqrt{2\pi}\sigma} e^{-(v-\mu)^2/2\sigma^2} dv$$

$$\int_{-\infty}^{\mu} \frac{1}{\sqrt{2\pi}\sigma} e^{-(v-\mu)^2/2\sigma^2} dv = 1/2$$

$$\int_{\mu}^{\infty} \frac{1}{\sqrt{2\pi}\sigma} e^{-(v-\mu)^2/2\sigma^2} dv = 1/2$$

$For\ \mu^2/2\sigma^2 >> 1\ (ave\ speed\ far\ from\ zero)\ then \int_0^\infty \frac{1}{\sqrt{2\pi}\sigma} e^{-(v-\mu)^2/2\sigma^2} dv \sim 1$

***World Nuclear Arsenal*** – We compare impact energies to the "World Nuclear Arsenal". This term needs clarification, as it is set by both nuclear weapons and delivery system treaties (NPT, TPNW, START etc) and hence time dependent as treaties are renegotiated and by weapons held in reserve or components stored that could be used in the future. In this paper the term "world nuclear arsenal", is used in a more general context and is not only the currently agreed to and deployed weapons systems but includes a "reasonable" estimate of weapons that could rapidly be put together based on "reserve, retired and stored deactivated components". The actual currently (treaty) agreed to and deployed nuclear arsenal is about 1.5-3 Gt, with the vast majority (~90%) being controlled by the US and Russia. In the not too distant past, the total available arsenal was in excess of 20 Gt in the US alone with the US peaking in 1966 and Russia peaking in about 1988 with about 30,000 warheads each. Treaties have drastically reduced this by about 90%. In this paper we use 6 Gt as an estimate of the total world nuclear arsenal. This number is not critical and is used only as a comparison metric.

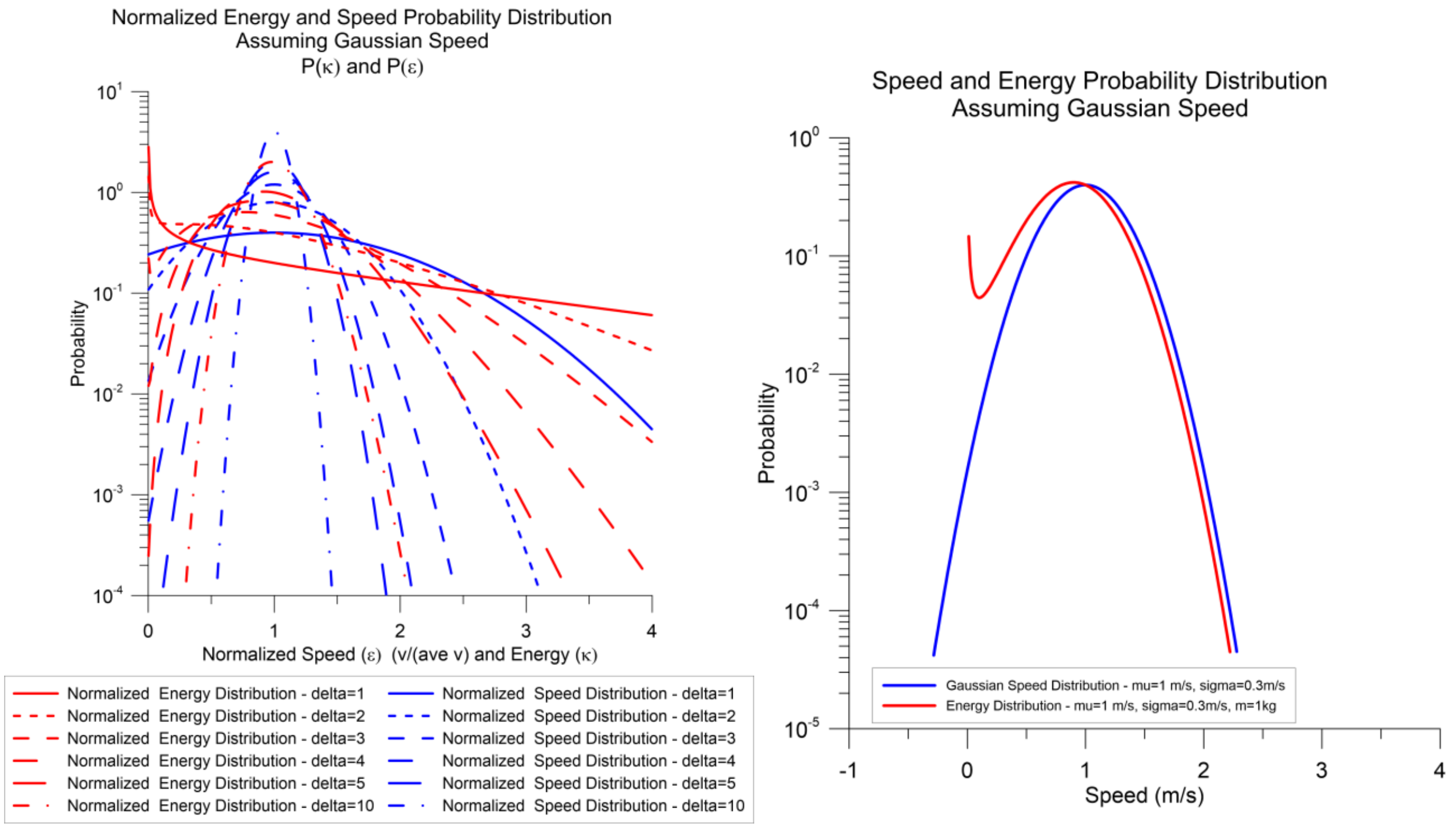

**Figure 9** – Left: Normalized speed and energy distribution (PDF) vs normalized speed for “delta from 1 to 4. Right: Speed and energy distribution (PDF) for the case of average speed of μ=1 m/s and a σ =0.3 m/s with a reference fragment mass of 1 kg. Note the large PDF tail for speed and energy for large speed dispersion (low δ) and corresponding small PDF tail for low speed dispersion (large δ).

***Multimodal capability -*** There are six basic modes of operation for this system allowing a natural growth from tackling modest threats to existential ones and from complete fragmentation to deflection. These modes depend on a combination of warning time and threat magnitude. The product of the intercept time $\tau_{impact}$ and the disruption speed $v_{dis}$ is the key mitigation metric for a given threat. Larger disruption speed allows for shorter intercept time with essentially $\tau_{impact} \alpha\ 1/\ v_{dis}$. In general, we assume a very conservative $v_{dis}=1m/s$ in calculating the intercept times for all cases unless otherwise indicated. An example if this trade-off is the (literally) last minute intercept for a 20m diameter (Chelyabinsk class) threat where with 10m/s we achieve a 100s intercept while with 1m/s we achieve 1000s intercept. Since the required disruption energy is quadratic in the disruption speed for disruption KE dominated cases (d< 1km) there is a trade-off in required energy for higher disruption speed and hence shorter intercept. In general, we use stony densities (2.6 g/cc) for most threat analysis though we explore a large range of densities and cohesive strengths in detail later in this paper.

***1) Short time warning – minutes to days intercept – terminal defense (15m to 100m diam threats)***

- *Fragment to <15m and use Earth's atmosphere as body armor – shock waves de-correlated*
- *Sum of all optical pulses below combustion limit – no fires*
- *Shock waves de-correlated – virtually no damage – possibly some minor window damage*
- *Ex: 100m diam (~100 Mt >> Tunguska) can be mitigated with 1 day intercept*
- *Ex: 20m (0.5 Mt - Chelyabinsk) can be mitigated with 100 sec intercept (10m/s disruption)*

***2) Moderate time warning – 10-60 day intercept (100 – 500m – Apophis, Bennu)***

- *Fragment to <15m and spread fragment cloud over large area on Earth (~ 1000 km radius)*
- *Earth's atmosphere is used as body armor*
- *Ex: 350m diam (~Apophis) (~ 4 Gt ½ Earth nuclear arsenal) can be mitigated with 10 day intercept*
- *Ex: 500m diam (~Bennu) (~ 8 Gt > Earth nuclear arsenal) can be mitigated with 20 day intercept*

***3) Longer time warning (>75 day intercept) (600-1000m threats)***

- *Fragment ideally to <15m but less restrictive*
- ***Fragment cloud spreads to be larger than the Earth - Virtually all fragments miss the Earth***
- *Residual fragments that will hit the Earth smaller than 15m are not a threat – atmosphere mitigates*
- *Residual fragments > 15m can be dealt with as in option 1) terminal defense IF needed*

***4) Long term warning and existential threat (>100 day intercept and >1& <15km diameter)***

- *Fragment using NED penetrator array – pure fission (eg W82 class NED) looks feasible - based on nuclear artillery technology already designed, developed and tested. Sequential penetrator option allows better NED effectiveness and possibly thermonuclear class penetrators if needed. These are internal and NOT standoff detonations.*
- *Possible use of "sequential following penetrators" to allow "hole drilling" for better NED coupling and lower "g" forces for devices such as B61-11 NED physics package – 3 kt/kg @ 360 Kt yield*
- *Fragment cloud spreads large enough to miss Earth for virtually all fragments.*

***5) Long term warning (> 1 year) – asymmetrical fragmentation/ enhanced deflection option***

- *Asymmetrical fragmentation to get extreme deflection enhancement - use energy not momentum.*
- *Blast off part of target to use it ("push") against itself – NOT like current deflection techniques but synergistic. See mode 6 below.*
- *Depending on target size can use kinetic only, kinetic with conventional explosives or NED penetrators for extreme threats.*
- *Use of penetrator to drive mass ejection via an induced "rocket exhaust" mode where the high temperature and pressure vaporized/ plasma created inside the bolide exists through the penetrator initiated "exhaust nozzle" to form a rocket engine using the bolide as "fuel".*

***6) Long term warning (> 1 year) – classical deflection***

- *If desired the same system can also be used as a classical deflector -complete multimodal use.*
- *Simple penetrator reconfiguration allows the same system to be used as a classical deflector.*
- *In general this mode is not needed as mode 5 (enhanced deflection) is superior but it is available.*

***This allows one system to be used from terminal defense to classical deflection mode.***

***The Earth's Atmosphere as a Shield*** -The Earth's atmosphere may seem like a tenuous and thus ineffective shield against asteroids and comets but it has a column depth or mass per unit area of 10,000 $kg/m^2$ which is equivalent to 10m of water, 4m of concrete or 1.3m of iron. It forms a literal Iron dome around the Earth. However, it is a much better shield than if it were solid iron since the long path length (~ 100 km) of the atmosphere allows for extreme heating and fragmentation of the fragments as well as rapid "self-healing". When thought of in the way, it is clear why our technique works so well. When the bolide fragments are small enough the atmosphere is an extraordinary capable shield with the primary remaining metric being the resulting shock waves and optical signatures generated from all the fragments in passing through the atmosphere. The difficulty of returning astronauts makes it clear just how difficult atmospheric entry is.

***Physics of Air Bursts and Shock waves*** **-** The physics of chemical and nuclear detonations and the generation of the resultant atmospheric blast including high speed winds and shock waves has been well studied theoretically, as well as measured experimentally. The term "blast wave" technically refers to both a sudden change in pressure (shock wave) and subsequent winds behind the shock front. The term "air blast or simply blast" often refer to both effects combined. The term "blast wave", "air blast", "blast" and "shock wave" are often used interchangeably. We will discuss both the shock wave which is the dominant destructive element at large ranges (weak shock regime) as well as the dynamic pressure due to the blast induced winds behind the shock front. The dynamic pressure due to the wind is sub dominant in terms of damage and energetics at large distances. The detonation energy levels for the relevant cases of asteroids impacting the Earth are much closer to the yields of nuclear weapons than to the chemical explosives. As shown above, for the case of asteroids in the 1-100m diameter class moving at relevant speeds of 15-50km/s, the energies involved are in the low kt to high Mt range, or

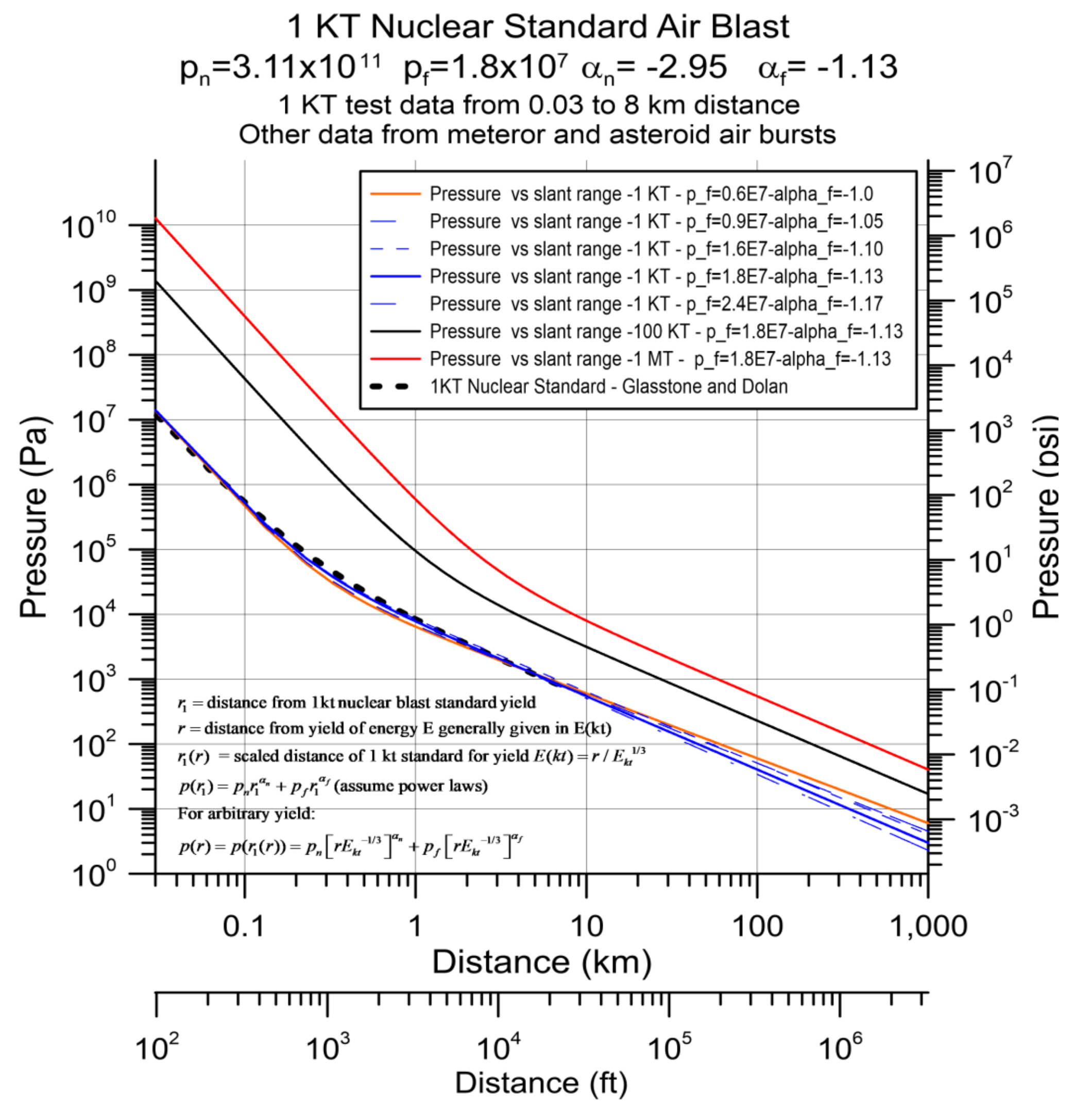

**Figure 10** – Sea level nuclear air burst "1kt standard" with scaling to 1Mt nuclear yield. Pressure vs distance from the point of detonation. Power law fits are shown with scaling to arbitrary yield and distance. For asteroids, we are not generally concerned with ranges less than 30km. A 10m diameter asteroid (20km/s, 2.6g/cc, 45deg angle) with 48kt blast yield is shown as a 100kt nuclear equivalent assuming approx. 50% nuclear yield to blast yield. Multiple power law models including P(r)~1/r in the far field (orange) are shown. Best fit shown in header.

much above common chemical explosive devices. The relevant energy ranges span the regime from small tactical to the largest strategic nuclear devices ever tested. Some of the early solutions to shock waves are discussed in the 1947 LANL paper of Bethe et al [20], as well as in papers by Sedov, von Neuman, and Taylor [20]–[23], and by Zeldovich in the USSR [24]. They dealt with the energy singularity problem relevant to compact explosive devices such as nuclear weapons. Some of the experimental data is summarized in the 1977 Effects of Nuclear Weapons by Glasstone and Dolan [2], wherein a 1kt "standard" is given which allows scaling to arbitrary yields. While much of the nuclear weapons test data is primarily focused on "close in" strong shock wave effects, we can scale to larger distances. While at distances close to the detonation, the shock wave speed is extremely supersonic the shock rapidly decays to near sound speed as we discuss below. Most of the relevant cases of interest in this paper are for long range effects where the peak over pressure is expected to scale ~1/*r* if there were no atmospheric absorption, no wind channel etc. We discuss absorption in the atmosphere below. As shown earlier, the typical fraction of nuclear weapon yield that goes into the shock wave and wind for air burst detonations is about 50% initially. The formation of the shock wave is essentially adiabatic due to the extremely short time scales relative to the thermal diffusion time scales. The shockwave from an asteroid air burst starts at extremely high speeds (at the speed of the bursting asteroid/fragment, which is typically > Mach 30 at production), but then, like the case of a nuclear weapon, it rapidly decreases with increasing distance and decays to sound speed (~340m/s). We model the peak over pressure from the shock wave as two power laws covering the "near" and "far" distance regimes however we note that the peak pressure generated by a bolide differs from that of a nuclear weapon airburst in that the "near" distance regime is very different due to the

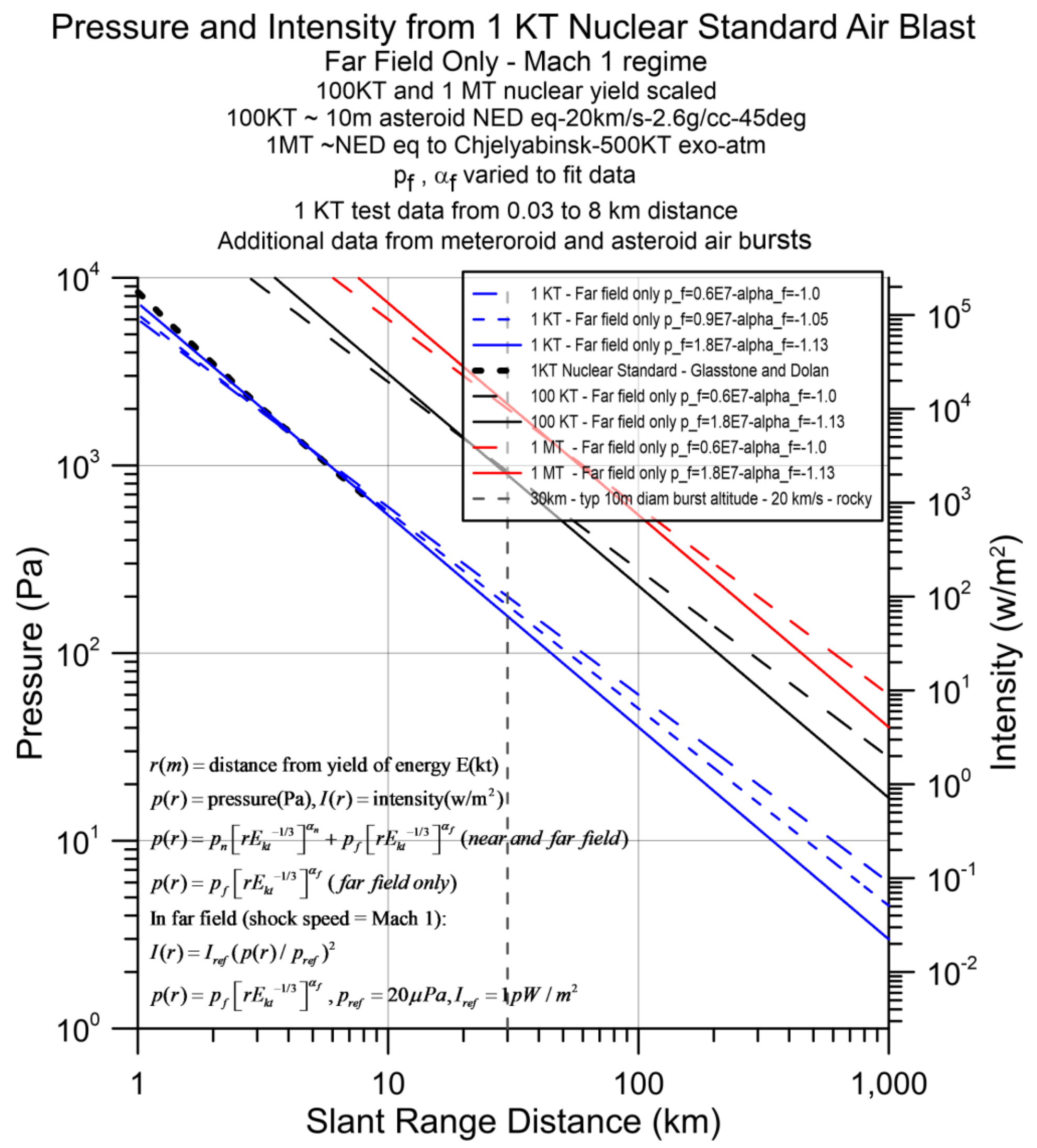

**Figure 11 –** Far field pressure and acoustic intensity vs slant range with multiple power fits for 1,100 and 1000 KT nuclear yields. Note that p(r)~1/r is a reasonable fit to the 1 KT NED standard which tracks pressure to about 700Pa (0.1 psi) at 8 km range though p(r)~ $1/r^{1.13}$ is a better fit. The critical range for our program is about 30km (directly under fragment for ~10m diam) to 600km (acoustic horizon for 30km altitude burst). The scaled cases for 100 KT (~ equivalent NED for 10m diam fragment with density 2.6g/cc and 20km/s closing speed) as well as a 1 MT case (~equivalent NED for Chelyabinsk class event with 20m diam, density 3g/cc and 20km/s). For the purposes of the conclusions of this paper, the precise scaling is largely irrelevant. Note the acoustic intensity which scales a $p^2$. At a pressure of p=1 kPa where we begin to have concerns about poorly framed residential window cracking, the intensity (flux) is 2.5 kW/m$^2$ about 2.5x of Earth surface sunlight.

shock production physics. The bolide airburst is set by the mechanical disassembly of the bolide from the impinging atmosphere and the bolide is not a compact object at the burst position but rather a mechanically distorted (in our model a pancake) object. The atmospheric "ram pressure" during bolide disassembly is set by the compressive cohesion strength of the bolide which is typically between 0.1 MPa (1 bar) loosely bound rubble to 100 MPa (1000 bar) for a "iron core" bolide. It does not make sense to consider shock wave peak pressures that are much higher than the bolide compressive cohesive material strength and hence the "near" pressure regime, unlike that of a nuclear weapons airburst, is approximately capped at the bolide material strength. Additionally, the typical distance from bolide airburst altitude to the ground directly underneath the burst for the fragment sizes of relevance here (<20m) is about 30km which places us in the "far" distance regime for the equivalent nuclear airburst. The energy regimes of fragments of interest to us are generally less than 100 Kt nuclear weapon equivalent. Later will come back to the tensile cohesive strength which is generally vastly smaller than the compressive cohesion. The tensile (pulling) and compressive strengths are very different. Think of them as picking up (tensile) and compressing sand as an example. Often the term cohesion refers only to the tensile term. We fit the NED shock wave data as follows:

$r_1(m) =$ distance from 1 kt nuclear blast standard yield

$r(m) =$ distance from yield of energy E given in E(kt)

$r_1(r)$ = scaled distance of 1 kt standard for detonation of yield $E(kt) = r / E_{kt}{}^{1/3}$

$p(r_1) = p_n r_1{}^{\alpha_n} + p_f r_1{}^{\alpha_f}$ (assume power laws for near (strong shock) and far (weak shock) regimes)

For arbitrary yield:

$$p(r) = p(r_1(r)) = p_n\left[rE_{kt}{}^{-1/3}\right]^{\alpha_n} + p_f\left[rE_{kt}{}^{-1/3}\right]^{\alpha_f}$$

A good fit to the measured data from atmospheric nuclear tests is:

$p_n = 3.11x10^{11}(Pa)$

$\alpha_n = -2.95$

$p_f = 1.8x10^{7}(Pa)$

$\alpha_f = -1.13$

Transition distance between near and far distances in power law:

$$r_{tr} = E^{1/3}\left[\frac{p_n}{p_f}\right]^{1/(\alpha_f-\alpha_n)}$$

$$p_{tr} = p_n r_{tr}{}^{\alpha_n} = p_f r_{tr}{}^{\alpha_f}$$

A reasonable to the 1 KT NED standard fit in the far field with p(r)~1/r is show below.

While not as good as the fit above, it allows a simple scaling interpretation.

$p_n = 3.11x10^{11}(Pa)$

$\alpha_n = -2.95$

$p_f = 6.0x10^{6}(Pa)$

$\alpha_f = -1.0$

**Shock and Blast Wave Notation** – The terms "shock wave" and "blast wave" and "blast" are at times used interchangeably in the literature and in this paper. In an atmospheric explosion there is both an acoustical wave or shock wave generated in addition to atmospheric heating and bulk air motion or "wind". All of this is at times simply called the "blast" effects or at times "blast wave" as it is both spatially and temporally dependent. We reserve the term "shock wave" in this paper to refer to the propagating acoustical wave though we will lapse occasionally into calling it the "blast wave" as this term is often used in the literature.

***Shock wave speed* - peak over pressure and distance from shock initiation** – The speed of the shock wave generated in an air burst depends on the peak pressure of the blast. The shock process is largely adiabatic since the time scale for thermal mixing is much longer than the relevant acoustical time scales. We calculate the speed vs pressure ratio, as well as vs distance. The latter is based on the peak pressure calculated above from nuclear weapons tests. Combining the measured weapons test data with the theoretical shock speed vs pressure ratio allows the calculation of shock speed vs distance. In general, we see the shock speed decays from highly supersonic to the speed of sound relatively quickly.

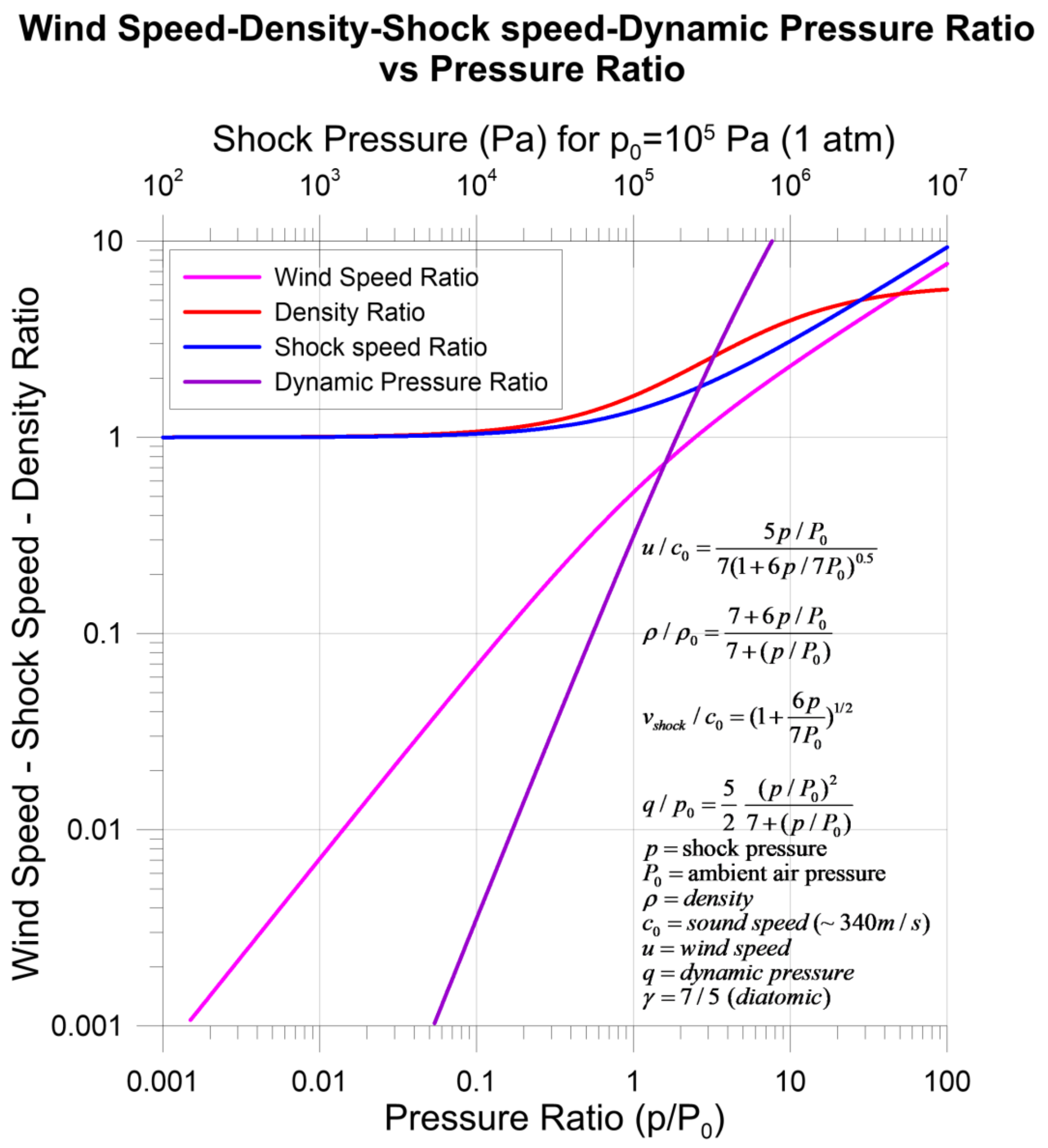

**Figure 12 –** Wind speed, density and shock speed ratios as a function of the peak pressure $p$ over ambient pressure $P_o$. At low pressure ratios, the shock speed is close to the sound speed $c_o$. At high pressure ratios, the shock wave is highly supersonic. However, typically the blast wave peak pressure is $p \ll P_o$ for the fragment size and distance to the observer in cases of interest here. This results in the shock speed being close to sound speed. At low peak pressures the wind speed scales proportional to peak pressure and hence is low (~1 m/s @ p=1 kPa). The scaling of the density behind the shock front and the dynamic pressure are also plotted. All cases are for $\gamma$=7/5 or for diatomic molecules.

**Pressure notation** – We generally use a lower-case p for the overpressure in the shock wave and the upper case P for the un-shocked atmospheric pressure with $P_0$ generally referring to the ambient ground pressure. The shock overpressure $p(\bar{x},t) \equiv \Delta P(\bar{x},t)$ is a function of space and time which can add additional confusion. In referring to shock or blast wave pressure, if not additionally qualified then the assumption is that this is the "peak or maximum overpressure" at a given spatial location. In other parts of the paper we also use upper case P for power when discussing the optical signature. The distinction is clear within the section topic.

$$v_{shock}/v_{sound} = (1+\frac{\gamma+1}{2\gamma}\frac{p}{P_0}) = (1+\frac{6p}{7P_0})^{1/2} \rightarrow 1 (p/P_0 << 1)$$

$p =$ shock overpressure $\equiv$ "shock pressure", $P_0 =$ ambient air pressure $(10^5\ Pa\ at\ sea\ level)$
$\gamma = c_p / c_v = 7/5\ for\ diatomic\ gas\ (5DOF\ low\ temp\ regime)$
$v_{sound} = c_o = ambient\ sound\ speed\ (temperature\ dependent \sim 340 m/s)$

A summary of related shock phenomenon from the Rankine-Hugoniot equations we will use later:

**The peak wind speed (u) behind the shock front for the peak pressure p:**

$$u/c_0 = \frac{pc_0}{\gamma p_0 (1+\frac{\gamma+1}{2\gamma} p/P_0)^{0.5}} = \frac{5p}{7p_0(1+6p/7P_0)^{0.5}} \quad (\gamma = 7/5) \rightarrow \frac{5}{7} p/P_0 \ (p/P_0 << 1)$$

**Strong shock regime (p>>p₀)** $u = c_0 \frac{5}{\sqrt{42}}\sqrt{\frac{p}{P_0}} = 0.772\sqrt{\frac{p}{P_0}}$

**Equal pressure regime (p=p₀)** $u = c_0 \frac{5}{\sqrt{91}} = 0.524 c_0$

**Weak shock regime (our case obsserver on the ground) (p<<P₀)** $u = c_0 \frac{5}{7}\frac{p}{P_0}$

Note the difference in scaling between the weak shock regime ($p << P_o$) which is our case on the ground) where the wind speed scales as $u = c_0 \frac{5}{7}\frac{p}{P_0} \propto \frac{p}{P_0}$ and the strong shock regime ($p >> P_o$ ) where the wind speed scales as $u = c_0 \frac{5}{\sqrt{42}}\sqrt{\frac{p}{P_0}} \propto \sqrt{\frac{p}{P_0}}$

**Density of air ρ behind shock front compared to ambient density ρ₀ :**

$$\rho/\rho_0 = \frac{2\gamma P_0 + (\gamma+1)p}{2\gamma P_0 + (\gamma-1)p} = \frac{7+6p/P_0}{7+p/P_0} \ (\gamma = 7/5) \rightarrow 1 \ (p/P_0 << 1)$$

**Dynamic pressure q (and KE/volume behind shock) definition:**

$$q = 0.5\rho u^2 = \frac{p^2}{2\gamma P_0 + (\gamma-1)p} = \frac{5}{2}\frac{p^2}{7P_0 + p} \ (\gamma = 7/5)$$

$$q/p_0 = \frac{(p/P_0)^2}{2\gamma + (\gamma-1)(p/P_0)} = \frac{5}{2}\frac{(p/P_0)^2}{7+(p/P_0)} \ (\gamma = 7/5) \rightarrow \frac{5}{2}\left(p/P_0\right)^2 \ (p/P_0 << 1)$$

**Head-on (normal incidence) shock wave on flat surface instantaneous peak reflected pressure pᵣ:**

$$p_r = 2p + (\gamma+1)q = 2p\frac{7P_0 + 4p}{7P_0 + p} \ (\gamma = 7/5)$$

$Strong\ shock\ (p >> P_0) \rightarrow p_r = 8p$

$Weak\ shock\ (our\ case\ on\ ground\ \ p << P_0) \rightarrow p_r = 2p$

**Shock Wave Intensity and Dynamic (Wind) Intensity at Distance r**

$$I_{dyn}(r)(w/m^2) = q(r)u(r) = 0.5\rho(r)u(r)^3$$
$$I_{shock}(r)(w/m^2) = p(r)^2/(\rho_{shock}(r)v_{shock}(r)) = p(r)^2/Z_{shock}(r)$$
$$Z_{shock}(r) \equiv \rho_{shock}(r)v_{shock}(r) = acoustic\ impedance \rightarrow Z_0 \equiv \rho_0 c_0 \sim 415 Pa/m/s\ (Mach 1, sea\ level)$$
$$Note\ acoustic\ impedance\ is\ similar\ to\ the\ vacuum\ impedance\ in\ EM\ where\ S = E^2/Z_{vac}\ \ Z_{vac} = \mu_0 c \sim 377\Omega$$

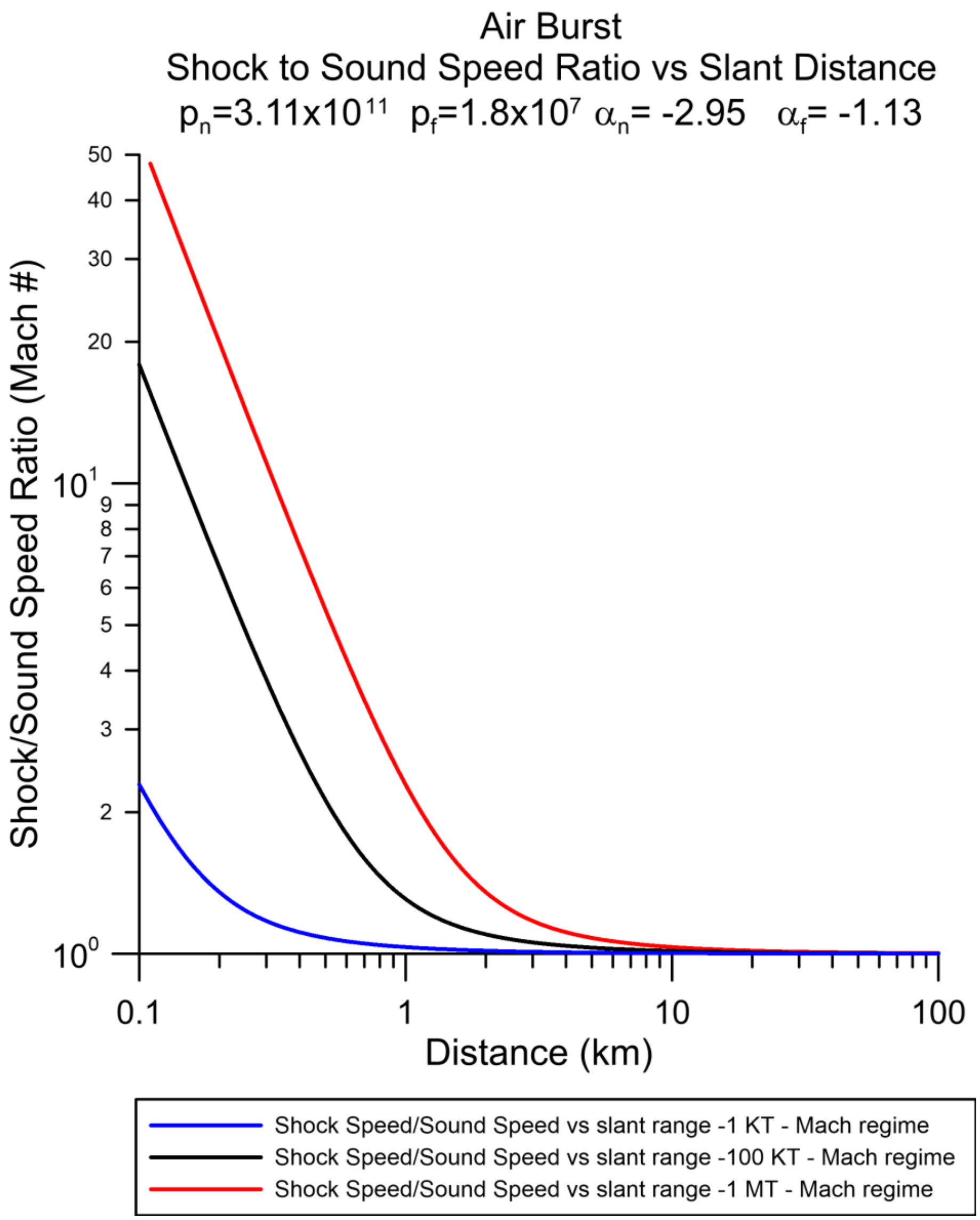

**Figure 13** – Shock wave Mach # vs distance to observer for a 1KT (blue) and 1MtT(red) equivalent nuclear detonation. Recall that about ½ of the energy of an air burst nuclear detonation goes into the shock wave. This means that the red curve (1 MT NED) corresponds to an asteroid shock wave energy of 0.5MT and the black curve (100KT NED) corresponds to a 50KT asteroid shock wave. In our mitigation technique, the asteroid fragments (~10m max diam.) have an energy of order 50KT or less and thus the cases of interest in this paper typically lay between the blue (1KT) and black (100 KT) curves. A 10m diameter asteroid (20km/s, 2.6g/cc, 45deg angle) with 48kt blast yield is shown as a 100kt nuclear equivalent assuming approx. 50% nuclear yield to blast yield. From this we observe that the shock wave rapidly decays to a speed of near Mach 1 (sound speed) at the relevant slant distances (>10km) since the final burst altitude is typically above 30km.

***Pressure – Flux and Sound Pressure Level*** – The relationship between the shock wave pressure *p* (Pa) and the acoustical flux *I* (W/m$^2$) as well as what is referred to as the Sound Pressure Level (SPL) is calculated below. The limit of nominal human hearing is 0 db SPL or 20 μPa by definition.

$$SPL(db) \equiv 20\log(p/p_{ref})$$
$$=10\log(I/I_{ref})$$
$$p/p_{ref} = 10^{SPL(db)/20}$$
$$p = \text{shock (acoustic) pressure}$$
$$p_{ref} = \text{ref shock pressure } (20\mu\text{Pa})$$
$$I(w/m^2) = \frac{{p_{amp}}^2}{2\rho v} = \frac{{p_{rms}}^2}{\rho v}$$
$$= I_{ref}(p/p_{ref})^2 = I_{ref} * 10^{SPL(db)/10}$$
$$I/I_{ref} = (p/p_{ref})^2$$
$$p_{amp} = pressure\ amplitude = \sqrt{2}\,p_{rms}$$
$$\rho = \text{air density } (1.2 \text{ kg/m}^3)\,@\,sea\,level$$
$$v = \text{ sound speed (\textasciitilde 330 m/s)}$$
$$p_{rms} = 20\mu\text{Pa} \rightarrow I = 1.0\,pW/m^2$$
$$I_{ref} = \text{reference flux } = 1.0\,pW/m^2$$

**Conversion to Nepers ($N_P$)**

$$SPL(db) \equiv 20\log(p/p_{ref})$$
$$\rightarrow p/p_{ref} = 10^{SPL(db)/20}$$
$$SPL(N_P) \equiv \ln(p/p_{ref})$$
$$\rightarrow p/p_{ref} = e^{SPL(N_P)}$$
$$e^{SPL(N_P)} = 10^{SPL(db)/20}$$
$$\rightarrow SPL(db) = 20\log(e)SPL(N_P)$$
$$SPL(N_P) = \frac{SPL(db)}{20\log(e)} = 0.1151 SPL(db)$$
$$1N_P = 8.686db$$

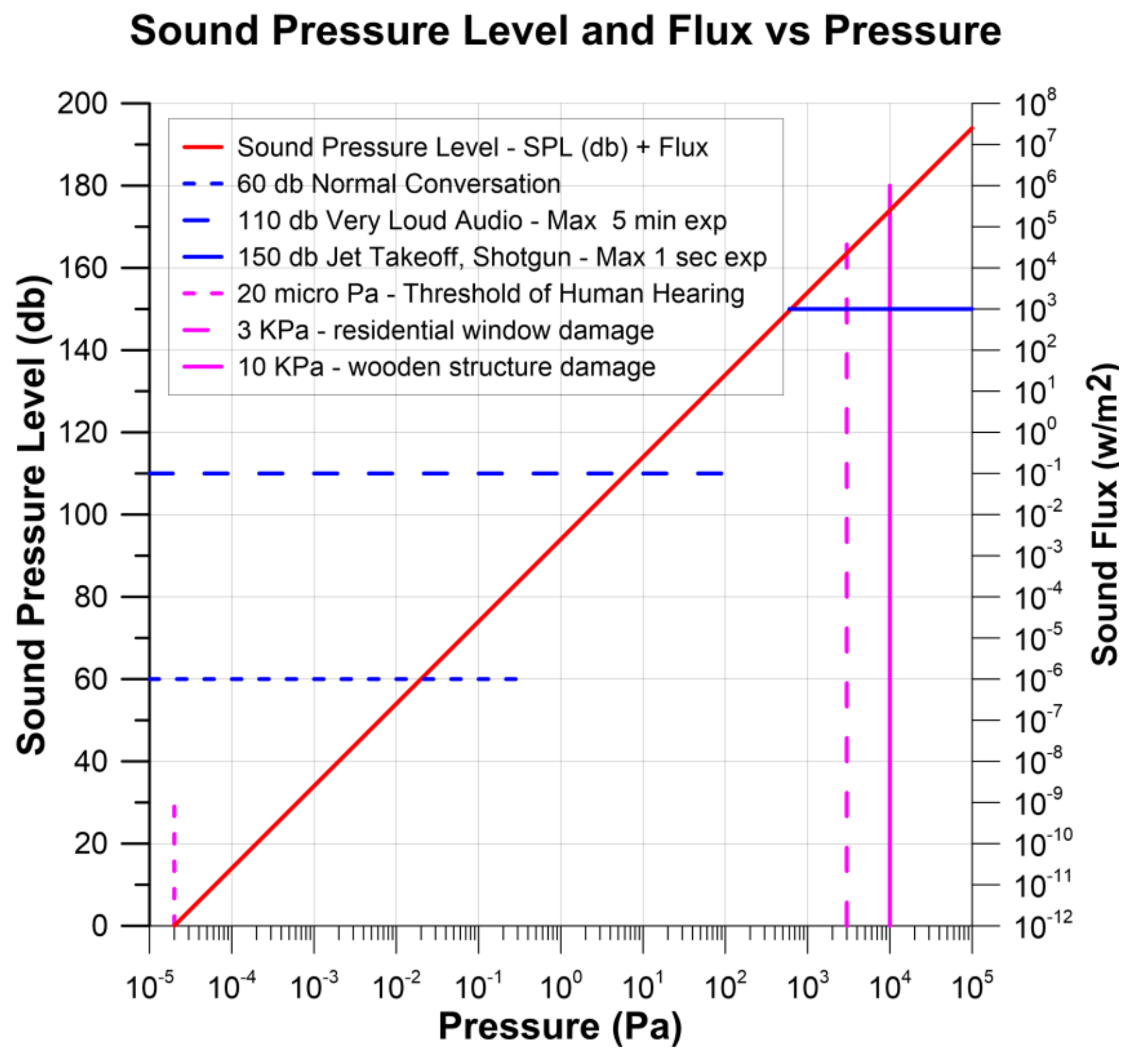

**Figure 14-** Relationship between sound pressure (SPL) and sound flux. Maximum exposure limits are the time before hearing damage occurs for unprotected hearing. *p*=3kPa (0.03 bar) is the pressure where window breakage can begin and *p*=10kPa (0.1 bar) is the limit where wood frame building damage begins.

***Hearing Damage*** – The threshold for hearing damage varies with the frequency and pulse shape. Typically, even short exposures to an SPL of >150db can cause hearing loss. An SPL of 150db corresponds to a peak pressure of approximately 1kPa (lower limit of residential window damage), which corresponds to an acoustical peak flux of a few kW/m$^2$. Simple mitigation includes using your hands/fingers to “close your ears” as well as simple ear plugs.

***Atmospheric Attenuation of Acoustical Waves***– There is significant absorption of acoustical shock waves in our atmosphere. The nature of the absorption is well understood from both a theoretical and experimental standpoint [25]–[31]. We summarize the relevant classical (viscous) and quantum mechanical corrections (molecular relaxation of vibration/rotation) for the primary species in our atmosphere (namely nitrogen, oxygen, and water vapor). The theory of absorption of the shock wave vs frequency and distance is shown in detail below. High frequencies are severely attenuated over the long ranges relevant here for the fragment created shock waves. Note than the exponent for “*e*-folding” scales as 1/v$^2$ at low frequencies with absorption dominated by $N_2$ below 1 kHz and $O_2$ above 1 kHz. At 100Hz, the absorption is about 100db/100km; at 10Hz the absorption is about 1db/100km; and at 1Hz it is ~ 0.01db/100km. The higher frequency components (>10Hz) are rapidly absorbed at long ranges (>100km), while the low frequency components from 1-10Hz are only mildly attenuated. The net effect of the atmospheric absorption at long ranges is to attenuate the high frequency components and thus act as a "low pass" spectral filter that changes the acoustical signature including overall reduction of the shock wave and its damage. This effect is also responsible for shock wave "pulse spreading/ stretching" that increases with time and distance the shock traveled from its source. This also why the thunder from a lightning strike nearby has a high frequency "sharp clap" while the thunder from distant lightning has a low frequency "rumble". Similarly, distant and particularly low on the horizon fragment detonations will sound like a low frequency "rumble" while relatively nearby fragment detonations will have more high frequency components that are more likely to cause damage.

Note that the absorption vs frequency below is given for pressure. The absorption vs frequency for intensity I(w/$m^2$) α $p^2$. The detailed mathematics are given below.

Atmospheric Acoustic Wave Absorption and Transmission vs Frequency

$\alpha(v,T,h)(db/km) = acoustical\ wave\ pressure\ absorption\ vs\ frequency\,(db/km)@1atm$

$Note\ that\ capital\ P\ is\ atmopsheric\ pressure\ while\ lower\ case\ p\ is\ shock\ overpressure$

$P(0) = P(z=0) = 1atm\ below\ where\ z = altitude$

$v = freq(Hz),\ T = temperature(K),\ h = humidity\ fraction\,(0-1)$

Converting to nepers ($N_P$):

$\alpha(v,T,h)(N_P/km) = \alpha(v,T,h)(db/km)/(20\log(e)) = 0.1151\alpha(v,T,h)(db/km)$

The absorption scales as the atmospheric density.

For different atmospheric densities $\rho$(z) and pressure P(z) where z= altitude assume:

$\alpha_{p(z)}(v,T,h) = \alpha(v,T,h)\rho(z)/\rho(0) = \alpha(v,T,h)P(z)/P(0)$

For an isothermal atmosphere $\rho(z)/\rho(0) = P(z)/P(0)$

$p_v dv = total\ acoustic\ pressure\ at\ freq\ v\ \&\ range\ dv,\ I_v dv = total$ intensity $at\ freq\ v\ \&\ range\ dv$

$I_v(x)/I_v(0) = [p_v(x)/p_v(x=0)]^2 = [p_v(x)/p_v(0)]^2$

where x=slant range where x=0 means no atmosphere absorption since slant distance=0 integrated over slant range - see next section on isothermal atmospheres

$$p_v(x)/p_v(0) = 10^{-\int_0^x [\alpha(v,T,h)(db/km)/20]dx(km)} = e^{-\frac{1}{20\log(e)}\int_0^x \alpha(v,T,h)(db/km)dx(km)} = e^{-0.1151\int_0^x \alpha(v,T,h)(db/km)dx(km)}$$

$$I_v(x)/I_v(0) = 10^{-\int_0^x [\alpha(v,T,h)(db/km)/10]dx(km)} = e^{-\frac{1}{10\log(e)}\int_0^x \alpha(v,T,h)(db/km)dx(km)} = e^{-0.2302\int_0^x \alpha(v,T,h)(db/km)dx(km)}$$

$p_v(x)dv = acoustic\ pressure\ at\ freq\ v\ \&\ range\ dv\ after\ going\ slant\ range\ x$

$$\alpha(v,T,h)(1atm) = 8690v^2 \begin{bmatrix} 1.84x10^{-11}\left(\frac{T}{T_0}\right)^{1/2} (classical) \\ +\left(\frac{T}{T_0}\right)^{-5/2} * \left[0.01275\frac{e^{-2239.1/T}}{F_{r,O}+v^2/F_{r,O}}\right](O_2) \\ +\left(\frac{T}{T_0}\right)^{-5/2} * \left[0.1068\frac{e^{-3352/T}}{F_{r,N}+v^2/F_{r,N}}\right](N_2) \end{bmatrix}$$

$$F_{r,O} = 24 + 4.04x10^4 h\frac{0.02+h}{0.391+h} = Oxygen\ relax\ freq$$

$$F_{r,N} = \left(\frac{T}{T_0}\right)^{-1/2}\left(9 + 280he^{\left(-4.17\left[\left[\frac{T}{T_0}\right]^{-1/3}-1\right]\right)}\right) = Nitrogen\ relax\ freq$$

$h = humidity\ fraction\ (0-1),\ T_0 = 293.15K$

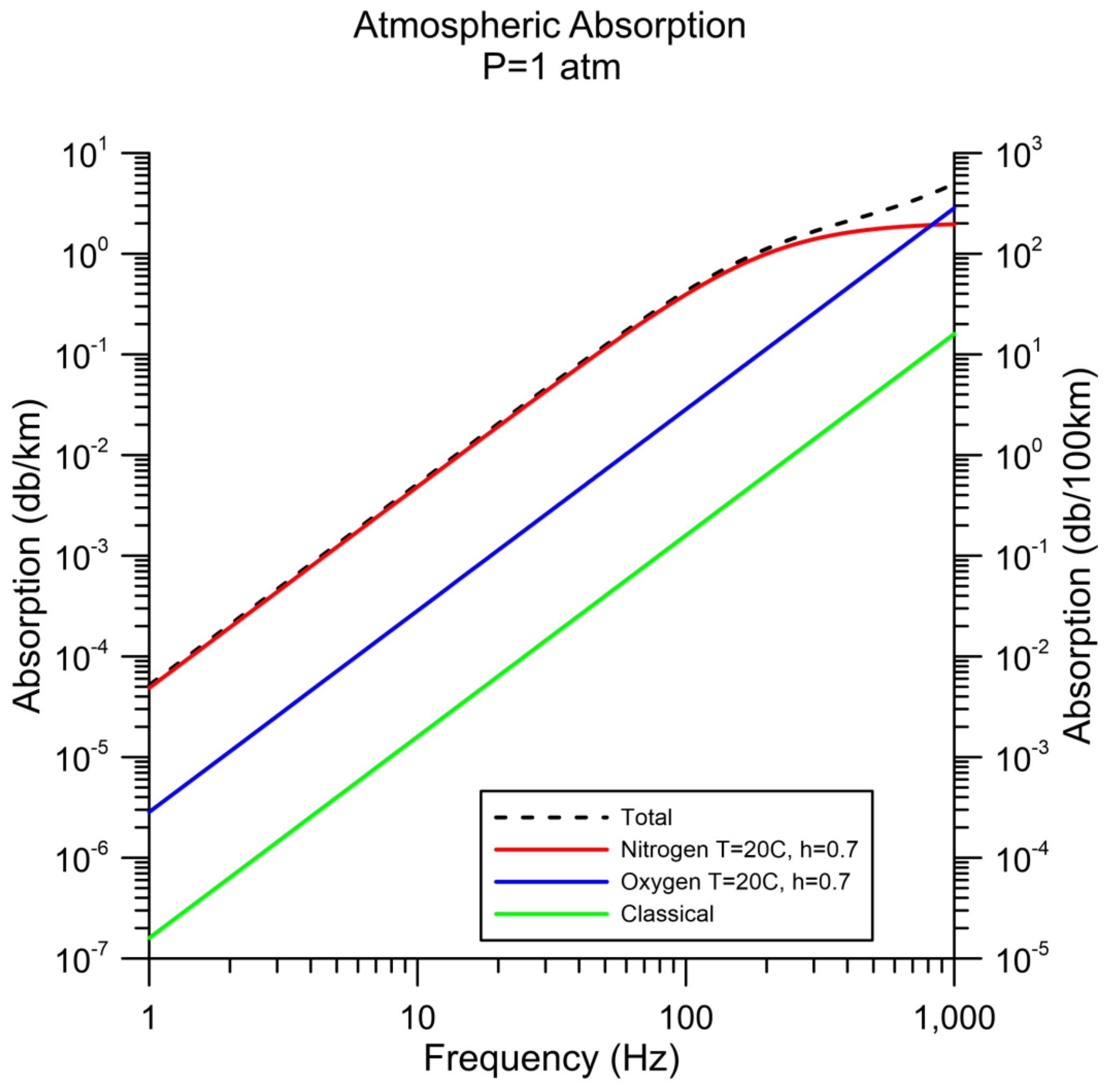

**Figure 15 -** Atmospheric absorption coefficient for pressure of sound waves vs frequency from 1 to 1000Hz. The classical (Green) and QM correction for $O_2$ (Blue) and $N_2$ (Red) molecules and the total absorption (black dashed) are shown. Note the extremely high absorption of frequencies above 100Hz at relevant distances. Note that for pressure ratios 20db=10x difference while for intensity ratios 10db=10x difference since intensity $I \alpha P^2$.

***Pressure wave attenuation over slant range from burst at high altitude*** – For a burst above ground level, we need to integrate over the path from the burst to the point of observation. **We assume an isothermal atmosphere for simplicity**.

$$r_g = \text{ground distance from ground zero}$$
$$r_s = \text{slant distance from burst}$$
$$r_b = z_b = \text{burst altitude}$$
$$H = e\ folding\ scale\ height\ (\sim 8km\ for\ air, less\ for\ water\ vapor\ which\ is\ \text{sub dominant})$$
$$r_s = \left(r_g^{\ 2} + z_b^{\ 2}\right)^{1/2} = z_b\left(1+\frac{r_g^{\ 2}}{z_b^{\ 2}}\right)^{1/2}$$
$$p(v) = \text{blast wave pressure with atmopsheric attenuation}$$
$$p_0(v) = \text{blast wave pressure without atmopsheric attenuation}$$

To make the problem analytically tractable we assume h is constant along the path

and that the absorption coef (in nepers/length) is proportional to the atm density along the path

$$\alpha(v,T,h,\rho(s)) = \alpha(v,T,h,\rho(z=0))*\rho(z)/\rho(0) = \alpha(v,T,h,\rho(0))e^{-z/H} \text{ and that } ds = dr_s = \frac{r_s}{r_b}dz$$

$$P(v)/P_0(v) = e^{-\int_{psth}\alpha(v,T,h,\rho(s))ds} \sim e^{-\frac{r_s}{r_b}\alpha(v,T,h,\rho(0))\int_0^{z_b} e^{-z/H}dz} = e^{-\frac{r_s}{r_b}\alpha(v,T,h,\rho(0))\mathrm{H}(1\text{-}e^{-z_b/H})} = e^{-\alpha_{eff}(v,T,h,\rho(0))\mathrm{H}}$$

$$\alpha_{eff}(v,T,h,\rho(0)) = \frac{r_s}{r_b}\alpha(v,T,h,\rho(0))(1\text{-}e^{-z_b/H}) = \alpha(v,T,h,\rho(0))\left(1+\frac{r_g^{\ 2}}{z_b^{\ 2}}\right)^{1/2}(1\text{-}e^{-z_b/H})$$

Whatever units $\alpha(v,T,h,\rho(0))\ is\ measured\ in\ (eg: nepers/m,\ db/km\ etc)\ then:$

$$\alpha_{eff}(v,T,h,\rho(0)) = \alpha(v,T,h,\rho(0))\left(1+\frac{r_g^{\ 2}}{z_b^{\ 2}}\right)^{1/2}(1\text{-}e^{-z_b/H}) \sim \alpha(v,T,h,\rho(0))\left(1+\frac{r_g^{\ 2}}{z_b^{\ 2}}\right)^{1/2}\quad (z_b >> H)$$

Note that for a low altitude burst where $z_b << H \rightarrow (1\text{-}e^{-z_b/H}) \sim z_b / H$

$$\rightarrow \alpha_{eff}(v,T,h,\rho(0))\mathrm{H} = \alpha(v,T,h,\rho(0))\left(1+\frac{r_g^{\ 2}}{z_b^{\ 2}}\right)^{1/2}(1\text{-}e^{-z_b/H})H \sim \alpha(v,T,h,\rho(0))\left(z_b^{\ 2}+r_g^{\ 2}\right)^{1/2}$$

$$= \alpha(v,T,h,\rho(0))r_s \sim \alpha(v,T,h,\rho(0))r_g\ (\,for\ z_b << r_g \text{ as expected})$$

$$Our\ typical\ burst\ altitude\ for\ 10m\ diameter\ class\ fragments\ is\ z_b \sim 30-40km \rightarrow z_b / H \sim 4-5$$

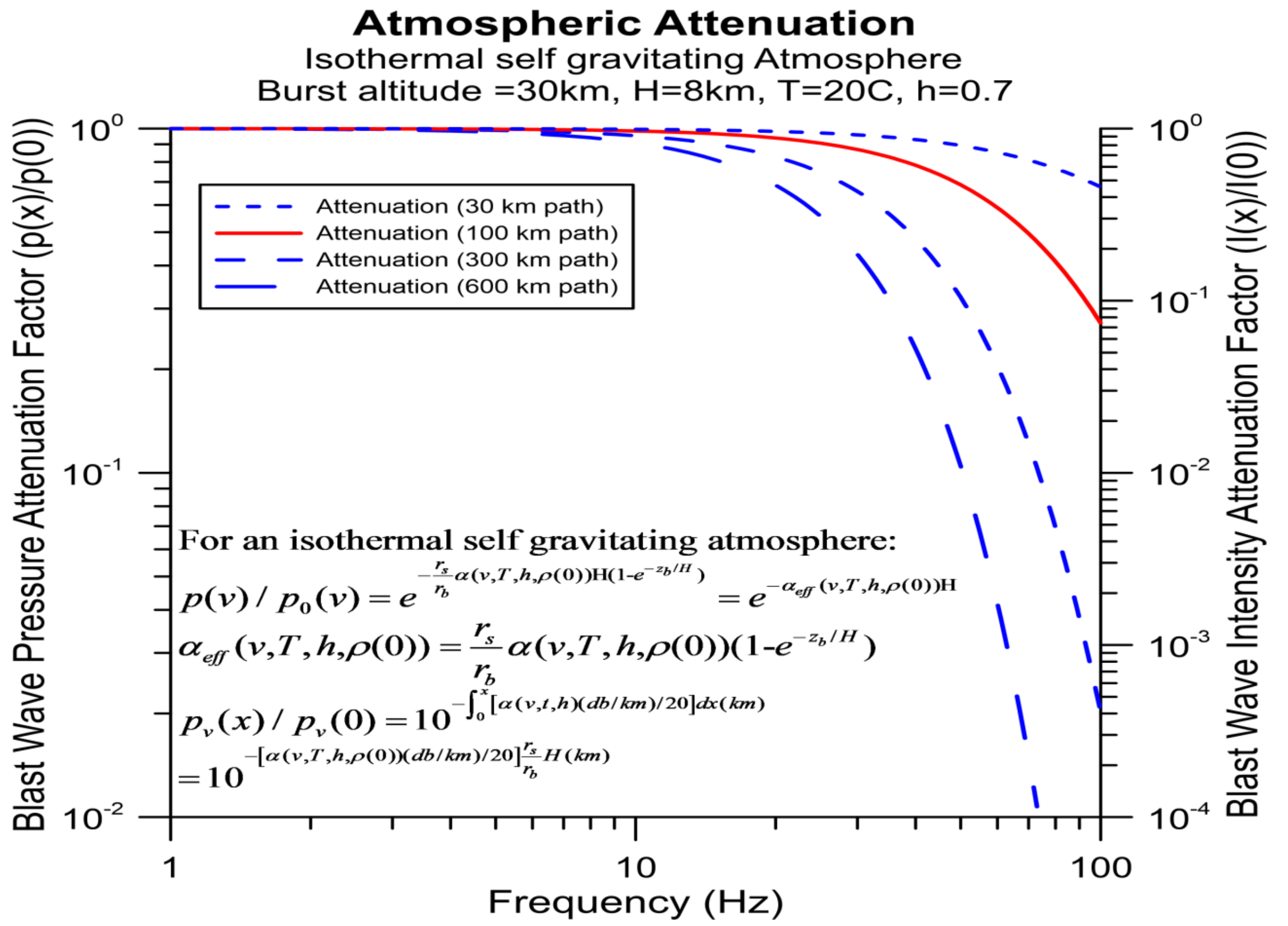

**Figure 17 –** Pressure and intensity attenuation for isothermal self-gravitating atmosphere case. $r_b$=$z_b$ = burst altitude.

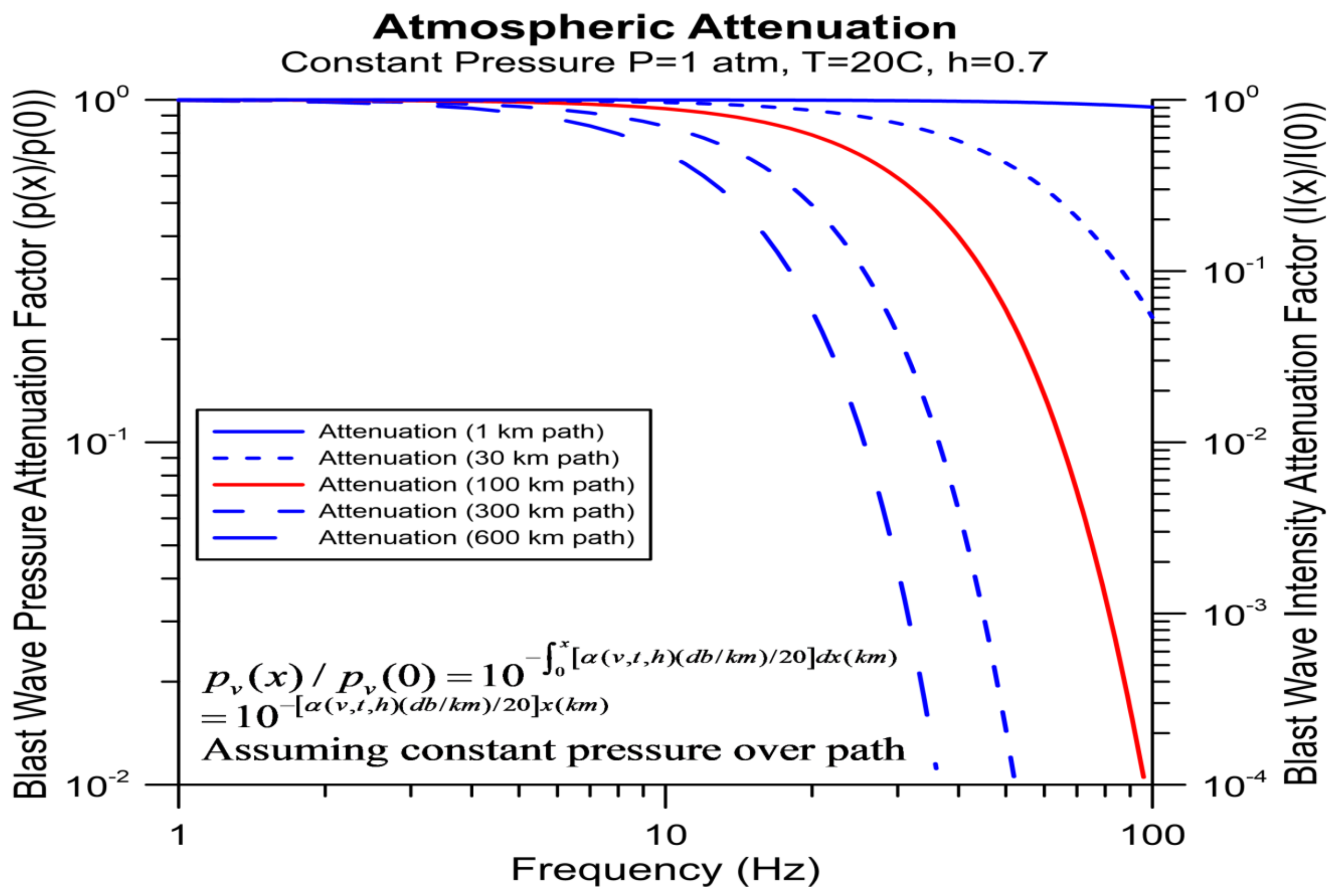

**Figure 16-** Blast wave pressure and intensity attenuation vs frequency and slant distance for constant pressure.

***High Altitude Detonation Effects and Data –*** There are relatively few high altitude-controlled detonations to cross check. While there is a wide variety of re-entry vehicles and hypersonic vehicle tests, the local energy depositions are far from the regime of interest to us. The general question that arises is "what is the effect of the altitude of the energy deposition on the ground effect of the shock wave"? Clearly, in the limit of detonation in a vacuum, there will no zero ground shock wave from direct coupling. If we look at the mean free path length (MFP) for atmospheric molecular collisions, we gain some insight into the general physics of the problem. For example at sea level the MFP ~0.1 micron, at 30 km MFP ~ 10 microns, at 50km MFP ~ 0.2mm at 100km MFP ~ 0.3m while at the altitude of the ISS (~ 400 km) the MFP ~ 60km. The MFP is relevant to the "tamping" efficiency for blast coupling to air. This is discussed in greater detail later in this paper when we analyze the ablation and energy deposition for hypersonic entry relevant to bolides. The details of the energy injection mechanisms become important in understanding how the energy will be coupled to the atmosphere. In the case of a hypersonic bolide the coupling to primarily through mechanical interaction between the bolide and the atmosphere whereas in a nuclear weapon there are long range and highly energetic processes from high energy photons (gamma) and from energetic neutrons both of which have long range interaction MFP's particularly in a low-density atmosphere at high altitude.

**Hardtack Teak and Orange High Altitude NED Tests** – There are a few high-altitude NED tests that are relevant in understanding the effects of detonation altitude on the ground effects. In particular, the Operation Hardtack series had two relevant high-altitude tests with the same 3.8 Mt yield. On August 1, 1958 the Teak test detonated at an altitude of 77 km nearly directly overhead at the Johnston Atoll in the Pacific and on August 12, 1958 the Orange test was detonated at an altitude of 43 km, horizontal offset of 40km and a range of 59 km from Johnston Atoll. Both of these tests were done in early ICBM tests using a Redstone rocket. Ground pressure at the Atoll were measured for both tests yielding valuable data for us to cross check our predictions. The data and numerical prediction scaled from ground level tests are shown. In both tests our numerical prediction over predict the peak shock pressure on the ground with the lower altitude Orange (measured ~25% lower than our prediction) test being closer than the Teak which measured ~ 40% lower than our prediction. **For the purposes of this paper and the relevant high altitudes and energies of bolides, this agreement is "good enough".** While the yield (3.8 Mt) was higher than our typical fragment energy range of interest (10-100 KT), the ground pressure near 1 kPa is ideal. As discussed later, while a spherical approximation of a bolide event is an approximation of the complex bolide detonation geometry, it is shown to be a reasonable one for our purposes.

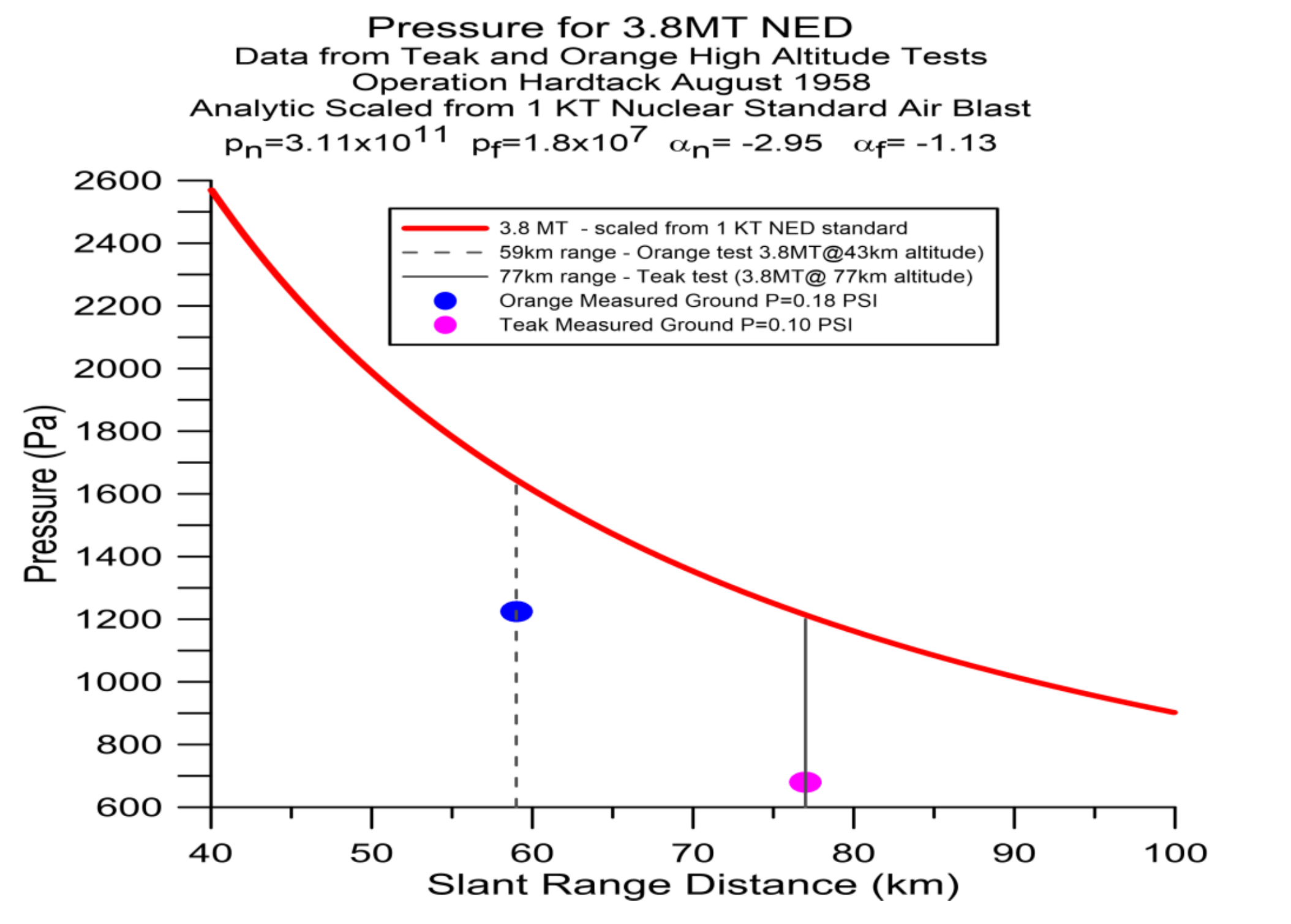

**Figure 18 –** Ground peak-overpressure from Operation Hardtack for test Teak (magenta) and Orange (blue) shown vs our model predictions (red curve). We over predict the ground pressure

**Yucca test** – On Aril 28, 1958 there was a test of a 1.4 KT device lofted to 26km using a large balloon but this has less relevant data unfortunately due to the low yield.

***Relating Blast Pressure to Common Phenomenon*** – It is useful to relate the observer maximum blast pressures we will be designing the mitigation to in terms of common phenomenon. A simple example that almost anyone reading this paper can experience is that of the pressure on a sea level car travelling at highway/ freeway speeds. If you hold your hand outside the window in a car travelling at 100 km/h (62 MPH) then the ram pressure $P = \rho v^2$ =0.96 KPa or ~1KPa (0.01 bar). Clearly your windshield does not crack, however car glass is laminated with plastic and secured much better than typically "single strength" (2.4mm thick in US) single pane residential glass. The same ram pressure is what would be experience by a house in a 100 km/h wind which is categorized as a tropical storm and below the speed of a Category 1 (lowest) hurricane. In such winds, your house windows typically do not break unless impacted by debris. If we increase the car speed to 200 km/h (124 MPH). the ram pressure being quadratic in speed, increases to ~ 4 KPa (0.04 bar). The car wind shield still does not crack, however a house buffeted with 200 km/h winds (Category 3 hurricane) may experience window and roof damage depending on the window construction. "Normal residential construction" houses will generally not be toppled in a 200 km/h wind though trees may be felled and "mobile homes not properly anchored" may topple. If we increase the speed to 300 km/h (~ 200 MPH) (Category 6 hurricane) the ram pressure is 9 KPa (0.1 bar) and while a properly designed car windshield will not break, residential windows will generally shatter and wood frame construction homes will be seriously damaged.
**A VERY significant difference between a storm/ hurricane and a shock wave is the short duration (~ seconds) of the shock wave vs long term storm conditions. Inertial mass effects on structures for short pulses often dominate building and tree resistance to damage.**

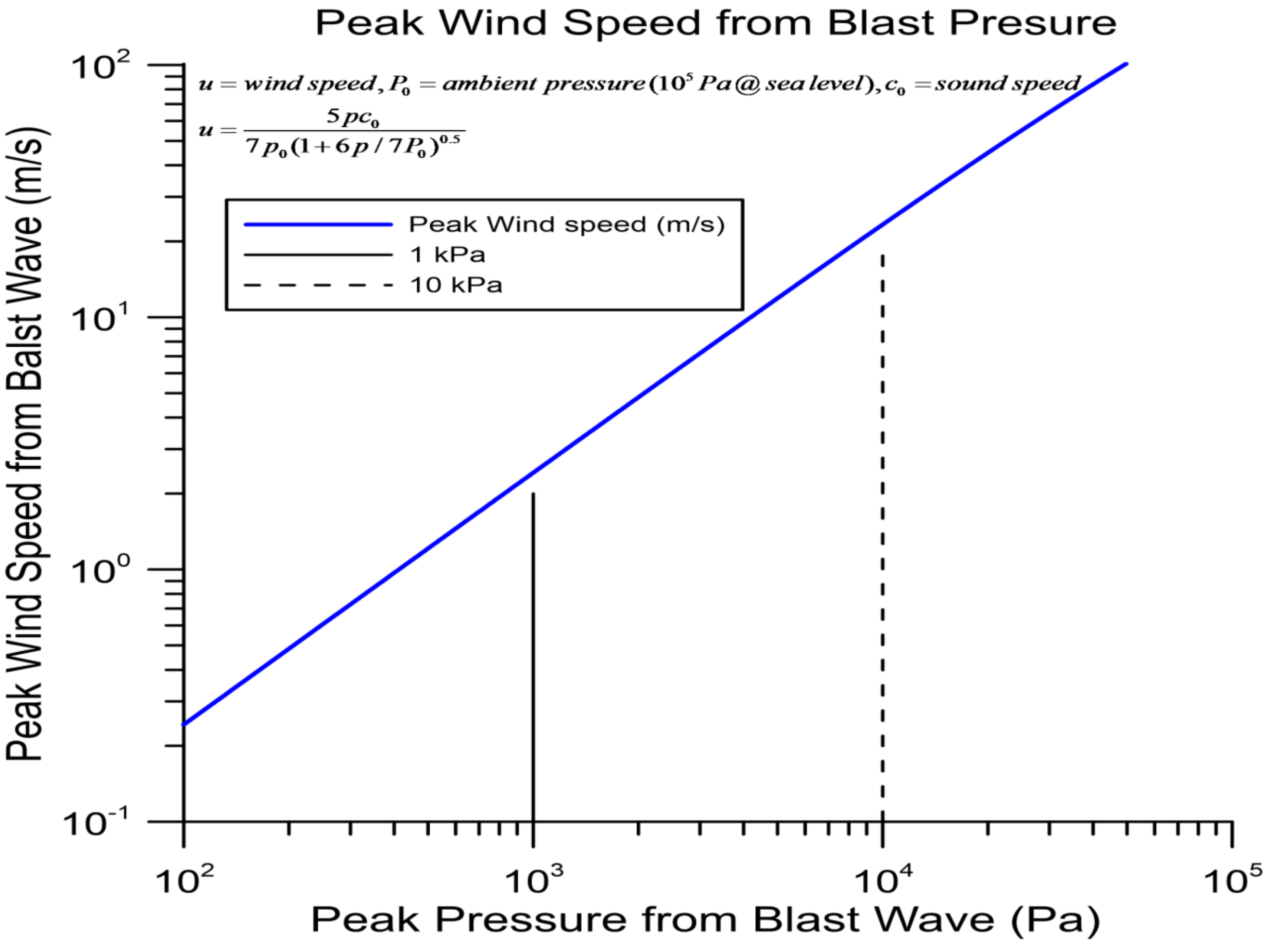

**Figure 19** – Peak wind sped created by blast wave vs peak pressure. Note the blast created wind speed is much smaller than the equivalent ram pressure from the bulk motion of atmosphere. 1 kPa (window cracking threshold) and 10 kPa (wood frame structure damage threshold) shown.

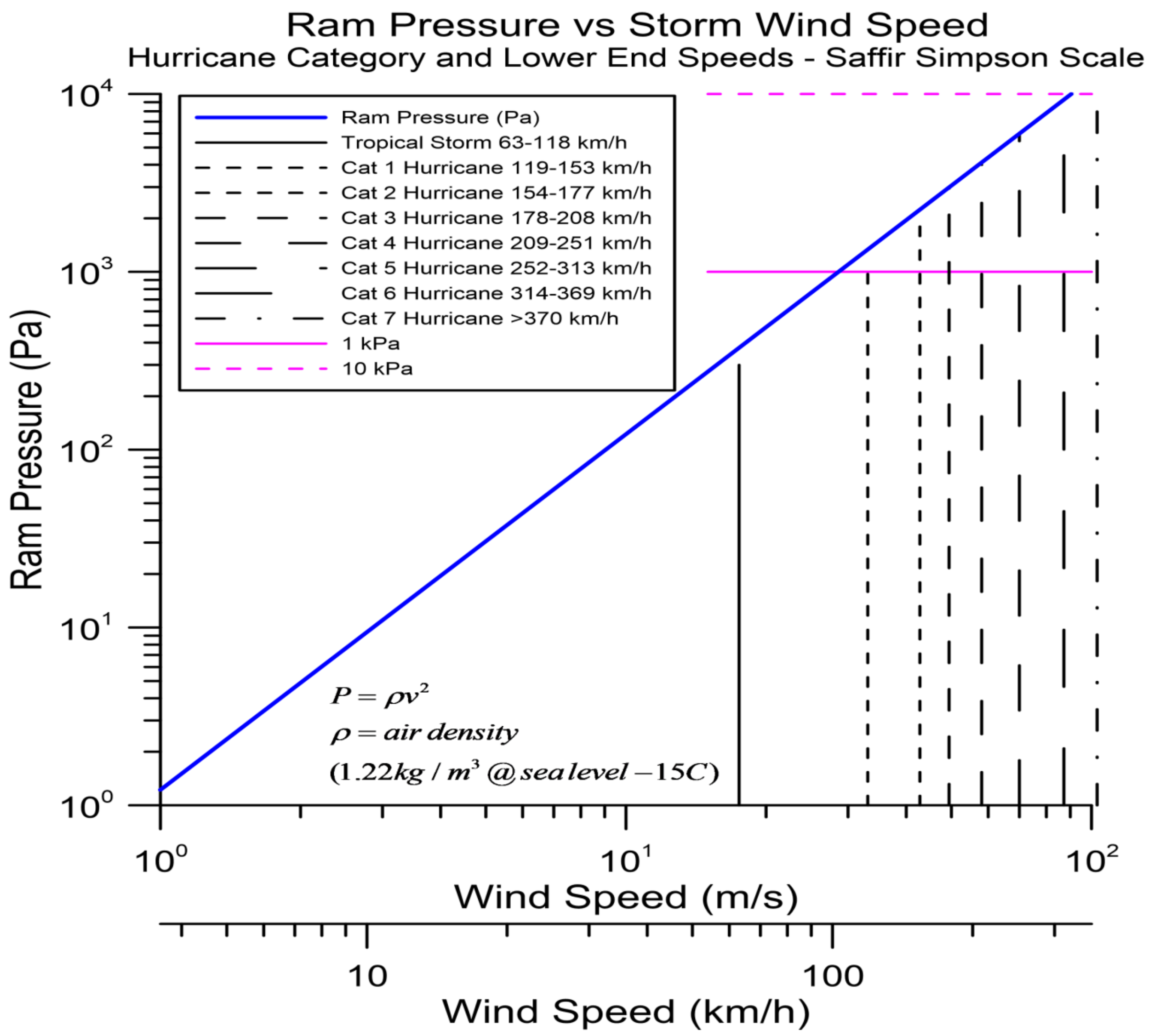

**Figure 20-**Ram pressure vs wind speed from storms with various storm categories. Also shown are 1 kPa (window damage threshold) and 10 kPa (wood frame structure damage threshold)

***Window Breakage*** – The breaking of windows and structures in general is a complex and less than satisfactory area for precise modeling due to the inhomogeneous nature of these elements. For windows the type of glass, size and thickness are critical as are the presence of defects, the edge mounting (plastic, metal, wood) and strength of the edge joint, age of the glass, previous history of shocks, angle relative to the blast source are all sources for which precise prediction are largely not feasible. We resort to both modeling and some large scale data such as nuclear airbursts, natural gas explosions, munition detonations both intentional and accidental. The issue of what constitutes a "failure" is also subjective in the literature. For example, is a slight crack induced by a shock wave of the same danger as ballistic shattering (flying glass) of the window? The studies of window damage in atmospheric nuclear tests such as discussed in Glasstone and Dolan gives peak pressure value for "window failure" of about 3 kPa while the work of Reed (1992) which studied the 1963 accidental HE explosive "Medina Incident" near San Antonio, Texas gives a probability vs peak pressure from 0.1 to 10 kPa for typical US residential windows of the 1960's. The conclusion of Reed's studies which looked at insurance claims for window replacement concludes "some damage – cracking for example" can begin at several hundred Pa.

On Friday the 13th, November 1963 at Site K bunker 572 at the Medina Disassembly Facility on Lackland Air Force Base in Texas, in a nuclear weapons assembly and disassembly area, Mrk 7 fission weapons were being disassembled with the conventional HE explosives stored for later incineration. At about 10 AM a small fire started in one of the storage areas that subsequently triggered the detonation of about 60 tons of HE from 209 explosive assemblies in bunker 572 that previously surrounded the fission cores of the disassembled weapons. The resulting explosion caused extensive damage to the local weapons storage facility though fortunately, and miraculously, the three technicians working in the storage bunker managed to escape largely unharmed. The blast wave from this explosion damaged a large number of windows on the base and in the adjoining city of San Antonino, which is what the Reed 1992 data is based on. In the adjoining plot below the term "window breakage" is largely based on the resulting insurance claims and thus not very quantitative in terms of the type of window damage. What is extremely useful however is the large number of windows damaged and the rather precise ability to calculate the peak pressure vs distance of the windows from the detonation site. There are detailed reports from a variety of schools, shops and homes at various distances with a wide variety of window sizes. For example, reports at 1.6km (significant window damage) and 17 km (large plate glass store window pushed in but did not burst) and at 20km where a large plate glass window broke. The expected pressure with no reflections (reflections can significantly increase in the local pressure) are shown in the accompanying figure.

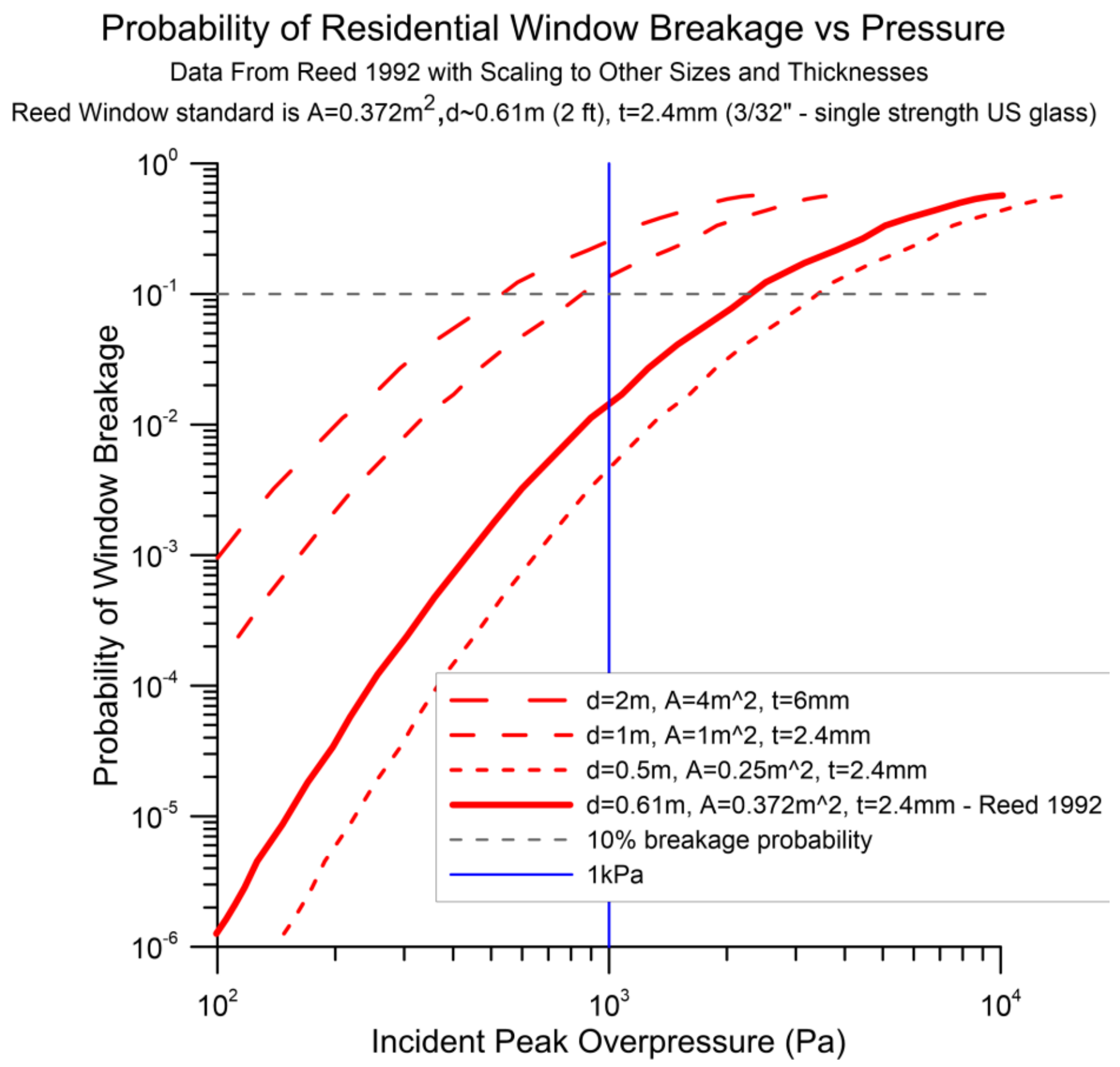

**Figure 21** – Probability of window breakage vs incident peak pressure from 1963 Medina Incident. Based on Reed 1992.

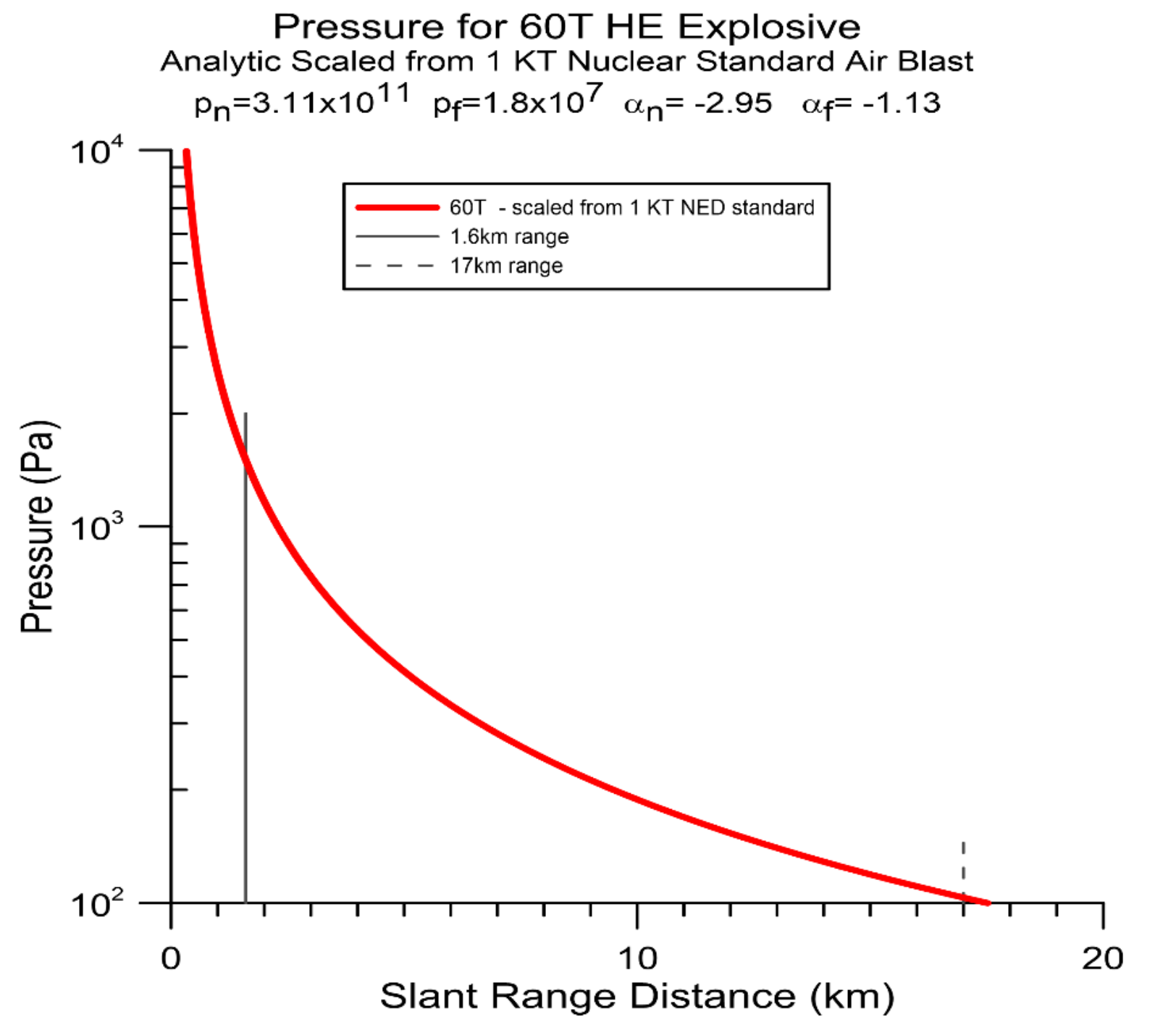

**Figure 22 –** Estimated pressure vs distance from Medina bunker.

For application to planetary defense simulations, additional complexities come from the varying size and thickness of glass windows as well as the edge mounting used in windows around the world. Making a precise statement about "how many windows will break" and to what extent will people be injured is even more complex to answer with any precision. Our approach, like that of others, is to bound the problem in a conservative manner. Comparing the Reed 1992 data with atmospheric nuclear tests, bomb damage in WW II and other conflicts, natural gas explosions, bolide shock wave such as the Chelyabinsk event all lead to reasonably consistent conclusions namely that above 3 kPa there is significant damage to typical residential glass windows and below 0.3 kPa there is little damage. As mention in other areas of this paper, reasonable simple civilian defense measures such as closing outside shutters, putting tape on windows, staying away from windows, closing blinds and curtains is generally sufficient to prevent significant harm to people even at 3 kPa peak pressure, though windows may break. A similar statement about staying away from unreinforced brick walls and poorly constructed building would also be common sense as it is in any natural event such as a hurricane or severe storm. In general, modern commercial building windows are significantly stronger than residential windows. Our analysis is general though we focus on residential windows.
The scaling of the size and thickness of the window glass and pressure for equal probability of damage is as follows with thinner and larger windows breaking more readily:

Pressure Scaling for Equal Probability of Breakage

$$p(A,t) = \text{pressure for given break probability}$$

$$p_0(A_0,t_0) = \text{pressure for given break probability in Reed data}$$

$$A = \text{window area} \;,\; t = \text{window thickness}$$

$$p(A,t) = \frac{t/t_0}{A/A_0}\, p_0(A_0,t_0)$$

$$A_0 = 0.372m^2\,(d = 0.61m\,(2\,ft)), t_0 = 2.4mm = 3/32\text{" US standard single strength (Reed 1992 data)}$$

**Angular Dependence of Window Breakage with Orientation** – The probability of window breakage also depends on the orientation of the window relative to the location of the generation of the shock wave. Based on both theoretical modeling as well as test data we can model the orientation dependence as shown below. It is important to note that window breakage is also highly dependent on the building specifics as well as nearby buildings that cause shadowing and reflection that can both decrease and increase the breakage probability. As discussed above the pressure required to break a window depends on the size (area) and the thickness to first order. Orientation dependence is a modest correction factor for a given window that we express as an "equivalent pressure multiplier" for a given shock wave incident pressure vs orientation with windows on the structure directly facing the direction of the source of the blast causing the shock breaking with higher probability than windows on the side and back of the structure. The presence of other structures and local topography (tress, hills, mountains etc)

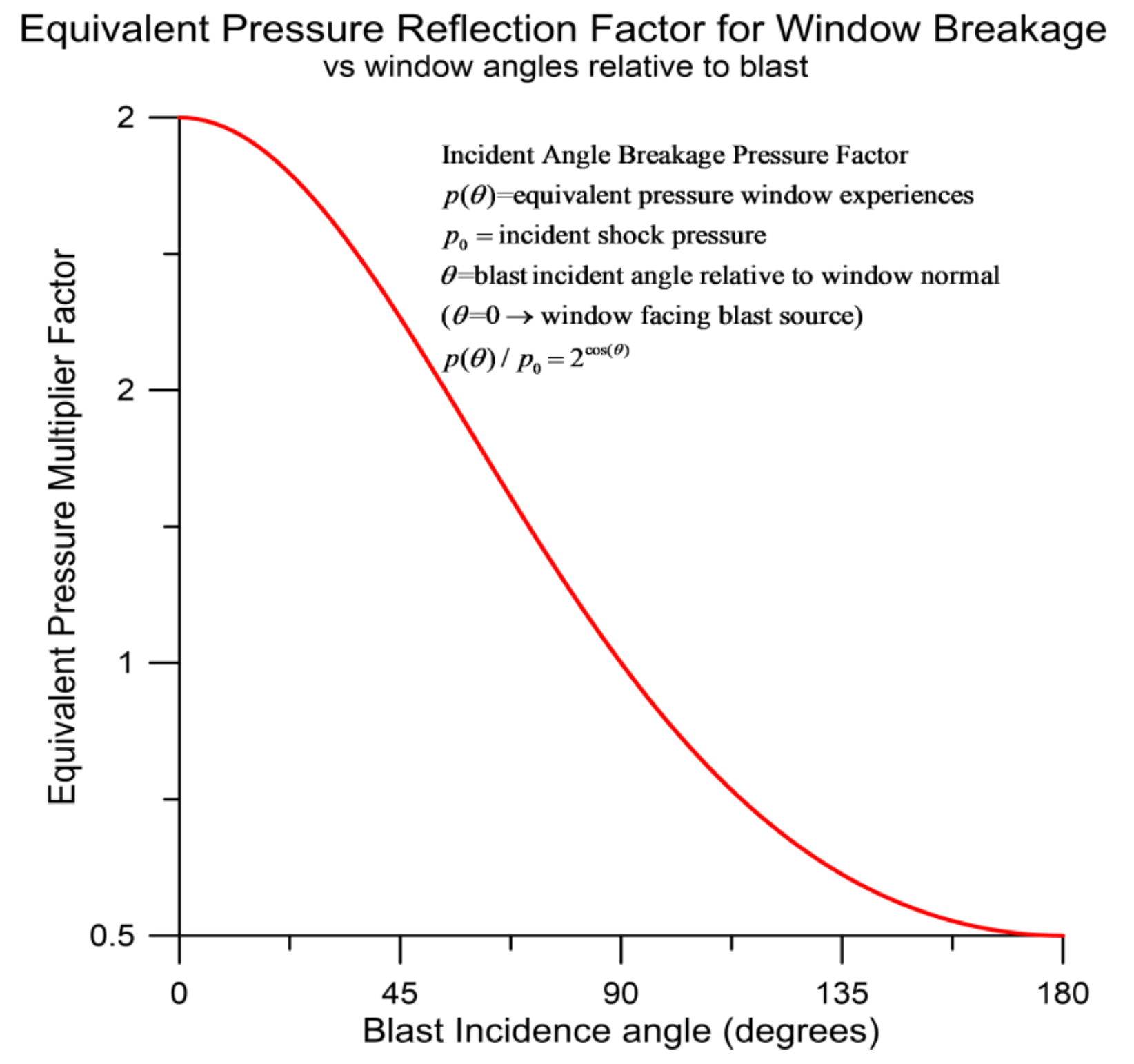

**Figure 23** – Pressure reflection enhancement factor for window breakage as a function of window orientation relative to shock source with incidence angle being the shock source direction relative to the window normal. Back side facing windows have a 4x lower effective pressure than windows on the front facing side of the structure and side facing have 2x lower eff pressure both relative to incident pressure.

nearby greatly complicates our predictive abilities.

Incident Angle Breakage Pressure Factor

$p(\theta)$=equivalent pressure window experiences

Note that in the weak shock regime the

"reflected pressure" is 2x incident pressure

$p_0$ = incident shock pressure

$\theta$=blast incident angle relative to window normal

($\theta$=0 $\rightarrow$ window facing blast source)

$$p(\theta)/p_0 = 2^{\cos(\theta)}$$

$\cos(\theta) = \hat{n}_w \bullet \hat{n}_{blast}$ *where* $\hat{n}_w$ = *window normal*, $\hat{n}_{blast}$ = *blast source direction*

Window in front experiences 4x more eff pressure than window in back of structure

Azimuthal average for random window orientation:

$$\langle p(\theta)/p_0 \rangle_{azimuth} = \int_0^{2\pi} [p(\theta)/p_0] d\theta / 2\pi = \int_0^{2\pi} 2^{\cos(\theta)} d\theta / 2\pi \sim 1.12$$

Note that airbursts will not generally be horizontal sources.

Assuming random burst distribution but equal burst magnitude in upper plane ($2\pi$)

$$\langle p(\theta)/p_0 \rangle = \int_{2\pi} [p(\theta)/p_0] d\Omega / 2\pi = \int_{2\pi} 2^{\hat{n}_w \bullet \hat{n}_{blast}} d\Omega$$

*Assume window normal horizontal along x axis (horizon facing window):*

$\hat{n}_w = (1,0,0)$

$\hat{n}_{blast} = (\sin(\theta)\cos(\phi), \sin(\theta)\sin(\phi), \cos(\theta))$

$\theta$ = *angle from vertical (z axis),* $\varphi$ = *angle from x axis (xy = horizontal plane)*

$$\langle p(\theta)/p_0 \rangle = \int_{2\pi} [p(\theta)/p_0] d\Omega / 2\pi = \int_{2\pi} 2^{\hat{n}_w \bullet \hat{n}_{blast}} d\Omega = \int_0^{2\pi}\int_0^{\pi/2} 2^{\sin(\theta)\cos(\phi)} \sin(\theta) d\theta d\phi / 2\pi \sim 1.08$$

*Assume window normal vertical along z axis (upward facing window):*

$\hat{n}_w = (0,0,1)$

$\hat{n}_{blast} = (\sin(\theta)\cos(\phi), \sin(\theta)\sin(\phi), \cos(\theta))$

$\theta$ = *angle from vertical (z axis),* $\varphi$ = *angle from x axis (xy = horizontal plane)*

$$\langle p(\theta)/p_0 \rangle = \int_{2\pi} [p(\theta)/p_0] d\Omega / 2\pi = \int_{2\pi} 2^{\hat{n}_w \bullet \hat{n}_{blast}} d\Omega = \int_0^{2\pi}\int_0^{\pi/2} 2^{\cos(\theta)} \sin(\theta) d\theta d\phi / 2\pi = 1/\ln(2) \sim 1.44$$

**Weak Shock Regime for Small to Large Explosives** – For comparison we compute the shock overpressure and intensity vs distance in the weak shock regime (here <0.1 bar) for explosives from 1g to 1 Mt. This regime is of primary interest to us for planetary defense (typ 10-100 Kt) but it also shows the connection to small explosive regimes such as large firecrackers like an "M80" which contains 3-5g of explosive and to a stick of dynamite used in construction/ demolition which contains ~ 190g of explosive. For reference a typical "firecracker" only contains about 50mg of explosive. A reasonable approximation to the peak over pressure in the weak shock (far field) regime is $P(kPa) \sim 6E(kt)^{1/3} / r(km)$ for P<10 kPa where we have forced a 1/r dependence. This works reasonably well even up to a few 10's of kPa.

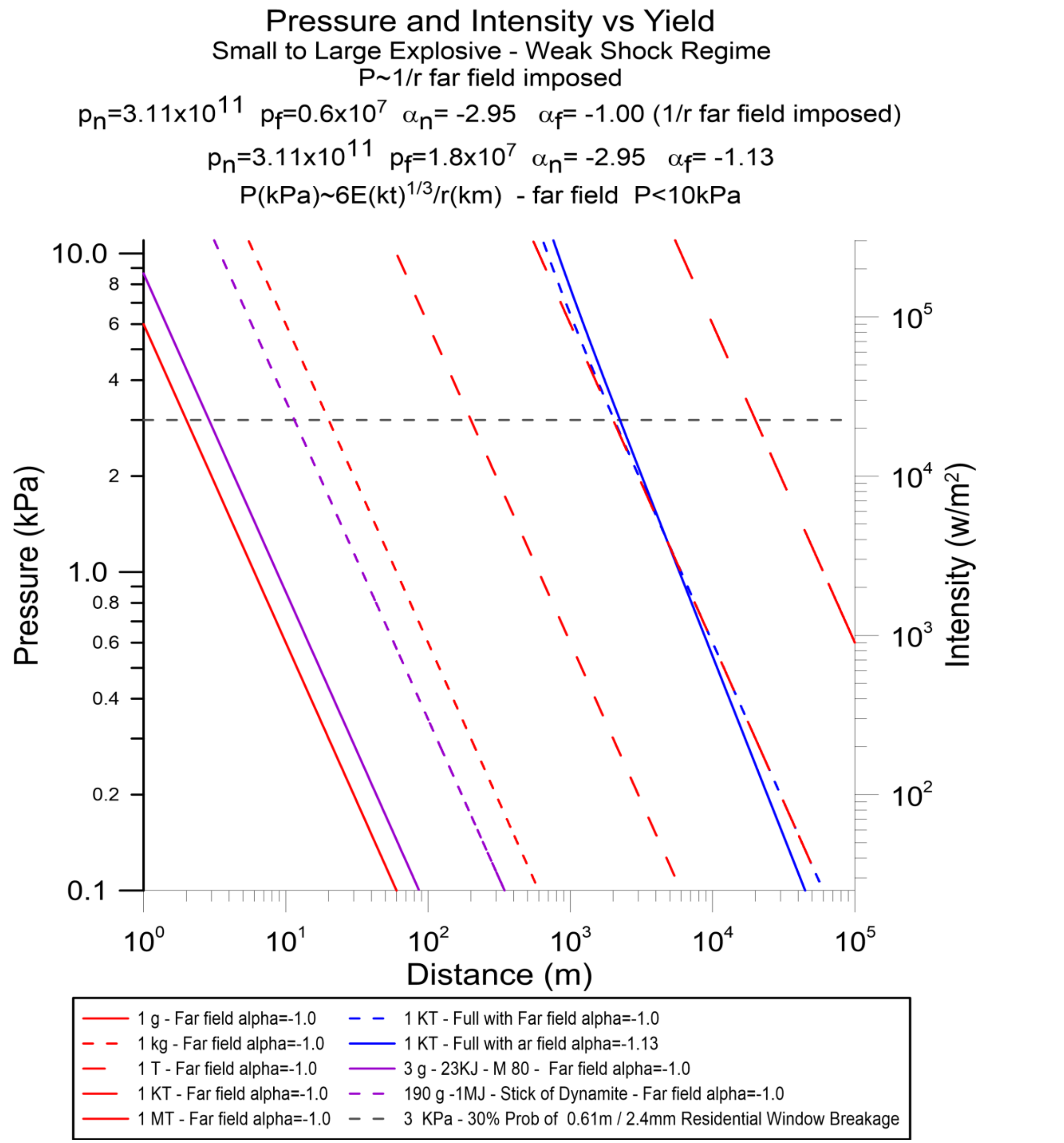

**Figure 24** – Peak overpressure and shock acoustic intensity vs distance from explosions in the weak shock regime for explosives from 1g (large firecracker) to 1 Mt (thermonuclear NED).

***Horizon Distance from Burst Altitude*** – The “limb” of the Earth (the horizon), or “radar horizon,” as seen from the burst altitude is computed and shown below. The horizon distance for a 30km burst, typical for a 10m diameter stony asteroid fragment, is roughly 600km from ground zero. Acoustical signatures from the shock wave can still be detected beyond the horizon due to diffraction and atmospheric “ducting” effects, but the primary effects of the shock wave are greatly diminished beyond the horizon in general. Depending on the intercept time prior to impact, the fragment spread can be greater than the horizon. **Note that h=height of asteroid burst below (referred to as $z_b$ in other sections) is NOT the h=humidity fraction used in the previous section on acoustical atmospheric absorption.**

$$D_h = \left(2hR_e + h^2\right)^{1/2}$$

$$D_h = \text{"radar horizon"=distance to "surface" of Earth when pointing horizontally}$$

$$h = \text{altitude of asteroid burst above surface}$$

$$R_e = \text{radius of Earth} \sim 6370 \text{ km}$$

$$D_h = \left(2hR_e + h^2\right)^{1/2} \sim \left(2hR_e\right)^{1/2} \sim 113h^{1/2} km\ for\ h\ in\ km \quad (h << R_e)$$

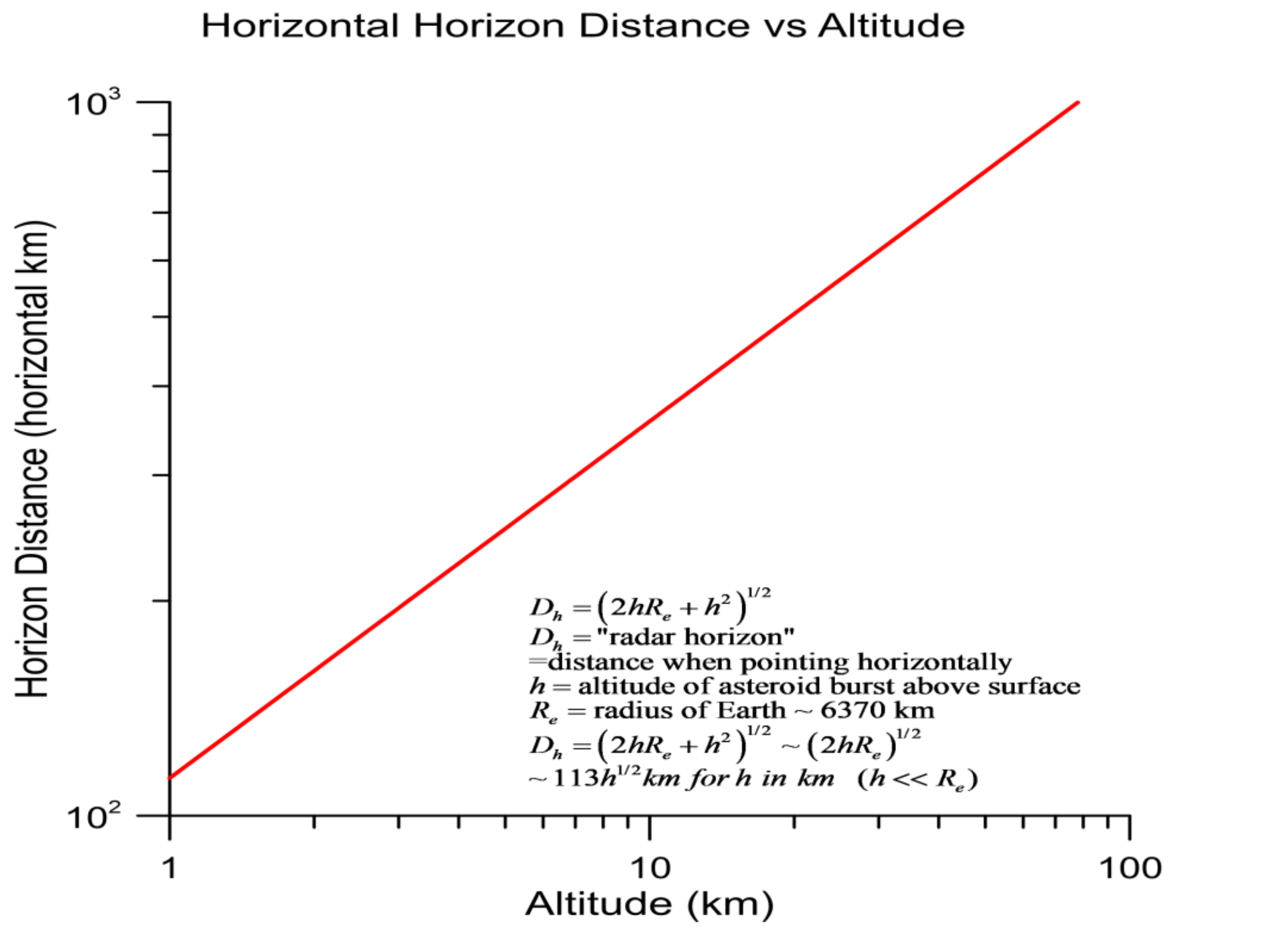

**Figure 25**– Horizon distance (horizontal range) vs altitude. The typical burst altitudes of 10m diameter fragment fragments is about 30km, yielding a horizon distances of about 600km. By this distance, the shock wave peak pressure will have diminished greatly and would not be hazardous.

## 3. Atmospheric Loading Capability

The ability of the atmosphere to act as a “beam dump” or a “bullet proof vest” gives us the ability to absorb an enormous amount of energy without significant damage IF that energy is distributed spatially and temporally. In our simulations, we explore these limits in detail. Some examples are helpful here to put this into perspective. The largest human-made explosion (impulse) was the Soviet test of the *Czar Bomba* NED on October 30, 1961. Detonated at an altitude of 4 km with a de-rated yield of 50Mt ($2x10^{17}$ J) down from the max design of 100Mt by replacing the last $^{238}U$ tamper with Pb, it was by far the largest human explosion. The shock wave including ground reflections broke windows up to 900 km away. For reference, it was about 4-5 times the likely yield as the 1908 Tunguska effect some 53 years earlier. Besides the radioactive fallout (it was one of the "cleanest" weapons ever detonated due to the large fusion yield), there was only modest surface damage. The 2013 Chelyabinsk asteroid event deposited about 0.5Mt of energy, largely into shock waves and produced no significant damage to the atmosphere (it did harm people below). Note that the Chelyabinsk event was about 100x less energy than the Tzar Bomba NED. The KT boundary extinction event some 66 million years ago had a yield of some 100 TeraTons (100Tt = $10^8$ Mt) and there is no evidence that it destroyed the atmosphere, though it certainly extinguished a massive amount of life with about 75% of life species extinguished [32]. The atmosphere is incredibly robust and though we would not want to try a 100Tt energy dump, we use the atmosphere to our advantage in dealing with asteroid threats. The key is to spatially disperse the energy of the asteroid if there is no time to deflect it.

***Mass of the atmosphere*** – The mass of the atmosphere can easily be computed from the pressure at ground level ($P_{ground}$~100kPa) and the surface area of the Earth ($A_{Earth}$~5x$10^{14}$m$^2$) giving $m_{atm} \sim P_{ground} A_{Earth}/g$ ~5x$10^{18}$ kg. Compared to the mass of the Earth of $m_{earth}$~6x$10^{24}$ kg this gives $m_{atm}/m_{earth}$~8x$10^{-7}$. The atmospheric mass is non-trivial at about 1ppm of the mass of the Earth. For reference, note that the mass of the asteroid that likely killed the dinosaurs 65Myr ago was roughly $10^{15}$kg, or three orders of magnitude less mass than the atmosphere, but still extremely large.

***Heat Capacity of the Atmosphere*** – The specific heat of air at 300K is measured to be $C_p$ =1.00 kJ/kg-K, or $C_v$ =0.718 kJ/kg-K; very close to that of a theoretical diatomic ideal gas (5 DOF → $C_v$=5/2$R$ J/K-mole ~0.7kJ/kg-K) with $\gamma = C_p/C_v$=7/5. **This gives a total heat capacity of the Earth's atmosphere of about 5x$10^{21}$ J/K or a temperature rise of ~0.8mK/Gt**. To put this in perspective, the total Earth's nuclear arsenal of ~ 6Gt would raise the Earth's atmosphere on average by 5mK while the total KE of a 10km diameter asteroid (2.6 g/cc) impacting at 10km/s would have KE~ 4x$10^{22}$ J, while at 20km/s speed we have KE~ 1.6x$10^{23}$ J. If all of this energy were deposited in the atmosphere as heat, it would cause a temperature rise of about 8K@10km/s and about 32K@20km/s. IF the KT extinction event injected 100Tt of energy, then the temperature rise would be about 80K. This is significant and is part of the existential threat as was the Earth's surface impact of the KT extinction event [32]. Since our goal is to disassemble the threat into fragments that are typically less than 10m in diameter, the detonation air burst will occur high in the atmosphere, typically above 30km where the air density is 1% or less compared to sea level. The local heating temperature injection will have a long thermal diffusion time scale, and hence the effects on air at low altitudes where people live will be minimal. The exception to this high-altitude energy injection is the shock waves and any thermal radiation which will have long range energy transfer.

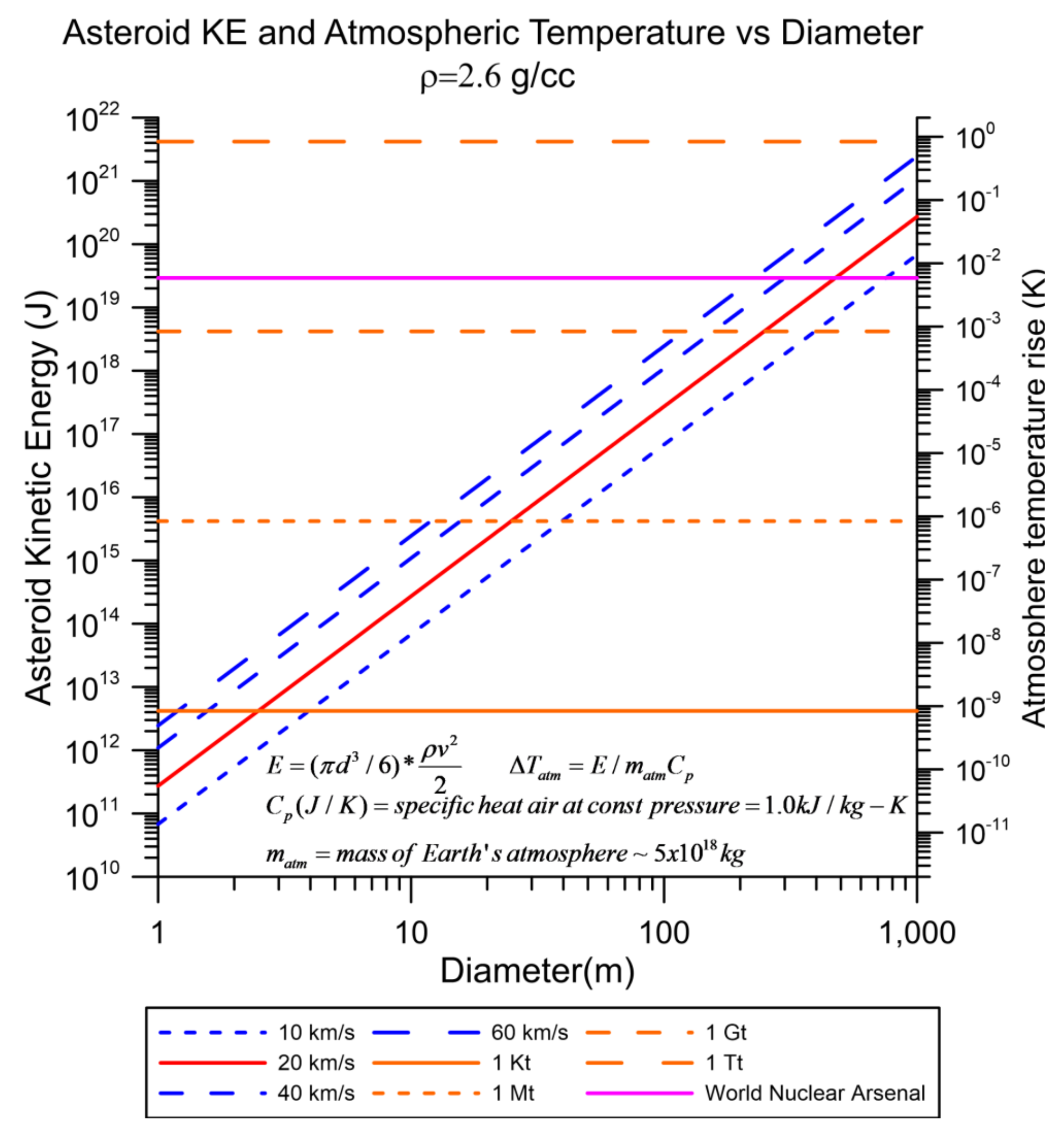

**Figure 26**– Asteroid exo-atmospheric KE vs diameter from 1 to 1000m for density 2.6g/cc. The right hand $y$-axis shows the temperature rise of the Earth's atmosphere assuming equilibrium. To first order, this plot is independent of fragmentation. Up to 1km diameter, the average temperature rise is small. The detailed issues of thermal diffusion and energy dissipation channels such as shock wave acoustics are discussed in the text.

***Local vs wide scale distributed atmospheric heating*** - In reality, any atmospheric impact will produce extreme local heating with significant shock wave generation that will eventually dissipate. Frictional heating due to asteroid surface heating and ablation, as well as local air heating due to adiabatic compression and shock heating are relatively local phenomenon with long diffusion time scales, except for the radiation transfer from asteroid surface heating. The acoustical shock waves generated, which does most of the damage from an air burst, rapidly spreads the shock energy out at supersonic speeds with slow dissipation. This is a complex heat transfer problem with very large intense local heating of the air with both slow diffusion times scales and fast time scales from the shock waves and the optical pulse which diffuse the energy rapidly. The fraction of energy that goes into local heating vs the large-scale distribution from shock waves depends on the specifics of the impact, but for the typical fragment sizes of 5-10m diameter, the fraction that goes into local vs shock wave is of order ½. The shock waves are by far the more destructive of the two dissipation modes.

If we look at a more modest possible large asteroid threat, Apophis for example, this has a mean effective diameter of 330-370m and a mass ~$6x10^{10}$ kg [33]. Assuming an impact speed of 10km/s, this would have a KE~$3x10^{18}$J , which would give a mean atmospheric temperature rise of about 0.6mK, which is negligible. A real "intact impact" of Apophis would deposit only a small fraction of its kinetic energy passing through the Earth atmosphere into the Earth's atmosphere and instead it would deposit most of its energy on the Earth's surface, which is precisely where we do not want it. It would be best if we could deposit the energy in the Earth's atmosphere instead. How do we prevent the impact energy from reaching the ground and instead, if impact is inevitable, direct the energy into the Earth's "body armor" of an atmosphere? The answer is to fragment prior to impact. This is precisely the point of this paper.

***Air Burst vs Ground Burst Heat Transfer*** – In an air burst with stony bolides with less than about 80m diameter, the energy transfer is largely slow thermal diffusion in the highly heated column of air along the path, as well some very modest light and thermal radiation. Most of the large-scale distribution of energy is via shock waves with virtually all the damage coming from the shock waves. In a ground burst (>100m diam.), there are two additional phenomena that are extremely damaging: the extreme thermal radiation from the "fireball" due to the ground hit, and the massive air and ground shock waves that are generated. The thermal radiation from the fireball and the ground shock wave causes a rapidly propagating and extremely dangerous though rapidly dissipating energy transfer. There is virtually no fireball from an air burst, though there is light generation. Given the choice, an air blast is preferable, and our method effectively prevents the ground burst.

***Granularity and Energetics*** – Much of our work in this paper can be thought of as analyzing the fragmentation granularity trade space between "doing nothing" at one extreme and "complete disassembly down to the molecular level" (vaporization) at the other extreme. Doing nothing for large threats that will impact is generally a bad idea but even this is a quantitation area. For example, a 50m threat whose "ground zero" is known to be in the middle of an ocean or sparsely populated region is a good example of "do nothing" (evacuate of course) while a 1km threat with "ground zero" in the same ocean or sparsely populated area is generally a very bad example of "do nothing".

There is an energetics trade space in achieving small scale fragmentation granularity. While smaller fragments are essentially always "better", it is generally sufficient to fragment to a size scale of less than 15m with a preference to be less than 10m for rocky bolides. There is also a dependence on internal cohesive strength of the threat that we explore in detail later. Up to the km diameter scale there is little concern that "atmospheric overloading" will be an issue, since even a 1km diameter threat at 20km/s and density 2.6 g/cc that is properly fragmented (10m scale) will only raise the average atmospheric temperature by less than 0.1K. **Our atmosphere has high enough heat capacity to absorb large but fragmented threats without** **significant damage as shown in the figure above.**

It is useful to compare the power and energy from impact threats to that of sunlight to gain some insight into the fundamental issues we face. For example, the average (essentially isotropic) “loading” of 100 tons/day or dm/dt=1.2 kg/s of small meteoritic debris has a power of $P_{debris} = 0.5v^2$ dm/dt assuming a constant closing impact speed of v. If we choose v=20km/s as being a reasonable impact speed, we get $P_{debris}$ ~ 240 MW. Comparing this to the solar insolation of 1350 W/m$^2$ and a projected Earth surface area of $A_{Earth\text{-}proj}= \pi R_E$ =1.3x10$^{14}$ m$^2$ gives a total Earth illuminated solar power of $P_{solar}$ =1.8x10$^{17}$ W (180 PW). This gives $P_{solar}/ P_{debris}$= 7.4x10$^8$ thus the average debris impact power is miniscule compared to the solar power illuminating our atmosphere. If we raised the debris rate from 100 tons/day such that $P_{debris}= P_{solar}$ we would need to have an average debris impact rate of 7.4x10$^{10}$ tons/day or 8.6x x10$^8$ kg/s. This would correspond to the impact equivalent (per day) of a bolide of diameter $d=(6m/\pi\rho)^{1/3}$ where m is the mass per day. For a density 2.6 g/cc asteroid this corresponds to an impact threat per day of diameter d= 3.8 km or the equivalent of an 86 m diameter asteroid per second! Clearly the atmosphere is very robust. Note that only about 30% of the solar illumination is actually absorbed in the atmosphere on a clear day but this is a small correction to the overall conclusion that our atmosphere is capable of absorbing the energy from an extremely large threat if spread both spatially and temporally.

In the trade space from “doing nothing” to “fine granulation” it then becomes a matter of understanding what is “good enough” for planetary defense. For example, an impact of Apophis on a large city would be devastating and even if had time to evacuate people from the city the cost of doing so and the disruption would be enormous. The “cost” in both resources and disruption to society would almost certainly be vastly less if we chose to mitigate. We will not do a costing analysis, except to note the bounds on launch vehicles for interception for various threat scenarios later in the paper.

# 4. Existential Threats – large asteroids and comets

***Large asteroids*** – Asteroids with diameters greater than 1km pose threats which approach existential levels. The KT extinction event is consistent with a 10km diameter asteroid with a kinetic energy of order 100Tt TNT, or more than 15,000 times the world's nuclear arsenal. As mentioned above, IF we are able to break such a future threat up of 1 km diameter into small fragments (~ 10m), then even in the worst case of "complete absorption" in the Earth's atmosphere of all the fragments, and hence all of the kinetic energy, we can still survive. We do not know of such a threat that is currently on the "horizon," and generally, with some exceptions such as sun grazing comets, we can track large threats and have plenty of warning time to either deflect them from the Earth or intercept and break them up with sufficient time that most fragments will miss the Earth. Extremely large asteroids that have diameters of hundreds of km fall into another category that need to be dealt with separately as we cannot afford to absorb the energy even in fragments unless the fragments are widely spread both spatially and temporally. In general, we would not want to "accept a complete hit of all the fragments" from anything much larger than a few km diameter and all known threats of such size would have years of warning and would allow us to fragment the threatening object, so only a small fraction would actually intercept the Earth (or deflect it completely).

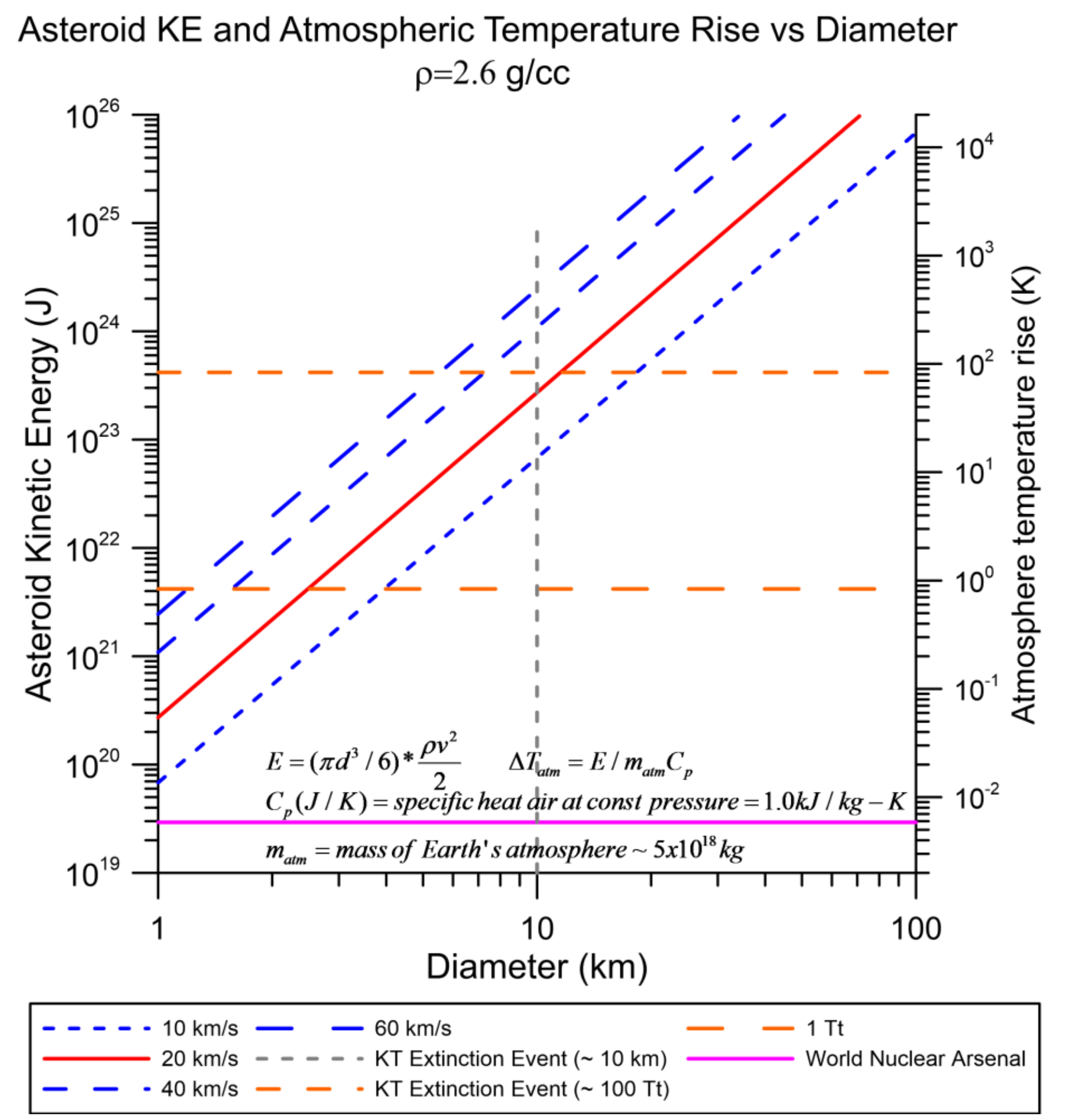

**Figure 27**- Asteroid exo-atmospheric KE vs diameter from 1 to 100 km (extremely large asteroids) and speed for density 2.6 g/cc. The right hand *y*-axis shows the temperature rise of the Earth's atmosphere assuming equilibrium. To first order, this plot is independent of fragmentation. Above a few km diameters, the effects on the Earth's atmosphere become serious, and critical above 10km. In these cases, either the interception for fragmentation must be done early enough to spread the debris cloud to be large enough so that most of the fragments miss the Earth, or complete diversion of the asteroid to miss the Earth must be implemented. Fortunately, very large asteroids are both extremely rare (though they have hit the Earth in our geologic history) and we would have sufficient notice to take action long before impact. The KT event we believe that caused the mass extinction 65 MYr ago is shown at 10km diameter based on the Iridium in the KT boundary clay layer.

***Comets*** – Comets pose unique threats as they are generally much faster (~ 60km/s) than asteroids and can be extremely large. Terminal defense (short time scale) vs long time scale deflection needs to be compared. Just as in the large asteroid case, we cannot in general afford to absorb a direct hit even when fragmented due to the extremely large KE input to the atmosphere for comets larger than a few km in diameter. Fragmentation at early enough times such that the majority of the fragments spread sufficiently that most will miss the Earth is one option, as is complete deflection. We have extensively explored directed energy ablation schemes for comets as another option [16], [18], [19], [34]–[36].

***High Density High Yield Strength Asteroids*** – The vast majority of asteroids are believed to be stony with densities approximately 3g/cc and low cohesion. There are a class of asteroids that may come from the breakup of protoplanets. One example of such an asteroid is 16 Pysche, which has a diameter of over 200km and a density of about 4g/cc. It is thought to be made primarily of "metals," but this does not mean it is a "solid" nickel-iron asteroid such as the famous Hoba meteorite [37]. 16 Psyche is not a current threat. It is not clear what the mesoscale yield strength is, though from the density we would estimate about 1MPa. This is a very modest yield strength (~ typ. Earth soil) and could be readily fractured. A much more challenging, though apparently very rare, asteroid would be like the 60,000kg Hoba meteorite (2.7 x 2.7 x 0.9m) with a density of approximately 9g/cc. This is a nearly homogeneous object with a vastly higher yield strength of about 50-100MPa. It is unknown if there are large asteroids with such densities and yield strengths like the Hoba case. Even if there were, we could still fragment them, though we would generally want to break them into small fragments, perhaps of order 3m in diameter. In the case of the ~ 3m Hoba meteorite, due to its high yield strength, it did not air burst and it is estimated that the atmospheric drag slowed the fragment down to about Mach 1 before it hit the ground and thus there was no significant fireball and little ground damage. There was likely only a modest atmospheric shock wave generated [38].

## 5. Nuclear vs Kinetic Penetrators vs Explosive Filled Penetrators

Nuclear weapons offer unique capabilities and unique challenges to planetary defense. In addition to political considerations, there are practical and technical issues. Nuclear deflection of asteroids has been discussed and simulated at length. In the US, this work has primarily been done at LLNL and LANL. The practical ways that nuclear weapon-based planetary defense is implemented is very different than "the movies" with a non-contact "standoff" approach where the device is detonated prior to impact with the target and the X ray and neutron yield is used to ablate the surface as the deflecting impulse. The reason for this approach is that no current nuclear weapons could remain functional with the extreme deceleration and casing destruction that occurs during a direct intercept impact penetration. No current materials are known that can withstand the impact at speeds greater than about 1km/s. This is simply a materials issue as well as a device integrity issue. Nuclear "bunker buster" Earth penetrators all operate at vastly lower impact speeds than those of a typical 10+km/s asteroid or comet. New approaches in this area would be an interesting possible direction. Slowing down the nuclear interceptor prior to impact so that it can survive the penetration phase is another, though this is complicated by the high relative speeds (>10km/s) and the relatively low Isp (typ <450s) of chemical rockets. In addition to non-contact nuclear deflection, asteroid landing and boring into the surface are conceivable, but still the "stuff of movies," though it is not impossible to consider. Unconventional approaches to nuclear detonation using the extreme impact speeds to "controllably" compress the primary/secondary is another area to ponder/simulate for the future, but the real problem here is the "controllable" compression from a high-speed impact with a highly heterogeneous target. The impact speed (>10km/s) are actually larger than the explosive detonation speed (<10km/s) for modern explosives such as RDX, so conceivably this could be used to advantage. The problem is that nuclear weapons only work with highly controlled explosive compression in both the primary (chemical driven implosion, ~ 10km/s) and the secondary (ablation/radiation driven implosion, ~ 100 km/s, from X rays from the primary). Neither of these compressions as currently implemented would work in conventional implosion nuclear devices, such as the B61-11 ground penetrator, that are designed for approximately 0.5km/s Earth impacts and penetration as they would be disabled and destroyed in an asteroid impact at >10km/s. While current NED's are not suitable, new designs may be viable. "Gun type as opposed to gun launch" fission only NED's may be a better approach for nuclear delivery, though small implosion devices were also developed. There has been considerable past work on gun launched nuclear munitions for artillery delivery with fission weapons that were designed for very high "gee" loading that could be good models for NED penetrators. Examples in the US were the W48 NED with a yield of approximately 72 tons TNT with a mass of approximately 54kg. It was tested to 12,000g successfully. A more modern device was the W82 with a design yield of 2kt, a mass of 43kg include the XM122 rocket booster (increased range), W82 NED and proximity fusing, and a bore size of approximately 155mm. The bare "physics package" W82 NED mass is

~30kg for 70 tons TNT/kg. Cancelled in 1991 at the end of the "Cold War". Another option, when combined with passive penetrators used first for "hole drilling", is the "ruggedized" W61 EPW (Earth Penetrating Warhead – cancelled 1992)) was designed for high "g" loading penetration into ~3m rock/ concrete. This unit was designed to take ~ $10^4$ g prior to detonation. It was replaced with the ruggedized B61-11/ B61-11 EPW. **Such NED's would be very attractive given their low mass and could be used for any threat with the disruption technique we propose, but the foreseeable future, NED intercepts will be reserved for very large threats as discuss later** [39, p. 82]**.**

The use of nuclear "standoff" devices is also conceivable for fragmentation, but only a modest fraction of their energy is effectively couple to the asteroid vs a true nuclear penetrator. For targets that miss the Earth, the radiological issues are essentially irrelevant. For targets that fragment and do hit the Earth, the radiological effects need to be considered. However, these are generally low due to the high-altitude breakup as well as the temporal radiological timeline. The large number of airburst nuclear tests show (as predicted) relatively low levels of residual radioactivity and critically. There are also no "prompt neutron" issues since the delay between intercept detonation and impact with the Earth's atmosphere is distributed both spatially and temporally. Some residual radioactive material due to both internal nuclear components (Pu, U tamper, etc.) as well as activated material in the target and bomb casing would be small and spread out over vast distances upon Earth impact.

Explosive filled penetrators are another area of exploration, but again there is the issue of the target intercept approach speed being significantly larger (>10km/s) than the best chemical explosive detonation speed (<10km/s). Detonation just prior to or at impact (shaped charge) (analogous, but different than the nuclear case) is one approach, as are creative penetrator designs. One advantage of chemical explosives compared to nuclear explosives is that chemical detonation is much more forgiving to asymmetries than are nuclear explosives. Explosive filled penetrators are an area that requires both extensive simulation and testing. Chemical penetrators can be tested on Earth while nuclear ones cannot. This is a problem with nuclear devices in that new iterations and generations of devices generally make only incremental changes to well tested older devices. This is a policy issue and not a fundamental issue. One path forward is to test hypersonic penetrators from chemical and gas gun driven sources up to 5-10km/s (depending on scale size) and to determine if a suitable design with acceptable deceleration parameters inside the unit is feasible. If suitable hypervelocity designs can be generated, then they could be tested without any payload on small asteroids that come close to the Earth by monitoring the deceleration and other parameters inside the penetrator vs time to determine if existing nuclear devices would remain viable inside. In any "real scenario" of planetary defense, using untested devices, whether passive, chemical, or nuclear would be foolish given the "stakes" involved. Fortunately, we have many opportunities to test the strategies on a wide variety of close approach asteroids. It is useful to remember that we take for granted the ability to have successful orbital re-entry as well as ICBM re-entry, yet in 1950 this was unclear. The development of ablative shields was able to solve a problem that at first appeared intractable due to the large amount of KE that had to be dissipated. Orbital re-entry is only about a factor 2x lower speed than asteroid impact speeds, BUT the thin air is vastly different than the dense asteroid material. It becomes a "dissipation timescale and power" issue as the asteroid KE dissipation (~4-10 higher KE/mass) must happen in a small fraction (0.001-0.1) of a second compared to the minutes of orbital re-entry. This means the dissipation power and deceleration levels are 4-6 orders of magnitude higher than that of a re-entry vehicle. In addition, the radiation channel for energy dissipation is largely blocked. The only "good news" is that testing is easier and can be done very rapidly to iterate designs. It is not "hopeless" to try to solve this problem.
A simple calculation of the deceleration is useful.

$$E = mv^2/2 = m*a*L$$

$$E/m = v^2 = aL \rightarrow a = v^2/L$$

$$Ex: v = 20km/s,\ L = 100m \rightarrow a = 4x10^6 m/s^2 = 400,000\,gee$$

This is a high "g" loading, but not without precedent. Electronics in artillery shells are routinely used at >15,000 gee and the acceleration in a common ultra-centrifuge is about 400,000 gee. The W48 tactical nuclear artillery shell was tested to >12,000 gee and was successfully fired. The follow on W82 (2kt) artillery shell was similar, but higher yield. The deceleration is not the primary issue, but rather it is energy management to allow a penetrator (or part of a penetrator) to survive and deliver its payload IF a nuclear option is chosen. A possible solution to the high gee loading is to use weakly coupled penetrators. We show that the nuclear option is generally not required except for extremely large threats.

***Testing Options*** – There are a wide variety of testing options for the kinetic and explosive filled kinetic penetrators. While getting to 20km/s at scale on the Earth is not generally feasible, it is possible to get to several km/s with explosive-driven (artillery) systems as well as railgun and gas gun options. Full up tests with "Earth-built" synthetic asteroids and high-speed (1-few km/s) penetrators is a very feasible option and would yield a wealth of information and feedback and would provide an opportunity for validation of simulations. Using the Moon as a "test target" is another option once Earth-based testing and development is well developed. Going after "real asteroids" to test the efficacy of the system is also a feasible and necessary element of a realistic long-term program. All of these (Earth, possible lunar, and actual asteroid testing) are not only critical, but feasible to pursue.

***Public Support*** – It is extremely likely that there would be enthusiastic worldwide public support for such mitigation programs. This is critically important to actualization of a functional Earth defense system. It could be seen as a truly large scale "environmental" program of help to all of the Earth and thus it would transcend national boundaries and national interests. This would also be a compelling opportunity for the private sector to participate in a "large scale human program" to benefit all of humanity. Additional input from the public in the form of crowd-funded efforts to both aid in the education of extraterrestrial threats and to support the R&D needed would be a likely response. Coupling to existing public programs such as Earth Day would also be logical. Such a program would not only be seen as altruistic, but could also help to bring countries together in a common effort against a common threat.

## 6. Detection Improvements

Our current detection of both asteroids and comets comes from passive visible light imaging from large scale astronomical surveys. Future space-based mid wave (3-5 microns) and long wave (8-15 microns - thermal IR) will improve our abilities significantly. Long range radar is complimentary, but the small target size and long-range capability that is needed will generally make radar a "follow up" capability. Another approach is to use long range laser active imaging, or LIDAR. We are beginning to possess the technological capability to develop large scale laser phased arrays in the near-IR that may offer a solution to the problem of finding and tracing small (<100m) objects that pose a threat. Using an orbital or lunar based system may finally allow us to tackle this critical part of the threat determination process, and the same system can also be used as a laser target designator for interception of the type we propose. As an example, in [35] we propose using a laser phased array at 1.06 microns to enable such a method.

## 7. Planetary Offense vs Planetary Defense - Proactive Mitigation

Sometimes the best defense is a good offense. This enters into an area of discussion related to passive vs proactive control of our environment. It is almost certainly a controversial area, but one that should be discussed. By proactive mitigation we refer to preemptively destroying the future threat before it becomes a current threat. Many asteroid threats, such as Apophis, approach the Earth many times before an impact. One option is to fragment such a threat on one of its close encounters. Destroying asteroids and even comets that could be future threats is a conceivable long-term strategy that should be pondered, even if it is an uncomfortable discussion. The analogy with biological threats is similar, but generally not controversial (depending on the level of proactive response). Examples of such preemptive biological threat responses include vaccination against or inoculation for or destruction of infected pathogens. Taking "control" of our extraterrestrial threats by preemptively fragmenting asteroids and comets is a thought to be pondered. Uncomfortable as it may be to have this discussion, a direct hit of a large asteroid or comet would be significantly more uncomfortable, assuming anyone is left to discuss it afterwards.

## 8. Asteroid – Earth Impact Physics

There have been numerous studies of the interaction physics of the high-speed Earth impacts of asteroids and comets. To enable rapid computation of a large number of scenarios, we adopt the formalism of [40], [41]. The goal of our mitigation strategy is to prevent ground impact by fragmenting the parent asteroid, and thus we focus on the physics of air bursts. The maximum size of the fragment depends on the details of its speed, angle of attack, density, and yield strength. For simplicity, we will model the fragments as being homogeneous with uniform density and uniform yield strength. Real asteroids are likely far from homogeneous. The maximum size to prevent a ground impact is roughly 100m for stony asteroids. However, this is much too large as the goal is to prevent large scale destruction. In practical terms this requires us to break up the parent asteroid into fragments whose maximum size is typically no larger than 10-15m in diameter for stony asteroids.

We use the six input parameters from [40], [41], but we do not allow ground impact and hence the last parameter (Earth impact composition) is not used in our simulations as we only allow air bursting of the disassembled fragments. While the general emphasis here is on asteroid fragmentation and their subsequent air bursts, the same phenomenology can be used with comet fragmentation and their subsequent air bursts. The term "ground zero" is used to refer to the point on the Earth's surface directly beneath the fragment air burst. We assume an isothermal atmosphere with scale height $H$ (~ 7-8km). The basic destructive process during atmospheric entry of the fragments occurs when the high speed "ram pressure" (or stagnation pressure) eventually exceeds the yield strength of the asteroid. We refer to this point as "breakup". After breakup, the asteroid undergoes a plastic deformation and begins to "pancake" or flatten and expand in size. With increased size, the drag due to the air will cause rapidly increasing flattening and expansion that will ultimately result in a runaway process leading to detonation or "bursting" of the pancake fragment. This typically happens when the flattened diameter exceeds about 7 times the original fragment diameter. Since the asteroid is hypersonic upon entry, a shock wave is created along the entire entry track until the burst phase at which point we assume all the remaining energy is dissipated into additional shock waves.

The fundamental channels of energy dissipation are:

1) Heating of the air due to friction with the asteroid
2) Heating of the asteroid due to air friction
3) Ablation of the asteroid and resulting fragments during entry (material-dependent – discussed later)
4) Production of acoustical shock waves
5) Radiation from the heated outer surface as well as from fragments ablated/removed
6) Production of light during the hypersonic interaction with the air – plasma – recombination emission

| **Parameter** | **Chelyabinsk Russia Feb 15, 2013** | **Tunguska Russia June 30, 1908** | **Arroyomolinos de Leon – Spain Dec 8, 1932** | **Indian Ocean (Prince Edward Isl) Aug 3, 1963** | **Bering Sea Kamchatka, Russia Dec 18, 2018** | **Qinghai, China Dec 22, 2020** |
|---|---|---|---|---|---|---|
| Exo-atmospheric Energy (MT) | 0.55 | ~10 | 0.19 | 0.26 | 0.173 | 0.01 |
| Diameter (m) | 19.5 | 50-100 | ~18 | | ~10 | 8 |
| Density (g/cc) | 3.3 | 3 | | | | |
| Speed (km/s) | 19.2 | 20 | | | | 13.6 |
| Angle rel horizon (deg) | 20 | ~45 | | | | |
| Yield Strength (MPa) | 2 | 1 | | | | |

**Table 1 –** Recent larger asteroid strikes. Since 2005 when the JPL Fireball and Bolide report became active, > 500 airbursts have been recorded with an average of about 35 significant events (>0.1 Kt) per year. Data prior to the use of atmospheric infrasound detectors (1960's) is generally sporadic.

Of all the energy dissipation modes, the two most observable are the light production (optical signature) from interaction with the atmosphere and the shock waves (acoustic signature). The light from meteorite impacts, or "shooting stars," are a common example of this for very small meteorites (typ. mm to cm sized) with the acoustical signature not noticeable except by specialized atmospheric infrasonic acoustical sensors that form a part of the network used in tracking of impacts (Fig 2). With a rough power law distribution in size (more at smaller sizes), the transition between meteorite and asteroid is arbitrary. When the size of the asteroid exceeds a few meters in size, the acoustic signature is noticeable with unaided human hearing near ground zero, and for sizes around 10m (for stony asteroids) the optical and acoustical effects are very apparent, though the damage is minimal in general. Once the size approaches 20m, the acoustical damage (window damage) becomes significant, as was seen in the Chelyabinsk event.

Clearly this is a complex process whose details are very dependent on the actual and complex structure of the asteroid fragment. To allow rapid approximate computation, we simplify this using the phenomenology in [40], [41], as well as the measured nuclear and conventional weapons air burst data [2]. We note that the precision of the details is much less important to the success of our program than are the qualitative conclusions. Any real asteroid interdiction scenario will have many unknowns as to the internal structure of the asteroid, and thus precision metaphysics simulations are only as good as the knowledge of the target asteroid, which is generally poor as to heterogeneity inside. However, over the very broad range of assumptions we have explored, the conclusion remains that the fragmentation of asteroids and comets, even relatively large ones on relatively short time intercept times scales (days for example), can result in vast reduction of the threat and should be seriously considered. The data from the Chelyabinsk and Tunguska events are the most comprehensive, allowing estimates of density, speed, angle of attack, and bolide material yield strength. Our results are consistent with the measured damage from the Chelyabinsk event.

***Limitations of the model*** – One of the primary limitations of the model is the ignoring of the ablation terms. Depending on the size of the bolide, this can be very important as it is for small bolides, or negligible as it is for large bolides. Ablation is also very composition- and structure-dependent. Complex issues such as fracturing during atmospheric entry are also ignored as they are highly dependent on the internal structure of the bolide. A precision simulation would require *a priori* detailed knowledge of the bolide, which we do not have. In addition, most critically, we are not concerned with the precise details of any individual entry as we want to design a defense system which is agnostic to the details of each bolide and concentrate on a general-purpose system that is conservative in its design. For example, ablation will tend to reduce the mass and thus the acoustical signature (lower peak pressure in the shock wave). Thus, ignoring ablation is a more conservative approach. Ablation does affect the optical signature by reducing the outer skin temperature. We come back to this later when we discuss the optical signature.

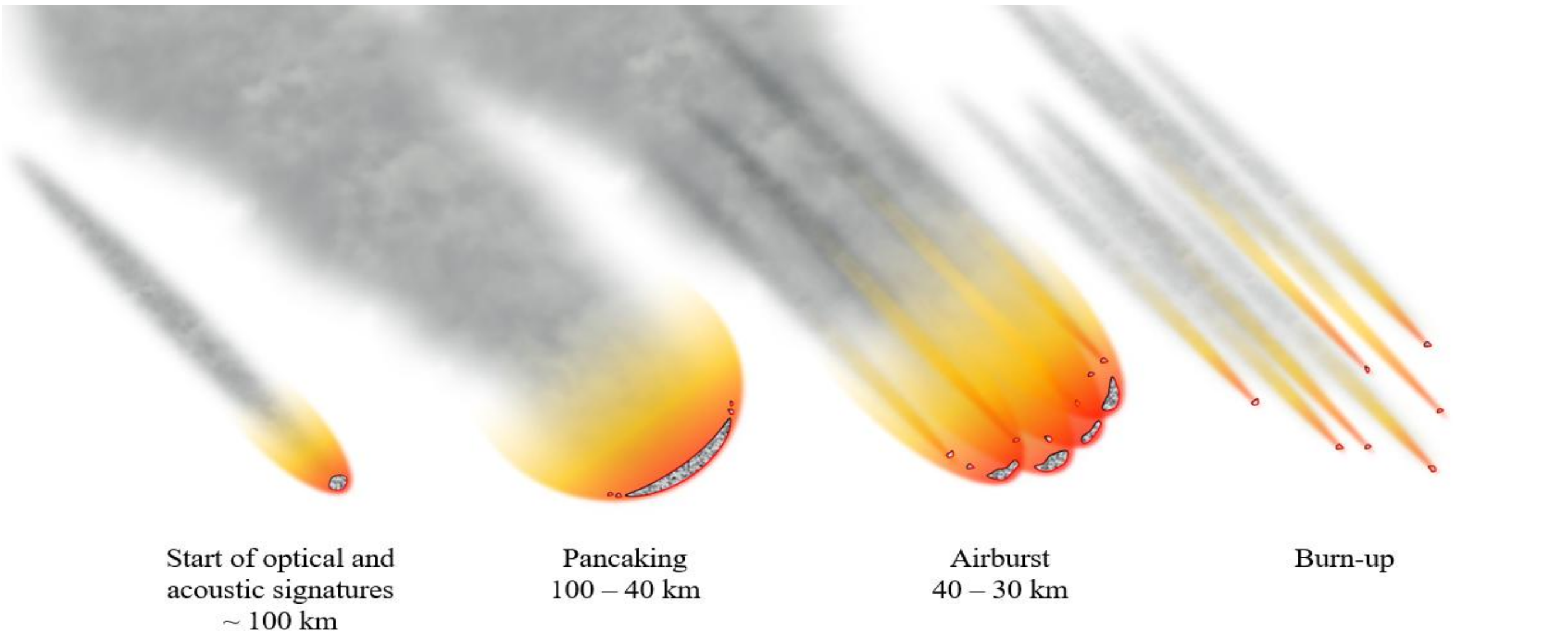

**Figure 28 –** Pancake model where atmospheric ram pressure causes plastic deformation of bolide until airburst detonation.

*Equations and Parametrization of Analytic Airburst Model*

**The input parameters to the model we use are as follows:**

1) $L_0$ =diameter of fragment
2) $\rho_i$ = fragment density
3) $v_0$ = exo-atmospheric parent asteroid speed
4) $\theta$ = asteroid velocity vector angle relative to horizon
5) $r$ = slant distance away from air burst
6) $\rho_t$ = Earth impact area target density - sedimentary, crystalline, marine – IF ground impact (we fragment so that all fragments will air burst and no intact fragment can hit the ground so this parameter is not relevant)

**Secondary inputs and assumptions are:**

1) $H$ =atmospheric scale height (typ = 8 km)
   a. $\rho(z) = \rho_0 e^{-z/H}$ where $\rho(z)$ is the atmospheric air density at altitude z
2) $C_D$ = asteroid/ fragment drag coef (typ = 2)
3) Assume a relationship between mean asteroid density and cohesion (yield strength= $Y_i$ )
   a. $log_{10} Y_i = 2.107 + 0.0624\sqrt{\rho_i}$ where $Y_i$ = asteroid yield strength (Pa)
4) Assume asteroid begins to breakup when stagnation pressure = asteroid yield strength
   a. $Y_i = \rho(z_*)v^2(z_*)$ where $z_*$ =breakup altitude

**From these inputs we calculate the following outputs:**

1) Asteroid exo-atmospheric KE, density and mass
   a. $E = \frac{1}{2} m_i v_0^2 = \frac{\pi}{12}\rho_i L_0^3 v_0^2$   $\rho_i$ = asteroid density
   b. $m_i = \frac{\pi}{6}\rho_i L_0^3$ =asteroid mass
2) Number of near-Earth asteroids with a diameter greater than $L_0$
   a. $N(>L) \approx 1148 L_{\text{km}}^{-2.354}$
3) Recurrence interval in years versus the impact energy
   a. $T_{\text{RE}} \approx 109 E_{\text{Mt}}^{0.78}$
4) Speed of the asteroid as a function of altitude z prior to breakup
   a. $v(z) = v_0 \exp[-\frac{3\rho(z)C_D H}{4\rho_i L_0 sin\theta}]$
5) Breakup altitude analytic approximation
   a. $z_* \approx -H[ln(\frac{Y_i}{\rho_0 v_i^2}) + 1.308 - 0.314 I_f - 1.303\sqrt{1-I_f}\,]$

b. Where $I_f = 4.07 \frac{C_D H Y_i}{\rho_i L_0 v_i^2 sin\theta}$

6) Speed at breakup

a. $v(z_*) = v_0 \exp[-\frac{3\rho(z_*) C_D H}{4\rho_i L_0 sin\theta}]$ where

$$z_* \approx -H[ln(\frac{Y_i}{\rho_0 v_i^2}) + 1.308 - 0.314 I_f - 1.303\sqrt{1 - I_f}]$$

7) After breakup but prior to final burst the breakup fragment the fragment size (pancake diameter) is

a. $L(z) = L_0 \sqrt{1 + (\frac{2H}{l})^2 (\exp[\frac{z_* - z}{2H}] - 1)^2}$

b. $l = L_0 sin\theta \sqrt{\frac{\rho_i}{C_D \rho(z_*)}}$ = “dispersion length”

c. $v(z) = v(z_*) \exp(-\frac{3}{4} \frac{C_D \rho(z_*)}{\rho_i L_0^3 sin\theta} \int_z^{z_*} \exp((z_* - z)/H) L^2(z) dz)$ = speed from breakup to burst

8) The altitude at burst is:

a. $z_b = z_* - 2Hln[1 + \frac{1}{2H}\sqrt{f_p^2 - 1}]$

b. $f_p$ =”pancake factor” = ratio of size of final “pancake” (at burst)/$L_0$

c. We will tyically use $f_p$ =7 at burst

9) The speed at burst is calculated using (in 7c)

a. $\int_{z_{\text{burst}}}^{z_*} \exp((z_* - z)/H) L^2(z) dz = \frac{l L_0^2}{24} \alpha [8(3 + \alpha^2) + 3\alpha \frac{l}{H}(2 + \alpha^2)]$

b. $\alpha \equiv \sqrt{f_p^2 - 1}$

c. $\int_0^{z_*} \exp((z_* - z)/H) L^2(z) dz = \frac{H^3 L_0^2}{3 l^2}(3[4 + (\frac{l}{H})^2] \exp(z_*/H) + 6\exp(2z_*/H) - 16\exp(3z_*/2H) - 3(\frac{l}{H})^2 - 2)$

d. $v(z) = v(z_*) \exp(-\frac{3}{4} \frac{C_D \rho(z_*)}{\rho_i L_0^3 sin\theta} \int_z^{z_*} \exp((z_* - z)/H) L^2(z) dz)$

e.
$$v(z_b) = v(z_*) \exp(-\frac{3}{4} \frac{C_D \rho(z_*)}{\rho_i L_0^3 sin\theta} \int_{z_b}^{z_*} \exp((z_* - z)/H) L^2(z) dz)$$
$$= v(z_*) \exp(-\frac{3}{4} \frac{C_D \rho(z_*)}{\rho_i L_0 sin\theta} \frac{l}{24} \alpha [8(3 + \alpha^2) + 3\alpha \frac{l}{H}(2 + \alpha^2)])$$

f. $l = L_0 sin\theta \sqrt{\frac{\rho_i}{C_D \rho(z_*)}} = "dispersion\, length"$

10) The energy that appears in the shock wave is computed from the energy lost between the initial KE at exo-atmospheric entry and the KE at burst:

a. $E_{blast} = \frac{m_i v_0^2}{2} - \frac{m_i v(z_b)^2}{2} = E*[1-(v(z_b)/v_0)^2]$

$E = \frac{m_i v_0^2}{2} = \text{exo-atmopsheric KE}$

11) The peak overpressure p (r) at slant distance r due to the energy in the shock wave is claculated using the measured nuclear weapons atmospheric tests as follows:

$p(r) = \text{pressure (Pa) (peak overpressure) at distance r}$

$p_n = \text{pressure for 1kt yield - near field= } 3.11\text{x}10^{11}(Pa)$

$\alpha_n = \text{power law index for - near field= -2.95}$

$p_f = \text{pressure for 1kt yield - far field= } 1.8\text{x}10^7(Pa)$

$\alpha_f = \text{power law index for - far field= -1.13}$

$r_1 = \text{distance from 1kt nuclear blast standard yield}$

$r = \text{distance from yield of arbitrary energy E given in kilotons TNT} = E(kt) = E_{kt}$

$r_1(r)$ = scaled distance of 1 kt standard for detonation of yield $E(kt): r_1(r) = r / E_{kt}^{1/3}$

$p(r_1) = p_n r_1^{\alpha_n} + p_f r_1^{\alpha_f}$ (assume power laws)

For arbitrary yield the pressure at distance r is:

$p(r) = p(r_1(r)) = p_n \left[ rE_{kt}^{-1/3} \right]^{\alpha_n} + p_f \left[ rE_{kt}^{-1/3} \right]^{\alpha_f}$

$\varepsilon_{blast}$ =fraction of 1 kt standard weapon that goes into blast wave (typ 0.5)

Pressure p(r) for an asteroid that has energy (in kt) of $E_{ast-kt}$ in the blast wave is calculated from the equivalent energy of a nuclear weapon with $E_{nuc} = E_{ast-kt} / \varepsilon$

$\rightarrow p(r) = p_n \left[ r(E_{ast-kt}/\varepsilon)^{-1/3} \right]^{\alpha_n} + p_f \left[ r(E_{ast-kt}/\varepsilon)^{-1/3} \right]^{\alpha_f}$

The reason for this is that only a fraction $\varepsilon$ of the nuclear weapon energy goes into the blast wave while we are assuming all of the asteroid energy $E_{ast-kt}$ calculated to be in the blast wave is actually fully in the blast wave.

12) The peak wind speed u for the peak pressure p is calculated as:

$u = wind\ speed,\ p_0 = sea\ level\ pressure\ (10^5 Pa), c_0 = sound\ speed$

$$u = \frac{pc_0}{\gamma p_0 (1 + \frac{\gamma+1}{2\gamma} p/p_0)^{0.5}} = \frac{5pc_0}{7p_0(1+6p/7p_0)^{0.5}}$$

$\gamma = c_p / c_v = 7/5\ diatomic\ 5DOF$

13) The sound pressure level (SPL) for peak pressure p(Pa) is:

a. SPL(db)=20log(p(r)/0.00002)
b. 20 μPa corresponds to 0 db SPL by definition of the nominal human acoustic threshold

14) The shock wave acoustic flux I(w/m$^2$) is: $I(r)(w/m^2) = 10^{-12} * [p(r)(Pa)/2x10^{-5}]^2$ I(w/m$^2$)

15) The shock speed is: $U(r) = c_0(1 + \frac{\gamma+1}{2\gamma}\frac{p(r)}{p_0}) = c_0(1+\frac{6p(r)}{7p_0})^{0.5}$ This starts out at the fragment speed and and rapdily decays asymtotically to the speed of sound (Mach 1).

16) The shock wave arrival time at distance r is: $T_b = \int_0^r \frac{dr}{u(r)}$ ~r/c₀

17) The shock wave time evolution for pressure p(t) is assumed to be a Friedlander waveform.

$$p(t) = p_0 e^{-t/t_1}(1-t/t_1)$$
$$Acoustic\,Flux\ (w/m^2):\ I(t) = I_0\left[e^{-t/t_1}(1-t/t_1)\right]^2$$
$$I_0 = (p_0/p_{ref})^2 I_{ref}$$
$$p_{ref} = 20\mu Pa,\ I_{ref} = 1pW/m^2\ (sea\,level)$$
$p_0(Pa) \equiv p(t=0) =$ peak overpressure at t=0

$I_0 =$ peak flux at t=0

$t_1 =$ zero crossing time when $\mathrm{p}(t_1), I(t_1) = 0\ (ambient)$

*Integrated blast wave energy per area*

$$\int_0^\infty I(t)dt = I_0\int_0^\infty \left[e^{-t/t_1}(1-t/t_1)\right]^2 dt = I_0 e^{-2t/t_1}\left[\frac{-t^2}{2t_1} - \frac{t_1}{4} + \frac{t}{2}\right]_0^\infty = I_0 t_1/4$$

18) For N fragments the shock wave time evolution of pressure and flux at any observer position is as follows where t=clock time that is the same (synchronized) at all location on Earth and in space. In this case t=0 is arbitrary:

$N = number\ of\ fragments$

$i = fragment\ number$

$\overline{x} = observer\ position$

$\overline{x}_{ast,i} = i_{th}\ fragment\ position\ at\ burst$

$v_s = speed\ of\ sound(\sim 340m/s)$

$p_{0_i}(\overline{x}) = peak\ pressure\ from\ fragment\ i\ at\ position\ \overline{x}$

$I_{0_i}(\overline{x}) = peak\ flux\ from\ fragment\ i\ at\ position\ \overline{x}$

$t_i(\overline{x}) = arrival\ time\ of\ blast\ from\ fragment\ i\ at\ position\ \overline{x}$

$t_{i-b}(\overline{x}_{ast,i}) = burst\ time\ of\ fragment\ i\ at\ position\ \overline{x}_{ast,i}$

$t_i(\overline{x}) = t_{i-b}(\overline{x}_{ast,i}) + \left|\overline{x}_{obs} - \overline{x}_{ast,i}\right|/v_s$

$t_{1i}(\overline{x}) = "t_1"\ from\ Friedlander\ blast\ evolution\ eq\ for\ fragment\ i$

$at\ position\ \overline{x} =$ zero crossing time for $\mathrm{p}(t,\overline{x}), I(t,\overline{x}) = 0$

$(t_{1i}(\overline{x})\ is\ slightly\ dependent\ on\ i\ and\ \overline{x}\ due\ to\ blast\ energy\ and\ atm\ freq\ dependent\ absorption)$

$$p(t,\overline{x}) = \sum_{i=1}^{N} p_{0_i}(\overline{x}) e^{-(t-t_i(\overline{x}))/t_{1i}(\overline{x})}\left[(1-(t-t_i(\overline{x}))/t_{1i}(\overline{x}))\right] \quad for\ t \ge t_i(\overline{x})$$
$$I(t,\overline{x}) = \sum_{i=1}^{N} I_{0_i}(\overline{x})\left[e^{-(t-t_i(\overline{x}))/t_{1i}(\overline{x})}\left[(1-(t-t_i(\overline{x}))/t_{1i}(\overline{x}))\right]\right]^2 \quad for\ t \ge t_i(\overline{x})$$
$$I_{0_i}(\overline{x}) = (p_{0_i}(\overline{x})/p_{ref})^2 I_{ref}$$
$$p_{ref} = 20\mu Pa,\ I_{ref} = 1pW/m^2\ (sea\,level)$$

19) Optical Pulse Modeling – We adopt a simple gaussian distribution vs time model for the power vs time coming from each fragment. The actual power vs time and distance along the bolide atmospheric path is better approximated by a spatially and temporally varying line emitter but the gausssian time varying point (at burst position) is a reasonable approximation with the gaussian time dispersion (sigma) of order a second depending on the fragment size. The time dispersion depends on the complex interaraction of each fragement with the atmosphere and will vary with fragment size and composition, speed, angle of attack among other systmetics. Since the dominant portion of the optical pulse occurs over a relatively short time scale with the power being strongly peaked in time, the optical damage (fire, eye and skin) metric is the deposited optical energy rather than the power. The time scale for cooling in the case of ignition is generally much longer the optical pulse dispersion time. As discussed later we will make an extremely conservative assumption about the effect of multiple fragment optical pulse ignition and **assume that the optical energy of ALL fragments as seen by any observer is simply the sum of the optical energy of each pulse**. This is uneccesarily conservative in general since the time between fragment burts is generally much larger than the optical dispersion time of any given fragment and the time between fragments can be long compared to the thermal cooling time scale that is relevant for ignition. Nonetheless, we start with this conservative assumption as it allows us to rapidly compute the optical pulse damage threshold for ignition (typ 0.2 $MJ/m^2$ for dry grass and loose paper). This is discussed in much greater detail later. We also assume that the optical energy from a fragment is related to the acoustic energy as both result from interaction of the high speed fragment with the atmosphere. We also discuss an analytic reltionship that is often used between the exo-atm energy and optical energy. The conversion of the kinetic energy into optical energy is highly dependent on the detailed fragment structure and internal cohesive (binding) energy/ strength and is poorly understood in general. We model this in multiple ways and the resort to measured optical data, primarily from DoD satellite observations of a small number of relevant bolide sizes of interest to us (typ 1-15m diam). While some use a fraction of the total exo-atmospheric energy to describe the optical energy, we also use a fraction of the total shock wave energy. We have modeled this in both ways. Based on the limited data available and the highly variable and complex nature of the optical pulse formation, we find a reasonable fit to (admittedly limited) data that is consistent with about 10% of the shock wave energy appearing as optical energy for some mesurements and about 10% of the exo-atm energy in other cases. The errrors on this assumption are large but the overall effect on our conclusions is minimal with the acoustic pulse causing the majority of the damage. This is discusssed in greater detail later in this paper.

For the propagation of the optical pulse through the atmosphere we use a full radiation transfer model to compute the optical power flux from each fragment at each observer. Since the optical propagation is occuring at very close to the speed of light in vacuum (300 km/ms) and since the relevant distance scale from the fragment to the observer are of order of tens to hundreds of km the light propagation time scale is very short. With the speed of light being approxomately $10^6$ time that of the acoustic propagation speed, the optical pulse can be well approximated as happening nearly simultaneously at all observer points with the optical pulse arriving very shortly after the fragment burst while the acoustic pulse (shock wave) arriving typically at least 100 sec after the arrival of the optical pulse from any given fragment.

$N = number\ of\ fragments$
$i = fragment\ number$
$\bar{x} = observer\ position$
$\bar{x}_{ast,i} = i_{th}\ fragment\ position\ at\ burst$
$t_{i-b}(\bar{x}_{ast,i}) = burst\ time\ of\ fragment\ i\ at\ burst\ position\ \bar{x}_{ast,i}$
$\sigma_i = optical\ power\ dispersion\ time\ of\ fragment\ i$
$p_{opt-0_i} = peak\ optical\ power\ from\ fragment\ i$
$p_{opt,_i}(t) = optical\ power\ vs\ time\ from\ fragment\ i$
$p_{opt,_i}(t) = p_{opt-0_i} e^{-(t-t_{i-b}(\bar{x}_{ast,i}))^2/2\sigma_i^2}\ (assumed\ gaussian\ time\ distribution)$
$F_{opt-power,i}(\bar{x},t) = optical\ power\ flux\ (w/m^2)\ at\ obs\ position\ \bar{x}\ at\ time\ t\ from\ fragment\ i$
$E_{opt,i} = total\ optical\ energy\ from\ fragment\ i$
$E_{opt,i} = \int_{-\infty}^{\infty} p_{opt,_i}(t)dt = p_{opt-0_i}\int_{-\infty}^{\infty} e^{-(t-t_{i-b}(\bar{x}_{ast,i}))^2/2\sigma_i^2} dt = p_{opt-0_i}\sqrt{2\pi}\sigma_i$
$E_{blast,i} = total\ energy\ in\ blast\ wave\ from\ fragment\ i$
$E_{exo-atm,i} = total\ exo-atm\ energy\ of\ fragment\ i$
$\alpha_{conv,i} = fraction\ of\ exo-atm/\ blast\ wave\ energy\ (\text{model dep})\ to\ optical\ energy\ for\ fragment\ i\ (0-1)$
$E_{opt,i} = \alpha_{conv,i} E_{blast,i} = p_{opt-0_i}\sqrt{2\pi}\sigma_i \rightarrow p_{opt-0_i} = E_{opt,i}/\sqrt{2\pi}\sigma_i = \alpha_{conv,i} E_{blast,i}/\sqrt{2\pi}\sigma_i$
$F_{opt-power,i}(\bar{x},t) = \alpha_{opt-trans}(\bar{x}_{ast,i},\bar{x}_{obs}) p_{opt-0_i} e^{-(t-t_{i-b}(\bar{x}_{ast,i}))^2/2\sigma_i^2} / (4\pi|\bar{x}_{obs}-\bar{x}_{ast,i}|^2)\ (assumed\ isotropic)$
$\alpha_{opt-trans}(\bar{x}_{ast,i},\bar{x}_{obs}) = optical\ transmission\ from\ burst\ for\ fragment\ i\ to\ observer\ (0-1)$
A resonable approximation (below) from fitting convolved BB source spectra is:
$\alpha_{opt-trans}(\bar{x}_{ast,i},\bar{x}_{obs}) = \alpha(r) = be^{-ar}$ with $r = |\bar{x}_{obs} - \bar{x}_{ast,i}|$
where a,b are determined from an atm model fit with a BB spectra (a,b fits shown later in paper)
A resonable approximation from fitting convolved BB source spectra is:
$\alpha(r) = be^{-ar}$ where a,b are determined from an atm model fit with a BB spectra

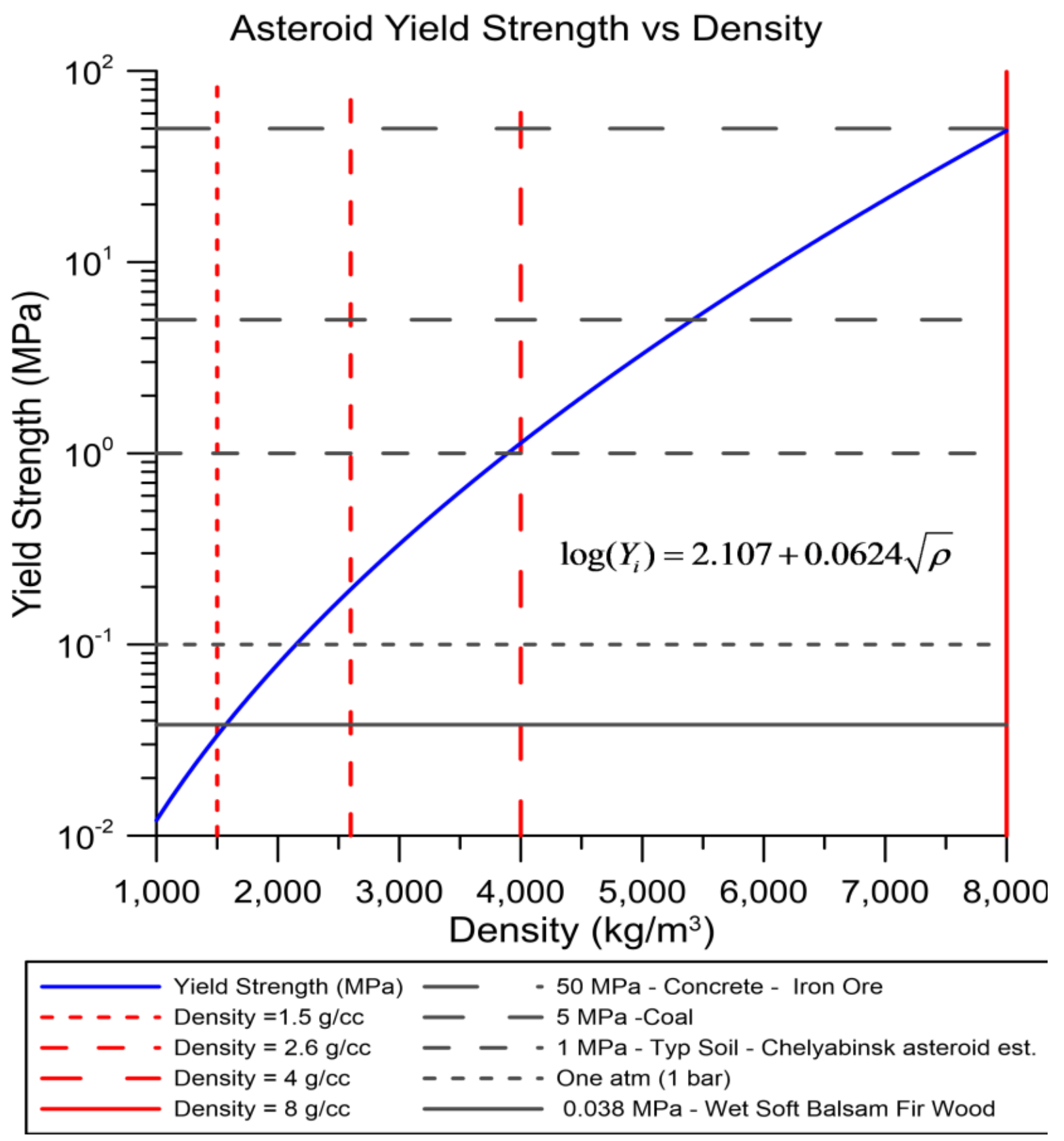

**Figure 29**– Functional form of asteroid yield strength vs density. This is only an approximate relationship between density and cohesion (strength).

We have run a large number (thousands) of simulations to test various fragmentation scenarios and summarize some of the saliant conclusions below. We have set two basic threat thresholds. The first is at a peak over pressure of 1kPa, which is the lower threshold where ordinary residential windows crack, and the second is at 10kPa, where wood frame and unreinforced brick residential building are threatened with significant damage or collapse. There have been a large number of studies of damage thresholds for shock waves, primarily in studies from the atmospheric nuclear tests of the 1940-1960's, as well as numerous conventional munition studies of shock wave hazards. As building damage thresholds are highly dependent on the details of both the construction and reflection due to local geometries relative to the blast site, we use conservative data from damage thresholds. In nuclear weapons tests, the typical threshold for standard residential (large) window breakage was roughly 4kPa, but there is a range depending on the prior history of the window, as well as the details of the size, mounting, acoustical resonant states excited, and local acoustical reflection issues. **In any real planetary defense scenario, there would be some public notification prior to intercept that would allow the public to avoid being near windows, or even simple measures such as using tape on the windows to miitigate potential damage. In general, our approach is to minimize any significant damage by fragmenting the parent asteroids sufficienctly such that the subsequent acoustical signature is small enough to avoid large scale damage of any kind.**

***Shock pressure time evolution*** – To model the time evolution of the shock wave, we use a Friedlander functional form. This describes the time evolution with two free parameters which are the peak pressure and a zero crossing time scale $t_1$. This also allows us to compute the time evolution of the acoustical pressure and flux. The time t=0 is when the shock wave first arrives at the observer NOT when the fragment bursts. See Equation set 18) above for a generalized time t which can be a common "clock time" not tied to the arrival time of the shock wave.

$t > t_1$ *below*

$(t) = p_0 e^{-t/t_1}(1 - t/t_1)$

$I(t) = I_0 \left[ e^{-t/t_1}(1 - t/t_1) \right]^2$

$I_0 = (p_0 / p_{ref})^2 I_{ref}$

$p_{ref} = 20\mu Pa,\ I_{ref} = 1 pW/m^2\ (sea\,level)$

$p_0 =$ peak over pressure at t=0

$I_0 =$ peak flux at t=0

$t_1 =$ time when crosses zero p$(t_1), I(t_1) = 0\ (ambient)$

*Integrated blast wave pressure momentum impulse flux from* $t_1$ *to* $t_2$ $(N - s/m^2)$

$$\int_{t_a}^{t_b} p(t)dt = p_0 \int_{t_a}^{t_b} e^{-t/t_1}(1 - t/t_1)dt = p_0 \left[ te^{-t/t_1} \right]_{t_0}^{t_b} = p_0 (t_b e^{-t_b/t_1} - t_a e^{-t_a/t_1})$$

*Positive pressure impulse*:

$$\int_0^{t_1} p(t)dt = p_0 t_1 / e$$

*Negative pressure impulse*:

$$\int_{t_1}^{\infty} p(t)dt = -p_0 t_1 / e$$

*Total pressure impulse*:

$$\int_0^{\infty} p(t)dt = 0$$

*Total Integrated blast wave energy flux* $(J/m^2)$

$$\int_0^{\infty} I(t)dt = I_0 \int_0^{\infty} \left[ e^{-t/t_1}(1 - t/t_1) \right]^2 dt = I_0 e^{-2t/t_1} \left[ \frac{-t^2}{2t_1} - \frac{t_1}{4} + \frac{t}{2} \right]_0^{\infty} = I_0 t_1 / 4$$

*For N fragments the time "t" above is replaced (as in 18 above as)*:

$t_i(\overline{x}) =$ *arrival time of blast from fragment i at position* $\overline{x}$

$t_{i-b}(\overline{x}_{ast,i}) =$ *burst time of fragment i at position* $\overline{x}_{ast,i}$

$t_i(\overline{x}) = t_{i-b}(\overline{x}_{ast,i}) + \left| \overline{x}_{obs} - \overline{x}_{ast,i} \right| / v_s$

$t(above) \rightarrow (t - t_i(\overline{x}))$

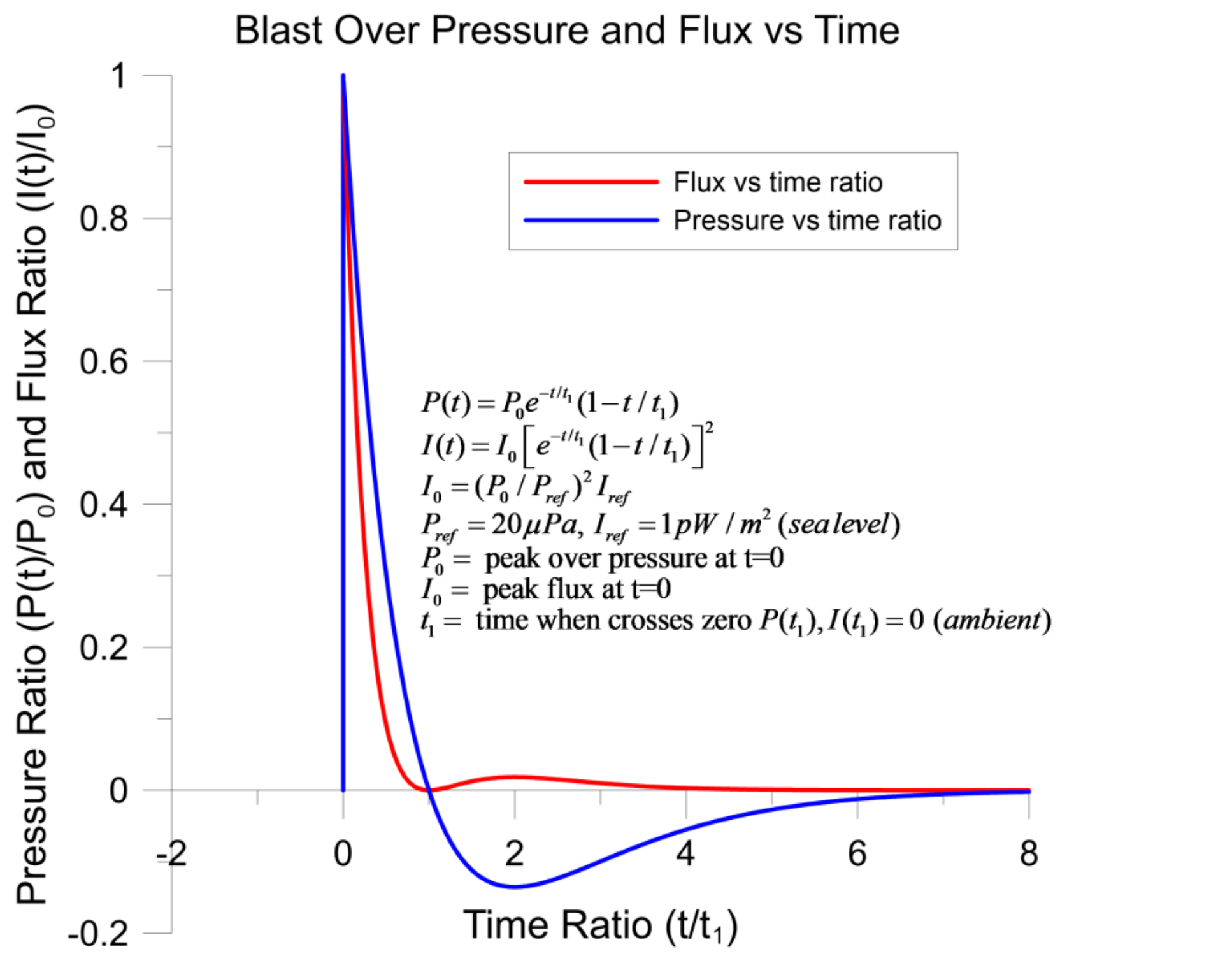

**Figure 30**– Friedlander functional form of blast wave pressure and flux vs time where $t_1$ is a parameter for the zero cross time when the pressure goes below ambient. The parameter $t_1$ is dependent on the detonation specifics, atmospheric specifics, and distance to the observer. None of our conclusions depend critically on $t_1$.

***Shock wave Caustics*** – Caustics form when there is a high contrast caused by interfering phenomenon. This does not require coherence in the normal sense. This is commonly seen in optical phenomenon. A relevant example is the caustics seen in the bottom of a pool of water with small surface waves on the surface causing changes in the refraction of the light which then causes the formation of a web of bright and dark regions which are the caustics. Ultimately this can be visualized as a "time of flight" effect. Similar phenomenon can occur in acoustics as well. A good analogy is the intersection when "blowing" bubbles. In this case the 3D intersection of the expanding bubbles is a plane. The projection onto the Earth of the expanding bubbles is a circle and the planar intersection is a line. The expanding bubbles in our case are the shock waves.

The caustics of interest are the pressure caustics caused by shock waves from multiple fragments that arrive at an observing point at the same time. A key difference for us is that the acoustic shock waves are pulses. This means the caustics will evolve spatially with time. In our case the shock waves from each fragment are emitted at slightly different times due to the longitudinal dispersion of the fragment arrival times. This latter effect will shift the caustic position on the surface of the Earth.

A simple way to visualize the acoustic caustic is to assume all of the fragments burst at the same time. For a given "clock time" that starts when the fragments burst. Choose a length of string equal to the shock wave travel time assuming the shock wave travels at the speed of sound (ignore the corrections from hypersonic to sonic). Make the string diameter inversely proportional to its length to indicate the shock peak pressure scaling roughly as 1/r where r=slant range=string length. As time increases the length of the string increases and the diameter decreases. Attach a string to each fragment and "see where the strings overlap on the surface of the Earth which can be assumed to be a plane for simplicity. Initially (with time) no strings overlap as the shock wave has not arrived at the Earth's surface. With increasing time and thus increasing string length and decreasing diameter, the "pulled taut" sum of the strings will begin to overlap. At infinite time all of the strings will evolve to a circle of infinite radius but with zero shock pressure (zero string diameter). Mathematically as we discussed the total pressure vs time and position of the observer and the fragments is given by:

$$p(t,\bar{x}) = \sum_{i=1}^{N} p_{0_i}(\bar{x}) e^{-(t-t_i(\bar{x}))/t_{1i}(\bar{x})} \left[(1-(t-t_i(\bar{x}))/t_{1i}(\bar{x}))\right] \quad for\, t \ge t_i(\bar{x})$$

$$p(t,\bar{x}) = \sum_{i=1}^{N} p_i(\bar{x},t) \quad for\, t \ge t_i(\bar{x})$$

$$p_{0_i}(\bar{x}) = Peak\, blast\, pressure\, from\, fragment\, i\, at\, position\, \bar{x}$$

$$p_i(\bar{x},t) = Blast\, wave\, pressure\, vs\, time\, from\, fragment\, i\, at\, position\, \bar{x}\, at\, time\, t \ge t_i(\bar{x})$$

$$p_i(\bar{x},t) = p_{0_i}(\bar{x}) e^{-(t-t_i(\bar{x}))/t_{1i}(\bar{x})} \left[(1-(t-t_i(\bar{x}))/t_{1i}(\bar{x}))\right] \quad for\, t \ge t_i(\bar{x})$$

$$t_i(\bar{x}) = arrival\, time\, of\, blast\, from\, fragment\, i\, at\, position\, \bar{x}$$

$$t_{i-b}(\bar{x}_{ast,i}) = burst\, time\, of\, fragment\, i\, at\, position\, \bar{x}_{ast,i}$$

`

$$t_i(\bar{x}) = t_{i-b}(\bar{x}_{ast,i}) + \left|\bar{x}_{obs} - \bar{x}_{ast,i}\right| / v_s$$

$$t_{1i}(\bar{x}) = "t_1"\, from\, Friedlander\, blast\, evolution\, eq\, for\, fragment\, i$$

$$\bar{x} = observer\, position$$

$$\bar{x}_{ast,i} = i_{th}\, fragment\, position\, at\, burst$$

The caustics form where multiple shock waves overlap in space and time. Caustics form a line in projection onto a flat surface such as the ~locally flat Earth's surface near ground zero for each fragment.

***Dependence of Blast Time Scale with Blast Yield*** – The positive pressure shock wave duration or $t_1$ in the Friedlander parametrization of the shock wave time evolution is dependent on a number of complex issues including blast yield, altitude of the blast source, distance from blast and atmospheric absorption vs frequency decomposition (FFT) of the shock wave. A first order fit to both a visual analysis of nuclear weapons air burst tests as well as chemical explosives gives $t_1(\sec) = 0.3E_{Kt}^{1/3}$ where $E_{Kt}$ is the total energy yield in kilotons. Recall in the case of nuclear air bursts that about ½ of the total energy goes into the shock wave and related atmospheric effects such as wind, and hence we roughly double the shock wave yield of an asteroid fragment to get the equivalent nuclear yield that would produce the same shock wave. Factors of "2" here are largely irrelevant to the larger conclusions. Much more detailed calculations of $t_1$ are given below.

**As an example, for a 10m diameter stony fragment with density 2.6 g/cc with a blast yield of about 25 Kt for an equivalent nuclear yield of 50 Kt (1/2 of this goes into the blast) giving $t_1$ ~ 1.1s (assuming $t_1(\sec) = 0.3E_{Kt}^{1/3}$ ). Note, in the nuclear airburst tests, the relevant distances for damage assessment were typically in the near field (strong shock regime) where the pressures were much higher (for a given yield) and thus the $t_1$ times were shorter and thus the measured relation of $t_1(\sec) = 0.3E_{Kt}^{1/3}$** is an under estimate for the far field weak shock regime of interest to us. We will see this in our calculations below. **Our simulations are not very sensitive to the value of $t_1$ in assessing the conclusions of blast damage.** Note that this method (fitting to observed weapons tests is difficult to fully understand systematic errors from the visual blast effects. This is particularly an issue since most of the visual effects data from nuclear and conventional explosive tests is done in the "near field" which is not as relevant to our case which is typically in the far field. We now compute the Friedlander equation time scale **$t_1$** from a "first principles" point.

**Shock wave Duration Time and Conservation of Energy – Scaling Laws**

We can relate the integrated shock wave flux to the total blast energy vs distance to derive the shock wave time constant $t_1$, The vast majority of the total input energy yield however is NOT in the shock wave in the far field but rather is mostly in the KE of the wind and umtimately dissipated as heat. In the near field the situation can be quite different with a significant portion of the total input yield going into the shock wave but we are not in the near field in our case as the ground effects are far from the detonation (typ>30km) as our gragments are small (~10m). The dissipation of energy into a combination of shock wave, wind and thermal dissipation yields a complex relationship of shock wave total energy vs distance and initial energy input.

If we assume an isotropic shock wave we can integrate the total energy in the shock wave and compare this to the initial energy input. The total shock wave energy fraction decreases significantly with distance as we will see, This will yield a scaling law for the shock wave duration time constant. Note that the following "first principles" analysis yields a reasonable agreement with observed blast effects from visual analysis of airburst weapons tests, thought it is somewhat larger. Below we get $t_1(\text{sec}) \sim 0.4E_{Kt}^{1/3}$ in the beginning of the far field whereas our visual fit to the weapons tests above was $t_1(\text{sec}) = 0.3E_{Kt}^{1/3}$. Remarkably they both have precisely the same functional form in energy dependence though we had no preconceived functional form in fitting the observed blast data. The positive pulse duration time $t_1$ is important in the multiple fragment shock wave acoustic caustic formation.

**Measurements and Estimations of $t_1$** – From numerous nuclear airburst tests and from some measured bolide shock wave pulse signatures, there are reasonable measurements of the positive pulse time scale ($t_1$). In the work of Glasstone and Dolan (1972) for the 1KT standard and scaling to larger yields and in Brode (1964) for measurements of 1 Mt class yields, there is some discrepancy though modest. Glasstone and Dolan measure $t_1(\text{sec}) \sim 0.4E_{Kt}^{1/3}$ at a 10 kPa pressure regime which would imply $t_1(\text{sec}) \sim 4E_{Mt}^{1/3}$ with Brode gives the positive pulse time scale closer to 3s for a 1 Mt yield @ 10kPa pressure. There are a wide range of technical issues related to the detonation specifics (altitude etc) with vertical (beneath burst) vs horizontal being slightly different. High altitude ( >20 km) data is limited. We have simulated various time scales consistent with the different measurements and while higher order blast caustics can form with much larger values of $t_1$, we find the "bottom line" ground blast effects to be weakly dependent on reasonable values of $t_1$. Measured data show clear pulse spreading from the near to far field with the strong shock high pressure regime being shorter $t_1$ and the weak shock low pressure regime having longer $t_1$.

$$p(t,r) = p_0(r)e^{-t/t_1}(1-t/t_1)$$

$$\text{Acoustic Power Flux } (w/m^2):\ I(t,r) = I_0(r)\left[e^{-t/t_1}(1-t/t_1)\right]^2$$

$$I_0(r) = \frac{p_{amp}^{\ 2}}{2\rho v} = \frac{p_{rms}^{\ 2}}{\rho v} = I_{ref}(p/p_{ref})^2 = 2.5x10^{-3}p^2(Pa^2)$$

$$p_{amp} = \sqrt{2}p_{rms}$$

$$p_{ref} = 20\mu Pa,\ I_{ref} = 1pW/m^2\ (sea\,level)$$

$p_0(r) =$ peak over pressure at t=0 at distance r

$I_0(r) =$ peak flux at t=0 at distance r

$t_1 =$ zero crossing time when $p(t_1,r), I(t_1,r) = 0\ (ambient)$

$\varepsilon_{blast}$(r)=fraction of equivalent weapon yield in blast wave measured at distance r (yield and pressure dependent)

$\alpha_{atm-trans}(r) =$ atmospheric transmission of blast wave intensity/energy (0-1) (integrated over full blast intensity (w/m$^2$) pulse)

*We can include $\alpha_{atm-trans}(r)$ in $\varepsilon_{blast}$(r) for simplicity*

*Integrated blast wave energy per area*

$$\int_0^\infty I(t,r)dt = I_0(r)\int_0^\infty\left[e^{-t/t_1}(1-t/t_1)\right]^2 dt = I_0(r)e^{-2t/t_1}\left[\frac{-t^2}{2t_1} - \frac{t_1}{4} + \frac{t}{2}\right]_0^\infty = I_0(r)t_1/4$$

At distance r from the initial detonation this gives a total energy:

$$E_{blast-wave}(r) = 4\pi r^2\int_0^\infty I(t,r)dt = \pi r^2 I_0(r)t_1$$

$$\varepsilon_{blast}(\text{r}) \equiv E_{blast-wave}(r)/E_{total-yield} = E_{blast-wave}(r)/\eta E_{kt} = \pi r^2 I_{acoustic-energy-flux}t_1/\eta E_{kt} = \frac{\pi r^2 t_1 p(r)^2}{\rho_{shock}\text{v}_{shock}\eta E_{kt}}$$

$$\varepsilon_{blast}(\text{r})/t_1 = \frac{\pi r^2 p(r)^2}{\rho_{shock}\text{v}_{shock}\eta E_{kt}} = \frac{\pi r^2\left[p_n\left[rE_{kt}^{-1/3}\right]^{\alpha_n} + p_f\left[rE_{kt}^{-1/3}\right]^{\alpha_f}\right]^2}{\rho_{shock}\text{v}_{shock}\eta E_{kt}}$$

$$t_1 = \frac{\rho_{shock}\text{v}_{shock}\eta\varepsilon_{blast}(\text{r})E_{kt}}{\pi r^2 p(r)^2} = \frac{\rho_{shock}\text{v}_{shock}\eta\varepsilon_{blast}(\text{r})E_{kt}}{\pi r^2\left[p_n\left[rE_{kt}^{-1/3}\right]^{\alpha_n} + p_f\left[rE_{kt}^{-1/3}\right]^{\alpha_f}\right]^2}$$

$$p(r) = p_n\left[rE_{kt}^{-1/3}\right]^{\alpha_n} + p_f\left[rE_{kt}^{-1/3}\right]^{\alpha_f}\ (near\ and\ far\ field)$$

$$p(r) = p_f\left[rE_{kt}^{-1/3}\right]^{\alpha_f}\ (far\ field\ - \text{v}_{shock} = v = c_o = speed\ of\ sound, \rho_{shock} = \rho_0 = ambient\,(sea\,level)\,air\,density)$$

For our cases of small fragments "detonating" in the atmosphere, the ground observer is in the far field

Far Field case:

$$\varepsilon_{blast}(\text{r})/t_1 = \frac{\pi r^2 p(r)^2}{\rho_0 \text{v}\eta E_{kt}} = \frac{\pi r^2\left[p_f\left[rE_{kt}^{-1/3}\right]^{\alpha_f}\right]^2}{\rho_0\text{v}\eta E_{kt}\eta E_{kt}},\ t_1 = \frac{\rho_{shock}\text{v}_{shock}\eta\varepsilon_{blast}(\text{r})E_{kt}}{\pi r^2 p(r)^2} = \frac{\rho_0\text{v}\eta\varepsilon_{blast}(\text{r})E_{kt}}{\pi r^2\left[p_f\left[rE_{kt}^{-1/3}\right]^{\alpha_f}\right]^2}$$

$$IF\ \alpha_f = -1.0 \rightarrow p(r) \sim 1/r, t_1 \sim E_{kt}^{1/3} \rightarrow \varepsilon_{blast}(\text{r})/t_1 \sim E_{kt}^{-1/3} \rightarrow \varepsilon_{blast}(\text{r}) \sim indep\,of\ r\ \&\ E$$

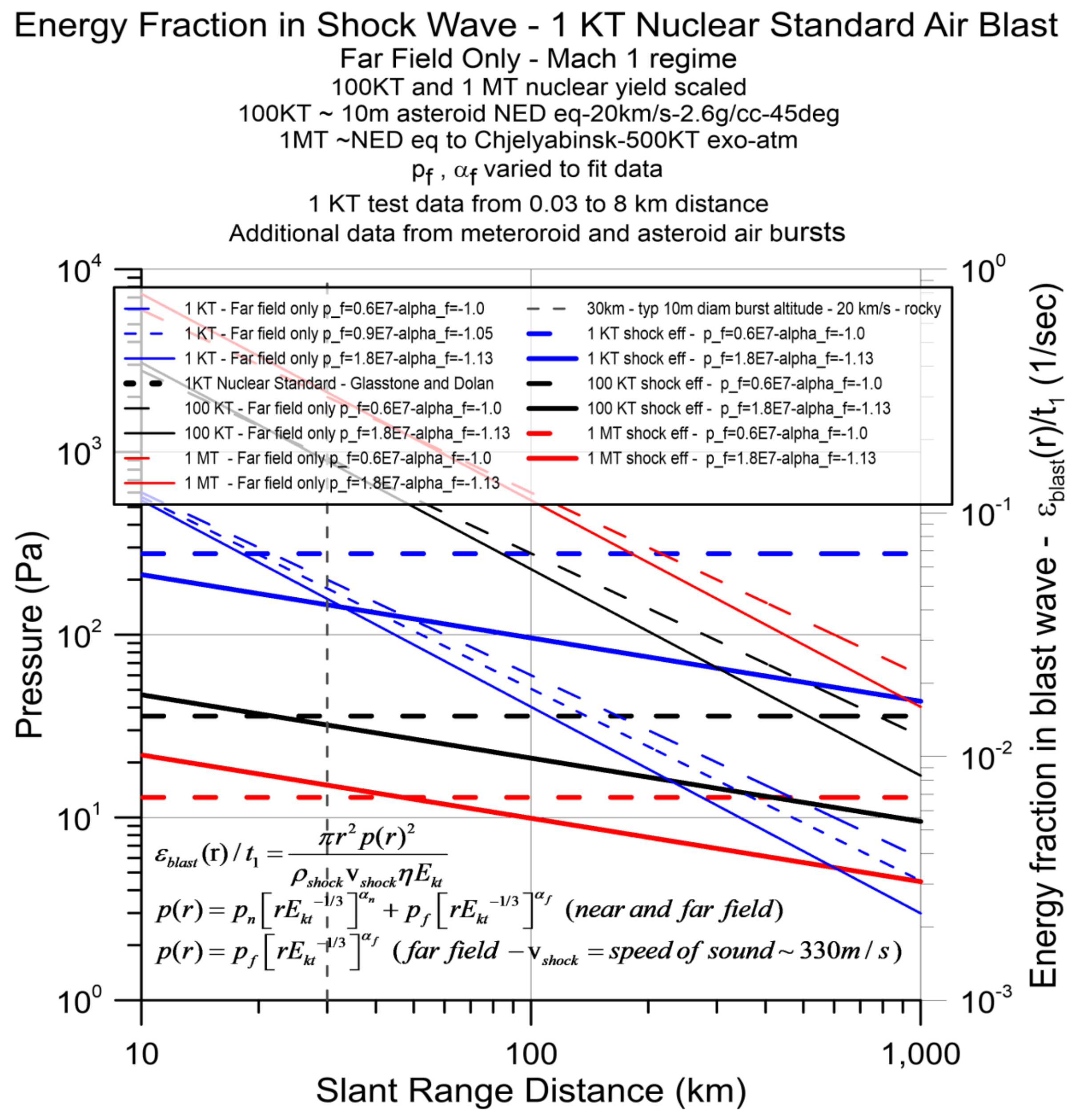

**Figure 31** – Pressure and energy fraction in blast wave normalized to $t_1$ for various fits to p(r). Note that for p(r)~1/r scaling in the far field then the energy fraction/$t_1$ is independent of distance and that if $t_1$ scales as $E_{kt}^{1/3}$ (which is normally assumed independent of P scaling with r, then the energy fraction is independent of r and E.

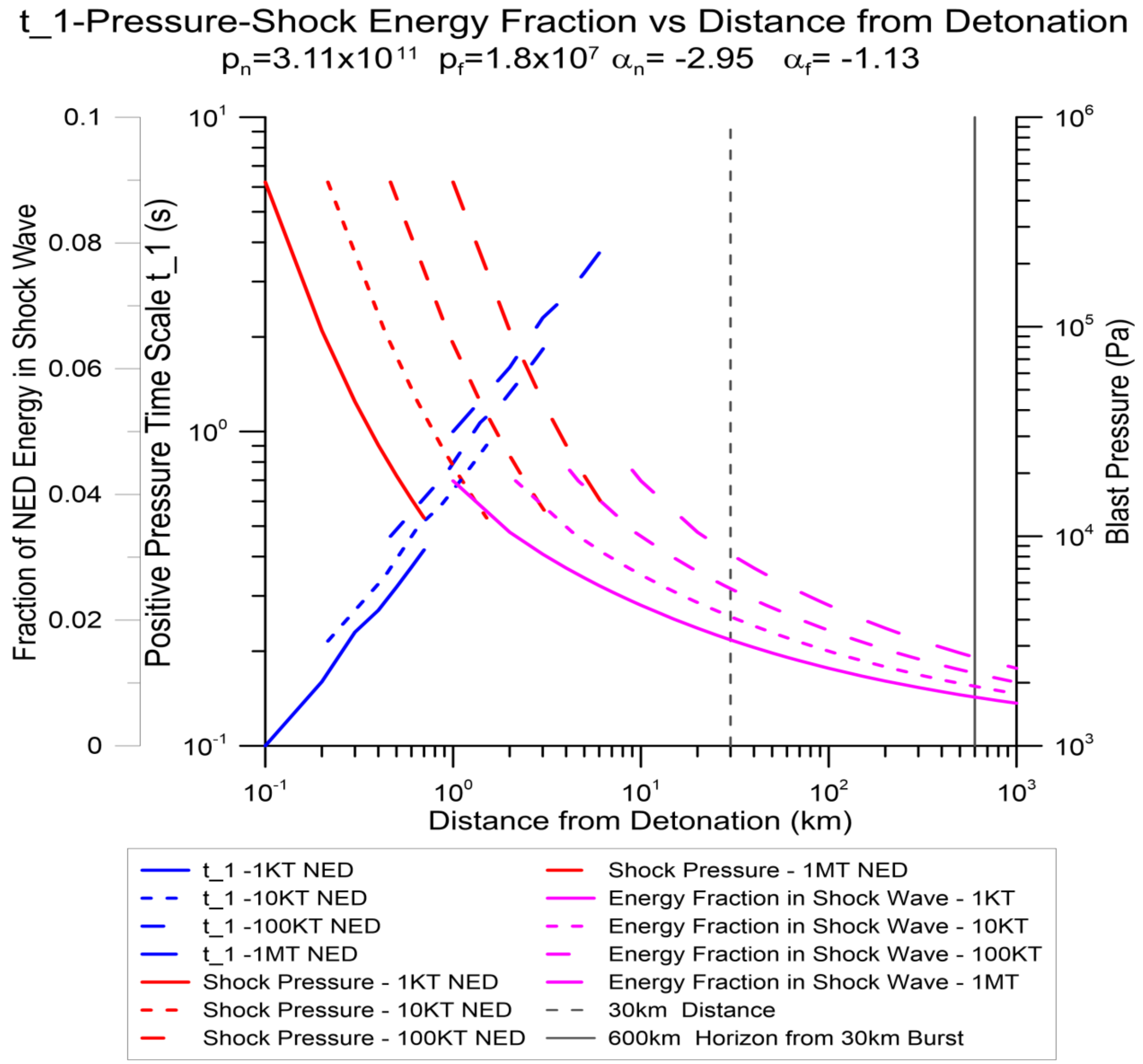

**Figure 32 –** Positive pressure time ($t_1$), peak blast pressure and blast wave fractional energy of total NED yield vs distance based on measured positive pressure time for nuclear weapon airburst tests. Note the pulse stretching ($t_1$ increases with distance from near to far field) The fraction of energy in the blast wave relative to the total weapon yield assumes $t_1$ is the measured value for the 1 KT standard at a distance where the blast pressure is about 10 KPa or 1/10 of ambient pressure. This is the regime where far field behavior (relevant to us) begins. For this regime $t_1(s) = 0.42 E_{kt}^{1/3}$. We assume a best fit pressure scaling in the far field of $\alpha_f$=-1.13. In the relevant fragment energy regimes of 10-200 KT for fragments of order 3-15m for nominal closing speeds of 10-20 km/s with about 50% of the initial energy in the equivalent NED going into the initial (early near field) blast wave, the fraction of blast wave energy in the relevant far field distance range of 30-800 km, is typically of order 1%. To convert from bolide fragment total blast wave energy deposition (near field) $E_{bolide\text{-}blast\text{-}wave}$ to the equivalent NED yield $E_{ned}$ we have $E_{ned} = E_{bolide\text{-}blast\text{-}wave}/\varepsilon_{ned\text{-}initial\text{-}blast}$ where is the fraction of the NED yield converted into the initial (very near field) blast wave energy which is typically about $\varepsilon_{ned\text{-}initial\text{-}blast}$=0.5 – ie the NED yield is double the bolide fragment total blast energy. For example, if the fragment has 20KT of energy put into the blast wave initially then the equivalent NED yield for computing pressure, $t_1$, energy fraction etc in far field corresponds to an initial equivalent yield of $E_{ned}$ =40 KT.

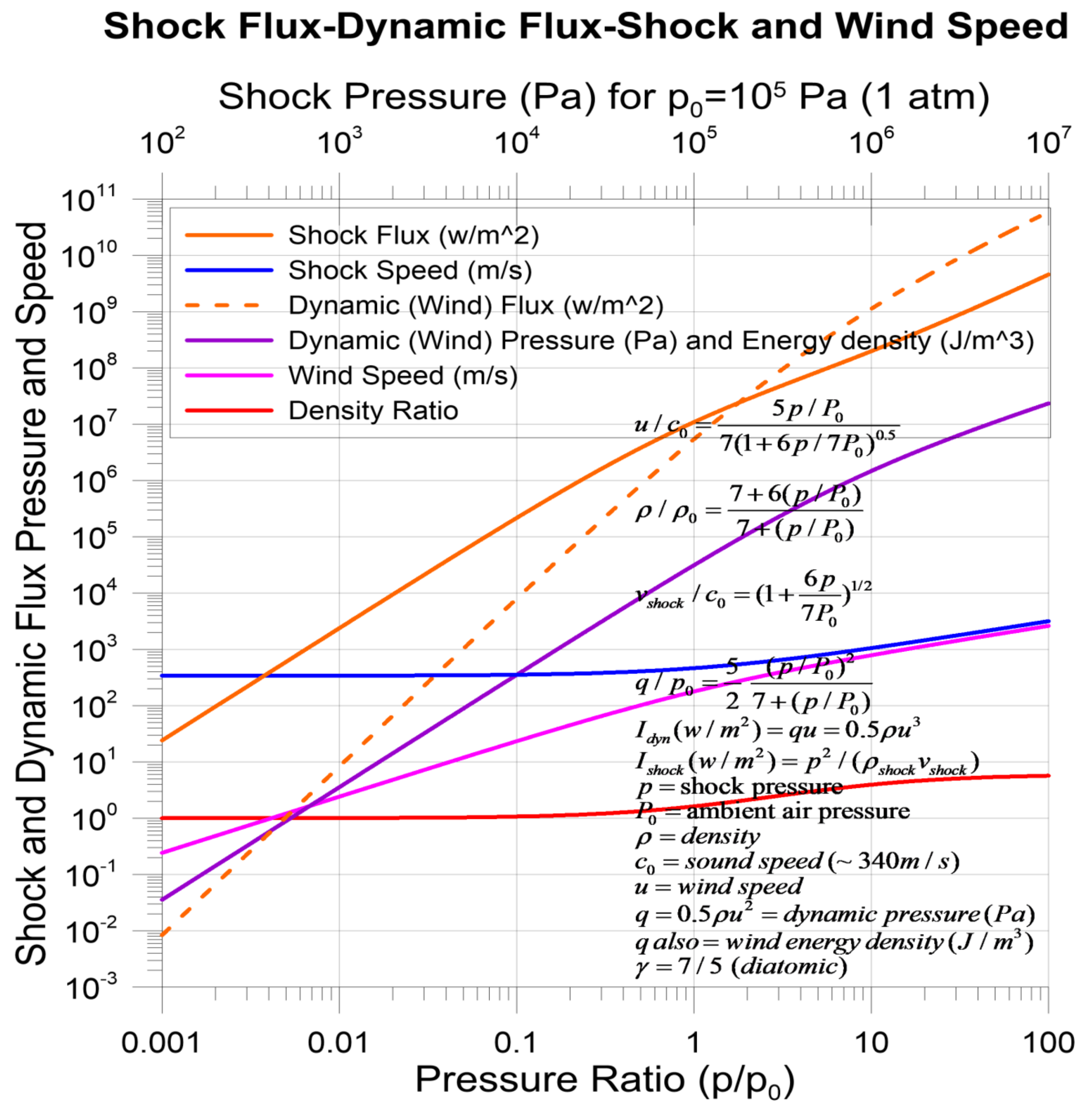

**Figure 33** – Comparing the shock wave flux and speed to the blast dynamic (wind) flux and speed. For overpressure > $P_0$ (ambient pressure – strong shock regime) the dynamic (wind) flux is greater than the shock flux with most of the energy going into the wind while at lower pressures (weak shock regime) the shock wave flux is generally much greater than the dynamic (wind) flux with wind speeds being quite low at low pressure ratios (large distance).

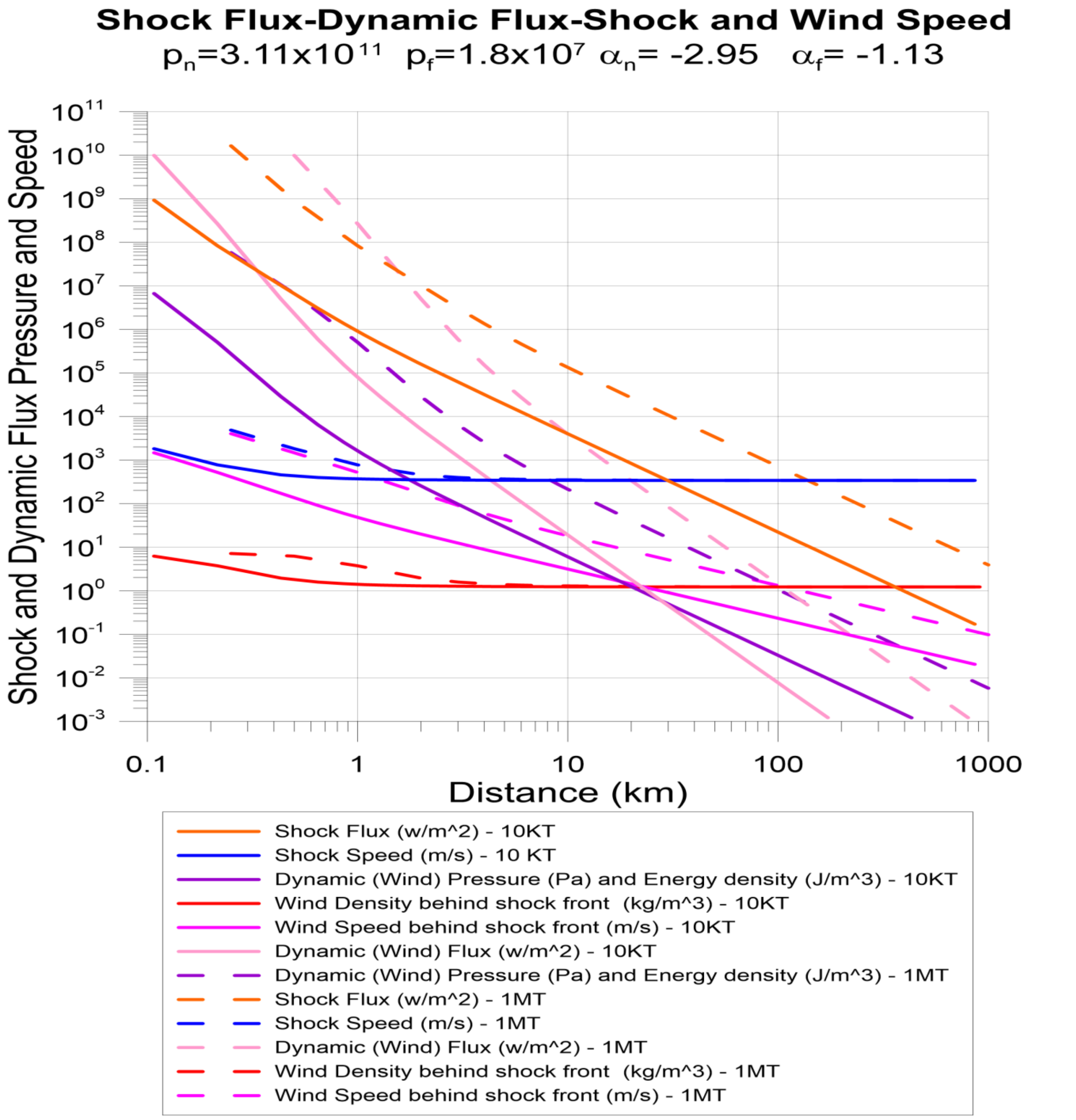

**Figure 34 –** Shock and Dynamic (Wind) Flux and Wind Density, Shock and Wind Speed and Wind Density vs range for 10KT and 1 MT equivalent NED yield. Note that the Wind flux (pink) is much less than the shock flux (orange) in the far field but the wind flux exceeds the shock flux in the near field. Since our fragments all burst at altitudes above about 25km (depending on size and density), we are essentially always in the far field.

$Far\ field - Forcing\ 1/r\ scaling:$

$$t_{1-a}(s) = \eta \varepsilon_{blast}(r) E_{kt}^{1/3} \frac{p_{ref}^{2}}{\pi I_{ref}\, p_{f-a}^{2}}$$

$A\ reasonable\ fit\ for\ forcing\ p(r) \sim 1/r \rightarrow p_{f-a} = 6x10^{6}\ (Pa), \alpha_f = -1 \quad p(r) \sim 1/r\ in\ far\ field)$

$\rightarrow t_{1-a}(s) = 14.8 \varepsilon_{blast}(r) E_{kt}^{1/3} \quad (note\ r\ dependence\ from\ \varepsilon_{blast}(r)\ which\ may\ also\ be\ P(r)\ dependent)$

$General\ form - near\ and\ far\ field:$

$$t_{1-b}(s) = \eta \frac{\varepsilon_{blast}(r)\, p_{ref}^{2}}{\pi I_{ref}\, p_{f-b}^{2}} r^{-(2\alpha_f + 2)} E_{kt}^{1+2\alpha_f/3} \quad (general\ form)$$

$$\frac{t_{1-b}}{t_{1-a}} = \left[\frac{p_{f-a}}{p_{f-b}}\right]^{2} r^{-2(\alpha_f+1)} E_{kt}^{2(1+\alpha_f)/3} \quad (far\ field)$$

$$p_0(r) = p_n \left[r E_{kt}^{-1/3}\right]^{\alpha_n} + p_f \left[r E_{kt}^{-1/3}\right]^{\alpha_f} \rightarrow p_f \left[r E_{kt}^{-1/3}\right]^{\alpha_f} \quad (far\ field)$$

Nominal fit to nuclear test data with $E_{kt}$ = NED yield in kt

$p_n$ = 3.11x10$^{11}$ $(Pa)(near\ field), \alpha_n$ = -2.95 (near field),

$p_f$ = 1.8x10$^{7}$ $(Pa)(far\ field), \alpha_f$ = -1.13 $(far\ field)$

$\varepsilon_{blast}$(r)=fraction of weapon yield that goes into blast wave (distance dependent)

$$\eta = 4.184x10^{12}\ J/kt,\ \eta \frac{p_{ref}^{2}}{\pi I_{ref}} = 5.33x10^{14}$$

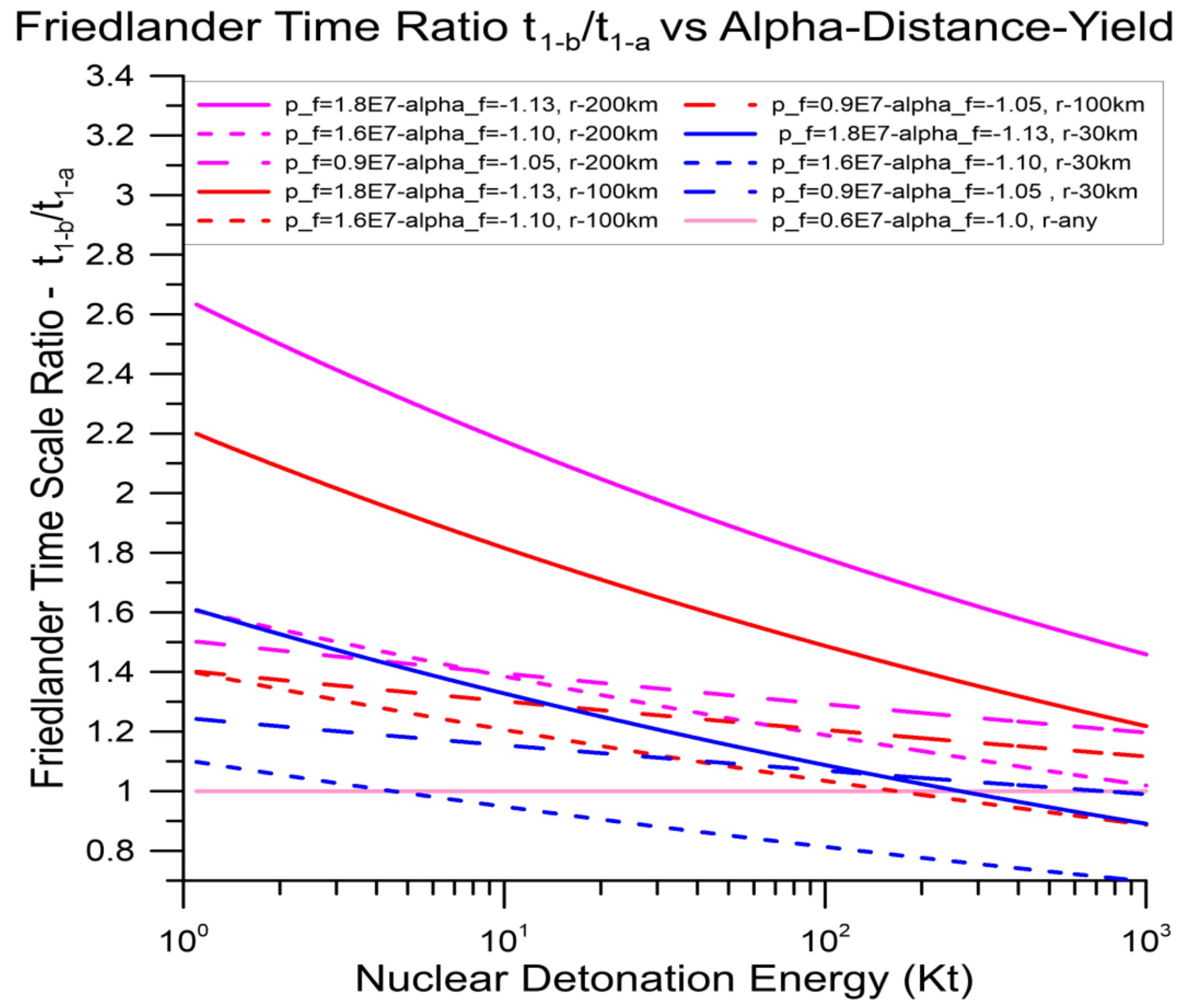

**Figure 35 –** Ratio of positive pressure times ($t_1$) for various far field fits from $\alpha_f$ =-1 (p(r)~1/r) to $\alpha_f$ =-1.13 (best fit). Note that for $\alpha_f$ =-1, the time ratio is independent of yield energy, again for the far field. If the conversion efficiency ($\alpha_{blast}$) is independent of distance in the far field then $t_1$ is also independent of r in the far field.

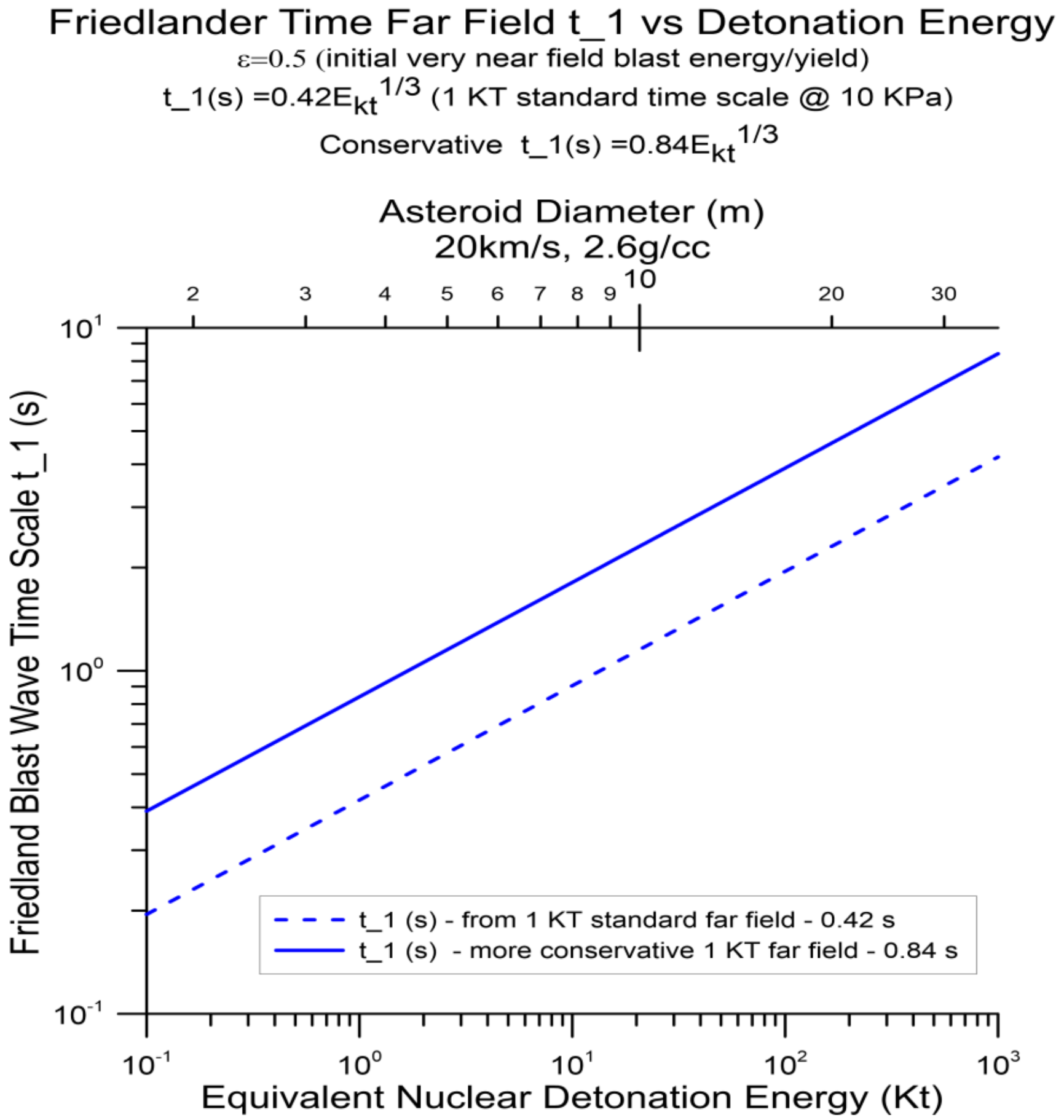

**Figure 36** – Time scale ($t_1$) for positive pressure phase of shock wave vs equivalent detonation energy at a pressure of about 10 kPa (Glasstone and Dolan, Brode) . This time scale is known as the Friedlander $t_1$ parameter. The relationship is approximate and influenced by many factors including atmospheric acoustic absorption of high frequency (short time scales). The equivalent asteroid diameter from 1.5 to 35m for a typical stony bolide (2.6 g/cc) is also shown. For a 10m asteroid the time scale is about one second. Note that pulse stretching will introduce a $t_1$ term that is pressure and thus distance dependent.

***Pulse Stretching – $t_1$ Pressure-Distance and Yield Dependence – Extrapolation to lower over pressure regime.*** As discussed above there is clear evidence from NED air-burst testing of the stretching (increase) of the positive pressure time ($t_1$) with increased distance and thus decreasing over-pressure. We have fit the data from various NED tests as summarized in Glasstone and Dolan (1972 and earlier versions) and in Brode (1964). We summarize the transition from strong shock ($\Delta P=p>P_0$) to the weak shock regime ($\Delta P=p<P_0$). In both regimes there is a pressure dependence on $t_1$ with the weak shock regime shock a clear trend of $t_1$ increasing at lower over pressure. In particular, we show the data from Brode for a 1 Mt yield and focus on the transition from strong to weak shock. We lack good data at low over-pressure, particularly in the regime of 0.1 -10 kPa. We extrapolate the best fit from the regime from 10-100 kPa where there is good data and assume the trend continues to lower over-pressures. We are in the process of running multiple different 3D hydrocodes using CTH, ALE3D and others to check the validity of this. We feel the extrapolation is a conservative one. Note that this does NOT affect the shock pressure but only the positive pressure regime and thus is relevant to caustic formation. We have also run a number of simulations over various $t_1$ times relevant to our yields and shock pressure regimes **and find only modest final effect on the ground "damage" assessment**. Our extrapolation to longer $t_1$ at greater ranger and lower pressure makes the caustic slightly larger and thus leads to even more conservative conclusions **with the bottom line being there is no significant change to our conclusions.**

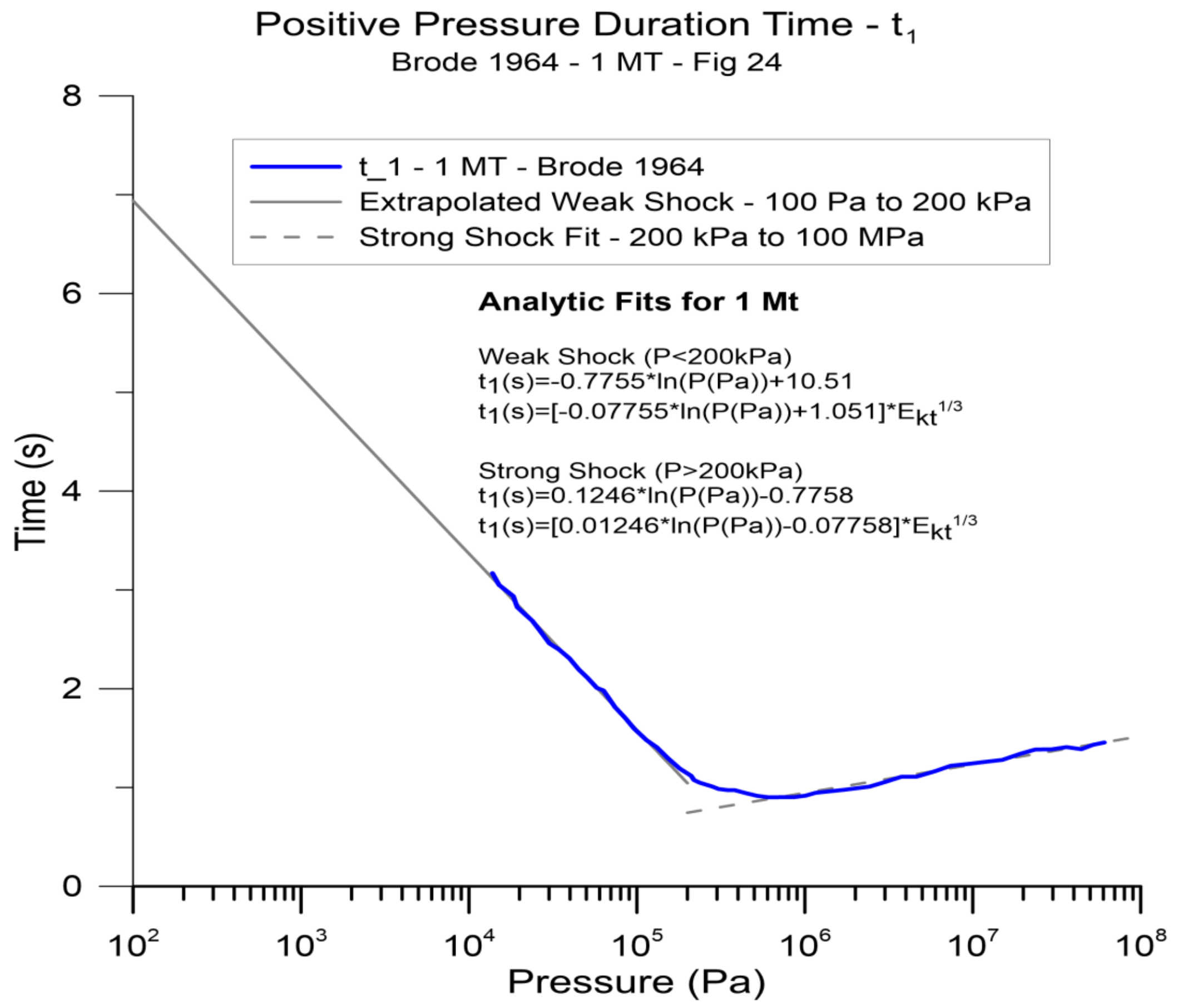

**Figure 38** – Positive pressure time ($t_1$ )vs over–pressure for a 1MT low altitude NED with fits to low pressure (weak shock) and strong shock regimes. Note that the positive pressure time scales with both pressure and energy.

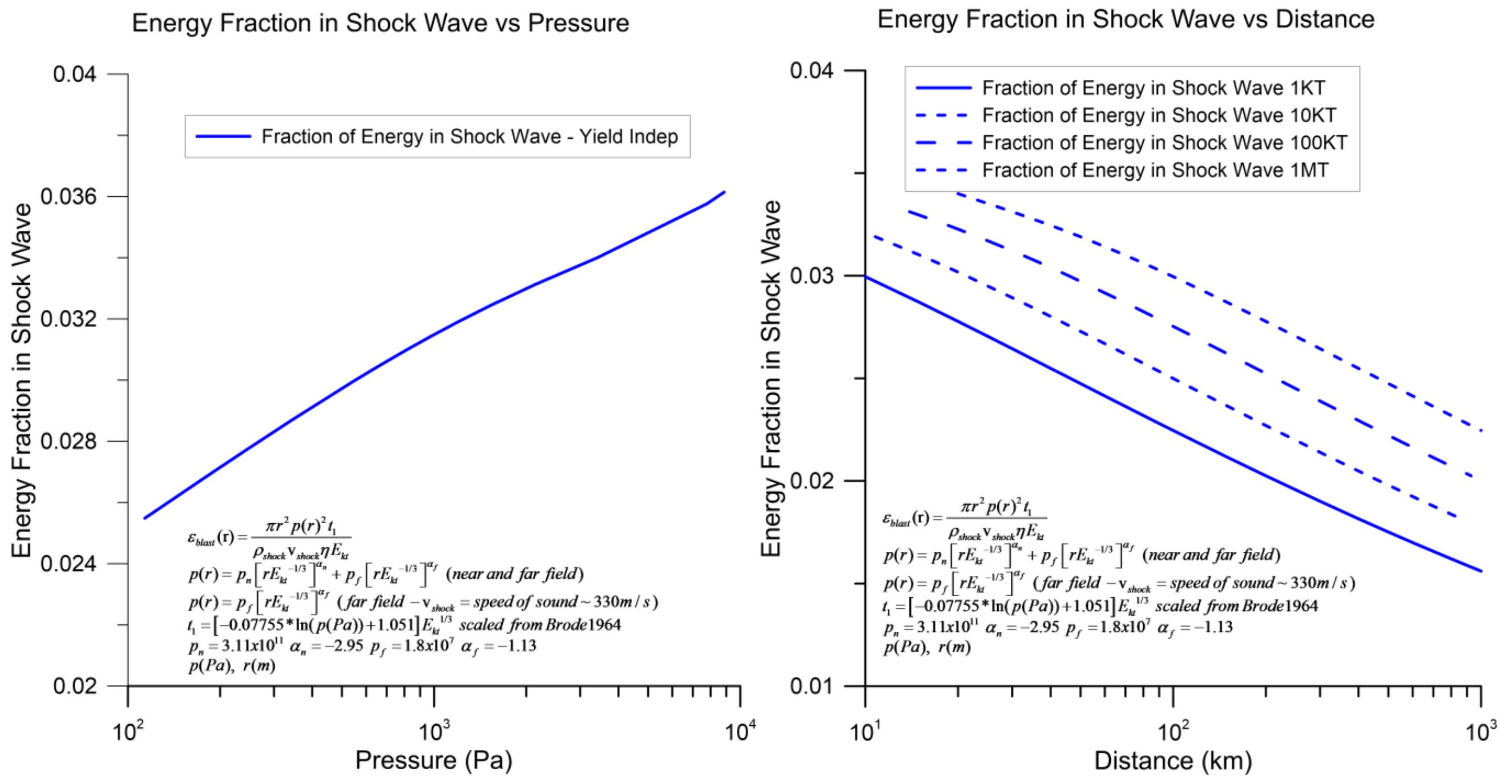

**Figure 37 – L:** Fraction of energy in shock wave **vs peak over-pressure** using $t_1$ analytic extrapolation. Independent of energy yield. **R:** Fraction of energy in shock wave **vs slant range distance** using $t_1$ analytic extrapolation

***Energy Fraction vs Pressure – Independent of Yield*** – For a given pressure, the energy fraction going into the shock wave is independent of the yield energy as shown in the accompanying figure. This is due to the assumed 1/3 power scaling with energy. As shown the figure, the energy fraction is a very mild function of pressure. In the range of general interest to us for pressures from about 0.1-10 kPa the energy fraction is about 3%.

$$\varepsilon_{blast}(\mathrm{r}) = \frac{\pi r^2 p(r)^2 t_1}{\rho_{shock} \mathrm{v}_{shock} \eta E_{kt}}$$
$$p(r) = p_n \left[ rE_{kt}^{-1/3} \right]^{\alpha_n} + p_f \left[ rE_{kt}^{-1/3} \right]^{\alpha_f} \quad (near\ and\ far\ field)$$
$$p(r) = p_f \left[ rE_{kt}^{-1/3} \right]^{\alpha_f} \quad (far\ field\ - \mathrm{v}_{shock} = speed\ of\ sound \sim 340 m/s)$$
$$t_1 = \left[ -0.07755 * \ln(p(Pa)) + 1.051 \right] E_{kt}^{1/3} \quad scaled\ from\ Brode 1964$$
$$p_n = 3.11x10^{11} \quad \alpha_n = -2.95 \quad p_f = 1.8x10^7 \quad \alpha_f = -1.13$$
$$p(Pa),\ r(m)$$

For a given pressure P(r) and given $p_n,\ \alpha_n,\ p_f,\ \alpha_f$ the product $rE_{kt}^{-1/3}$ is fixed.
Thus r scales as $r \sim E_{kt}^{1/3}$ This is the usual energy to the 1/3 power scaling.
If we also assume that $t_1$ scales as the 1/3 power of energy then $\varepsilon_{blast}(\mathrm{r}) = \varepsilon_{blast}(\mathrm{P(r)})$ is independent of energy.

$$\varepsilon_{blast}(\mathrm{r}) = \frac{\pi r^2 p(r)^2 t_1}{\rho_{shock} \mathrm{v}_{shock} \eta E_{kt}} \propto \frac{r^2 p(r)^2 t_1}{E_{kt}} \propto \frac{E_{kt}^{2/3} p(r)^2 E_{kt}^{1/3}}{E_{kt}} \propto p(r)^2 \quad (independent\ of\ E_{kt}\ in\ both\ near\ and\ far\ field)$$

***Energy Fraction vs Distance – Scaling with Distance and Yield*** – For a given yield from 1-1000 Kt the energy fraction going into the shock wave is a relatively mild function of both yield and distance with a typical energy fraction of 2-3% from 10-1000 km distance and 1-1000 Kt yield

***Shock Wave Surface and Local Topography Reflections*** – The shock wave time evolution that any observer measures are influenced by a number of factors including surface reflections of the primary shock wave on the Earth's surface and the local land and building topography. Much like radio reception has nulls and peaks due to the interference effects from local reflections, so too do acoustical shock waves have interference effects. This has been noted in a number of conventional, nuclear and some measured asteroid acoustic measurements. In general, the blast source we are interested in occurs at high altitude typically around 30 km and for typical relevant distances away from "ground zero" from any fragment the observer is normally in Mach regime where reflections are relevant. Constructive interference has been noted in the Chelyabinsk event where window breakage appeared correlated with possible multiple reflections yielding constructive interference in some cases and destructive interference in other cases. This is a site-specific issue (local topography, multiple buildings etc) and this not one that we can take account of in general.

***Scaling Relations*** – It is useful to understand the scaling of the slant rang (r), shock wave peak pressure P, shock wave duration ($t_1$), equivalent detonation energy ($E_{kt}$) and bolide diameter (d).

$$t_1 \propto (E_{kt})^{1/3} \propto (d^3)^{1/3} = d$$
$$IF\ we\ approximate\ 1/r\ scaling\ for\ p(r)\ in\ far\ field\ (not\ the\ best\ fit)\ then:$$
$$\rightarrow p(r, E_{kt}) \propto r^{-1} (E_{kt})^{1/3} \propto d/r \rightarrow p(r, E_{kt}) \propto t_1 / r$$
$$As\ we\ showed\ above\ a\ better\ fit\ for\ t_1\ involves\ pressure\ corrections:$$
$$t_1 = \left[ -0.07755 * \ln(p(Pa)) + 1.051 \right] E_{kt}^{1/3}$$
$$with\ p(r) = p_f \left[ rE_{kt}^{-1/3} \right]^{\alpha_f} \quad p_f = 1.8x10^7 \quad \alpha_f = -1.13\ though\ this\ is\ close\ to\ p(r) \propto 1/r\ scaling$$

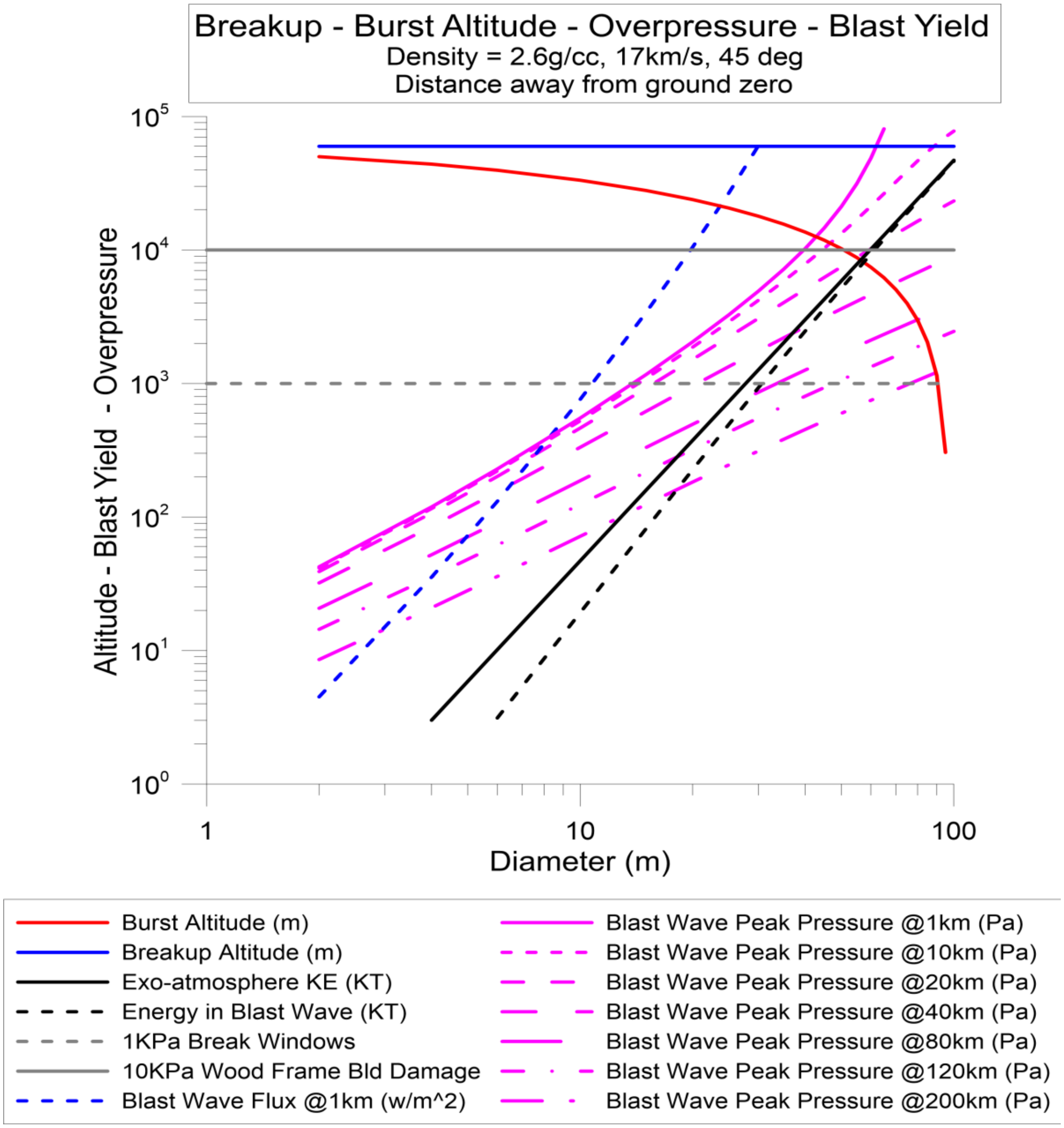

**Figure 39**– Summary of our modeled breakup and burst altitude as well as the blast wave peak pressure and flux and total blast wave energy vs asteroid/fragment diameter for 17 km/s entry, as well as horizontal distance from ground zero (point on the Earth directly below the final burst) for diameters from 1 to 100m. Beyond ~100m diameter, the asteroid will hit the ground, resulting in massive local damage. The curves are shown for a density 2.6g/cc (stony) asteroid at 17km/s and 45-degree impact (horizon angle). The lines for common residential and commercial building glass breakage and the beginning of serious residential structural damage are shown in grey dashed and solid lines, respectively. Breakup altitude is virtually independent of diameter, while burst altitude is critically dependent on diameter. As is seen, if we can keep the fragment size below 10m diameter for rocky asteroid (~ 5m for higher density "iron rich" asteroids) then the damage even directly below the burst (1km distances from ground zero, for example) is below the glass breakage threshold (~ 1kPa). **This is a critical take away from this plot. *Fragmentation into reasonably small fragments prevents serious damage.* The fragmentation maximum diameter to be allowed does depend strongly on the asteroid density and angle of attack and weakly on the speed. The latter weak dependence on speed is not obvious, but is due to high altitude breakup at high speed and energy dissipation into modes other than shock waves.**

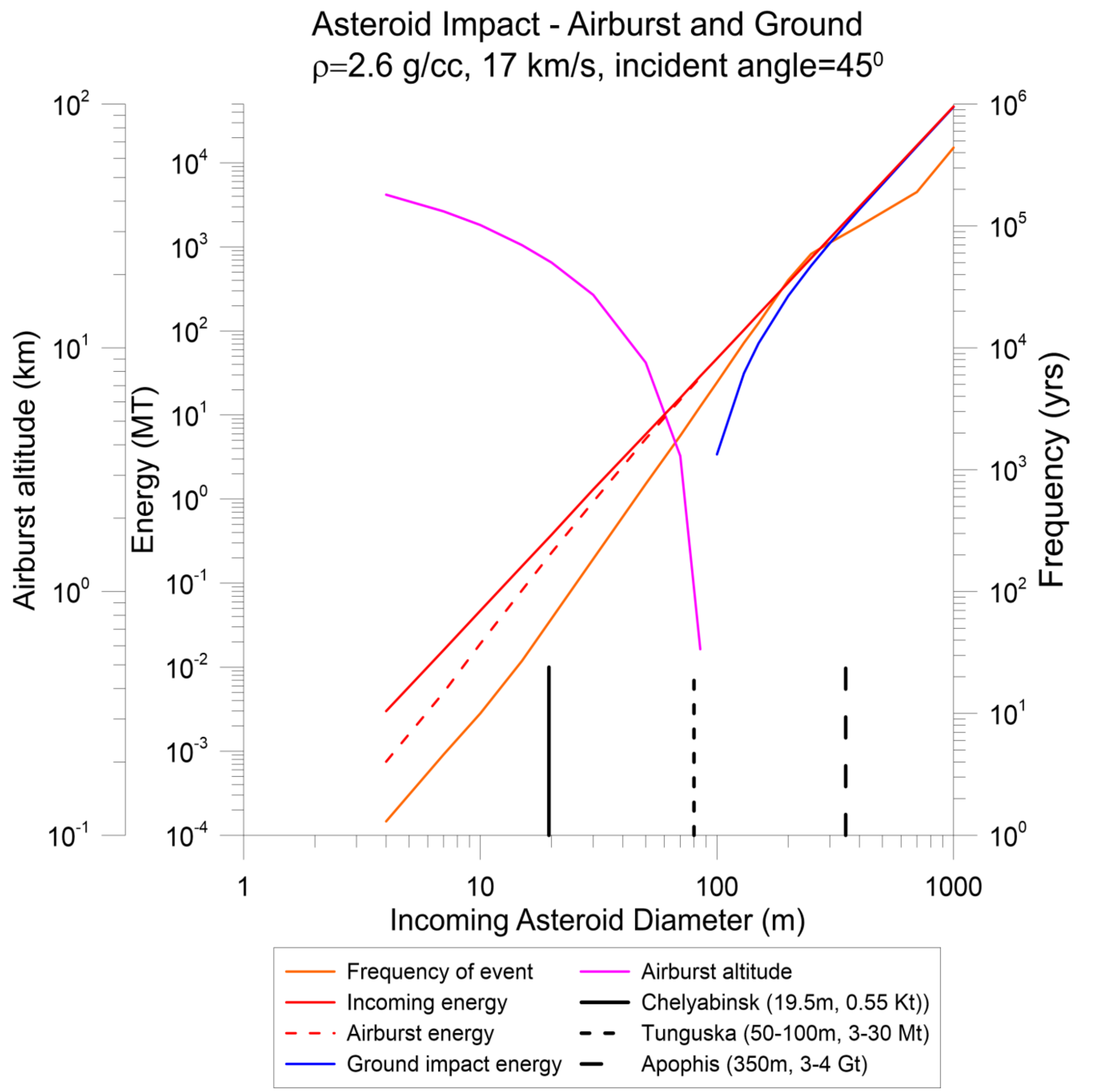

**Figure 40**– Frequency of impacts, exo-atmospheric energy, and air burst/ground impact energies for asteroids from 1 to 1000m diameter – density 2.6. **Below ~90m diameter the asteroid will air burst**, **and above 100m the asteroid will impact the ground.** The energies in Mt (TNT) are shown for incoming, air burst (shock wave), and ground impact energies. **Preventing ground impact is critical.**

***Mass of atmosphere traversed by bolide prior to burst*** – We compute the air mass a spherical bolide traverse's through before it air bursts and compare this to the mass of the bolide. This air mass is the mass of the column of air the bolide goes through prior to bursting. We treat the bolide as a sphere and the air column as a cylinder with diameter equal to the diameter of the bolide. Atmosphere is modeled as an isothermal self-gravitating sphere with scale height H ~ 8km.

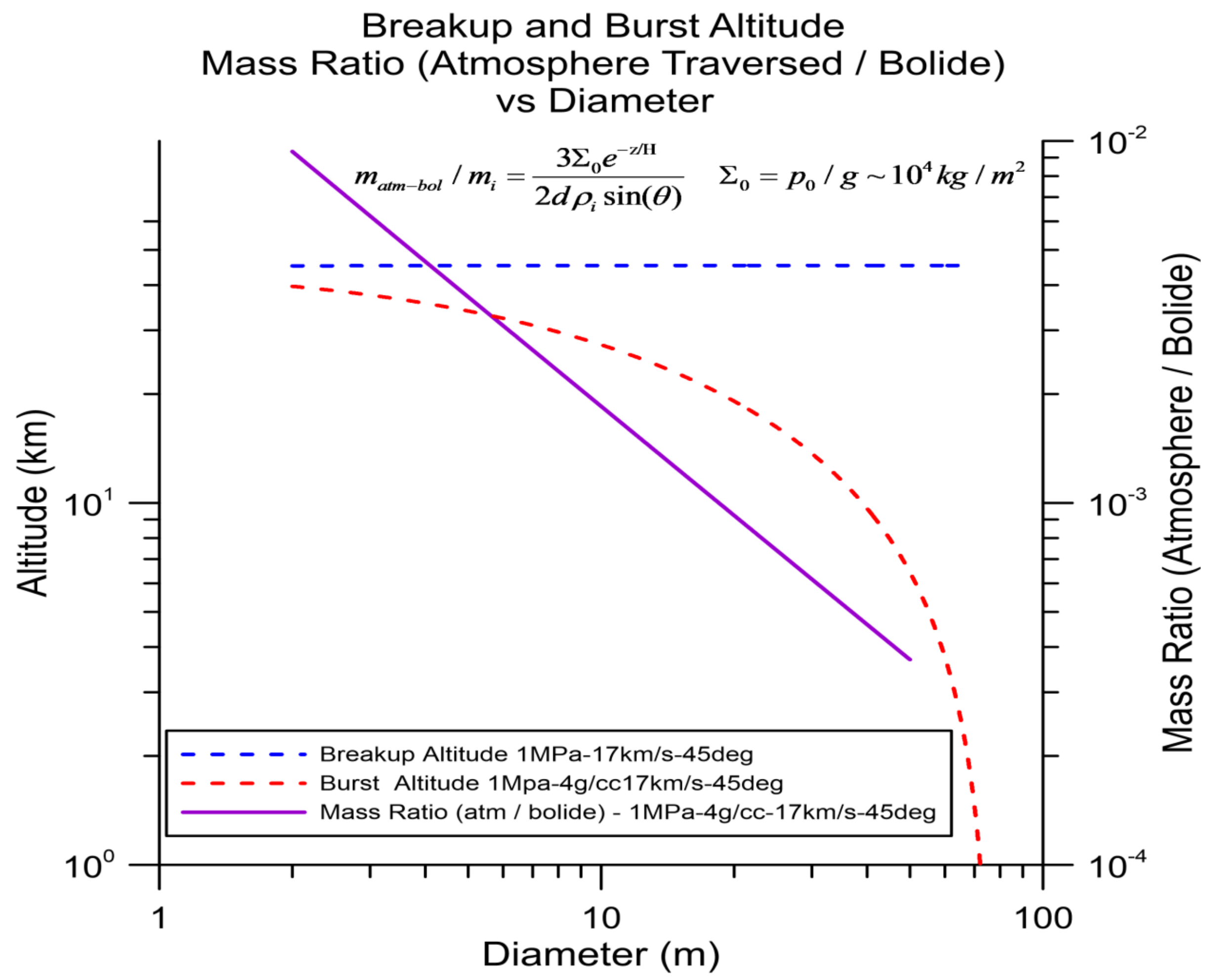

**Figure 41** – Ratio of mass of atmosphere traversed before burst over bolide mass for case of 17 km/s, 4 g/cc bolide with about 1 MPa yield strength and attack angle of 45 degrees. Also shown are breakup and burst altitude.

Pressure p(z) at altitude z for isothermal atmosphere with scale height H (~8km):

$$P(z)/P_0 = \rho(z)/\rho_0 = e^{-z/H}$$

Atmosphere mass/area ($\Sigma(z)$) above height z:

$$\Sigma(z) = P(z)/g = \Sigma_0 e^{-z/H} \quad where\ \Sigma_0 = P_0/g \sim 10^4 kg/m^2$$

Mass of atm $m_{atm-bol}$ that bolide of diameter d traverses above altitude z:

$$m_{atm-bol} = (\pi d^2/4)\Sigma(z)/\sin(\theta) = \frac{\pi d^2 \Sigma_0 e^{-z/H}}{4\sin(\theta)} \quad where\ \theta = attack\ angle\ rel\ horizon$$

$$m_i = mass\ frag\ i = \frac{\pi}{6}\rho_i d^3$$

$$m_{atm-bol}/m_i = \frac{3\Sigma_0 e^{-z/H}}{2d\rho_i \sin(\theta)}$$

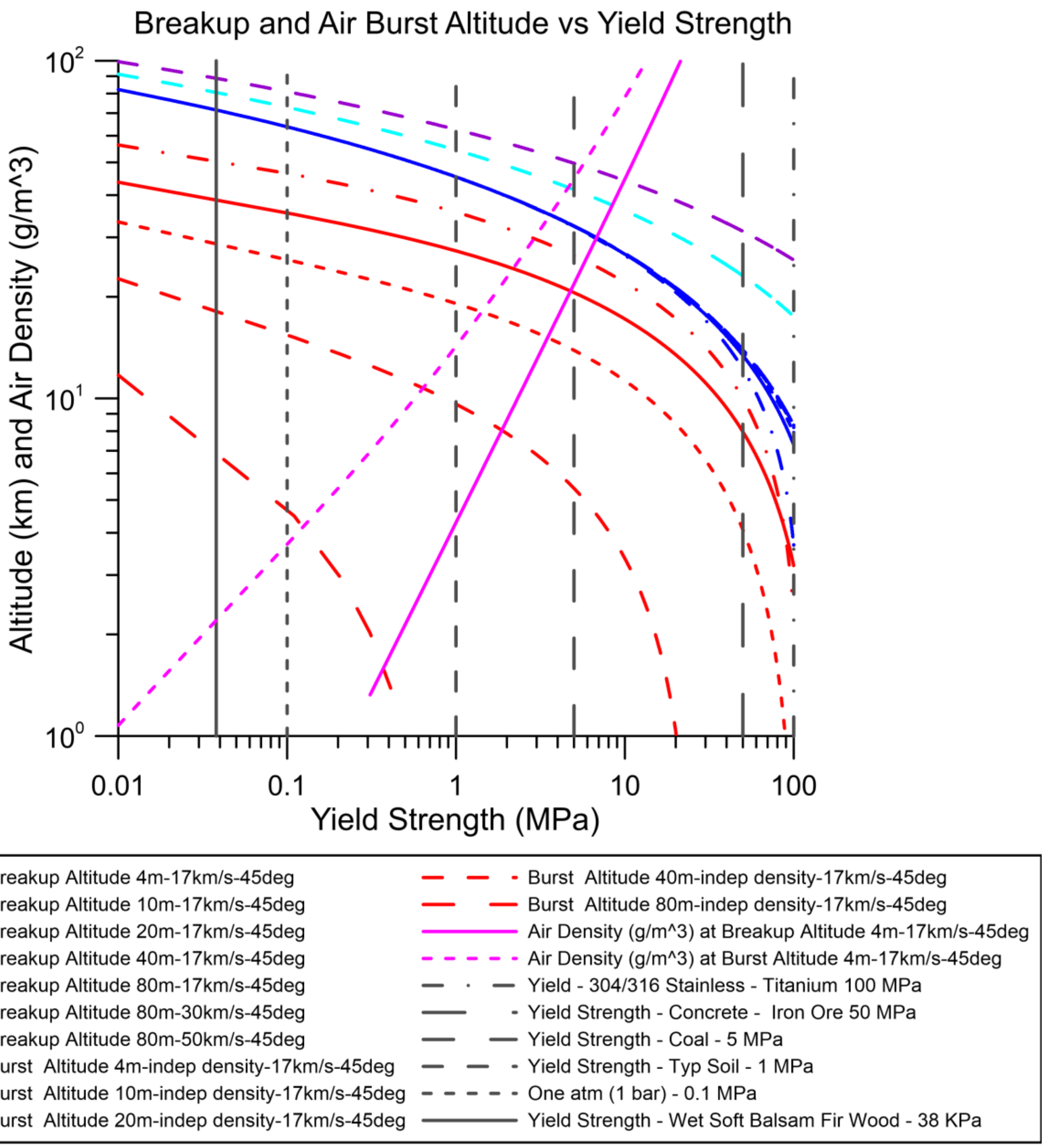

$$z_* = -H\left[\ln\left[\frac{Y_i}{\rho_0 v_0^{\,2}}\right] + 1.308 - 0.314 I_f - 1.303\sqrt{1-I_f}\right] \qquad I_f = 4.07\frac{C_D H Y_i}{\rho_i L_0 v_0^{\,2}\sin(\theta)}$$

$$z_b - z_* = -2H\ln\left[1 + \frac{l}{2H}\sqrt{f_p^{\,2}-1}\right] \qquad f_p = L(z_b)/L_0 = \mathit{breakup\ spread\ ratio}\,(\sim 7\ \mathit{works\ well})$$

$$z_* = \text{breakup altitude} \quad z_b = \text{burst altitude} \qquad (z_* > z_b)$$

$$l = L_0\sin(\theta)\sqrt{\frac{\rho_i}{C_D\,\rho(z_*)}} = \mathit{dispersion\ length} \qquad \rho_i = \left[\frac{\log_{10}(Y_i) - 2.107}{0.0624}\right]^2 = \mathit{impactor\ density}$$

**Figure 42**– Breakup and burst altitude vs asteroid yield strength for our model. Air density at breakup and burst is also shown. Comparison to various common material yield strengths is also shown. While there are no known bolides with large scale yield strength, we include high strengths for completeness. While some fragments may have high yield strength, our simulations show these can be fragmented sufficiently small as to be of little concern.

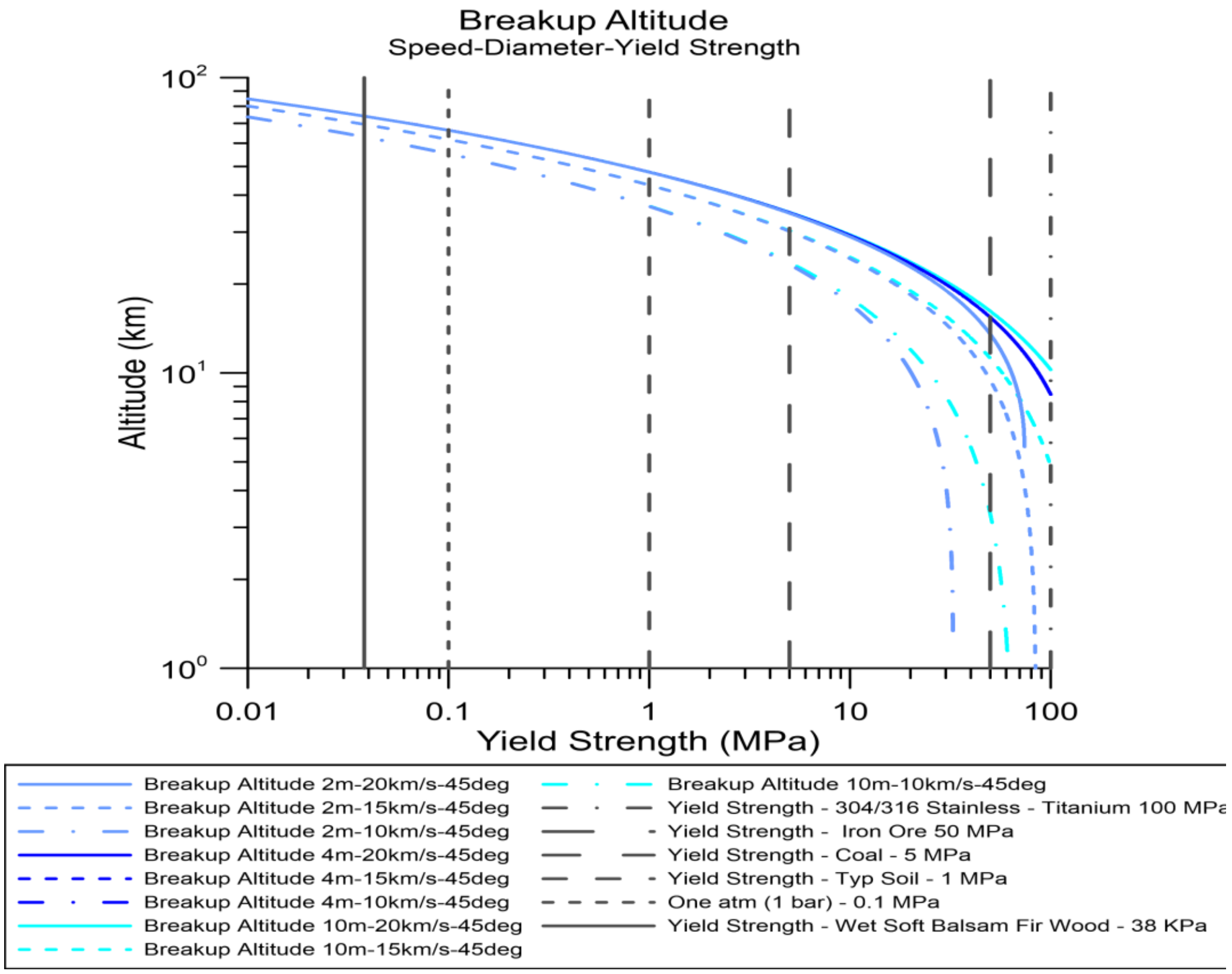

**Figure 43 –** Breakup altitude vs yield strength for 2,4 and 10m diam fragments. Note that slow very high yield strength fragments can penetrate the atmosphere.

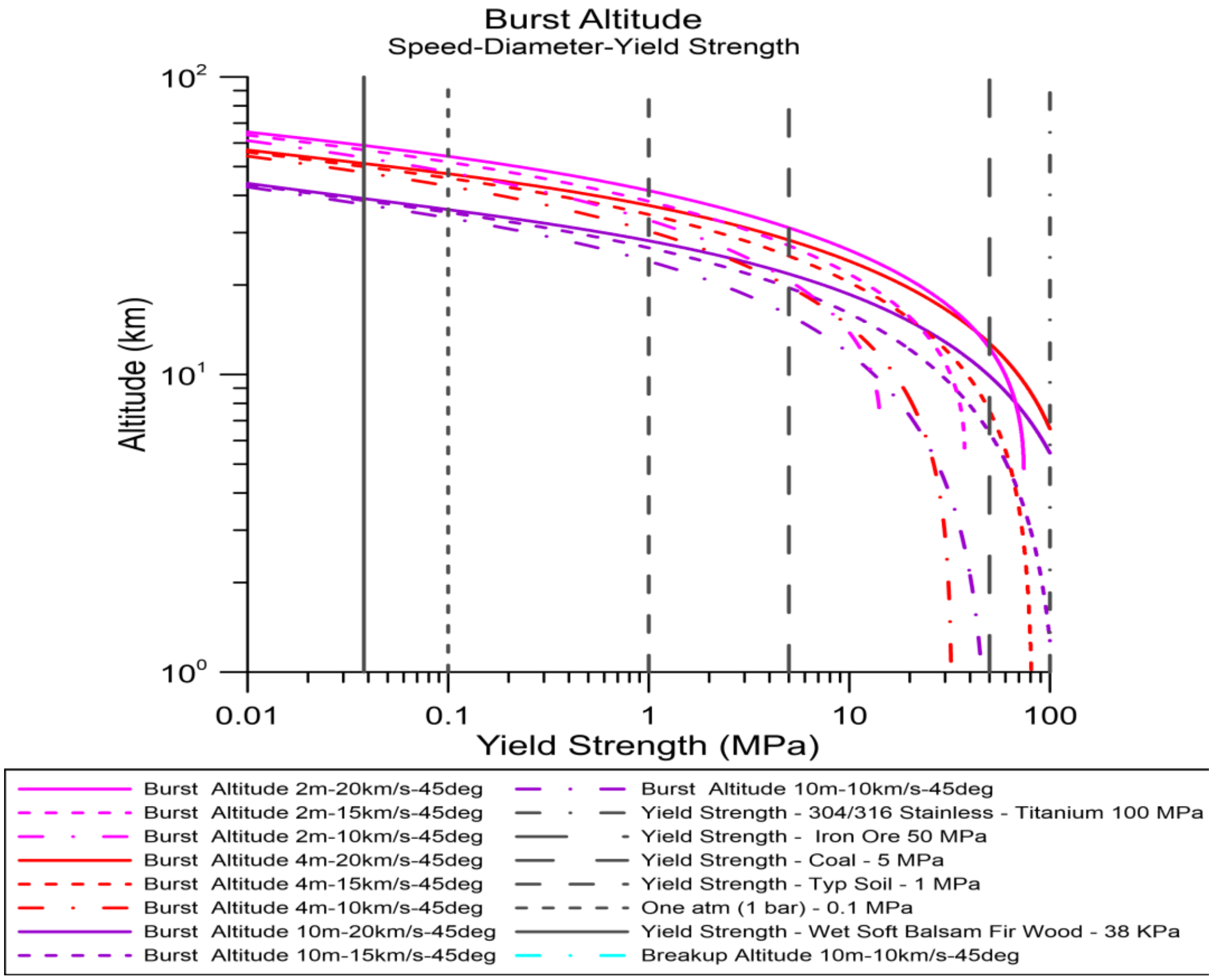

**Figure 44** – Burst altitude vs yield strength for 2,4 and 10m diam fragments. Note that slow very high yield strength fragments can penetrate the atmosphere due to low atmospheric ram pressure.

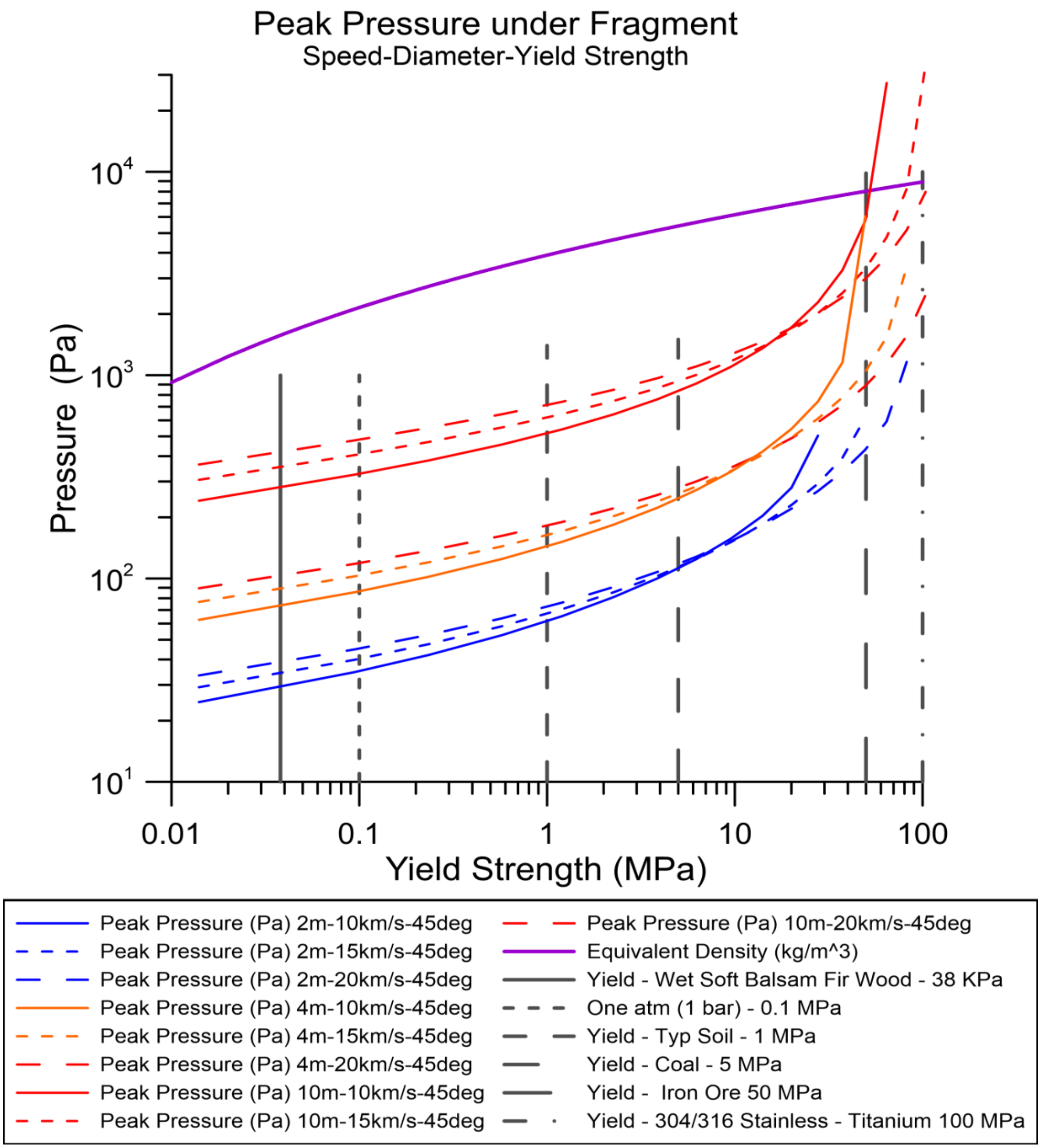

**Figure 45 –** Peak pressure directly under fragment (ground zero) vs yield strength, diameter and speed. Note the strong dependence of the peak pressure vs diameter and yield strength (which approximately correlates with density) as expected. Various common material yield strengths are shown for reference. The peak blast pressure is a function of both the fragment energy (speed and density dependent) but also the burst altitude above the ground with higher yield strength fragments penetrating deeper in the Earth's atmosphere but also related to exo-atmospheric fragment speed with a complicated relationship between speed, penetration depth and yield strength. Slower fragments can penetrate more deeply due to their smaller drag and smaller ram pressure. High yield strength slower fragments can penetrate very deeply which can cause higher peak pressures as they can burst closer to the ground. This is seen clearly for very high strength low speed fragments. Some very high strength slow speed fragments, which cause relatively little damage, can penetrate all the way to the ground. An example of this is the Hoba nickel-iron meteorite which is about 3x3x1m that is estimated to have hit the ground at about 300m/s having slowed down greatly in the Earth atmosphere and not having burst due to it high yield strength and apparently relatively low entry speed giving insufficient ram pressure for fragmentation to occur. For reference, both the 2013 Chelyabinsk and 1908 Tunguska events are estimated to have had yield strengths of approximately 1 MPa with less confidence for Tunguska.

***Intercept Safety Decision Logic –*** In running the simulations we implement a logic flow to test whether a simulation meets a safety metric that is based on an "acceptable level of ground level damage". We set this "safety metric" primarily based on two observables for all locations on the Earth's surface. These observables are discussed in more detail in the many sections below but they are summarized as:

1) Keeping the peak acoustical shock wave pressure below a level that breaks nominal residential size and thickness windows. We will assume an allowed peak pressure of ~2 kPa at all observers. Note that while preventing the breakage of glass is desirable, we consider a reasonable trade space to allow for some glass breakage.

2) Keeping the optical energy flux below the level that could start a fire from dry combustibles such as dry grass or loose paper. We will assume a maximum time integrated optical energy flux from ALL fragments summed of 0.2 $MJ/m^2$ at all observers. This is extremely conservative as we assume the optical energy from ALL fragments arriving from all entry times and all direction to be simply summed. We do compute the optical atmospheric absorption for every observer path and every fragment.

**Both of these are very conservative safety metrics. In both cases we propagate the acoustic and optical signature through the atmosphere and include atmospheric absorption for every fragment and every observer position. In the optical case we assume there is no cooling time between optical pulses (worst case).**
**We proceed in the decision-making process for any given threat as follows:**

1) Threat Detection and Trajectory Analysis – Determine if threat mitigation is warranted. Determine threat parameters

2) *Parent Threat Parameters*
$L_0, m_o, v_o, \langle \rho \rangle$

3) Parent Bolide - Impact Date, Time, Latitude, Longitude, Attack Angle

Intercept Parameters

4) $\tau$ (Time between intercept and impact)
$v_{ave}$ (Average frag dispersal speed)
$L_{\max}$ (Largest fragment Diam Accepted)
$N$(# fragments) = $\left(L_0 / L_{\max}\right)^3$

Fragment Properties

5)
$\bar{v}_o = speed\ frag\ cloud\ at\ impact - assume\ all\ frag\ same\ longitudinal\ speed$

$\bar{x}_{ast,i} = i_{th}\ frag\ position\ incl\ burst\ alt\ z_b$

$\theta_i = i_{th}\ frag\ attack\ angle$

$L_i = i_{th}\ frag\ diam$

$\rho_i = i_{th}\ frag\ density$

$t_i = i_{th}\ frag\ burst\ time$

$E_i = i_{th}\ frag\ entry\ KE$

$E_{i-acoustic} = i_{th}\ frag\ acoustic\ energy\ release$

$E_{i-optical} = i_{th}\ frag\ optical\ energy\ release$

Pressure from Acoustical Pulse (Blast Wave) and Optical Pulse at Observer

6)
$\bar{x}_{ast,i} = i_{th}\ frag\ position\ at\ burst\ z = z_b$

$\bar{x} = observer\ position$

$\bar{r}_{slant-i} \equiv (\bar{x}_{ast,i} - \bar{x}) = slant\ vector\ from\ i_{th}\ frag\ burst\ to\ observer$

$r_{slant-i} = |\bar{r}_{slant-i}| = slant\ dist\ from\ from\ i_{th}\ frag\ burst\ to\ observer$

$p_o(\bar{x}, \bar{x}_{ast,i}) = peak\ pressure\ at\ observer\ from\ i_{th}\ frag\ including\ atm\ acoustic\ absorption$

$E_{optical}(\bar{x}, \bar{x}_{ast,i}) = peak\ optical\ energy\ flux\ at\ observer\ from\ i_{th}\ frag\ including\ atm\ optical\ absorption$

$t_i(\bar{x}) = arrival\ time\ of\ blast\ from\ fragment\ i\ at\ observer\ position\ \bar{x}$

$t_{i-b}(\bar{x}_{ast,i}) = burst\ time\ of\ fragment\ i\ at\ position\ \bar{x}_{ast,i}$

$t_i(\bar{x}) = t_{i-b}(\bar{x}_{ast,i}) + |\bar{x} - \bar{x}_{ast,i}| / v_s$

$t_{1i}(\bar{x}) = "t_1"\ from\ Friedlander\ blast\ evolution\ eq\ for\ fragment\ i$

$at\ position\ \bar{x} = \text{zero crossing time for } p(t,\bar{x}), I(t,\bar{x}) = 0$

$(t_{1i}(\bar{x})\ is\ slightly\ dependent\ on\ i\ and\ \bar{x}\ due\ to\ blast\ energy\ and\ atm\ freq\ dependent\ absorption)$

$$p(t,\bar{x}) = \sum_{i=1}^{N} P_{0_i}(\bar{x}) e^{-(t-t_i(\bar{x}))/t_{1i}(\bar{x})} \left[(1-(t-t_i(\bar{x}))/t_{1i}(\bar{x}))\right] \quad for\ t \ge t_i(\bar{x})$$

$p(t,\bar{x}) = sum\ of\ all\ blast\ waves\ at\ observer\ (\bar{x}) - includes\ decorrelation\ from\ time\ evolution$

$$E_{opt-sum} \equiv \sum_{i=1}^{N} E_{optical}(\bar{x}_{obs}, \bar{x}_{ast,i}) = sum\ of\ optical\ energy\ flux\ from\ ALL\ fragments\ (worst\ case)$$

7) Check if intercept time before impact τ and intercept parameters (dispersion speed, number of penetrators, fragment size etc) allows all observers (all position on Earth surface) to be safe

- Check peak acoustic pressure $P_o$ at observer from each fragment is less than acceptable damage threshold (~ 2 KPa)

- Check total pressure from all fragments (P(t,x)) including time evolution decorrelation is less than acceptable damage threshold (~2-3KPa) for peak pressure

- Check $E_{opt-sum}$ summed peak optical energy flux at observer from ALL fragments is less than acceptable damage threshold 0.2 MJ/m$^2$)

**Decision – if peak pressure (P(t,x)) or sum of peak optical energy flux is not acceptable at any observer then increase $\tau * v_{avg}$ (make intercept earlier and/or dispersal speed larger) and repeat process until converged**

***Intercept time – Dispersal Speed product and Dispersal energy dependence* – For a given mitigation outcome in terms of the Earth surface affects, it is the product of the intercept time prior to impact $\tau$ and the dispersal speed $v_{avg}$ that determines the damage outcome to first order.** This is due to the fact that the fragment cloud radius $\mathbf{R_{cloud} = \tau * v_{avg}}$ is what determines both the acoustic and optical damage to first order with larger radius being desirable. We can increase the fragment could radius by either increasing the time $\tau$ and/or increasing the dispersal speed $v_{avg}$. Since the dispersal kinetic energy scales as $v_{avg}^2$ there is a dispersal energy penalty for increasing $v_{avg}$ while increasing $\tau$ requires us to launch earlier or go faster to the target. There is a trade space here that is target and situation dependent.

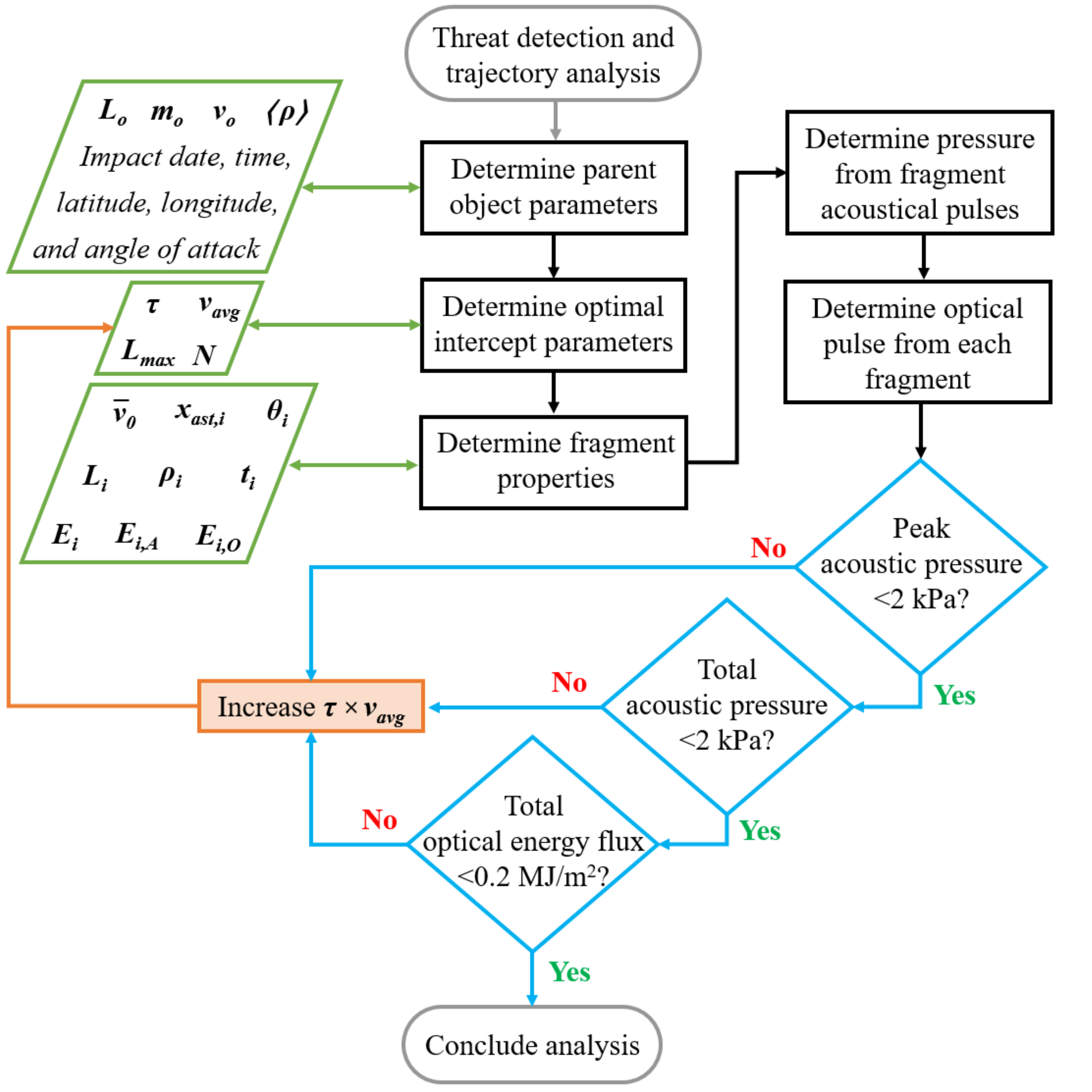

**Figure 46**– Threat Mitigation Decision Diagram from threat detection to successful mitigation with a threat metric being both acoustic and optical signature thresholds.

***Smaller Fragment Size is Better –*** Due to the strong connection between fragment size and the peak blast pressure at "ground zero" (directly below the fragment burst), the minimum damage on the ground for a given parent bolide is when the fragments are as small as possible. There is a practical trade-off between achieving small fragment size, the disruption strategy, and the energy required. Our approach in this paper is to adopt a conservative strategy. We have run a large number of simulations over a wide range of fragmentation strategies and generally adopt a very conservative mean fragment diameter of 10m for stony bolides, which yields an acceptable mitigation strategy. We show this vividly below where we compare a 5m and 10m mean fragment size with sigma indicated. In these histograms, the blast pressure is the maximum pressure right under the fragment (ground zero – worst case). **This is a "zero time" intercept with no spatial spread, and thus it represents the worst case possible as the observer is always directly under each fragment**. Even so, in this case the peak blast pressure is below 2kPa for all fragments. At 2kPa peak pressure, some residential windows may break, but large-scale blast damage is avoided. If we fragment to a 5m mean diameter instead of 10m mean diameter, then the peak blast pressure is less than 1kPa everywhere, and thus virtually no window breakage should occur. **However, as ground and building reflections of the acoustical wave do occur, there are some limited sites where constructive interference may cause window breakage.**

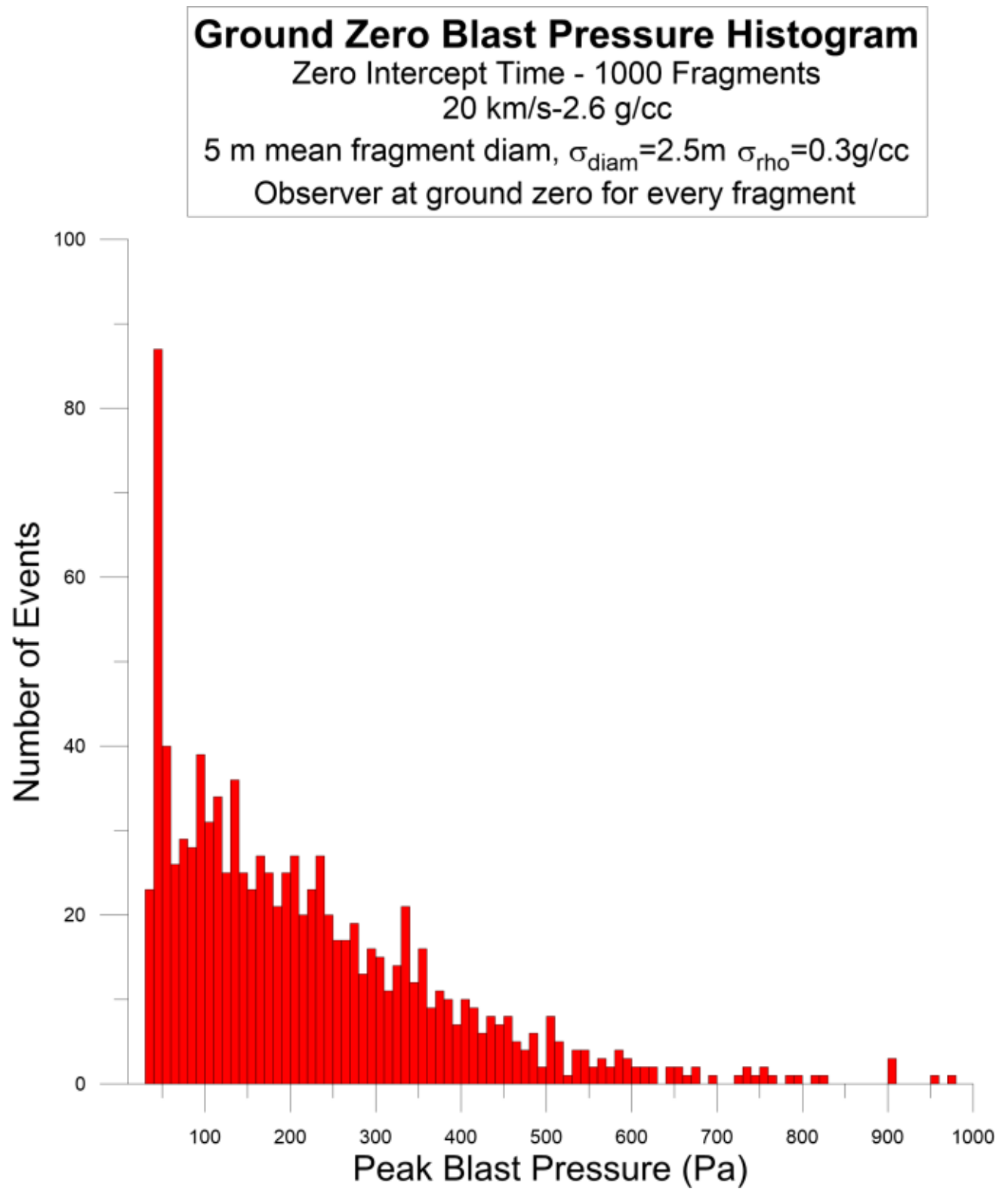

**Figure 47 –** Histogram of ground zero peak pressure for 1000 fragments with mean diameter of 5m at 20km/s. The blast wave is below 1kPa for all fragments thus mitigating any serious damage. **Angle of attack is 45 deg unless specified.**

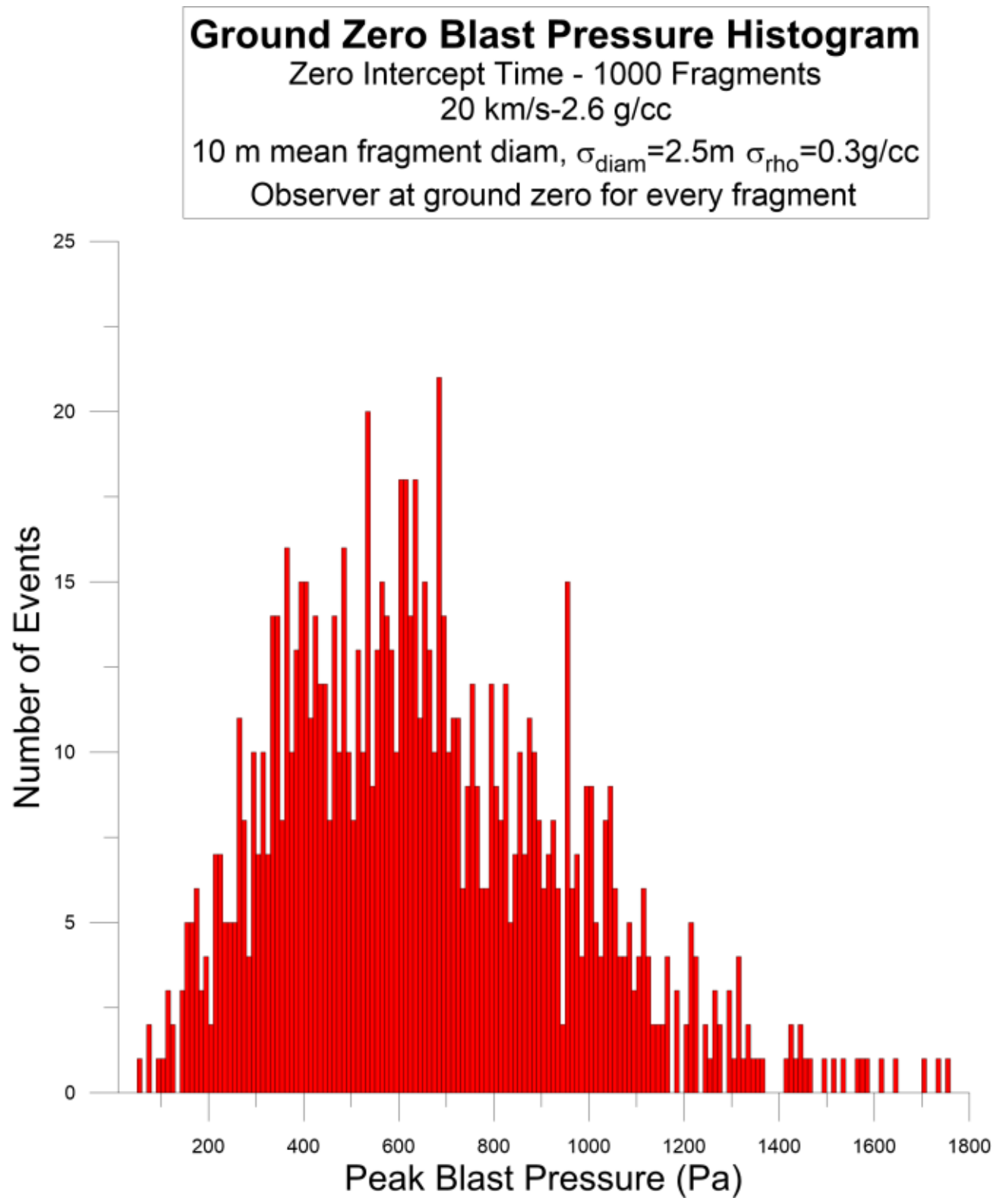

**Figure 48 –** Histogram of ground zero peak pressure for 1000 fragments with mean diameter of 10m at 20km/s. The blast wave is below 2kPa for all fragments mitigating any serious damage though some windows may break.

***Shock wave pressure and travel time vs observer position*** – In preparation for the computation of multiple fragment shock waves, we first compute the shock wave and travel time for any position on the ground. In the example below, we compute the maximum blast pressure and travel time for a 10m diameter fragment with density 2.6g/cc and an angle of attack of 45 degrees, for an observer grid of ±100km in $x$ and $y$. The asteroid/fragment bursts at approximately 33km altitude at $x=y=0$. Entry speed is 17 km/s.

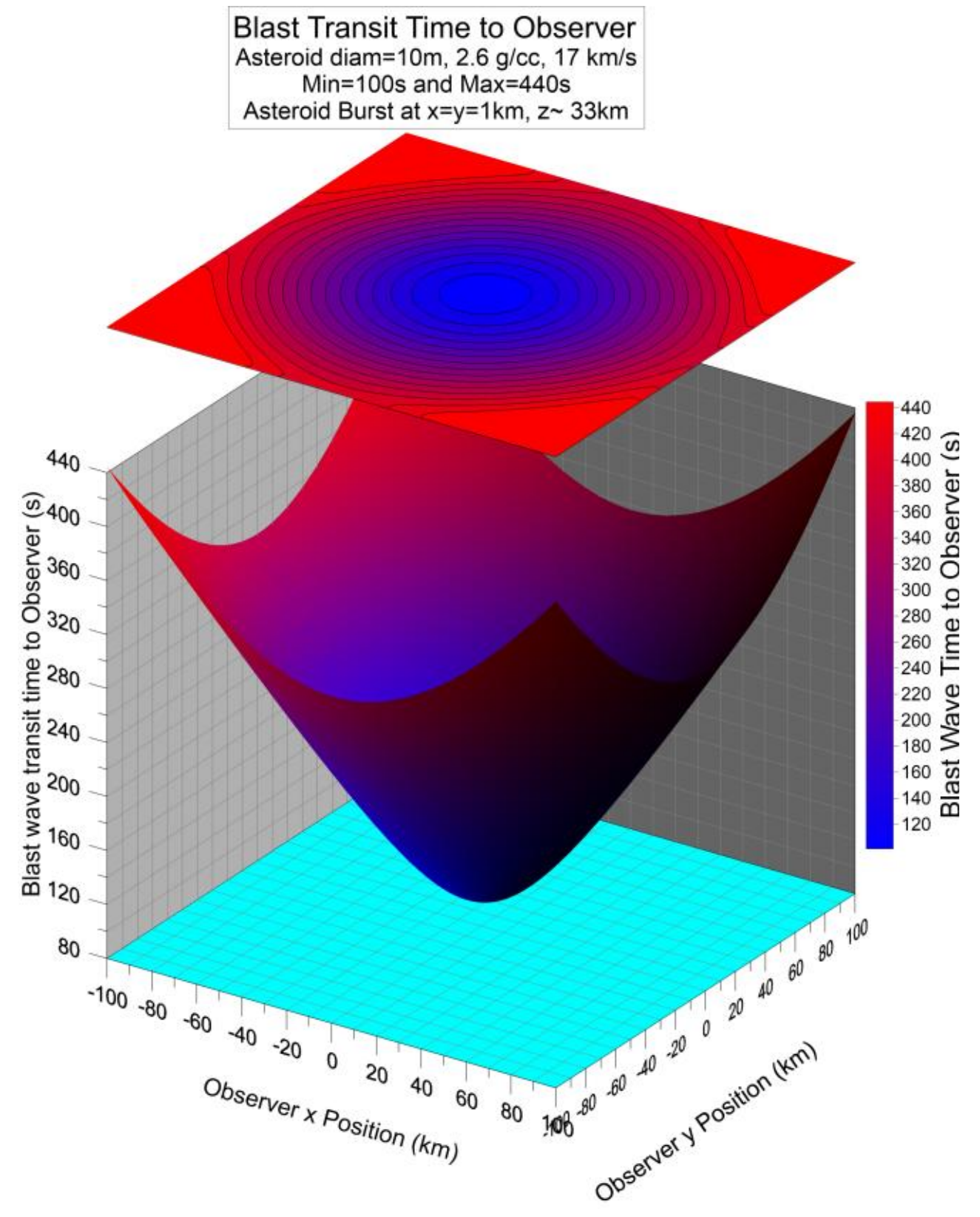

**Figure 50** – Shock wave transit time from burst to observer vs observer position on a 200x200km ground grid for a 10m diameter asteroid/fragment at 17km/s and density 2.6g/cc.

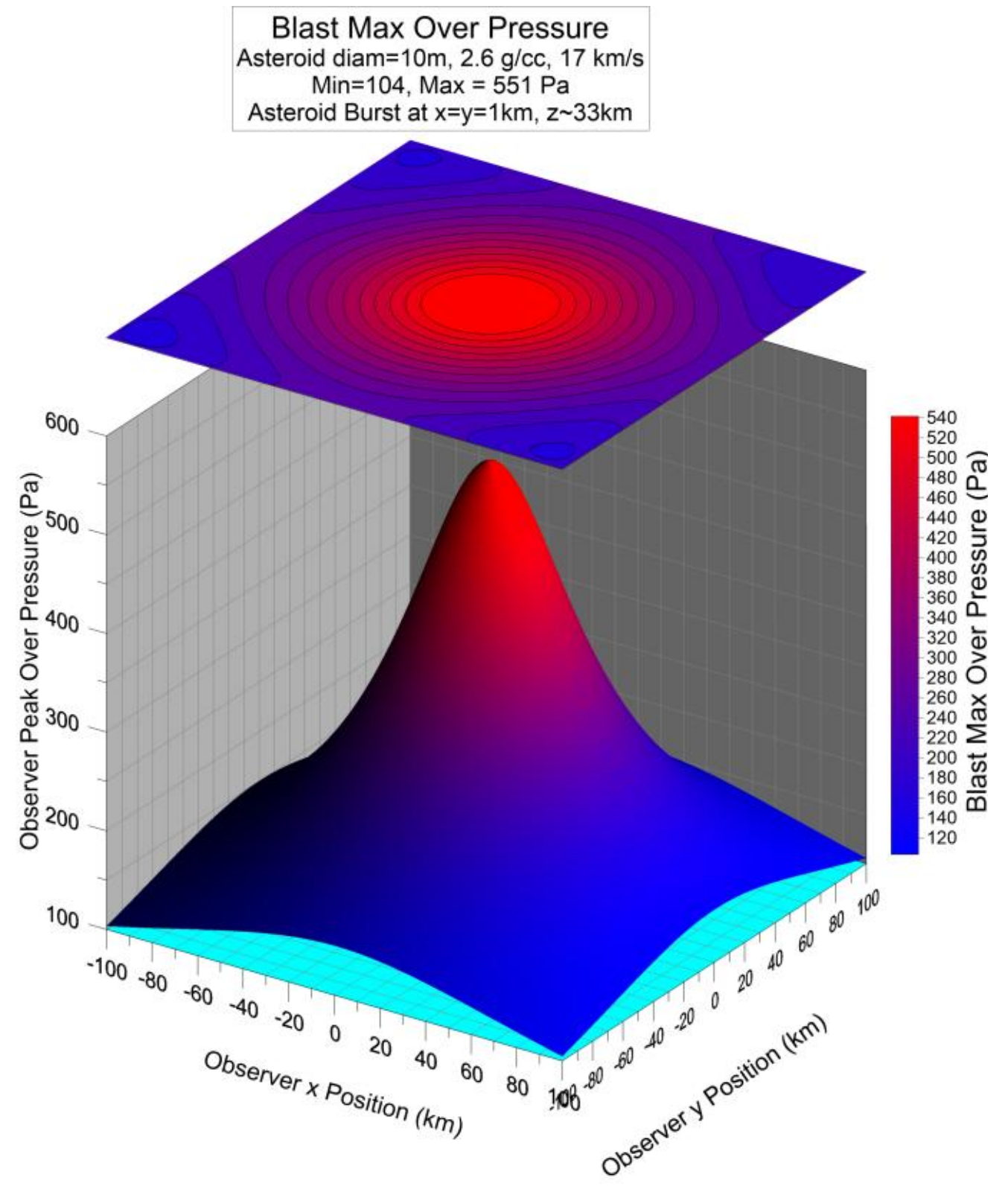

**Figure 49**– Shock wave max pressure vs observer position on a 200x200km ground grid for a 10m diameter asteroid/fragment at 17km/s and density 2.6g/cc.

***Target Interception Distance and Fragment Cloud Spread vs Intercept time prior to Impact*** – We compute the intercept distance vs time of intercept prior to impact for bolides of various speeds from 10 to 60km/s. We then compute the fragment cloud diameter upon arrival at the Earth for fragment dispersal speeds of 1, 3, and 10m/s. The fragment dispersal is key to the success of the program.

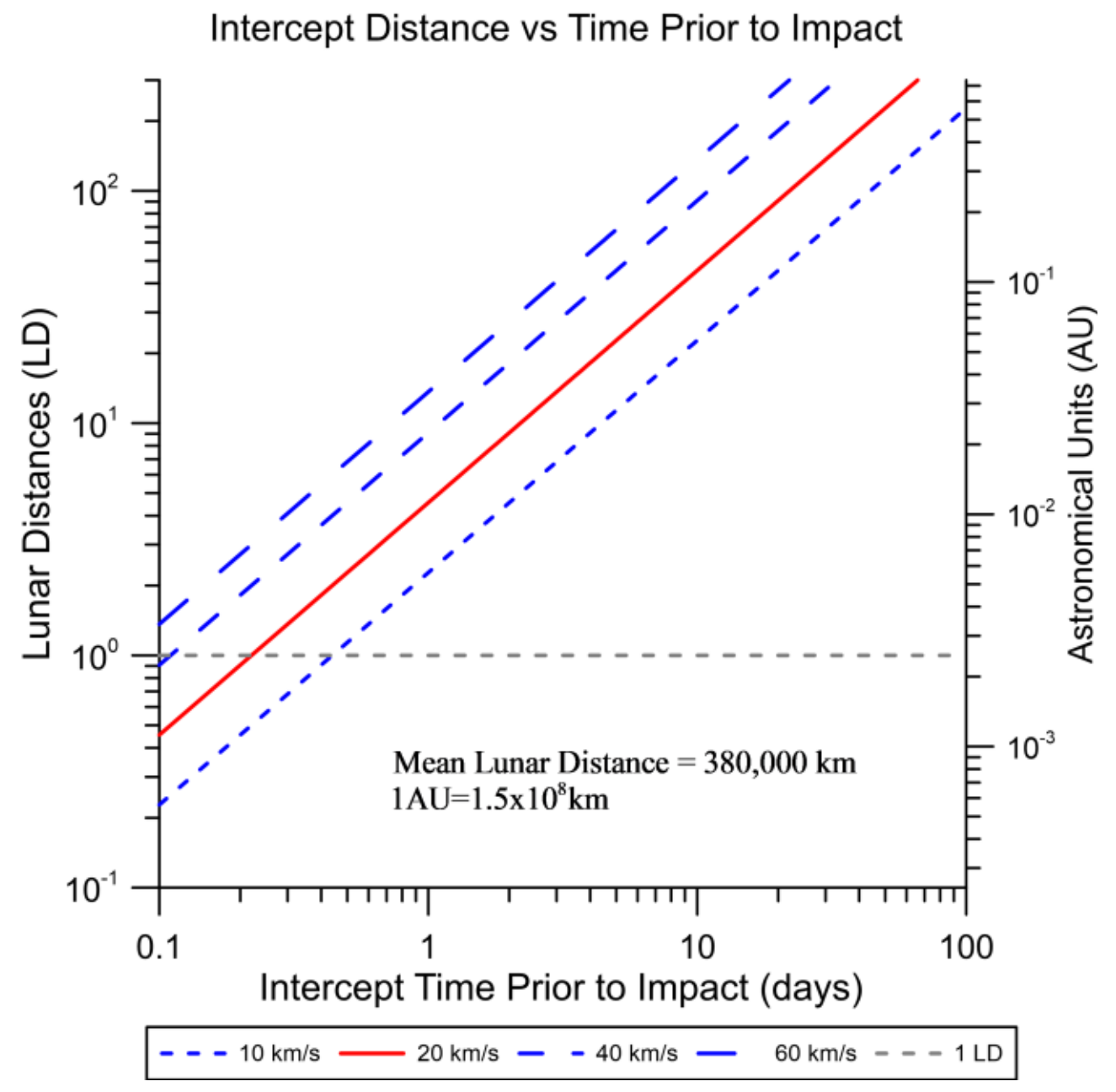

**Figure 51**– Distance to intercept vs intercept time prior to impact.

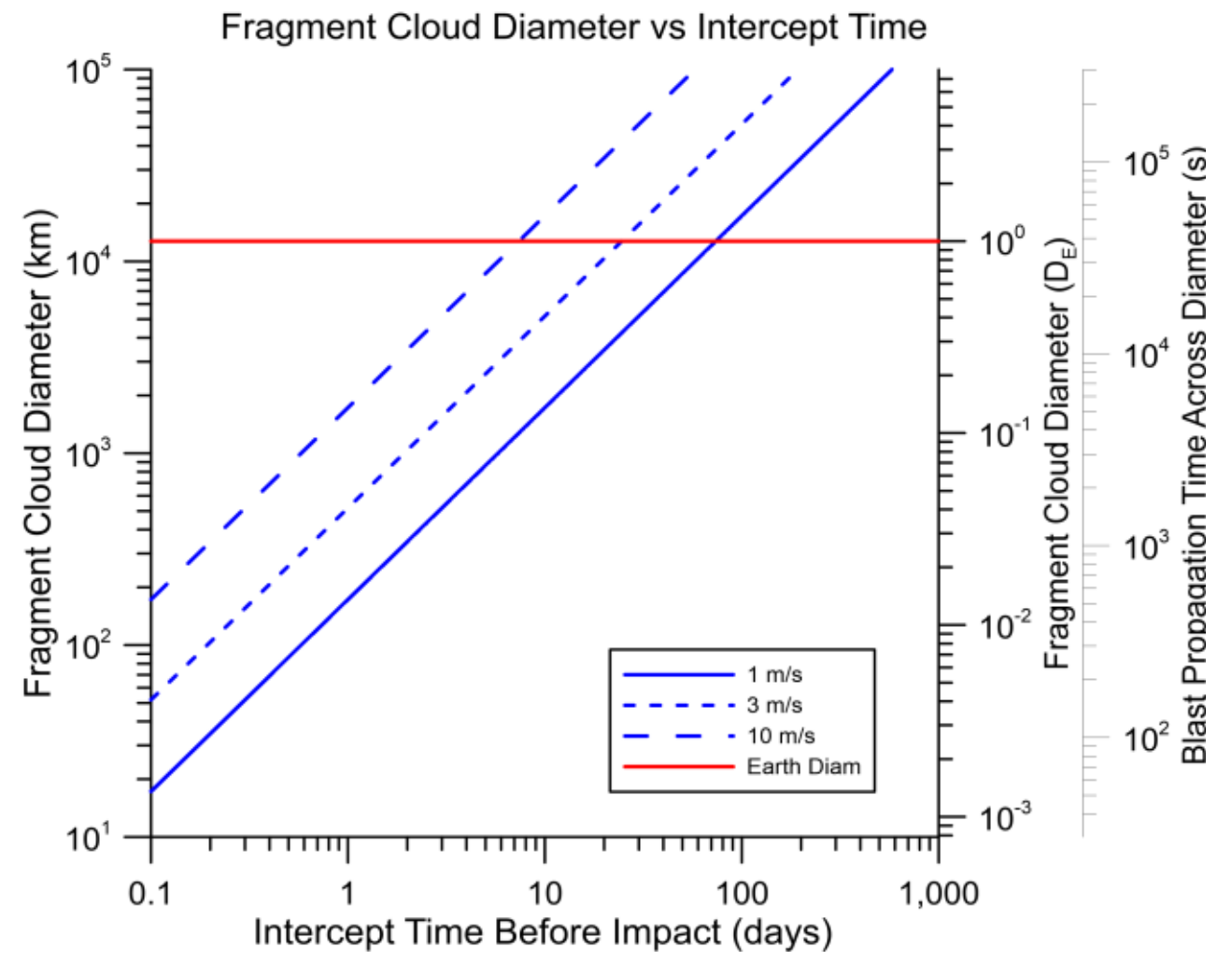

**Figure 52**– Fragment cloud diameter vs time before impact assuming the fragments have a 1m/s average speed relative to the parent asteroid center of mass.

***Mitigation of 100m diameter Bolides*** – A 100m diameter asteroid would cause significant damage to a populated area if struck. In general, a 100m diameter asteroid that was intact would hit the ground and cause a very large area of destruction. At a typical 20km/s, the energy of such a bolide would be of order 100Mt. For comparison, the largest thermonuclear weapon ever tested was the *Czar Bomba*, with a yield of around 50Mt. We show the results of a number of simulations using our method to fragment the parent bolide into 1000 fragments, each of which has an average diameter of 10m. The results below show one of these simulations where we have taken an extreme short warning time with an intercept of 100ksec (1.2 day) prior to Earth impact. The simulation extends out ±180km in *x* and *y*, where the origin is where the asteroid would have struck if it were not mitigated. We can place the observer at any position, but for this simulation we placed the ground observer in the worst position, namely right under the fragment "ring" at *x*=100km, *y*=0km. In this simulation, we assume a nominal incoming speed of 20km/s, a mean density of 2.6g/cc, and a mean fragment dispersal speed of 1m/s. We model a heterogeneous target by introducing density variations with σ=0.3g/cc. We also allow for a large range of fragments with σ=5m around a mean of 10m, truncated on the "small side" at 2m. We also introduce dispersion in the fragment speed with σ=0.3m/s. All of these are free parameters allowing us to rapidly run any reasonable scenario. The plots below show the fragment cloud distribution in space upon entry into the Earth's atmosphere, the distribution in fragment diameters, blast wave pressure vs arrival time distribution for all 1000 fragments, and the blast wave pressure vs time for the first five blast waves to arrive at the observer. We chose a simulation to show a relatively rare event in overlapping blast waves from different fragments, as is seen in the first two blast waves to "hit" the observer. Note that even in this relatively extreme mitigation of 1.2 days prior to impact, there is virtually no damage as very few of the blast waves at the observer are above 1kPa where residential glass breakage begins. When we include blast wave caustic formation we find some small fraction of cases with overlapping "2 point caustics" with pressures that are about 2x higher in these small fraction of spatial locations.

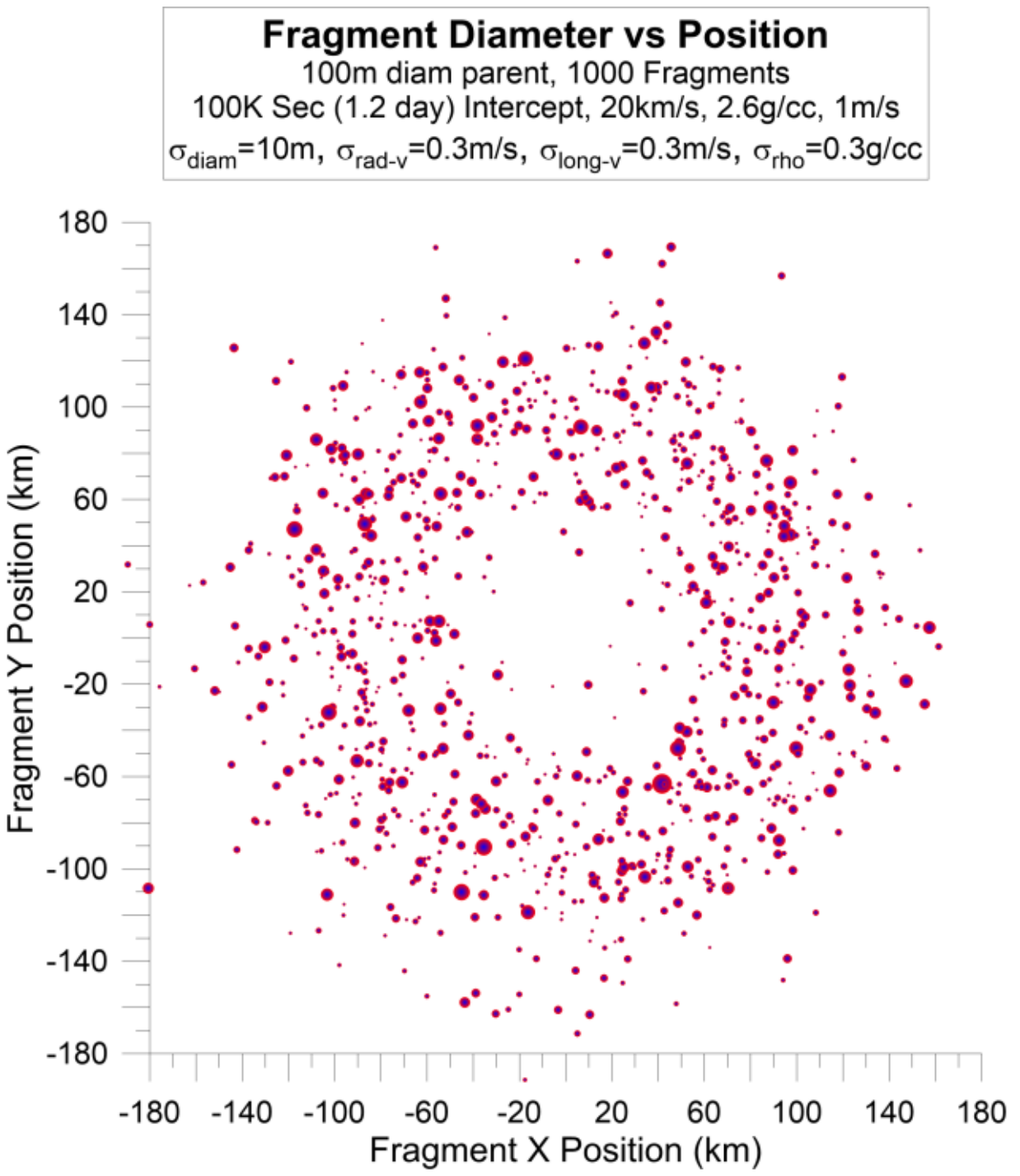

**Figure 53**– Fragment positions and diameters for a 100m diameter parent asteroid fragmented into 1000 pieces (10m diam. avg.) traveling at 20km/s intercepted 100ksec (1.2 days) prior impact. Fragments have a 1m/s average speed relative to the parent asteroid center of mass.

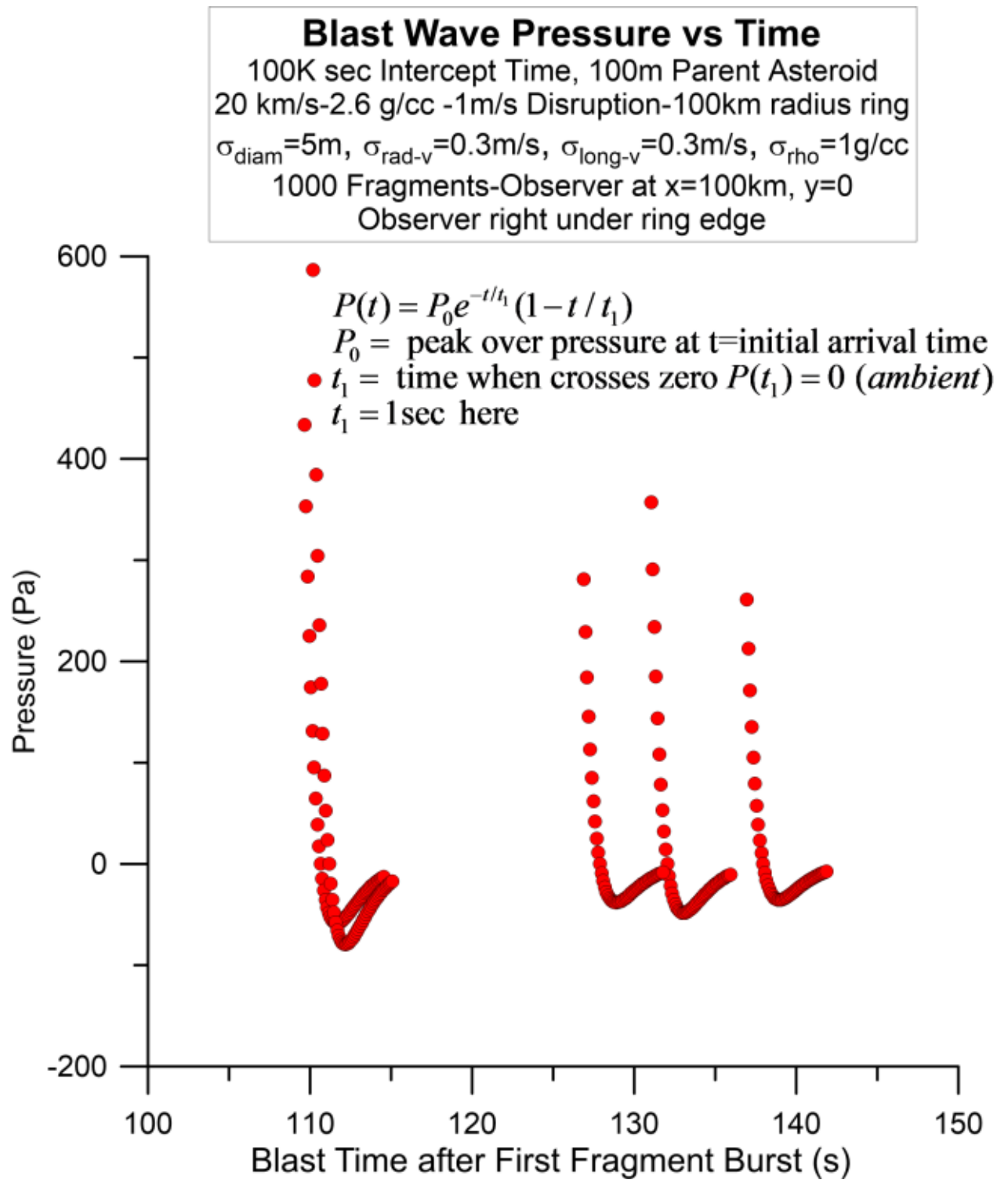

**Figure 54**– Fragment shock wave pressure vs time for a 100m diameter parent asteroid fragmented into 1000 pieces (10m diam. avg.) 100ksec (1.2 days) prior to impact. Fragments have a 1m/s average speed relative to the parent asteroid center of mass. First five waves shown. Worst case with observer right under ring.

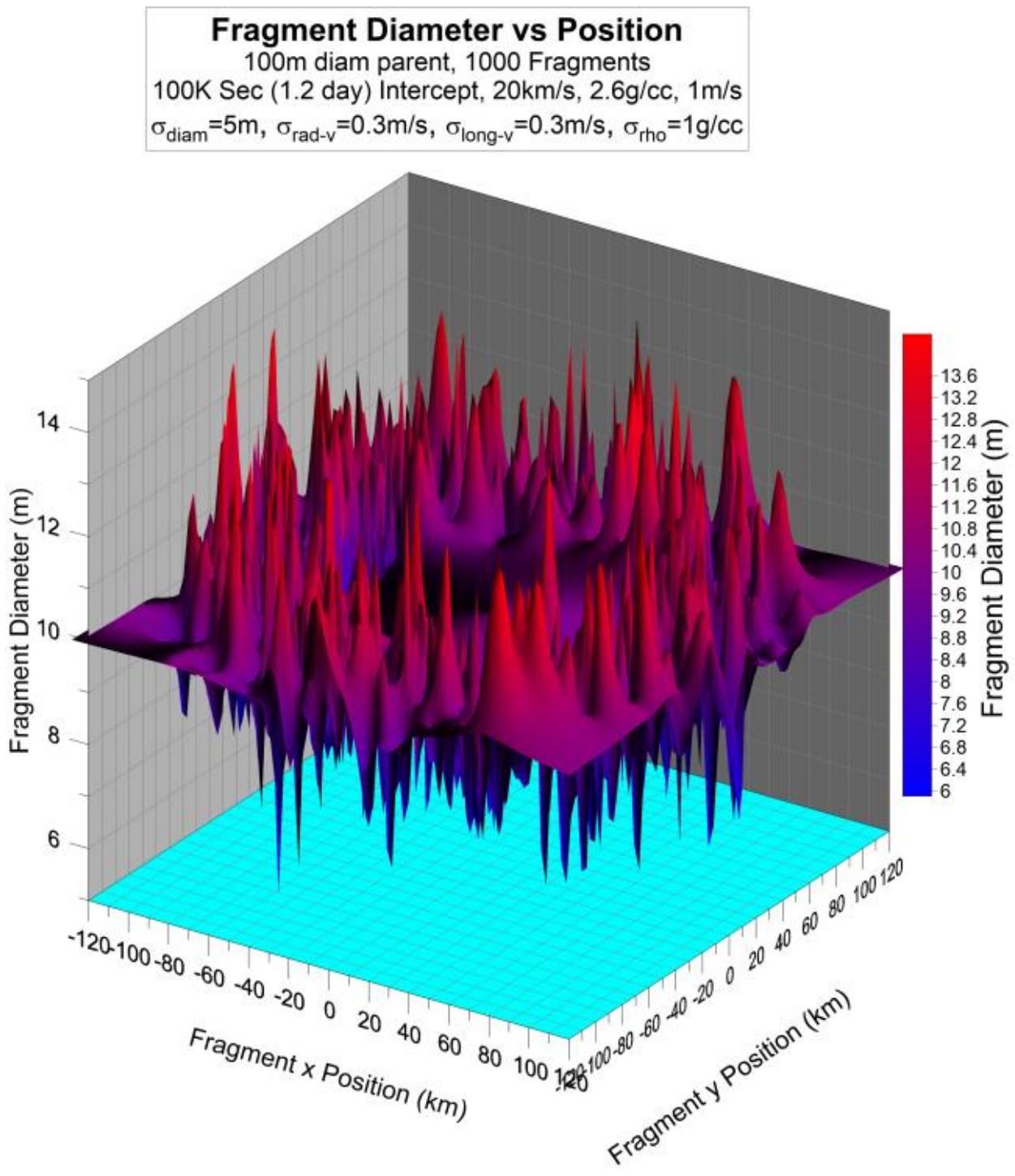

**Figure 55**– Fragment diameters and positions for a 100m diameter parent asteroid fragmented into 1000 pieces (10m diam. avg.) traveling at 20km/s intercepted 100ksec (1.2 days) prior to impact. Fragments have a 1m/s average speed relative to the parent asteroid center of mass.

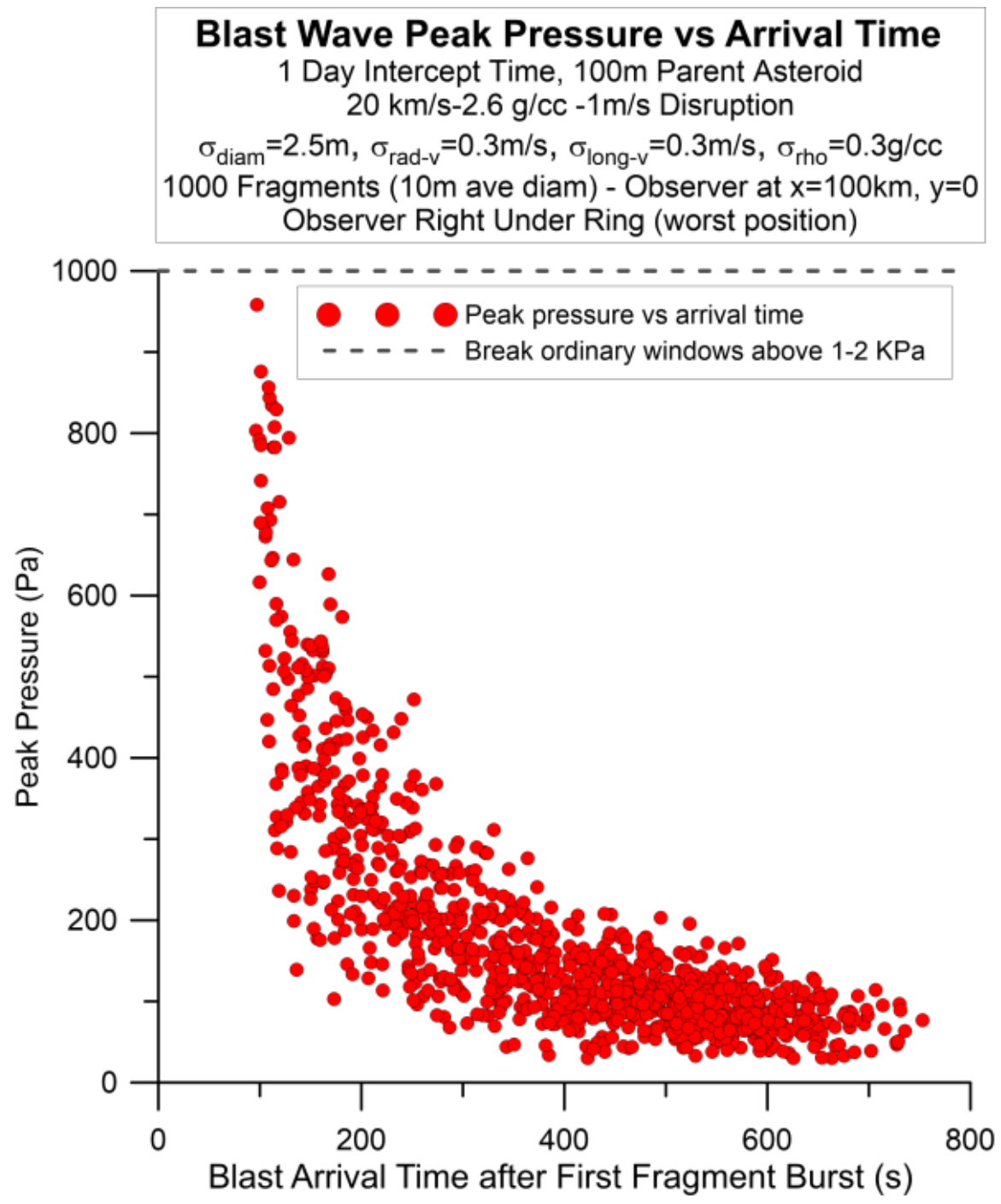

**Figure 56**– Fragment shock wave pressure distribution vs blast arrival time for a 100m diameter parent asteroid fragmented into 1000 pieces (10m diam. avg.) 1 day prior to impact. Fragments have a 1m/s average speed relative to the parent asteroid center of mass. Worst case with observer right under ring.

***Mitigation of Apophis*** – We chose the specific case of Apophis as it is a potential Earth hazard. Apophis has an orbital period of about 0.9 years and frequently comes close to the Earth. The next "near miss" of Apophis is expected to be Friday April 13, 2029. It is estimated to come within 31,000km of the Earth's surface, which is closer than the geosynchronous satellite belt. This is truly a close call. Other encounters in 2036 and 2068 will miss the Earth. Apophis (99942) has an estimated mean diameter of 330-370m and a density of approximately 2.6-3.2g/cc and an estimated mass of $m=6x10^{10}$kg . The equivalent diameter d for a solid sphere is $d=(6m/\pi\rho)^{1/3}$ or 350m for a density of 2.6g/cc if we use the estimated mass of $6x10^{10}$kg. We assume a conservative 350m diameter with density 2/6g/cc in general when we use the term "Apophis". We should gain much more information about Apophis this decade when it passes close by. The kinetic energy of Apophis depends on the particular pass and orbital phase of both it and the Earth it makes. Typical known orbital information yields an "impact speed" of 11-18 km/s IF it were to impact Earth. A reasonable conservative estimate is to assume about 20km/s which gives a KE of about 3Gt. To be extremely conservative in some of our calculations, we will also assume 4Gt total yield later in the paper. For reference, the Earth's total arsenal of nuclear weapons is approximately 6Gt (including both deployed and not deployed – at peak in ~1960 it was over 20Gt), so a direct hit by Apophis would be a "bad day". We show the results below of an extreme case of a 1 day prior to impact mitigation, as well as a 10 day prior to impact with the same basic parameters as for the 100m case, but we break Apophis into 30,000 fragments with a mean size of 12m diameter. We also introduce parameter perturbations in density, fragment speed, and fragment size to give a more realistic simulation. A one day prior to impact strategy is a really bad idea here, though the blast wave distribution shows relatively minor damage, but the optical signature discussed in later sections could be a hazard. The case for 10 and 30 days prior to impact are acceptable in both blast wave and optical signature mitigation with 30 days or longer being preferred for an object like Apophis. What is truly remarkable about these results is that even large asteroids like Apophis could be mitigated by PI with relatively short intercept times (10 day for the Apophis example). In addition, as discussed below, frequent Earth-crossing asteroids provide ample "target practice" for testing the system. It is best to start with smaller targets first, however!

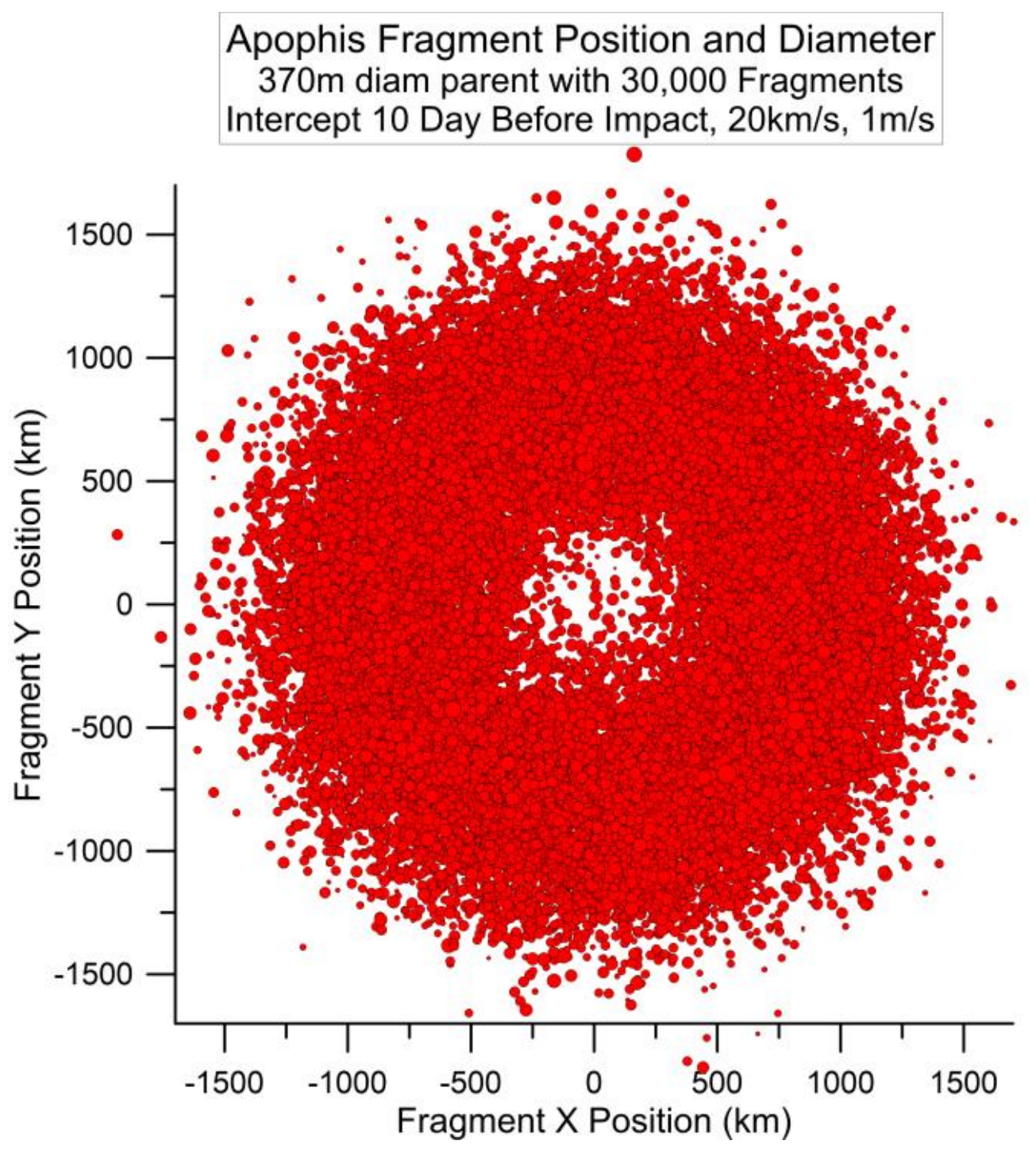

**Figure 60**– Fragment positions and diameters for Apophis fragmented into 30,000 pieces (12m diam. avg.). 20km/s intercepted 10 days prior to impact. Fragments have a 1m/s average speed relative to the parent asteroid center of mass.

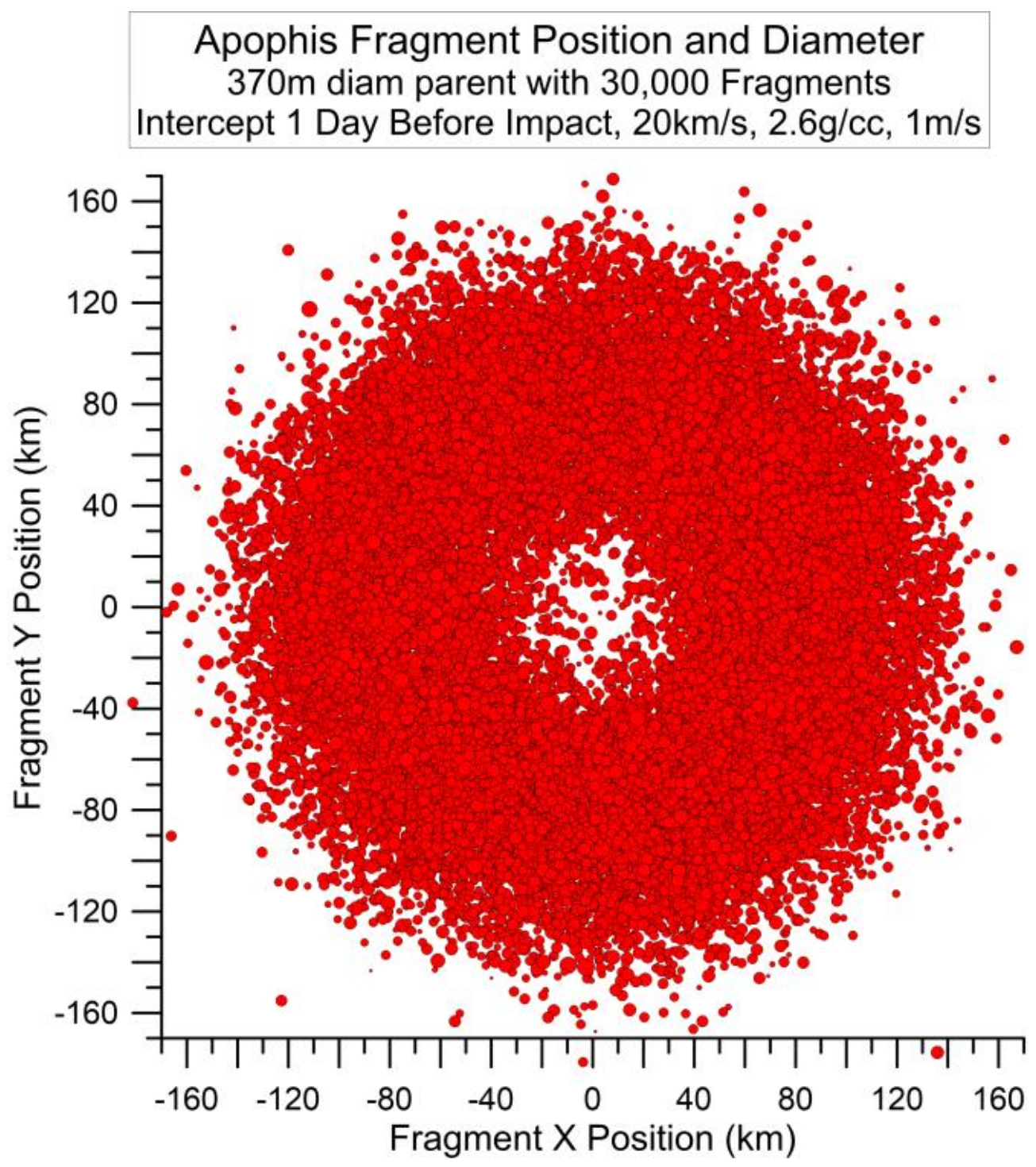

**Figure 58**– Fragment positions and diameters for Apophis fragmented into 30,000 pieces (12m diam. avg.), 20km/s intercepted 1 day prior to impact. Fragments have a 1m/s average speed relative to the parent asteroid center of mass.

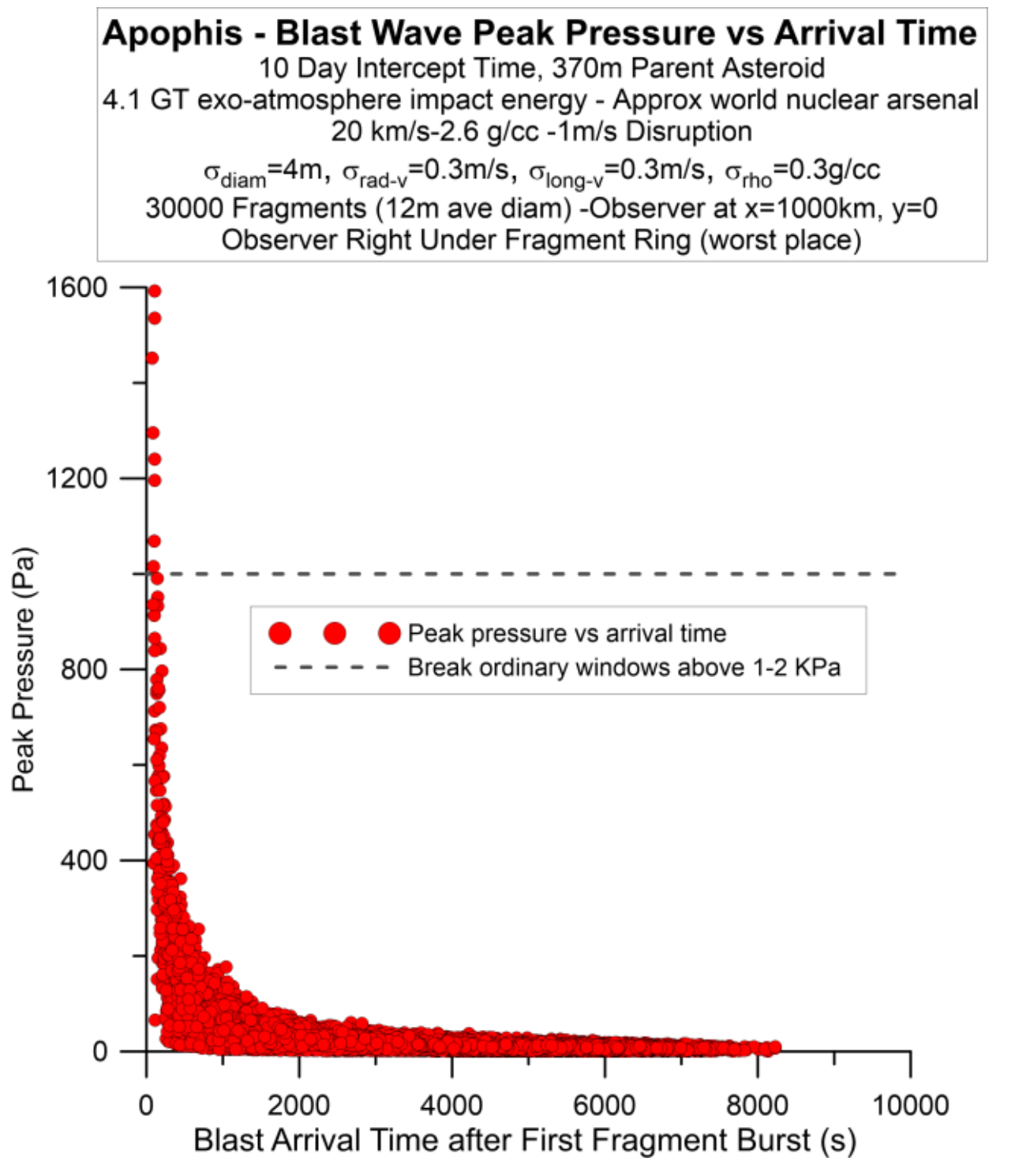

**Figure 59**– Fragment blast wave pressure distribution vs blast arrival time for Apophis fragmented into 30,000 pieces (12m diam. avg.) 10 day prior to impact. Fragments have a 1m/s average speed relative to center of mass.

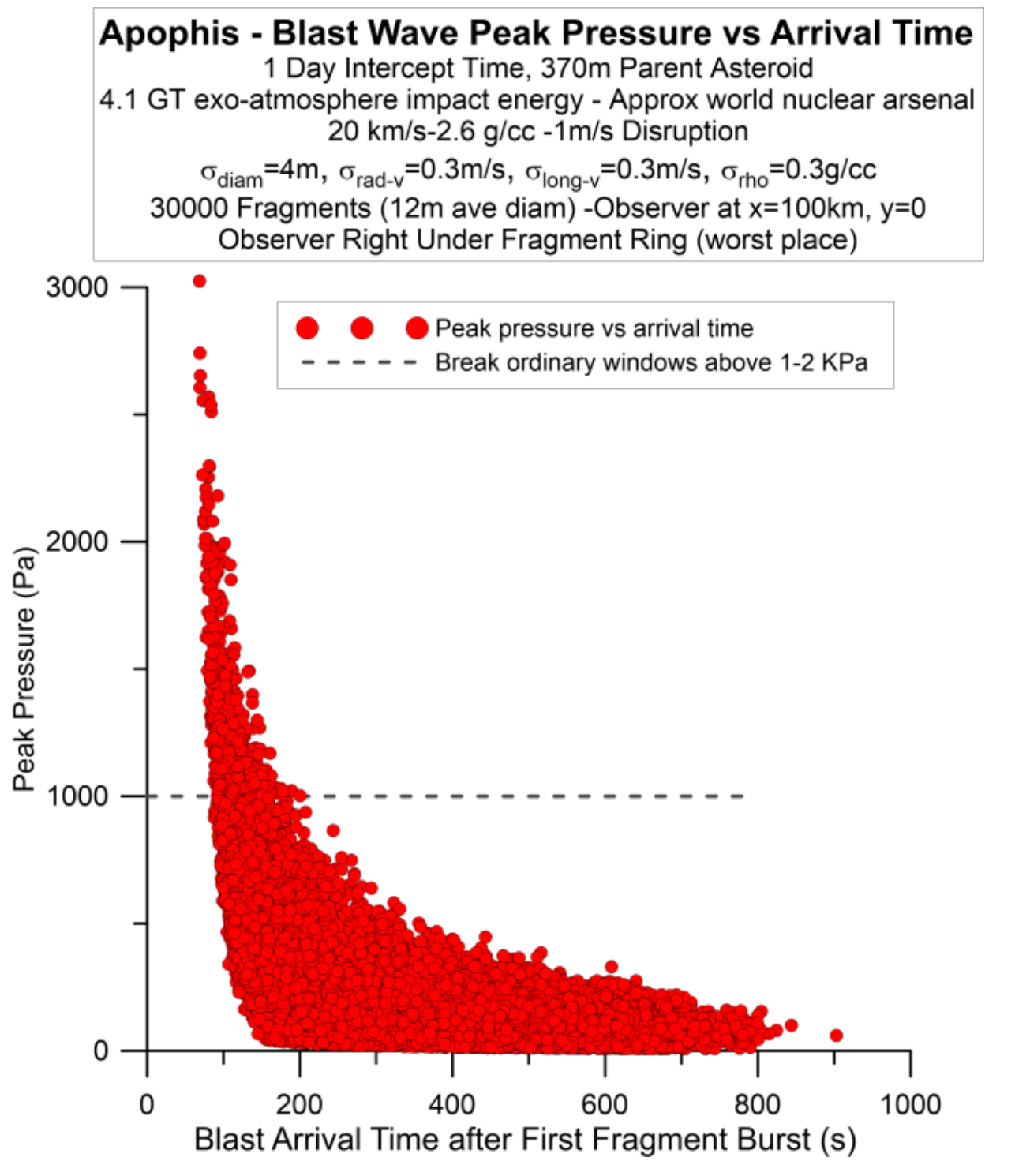

**Figure 57-** Fragment blast wave pressure distribution vs blast arrival time for Apophis fragmented into 30,000 pieces (12m diam. avg.) 1 day prior to impact. Fragments have a 1m/s average speed relative to center of mass.

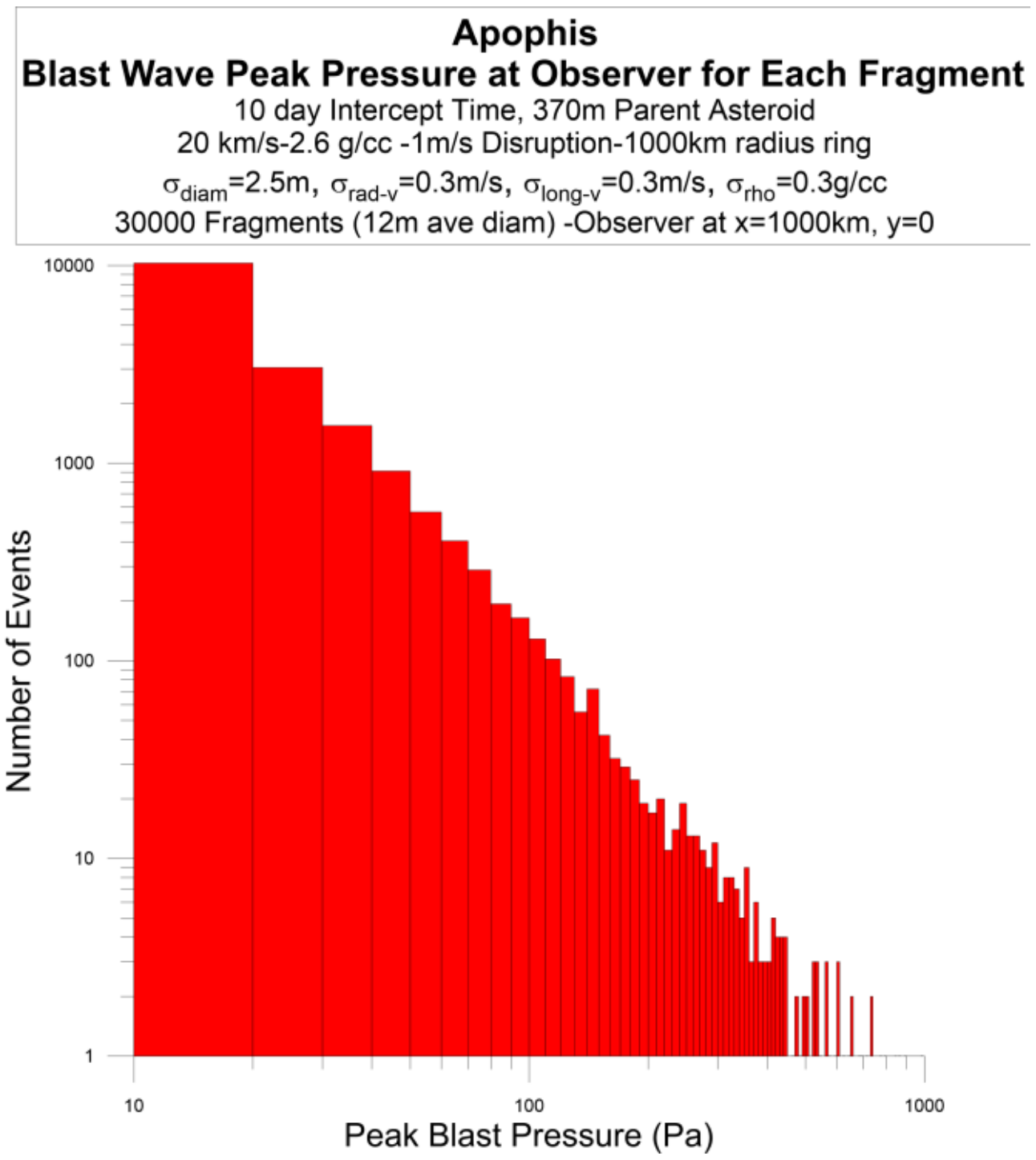

**Figure 63 –** 10-day intercept of Apophis in 30,000 fragments. Histogram of peak pressure from each of the fragments at the observer which is under the fragment cloud ring (worst position) at x=1000km. 30,000 fragments with 12m diam. avg. and 20km/s. Fragments have a 1m/s average speed relative to the Apophis center of mass. The blast waves at the observer are generally very small due to the large spread of the fragment cloud for a 10-day intercept**.**

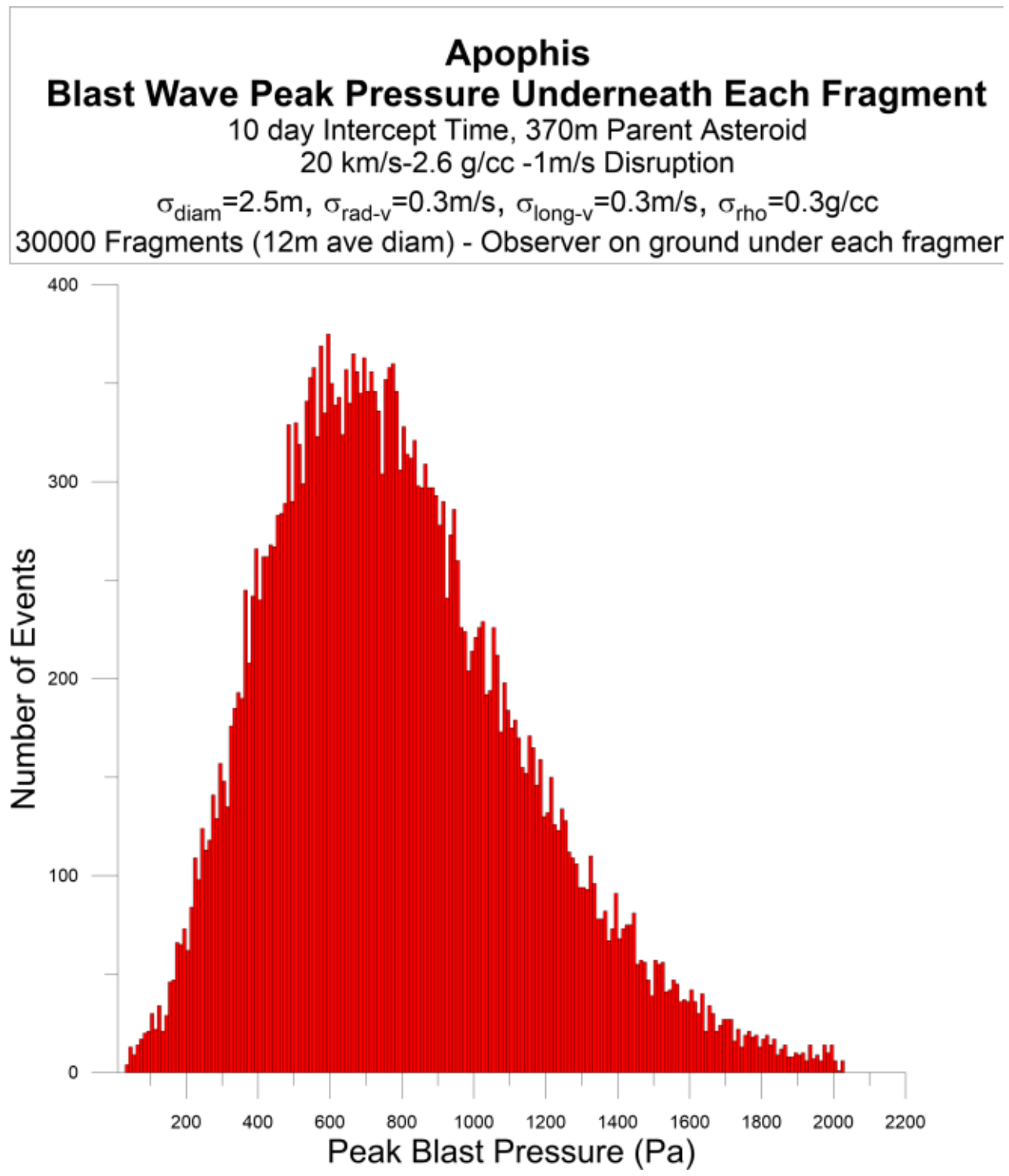

**Figure 62**– 10-day intercept of Apophis in 30,000 fragments. Histogram of peak pressure underneath each of the fragments (worst place). 30,000 fragments with 12m diam. avg. and 20km/s. Fragments have a 1m/s average speed relative to the Apophis center of mass. Effectively this shows what 30,000 observers, each directly under each fragment, would experience from "their fragment"/.

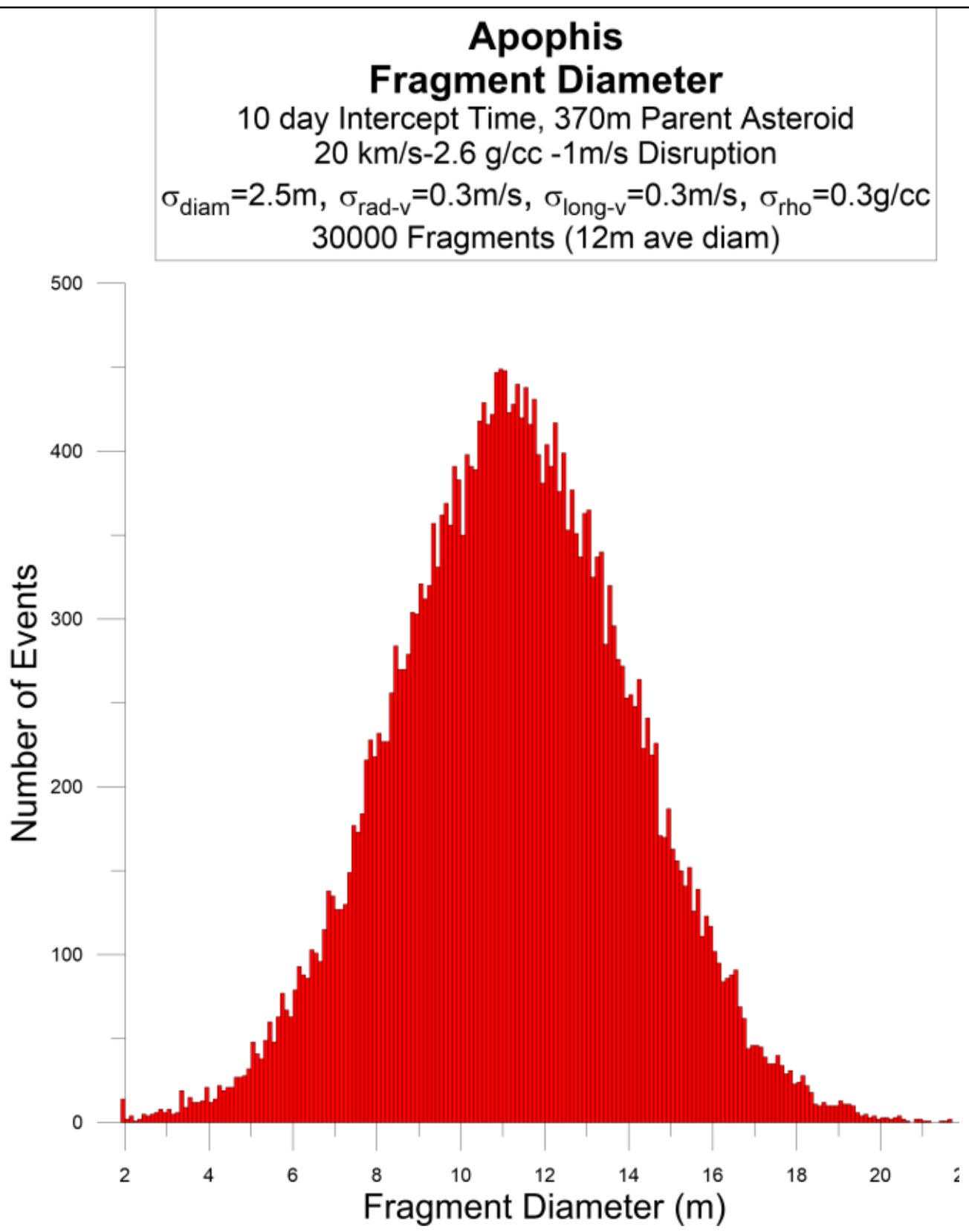

**Figure 64** – 10-day intercept of Apophis in 30,000 fragments. Histogram of fragment diameters.

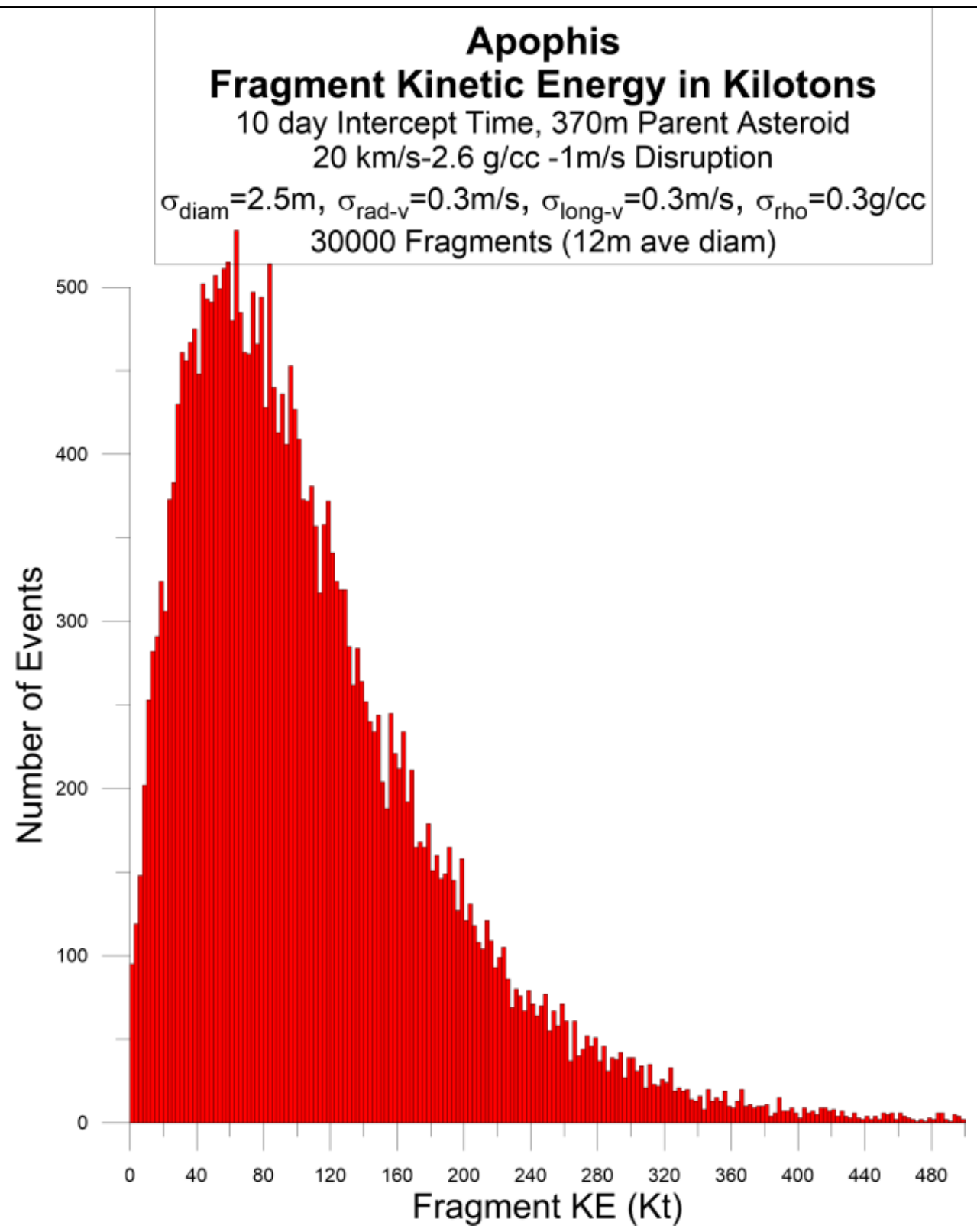

**Figure 61** – 10-day intercept of Apophis in 30,000 fragments. Histogram of fragment exo-atm. kinetic energy.

## 9. Coupling of penetrator energy to disassembly energy

The coupling between the penetrator KE and the fragment KE is a key part of the implementation of this program. We compute the mass of the penetrator required to deliver the needed disassembly KE for a 1m/s fragment speed for different intercept speeds, as well as the conversion efficiency of penetrator KE to disassembly KE for a possible future 100 ton delivered penetrator array capability. A 100-ton delivery capability of an interceptor array is possible with upcoming heavy lift options such as the Space X Starship and two NASA SLS launchers. The required efficiency is very low (<1%) even up to Apophis-class asteroids at 20km/s. This is very encouraging. The specific nature and workings of the interceptor to achieve this efficiency requires detailed design, simulations, and ground testing of both passive and explosive-filled penetrators. While the addition of explosives to the penetrator does not add significantly to their overall energy, it may add significant to their efficiency of scattering the fragments. Other options to explore include slowing the penetrator using ablation and "front side" staged explosives, as well as filling the penetrator with fluids/solids that will vaporize and "push" the fragments away. Earth testing of various configurations is critical in iterating the design. Testing of designs on actual smaller asteroid that come close to the Earth will allow for iterations on a variety of real targets.

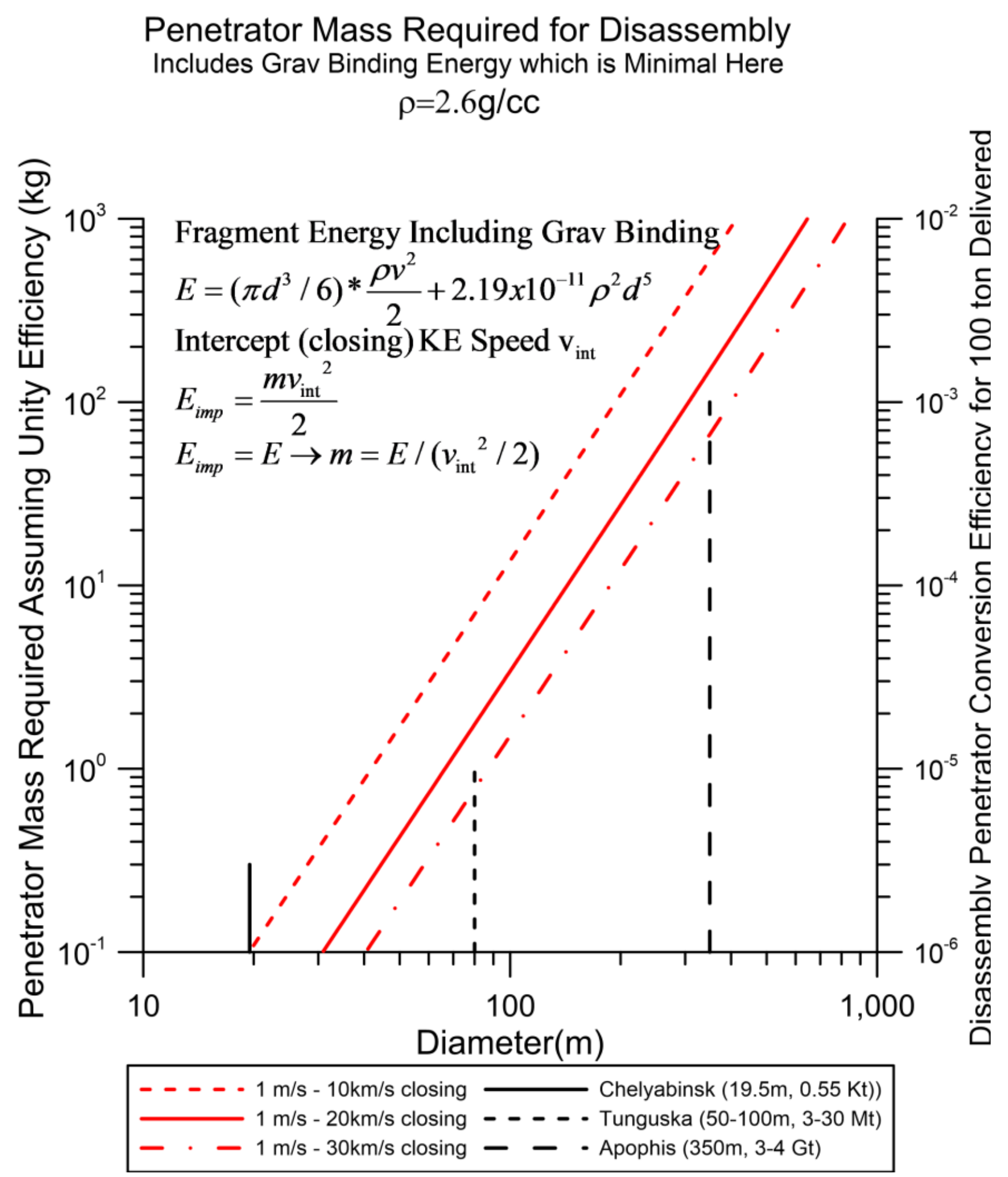

**Figure 65** – Penetrator mass required to disassemble asteroid with 1 m/s residual average fragment speed relative to the center of mass. . Mass assuming unity conversion efficiency of penetrator KE to fragment KE as well as fractional conversion efficiency for 100 ton delivered penetrator load (e.g. Starship).

***Mass of Penetrators vs Mass of Bolide*** – Assuming unity coupling of the KE of the penetrator in the frame of reference of the bolide to the KE of the fragments allows us to compare the ratio of penetrator mass $m_p$ to bolide mass M.

Bolide mass M, disperal energy incl grav Binding $E_{disp}=E_{KE-frag}+E_{BE}=Mv_{disp}^2/2+3/5\frac{GM^2}{R}$

Penetrator mass m, intercept (closing) speed $v_{int}$ $\quad E_p=mv_{int}^2/2$

Assuming unity coupling $\rightarrow E_p=E_{disp}\rightarrow m/M=[(v_{disp}/v_{int})^2+(6GM/5R)/v_{int}^2]$

$=[(v_{disp}/v_{int})^2+(3/5)(v_{esc}/v_{int})^2]$

$v_{esc}=\sqrt{2GM/R}$ $\quad$ For $E_{KE-frag}>>E_{BE}$ $(smaller\ bolide)\rightarrow m/M=(v_{disp}/v_{int})^2$

$Ex: v_{disp}=1m/s,\ v_{int}=20km/s\rightarrow m/M\sim 2.5x10^{-9}$

***Energy per mass*** – The energy per fragment mass to achieve speed *v* relative to the asteroid CM (ignoring gravitational binding) is: $v^2/2 = 1/2J/m(kg)-(m/s)^2 = 1.2x10^{-7} kgTNT/m(kg)-(m/s)^2$, as 1kg TNT is 4.2MJ, or 1g is 4.2kJ. Testing the coupling efficiency of energy deposition, whether via passive penetrators or via explosive penetrators, is critical to assess and design the system. Conventional and nuclear explosives deliver a significant fraction of their energy into the blast wave via gas coupling for terrestrial detonations. Airbags in cars are another example of this. Designing an efficient energy conversion system for PI will be a critical step in its success. Ground testing in a large vacuum chamber, if necessary, could be done. To put this in perspective, imagine two 5000kg railcars that are oriented back-to-back with an explosive charge between them of 1g of TNT. This would achieve a speed of each car of

$$v = (2E/m)^{1/2} = 0.92m/s$$

$$E = 4184J \ (1g\,TNT), \ m = 10{,}000kg \ .$$

This is essentially identical to the speed we have been simulating for fragment dispersion IF all of the energy of detonation could be coupled into the kinetic energy of the motion of the railcars. Some of the detonation energy will clearly go into heating of the railcars, which is not a useful form of energy for us. Coupling the detonation energy efficiently into kinetic energy will be a critical element of the design. This can be ground-tested by using a small rail or air table suspension system. This allows very practical testing and optimization to rapidly proceed at low cost. The testing of passive kinetic penetrators is similar, though more complex in nature. Unfortunately, there are no ground assets that allow 20km/s testing at appropriate scales. Lower speed ground testing to cross check our hypervelocity penetrator simulations codes is one option we are exploring.

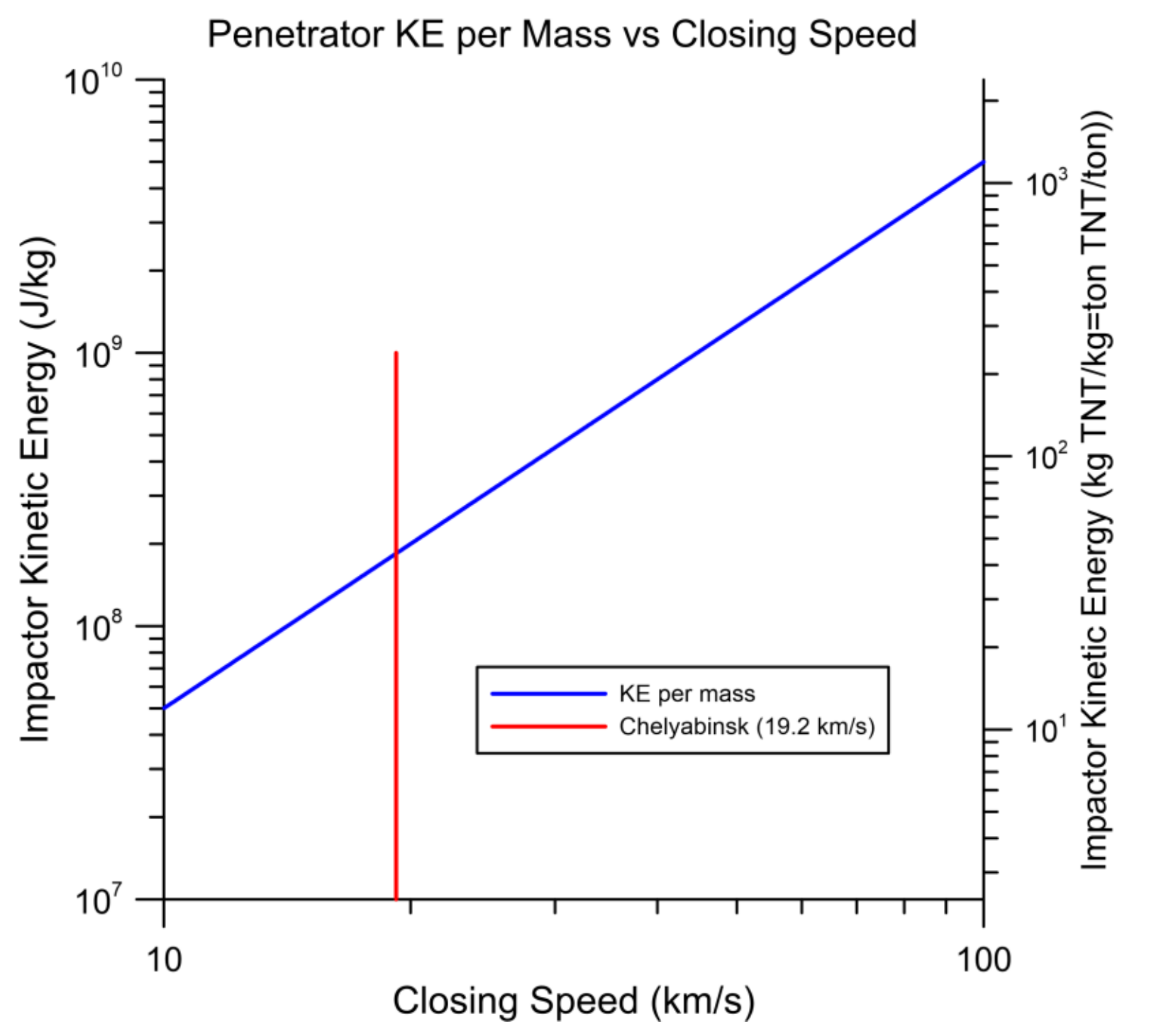

**Figure 66** – Passive (no explosives) penetrator energy per unit mass vs closing speed onto asteroid. Due to the high closing speed, the amount of kinetic energy per unit penetrator mass is far greater that the equivalent energy of explosives. For a Chelyabinsk-like event with a speed of 19.2km/s, the KE is about 44 times higher than the equivalent mass of TNT. As an example, if the delivered penetrator mass is 100 ton (Starship) then the total energy would be 4.4kt or equal to a tactical nuclear weapon.

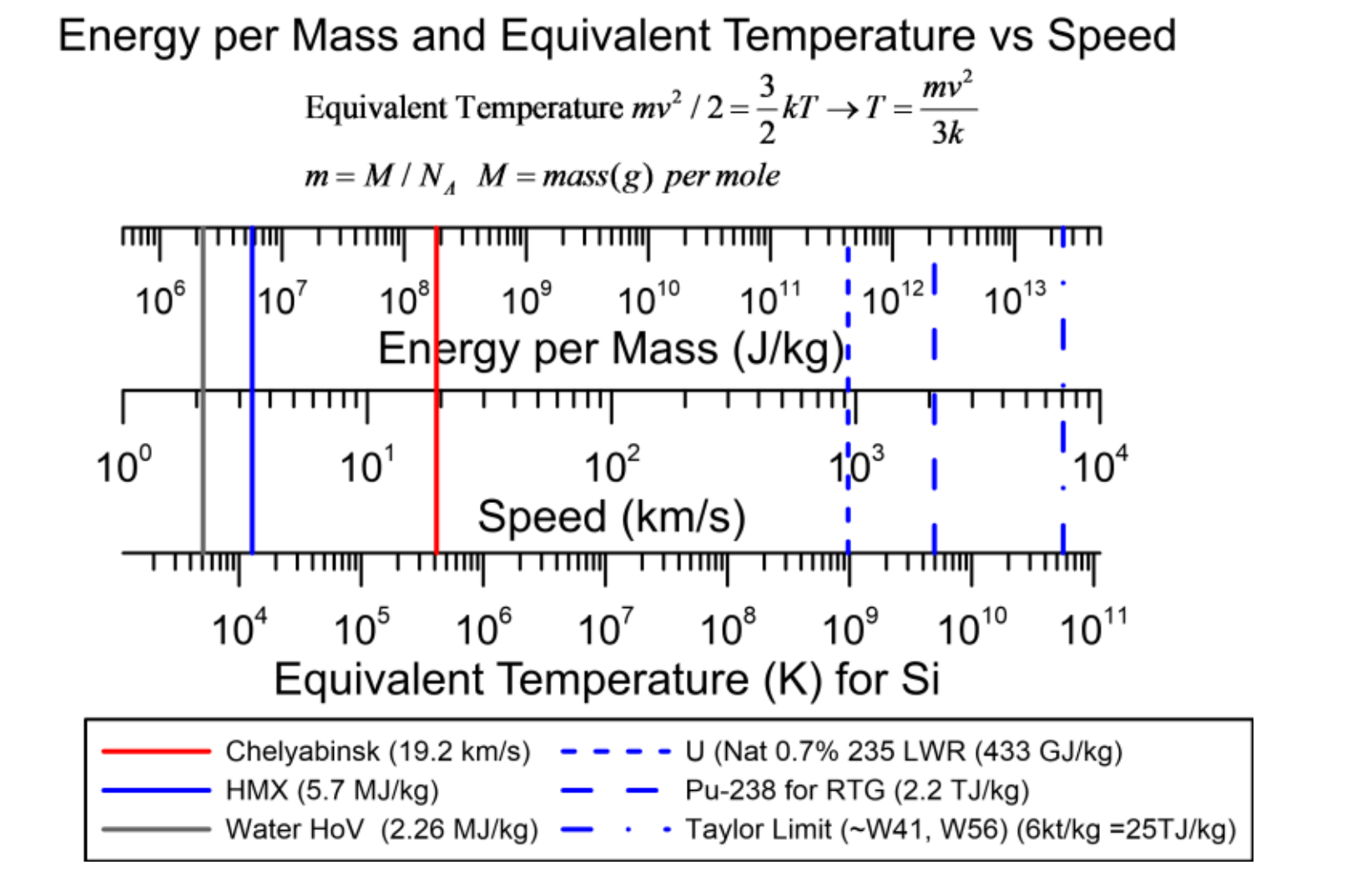

**Figure 67** – Comparison of energy per unit mass, speed and equivalent temperature of Si. Relevant for comparison of chemical, kinetic penetrators, and nuclear explosives.

## 10. Penetrator Physics and Depth

We model the penetration depth for a hypervelocity penetrator by assuming a model in the strong shock limit where the speed of the penetrator greatly exceeds the sound speed in the material being penetrated, as well as in many cases of the speed of sound "inside" the penetrator. We also assume that the speeds of impact are sufficient to compress and vaporize the material ahead of the penetrator. This is similar to the Earth-penetrating devices used to mitigate underground targets or bunkers [42]. Note that the speeds of the penetrators we are considering for asteroid mitigation (typ. > Mach 60) are far in excess of those used in "Earth-penetrating munitions" (typ. ~ Mach 1-3), and thus the basic physics will need to be refined to give accurate results. We use the existing work on Earth-penetrating munitions as a guide and focus on the approximate penetration depths and leave open the use of chemical, nuclear, or no (kinetic only) additional energy deposition. The penetrator will generally have a length $L$ to diameter $d$ ratio $L/d$>>1 – i.e. a "rod". We follow the approach of [42] in computing the penetration depth for a given penetrator design. Comparing experimental penetration depths to the theoretical depths shows that the theoretical depths are typically an over-estimate of the actual measured depths by a factor of a few. More realistic simulations are needed, but the theoretical estimates give us a good starting point. The yield strength of the penetrator material will be a critical area of work to optimize. Various high strength alloys (steels, W, Ti, for example) can have compressive yield strengths of about 1GPa, while some ceramics and glasses can exceed 4GPa ($Al_2O_3$, for example). However, high yield strength must also be accompanied by the ability to survive fracturing.

***Penetrator strategy for asteroid fragmentation*** – Numerous high-fidelity simulations are needed to determine optimized fragmentation strategies. Currently, we feel an "onion peeling" strategy is appealing in that the asteroid outer layers are removed first and then successive penetrators work inwardly. This also greatly reduces the penetration depth needed for the outer penetrators, and hence a heterogeneous penetrator array optimized for this strategy may be helpful.

***Similarity to Asteroid Penetration of the Earth's Atmosphere*** – The asteroid penetrators for disruption have similarities to the penetration of the Earth's atmosphere by low strength asteroids (the majority), where the asteroid begins to break up at high altitude when the yield stresses are exceeded and then continues to lower altitudes before detonation. This is an example of objects going far beyond their yield point and still carrying significant energy and momentum.

***Similarity to Explosively Formed Penetrator (EFP)*** – The penetrator rods can be designed to form an explosively formed penetrator (EFP) using internal explosives to liquefy a metal penetrator that may increase penetration depth.

***Explosive tipped penetrators*** – One option for increased penetration has been to use explosive or shaped charge tipped penetrators. This may be an option, but one serious problem is that the detonation speed of explosives (max 10km/s) is slower than the asteroid speeds of interest (>10km/s) and hence a conventional explosive or shaped charge tipped penetrator would not generally work as it does for terrestrial applications. Having multiple penetrators that follow each other in so that the initial entry "clears the way" for the subsequent penetrators may be an option if needed.

***Low Yield Strength Asteroid*** – Low density asteroids may be quite low yield strength, particularly on larger macroscopic levels where cohesion is extremely low. For example, it appears that low density stony asteroids have macroscopic yield strengths under 1MPa. We have a very limited dataset to go from on this and much more work in characterizing them is needed. One strategy is to design "smart" penetrators that can measure the penetration length vs time since entry and then detonate if needed before exiting.

***Realistic Asteroid Composition*** – A realistic asteroid is generally thought to be significantly heterogeneous, and thus complex in nature. This will complicate our efforts in some ways, but may also make asteroids easier to break apart in other ways. Preliminary missions to probe a variety of asteroids using penetrators would give us a great amount of structural information. Measurements of asteroids using penetrators as probes would be an extremely interesting mission space. Interior structural information could, in theory, be obtained from a smart penetrator that was instrumented with accelerometers that could record and relay acceleration vs time at high cadence (MHz rates, for example) to give sub-meter structural resolution.

***Heterogeneous Penetrators*** – There are a wide variety of options in designing penetrators for maximum "pulverization" of the asteroid being targeted. Since the impact speed of the penetrator ensures pressures above the structural strength of known materials, structural failure is virtually assured. This is analogous to the case of the asteroid impacting the Earth's atmosphere that then fragments due to structural failure. Just as the asteroid in the Earth's atmosphere undergoes catastrophic detonation, so too could the catastrophic failure of the penetrator be useful in fragmenting the asteroid. Penetrators specifically designed to fail and release either explosive charges, or a load of ceramic, or other type of sub-penetrators could be extremely useful in transferring momentum and disassembly energy into the asteroid. A type of "compressively formed penetrator" analogous to EFP's could be useful in this regard. Heterogeneous penetrators with hardened tips and explosive interiors with clusters of internal penetrators and explosives is another option.

***Explosive Optimization*** **–** Part of the problem in optimizing the conversion of chemical or passive hypervelocity penetrators into useful kinetic energy for fragment dispersion is the large mismatch in speeds. Chemical detonations occur at speed of approximately 3-10km/s, while the passive penetrator closing speed on the asteroid is 10-40km/s. The desired speed of the fragments relative to the asteroid CM is of order 1m/s. The ratio of the detonation (whether chemical or passive impact) speed to the desired fragment speed is of order $10^4$. It is this "impedance" mismatch that requires optimization. It is not as much an energy problem, but rather a momentum transfer problem. Another way to ponder the problem is to consider a completely inelastic collision between the penetrator/explosive and the fragment. The ratio of the momentum (useful work) to energy input is $P/E = mv/(mv^2/2) = 2/v$ where $v$ is the penetrator or chemical explosive speed. The higher the speed $v$, the lower the momentum (per energy) transfer for an inelastic collision. For $v$=10km/s, this inelastic momentum per unit energy $P/E = 2x10^{-4} N-s/J \ (or\, N/w)$. This is the same relationship of *P/E* as for a rocket engine of specific impulse $I_{sp}$ *where* $v_{exh} = gI_{sp}$ *and* $P/E = 2/v_{exh} = 2/gI_{sp}$. In a very real sense, an explosive denotation, or a penetrator with speed $v$ for an inelastic collision, is equivalent to a rocket engine with the equivalent energy in fuel. In precise analogy to a rocket, the way to get higher momentum transfer *P* for the same fuel energy *E* is to use the fuel in a more efficient cycle (such as a Stirling or Carnot engine cycle, both of which use gas as the "momentum extraction working fluid") to extract more mechanically useful energy.

***Optimized detonation speed for pure inelastic collision*** - Since momentum/energy $P/E = 2/v$, we can model the final momentum in a fully inelastic collision as a simple model. If we assume a desired fragment speed $v_{frag}$ then only input particles from the detonation moving at speed $v \geq v_{frag}$ can do useful work on the fragment as slower detonation particles cannot catch up to the fragment and thus are not useful, while particles that move faster than the fragment "waste energy" in the inelastic collision as the excess energy is converted to heat. Thus, the optimum detonation speed is to match the speed of the fragment. If the collision is not inelastic and there is a way to store the "propulsion energy" in a mechanism such a spring (gas for example), then the detonation speed can be very fast compared to the final fragment speed as the spring is "slowly released" to push on the fragment, and thus efficiently transfer the detonation energy into fragment momentum.

Another complication is that part of the total energy of the penetrator/explosive is generated as heat, which is not generally useful for our purposes UNLESS we can use the heat to cause gas expansion, such as in an airbag or a confined (tamped) explosive. We can do much better if we effectively introduce a "spring" into the system where much of the energy is used to rapidly compress the spring and then the spring is slowly expanded to "push" the fragments away. In a sense, this is precisely what happens in a conventional or nuclear detonation where the initial explosive speeds for chemical explosives are ~ 10km/s, and for nuclear weapons they are of order 100-1000km/s. In the resultant air burst blast wave, the "spring" is the air itself when it is rapidly (adiabatically) compressed and then slowly expanded as the shock wave propagates ultimately at about 0.3km/s (speed of sound in air) far from the explosion. To efficiently couple the energy from the penetrator, we need this "spring". In both the passive penetrator and the explosively loaded penetrator, there is gas generated that can be useful in this. For the passive penetrator the gas/plasma comes from the vaporization of the asteroid/penetrator during created during penetration, while for the chemical explosive penetrator there is also the gas generated by the chemical explosive decomposition. The "ideal explosive" for our application would be one with a detonation speed of order 1m/s. While no such explosive exists, conventional explosives coupled to mechanical (gas, for example) systems do exist. This area has strong overlap with the mining and explosive construction industry. This is an area ripe for creative and testable designs with an emphasis on rapidly testable designs.

***Testing Paths*** – High fidelity ground and then space testing can be accomplished by a series of steps including:

- Ground-based tests using hypervelocity projectiles into custom designed and fabricated targets. One of the advantages of the approach we take is that robust ground testing is feasible and can be done at relatively low cost using many of the assets already in place such as railguns, gas guns, rocket assisted artillery shells, etc. This allows a rapid design, testing, and iteration program with both passive and explosive filled penetrators. The critical steps of optimizing the coupling of the large penetrator KE for passive penetrators and the option of explosive filled penetrators and possible "gas inflated" penetration options should be explored. There are numerous opportunities for space-based testing, though this may be politically complex, due to the large number of "flyby" asteroids that come close to the Earth. Proving out the system in actual intercept and fragmentation missions would give us confidence that we would be ready for "the real one" when needed.
    - Building parts of scaled heterogeneous targets that can be simulated.
    - Using hypervelocity rail gun or gas gun and EFP projectiles with and without explosives.
    - Building parts of full-scale targets (longitudinal slices) to test penetrators.
- Ground testing coupling efficiency of explosives on targets from 1kg to beyond 1000 tons (~ 10m).
    - Goal is to understand coupling efficiency of various explosive configurations.
    - Testing simulating zero gee by using horizontal (air table) rails to measure speed (stiction).
    - Testing simulating zero gee by using floating targets to measure acceleration (no stiction).
    - Test 1kg to 1000 ton targets with explosives from grams to multiple kg.
    - Feed explosive test data into explosive penetrator modeling and iterate.
- Lunar and space intercept tests.
    - Using the moon as a "target" where real-time sensor data (acceleration) could be monitored on the moon or on Earth would provide a wealth of data and build confidence. This data could also be used to probe the local interior of the lunar surface to compare with acoustic measurements.
    - Interception of small asteroids that come close to the Earth would provide numerous opportunities to test the system as it is developed.

***Penetrator Physics*** – We calculate the penetration depth of a high-speed penetrator below. The physics here is not valid above 2km/s [43]–[47]. The term "penetration depth" is the depth of the "hole" formed by the penetrator in the material. It can be considered as the effective "cavity depth" formed in the material by the penetrator.

Penetrator depth model for low speed (<2 km/s) entry

$D =$ penetration depth

$L =$ impactor length

$v_{p-init} =$ initial impactor speed

$v_{p-final} =$ "final" impactor speed when the ram pressure $\frac{1}{2}\rho_s v_{p-final}{}^2 = Y_t$

Note that actual "final" impactor speed is 0 when if come to rest.

$v_{p-final}$ is the impactor speed when the impactor can no longer "break" the target cohesion

$\rho_p =$ density of impactor

$\rho_t =$ mean density of target

$\rho_s =$ density of shocked target material ahead of impactor

$Y_p =$ compressive yield strength of impactor

$Y_t =$ compressive yield strength of target

For targets that are not "bulk molecularly bound or cohesionless or granular"

the range of is typically 0.01 to 1 MPa while well bound materials have ranges of 10-1000MPa

$\gamma =$ polytropic index of vaporized shocked material ahead of impactor

(typically vaporized material is monatomic gas $\rightarrow \gamma = 5/3$)

$$\rho_s = \rho_t \frac{\gamma+1}{\gamma-1} \rightarrow \rho_s = 4\rho_t (if\, \gamma = 5/3)$$

$v_{p-final} =$ "final" impactor speed when the ram pressure $\frac{1}{2}\rho_s v_{p-final}{}^2 = Y_t$

$$\rightarrow v_{p-final} = \sqrt{\frac{2Y_t}{\rho_s}} = \sqrt{\frac{Y_t}{2\rho_t}} \;(if\, \gamma = 5/3)$$

$$D/L = \frac{\rho_p}{\rho_t}\sqrt{\frac{\rho_t}{2Y_t}}(v_{p-init} - v_{p-final})$$

$$\rightarrow D/L = \frac{\rho_p}{\rho_t}\left[\sqrt{\frac{\rho_t}{2Y_t}}v_{p-init} - 1/2\right] = \frac{\rho_p}{\rho_t}\left[\frac{v_{p-init}}{2v_{p-final}} - 1/2\right](if\, \gamma = 5/3)$$

Note that for the speed of our impactors relative to the asteroid $v_{p-final} << v_{p-init}$

$$\rightarrow D/L \sim \frac{\rho_p}{\rho_t}\sqrt{\frac{\rho_t}{2Y_t}}v_{p-init} = \frac{\rho_p}{\rho_t}\frac{v_{p-init}}{2v_{p-final}} \;(if\, \gamma = 5/3)$$

Destruction of impactor

The impactor speed impose a severe constraint on the maximum impactor speed

Since the ram pressure on the impactor at very high speeds will cause deformation

and ultimately impactor failure when the ram pressure exceeds the impactor yield strength

$$P_{ram-imp-\max} = \frac{1}{2}\rho_t v_{p-\max}{}^2 = Y_p \rightarrow v_{p-\max} = \sqrt{\frac{2Y_p}{\rho_t}}$$

Replacing $v_{p-init}$ with $v_{p-\max}$ we get:

$$D/L \sim \frac{\rho_p}{\rho_t}\sqrt{\frac{\rho_t}{2Y_t}}v_{p-init} = \frac{\rho_p}{\rho_t}\sqrt{\frac{\rho_t}{2Y_t}}v_{p-\max} = \frac{\rho_p}{\rho_t}\frac{v_{p-\max}}{2v_{p-final}} \;(if\, \gamma = 5/3)$$

Example: $Y_p = 1GPa,\ \rho_t = 2000kg/m^3$ $(\sim stony\ asteroid), Y_t = 1MPa, \rho_p = 8000kg/m^3$

$$v_{p-\max} = \sqrt{\frac{2Y_p}{\rho_t}} = 1km/s$$

$$v_{p-final} = \sqrt{\frac{2Y_t}{\rho_s}} = \sqrt{\frac{Y_t}{2\rho_t}}\ (if\,\gamma = 5/3) \sim 22m/s$$

Replacing $v_{p-init}$ with $v_{p-\max}$ we get $v_{p-init}, v_{p-\max} >> v_{p-final}$:

$$D/L = \frac{\rho_p}{\rho_t}\left[\sqrt{\frac{\rho_t}{2Y_t}}v_{p-init} - 1/2\right] = \frac{\rho_p}{\rho_t}\left[\frac{v_{p-init}}{2v_{p-final}} - 1/2\right] \sim \frac{\rho_p}{\rho_t}\frac{v_{p-\max}}{2v_{p-final}}\ (if\,\gamma = 5/3) \sim 90$$

This is an extremely large penetration to impactor ratio and it is not clear how realistic this would be given measured penetration is often a factor of a few times less. As discussed with $v_{p-\max} >> v_{p-final}$ in general this would mean distortion and ultimately structural failure of the penetrator which is not necessarily bad.

Comparison to testing of B61-11 NED penetrator in frozen soil where the measured penetration was a maximum of 8 meters (*D/L*~3) when the drop was from 8000 feet (2.4km) with max impact speed (ignore air resistance) of $(2gh)^{0.5}$~ 220m/s. Using ice and two frozen soil examples as the target:

Example: $Y_p = 1GPa,\ \rho_t = 1000kg/m^3\ (ice), Y_t = 25MPa, \rho_p = 8000kg/m^3, v_{p-init} = 220m/s$

$$v_{p-final} = \sqrt{\frac{2Y_t}{\rho_s}} = \sqrt{\frac{Y_t}{2\rho_t}}\ (if\,\gamma = 5/3) \sim 110m/s$$

$$D/L = \frac{\rho_p}{\rho_t}\left[\sqrt{\frac{\rho_t}{2Y_t}}v_{p-init} - 1/2\right] = \frac{\rho_p}{\rho_t}\left[\frac{v_{p-init}}{2v_{p-final}} - 1/2\right](if\,\gamma = 5/3) = 4$$

Example: $Y_p = 1GPa,\ \rho_t = 2000kg/m^3\ (\sim\ frozen\,soil), Y_t = 10MPa, \rho_p = 8000kg/m^3, v_{p-init} = 220m/s$

$$v_{p-final} = \sqrt{\frac{2Y_t}{\rho_s}} = \sqrt{\frac{Y_t}{2\rho_t}}\ (if\,\gamma = 5/3) \sim 50m/s$$

$$D/L = \frac{\rho_p}{\rho_t}\left[\frac{v_{p-init}}{2v_{p-final}} - 1/2\right](if\,\gamma = 5/3) = 6.8$$

Example: $Y_p = 1GPa,\ \rho_t = 2000kg/m^3\ (\sim\ frozen\,soil), Y_t = 5MPa, \rho_p = 8000kg/m^3, v_{p-init} = 220m/s$

$$v_{p-final} = \sqrt{\frac{2Y_t}{\rho_s}} = \sqrt{\frac{Y_t}{2\rho_t}}\ (if\,\gamma = 5/3) \sim 35m/s$$

$$D/L = \frac{\rho_p}{\rho_t}\left[\frac{v_{p-init}}{2v_{p-final}} - 1/2\right](if\,\gamma = 5/3) \sim 11$$

This calculated *D/L* is in reasonable agreement with the measured penetration, though there is significant uncertainty as to the soil compressive yield strength for the B61-11 penetrator tests in Alaska frozen soil.

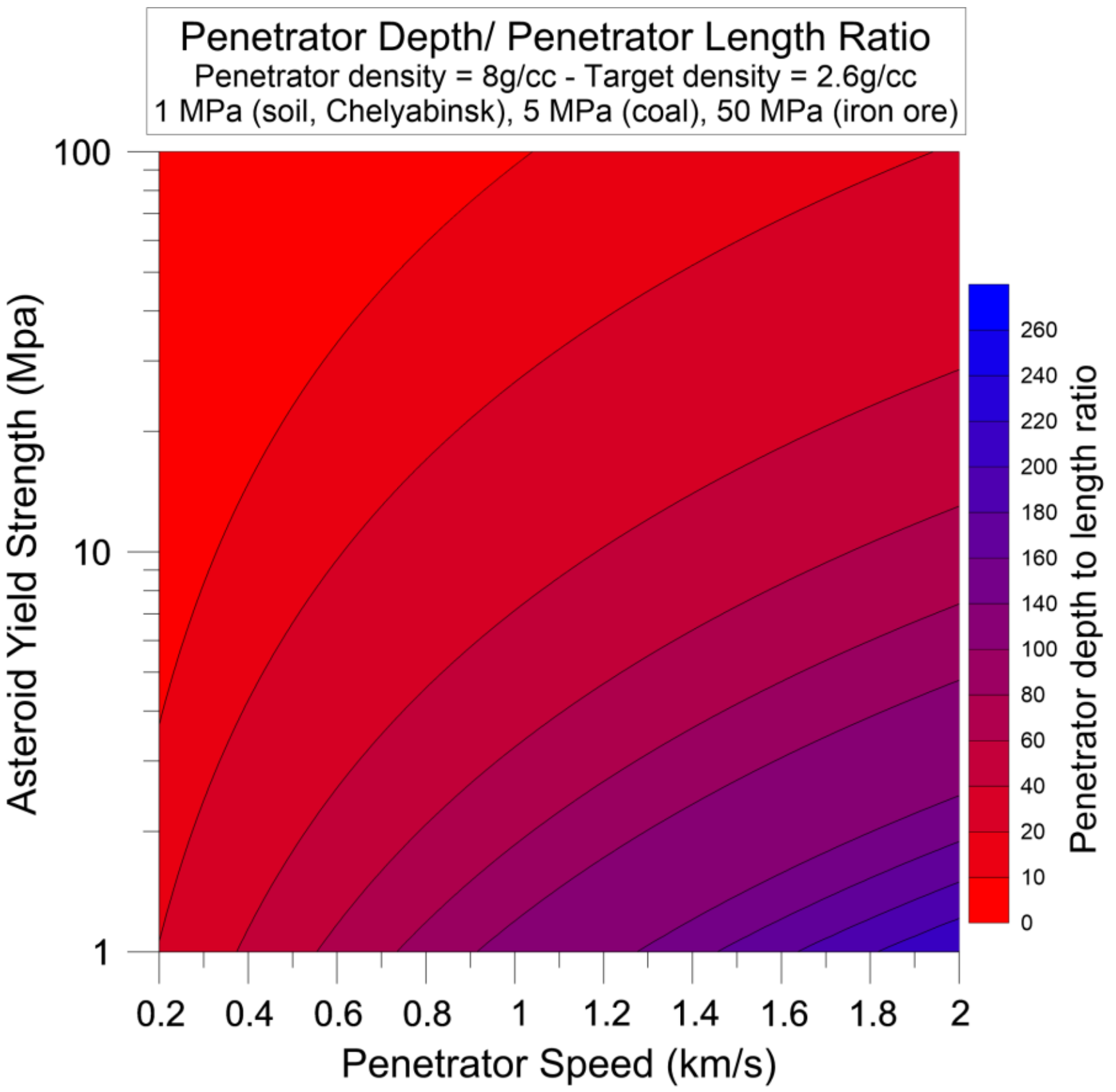

$$D/L = \frac{\rho_p}{\rho_t}\left[\sqrt{\frac{\rho_t}{2Y_t}}v_{p-init} - 1/2\right] for\, \gamma = 5/3$$

**Figure 68** – Estimated penetrator depth-to-length ratio for asteroid yield strength vs penetrator speed. This is a rough estimate based on relatively low speed penetrator physics. The physics for relevant speeds of 10's km/s requires the inclusion of penetrator vaporization and plasma effects that are not included here. It is encouraging that at low km/s speeds the penetration depths are potentially extremely large for modest yield strengths appropriate for stony asteroids.

## 11. Optical and Thermal signature

Air burst asteroids/fragments have an optical signature due to the extreme heating of the outer surface of the bolide as well as the air. As discussed above, the effective temperature of the impacting air molecules for typical impactors speeds of 10-60km/s is in the millions of degree range. However, there are a number of physical processes that will limit the actual physical temperature of both the outer skin of the bolide as well as the air. In many respects, this is analogous to the re-entry of spacecraft where the heat shield is the equivalent to the outer skin of the bolide and the plasma layer around the heat shield is similar to the extreme heating of the air around the bolide. This is also seen in the many "dash cam" videos of the Chelyabinsk event and the common "shooting star" optical trails of meteors. In meteor communications, the ionized trail of the meteor acts as a plasma mirror and reflects radio waves, much like the Earth's ionosphere.

***Atmospheric absorption*** - We can ignore atmospheric absorption in order to get a worst-case analysis. In reality, the atmosphere will significantly scatter and absorb the radiation. UV and some IR will be rapidly attenuated. Realistic models will need to take into account weather conditions as well as humidity, particulates, etc. The numbers below ignore all atmospheric absorption and are thus truly a worst case. **Our actual simulation does include atmospheric absorption for every fragment and observer.**

***Ablation Limited Temperature of Asteroid Skin*** – The extreme speed of the air molecules heats the outer surface of the asteroid until the asteroid begins to ablate, at which point there is an energy balance between air bombardment heating and ablation and radiation cooling. The ablation rate, and hence the cooling flux, is exponentially dependent on the temperature, and hence the temperature rises rapidly and then stabilizes. This is similar to boiling water [18], [34]–[36].

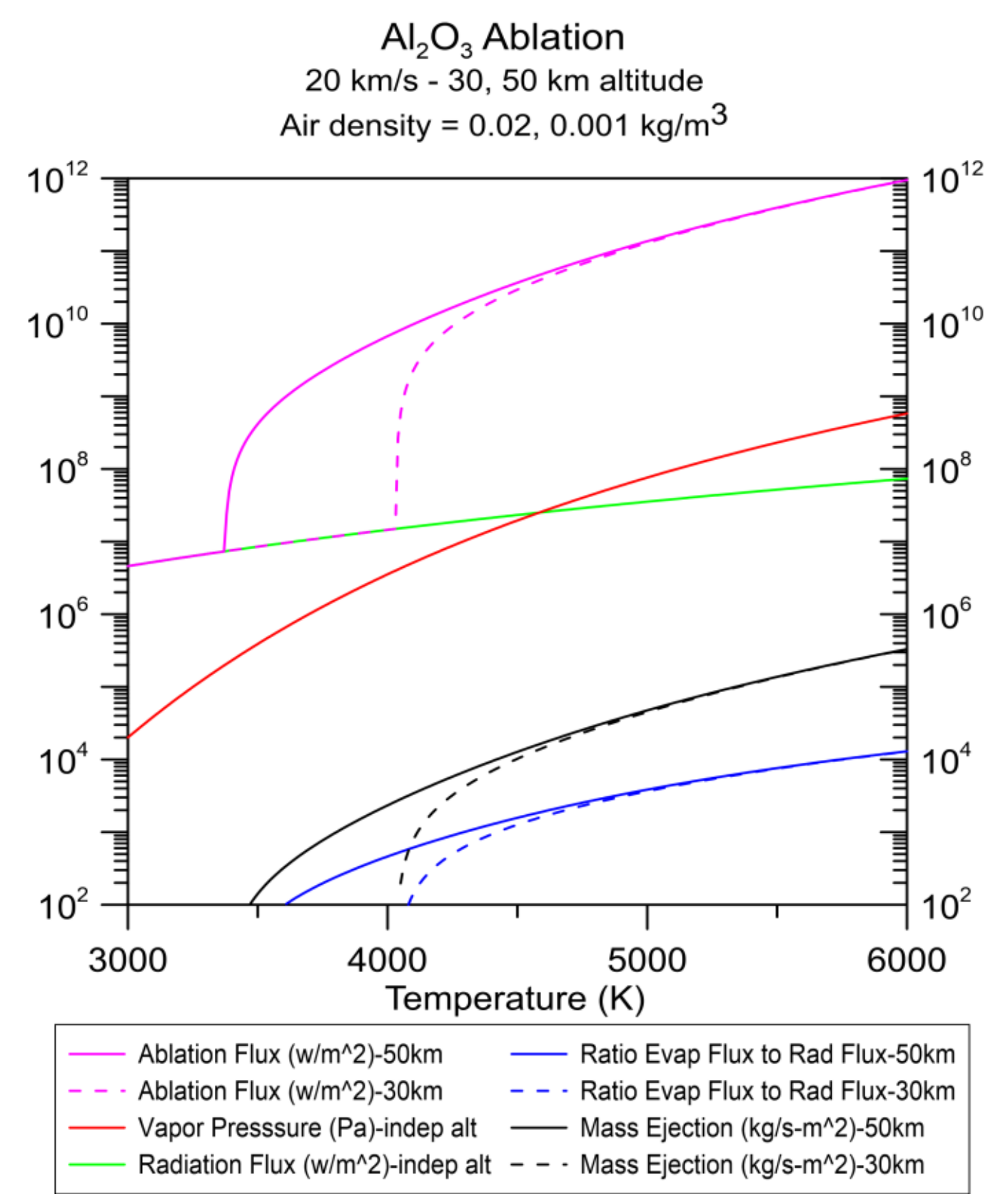

**Figure 69** – Ablation rate, total flux including radiation and other related parameters vs temperature for an example compound (Al2O3). Modeled for 20km/s asteroid (indep. size) at 30 and 50km altitude. Includes ram pressure suppression of ablation which is clear at lower temperatures.

The air heating flux can be approximated as $F_{air\ heated\ ablation}$ (W/m$^2$) = $\rho_{air}V_{bolide}{}^3/2$. The impinging air mass flux is $\rho_{air}V_{bolide}$ (kg/s-m$^2$). Note that this is related to the stagnation pressure $P_{stag} = \rho_{air}\ V_{bolide}{}^2$ we calculated above by $F_{air\ heated\ ablation} = 1/2\ V_{bolide}\ P_{stag}$. The real situation is far more complex due to the air/ plasma flow around the bolide. For the typical altitudes of the asteroid fragment size where burnup occurs (typ. 30-50km altitude), $\rho_{air}$ ~ 0.02-0.001kg/m$^3$. For $V_{bolide}$=20km/s, this gives an air bombardment flux of about $F_{air\ heated\ ablation}$(W/m$^2$) ~ 4-40GW/m$^2$ with a stagnation pressure $P_{stag}$ = 0.4-8MPa. While extremely high, this is within the range of laser ablation experiments. We have studied this for a variety of materials as a part of our previous laser ablation planetary defense program [18], [19], [34]–[36]. There are some significant differences, however, between laser ablation and atmospheric impact ablation. The area of ablation is highly material- and structure-dependent and is thus often unique to each case. We can make an approximation using some of our previous studies in this area. We show an example of $Al_2O_3$, which at 6000K has an extrapolated vapor pressure of 580MPa and an ablation rate of 3.3x10$^5$ kg/s-m$^2$. For a 10m diameter asteroid with density 2.6g/cc, the mass is 1.4x10$^6$ kg. The projected effective front surface area is 79m$^2$, and this gives a mass loss rate of 2.6x10$^7$ kg/s, which would imply the asteroid would not last long (<1s). Since the total ablation mass loss (kg/s) is dependent on the asteroid surface area, it scales as $r^2$, while the asteroid mass scales as $r^3$, where $r$ is the asteroid radius. For a constant ablation flux, this gives $dr/dt$=constant, and thus a finite time to complete evaporation. For the example above, the time to complete evaporation would be 0.16 seconds from the analysis below, assuming a constant ablation flux.

$$r_o = initial\ asteroid\ radius,\ \rho = density,\ \tau = time\ to\ evaporate,\ \Gamma(kg/s-m^2) = ablation\ \ flux$$

$$m = \rho\frac{4\pi}{3}r^3 \quad dm/dt = 4\pi\rho r^2(dr/dt) = \Gamma\pi r^2 \ \ (\pi r^2\ is\ projected\ area) \ \rightarrow dr/dt = \Gamma/4\rho\ \ \rightarrow \tau = \frac{r_o}{dr/dt} = \frac{4\rho r_o}{\Gamma}$$

In reality, it is much more complex as there are competing effects of slowing down the asteroid, slowed ablation rate from high density plume (back scattering), and atmospheric pressure [36]. The physics becomes quite complex due to the extreme air bombardment suppressing ablation (like in a halogen lightbulb), and thus the ablation cooling is suppressed allowing the temperature to rise even higher than shown. Based on the color spectrum of observed objects, we will nominally assume an equivalent temperature of 6000K blackbody, but note that the effective temperature is time dependent since the bolide is slowing down and fracturing, as well as material dependent and greatly increased effective surface area during the atmospheric breakup. This makes a first principles analysis only an approximation. We also analyze the cases of cooler 2000K and 3000K systems. We compare to measured data from satellites for a select case below.

We assume a blackbody flux with $F_{blackbody}$(W/m$^2$) = σT$^4$ ( σ=5.67x10$^{-8}$) which for

- T=6000K gives a bolometric flux (integrated over all wavelengths) of $F_{blackbody}$(W/m$^2$) = 74MW/m$^2$ or about 12GW for a 10m diameter asteroid/fragment assuming the forward half is glowing.
- T=3000K gives a bolometric flux (integrated over all wavelengths) of $F_{blackbody}$(W/m$^2$) = 4.6MW/m$^2$ or about 0.72GW for a 10m diameter asteroid/fragment assuming the forward half is glowing.
- T=2000K gives a bolometric flux (integrated over all wavelengths) of $F_{blackbody}$(W/m$^2$) = 0.91MW/m$^2$ or about 0.14 W for a 10m diameter asteroid/fragment assuming the forward half is glowing.

For a 6000K blackbody, the spectrum peak wavelength is about 0.5 micron, or in the peak of the human eye response. The visible portion is 38% of the total emission in the 0.4 to 0.7 micron visible band. For a 3000K blackbody, the spectrum peak wavelength is about 1 micron or in the near IR. The visible portion is 8% of the total emission in the 0.4 to 0.7 micron visible band. For a 2000K blackbody, the spectrum peak wavelength is about 1.5 micron or in the near IR. The visible portion is 1.7% of the total emission in the 0.4 to 0.7 micron visible band.

We model the optical radiation emitted power $P$ from the fragment as that of a hemisphere (front section of spherical bolide) radiating into the forward 2π.

$$L_{0_i} = \text{fragment diameter}$$

$$r_i = \text{distance to observer}$$

$$F_{blackbody} = \sigma T^4$$

$$P_{emitted} = F_{blackbody} 2\pi \left(L_{0_i} / 2\right)^2 = \frac{\pi}{2} L_{0_i}{}^2 \sigma T^4$$

$$F_{observer} = P_{emitted} / 2\pi r_i^2 = \frac{L_{0_i}{}^2 \sigma T^4}{4 r_i^2}$$

**The minimum distance to the observer is the distance from the burst altitude to the ground, which is typically about 30km** for the cases we design the system for (typ. 10m fragment size for stony asteroids) and also calculate the emitted power and observed flux for the 7x pancake at burst.

As an example, if we assume:

$$F_{observer} = P_{emitted} / 2\pi r_i^2 = \frac{L_{0_i}{}^2 \sigma T^4}{4 r_i^2}$$

$$Ex: L_{0_i} = 10m,\ T = 6000K,\ r_i = 30km \rightarrow F_{observer} = 2.4W/m^2\ P_{emitted} = 12GW$$

$$Ex: 7x\ pancake\ L_{0_i} = 7x10m,\ T = 6000K,\ r_i = 30km \rightarrow F_{observer} = 120W/m^2\ P_{emitted} = 600GW$$

$$Ex: L_{0_i} = 10m,\ T = 3000K,\ r_i = 30km \rightarrow F_{observer} = 0.15W/m^2\ P_{emitted} = 0.75GW$$

$$Ex: 7x\ pancake\ L_{0_i} = 7x10m,\ T = 3000K,\ r_i = 30km \rightarrow F_{observer} = 7.4W/m^2\ P_{emitted} = 37GW$$

$$Ex: L_{0_i} = 10m,\ T = 2000K,\ r_i = 30km \rightarrow F_{observer} = 0.03W/m^2\ P_{emitted} = 0.15GW$$

$$Ex: 7x\ pancake\ L_{0_i} = 7x10m,\ T = 2000K,\ r_i = 30km \rightarrow F_{observer} = 1.5W/m^2\ P_{emitted} = 7.4GW$$

All of these are an extremely small flux compared to sunlight (~ 1000W/m$^2$), but is similar to a modestly illuminated room. For reference, full moonlight flux on the Earth’s surface is about 7mW/m$^2$, while the thermal emission flux from the Moon on the Earth is only about $25mW/m^2$.
Even the extreme case of 30,000 fragments for the Apophis case where ALL of the fragments were in the same location right above the observer, the flux would be 72x solar flux for T=6000K, 4.5x solar flux for T=3000K ablation, or 0.9 solar flux for T=2000K. Note that IF we allow the “pancake 7x diameter expansion” then these ratios would increase by 49x (or 3500x solar flux) for T=6000K, 220x solar flux for T=3000K ablation, or 44x solar flux for T=2000K. **We would never allow the scenario of all 30,000 fragments for the Apophis case to be directly overhead as they would be dispersed depending on the interdiction time. See the sheet model below for the distributed fragment case.**

The bottom line is that the danger from the optical/thermal IR flux from a large number of spatially spread fragments is low relative to the blast wave. The optical illumination is also very brief (typ. ~ several seconds). The damage from the blast wave is vastly larger and more dangerous than the optical burst in general. **In addition, people can be warned to stay inside or be covered when the event occurs.**

***Scaling due to fragmentation*** – As the bolide is fragmented to finer and fine scales, the total surface area of the fragments increases, though only as the 1/3rd power of the fragmentation number. This increased fragmentation is offset by the increased altitude of the burnup of the finer fragments and hence increased distance from the optical flash to the observer.

$$L_{0_i} = \text{fragment diameter} = L_0 / N^{1/3}$$

$$P_{total-emitted} = NP_{one-fragment-emitted} = \frac{N\pi}{2}\left[\frac{L_0}{N^{1/3}}\right]^2 \sigma T^4 = \frac{\pi}{2} N^{1/3} L_0^{\,2} \sigma T^4$$

$$F_{total-observer} = P_{total-emitted} / 2\pi r_i^2 = \frac{N^{1/3} L_0^{\,2} \sigma T^4}{4 r_i^2}$$

***Plastic deformation and burst light increase*** – There are two other effects, though they are not dramatic.

- Increase in surface area as the asteroid/fragment plastically deforms and increases in area prior to burst. This increases the light yield as the forward surface area increases.
- Additional fragmentation upon breakup and final burst – both of these will also increase the effective surface area and thus can increase the light output.

***Heated Air Optical Signature*** – The optical signature of extremely hot air depends on the density of the air and the ionization state. For non-ionized air, the photon cross section is so low that the effective emissivity of the air is very low, and thus very little emission is observed. For highly ionized air, the optical depth can be quite small (think of turning on a Neon sign or a high-pressure Hg or Xe gas bulb). Thus, an optical signature can be significant. In addition, the ablating asteroid material can “seed” the heated air and thus provide a higher effective emissivity.

***Comparison to Chelyabinsk*** – The optical and the acoustical signatures have been analyzed for the Chelyabinsk event. While calibration of the optical signature in and near Chelyabinsk is difficult to do precisely, there are numerous reports of an intense flash of light with estimates being a peak flux of 30x solar with a thermal signature “felt” by people nearby, and there were are also reports of UV “sunburn” effects some 20 miles away, as well as some retina burns [1], [48], [49]. The UV burns would favor a higher effective blackbody temperature or UV line emission or short wavelength emission from the created plasma.

Modifying the above analysis of the 10m diameter to a 20m diameter for Chelyabinsk we get:

$$F_{observer} = P_{emitted} / 2\pi r_i^2 = \frac{L_{0_i}^{\ 2}\sigma T^4}{4r_i^2}$$

$$Ex: L_{0_i} = 20m,\ T = 6000K,\ r_i = 30km \rightarrow F_{observer} = 10W/m^2\ \ P_{emitted} = 46GW$$

$$Ex: L_{0_i} = 20m,\ T = 3000K,\ r_i = 30km \rightarrow F_{observer} = 0.6W/m^2\ \ P_{emitted} = 2.9GW$$

$$Ex: L_{0_i} = 20m,\ T = 2000K,\ r_i = 30km \rightarrow F_{observer} = 0.12W/m^2\ \ P_{emitted} = 0.57GW$$

None of these cases agree with optical signature observed (even if weakly calibrated) of 30x sun (30,000 $W/m^2$) at peak. Even the 6000K case at $10W/m^2$ is far too low to agree with 30x sun for the assumptions.

***Plastic Deformation and Increased Brightness*** - Our model assumes that the parent asteroid/fragment deforms due to the extreme atmospheric ram pressure with a "pancake diameter" limit of 7x asteroid/fragment diameter. This would give an increase in area of the forward surface by of order $7^2$=49 with an increase in both total power and received flux of ~ 50x. This gives:

$$F_{observer} = P_{emitted} / 2\pi r_i^2 = \frac{L_{0_i}^{\ 2}\sigma T^4}{4r_i^2}$$

$$P_{emitted} = F_{blackbody} 2\pi \left(L_{0_i}/2\right)^2 = \frac{\pi}{2} L_{0_i}^{\ 2}\sigma T^4$$

$$Ex: L_{0_i} = 7x20m,\ T = 6000K,\ r_i = 30km \rightarrow F_{observer} = 500W/m^2\ (1/2\,sun)\ P_{emitted} = 2.3TW$$

$$Ex: L_{0_i} = 7x20m,\ T = 3000K,\ r_i = 30km \rightarrow F_{observer} = 30W/m^2\ \ (1/30\,sun)\ P_{emitted} = 140GW$$

$$Ex: L_{0_i} = 7x20m,\ T = 2000K,\ r_i = 30km \rightarrow F_{observer} = 6W/m^2\ \ (1/170\,sun)\ P_{emitted} = 28GW$$

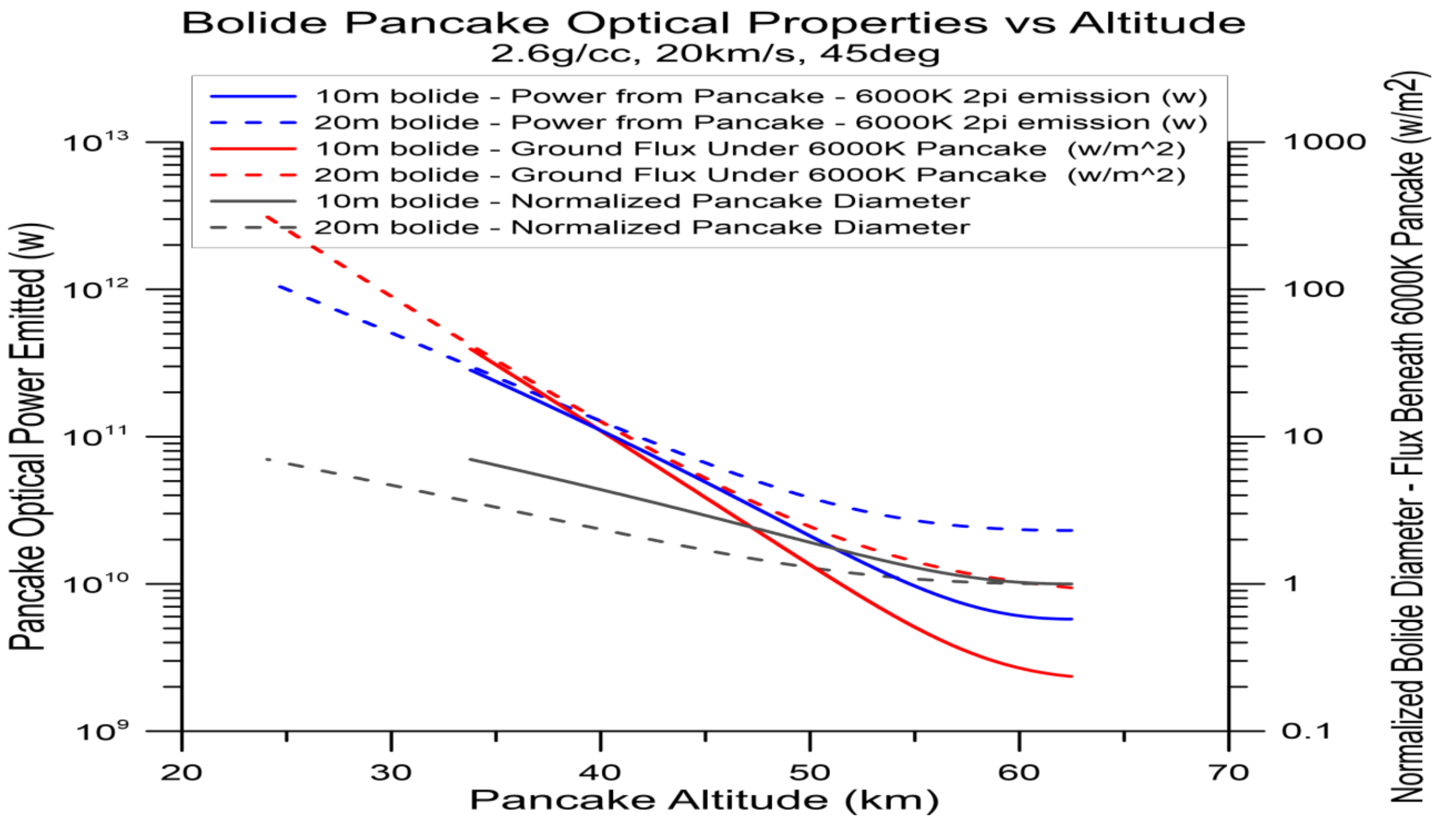

**Figure 70 –** Pancake model diameter, power emitted assuming a 2 Pi radiating 6000K flat surface and ground flux directly under bolide vs altitude during entry at an exo-atmospheric speed of 20 km/s, 45 degree angle of attack and 2.6g/cc. The breakup (plastic deformation) begins at about 62km and the burst (7x expansion) occurs at about 34 km for 10m and 24 km for 20m diam. A 2 Pi radiating hemispherical surface would have 2x the total power and 2x the flux. *A 4 Pi radiating flat surface would have 2x the total power and the same surface flux while a 4 Pi radiating spherical surface would have 4x the total power and 2x surface flux.* Comparing to the space measured case below (~10m – 4Pi) is reasonable given bolide unknowns, though there may be a larger radiating area possibly caused by fragmentation into small pieces during burst.

These number calculated are for the closest point directly under the burst. This still is far away (60x) from the ~ 30x sun (30,000W/$m^2$) reportedly "observed," but the 500W/$m^2$ would be felt as a mild thermal signature (1/2 sun). If the 30x is to be literally interpreted, then we are still a factor of 60x too small even for the T=6000K case. Possible resolutions of the calculated BB optical signature and Chelyabinsk estimation for the optical signal are:

- Our calculations are based on a static model. A real event may have short time scale flux peaks.
- Additional debris area generated at the burst which briefly glow brighter due to larger surface area.
- Line emission from the shocked and super-heated air/plasma
- The "30x sun" is qualitative and could be too high. However, the UV and some retinal burns need to be factored into a comparison.

***Comparison of Optical and Acoustical Signature* – Correlation vs De-Correlated Signatures** – The comparison between the optical and acoustical signatures is the key as to why our mitigation technique works. The speed of the optical signal ($3x10^8$ m/s) compared to the acoustical signature ($3x10^2$ m/s) means that the optical signal arrives one million times faster than the acoustical signal. To calculate the correlation lengths that are appropriate, we need to look at the two relevant distance scales.

- Spatial scale of production – asteroid size and maximum "pancake deformation" before burst. These are small with the asteroid fragments of interest being about 10m diameter and the maximum pancake ratio being about 7 for a pancake diameter of 70m. The optical light crossing time for these production scales is of over 0.1 microsecond while the acoustical cross time is of order 0.1s.
- Spatial scale of line of sight varies from a minimum of about 30km (typ. burst height) to the horizon scale for a 30km burst, which is about 600km.
- Time scale of production is of order 1 second – ablation heating time and acoustical shock wave production time, both set by the speed of the bolide in the atmosphere (~ 10-20km/s) and the atmospheric traversal distance during the critical production of the optical and acoustical signatures.

Comparing the optical and acoustical propagation time scales to the production time scale immediately shows that the optical signatures from all the fragments (out to the horizon) are correlated (differential travel times less than production time) while the acoustical signatures are de-correlated (differential travel times much greater than production time). This observation, as mentioned previously in a slightly different context, is why this program works. The acoustical signatures are de-correlated over the various travel paths and **hence peak pressures**, and the acoustical fluxes do not add while the optical flux is correlated in time and does add.

***Comparing Optical and Acoustical Flux at Observer*** – There are some well measured observations of the optical signature from a bolide. An example is the June 6, 2002 event shown in the figure below. It reached a peak optical power of about 9TW (4π) with a FWHM time of about ¼ second. If the effective temperature of this object was 6000K, which has an object surface flux of 74MW/m$^2$, this would imply an effective radiating surface area of about 1.2x10$^5$ m$^2$. This implies an effective spherical radiating diameter of about 200m. This is clearly not the physical size of the bolide, but rather represents an equivalent diameter of a sphere whose surface area would radiate 9TW if the spherical shell were at 6000K. This is very sensitive to the source effective temperature, which we do not know well. As shown above, the optical flux from a 10m diameter asteroid/fragment moving at 20km/s is only about 2.4W/m$^2$ at "ground zero" (shortest path) for T=6000K, 150mW/m$^2$ for T=3000K, and 30mW/m$^2$ for T=2000K, while for the 7x diameter pancake at burst the flux at ground zero is 120W/m$^2$ for T=6000K, 7.4 W/m$^2$ for T=3000K, and 1.5W/m$^2$ for T=2000K. The **peak acoustical flux** is about 1500W/m$^2$ at ground zero. For a 10m bolide with no pancaking, the acoustical to optical flux ratio is about 600x for T=6000K, 10,000x for T=3000K, and 50,000x for T=2000K. For a 7x pancake diameter, the acoustical to optical flux ratio is about 13x for T=6000K, 200x for T=3000K, and 1000x for T=2000K. **It is important to keep in mind that the acoustical flux (and pressure) is rapidly evolving due to the blast wave time evolution (Friedlander function), while the optical flux may be more slowing evolving**. Each pulse is unique to the fragment's parameters (speed, composition, attack angle…)

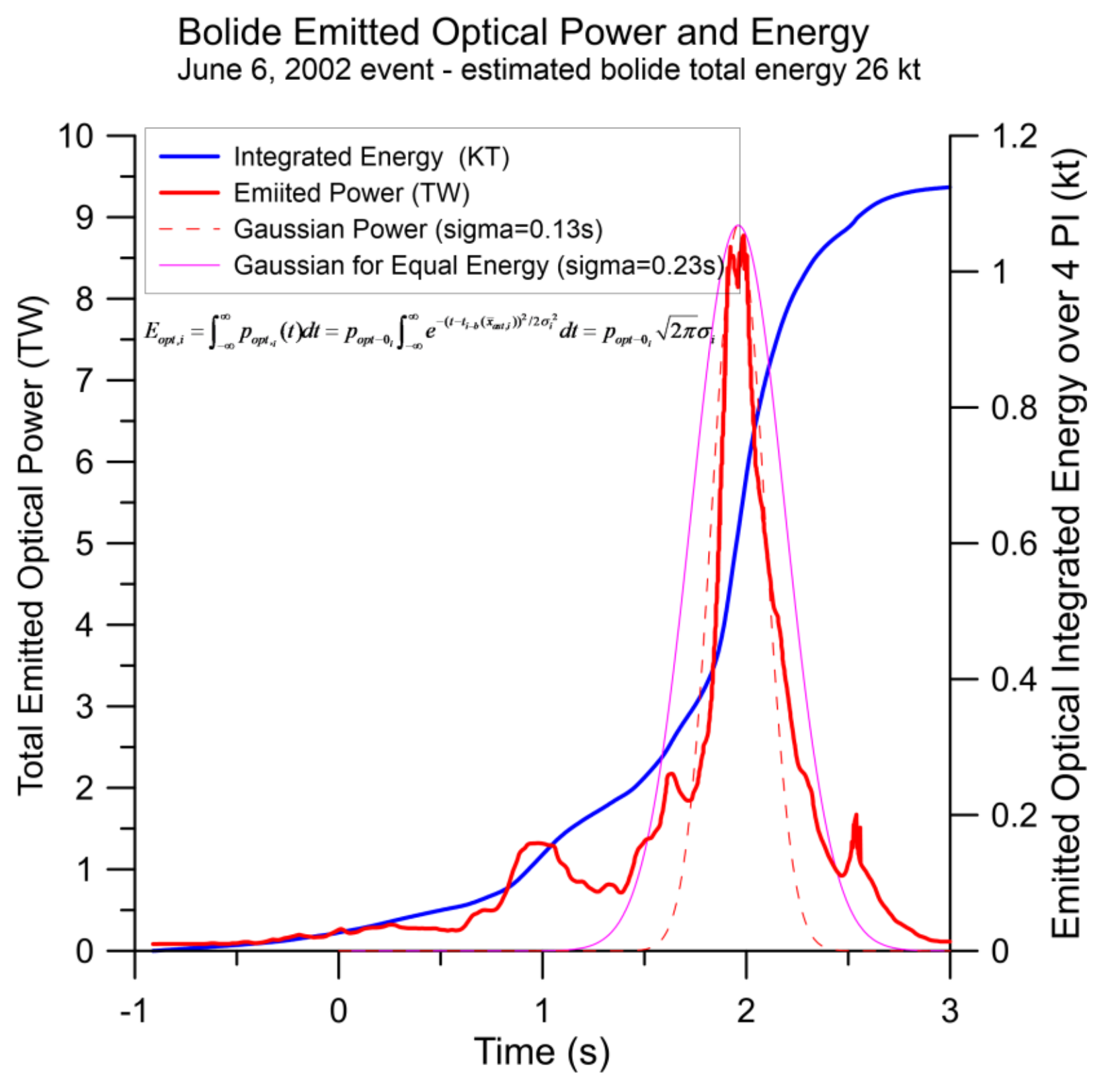

**Figure 71** – Optical signature measured from satellite observations for an event on June 6, 2002. With an estimated total yield of 26kt the optical integrated fraction is about 4% (1.1Kt). This is roughly consistent with a 7x diameter expansion pancake model and a 6000K pancake temperature. The optical output fraction will vary depending on many parameters including composition. The 26kt yield is comparable to a 10m diameter bolide. FWHM is about ¼ sec. Time is relative to 04:28:20 UTC. Fitted Gaussians shown. Note significant energy outside (before, after main pulse). Two fits shown.

A simple global estimate comparison of the acoustical to optical ratio is that blast energy from 10m diameter asteroid/fragment with 2.6g/cc density, 20km/s exo-atmospheric speed, and 45-degree attack angle is about 48kt, or 200TJ. With a nominal 1 second atmospheric traversal time, this gives an average acoustical power of about 200TW. Compare this to the optical power for the "7x pancake" of order 140GW (for T=3000K) or 2.3TW (for T=6000K) during this same ~ 1 second production time. The average acoustical power is about 1400 times larger than the optical power in this example for T=3000K and 90 times larger than the optical power for T=6000K. The ratio of acoustical to optical power and energy depends on the specifics of the asteroid/fragment size, density, speed, angle of attack, and material thermal/physical properties, but the general conclusion for all asteroids/fragments of relevance here is that the acoustical component is significantly larger and more dangerous. For comparison, in observations of smaller objects [1], [3], there is about **a 10x ratio of total asteroid KE to optical energy noted with significant variation.**

***Chelyabinsk event comparison*** - There are significant systematic errors in the measurements (both optical and acoustical) with 17% conversion from total exo-atmospheric energy to optical being reported from the Chelyabinsk event [48]. In this same paper, the estimated total optical yield is 375TJ with a peak optical power of 340TW. The total exo-atmospheric energy of this event is estimated to be approximately 450 - 550kt with 90kt being estimated to having gone into the optical signature (about 16-20% of exo-atmospheric energy). The peak over pressure is estimated to be 3.2±0.6kPa, which is in reasonable agreement with our estimation of the blast wave for the Chelyabinsk event assuming a single bolide of 19.5m diameter, 19.2 km/s speed, 20 degrees above the horizon attack angle, and density about 3.3g/cc. The density is not well known, but the yield strength is estimated to be about 0.7MPa at first breakup and then rising to about 4MPa at burst. It is possible that the outer layers were of lower density than the inner portion or that the bolide was heterogeneous enough to have a wide range of yield strengths. The bolide path came within about 40km (south) of the town of Chelyabinsk. The peak pressure estimates of 3.2kPa is for the town of Chelyabinsk which is about 40km North of the closet asteroid ground track. The comparison to the Chelyabinsk event is discussed further below [49]. Ground and building blast wave reflections were likely significant.

***Ground zero optical flux from June 6, 2002 Event*** – Another comparison is to use the June 6, 2002 event, which had a peak optical power of approximately 9TW with about 1.1kt going into the optical signature and an estimated total yield of 26kt. The 26kt total yield is subject to significant error. If we assume the 26kt total energy is correct, this would give a 4% conversion of total KE to light which is significantly lower than that measured for Chelyabinsk. This is not unexpected as the optical yield is dependent on the bolide composition and details of the breakup. **The flux at ground zero (assuming a 30km altitude burst) for the June 6, 2002 event is 0.8kW/m$^2$**. This is about 7x larger than the 120W/m$^2$ for T=6000K 7x pancake for a 10m fragment calculated above.

***Assumed conversion of blast wave energy to optical energy*** – Given the many uncertainties **we assume a nominal 10-20% conversion** of the energy in the acoustical signature (blast wave) to the energy in the optical signature as well as an analytic relationship between exo-atm KE and $E_{opt}$. **The overall conclusions of this paper do NOT depend on the details of the conversion of the acoustical to optical signature.**

***Comparing Optical and Acoustical Flux at Observer for Large Fragment Clouds*** – Since the optical signatures are time correlated while the acoustical one are de-correlated, we need to consider the cases where we create a very large number of fragments for very large asteroids. In the case of Apophis, for example, we split the parent asteroid into roughly 30,000 fragments, but the fragment cloud is spatially spread out. Hence, while the optical signatures are time correlated, the optical signature from each fragment that is distant is much lower in flux ($1/r^2$ dependence) and thus the overall optical flux is still very low. Once the fragment cloud is spread out, the acoustic signature is vastly mitigated as we showed in great detail earlier.

***Uniform Light Sheet Model*** – For the case of a large spread in fragments, a simple optical model is to use a uniform light sheet. If we assume the fragments are spread over a region that is large compared to the burst altitude (typ ~ 30km), then a first order assumption for optical emission is to model this as a uniform sheet of light with optical power per area (flux) of $F_{sheet}$ (W/m$^2$) illuminating both up and down. The observed optical flux on the ground is independent of the position on the ground for a large light sheet and it is also independent of the distance to the burst altitude (assuming the sheet is large compared to the burst altitude). This is the precise analogy to a uniformly charged dielectric with the electric field being $\sigma/2\varepsilon_0$ independent of the position of the observer, where $\sigma$ is the total charge per unit area. **In our case, the observed ground flux on is simply $F_{obs}$ (W/m$^2$)= $F_{sheet}$ (W/m$^2$)/2, where $F_{sheet}$ is the total optical flux (both sides).**

***Light Sheet Model and Horizon - Curvature of Earth*** – We have assumed a "flat Earth model" for simplicity. One justification in addition to simplicity is that the horizon scale for a typical 30km burst is approximately 600km (radius). This is 20 times larger than the 30km burst altitude. There are two additional effects that aid us.

- With the Earth's curvature there is an additional time delay for the fragments entering the atmosphere. This helps to de-correlate the optical pulse.

- The line-of-sight path from observer to any fragment is longer and traverses more atmosphere and hence has more absorption and scattering of the optical pulse.
- Fragments beyond the horizon for an observer will have the optical pulse blocked by the Earth.
- While local fragments near the observer are the largest contribution, the above effects do mitigate the total optical flux.

***The Optical Case for a 100m Diameter Asteroid*** – Assume a nearly worst-case scenario of intercepting a 100m diameter asteroid 1, 10, and 30 days prior to impact with a 1m/s fragment spread speed and breaking it into 1000 fragments. The 1000 fragments will be spread out over an approximately 200, 2000, 6000km diameter (sheet) for the 1, 10, 30 day cases, respectively. If we assume a peak power similar to the June 2002, 26kt case (close to a 10m fragment) shown in the previous figure, with a maximum power of 10TW optical for each fragment, the total power for the 1000 fragments (assuming they all arrived within one second of each other) would be 10PW.

- **1 day intercept** – fragment cloud is approx. 200km diameter with a peak sheet flux of $10PW/(200km)^2 = 250kW/m^2$ = 250x sun assuming no time spread. The ground flux is ½ this or ~ 125x sun. Even though the peak optical flash is less than 1 sec it is still very dangerous.
  - Optical time spread for 1 day intercept is about 10 sec for a 20km/s bolide with 1m/s fragments. This time spread helps reduce the total peak optical flux as the fragment optical pulses become de-correlated on time scale of order 1s.
- **10 day intercept** – fragment cloud is approx. 2000km diameter with a peak sheet flux of $10PW/(2000km)^2 = 2.5kW/m^2$ = 2.5x sun assuming no time spread. The ground flux is ½ this or ~ 1.3x sun.
  - Optical time spread for 10 day intercept is about 100 sec for a 20km/s bolide with 1m/s fragments. This time further reduce the total peak optical flux as the fragment optical pulses become de-correlated on time scale of order 1s.
- **30 day intercept** – fragment cloud is approx. 6000km diameter with a peak sheet flux of $10PW/(3000km)^2 = 280W/m^2$ = 1/4x sun assuming no time spread  The ground flux is ½ this or ~ 1/8x sun. Acceptable.
  - Optical time spread for 30 day intercept is about 300 sec for a 20km/s bolide with 1m/s fragments. This time spread significantly reduces the total peak optical flux as the fragment optical pulses become de-correlated on time scale of order 1s.

***The Optical Case for Apophis*** – Assume a nearly worst-case scenario of intercepting Apophis 1, 10 and 30 days prior to impact with a 1m/s fragment spread speed and breaking it into 30,000 fragments. The 30,000 fragments will be spread out over an approximately 200, 2000, 6000km diameter (sheet) for the 1, 10, 30 day cases, respectively. If we assume a peak power similar to the June 2002, 26kt case (close to a 10m fragment) shown in the previous figure with a maximum power of 10TW optical for each fragment, the total power for the 30,000 fragments (assuming they all arrived within one second of each other) would be 300PW. A formidable number!

- **1 day intercept** – fragment cloud is approx. 200km diameter with a peak sheet flux of $300PW/(200km)^2 = 7.5MW/m^2$= 7500x sun assuming no time spread. The ground flux is ½ this or ~ 3750x sun. Even though the peak optical flash is less than 1 sec it is still very dangerous.
  - Optical time spread for 1 day is about 10 sec for a 20km/s bolide with 1m/s fragments. This time spread helps reduce the total peak optical flux as the fragment optical pulses become de-correlated on time scale of order 1s.
- **10 day intercept** – fragment cloud is approx. 2000km diameter with a peak sheet flux of $300PW/(2000km)^2 = 75kW/m^2$ = 75x sun assuming no time spread. The ground flux is ½ this or ~ 38x sun. This is of concern though the short time pulse and time spread help.
  - Optical time spread for 10 days is about 100 sec for a 20km/s bolide with 1m/s fragments. This time further reduce the total peak optical flux as the fragment optical pulses become de-correlated on time scale of order 1s.

- **30 day intercept** – fragment cloud is approx. 6000km diameter with a peak sheet flux of $300PW/(3000km)^2$ = $8.3kW/m^2$ = 8.3x sun assuming no time spread The ground flux is ½ this or ~ 4x sun. For a short flash this is acceptable.
    - Optical time spread for 30 days is about 300 sec for a 20km/s bolide with 1m/s fragments. This time spread significantly reduces the total peak optical flux as the fragment optical pulses become de-correlated on time scale of order 1s.

The real ground illumination pattern will be closer to larger flux directly under a fragment and smaller in between fragments. The sheet model is a crude approximation to the complex fragment distribution. We do more detailed simulations that include various parent diameters, speeds, compositions and attack angles.

***Energy vs Power Flux – Optical Flash and Combustion*** – The total energy flux, peak power, and time profile are all important for understanding the effect of the optical signature on the ground. For short optical pulses, the total energy flux is typically the most important metric for calculating the danger of combustion. This is a well-studied area from nuclear testing. Typical energy fluxes for ignition are given in $cal/cm^2$ [2] with **ground values of 5 $cal/cm^2$ ~ 0.2 $MJ/m^2$ being sufficient to ignite easily combustible materials such as leaves and paper.**

***Apophis Optical Flash Comparison to Total Energy*** – Calculating the total average optical energy flux by calculating the exo-atmospheric yield and then making a "reasonable assumption" about the optical coupling fraction helps put this into perspective. Assuming conservatively that Apophis has an exo-atmospheric energy of 4Gt ~ $1.6x10^{19}$J, we calculate the energy flux (J/m) for the 1, 10, and 30 day intercepts. **We parametrize the 4π optical fraction of the total yield as $\varepsilon_{opt}$**

- **1 day intercept** – fragment cloud is approx. 200km diameter with 4Gt gives a ground energy flux of $\varepsilon_{opt}$ *$1.6x10^{19}$ J $/(200km)^2$ = ½ $\varepsilon_{opt}$ *400 MJ/m2 = 1000 $\varepsilon_{opt}$ x "combustion limit" of $0.2MJ/m^2$. Even for $\varepsilon_{opt}$ =0.01 giving 10x "combustion limit", this is still extremely dangerous.
    - Time spread for 1 day is about 10 sec for a 20km/s bolide with 1m/s fragments. This time spread helps reduce the total peak optical flux as the fragment optical pulses become de-correlated on time scale of order 1s and may allow even the one-day case to be reasonable. More simulations are needed.
- **10 day intercep**t – fragment cloud is approx. 2000km diameter with 4Gt gives an energy flux of $\varepsilon_{opt}$ *$1.6x10^{19}$ J $/(2000km)^2$ =½ $\varepsilon_{opt}$ *4 MJ/m2 = 10 $\varepsilon_{opt}$ x "combustion limit" of $0.2MJ/m^2$. This is small for reasonable values of $\varepsilon_{opt.}$
    - Time spread for 10 days is about 100 sec for a 20km/s bolide with 1m/s fragments. This time spread significantly reduces the total energy flux as the fragment optical pulses become de-correlated on time scale of order 1s.
- **30 day intercept** – fragment cloud is approx. 6000km diameter with 4Gt gives an energy flux of = $\varepsilon_{opt}$ *$1.6x10^{19}$ J $/(6000km)^2$ = ½ $\varepsilon_{opt}$ *0.4 MJ/m2 = 1 $\varepsilon_{opt}$ *x "combustion limit" of $0.2MJ/m^{2.}$ This is small for reasonable values of $\varepsilon_{opt.}$
    - Time spread for 30 days is about 300 sec for a 20km/s bolide with 1m/s fragments. This time spread significantly reduces the total energy flux as the fragment optical pulses become de-correlated on time scale of order 1s.

***100m diam Asteroid Optical Flash Comparison to Total Energy*** – Using the same analysis as the Apophis case, but noting that the total exo-atmospheric energy of a 100m diameter asteroid (x3.5 smaller diameter and ~ 40x less energy for the same speed and density), allows using the same numbers for the optical energy flux, but divided by 40. The conclusion is that even a 1-day intercept for the 100m diameter asteroid is acceptable in terms of the optical term. We also showed the blast wave distribution was acceptable for 1-day intercept with the same assumptions of 1m/s fragment spread.

***Atmospheric Attenuation and Blackbody Source*** – There is significant atmospheric attenuation that varies with the complex and time varying nature of the atmosphere. It is also highly wavelength-dependent due to the many absorption lines. The observed flux depends on the source spectral energy distribution (SED) that is wavelength-dependent, as well as the atmospheric attenuation. Just as we calculated the attenuation vs acoustic frequency for the blast wave, we now model in detail the atmospheric absorption of the optical signature vs wavelength. We model the atmosphere using MODTRAN 6 under various conditions. We assume a "worse case" scenario with no clouds (clear skies in all directions), though we do include nominal stratosphere aerosols as well as all relevant molecular line absorption. We do a full analysis including the curvature of the Earth's atmosphere, as well as nominal atmospheric pressure and temperature vs altitude. For this we use the 1976 atmospheric model. To calculate the attenuation of the (wavelength integrated – bolometric) optical signature we assume a blackbody spectrum of the source and have produced models from 3000-7000K under numerous atmospheric conditions. We summarize a few here as well as parametrize the attenuation to allow inclusion into our fragmentation model, allowing an integrated optical signature solution for both a vertical and horizontal plane at the observer. Our model allows placing the observer at any location. We calculation the optical absorption from 0.3μm (UV) to 5μm (mid IR) with high resolution (typ ~ 0.1 cm$^{-1}$ in wavenumber or an equivalent resolution $R= \lambda/\Delta\lambda \sim 10^5$ ). We can propagate the optical solution from any altitude, but focus here on the breakup altitude of (typ.) 60km and the burst altitude of (typ.) 30km. We propagate the optical emission from every fragment to the observer to get a total energy flux at the observer. The observed spectrum is modified from a blackbody due to the wavelength dependence transmission of the atmosphere. We convolve the atmospheric attenuation with the assumed blackbody spectrum to show the observed spectrum for a 200km slant path as an example.

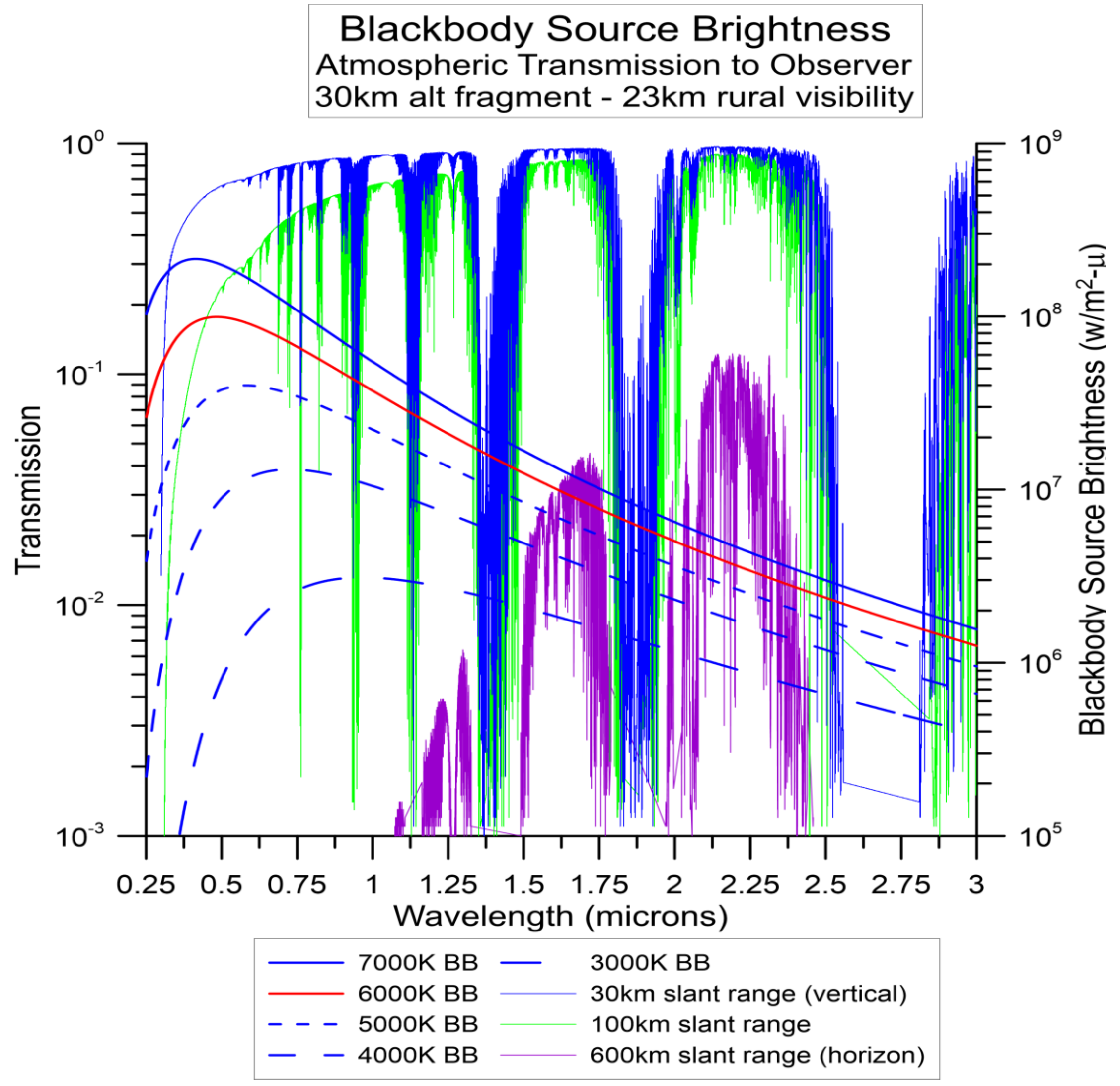

**Figure 72** – Blackbody source (fragment) brightness vs wavelength and temperature. Several slant length atmospheric transmissions are shown.

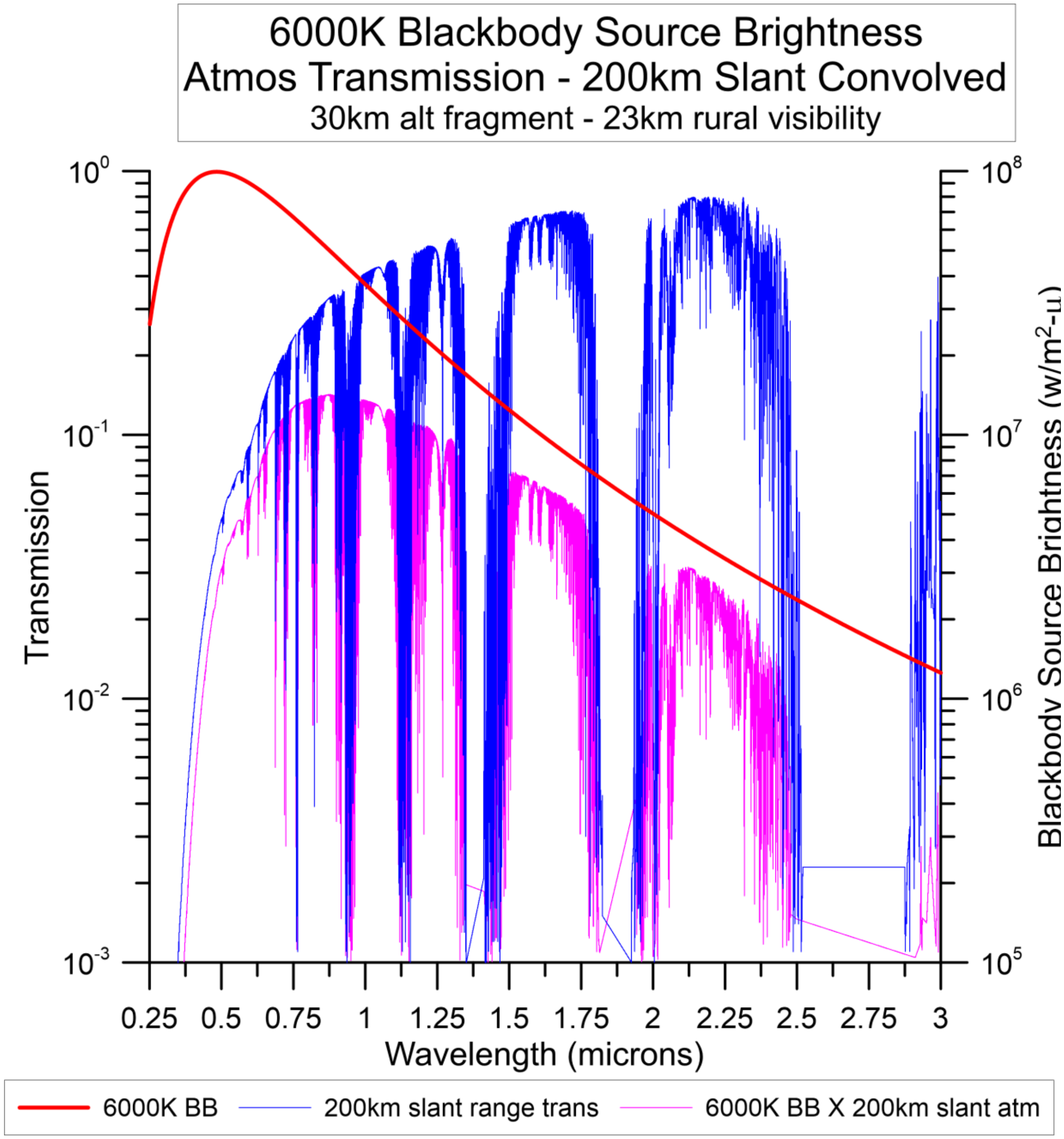

**Figure 73** – The observed spectrum for an assumed 6000K source spectrum at a slant range of 200km after passing through the atmosphere. The observed spectrum (magenta) is significantly reddened much like

$B(\lambda)(w/m^2-m)=$ source brightness (power per area per unit wavelength)

$P(\lambda)(w/m)=\int_{source\,area} B(\lambda)dA = B(\lambda)A_{source}$ source power per wavelength

Assuming $B(\lambda)$ is constant over source area

$\tau(\lambda)=$ transmision over path between source and observer

$r=$ distance from source to observer

$F(r)(w/m^2)=\dfrac{\int_0^\infty P(\lambda)\tau(\lambda)d\lambda}{4\pi r^2}$ = flux at observer assuming an isotropic source

Define the ratio between flux with and without atmospheric absorption:

$$\alpha \equiv F(r)/F_{no-atm}(r)=\frac{\int_0^\infty P(\lambda)\tau(\lambda)d\lambda}{\int_0^\infty P(\lambda)d\lambda}=\frac{\int_0^\infty B(\lambda)\tau(\lambda)d\lambda}{\int_0^\infty B(\lambda)d\lambda}$$

We then apply this as a correction factor to the "no atmosphere" case.

Note that $\alpha$ depends on the source brightness and not just the atmospheric transmission function

$F(r)(w/m^2)=\alpha F_{no-atm}(r)$

For a blackbody source:

$B(\lambda)d\lambda = B(v)dv$

*For both polarizations*

$$B(v)(\text{w/m}^2-\text{Hz})=\frac{1}{4}c\rho=\frac{2\pi h v^3}{c^2}\left[\text{e}^{hv/kT}-1\right]^{-1}=\frac{2\pi hc}{\lambda^3}\left[\text{e}^{hc/\lambda kT}-1\right]^{-1}$$

$$B(\lambda)(\text{w/m}^2-\text{m})=B(v)dv/d\lambda=-\frac{c}{\lambda^2}B(v)=\frac{2\pi hc^2}{\lambda^5}\left[\text{e}^{hc/\lambda kT}-1\right]^{-1}(\text{ignore minus})$$

$$=\frac{2\pi k^5}{h^4c^3}T^5x^5\left[\text{e}^x-1\right]^{-1}=6.028x10^{-7}T^5x^5\left[\text{e}^x-1\right]^{-1}$$

$$x=hv/kT=hc/\lambda kT=1.44x10^4(\lambda(\mu\text{m})T(K))^{-1}$$

$$B(\lambda)(\text{w/m}^2-\mu\text{m})=10^{-6}B(\lambda)(\text{w/m}^2-\text{m})=6.028x10^{-13}T^5x^5\left[\text{e}^x-1\right]^{-1}$$

$B(\lambda)$ peak $\lambda_p$ when $\text{d}B(\lambda)/d\lambda=0\rightarrow d(x^5\left[\text{e}^x-1\right]^{-1})/dx=0\rightarrow 5(1-e^{-x})=x\rightarrow x\sim 4.965$

$\rightarrow \lambda_p(\mu\text{m})T(K)\sim 2900\mu m-K$

$$S(v)(\text{w/m}^2-st-\text{Hz})=\frac{1}{4\pi}c\rho=\frac{1}{\pi}B(v)(\text{w/m}^2-\text{Hz})$$

$$S(\lambda)(\text{w/m}^2-st-\mu\text{m})=\frac{1}{\pi}B(\lambda)(\text{w/m}^2-\mu\text{m})=1.92x10^{-13}T^5x^5\left[\text{e}^x-1\right]^{-1}$$

$$\alpha \equiv F(r)/F_{no-atm}(r)=\frac{\int_0^\infty B(\lambda)\tau(\lambda)d\lambda}{\int_0^\infty B(\lambda)d\lambda}=\frac{\int_0^\infty \lambda^{-5}\left[\text{e}^{hc/\lambda kT}-1\right]^{-1}\tau(\lambda)d\lambda}{\int_0^\infty \lambda^{-5}\left[\text{e}^{hc/\lambda kT}-1\right]^{-1}d\lambda}$$

A resonable approximation (below) from fitting convolved BB source spectra is:

$\alpha(r)=be^{-ar}$ where a,b are determined from an atm model fit with a BB spectra

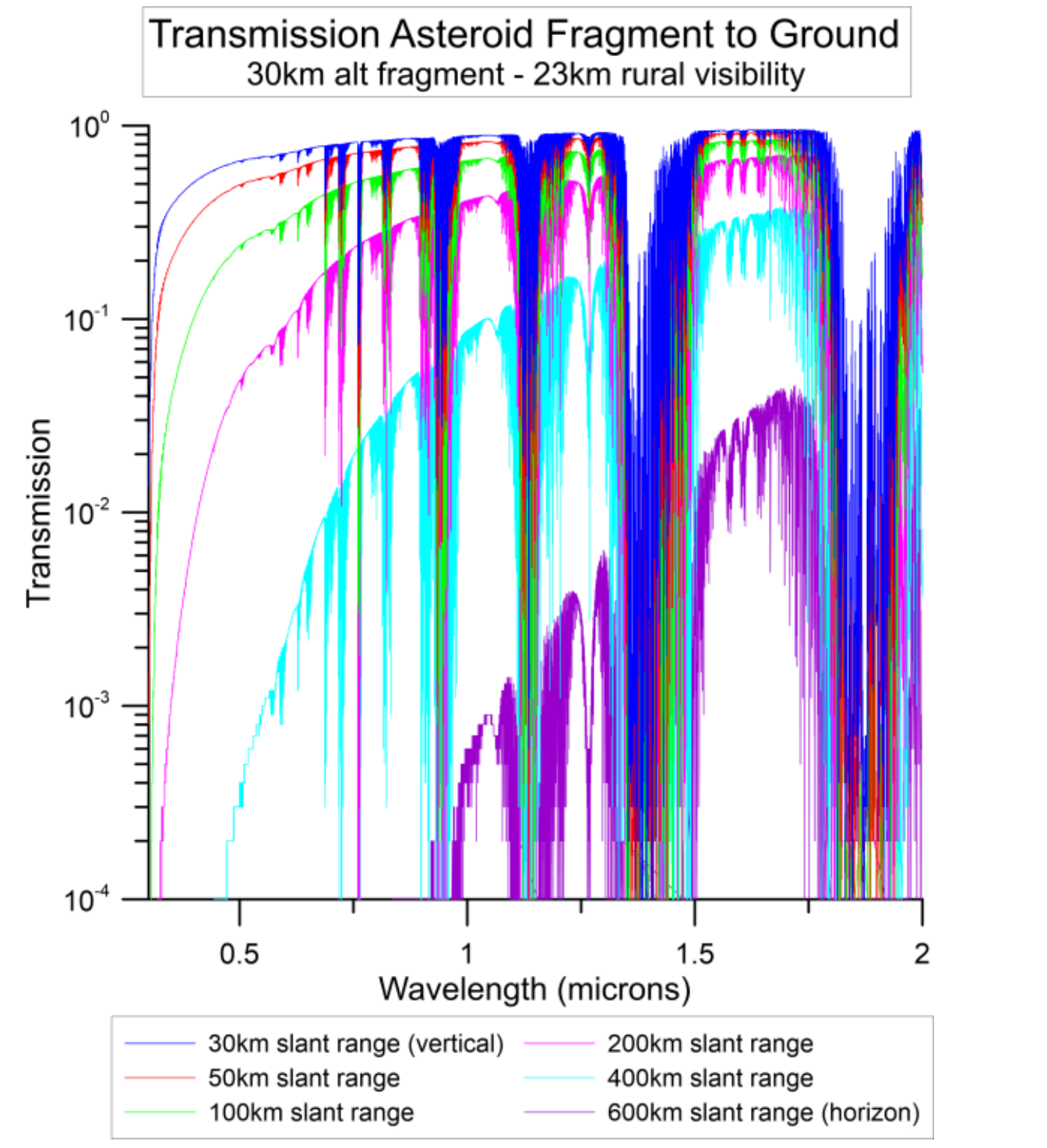

**Figure 76 –** Atmospheric transmission vs wavelength from 0.3 to 2 microns vs slant range for 30km alt. bolide.

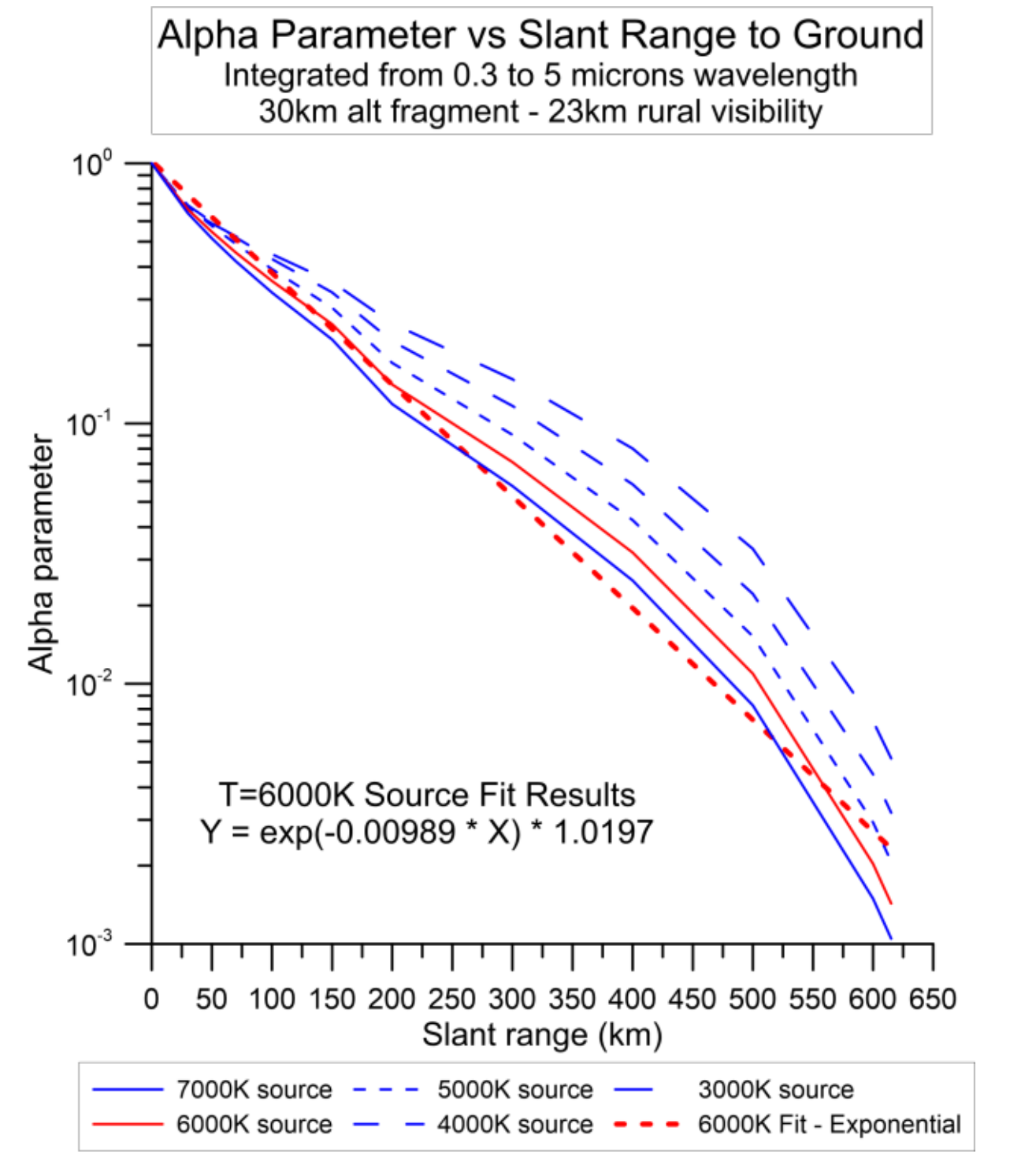

**Figure 74 –** Alpha parameter (BB convolved with atmosphere) vs slant range for 30km alt. bolide.

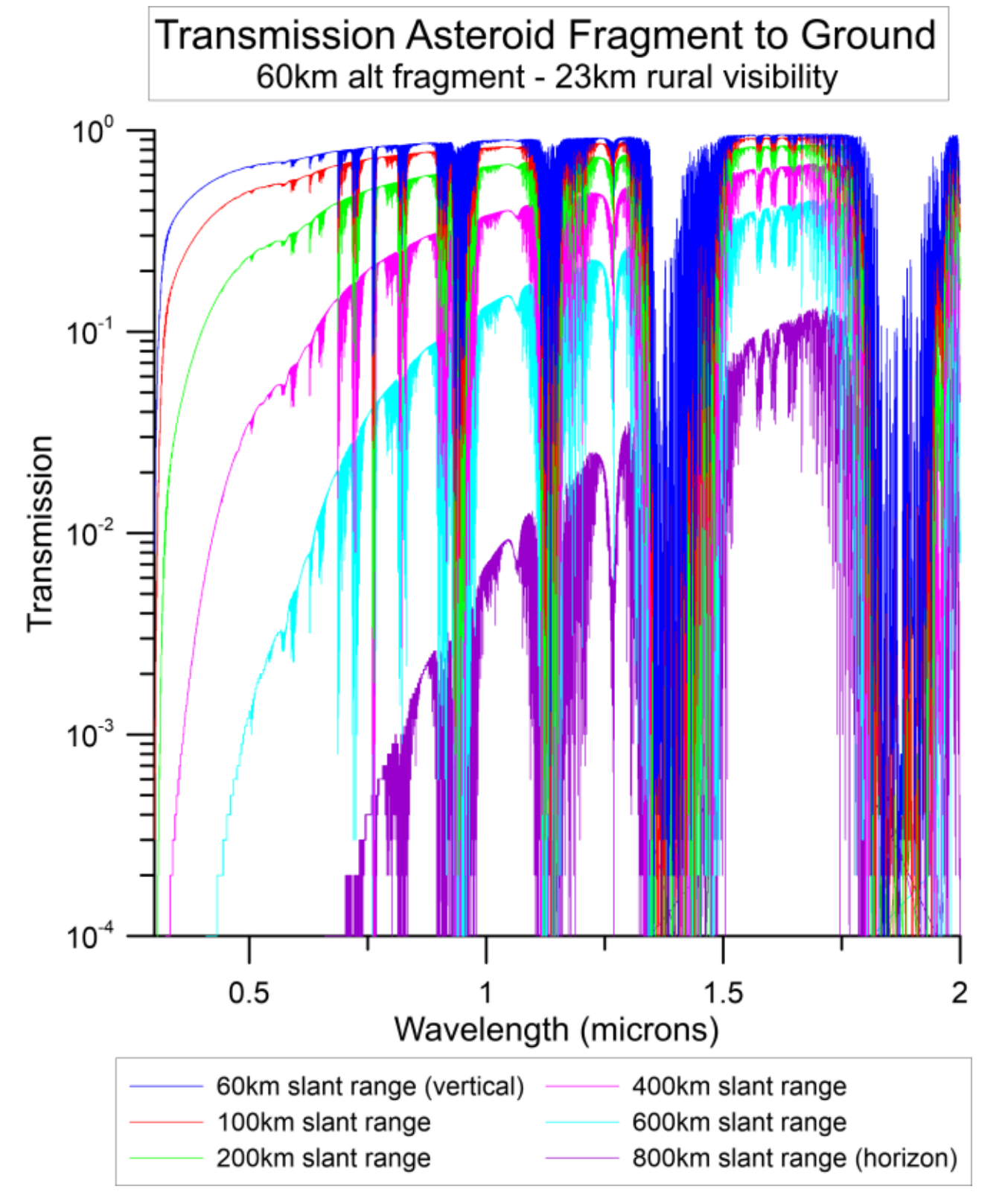

**Figure 77 –** Atmospheric transmission vs wavelength from 0.3 to 2 microns vs slant range for 60km alt. bolide.

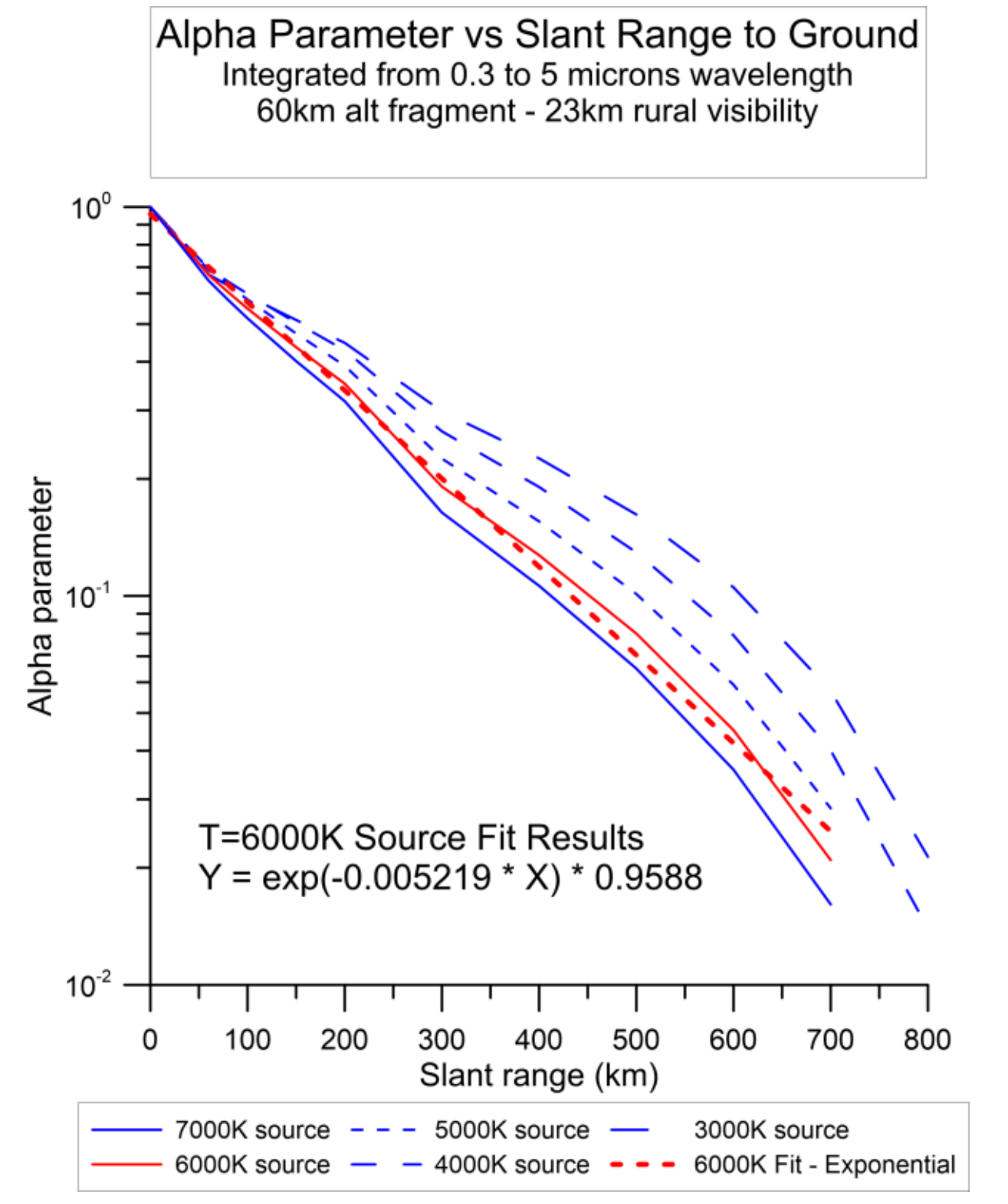

**Figure 75 –** Alpha parameter (BB convolved with atmosphere) vs slant range for 60km alt. bolide.

***Optical Signature Measurements from Orbital and Ground Based Assets*** – There are ground based and orbital assets equipped with optical sensors that are useful in measuring the optical signatures of impacting bolides. While much of the orbital data is not publically released, there is a public compilation maintained by NASA-JPL Center for Near Earth Object Studies (CNEOS Fireballs) that contained recently updated light curves. In this dataset are the measured bolometric total optical radiance (J) (over a limited pass band normalized to a 6000K blackbody) as well as the optical power (w/st) vs time, the latter being the "light-curves". Ideally we would have independent measurements of the bolide exo-atmospheric energy, total blast wave energy emitted as well as the measured optical energy and light-curve. In some instances, this data exists, while in others one parameter is inferred from the other rather than being independently measured. For example, the CNEOS Fireballs compilations infers the exo-atmospheric energy from the measured optical signature. In the case of Chelyabinsk, we have multiple datasets to cross compare including infrasound acoustics, optical data from both orbital assets as well as semi quantitative ground cameras (generally uncalibrated) as well as observed blast damage from broken windows and structural damage. In the case of the CNEOS measured optical signature the assumed relationship used to infer the exo-atm energy is shown below.

What is needed is to compare the measured acoustical data to the measured optical data to get a better understanding of the variability of the KE to optical conversion. Our conclusions do not require this but it would be extremely useful to the field to have this data. It is not currently available unfortunately.

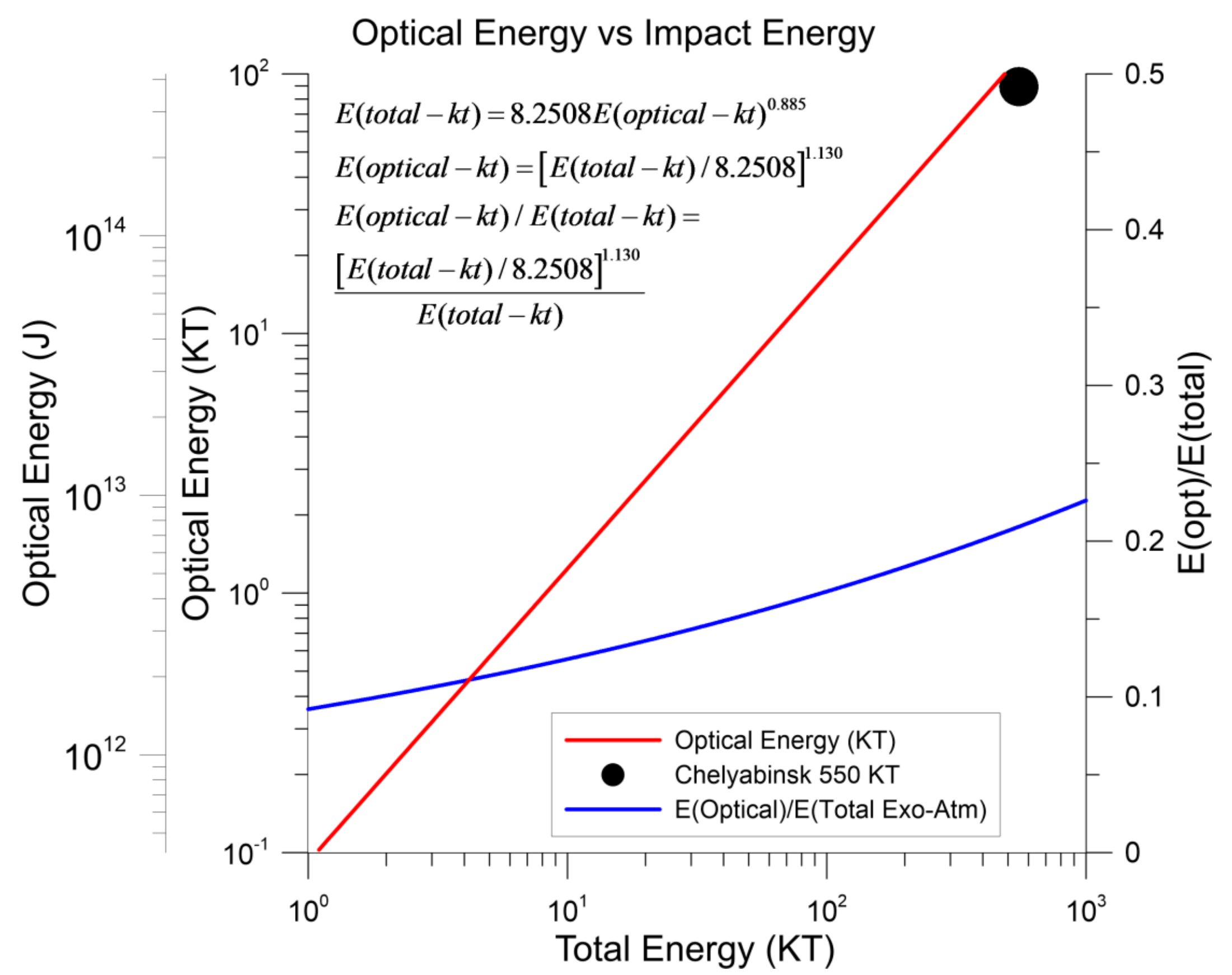

**Figure 78 –** Analytic relationship that is often used to relate the measured total optical energy in an atmospheric "fireball" to the inferred exo-atmospheric total kinetic energy (exo-atm "impact energy"). Note that the measurement is the total optical radiance integrated over the filter band with an inferred total optical signature assuming a T=6000K blackbody. The "total energy (exo-atm) is inferred analytically. The total energy is NOT measured but is inferred. In the range of most interest to us in this paper the inferred total exo-atm energy from 10-100 KT implies a conversion (from total to optical) of about 12-15%. This is about 3x higher than our typical simulation which assumes 10% of the blast wave energy into optical energy. We have run simulations that bound both lower and higher estimates of conversion into optical energy and they have little effect on the conclusions as we are already extremely conservative in assuming ALL the light from ALL the fragments arrive instantaneously at the observer as discussed below. There is a wide scatter of the limited observed conversion of estimated exo-atm energy into optical energy as the process of conversion is highly dependent on the composition, internal structure, speed, impact angle etc. Our overall conclusions are not dependent on these details.

$$E(total-kt) = 8.2508E(optical-kt)^{0.885},\ E(optical-kt) = \left[E(total-kt)/8.2508\right]^{1.130}$$

$$Energy\ fraction\ in\ optical: E(optical-kt)/E(total-kt) = \frac{\left[E(total-kt)/8.2508\right]^{1.130}}{E(total-kt)}$$

***Single vs Spatially and Temporally Distributed Optical Sources*** – The various optical impulse data relevant to ignition of ground materials virtually all assume a single source with a relatively short energy injection pulse time. In above ground nuclear testing for example, there is a single compact source that yields a sudden local energy injection whose thermal (optical/ radiative) pulse width scales roughly as the energy injection (yield) to the 1/3 to 1/2 power (Glasstone and Dolan – Chap VII) (similar to the $\tau \sim E^{1/3}$ for acoustical pulse width as we have discussed in detail in other parts of this paper). For example, a 1 KT "low altitude" (<30km altitude) airburst has a thermal pulse width of ~ 0.4s while a 10 Mt low altitude airburst has about a 20s thermal pulse width. Note that a nuclear detonation is not a good "physics model" for an asteroid (fragment) hypervelocity airburst. In analyzing the JPL CNEOS asteroid atmospheric airburst light curves from 2 – 440 KT total exo-atm kinetic energy (yield), there is a wide variation of light vs time as would be expected from a diverse population of asteroids with varying strengths, speeds, angle of entry etc but we find that a "reasonable" estimate of the light pulse duration is about 1 second. Coincidentally, this is similar to the optical pule width for a nuclear air burst of equivalent yield. For a detailed analysis of an event see section above on "Comparing Optical and Acoustical Flux at Observer". This number is not critical in our calculations below. In our case of spatially and temporally distributed fragments the optical power flux ($w/m^2$) at a given observer from each fragment arrives at different times, with varying pulse profiles and from different directions. The integrated **observed** energy flux ($J/m^2$) from each fragment arrives at different times, with different pulse duration and from different directions. These distributed source effects tend to greatly lessen the overall effect on ignition at the observer. In our analysis we take an extremely (and excessively) conservative view in that we simply add all the integrated energy flux from all fragments, after correctly for the atmospheric absorption for each optical path from fragment to observer. Since the fragment cloud is a 3 dimensional distribution on a curved (Earth) surface, there is a distribution of arrival times of each fragment over a time scale from the mean bottom to the mean top of the fragment cloud that is approximately given by the diameter of the fragment cloud when it hits the Earth's atmosphere divided by the closing speed:

$$Fragment\,Cloud\,Radius = \tau_{intercept-time-prio-to-impact} v_{disruption}$$

$$\tau_{fragment-arrival-time-dispersion} \sim 2\tau_{intercept-time-prio-to-impact} v_{disruption} / v_{parent-bolide}$$

$$Ex: \tau_{intercept-time-prio-to-impact} = 10\,days, v_{disruption} = 1m/s, v_{parent-bolide} = 20km/s \rightarrow \tau_{fragment-arrival-time-dispersion} \sim 80s$$

$$Ex: \tau_{intercept-time-prio-to-impact} = 1\,day, v_{disruption} = 1m/s, v_{parent-bolide} = 20km/s \rightarrow \tau_{fragment-arrival-time-dispersion} \sim 8s$$

Depending on the ignition target (dry grass, trees etc), there is cooling between each pulse arrival driven by heating from the pulse (typically from convection and radiation rather than conduction). Over the period of the example above for a 10 day intercept (Apophis example). 1m/s disruption and 20km/s closing speed (40s over the fragment pulses arrival times) there will be considerable cooling so that the effective ignition danger will be vastly less that than implied by the simple sum of all optical pulse energy fluxes. For the extremely case of a 100m asteroid with 1 m/s disruption speed and 20 km/s closing speed intercepted 1 day prior to impact the fragment time dispersion is about 8s. There will be modest cooling over 8 seconds and thus this is a relatively extremely case for such a large threat. Even with this extreme case (100m, 1day intercept) the maximum energy flux anywhere in the 200x200km ground patch, centered at ground zero, is only slightly higher than our goal of staying below $200KJ/m^2$ ($4.8\ cal/cm^2$) at every point on the ground. In this case assuming the analytic connection between the exo-atm energy and the total optical radiation as described above, with atmospheric attenuation included and assuming no cooling between pulses, there are some ground points slightly greater than $200KJ/m^2$ while for our assumed connection between the optical pulse and the total blast wave energy we are slightly lower than $200KJ/m^2$ . In all cases we have studied we do not achieve any significant ignition danger. Our cases are designed with the criteria of "keep both the acoustic and optical signature low enough to not cause any serious threat". If we were to take into account, the realism of the spatially distributed source of radiators (each fragment) with temporal arrival time variations and some realistic cooling then we stay well below any ignition even with the most extremely mitigation cases such as the 100m diam threat closing at 20 km/s and intercepted one day prior to impact.

To do a high fidelity ignition estimate would require detailed knowledge of the target material, nearby structure that might shade different directions, detailed cooling channel physics that depends on the target structure and surroundings etc. None of this is realistic in general so we do a worst case analysis of a simple sum. While we have computed projection effects for horizontal and vertical surfaces at the target for each fragment, we assume a worst case of full isotropic absorption. Hence our "summed energy flux ignition" limit for shredded paper/ loose dry grass of 200KJ/$m^2$ should be considered extremely conservative.

***Acoustic and Optical Horizons – Shadowing – Sky Glow***

There are a variety of phenomena that arise from the horizon effects. Most of these effects are sub-dominant to the primary line of sight acoustic and optical signature, but are worth noting.

***Primary line of sight horizon*** – As discussed previously, there is a line-of-sight horizon due to the altitude of the breakup and subsequent burst that affects both the acoustic and optical signatures. For the typical breakup altitude of 50km and burst altitude of 30km, the horizons due to the Earth's radius are about 800 and 600km slant range for the breakup and burst altitude, respectively. Both the acoustic and optical signature are greatly attenuated beyond the burst horizon. Secondary effects due to acoustic diffraction and atmospheric "ducting" allow some small over-the-horizon acoustic signature to remain, while the optical signature is severely attenuated with only minimal atmospheric scattering ("sunset effect") for ranges beyond the horizon.

***Acoustic and optical horizon propagation time scales*** – For a typical 600km burst altitude horizon, the acoustic propagation time is about 180 seconds, while the optical propagation time is about 2ms.

***Earth forward hemisphere fragment interaction for very large fragment clouds*** – For very large fragment cloud scenarios from early time interdictions that may be required for very large bolides, the entire forward hemisphere of the Earth relative to the bolide position is subject to fragment bombardment. Virtually all of this is beyond the local horizon due to the Earth's curvature. There is a slight increase beyond the forward $2\pi$ hemisphere due to the altitude of the fragment breakup and burst, but this is a small effect. The arc length at the edge of the forward hemisphere is $(\pi/2)^* R_E$ or about 10,000km giving an "over horizon" arc acoustic propagation of about 30,000 seconds and an optical propagation time of 33ms for an observer at the center of the ring (parent bolide velocity vector Earth intercept point), or twice the arc distance and propagation time for the "equator-to-equator" case. In this latter case. the arc length is $\pi/R_E$ or about 20,000km, giving an "over horizon" arc acoustic propagation of about 60,000 seconds and an optical propagation time of 66ms. This can be seen in some of the plots for blast wave peak pressure vs arrival time where the fragment cloud diameter is comparable to the diameter of the Earth,

Past large explosions such as the Krakatoa eruption and large nuclear tests as well as infrasound measurements of bolide strikes have shown acoustic signature with multiple Earth "acoustic circumnavigations". These effects would be measurable, though of minor consequence in our case. Another effect is caused by the fragment speed causing a time delay between the time difference of the observer "overhead" fragment atmospheric interaction and the fragment interaction in the atmosphere at the "equator relative to the observer," where there is a fragment traversal distance of $R_E$ resulting in a time difference of $\tau=R_E/v_o$ where $v_o$ is the parent bolide speed relative to the Earth. For a typical 20km/s bolide, this would be about $\tau=R_E/v_o$ =320 sec.

***Sunset and twilight atmospheric scattering effect on the optical signature***

Over the horizon optical signatures due to scattering in the atmosphere are a common occurrence every day at sunset and sunrise as well as in previous nuclear testing. In the case of large fragment cloud mitigation bombardment, this would be seen as a "spectacular light show" over a period of about $\tau=R_E/v_o$ =320 sec between the first fragments overhead and the last fragments at the "observer Earth horizon" for a 20km/s bolide. These would be seen as faint atmospheric glow from the many fragment hits. Just as in the atmospheric scattering sky glow past sunset, this effect becomes extremely dim (and reddened) for fragments that are interacting far below the local horizon. This is further diminished by the extremely long atmospheric path length that greatly diminishes the optical signature. A similar effect happens to the acoustic signature, though the atmospheric absorption is vastly less for very low frequency acoustic waves. Monitoring both the optical and acoustic signatures from multiple ground stations and the optical signature from space would be useful in post-strike analysis.

***Optical Signature Simulations*** – We have simulated a large parameter space of parent asteroids and intercept scenarios and present a few summaries here. In all cases, we fragment the parent asteroid and then follow each fragment from intercept to atmospheric entry through breakup and burst and through the optical signature phase, and then propagate the optical signature through the atmosphere to the observer. **We err on the conservative side and assume there is no cooling between optical pulses, and thus the energy flux adds. This is clearly excessively conservative, but allows us to access a worst-case scenario.** We treat the atmosphere as being "a clear day with no clouds," which is also a worst-case scenario. We look at two large asteroids in detail, a 100m diameter and Apophis (assumed to be 350m diameter). Each is treated as having a density of 2.6 /cc and an attack angle of 45-degrees. We cut the optical signature when the fragment is beyond the visible horizon for each case. **The bottom line is that even a one-day intercept of a 100m asteroid is acceptable, as is a 10-day intercept of Apophis. These are both remarkably short terminal defense intercepts for such large potential threats.**

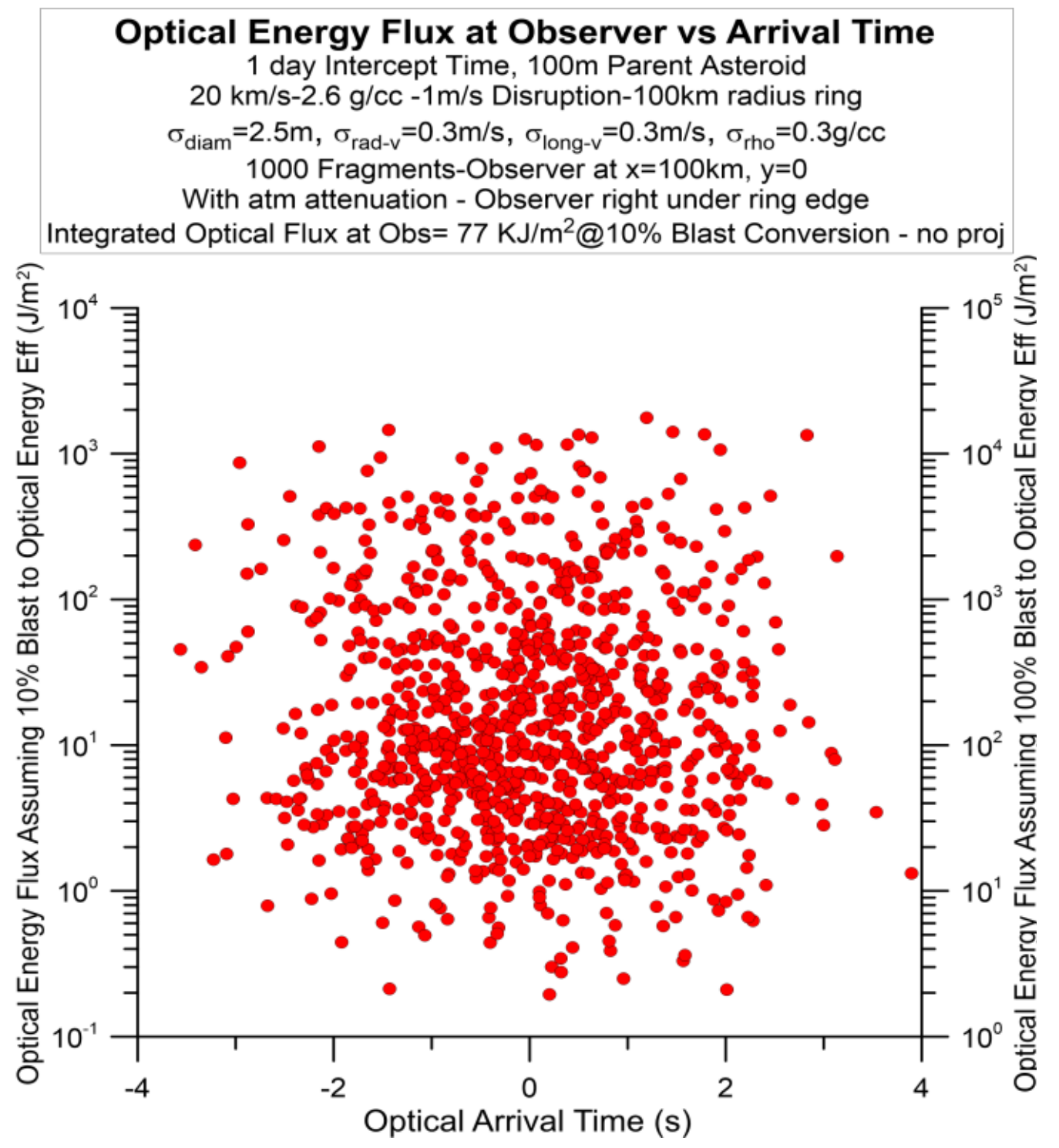

**Figure 79 –** Optical energy flux (J/$m^2$) at observer vs optical pulse arrival time for 1 day intercept of 100m bolide in 1000 fragments.

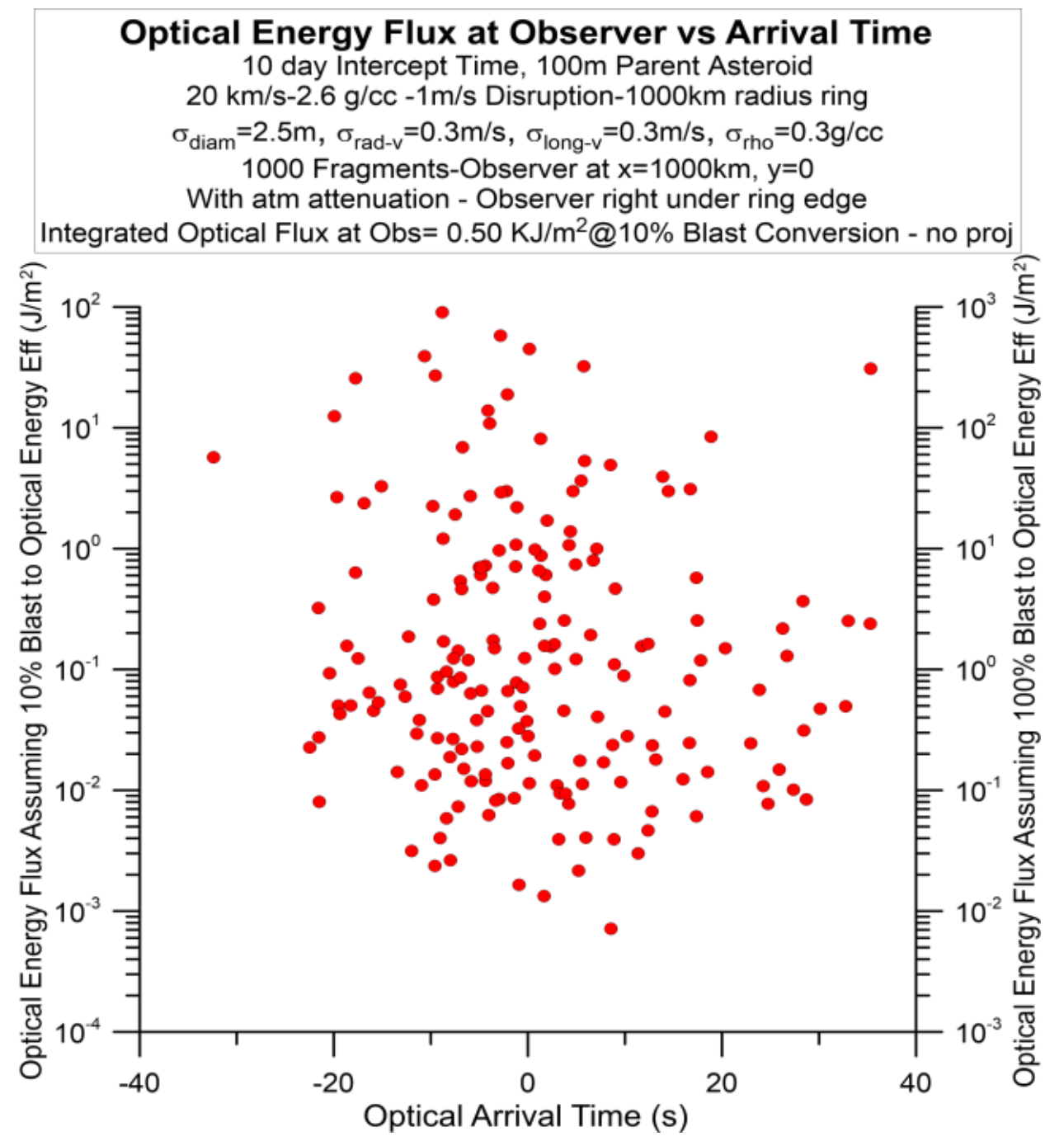

**Figure 80 –** Optical energy flux (J/$m^2$) at observer vs optical pulse arrival time for 10 day intercept of 100m bolide in 1000 fragments.

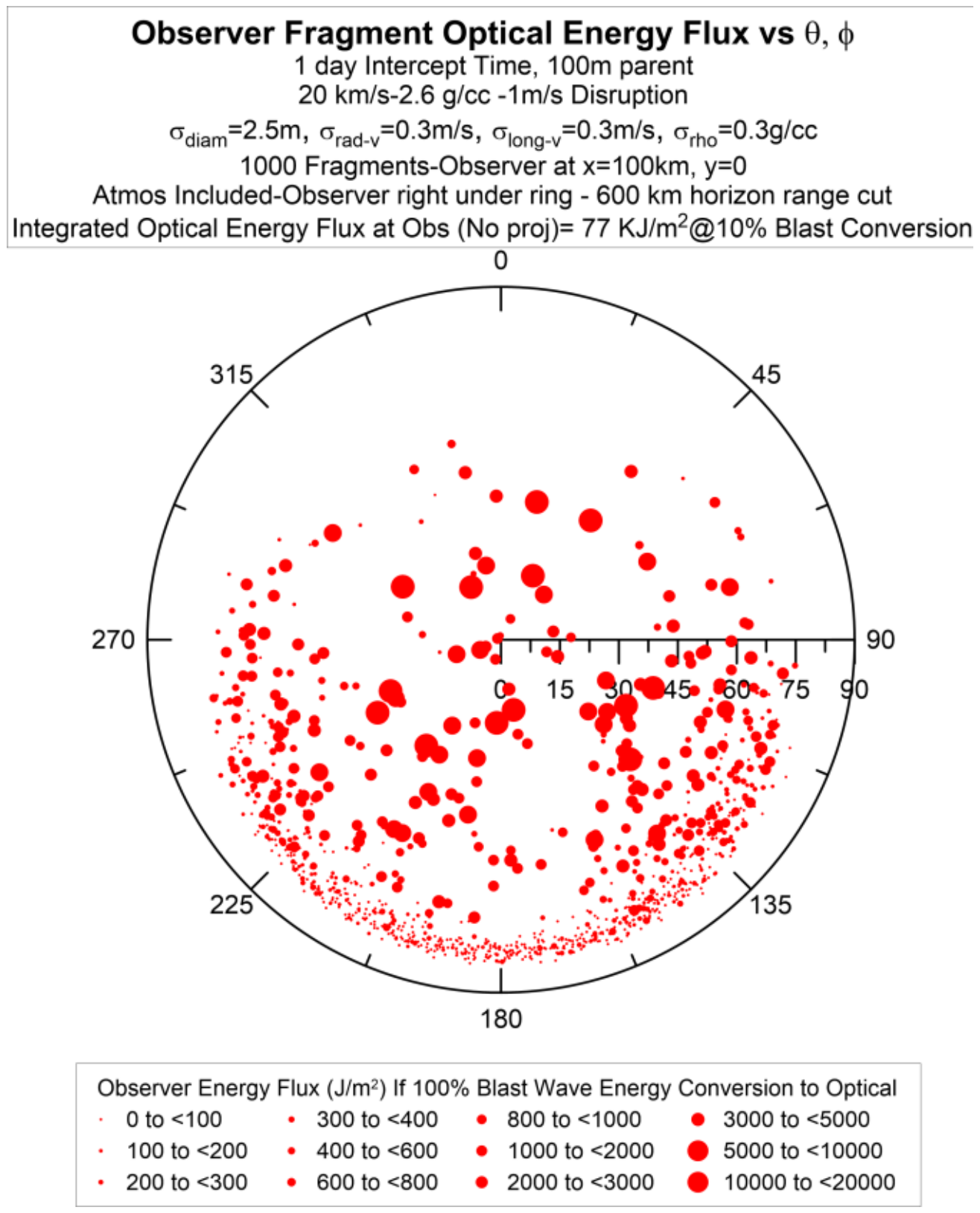

**Figure 82** – Observed fragment theta, phi for 1 day intercept of 100m bolide. Observer under edge of ring.

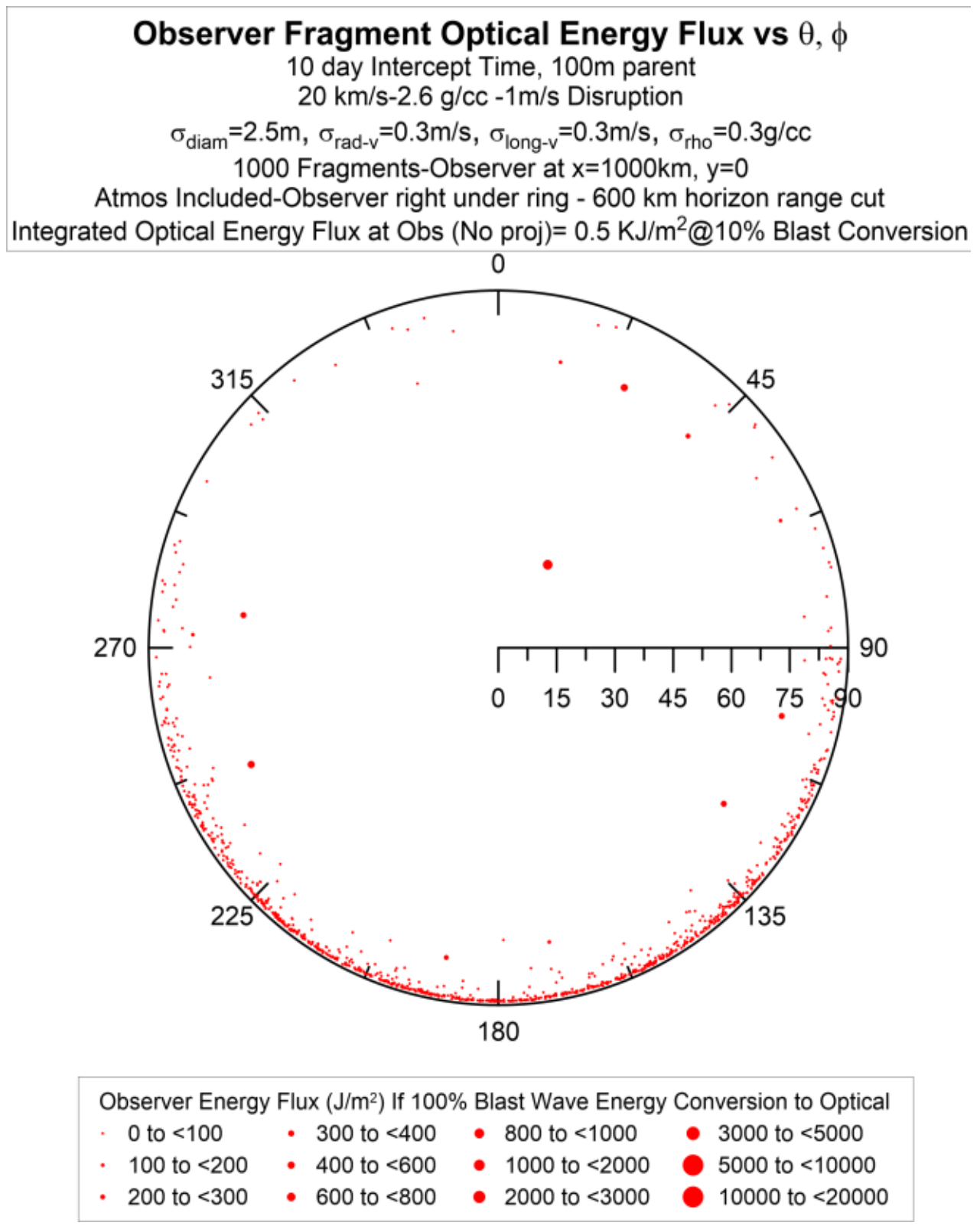

**Figure 81** – Observed fragment theta, phi for 10 day intercept of 100m bolide. Observer under edge of ring.

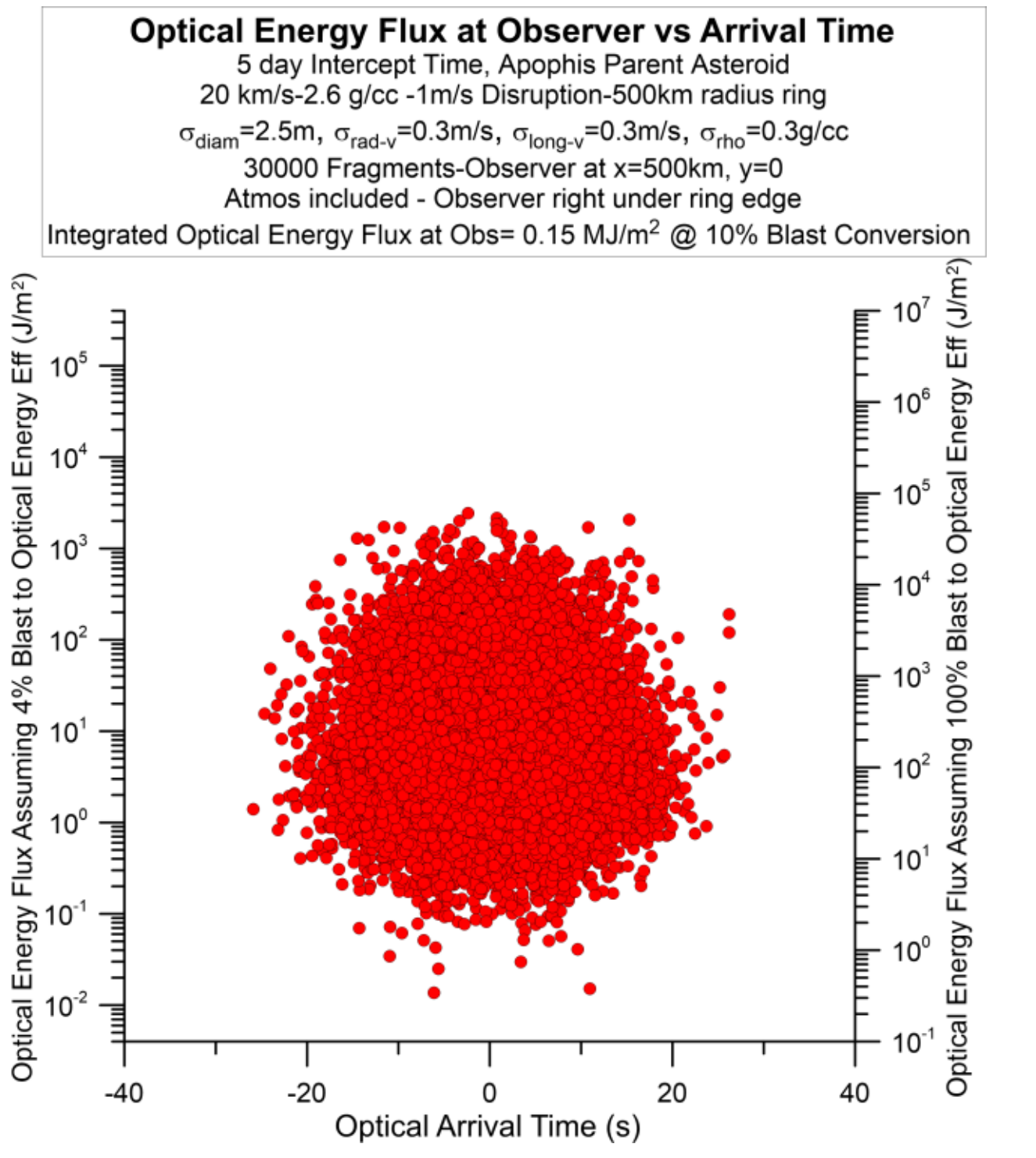

**Figure 83 –** Optical energy flux (J/m$^2$) at observer vs optical pulse arrival time for 5 day intercept of Apophis in 30000 fragments.

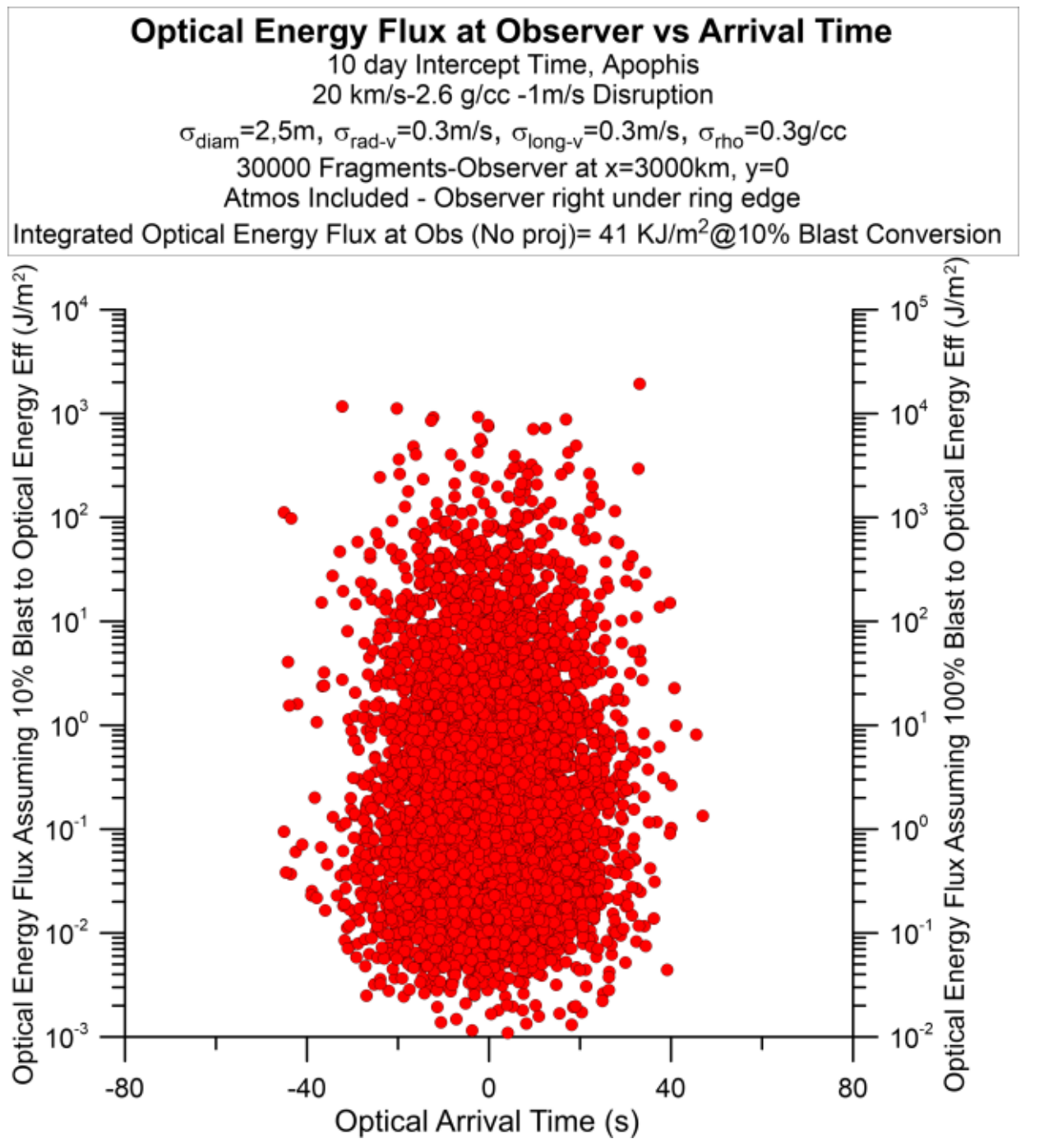

**Figure 84 –** Optical energy flux (J/m$^2$) at observer vs optical pulse arrival time for 10 day intercept of Apophis in 30000 fragments.

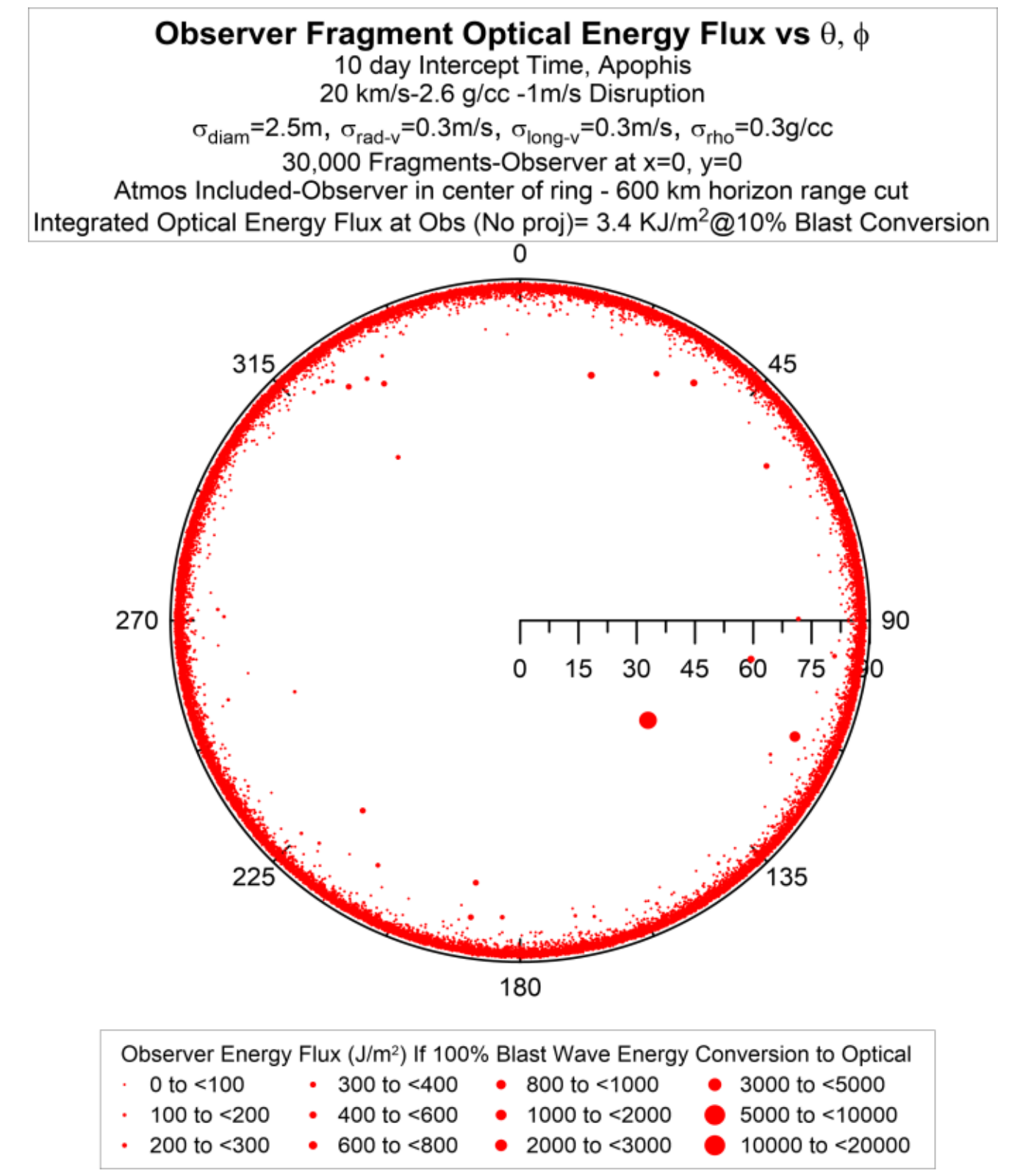

**Figure 85** – Observed fragment theta, phi for 10 day intercept of Apophis. Observer center of ring.

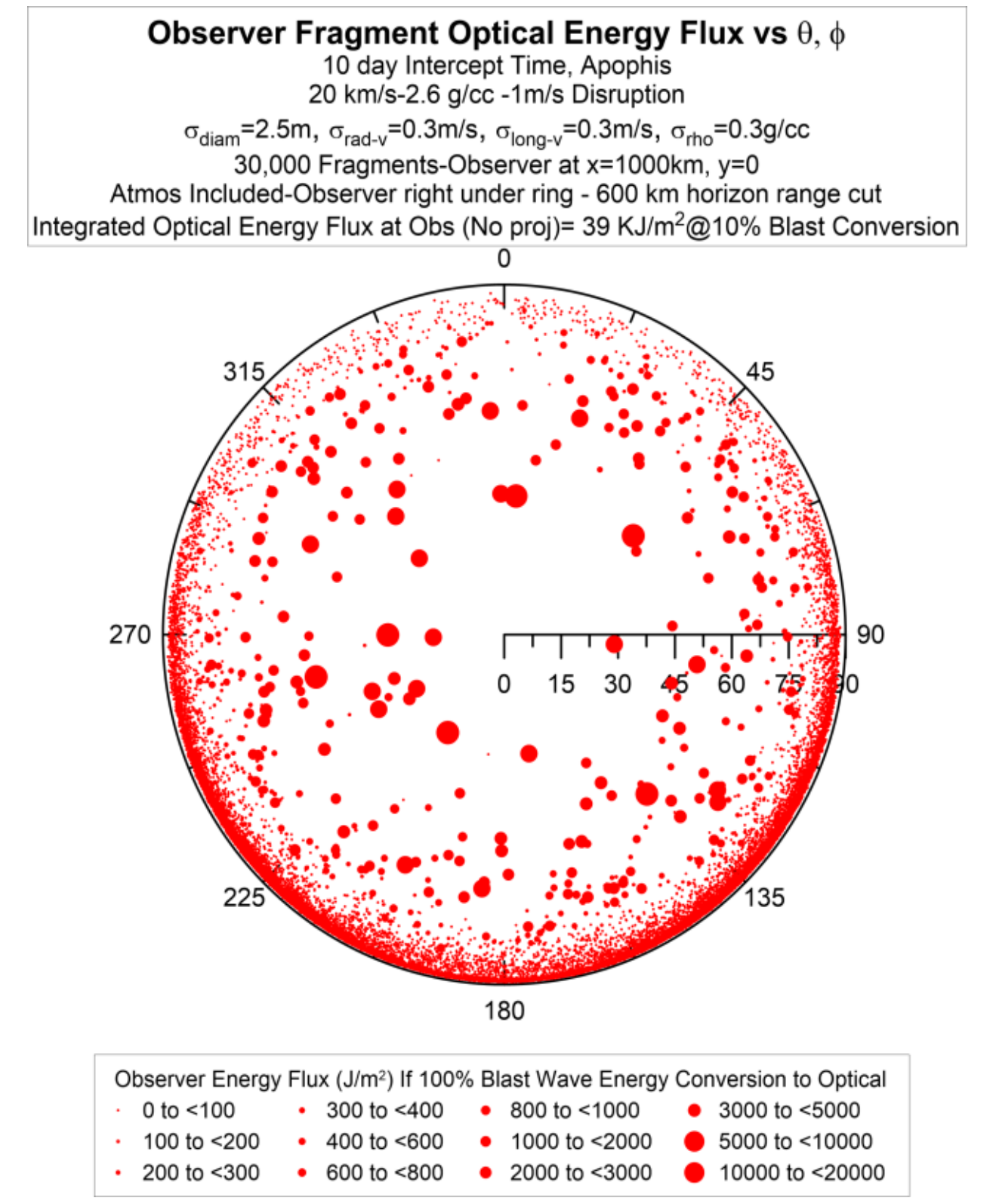

**Figure 86** – Observed fragment theta, phi for 10 day intercept of Apophis. Observer under edge of ring.

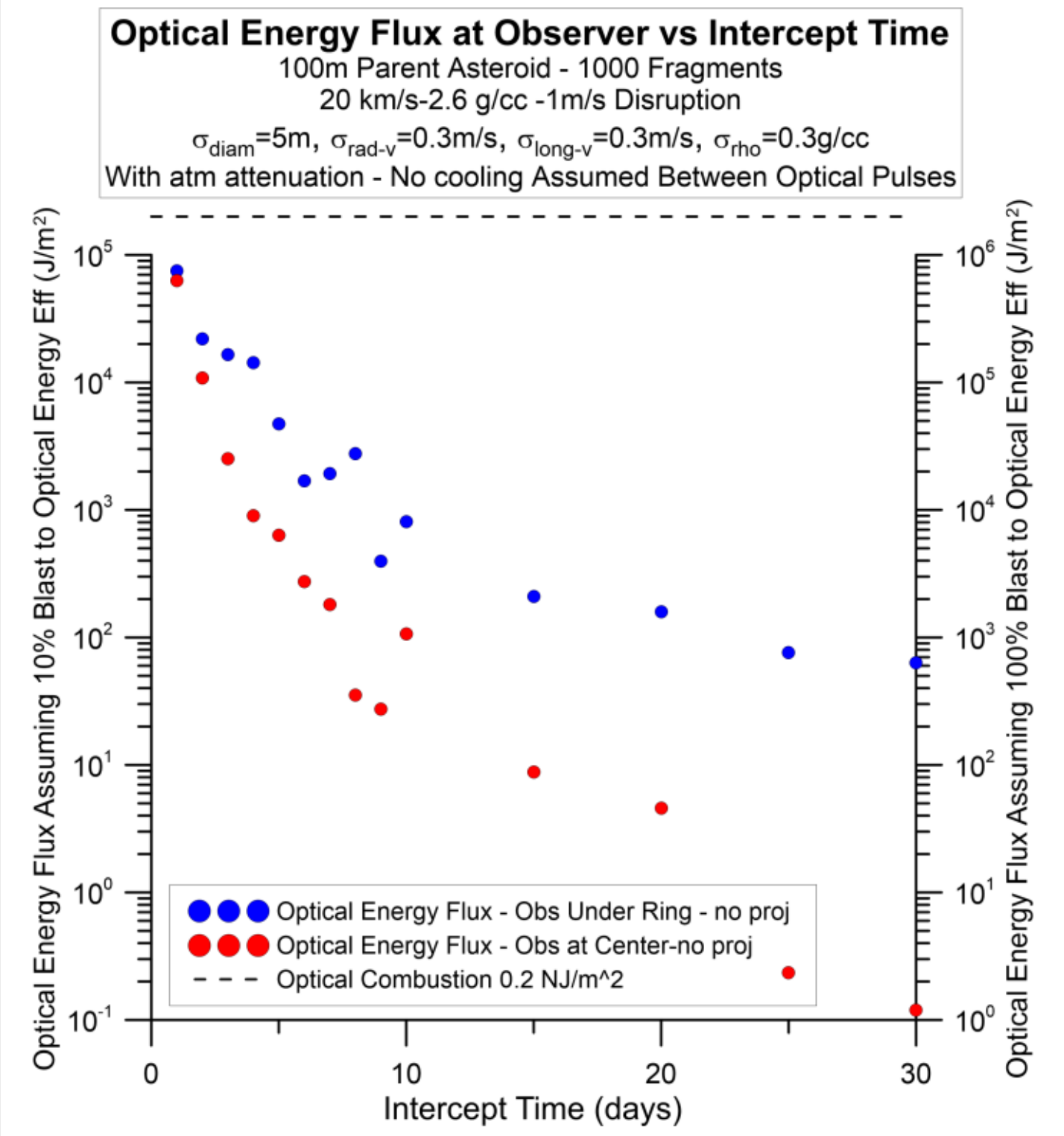

**Figure 88** – Optical energy flux (J/m$^2$) at observer vs intercept time for 100m bolide in 1000 fragments.

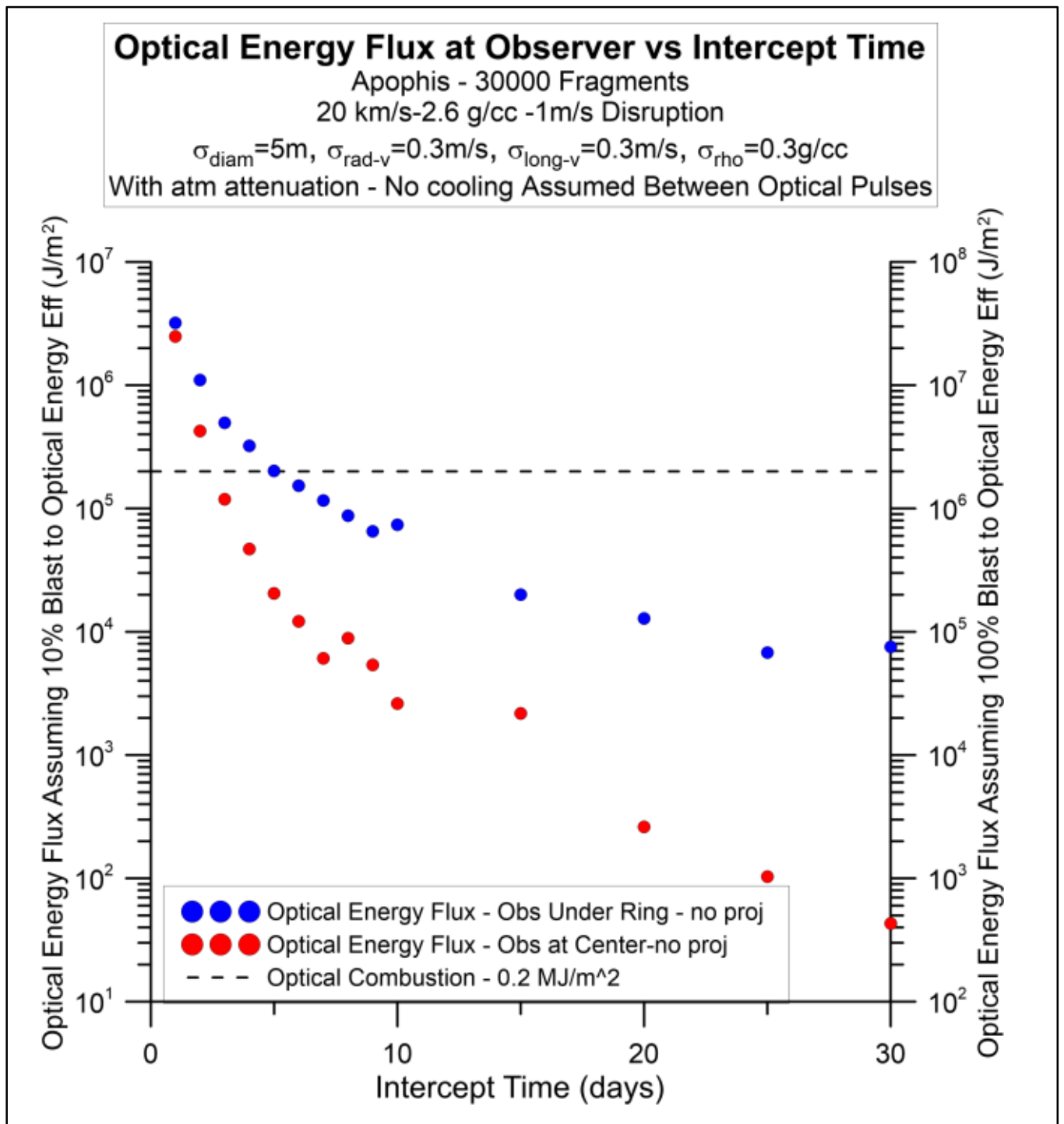

**Figure 87** – Optical energy flux (J/m$^2$) at observer vs intercept time for Apophis in 30000 fragments. Very conservative as no cooling between optical pulses is assumed.

***500m diameter Bolide Mitigation – Bennu***

The asteroid 101955 Bennu, with an equivalent diameter of 490m, is an object that poses a potential large-scale threat to Earth much like Apophis, with a yield, depending on the specific impact orbit. Its energy yield is ~ 4Gt at 20km/s, were it to hit Earth, is comparable to all the world's nuclear arsenal combined. With a period of 1.1955 years, it is nearly identical to Earth. It was discovered relatively recently in September 1999 and was visited by the OSIRIS-Rex mission, which reached it in December 2018, landed in October 2020, and will return samples to Earth in 2023. It is a carbonaceous asteroid with a relatively low average density of 1.26g/cc and a mass of ~ $7.8 \times 10^{10}$kg. Note we generally use a density of 2.6g/cc for "rocky asteroids". Bennu has about ½ the density we normally use for asteroids of this type. It is considered one of the two most hazardous asteroids in the solar system, the other being 1950DA. Bennu has the highest cumulative rating on the Palermo Impact Hazard scale. It is currently estimated to have a probability of impact through the year 2300 of about 1/1750, which will be refined with further observations. The next close approaches of less than 0.09 au will be September 30, 2054, and then September 23, 2060. We have explored numerous mitigations scenarios and have found that fragmenting into $10^5$ fragments 20 days prior to impact with 1m/s disruption is sufficient as shown below. Like Apophis, we are able to mitigate very real large-scale threats (like Bennu) with extremely short notice, though in general we will have significant advance notice of large threats.

***1km diameter Bolide Mitigation*** – Using the same analysis techniques, we have shown that a 1km bolide can be mitigated using our approach. We explored $10^5$, $5 \times 10^5$, and $10^6$ fragment disruption for the 1km diameter case with mean fragment size of about 21.6, 12.7, and 10m diameter respectively. The case of $5 \times 10^5$ fragments at 60 days and 1m/s disruption allows for a successful mitigation with minimal damage, though some broken windows in residential buildings close to ground zero for some of the larger fragments, while for the $10^6$ case the damage is even less, with pressures less than 1kPa virtually everywhere on the ground. Allowing for a 60-day intercept allows the fragment cloud to expand to nearly the size of the Earth. Even longer-term intercepts would allow virtually all of the fragments to miss the Earth entirely. This event would be an extremely large and catastrophic bolide if not mitigated. With a total exo-atmospheric energy of about 100Gt, it greatly exceeds the Earth's nuclear arsenal and would cause extensive damage to humanity if not mitigated. For a 1km diameter bolide disruption, an array of NED penetrators may be preferred to conventional penetrators. Note that the peak pressure scales (on average over the ensemble of fragments) approximately as $1/t$ where $t$ is the blast arrival time. This is consistent with our scaling of the peak blast pressure with slant range as $(r_slant)^{-1.13}$.

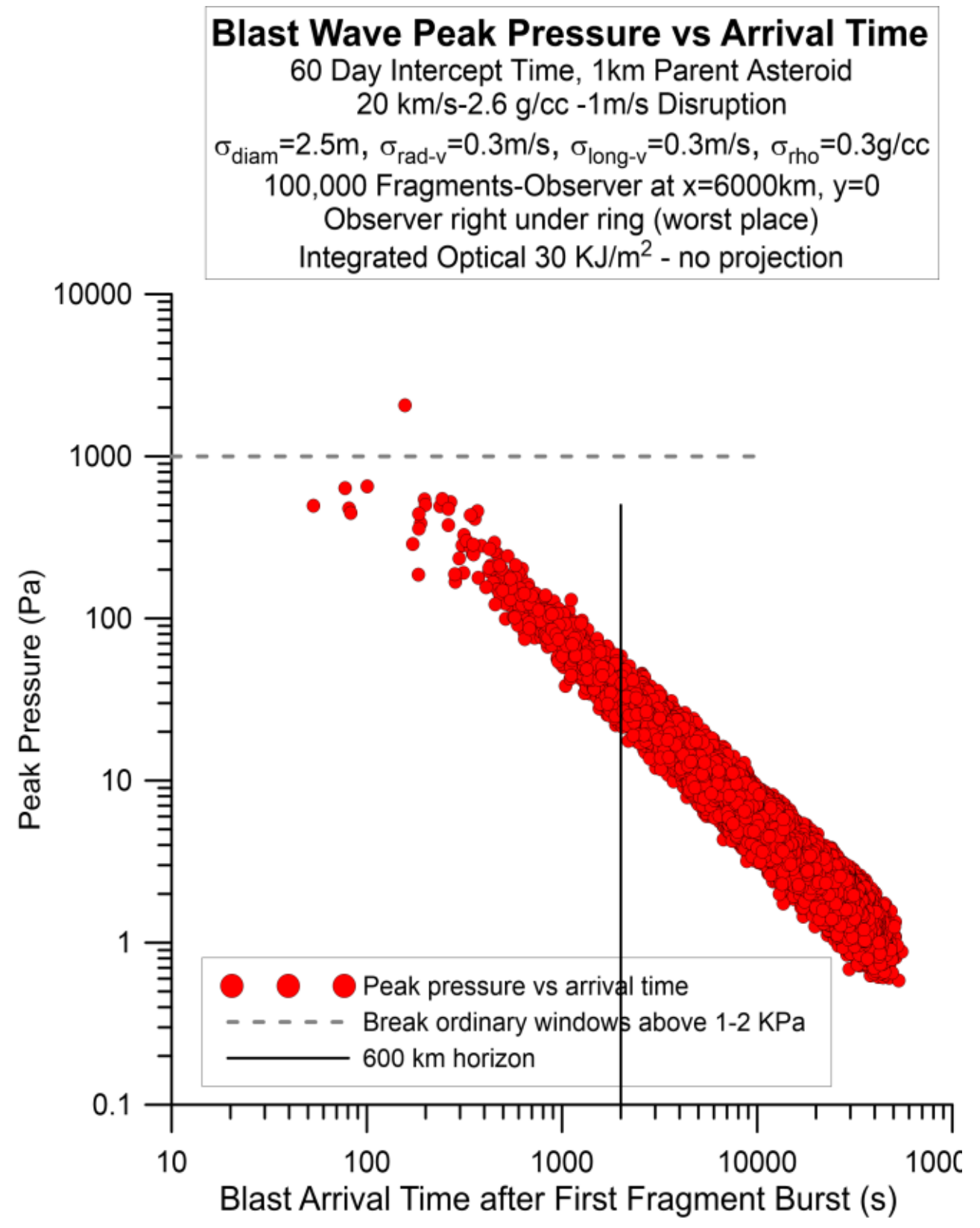

**Figure 89** - Fragment blast wave pressure distribution vs blast arrival time for a 1km bolide fragmented into 100,000 pieces (21.6m diam. avg.) 60 day prior to impact. Fragments have a 1m/s average speed relative to center of mass. The observer is right under the right (worst place) at 6000km from center of fragment ring The radar horizon at 600km is shown as an acoustical propagation time of approx. 2000 sec. We allow "over the horizon" acoustical signatures to propagate but their effect at the observer is minimal.

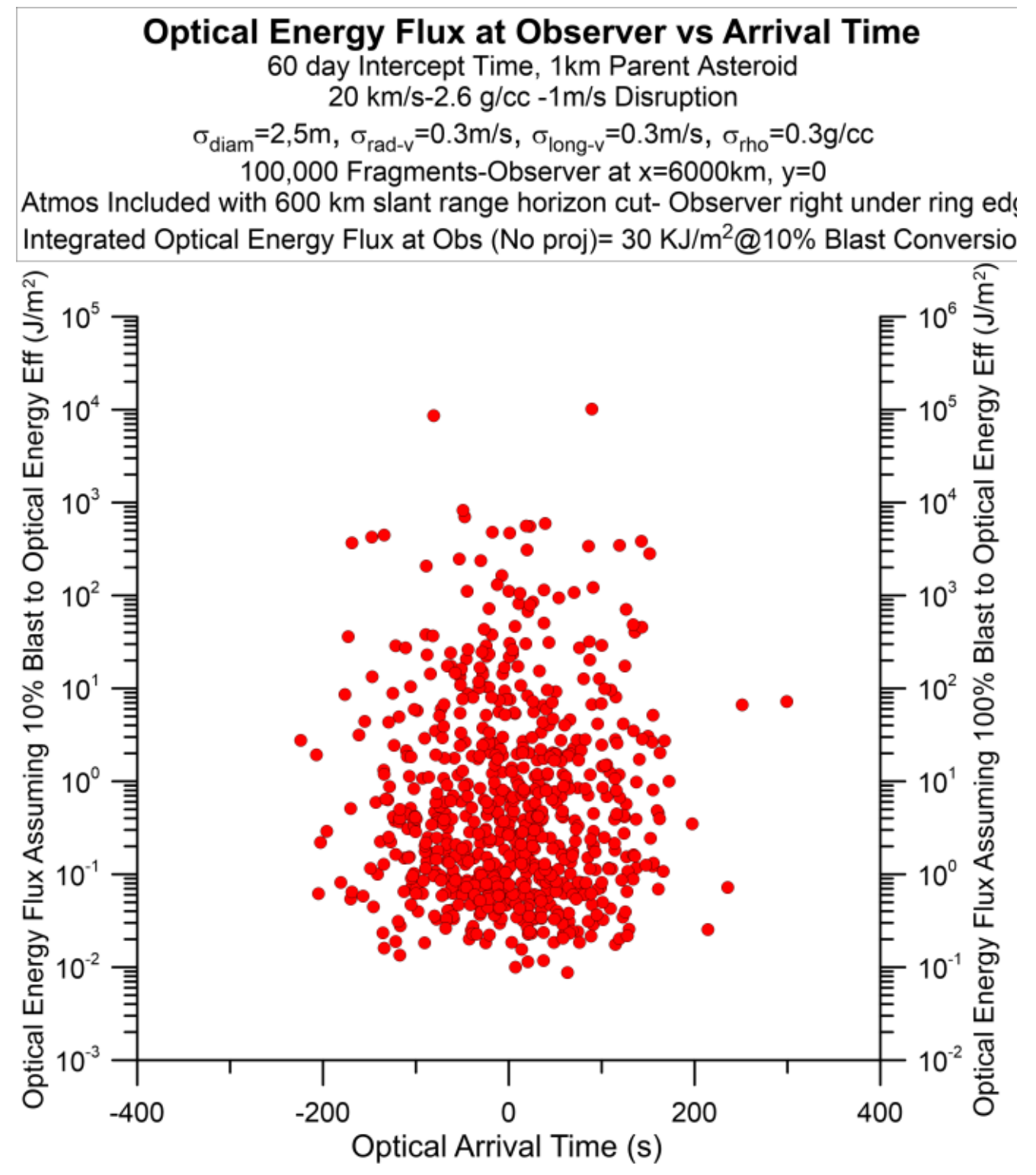

**Figure 90** - Optical energy flux (J/m$^2$) at observer vs optical pulse arrival time for 60 day intercept of 1km bolide in 100,0000 fragments. The observer is right under the right (worst place) at 6000km from center of fragment ring. Note the large optical arrival time range. This range will greatly reduce the effect of the optical pulse from each fragment.

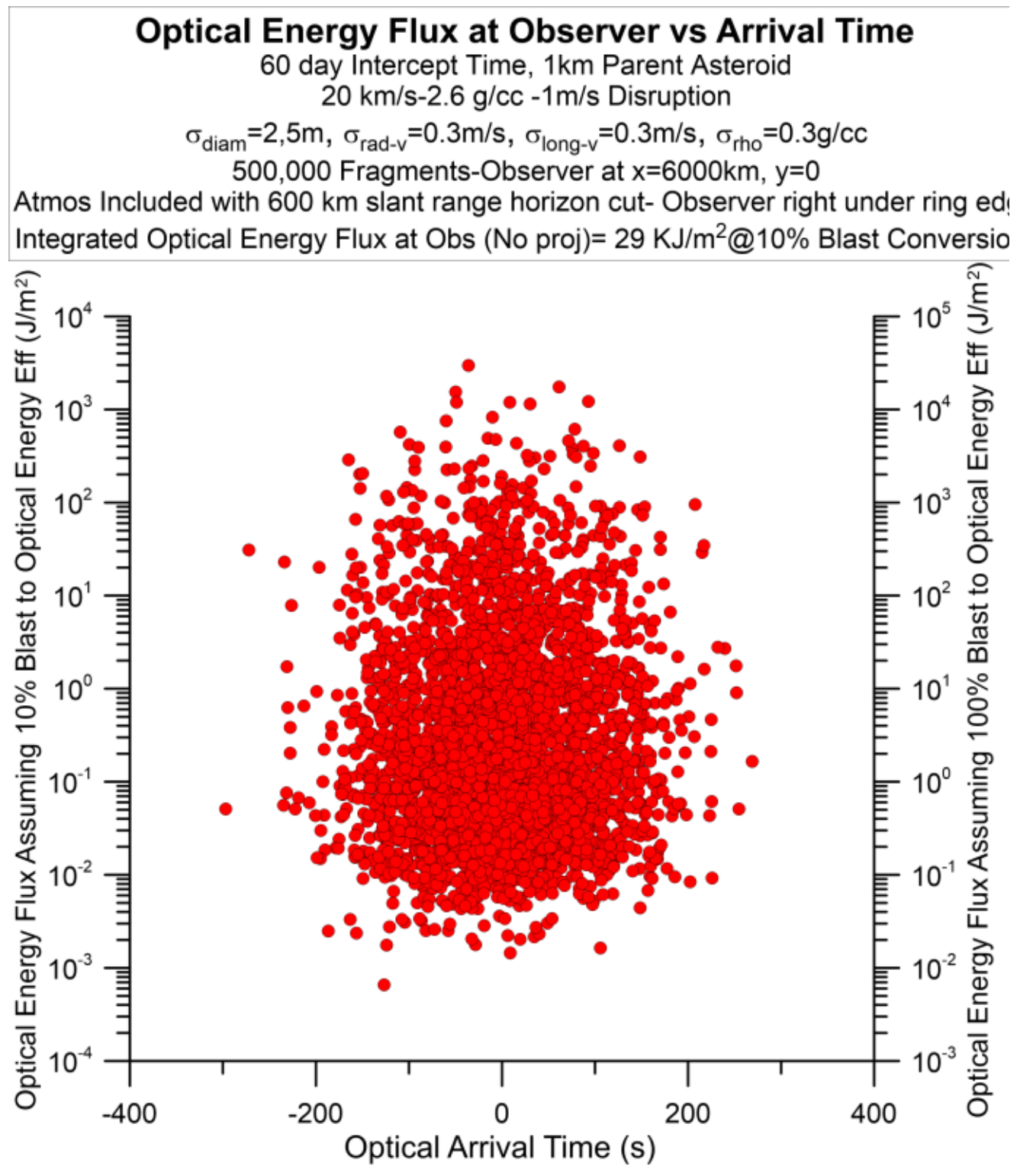

**Figure 92 -** Optical energy flux (J/m$^2$) at observer vs optical pulse arrival time for 60 day intercept of 1km bolide in 500,0000 fragments. The observer is right under the right (worst place) at 6000km from center of fragment ring. Note the large optical arrival time range. This range will greatly reduce the effect of the optical pulse from each fragment. Note the similarity to the 100,000 fragment case above but with a smaller mean fragment size (12.7m) due to the 5x increased number of fragments.

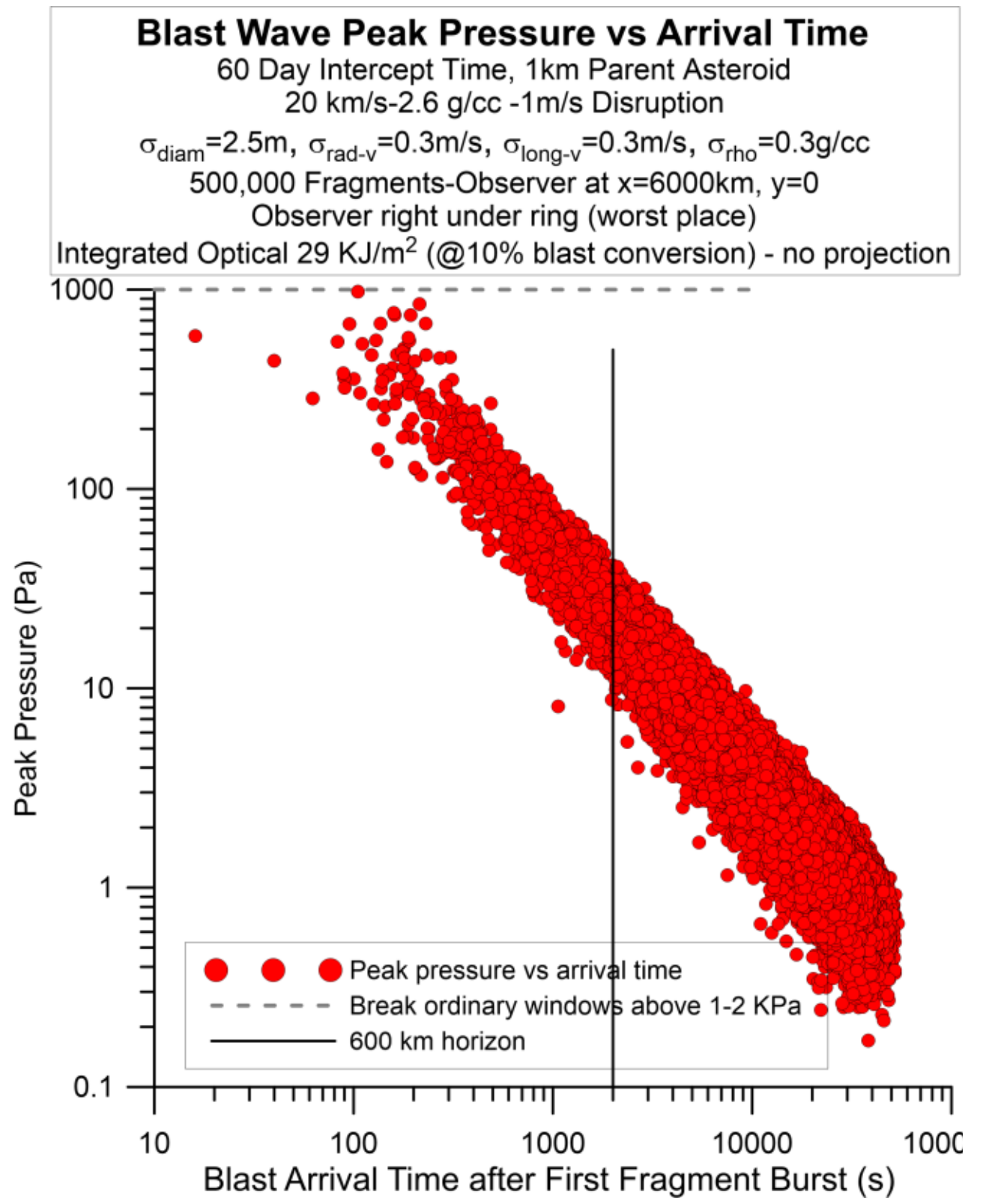

**Figure 91** - Fragment blast wave pressure distribution vs blast arrival time for a 1km bolide fragmented into 500,000 pieces (12.7m diam. avg.) 60 day prior to impact. Fragments have a 1 m/s average speed relative to center of mass. The observer is right under the right (worst place) at 6000km from center of fragment ring The radar horizon at 600km is shown as an acoustical propagation time of approx. 2000 sec. We allow "over the horizon" acoustical signatures to propagate but their effect at the observer is minimal. Note the similarity to the 100,000 fragment case above, but with a smaller mean fragment size due to the 5x increased number of fragments.

### *Existential threats – very large bolides – 10km case*

Truly existential threats to humanity can come in many forms. Extraterrestrial ones include impacts from large asteroids and comets with diameter of order 10km or larger. The biological and geological records indicate a number of mass extinction events, such as the death of the dinosaurs some 65Myr's ago. As mentioned, this event is consistent with a 10km diameter stony asteroid due to the observed Ir layer and shocked quartz. Some have speculated this extinction may have also be a comet strike or even a grazing comet. How the Ir layer would have formed from a typical comet is unknown, though some have also speculated that significant volcanic activity may have also been a part of this extinction period. While there is no known threat in the foreseeable future from such a large asteroid or comet, it is useful to ponder the question of whether or not our proposed technique could mitigate such an existential threat. Due to the $d^5$ dependence of the gravitational binding energy and the $d^3$ dependence of the disassembly kinetic energy on diameter, as shown, the gravitational binding energy now begins to dominate the disassembly energy budget for a 1m/s disruption. For example, with a 1m/s disruption and a 10km diameter, 2.6g/cc bolide, the total energy required is roughly 5Mt, while for a 1km diameter the energy is only about 0.2kt. The mitigation for the 10km diameter case calls for a nuclear disruption, which is possible, but the details become critical to understand. As discussed previously, NED penetrators using modern thermonuclear penetrators such as the B61-11 are highly problematic due to the extremely high "gee" loading on direct penetration, while pure fission NED's may be feasible, particularly those with a nuclear artillery heritage. The problem currently is partly political in that development is possible, but testing is not due to the CTBT that bans all nuclear weapons tests that have nuclear yield. This could always change, but currently this would limit NED options to the existing stockpile or to the development, but not testing, of NED's that do not require additional testing to ensure their performance. In any realistic scenario of an existential threat, presumable logic would prevail, at

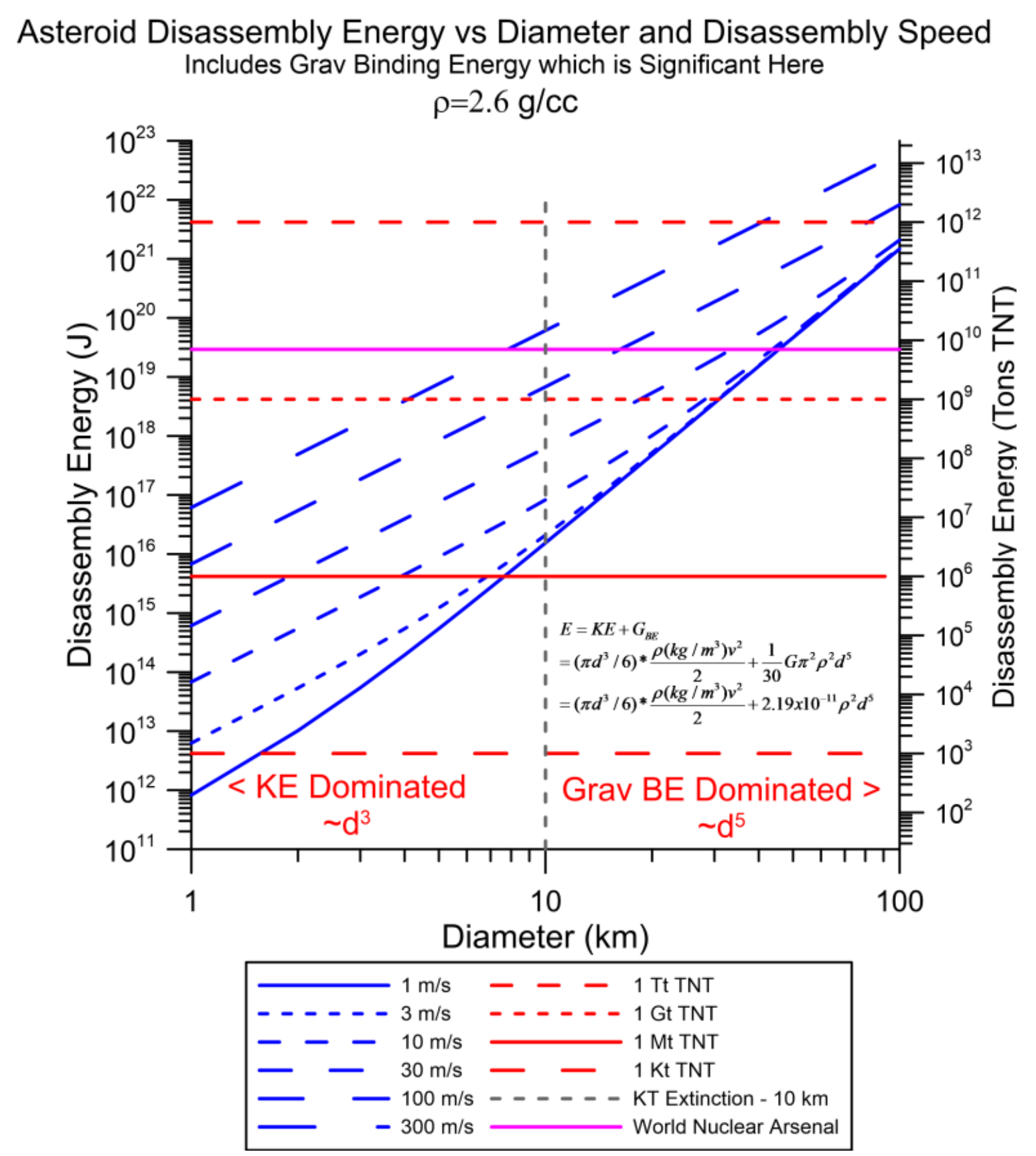

$$E = KE + G_{BE}$$
$$= (\pi d^3/6) * \frac{\rho(kg/m^3)v^2}{2} + \frac{1}{30}G\pi^2\rho^2 d^5$$
$$= (\pi d^3/6) * \frac{\rho(kg/m^3)v^2}{2} + 2.19x10^{-11}\rho^2 d^5$$

**Figure 94 -** Energy required to disassemble a large asteroid/comet at a given fragment speed relative to the center of mass for fragment speeds from 1 to 300m/s and parent asteroid diameters from 1 to 100km. Note that unlike the case of small bolides, the case of large bolides has very significant gravitational binding energy. **Note that above 40km diameter there is not sufficient explosive energy in the world current arsenal to even gravitationally de-bind the bolide.**

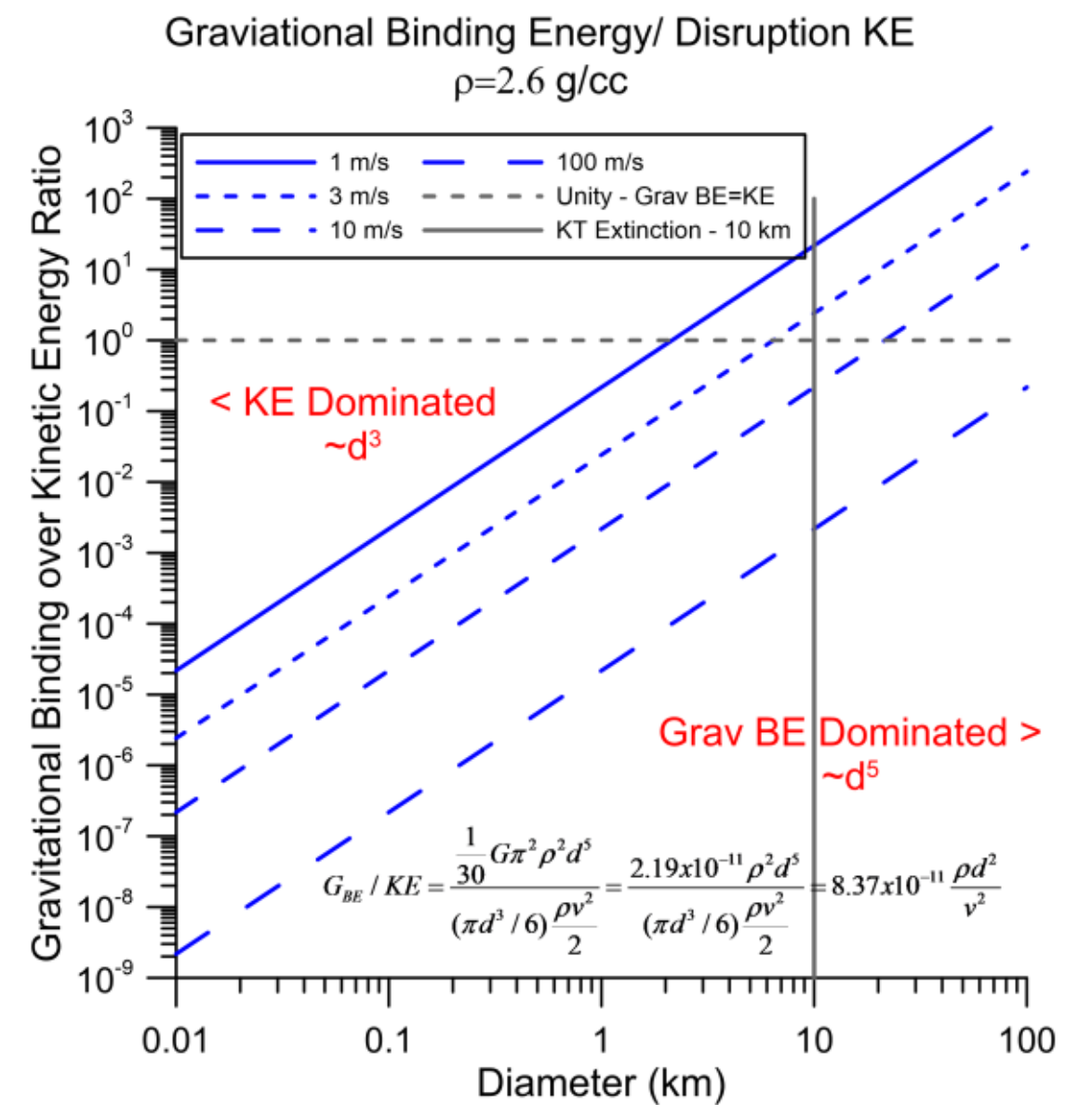

$$G_{BE}/KE = \frac{\frac{1}{30}G\pi^2\rho^2 d^5}{(\pi d^3/6)\frac{\rho v^2}{2}} = \frac{2.19x10^{-11}\rho^2 d^5}{(\pi d^3/6)\frac{\rho v^2}{2}} = 8.37x10^{-11}\frac{\rho d^2}{v^2}$$

**Figure 93** – Ratio of gravitational binding energy to fragment KE. Above 2km diameter gravity dominates for 1m/s disruption for density 2.6g/cc.

least one would hope. If we are confined to the existing stockpile or new untested devices which are "guaranteed" to work in full speed penetrators, then pure fission devices are possible. If we use the (cancelled) W82 pure fission NED (2kt yield 30kg bare NED mass) as a conservative example, this gives a yield per mass of 70t(TNT)/kg. Note that this is far less (~ 90x) than the Taylor limit for thermonuclear devices of 6kt/kg. Thermonuclear devices are far more mass efficient.

At 70t(TNT)/kg for the W82, we would still have about 1400x larger energy than purely kinetic impactors of the same mass for a 20km/s closing speed. If we could couple all of the W82 energy into disruption of the 10km bolide (5Mt minimum needed for 1m/s disruption including gravitational binding), we would require at least 2500 W82 NED's with a total mass of 75 tons. This mass is above the limit of a single SLS and at the limit of a flight-refueled Starship capability and we are already assuming unity coupling efficiency of NED yield into gravitation de-binding and KE of the fragments which is unreasonably optimistic. Pure fission devices for existential threats do not seem reasonable with thermonuclear devices such as a B61-11 "physics package" ~ 360 kt yield @ 3 kt/kg (physics) being preferred. **Sequential passive penetrators prior to the NED may allow this option. Note that above 40km diameter threats, even the entire world's current nuclear arsenal is not able to even gravitationally de-bind the bolide at unity coupling.**

***Nuclear Weapon Yield vs Mass – Taylor Limit*** – The yield of thermonuclear weapons that are particularly relevant to large threats including potential existential threats, are important to put into context. The largest weapon ever detonated was the Soviet RDS 220 "Tsar Bomba", "air dropped on a parachute" and detonated on October 30, 1961. It was more of a "show of force" than a practical weapon. It was designed for a yield of 100 MT but derated to 50 MT to mitigate fallout and to allow the aircraft to escape the blast wave. It had a total mass of 27 mt (including ~ 1 ton parachute) giving a design yield to mass ratio (YTM) of 3.7 kt/kg. Note this is a "total weapon" YTM NOT just the "physics package" YTM. Due to the enormity of the device, the two ratios are fairly similar. The "physics package" YTM for the B61-11 is similar at 3kt/kg while the total weapon mass is about 320 kg vs the physics package mass of about 130 kg giving a B61-11 total weapon YTM of about 1 kt/kg. The W61 EPW had a total weapon mass of ~ 375 kg or a YTM of 1 kt/kg/ The largest current US stockpile weapon is the B83 with a yield of 1.2 MT and weapon mass of 1100 kg for a YTM of 1 kt/kg. A B83 EPW has been studied. The most efficient (largest YTM) US weapon was the B41 with a design yield of 25 MT and a mass of 4850 kg giving a total weapon YTM of 5.2 kt/kg. This is the highest YTM weapon and close to the so-called "Taylor Limit" YTM of 6 kt/kg. Current stockpile weapons have smaller yield due to both targeting improvements and treaty limits, than earlier generations of extremely large yield weapons. In the discussion of NED's for planetary defense, it is important to distinguish between "what could be built/ what was once built" and those in the existing stockpile that could be rapidly accessed. In addition, for a PD mission requiring NED's the core "physics package" vs the existing "full weapon/ warhead" would need to be considered.

***Intercept Times for Very Large Bolides*** – While we have focused on the ability to fragment and de-bind the large bolides, it is interesting to note that for early interception where the fragment cloud diameter exceeds the Earth's diameter, then the threat becomes much less even if the fragments are larger. For example, simply splitting the threat in "half" and separating the "halves" by a distance significantly larger than the Earth's diameter avoids the problem of fragments in the atmosphere altogether. This statement is true of any threat. For 1m/s disruption speeds, intercept time of greater than 75-150 (depending on ground zero position) days or larger avoids any significant damage as the fragments miss the Earth. The specific fragment trajectories and the interaction with the Earth's gravity is of course relevant, though the bolide speed is generally much larger than the Earth escape speed, so the impact parameter is only modest.

***Smaller asteroid intercepts*** - While we have focused on large scale threats with 100m and Apophis as examples, we now discuss relevant impacts in the last century, namely the 20m diameter 2013 Chelyabinsk and the 50m diameter 1908 Tunguska events. For these smaller threats, the optical signature is much less important than the acoustical signature (blast wave). The very short required intercept times allow for significant flexibility in threat mitigation.

***20m diameter asteroid*** – The 2013 Chelyabinsk event was a 20m diameter stony asteroid with an exo-atmospheric energy of about 0.5Mt, similar to that of a large strategic thermonuclear weapon. It airburst near the town of Chelyabinsk. This class of event is expected about once every 50 years and could cause loss of life if the asteroid hits over a densely populated area.
If we use the same strategy, we have outlined and fragment the asteroid into 20 pieces, we can intercept with very short time scales. **With a 15 minute prior to impact intercept with a 1m/s disruption, the threat is effectively mitigated. Even a 100 s prior to impact mitigation (10m/s disruption) largely mitigates it.**

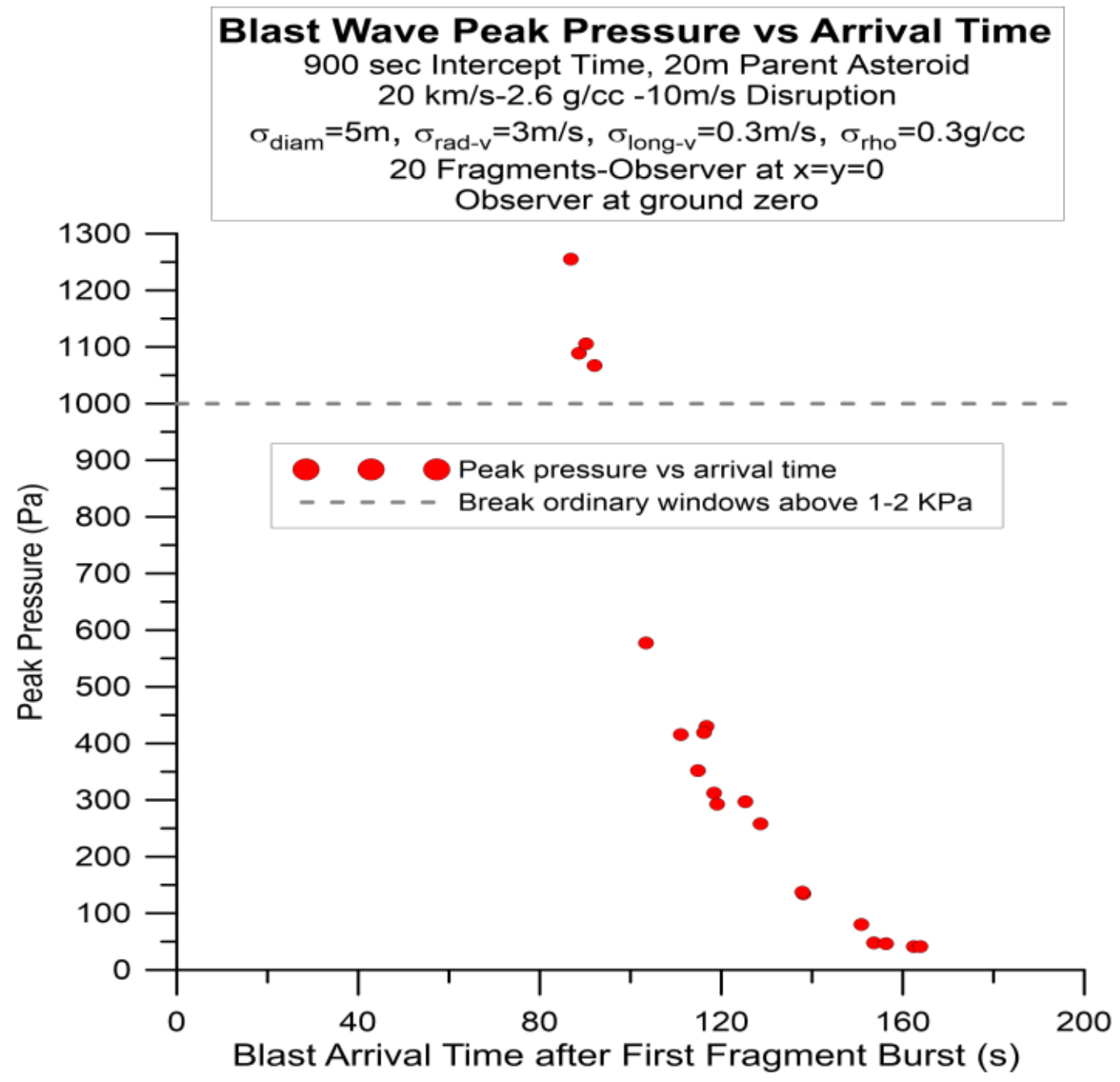

**Figure 95** – Blast wave peak pressure vs arrival time for 15 minute intercept of 20m bolide in 20 fragments.

***50m diameter asteroid*** – The 1908 Tunguska event was an approximately 50m diameter asteroid, or possibly a comet fragment, with an exo-atmospheric energy estimated to be about 10Mt, similar to that of the largest US strategic thermonuclear weapons tested. It is assumed to have been an airburst as it blew down more than ~2000km$^2$ of trees. This class of event is expected about once every several hundred years and would cause significant loss of life if the bolide were over a densely populated area. If we use the same strategy, we have outlined and fragment the asteroid into 100 pieces, **we can intercept with short time scales. For example, with a 5 hour prior to impact and a 1m/s disruption, the threat is effectively mitigated** though larger time intercepts are always preferred.

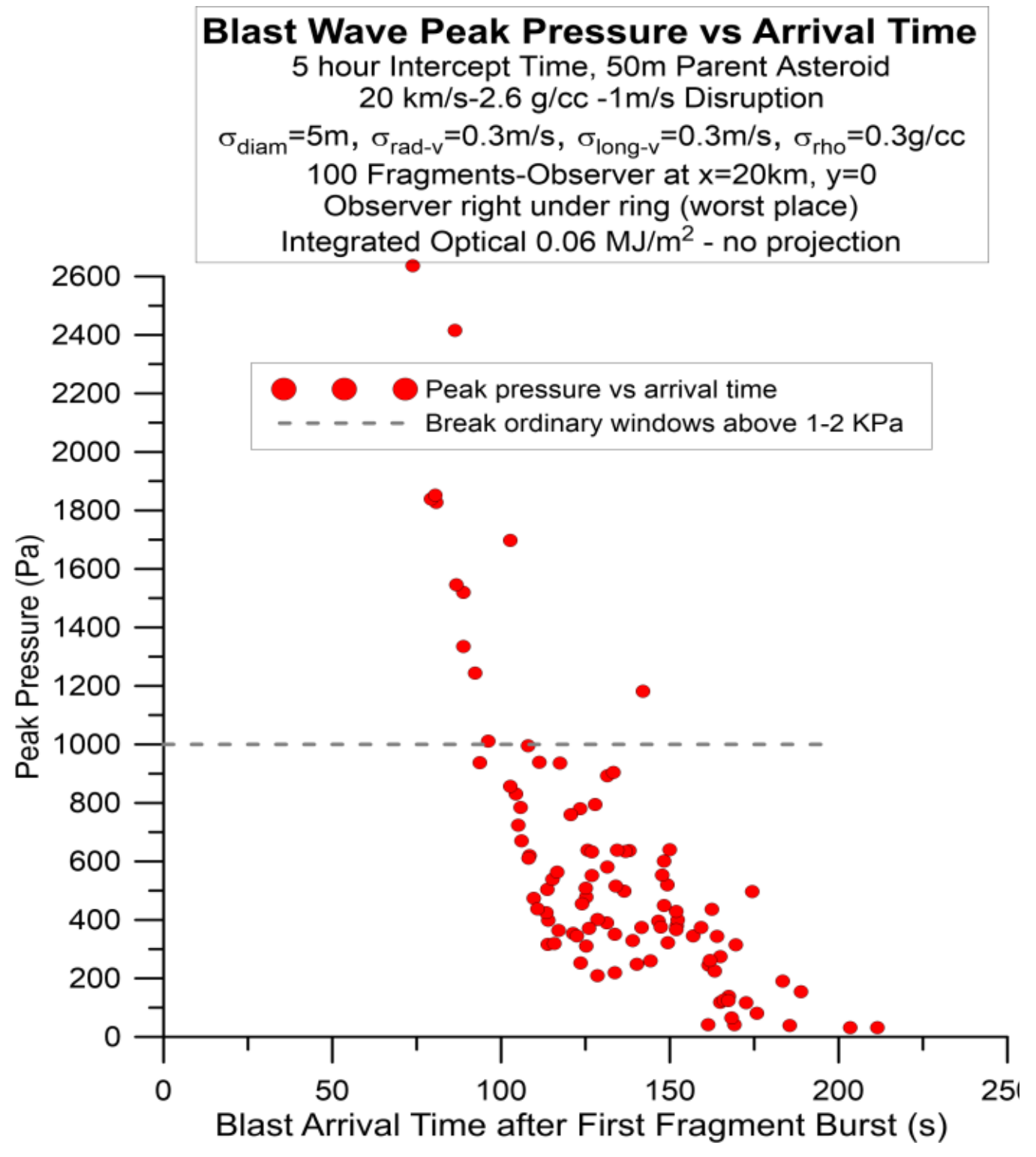

**Figure 96** – Blast wave peak pressure vs arrival time for 5 hour intercept of 50m bolide in 100 fragments.

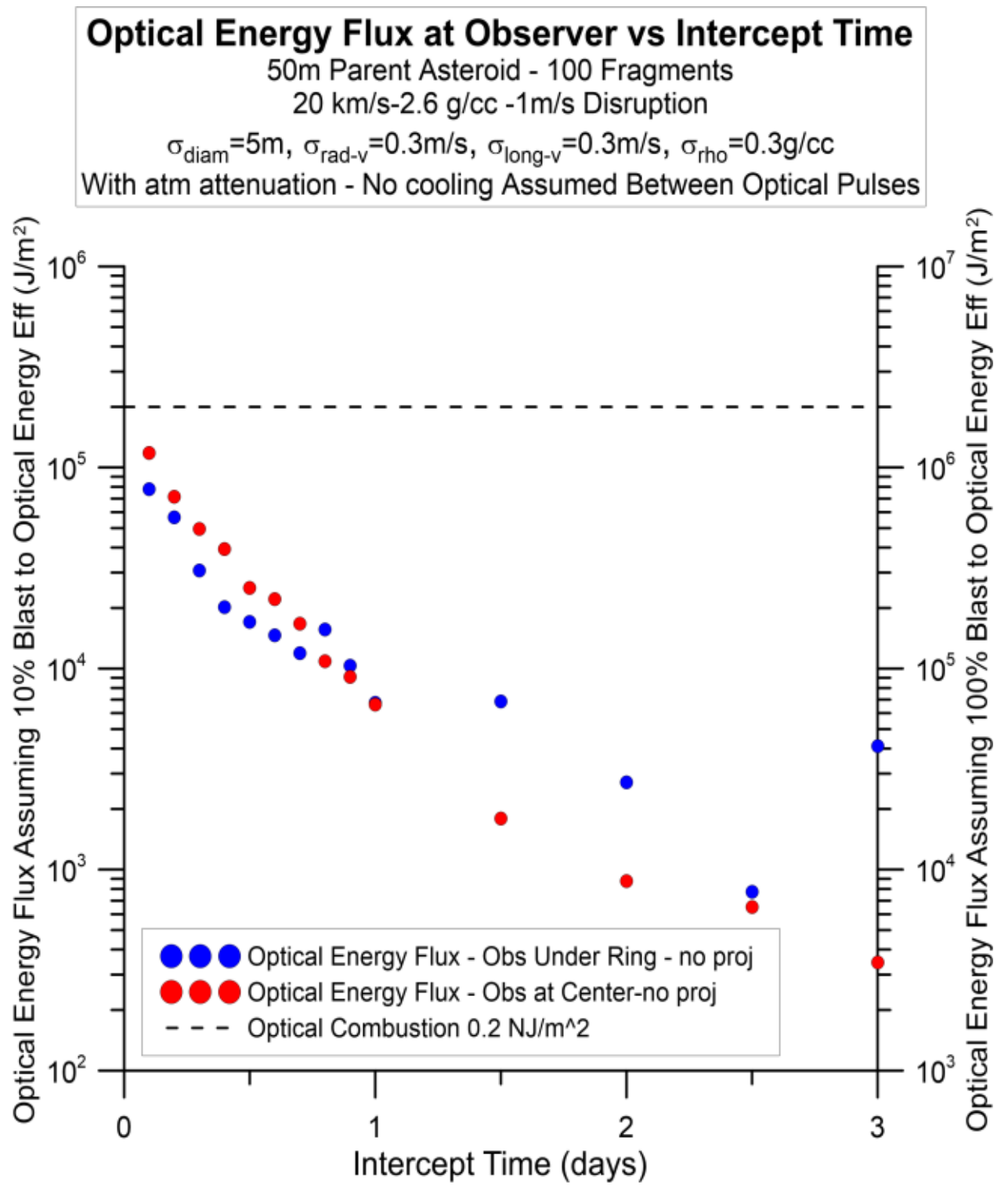

**Figure 97** – Optical energy flux (J/m$^2$) at observer vs intercept time for 50m bolide in 100 fragments.

### *Eye safety from optical signature (pulse)*

The optical signature can be bright enough from a single fragment to be of concern for retinal safety. The maximum permissible exposure (MPE) can be characterized in both a power and energy flux. The general conclusion is that energy fluxes in excess of $10J/m^2$ are of concern with the need to avoid looking at the optical pulse from fragments without special glasses (welding glasses). The pulsed MPE assumes a worst case of a point source being fully imaged onto the retina, though at very short (<400nm) and long wavelength (>1.3 microns) the light is greatly attenuated in the cornea, retina, or other parts of the eye before reaching the retina. Recall that the combustion threshold is $0.2MJ/m^2$ or vastly (20,000x) above the MPE. Just as we do not stare at the sun without special glasses, the same is true here. We should not stare at the optical pulse of the fragments without special glasses. The simplest solution is simply to stay indoors during an asteroid/comet strike or at least to close your eyes. For reference, during the Chelyabinsk event nearly 200 people reported "eye pain" with some diagnosed with mild retinal burns after having "stared" at the bolide. Approximately 70 people reported "flash blindness" [1], [48].

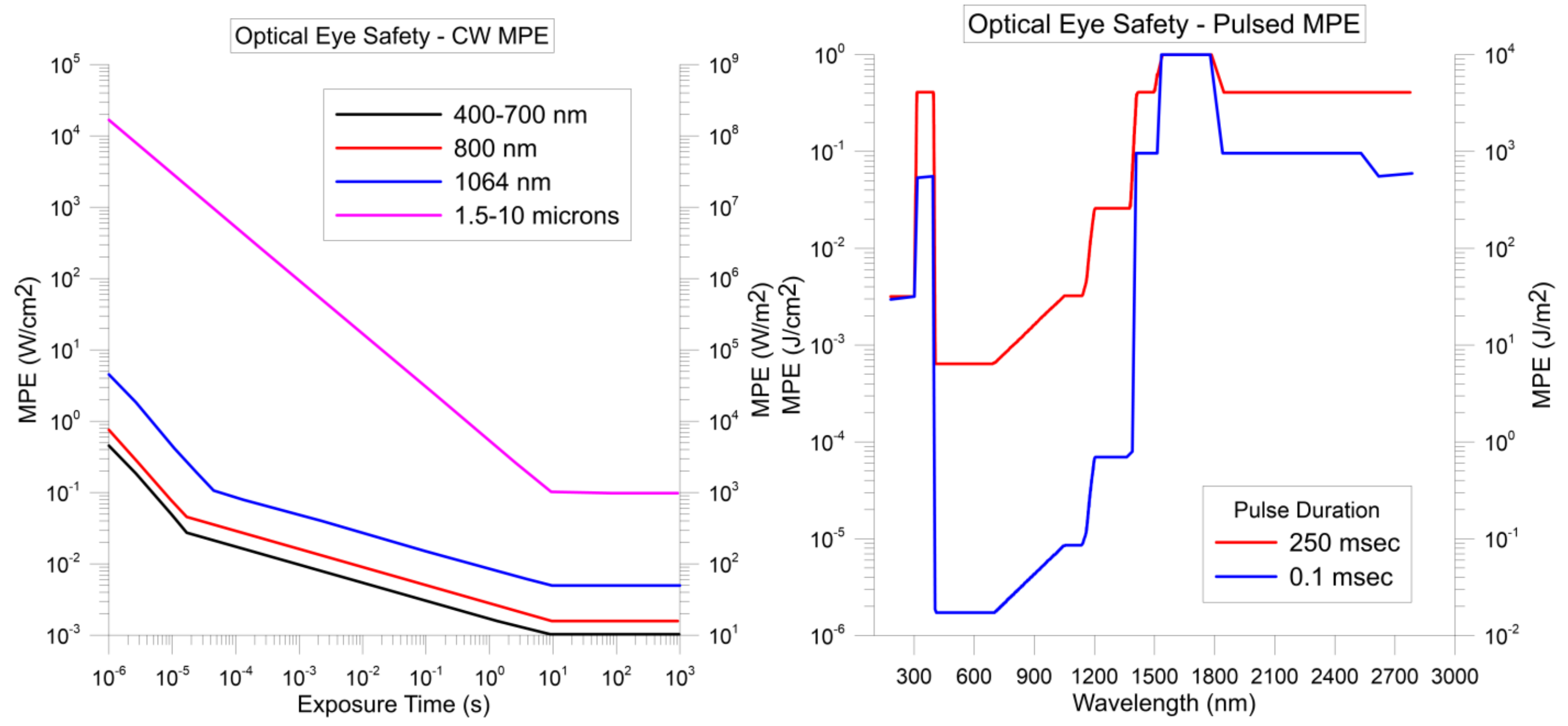

**Figure 98**- Left: CW MPE vs exposure time in power flux ($W/m^2$). Right: Pulsed MPE vs wavelength in energy flux ($J/m^2$). With a typical bolide optical pulse of about ¼ sec a conservative estimate of the Pulsed MPE is 10 $J/m^2$.

### *Effects on exposed skin – UV and thermal burns*

The optical signature can cause "sunburn-like" skin damage. This is dependent on the optical spectrum and exposure time just as it is for sunburn. For near UV (UVA) in the 315-400nm range, the permissible exposure limits are $1J/cm^2 = 10^4J/m^2$ for exposures less than 1000s (our case), or $1mW/cm^2 =10W/m^2$ for longer exposures (>1000s, which is not our case). Note that this is NOT the integrated optical spectrum, but rather the UVA portion of the spectrum. As we showed in the detailed calculation, we do include the atmospheric absorption for the light from each fragment which tends to "redden" the spectrum and block the UV portion of the spectrum. To be conservative we adopt a $10kJ/m^2$ integrated optical exposure as a suggested upper limit to minimize "sunburn". Recall that the combustion limit is 20 times higher, or $200kJ/m^2$ and this limit is the full spectrum limit and not the UV limit. For reference, during the Chelyabinsk event, 20 people subsequently were diagnosed with mild to moderate sunburn from the optical signature, likely from the UV and enhanced by snow reflection.

***Thermal burns*** to exposed skin are a related issue, but are largely a temperature effect where the outer layers of the skin are suddenly heated to excess levels causing a "burn". First, second, and third degree burns from thermal optical flash events are 10, 20, 32 $J/cm^2$ = 100, 200, 320 $kJ/m^2$ respectively. Note that these are approximately that of combustion, $200kJ/m^2$, and that the second degree burn level is equal to the combustion limit for dry leaves/paper. Flash burns are extremely serious and can be life threatening for large exposure fractions of the body. In the previous analysis of the optical signature, we keep the total thermal flash (optical signature) well below the combustion level of 200kJ/$m^2$ which allows us to avoid thermal flash burns in general, but it is critical to keep the UV exposure to a much lower level which we have set to be 10kJ/$m^2$ to avoid UVA "sunburn". As is the case for retinal effects, a simple course of action is to stay indoors during an event or if outside then seek shelter to shade the body and avoid exposed skin exposure.

***Chelyabinsk Comparison and Mitigation*** - In a detailed analysis of the Chelyabinsk event [48], the estimated total optical yield is 375TJ with a peak optical power of 340TW. The total exo-atmospheric energy of this event is estimated to be approximately 450 -550kt with 90kt being estimated to having gone into the optical signature (about 16-20% of exo-atmospheric energy). The peak optical ground flux at ground zero (under 30km burst) is 30kW/$m^2$ (30 sun) while the peak flux at 40km surface range is 11kW/$m^2$ (11 sun). The peak over pressure is estimated to be 3.2±0.6kPa, diameter of about 19.5m, an exo-atmospheric speed of 19.2km/s, 20 degrees above the horizon attack angle, and density about 3.3g/cc. The density is not well known, but the yield strength is estimated to be about 0.7MPa at first breakup and then rising to about 4-5MPa at burst. It is possible that the outer layers were of lower density than the inner portion if the bolide was heterogeneous enough to have a wide range of yield strengths. The bolide path came within about 40km (south) of the center of the town of Chelyabinsk. At these short distances, ground reflections of the blast wave can significantly enhance the its amplitude and to what extent this was an issue is unknown as the peak pressure is estimated based on a variety of inputs including (importantly) the breakage of glass and building damage in and around the town.

We have modeled a variety of simulations for Chelyabinsk with both a single bolide and a very short-term fragmentation mitigation at 0.01 day (864 sec <15 min@1m/s disruption) by breaking the parent bolide into 20 fragments with average diameter of about 7m. It is also possible to mitigate with a 100s intercept @10m/s disruption due to longitudinal dispersion in the burst altitude allowing blast wave de-correlation. These are extremely short mitigation times!

***As measured - Not Mitigated - Single bolide case***

At ground zero under the burst, we compute a peak pressure of 1.7kPa and a blast yield of 328kt (1.4PJ) with a total exo-atmospheric KE of 564kt. We get an estimate for the optical pulse of 140TJ with 10% blast-to-optical conversion and 240TJ with 17% blast to optical conversion. Both are in reasonable agreement with the "measured" total optical signature of 375TJ. The optical energy flux at the ground zero observer is approximately 6.7kJ/$m^2$ including atmospheric absorption. The optical energy flux is large enough to cause retinal burns and some UV "sun burn" consistent with what was observed.

At a distance of 40km, we get a peak pressure of 1.1kPa. Note that this is less than the estimated 3.2kPa from observed window damage in the town. A ground reflection could easily double this pressure, which our simulation does not include, nor local topography. Even so, the numbers are close enough for our simulations to be reasonably consistent with the "measured" values. The optical energy flux at 40km from ground zero is approximately 3.3kJ/$m^2$ including atmospheric absorption. The optical energy flux is large enough to cause retinal burns and some UV "sun burn" consistent with what was observed but far below the combustion limit.

***Mitigation case (20m, 20 fragments @ 864 sec intercept)***

With an intercept case below of 0.01 day = 864 seconds (<15 min) and a breakup speed of 10m/s, we intercept about 17,000km from the Earth's surface which is well inside the GEO orbit distance. We fragment the Chelyabinsk bolide into 20 fragments. The fragments breakup at an altitude of about 55km and burst at about 30km. The mitigation is sufficient to de-correlate the blast waves and significantly reduce ground damage as seen in the plot of all 20 fragment blast waves vs time. Even for this extremely short intercept time, the blast damaged is reduced by a factor of 3x. Longer time scale intercepts greatly reduce the blast wave even further. As mentioned above if we model longitudinal dispersion in the burst altitude allowing blast wave de-correlation even a 100 s mitigation is possible. In this case "zero-time mitigation" also works.

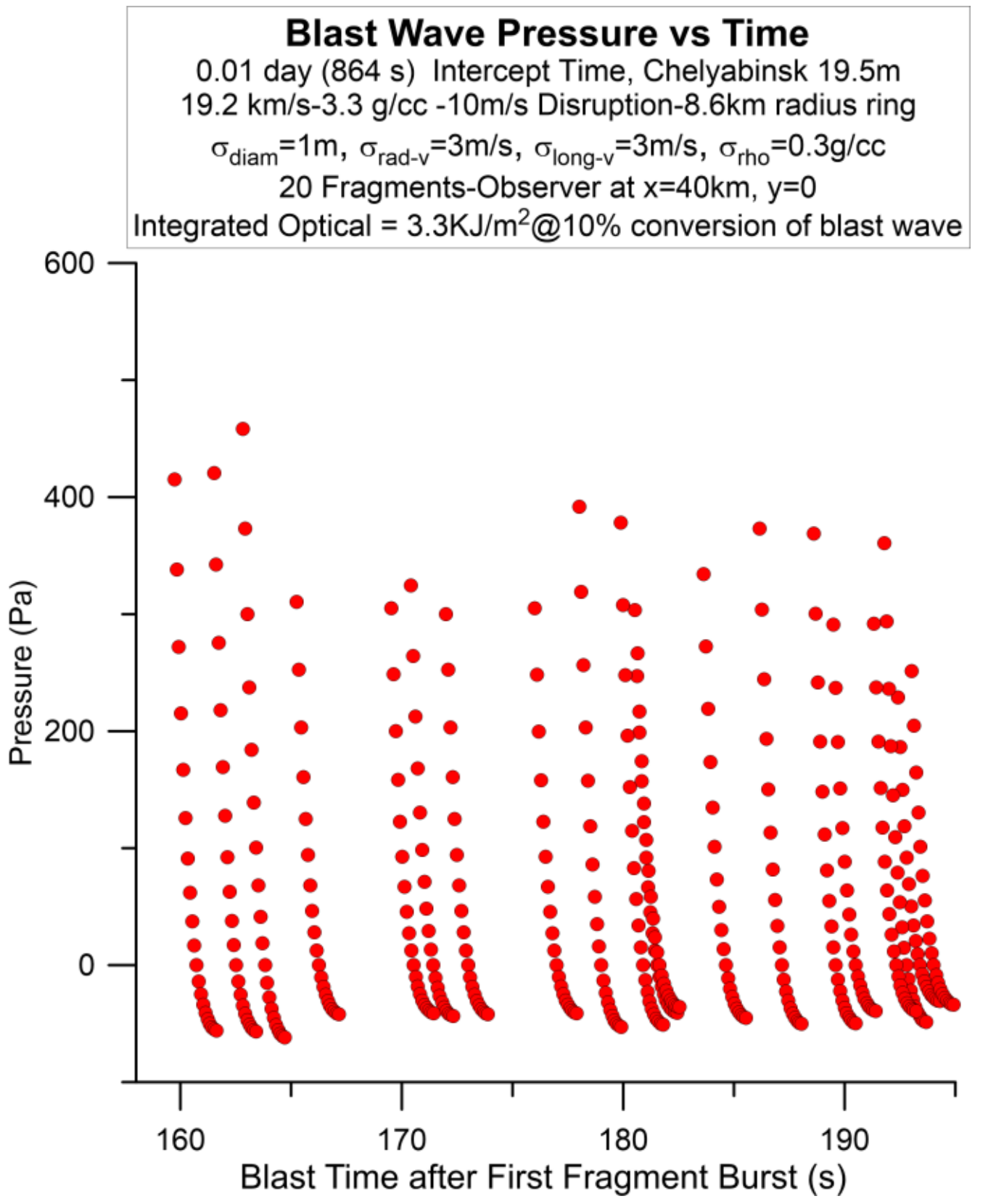

**Figure 100** – Time evolution of the 20 blast waves from the fragmented Chelyabinsk bolide at 40km surface distance.

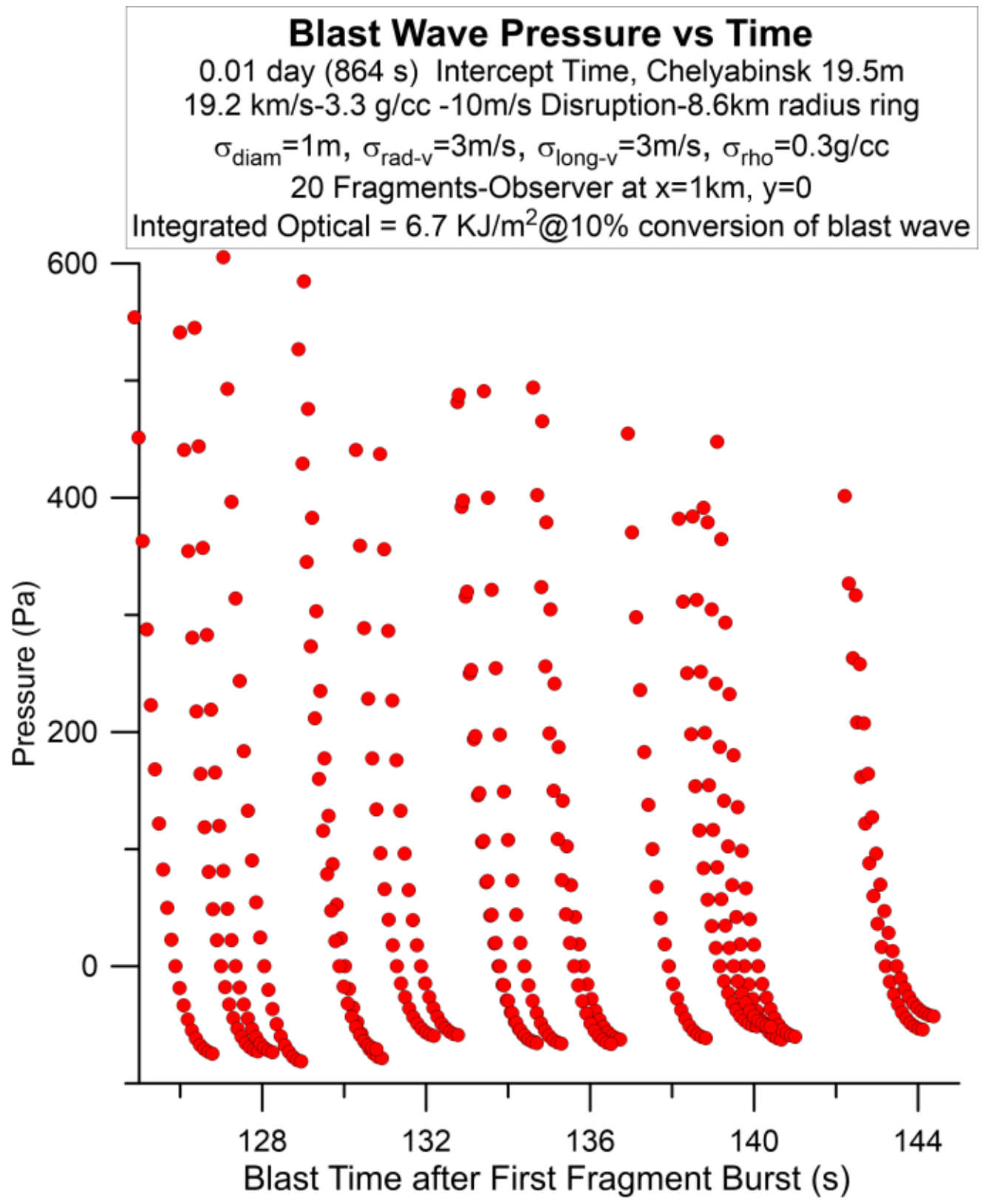

**Figure 99** – Time evolution of each of the 20 blast waves from the fragmented Chelyabinsk bolide at ground zero.

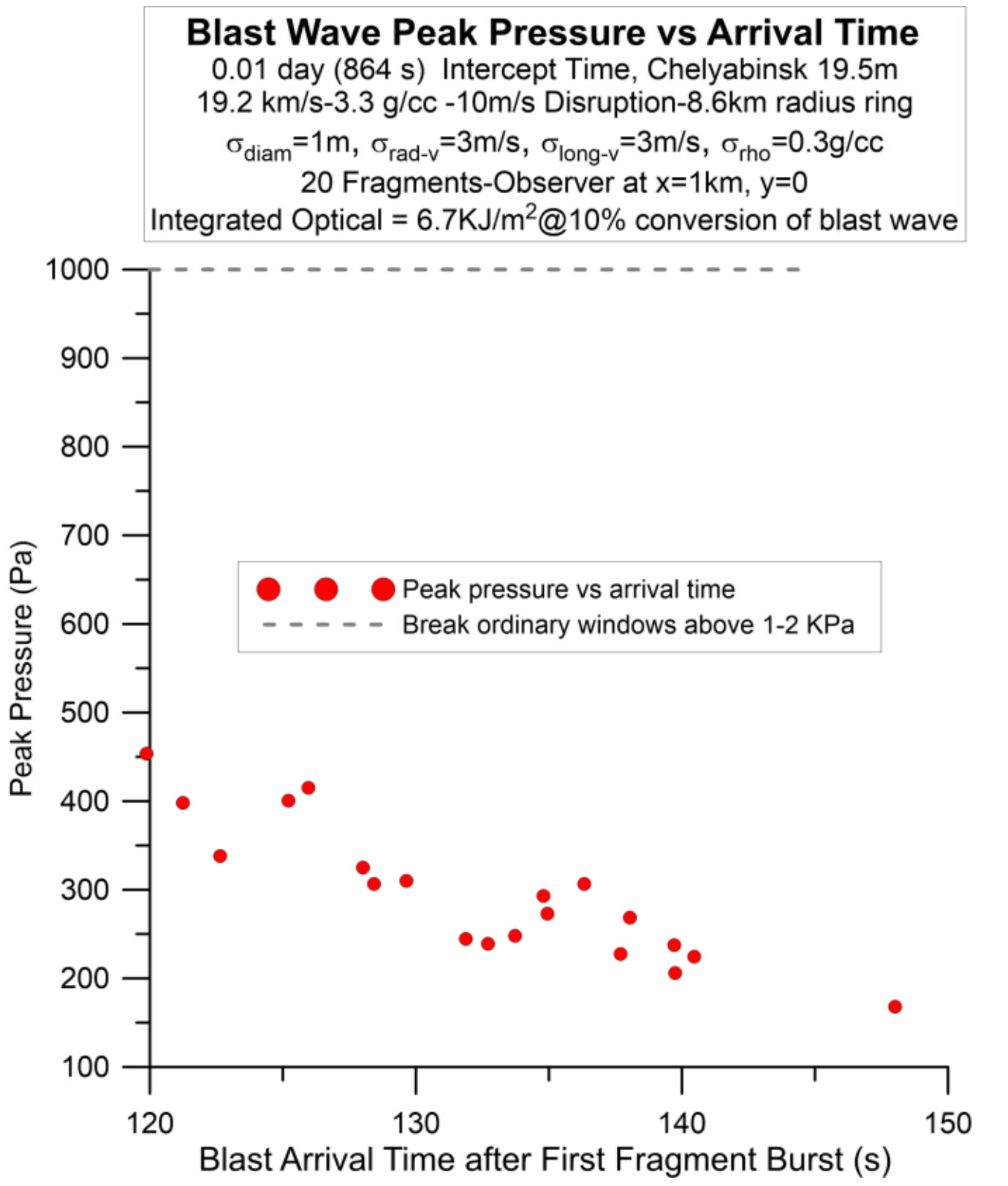

**Figure 102** – Peak pressure and arrival time of the 20 blast waves from the fragmented Chelyabinsk bolide at ground zero.

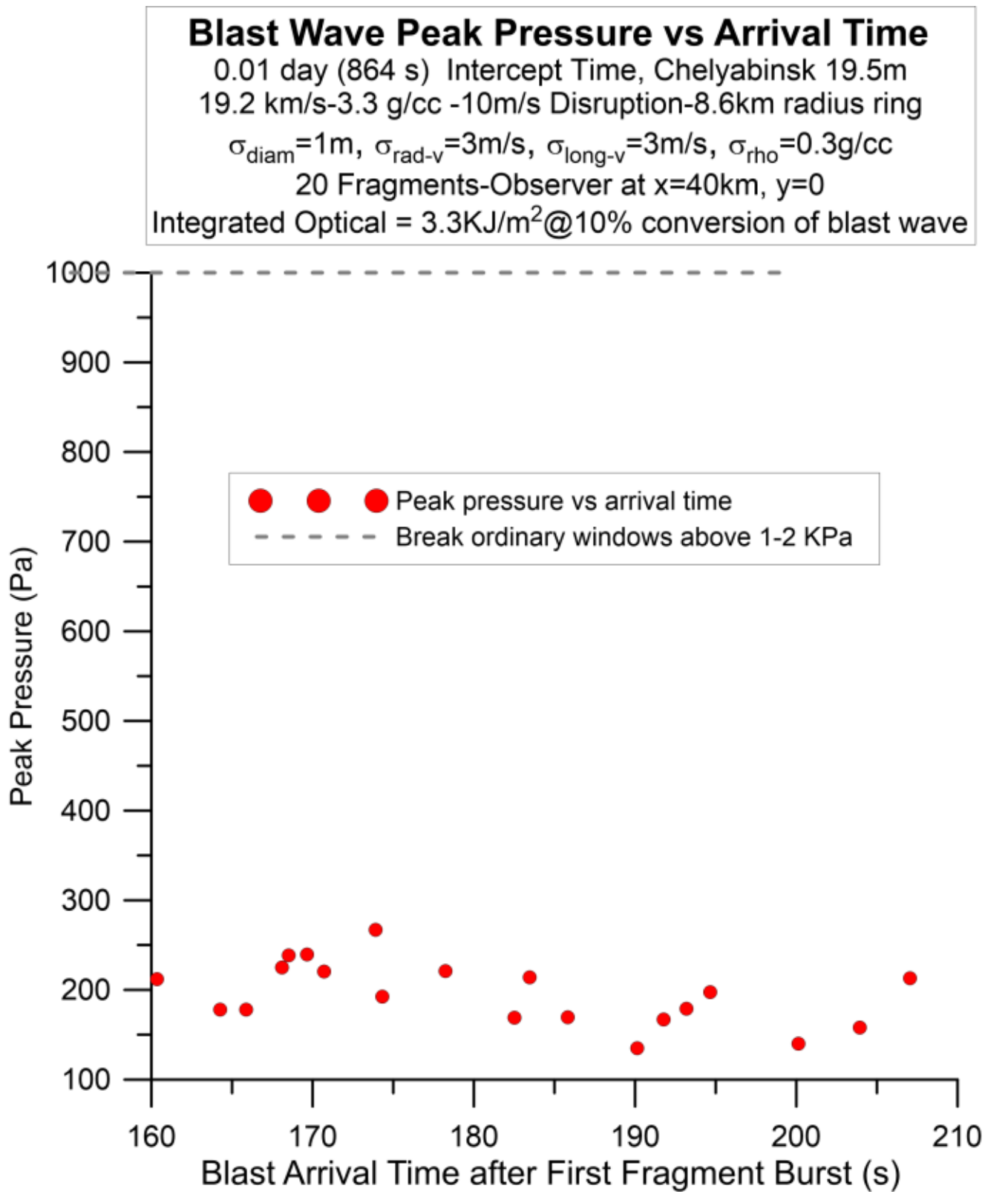

**Figure 101** – Peak pressure and arrival time of the 20 blast waves from the fragmented Chelyabinsk bolide at 40km surface distance from ground zero.

### *Scaling of intercept time and disruption energy*

The effects of the fragments bombarding the Earth are related to fragment cloud size upon entry into the Earth's atmosphere. The size of the fragment cloud is proportional to the intercept time and the mean disruption speed: $r_{frag_cloud} = \tau_{\text{intercept}} v_{disrupt}$ . This allows us to trade the intercept time and disruption speed so that shorter intercept times can be accomplished with the larger disruption speed. In this analysis, we need to consider both the blast wave and the optical signatures in deciding what is the minimum fragment cloud size. For smaller bolides (<50m), the optical signatures are small enough that the cloud radius is set by blast wave considerations and not optical considerations. This is both a practical question of intercept capability and the disruption energetics, the required disruption energy (ignoring the gravitational binding energy) being proportional to the square of the disruption speed and hence inversely proportional to the square of the intercept time. We have presented numerous scenarios with most of the detailed effects for a 1m/s mean disruption speed. If we were to use a 3m/s disruption speed, requiring about 10x larger disruption energy, we could lower the intercept time by 3x. At 10m/s disruption speed, we could achieve 10x lower intercept time, etc. This is an important trade space to consider. The disruption energy scales as the parent bolide mass, which scales as the cube of the bolide diameter (*d*), which allows us to have very short time intercepts for smaller bolides, which are more numerous.

$$E_{disrupt} = (\pi d^3/6) * \frac{\rho v_{disrupt}^{\,2}}{2} + G_{BE} = (\pi d^3/6) * \frac{\rho v_{disrupt}^{\,2}}{2} + \frac{1}{30} G \pi^2 \rho^2 d^5$$

$$= (\pi d^3/6) * \frac{\rho}{2} \left[ \frac{r_{frag_cloud}}{\tau_{\text{intercept}}} \right]^2 + \frac{1}{30} G \pi^2 \rho^2 d^5$$

As we showed, the gravitational binding energy is small compared to the disruption KE terms for bolides less than 1km diameter and for disruption speeds ~ 1m/s or larger giving

$$E_{disrupt} = \frac{\pi d^3 \rho}{12} v_{disrupt}^{\,2} = \frac{\pi d^3 \rho}{12} \left[ \frac{r_{frag_cloud}}{\tau_{\text{intercept}}} \right]^2 \sim d^3 / \tau_{\text{intercept}}^{\,2}$$

$$\tau_{\text{intercept}} = r_{frag_cloud} \left[ \frac{\pi d^3 \rho}{12 E_{disrupt}} \right]^{1/2} \sim r_{frag_cloud} d^{3/2}$$

$$v_{disrupt} = \left[ \frac{12 E_{disrupt}}{\pi d^3 \rho} \right]^{1/2} \sim \sqrt{E_{disrupt} / d^3}$$

In a threat mitigation analysis of a given bolide, our previous calculations allow us to set the minimum fragment cloud radius for an acceptable effect level. For small threats, the fragment cloud radius is smaller and hence the actual minimum intercept time is an even stronger function of the threat diameter.

Part of any serious planetary defense system would focus on layers of preparations for various scenarios, including fast reaction to smaller bolides. As we have shown, for stony bolides that are 10m and smaller in diameter, no action is required. In an operational interceptor system, we would need to have "at the ready" a strategy to handle a reasonably wide level of threats with a focus on the smaller threats (<100m) due to their increased rate. If we fix the maximum disruption energy capability of an interceptor, we can then calculate the minimum intercept time for a given threat. This is important in determining the overall system capability. This would allow us to design an overall system considering threat size detection capability and threat mitigation intercept time. Since we can ignore small bolides (<15m), this allows us to potentially have extremely short response times to smaller bolides. For example, the 20m class bolides (Chelyabinsk) that are difficult to detect at long range, could be mitigated with sub-hour interception if we have already designed the capability of the interceptor to mitigate 100m class bolides with a ~1-day intercept time. As a specific example, an interceptor that can impart a 1m/s disruption speed to a 100m diameter bolide could be capable of a disruption speed of $(100/20)^{3/2}$ ~11x larger, or 11m/s for a 20m threat. IF we wanted the SAME fragment cloud radius as for the 100m threat, then the intercept time could be 11x smaller for the 20m threat, BUT this is not correct as the fragment cloud radius for the 20m threat can be much smaller. For example, a 10km fragment cloud radius should be sufficient for a 20m stony bolide threat with an incoming speed of 20km/s. **A 10m/s disruption speed for the 20m threat gives an intercept time capability of 15 minutes with an intercept distance of 18,000km, or INSIDE the geosynchronous satellite belt. This would require a disruption energy of 170kg TNT equivalent. This is within reach of rapid launch interceptors. If we add the longitudinal dispersion of burst altitude with fragment size then a 100s intercept at 2000km altitude becomes possible, well within rapid launch interceptors.**

### *Zero-time intercept*

Can small asteroids (~20m like Chelyabinsk) be intercepted and mitigated immediately prior to impact? This would be a “zero time intercept”. At the extreme end of fragmentation, we could imagine a very fine granularity where the fragments are taken to the molecular level, which would be effectively vaporization of the threat. In theory this could be done, though it is not practical due to the large energy input required. At a somewhat coarser level, imagine we fragment to the “dust grain level” of 1-100 micron diameter. At very small fragment sizes, the interaction physics between the hypersonic grain fragments and the atmosphere become very different than the macroscopic level (meter-scale) we have been focusing on. The generation of both the acoustic and optical signatures would be very different. While not the focus of this paper, we note that in theory, extremely short intercept times are possible for any size threat, though the ramifications both in practical terms of energy requirements and understanding the details of the fine granular interactions would be critical to model. We can gain some insight into the general issue of the effects of going to smaller fragments by extending our current model down to the scale of 2m fragments. We compute the breakup and burst altitude, the peak blast pressure, and acoustic flux at ground zero (directly beneath the fragment) and the energy of the fragment in the accompanying figure. **From 2-20m diameter fragments, both the peak pressure and acoustic flux are well modeled by power law fits, as shown. It is still critical to de-correlate the blast waves at the observer to prevent correlated collective effects.** This is key to mitigation, as we have shown. There is a trade space between the energy requirements of higher speed fragment ejection to allow larger fragment spatial spread and hence temporal de-correlation of the arriving blast waves. As we have shown, there is significant advantage to reducing the fragment size where feasible. The case of the 20m diameter threat

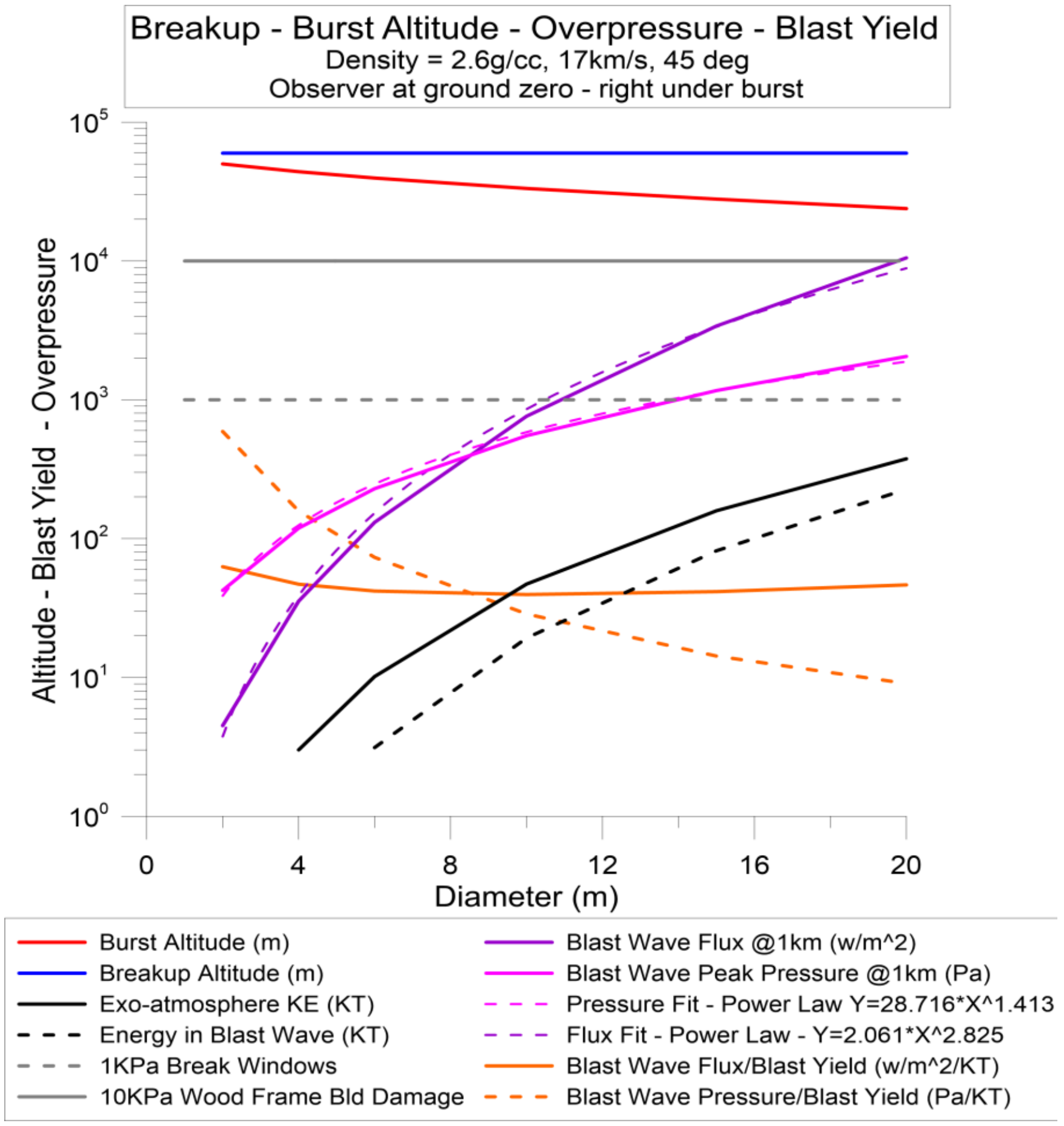

**Figure 103** – Breakup and burst altitude as well as peak pressure and acoustic flux at ground zero (under fragment) for fragment diameters from 2 to 20m. The peak pressure and acoustic flux are well fit by power laws, as shown. Blast wave flux per unit of blast yield is also shown. It is fairly constant over this fragment range. Blast wave pressure is shown at 1km ground distance from ground zero which is essentially equivalent to being at ground zero (right under fragment

(~Chelyabinsk) where we break the bolide into 20 fragments with a mean size of 7.4m allows for a 1000s intercept with a 1m/s disruption speed and 100s intercept with a 10m/s disruption speed. These are very short intercept times. If we shatter the 20m bolide into 1000 fragments with a 2m mean diameter, we can do even better, but the limit becomes the spatial spread of the fragments to allow sufficient de-correlation of the blast wave. To get even shorter intercept times, we increase the fragment speed which then becomes a disruption energy issue. To first order, the product of the intercept time and fragment speed, which then sets the fragment spatial scale, is a constant for a given bolide.

***Variation of burst altitude with diameter – longitudinal dispersion - implications for zero-time intercept***

Smaller fragments burst at higher altitudes, which then gives longer blast wave propagation times for these smaller fragments. The time delay is significant enough to assist in de-correlating the various fragment blast wave arrival times at the observer even for zero-time intercepts. In our mitigation scenario, fragments come off in different sizes and hence burst at different altitudes leading to varying propagation times, as shown in the accompanying figure. For example, the time difference for a 2m vs a 10m or a 5m vs a 15m fragment is roughly 50 seconds. **This longitudinal dispersion is largely independent of the interception time and is another source of blast wave de-correlation.** *For zero-time interceptions, this becomes the dominant time dispersion mechanism. As the intercept time increases, the transverse dispersion from the lateral dispersal of the fragments eventually becomes the dominant mechanism for blast wave time de-correlation.*

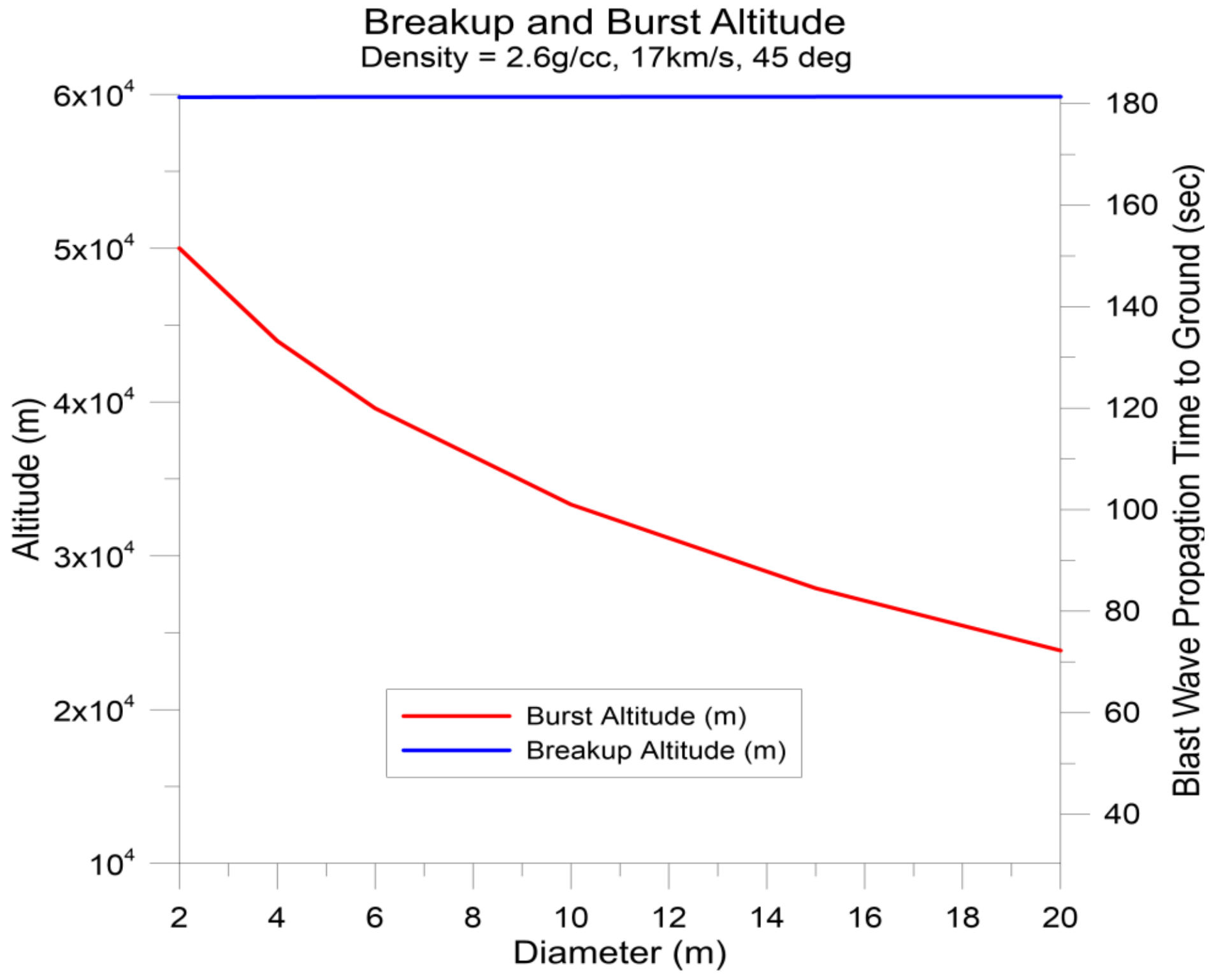

**Figure 104** – Breakup and burst altitude and blast wave arrival time at ground zero (directly under fragment) **vs** fragment diameter. This shows the large spread in arrival time caused by the range of burst altitudes for different fragment sizes. Smaller fragments burst at high altitudes and have shock waves that arrive later compared to larger fragments that burst at lower altitudes

### *Fragment and Blast Wave Self Shielding in Short Time Intercepts*

In very short time intercepts, the fragments do not have time to disperse very far in the transverse dimension and thus the potential for self-shielding of fragments with each other and their interference of their blast waves with each other becomes a concern. This interaction among fragments that are overlapping in projections would be a complex mixture of supersonic blast waves interacting with each other and with fragment breakup and may trigger an "acoustic cavitation" due to the partial vacuum creating behind the supersonic fragments. For example, at a 20km/s exo-atmospheric entry this corresponds to Mach #~ 60. With the fragment speeds being of such high Mach # upon entry, there will a significant vacuum created behind each fragment. The physics of the vacuum column behind the fragments is complex, but it can be simplified as a cylindrical volume collapsing at roughly the sound speed (~ 340m/s). The ratio of the vacuum column length $l$ to diameter $d$ can be approximated as roughly $l/d$=0.5M#, where the factor of 0.5 is from the radial inflow of the gas filling the vacuum column. In reality, the situation is much more complex due to turbulence and supersonic shock wave generation from the fragment. For a 20km/s bolide, this gives M#~60, which gives $l/d$~30. Thus, for a $d$=10m fragment, the length of the column is only about $l$=0.5d*M#~300m. This is small compared to the transit distance through the atmosphere, and hence the self-shielding of fragments along the longitudinal direction should be relatively small. To be conservative, we use a metric where $f$<<1 for breakup AND for burst even with pancake factor $\chi$=7. We see this in the calculations below where we choose as an example a 20m parent bolide fragmented into $N$=20 pieces. For an extremely short intercept time of 100 seconds, we choose a 10m/s disruption rather than a 1m/s disruption for this purpose. The alternative is to choose a longer intercept time of (say) 1000 sec, and then a 1m/s disruption is acceptable in terms of self-shielding concerns.

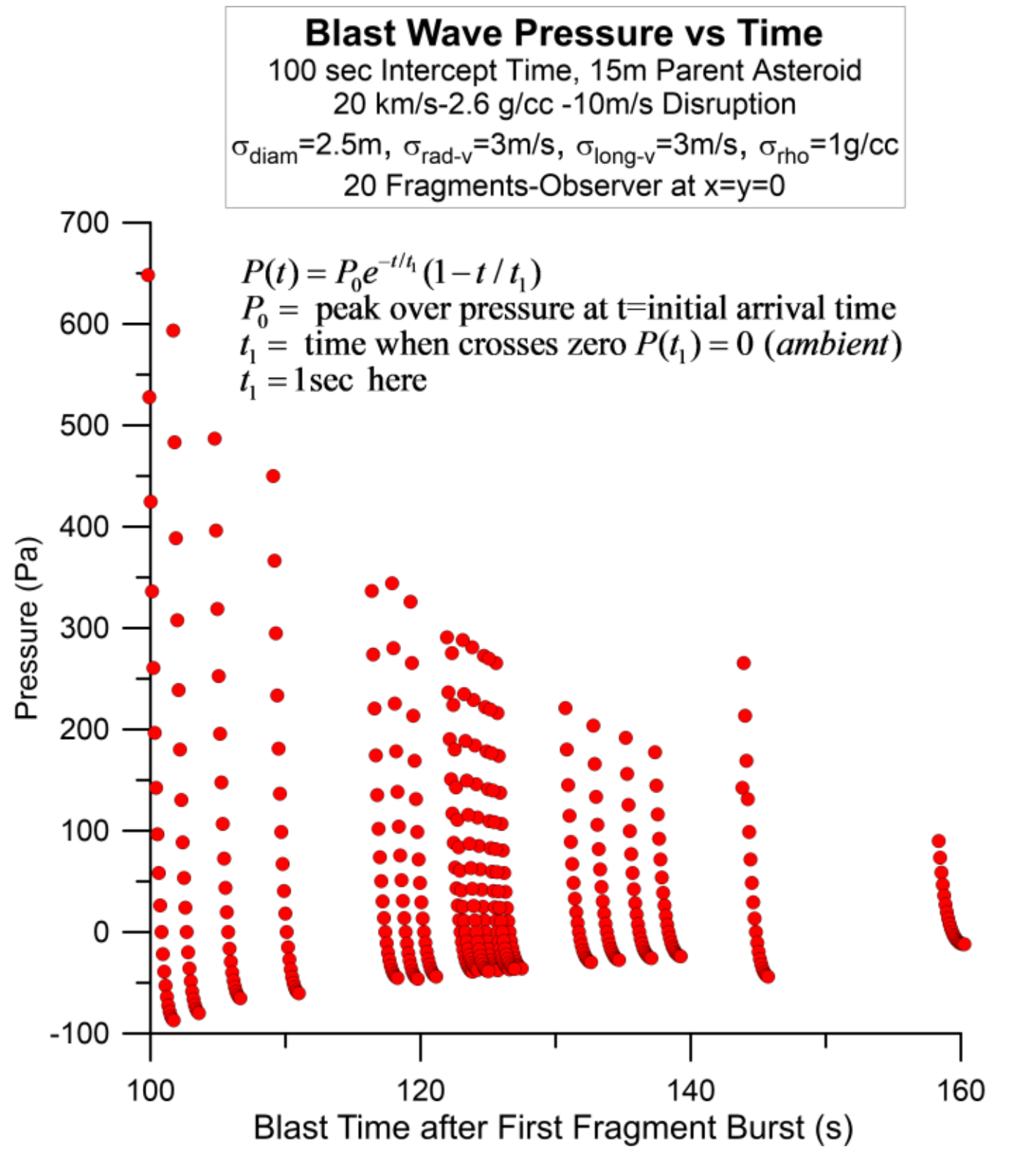

**Figure 105** – 100 sec intercept time evolution of each of the 20 blast waves from a fragmented 15m bolide at ground zero with 10m/s disruption.

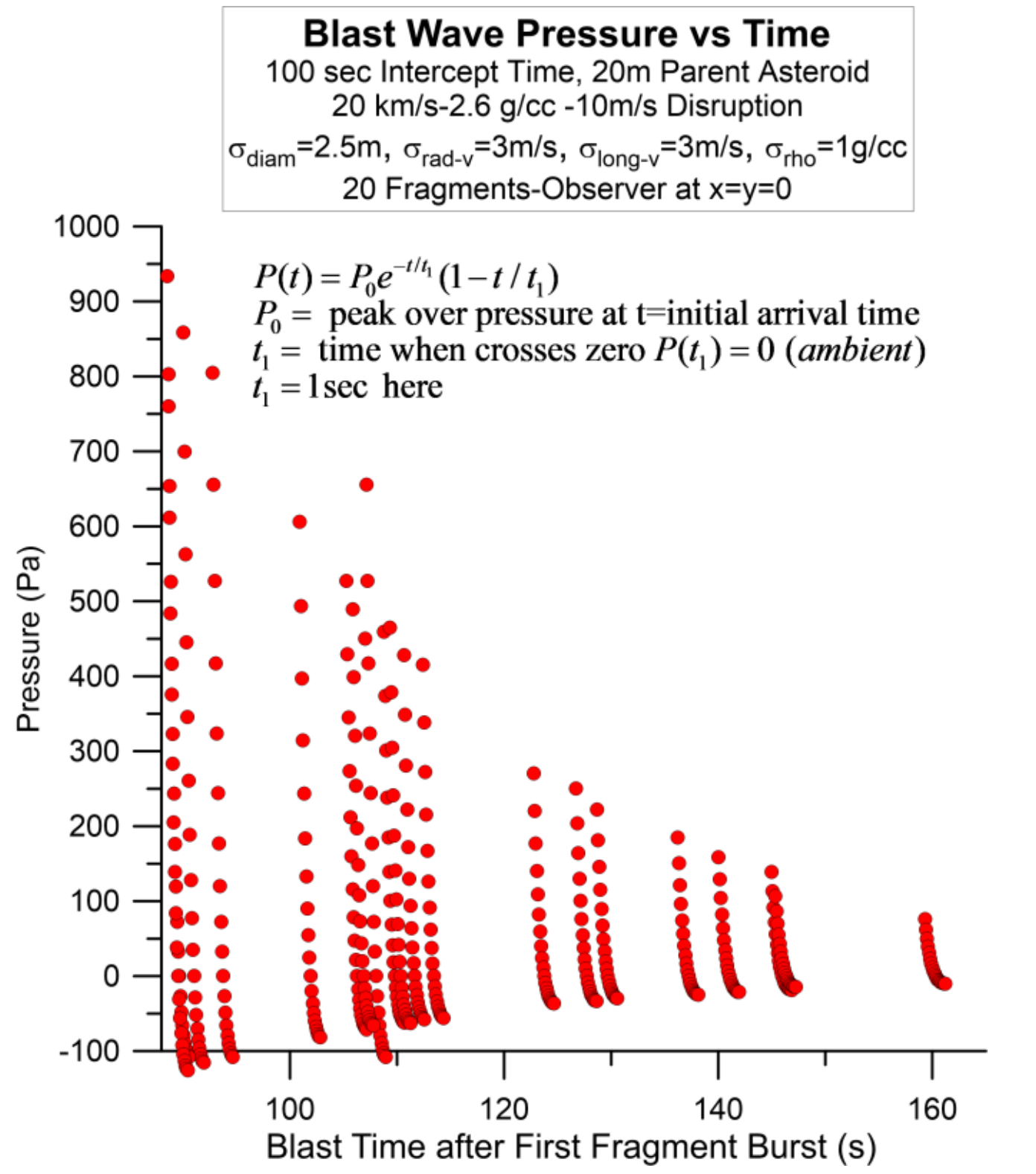

**Figure 106** – 100 sec intercept time evolution of each of the 20 blast waves from a fragmented 20 m (Chelyabinsk) bolide at ground zero with 10m/s disruption.

$\tau =$ time between intercept and impact

$L_i = \mathrm{i_{th}}$ fragment diameter

$N$=# fragments

$\chi$=fragment effective diameter multiplier (typ $\chi$=1 at breakup and $\chi$=7 at burst)

$\mathrm{v}_{ave} =$ average transverse disruption speed

$f$ =fractional overlap of fragments (self shielding fractor)

$$f = \left[ \sum_{i=1}^{N} \pi(\chi L_i)^2 / 4 \right] / \left[ \pi(\tau v_{ave})^2 \right] = \frac{\sum_{i=1}^{N} \pi\chi^2 L_{i^2}}{4(\tau v_{ave})^2}$$

$Ex:$

$L_0 = 20m, N = 20, \mathrm{v}_{ave} = 10m/s, \tau = 100s$

*Assume all fragments are the same size*

$L_i = L_0 / N^{1/3} = 2.71 \rightarrow L = 7.37m$

$$f = \frac{\sum_{i=1}^{N} \pi\chi^2 L_{i^2}}{4(\tau v_{ave})^2} = \frac{N\chi^2 L_i^{\,2}}{4(\tau v_{ave})^2} = 2.7x10^{-4}\chi^2$$

$= 2.7x10^{-4} \quad \chi = 1 @ breakup$

$= 1.3x10^{-2} \quad \chi = 7 @ burst$

We conclude that the self-shielding fraction is very small for this case (~ 1%) at burst and negligible at breakup. Note that had we used $\mathrm{v}_{ave} = 1m/s$with the same intercept time of 100s then f would be 100x larger yielding:

$f = 2.7x10^{-2} \quad \chi = 1 @ breakup$

$f = 1.3 \quad \chi = 7 @ burst$

*This implies small overlap at breakup but severe overlap at burst.*

Since the disruption energy scales as $v_{ave}^2$ there is a premium on earlier detection to allow longer intercept times if possible. We have already shown, however, that very short time scale intercepts (less than 1 hour) are very viable for smaller diameter late intercept bolides (<30m diameter) and that close to zero-time intercepts (τ~ few minutes) are feasible with 20m bolides. Obviously, we would prefer not to intercept at such short times if we could avoid it, but it is reassuring that it is possible if necessary.

### *Optical flux vs acoustical flux*

The two key observables are the acoustic blast wave signature and the optical signature. We compute this below for the example of a 100m diameter parent bolide broken into 1000 fragments. We show simulations of intercept times of 0.01, 1, and 5 days with the observer at the center and under the ring. The general trend of larger optical signature per acoustic signature (J/w) favors smaller fragments to lessen the optical signature. Smaller fragments also lessen the acoustic signature, so smaller is better. The range of fragments is 2 to 24m, with the vast majority being smaller than 15m. The dispersion is due to the simulated fragmentation distribution and the atmospheric absorption for the larger fragment cloud (larger intercept times), where many of the fragments are near the horizon.

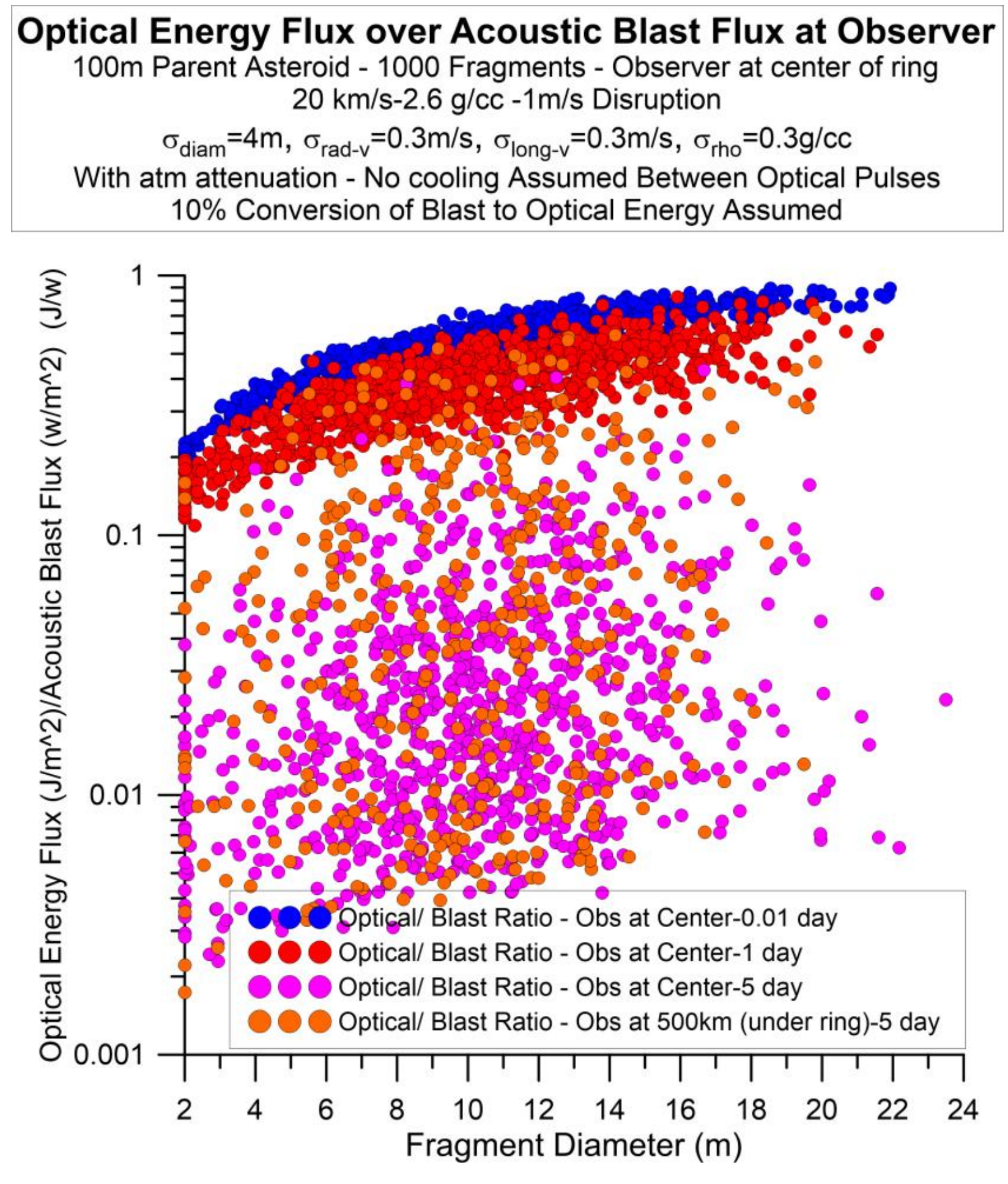

**Figure 107** – Ratio of integrated optical flux to peak acoustical (blast wave) flux. Specific case is for 100m diameter in 1000 fragments with observer at center of fragment cloud but ratio is generally indicative for a variety of scenarios.

The 0.01-day intercept has virtually all of the fragments nearly directly over the observer. The 0.01-day intercept case provides relevant data to general cases where the fragments are overhead. For fragments overhead, small fragments (2-5m diameter) have a value of about 0.2-0.3J/W, 10m diameter having a value of about 0.5J/W, and 15m diameter with a value of about 0.7J/W, with a generally increasing trend. Part of the scatter is also due to the atmospheric absorption for fragments that are low on the horizon relative to the observer.

### *Dust production from asteroid ablation and burst*

There is dust production from the asteroid fragments "burning up" in the atmosphere at high altitudes (typ. 30-60km). The finer the final particle size from burn up, the higher the effective surface area of the combined effect of all the dust grains. Like the injection of dust and debris from volcanoes, the stratospheric dust can persist for years and potentially cause a "nuclear winter" scenario. Is this a significant issue with our mitigation strategy? There is a tradeoff between forming dust at extremely small sizes (compared to the relevant wavelength of light from our Sun) and the effective cross section for light blockage and the larger grain sizes, where the cross section is essentially geometric, but there are fewer total grains. Since physical area of the sum of all the dust grains scales as $1/d$, where $d$ is the dust grain diameter, it would seem that the total area diverges as $d$ approaches zero. This scaling comes from the surface area of a grain scaling as $d^2$ and the volume (mass) of the grain scaling as $d^3$ and hence the sum of all the areas of the grains is $A_T \sim Nd^2$ where $N$ is the total number of grains with $N \sim (L_0/d)^3 \sim d^{-3}$ and $A_T \sim Nd^2 \sim (L_0)^3/d$. Of course, we are stopped at the molecular sizes of order1nm. However, there is another important effect when $d<\lambda$ where $\lambda$ is the typical wavelength of sunlight, namely $\lambda \sim 0.5\mu m$. For $d<\lambda$, the effective cross section scale as $(d/\lambda)^4$ and hence extremely small grains are largely ineffective in blocking sunlight, even though there are more of them. The optimal grain size for blocking sunlight is roughly d=λ. We will assume an optimal (worst case) grain size of d=1μm. Consider the case of a 100m diameter asteroid. If we convert this asteroid into optimal sunlight blocking grains of d=1μm we get $10^{24}$ dust grains with a total surface area of $10^{12}$ $m^2$. This is equivalent to a 1000x1000km sun shade! The Earth's surface area is about $5x10^8$ $km^2$ and hence this 1000x1000km sun shade would block 0.2% of sunlight. However, this is an absolutely worst-case analysis as grains at d=1μm would still have significant diffraction and thus the actual blockage would be significantly less. Of course, reducing ALL of a 100m diameter asteroid into 1μm dust grains is not reasonable for any ablation scenario.

As a comparison, note that the Krakatoa volcanic eruption of August 1883 is estimated to have hurled approximately $45km^3$ of debris into the stratosphere. This caused significant decreases (~1deg C) in global temperatures for about 5 years and did produce a measurable climate change (nuclear winter). In addition to the dust and debris, there was a significant amount of $SO_2$ formed in the stratosphere. By comparison, the volume of our 100m asteroid is about 45,000 times less material. There does not appear to be a significant "nuclear winter" event that would be caused by asteroids up to 1km using our mitigation method [50], [51].

***EM signature – EMP***

High altitude detonations of nuclear devices can cause severe EM signatures known as an EMP (Electromagnetic Pulse). This is causes by the high energy gamma rays from the nuclear reactions (primarily fission reactions) causing electrons to be ejected by air molecules via Compton scattering and the subsequent interaction of the high energy electrons and the Earth's magnetic field. This has been seen in numerous high-altitude tests (typ. >30km altitude) where the mean free path (MFP) is large enough to allow a very large electric field to be generated. The electric fields generated are typically up to 50kV/m. The phenomenon is a partially coherent effect due to the low frequencies involved. Detonations at low altitude cause relatively little EMP due to the short MFP.

Bolide impacts do not generate high energy photons and thus cannot generate the classic nuclear EMP. However, the air around the bolide is ionized and there can be small EM effects, similar to lightning. While related effects such as meteor trail communications due to the ionized trail have been long observed and utilized, there is little experience with large fragment bombardment. The physics of EM generation from hypersonic fragments is strongly against coherent and collective EM effects and thus very little damage is expected from fragmented bolides. For example, in the Chelyabinsk event, no significant EMP was noted. There is one Canadian event in January 2000 near Whitehorse in the Yukon region, where a 5m bolide is reported to have "knocked out" the power grid, but this event is not verified to have actually caused the power outage and there is significant skepticism as to its veracity, particularly since telecommunication equipment continued to function.

***Intercept time for mitigation vs target size***

We summarize the acceptable intercept times for mitigation where we keep both the acoustic and optical signatures at values low enough to avoid any large-scale damage. There is a trade space for any target as to what is an acceptable level of acoustic and optical signatures that we are willing to live with. Note that a 20km/s bolide with density that of stony asteroid (~ 2.6g/cc) whose exo-atmospheric energy roughly equals the total world nuclear arsenal, corresponds to a diameter of about 500m (Bennu) or not much larger than Apophis. This is a somber reminder that extra-terrestrial threats from asteroids and comets are at a scale that can far exceed current human weaponry, and yet we are now able to mitigate these threats using our current technology. This would truly be a remarkable achievement for humanity to be able to actualize planetary defense.

For comparison to our detailed simulations, we can make a simple analytic model for the minimum intercept time. The analytic model computes the intercept time based on the need to de-correlate the blast waves. To accomplish de-correlation, we need to ensure that there is acceptably small blast wave temporal overlap. As the intercept time increases, the fragment cloud spreads out further until it exceeds the local horizon for the observer, which greatly mitigates the blast wave and optical signature. With further increases in the intercept time, eventually the fragment cloud radius will exceed the Earth's radius with fragments no longer impacting the Earth.

The overall logic in fragmentation is as follows:

1) Choose the maximum fragment size $L_{max}$ so that the max blast pressure $P_0$ at ground zero is acceptably small to prevent significant damage. Typically, this gives $L_{max}$~10m for $P_0$~1kPa at ground zero.
2) $L_{max}$ yield the number of fragment $N$ for a given parent bolide diameter $L_0$ where $N=(L_0/L_{max})^3$.
3) Choose the average fragment speed $v_{ave}$ depending on the interceptor capability.
4) For a given $v_{ave}$ and $N$, we then need to ensure that the fragment blast waves at the observer are de-correlated. See equations below. This will then give the needed cloud spread radius $a_{ave}$.
5) Once the cloud radius is determined, this then gives the intercept time $\tau$ where $\tau=a_{ave}/v_{ave}$.
6) Check that optical signature at observer is acceptable.
7) Note the intercept time can always be made smaller by increasing the mean disruption speed $v_{ave}$

$\tau =$ intercept time
$t_1 =$ Friedlander blast time scale (0.1-1 sec)
$v_s =$ Sound speed
$v_{ave} =$ Ave fragment disruption speed
$z_b =$ burst altitude
$L_{\max} =$ Ave fragment diam desired (typ 5-10m)
$L_0 =$ Bolide diam
$a_i = \tau v_i = \tau(v_{ave} + \delta v_i) = \mathrm{i_{th}}$ fragment transverse distance
$v_i = \mathrm{i_{th}}$ fragment transverse disruption speed
$\delta t_i = (r_i - z_b)/v_s = \mathrm{i_{th}}$ blast wave arrival time at observer compared to burst directly overhead observer
$r_i = \mathrm{i_{th}}$ fragment slant range to observer$=(a_i^2 + z_b^2)^{1/2}$
$a_{ave} = \tau v_{ave} =$ average fragment transverse distance
$r_{ave} =$ average fragment slant range distance$=(a_{ave}^2 + z_b^2)^{1/2}$
$\delta t_{ave} = (r_{ave} - z_b)/v_s = \mathrm{i_{th}}$ blast wave arrival time at observer minus burst directly overhead observer
$\delta t_{ave} = (r_{ave} - z_b)/v_s > \alpha N t_1/2 =$ condition to de-correlate blast waves arring at observer
$N$=total number of fragments = $\left(L_0/L_{\max}\right)^3$, $\alpha =$ de-correlation multiplier (~typ 3-10)

$$\tau v_{ave} = a_{ave} \rightarrow \tau = a_{ave}/v_{ave} = \left(r_{ave}^2 - z_b^2\right)^{1/2}/v_{ave} \propto 1/v_{ave}$$

$\rightarrow \tau$ can always be shortened by making $v_{ave}$ larger BUT this requires more energy ($\propto v_{ave}^2$)

$$(r_{ave} - z_b)/v_s = \alpha N t_1/2 \rightarrow r_{ave} = \alpha N t_1 v_s/2 + z_b$$

$$\tau = a_{ave}/v_{ave} = \left(r_{ave}^2 - z_b^2\right)^{1/2}/v_{ave} = \left((\alpha N t_1 v_s/2 + z_b)^2 - z_b^2\right)^{1/2}/v_{ave} = \left((\alpha\left(L_0/L_{\max}\right)^3 t_1 v_s/2 + z_b)^2 - z_b^2\right)^{1/2}/v_{ave}$$

$$\tau = \left(r_{ave}^2 - z_b^2\right)^{1/2}/v_{ave} = \left[(r_{ave} - z_b)(r_{ave} + z_b)\right]^{1/2}/v_{ave}$$

$\tau_c \equiv \alpha t_1 v_s/v_{ave}$, $\tau_b \equiv z_b/v_{ave}$ Note that $\tau_c$~300 sec while $\tau_b$ ~ 30,000 sec for $v_{ave}$=1m/s

$$\tau(a_{ave} << z_b \rightarrow r_{ave} \sim z_b) = \left[(r_{ave} - z_b)2z_b\right]^{1/2}/v_{ave} = \left[\alpha N t_1 v_s z_b\right]^{1/2}/v_{ave} = \left[\alpha\left(\frac{L_0}{L_{\max}}\right)^3 t_1 v_s z_b\right]^{1/2}/v_{ave}$$

$$\tau(a_{ave} >> z_b \rightarrow r_{ave} >> z_b) = \alpha N t_1 \frac{v_s}{2v_{ave}} = \alpha\left(\frac{L_0}{L_{\max}}\right)^3 t_1 \frac{v_s}{2v_{ave}} \propto L_0^3 \propto m_0 \text{(total bolide mass)}$$

$$\tau = \left(\left[\frac{1}{2}\alpha t_1 v_s\left[\frac{L_0}{L_{\max}}\right]^3 + z_b\right]^2 - z_b^2\right)^{1/2}/v_{ave} = \tau_b\left(\left[1 + \frac{N}{2}\frac{\tau_c}{\tau_b}\right]^2 - 1\right)^{1/2}$$

$$\tau(N\tau_c << \tau_b \text{ short range intercept}) = \sqrt{N\tau_c\tau_b}$$

$$\tau(N\tau_c >> \tau_b \text{ long range intercept}) = N\tau_c/2$$

As shown in the figure, comparing the simulations we see that the analytic model is reasonably consistent. The intercept time for the smaller diameters (<40m) is dominated by the dispersion in the burst altitude ($z_b$), while the very large diameters (>300m) are dominated by the finite size of the Earth. In both extremes (small and large), the intercept time is expected to be lower than the analytical solution, consistent with the accompanying plot. Note that zero-time intercepts are possible for bolides less than 30m due to the shock wave arrival time dispersion from the burst altitude dispersion dominating over the lateral dispersion. **In general, we take a more conservative approach and do not use true zero-time intercepts, though these are possible.**

**Table 2 – Summary of Short Time Intercepts that meet both Shock wave and Optical Pulse Limits**

| $L_0$ (m) | τ (days) | N | $L_{frag}$ (m) | v (m/s), σ(m/s) | Obs (km) | Max P at Obs (Pa) | Max overpressure under fragment (Pa) | Observed Optical Energy (J/m$^2$) @10% conversion |
|---|---|---|---|---|---|---|---|---|
| 15 | $10^{-3}$ (86s) | 20 | 5.5 | 10, 3 | 0 | 629 | 900 | 1866 |
| 20 | $10^{-3}$ (86s) | 20 | 7.4 | 10, 3 | 0 | 746 | 953 | 4979 |
| 30 | $10^{-2}$ (864s) | 100 | 6.5 | 1, 0.3 | 0 | 739 | 1002 | 11923 |
| 40 | 0.03 (43min) | 100 | 8.6 | 1, 0.3 | 0 | 1142 | 1197 | 33375 |
| 50 | 0.2 (4.8 hr) | 100 | 10.8 | 1, 0.3 | 17 | 1193 | 1555 | 57205 |
| 100 | 1 | 1000 | 10 | 1, 0.3 | 100 | 999 | 1738 | 75437 |
| 370 | 10 | 30,000 | 11.9 | 1, 0.3 | 1000 | 1625 | 2894 | 94746 |
| 500 | 20 | 100,000 | 10.8 | 1, 0.3 | 2000 | 775 | 3004 | 43207 |
| 1000 | 60 | 500,000 | 12.6 | 1, 0.3 | 6000 | 1111 | 3889 | 43422 |

*N*= number of fragments, observer is directly under fragment ring – worst place. $L_{frag}$ = mean fragment size. *v*=disruption speed relative to center of mass. Density =2.6g/cc with 30% dispersion (low end clipped Gaussian distribution) for fragment size, disruption speed and density. Attack angle relative to horizon is 45 deg. The maximum pressure under any fragment is shown (an observer under every fragment for this). The "Obs" is the chosen worst case observer distance from "ground zero" for maximum damage. The observer position is chosen to be conservative and calculate damage threshold for worst case. The "Max P under fragment" column is the maximum blast wave pressure directly under ANY fragment. This is an extremely small region on the ground. This can be further mitigated if needed by going to small fragment size (for example limiting fragments to be less than 10m or smaller diameter).

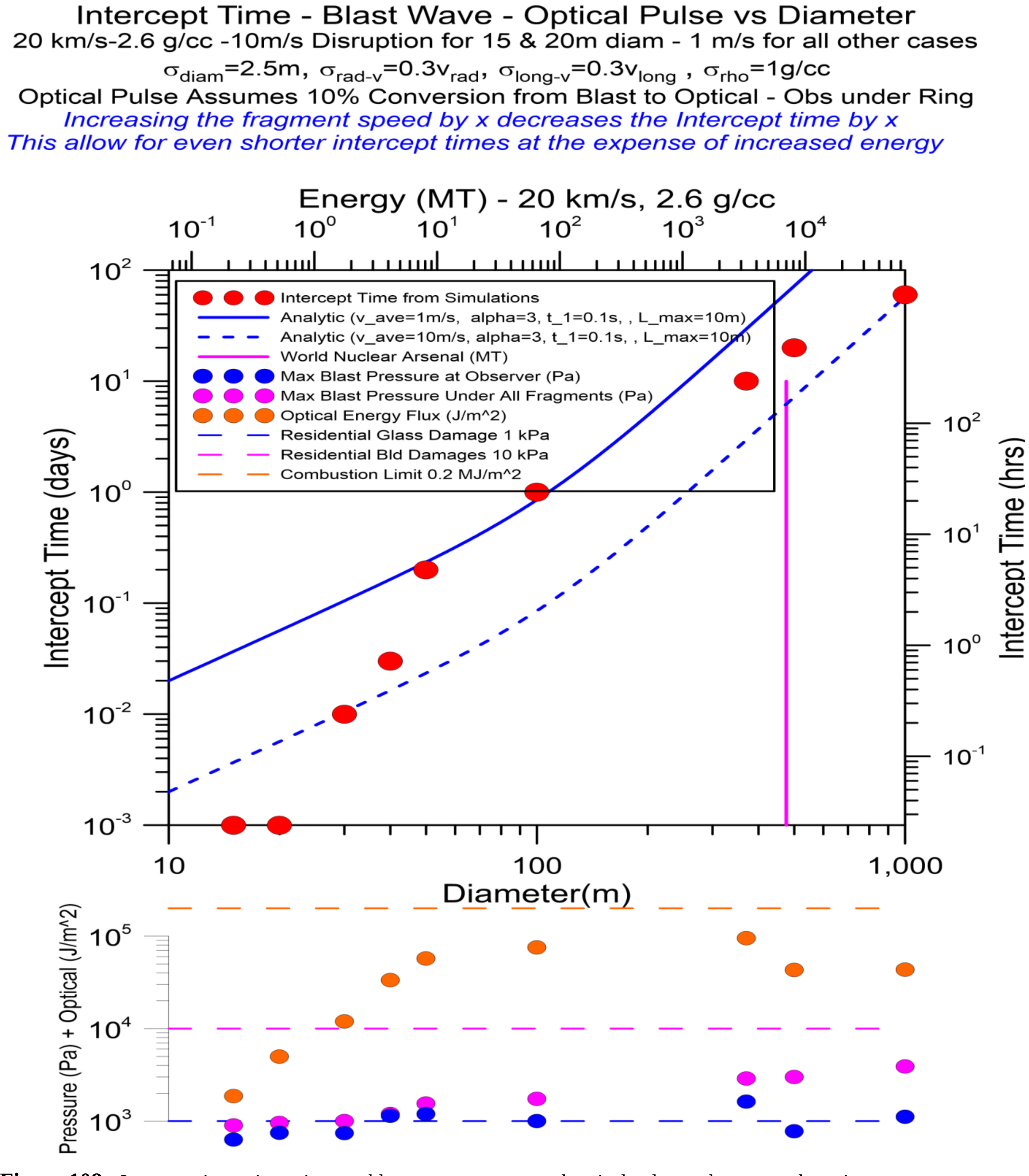

**Figure 108 -** Intercept time prior to impact, blast wave pressure, and optical pulse at observer and maximum pressure directly underneath all fragments vs bolide diameter. The simulations were done for a speed of 20km/s, density of 2.6g/cc and 45-degree angle of attack for 15, 20, 30, 40, 50 , 100, 370 (~**Apophis**), 500 (~**Bennu**) and 1000m diameters. Recall Bennu has a much lower density of about 1.26g/cc than our nominal rocky asteroid density of 2.6g/cc. The disruption speed dispersion is 30% of the mean speed and the density dispersion is 30% of the mean density. The disruption speed has a mean of 1m/s for all bolides except the 15 and 20m diameters, which have a mean disruption speed of 10m/s. The intercept time for the smaller diameters (<40m) is dominated by the dispersion in the burst altitude ($z_b$), while the very large diameters (>300m) are dominated by the finite size of the Earth. In both extremes (small and large), the intercept time is lower than expected analytically. **Note that zero-time intercepts are possible for bolides less than 30m due to the shock wave arrival time dispersion from the burst altitude dispersion dominating over the lateral dispersion. As mentioned, the intercept time τ can always be shortened by making the mean disruption speed ($v_{ave}$) larger, but this increases the disruption energy required. In general, even larger bolides have acceptable intercept times when the cloud diameter exceeds the Earth diameter. For 1m/s disruption, intercept times greater than 75 days are sufficient for bolides even larger than 10km.**

## 12. Mission Scenarios

### *Long Range Interceptors*

### *The Moon as a Possible Base for Detection and Interceptors*

The moon may be a future option for a longer-range program of planetary defense. The installation of both a long-range active detection based LIDAR system as well as a passive visible, NIR from scattered sunlight, as well as a possible thermal IR system would be a powerful augmentation to Earth and current space assets. A lunar based LIDAR system could also be as a "laser designator" for target locking. The low escape speed of 2.4km/s for the Moon vs 11.2km/s for the Earth is another significant advantage for a possible high speed interceptor program.

### *Short Range Interceptors*

As an example, assume we hit a 100m diameter asteroid with specific density 2g/cc travelling at $v_a$ ~ 10km/s relative to the Earth at a time prior to impact of $t_I$ ~ 1 day prior to impact with a small mission carrying 10 metric tons of penetrators. The total asteroid mass $m_a$ would be about $1\text{x}10^9$ kg. The impact would be at a distance of about $v_a * t_I \sim 8\text{x}10^8$ m or about 2 $R_{L,}$ where $R_L$ is the mean Earth-moon distance. Assuming we have a spacecraft speed of 5km/s, this gives a closing speed of 15km/s. If we use a 20x20 array of 25kg penetrators with a mean spacing of 5m, then the energy of each penetrator would be about $3\text{x}10^9$ J or ~ 1 ton TNT and the total kinetic energy of all the penetrators ($10^4$ kg) would be about $10^{12}$ J, or 250 tons TNT. The total gravitational binding energy is only about 1MJ, or $10^{-6}$ of the kinetic energy of impacts. To compute the effective speed of the debris field relative to the center of the asteroid, we would need to know the specifics of the material and binding distribution properties of the asteroid. This is dealt with in our simulations section for hyper velocity impacts. To be conservative, we assume a coupling between impact KE and debris KE of 1%. We would then have a debris KE of about $10^{10}$ J and get a mean debris speed $v_d$ ~ 4.5m/s. Note the gravitational escape speed is about 0.05m/s, so we are nearly 100 times the escape speed. In the remaining one day to impact, the debris cloud would spread out about $2v_d\, t$ where $t$=time to impact ~ $10^5$ s in this case (~ 1 day). This yields a debris foot print at impact of about $8\text{x}10^5$ m, or 800km. Note that there is both a lateral debris diameter ($v_d\, t$ ~ 800km in our case) as well as a longitudinal diameter that translates into a range of impact times. The latter is important in the context of atmospheric loading where there is both a spatial and a temporal component. If we assume the worst case of an instantaneous hit of all the debris that is spread out over 800km, we get an atmospheric flux of $1/2\ m_a v_a^2\ /(\ v_d\, t)^2 \sim 5\text{x}10^{16}\ \text{J}/6\text{x}10^{11} \sim 10^5\ \text{J/m}^2$ . This is equivalent to 100 seconds of Sunlight in terms of energy per unit area. This is a very small loading due to the spread of the debris cloud even in the case of a scenario where mitigation is one day prior to impact.

***Interceptor Energy Delivered*** – It is useful to understand the amount of energy delivered in the asteroid frame and compare this to the numerous calculations we have on the energy required to disassemble and spread out the fragments. We show a few representative cases:

- Passive penetrator - 20km/s asteroid →44 ton TNT equivalent per ton delivered.
    - 4.4kt for 100 ton delivered (Starship).
- Chemical explosive penetrators - 20km/s asteroid →45 ton TNT equivalent per ton delivered.
    - 4.5kt for 100 ton delivered (Starship).
- Nuclear explosive penetrators (assume B61-11 360kt ~130kg physics pkg) →3Mt per ton delivered (assuming 50% of mass is NED). For nuclear devices, especially thermonuclear devices, the ability to couple the NED energy efficiently into the bolide is complicated by the inability of standard thermonuclear devices to survive the high deceleration during direct penetration. Passive penetrator "hole drilling" prior to NED insertion is a possible solution. This is an area that needs future exploration IF NED's are needed for extremely large threats.
    - ~300Mt for 100 ton delivered (Starship) ~ 2% world arsenal. **Highly unlikely that such a mission of this type would be undertaken given the large yield within a single mission**

***Launch Vehicles and Multiple Independent Active Targeting Penetrators*** – After the initial intercept there may be residual large fragments that need to be additionally fragmented. For the case of stony bolides this would correspond to fragments with diameter larger than about 15m diameter. One option is to have a second interceptor array with independently targeting penetrators to "go after" these remaining large fragments. In this case the active penetrators would have small thrusters for lateral terminal guidance and also would be an intelligent swarm in that "they would communicate with each other" to decide on targeting priorities. Such a secondary system may not be necessary but would be a useful part of any robust strategy. In addition to lower mass launchers such as the Falcon 9 and others, we explore heavier lift launchers as shown in the table below. Note that for many smaller threats (<~150m diameter), the use of a "single launcher solution" using a modest lift launcher such as a Falcon 9 is a feasible path forward. **Since the development of launch vehicles will only "get better", the options we discuss in this paper, will rapidly be superseded by even more capable launchers in the future. The fact that we already have launchers capable of robust planetary defense using PI, is very encouraging. This is one area that we do not need to develop specifically for planetary defense as it is being driven extremely rapidly by other needs.**

**Table 3**. Parameters of various heavy lift launch vehicles in consideration for PI for large threats

| Parameter | Space X Starship | SLS Block 1,2 | Falcon Heavy | Delta IV Heavy |
|---|---|---|---|---|
| Payload Mass to LEO (mt – 1000 kg) | ~150 | 70 (1), 105 (1B) 130 (2) | 53 | 28.8 |
| Payload Mass to "deep space" (mt – 1000 kg) | ~100 with refueling? | 38 (1B), 46 (2)<br>27 TLI (1)<br>43 TLI (1B) | 26 GTO<br>16.8 TMI | 14 GTO |
| Cost per unit mass to LEO | Goal <1 k$/kg | 19 k$/kg | 1.9 K$/kg | 13 k$/kg |
| Fairing Diameter (m) | 9 Dx18 L | 8.4 B1 to 10 B2 | 5.2 | 5 |
| Status | Expected 2027 | Block 1 Flight proven | Flight proven | Flight proven |
| Unique features | In orbit refueling | | | |
| Could be Lunar Based | Possible | Unlikely | Possible | Unlikely |
| Could deploy lunar LIDAR tracking and targeting | Y | Y | Possible | Unlikely |

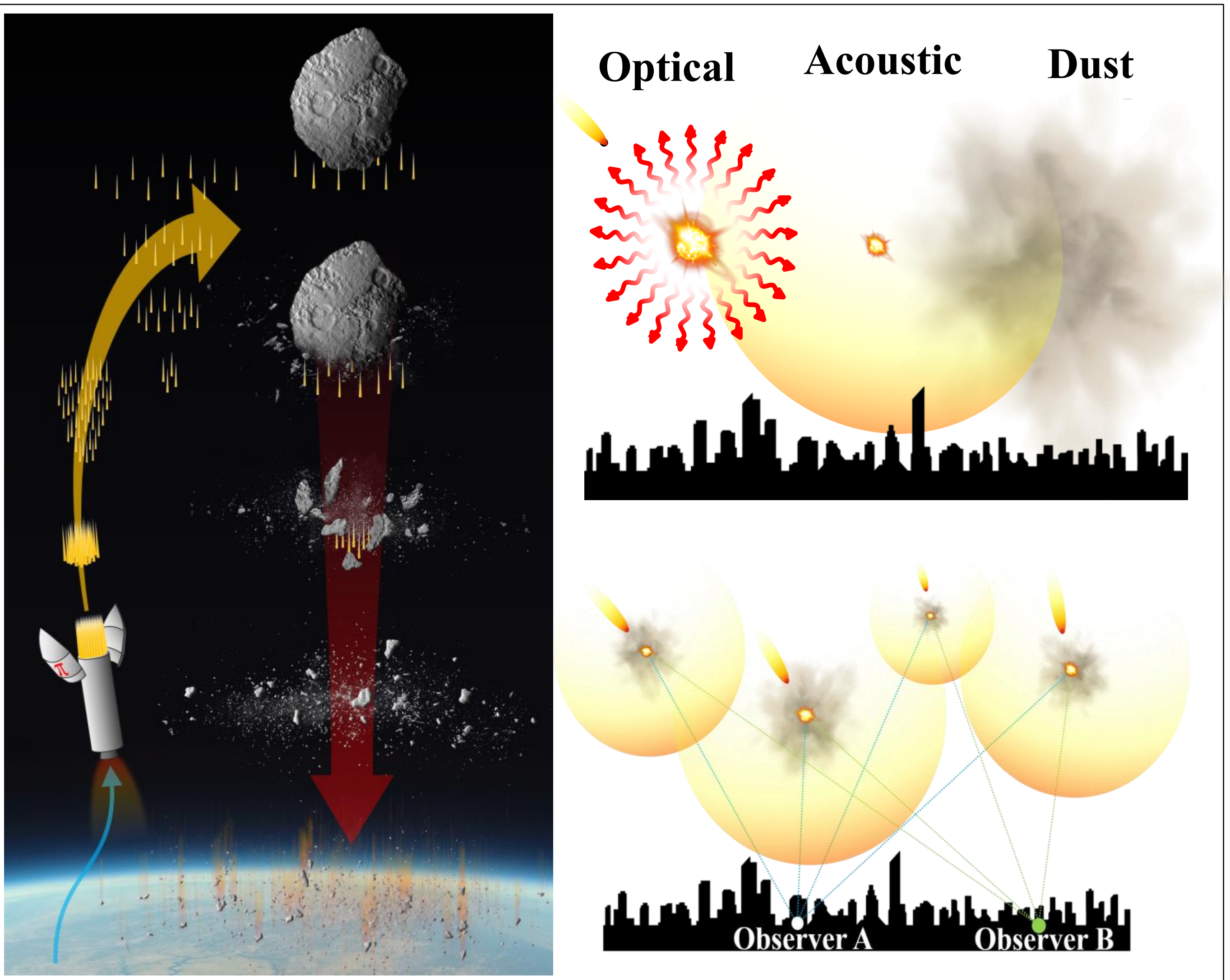

**Figure 109 –L:** Interceptor penetrator array deployment sequence. Note that the outer layers are ejected (peeled away) first and then successive penetrators peel away the inner layers in a conical pattern. R upper: Optical and Acoustic Signature from each fragment as well as minor contribution of dust from fragment disintegration. R Lower: Effects of multiple fragments on observer. The acoustic de-correlation is clearly seen from the varying acoustic path lengths from each fragment.

### *Tilt of Fragment Plane Relative to Observer and Finite Earth Radius Effects*

We have assumed a conservative model where there is no significant difference between the atmospheric entry times across the fragment cloud plane. There are several effects that can cause a significant difference in this. In the case of a large fragment cloud, where large is considered relative to the Earth's radius, this can be an issue for earlier time intercepts and/or large disruption speeds. This results in two primary effects:

- The angle of attack will vary significantly for the various fragments.
- There will be a significant time delay between fragments entering the atmosphere.
- The fragment flux will be reduced by the projection at larger angles.

In general, both of these effects will further de-correlate the acoustic blast wave signature and increase the spread of the optical signature arrival times. Both of these further mitigate the effects and further reduce the danger. Thus, we are conservative in that we do not take this into account in the current simulations. We note that this effect is only significant when the fragment cloud is very large and thus in general the danger is already greatly mitigated. Adding in the tilt correction only further reduces the effects This is computed below allowing for an arbitrary angle of attack of both the parent bolide as well as the effect on subsequent angle of attack of each fragment in the fragment cloud as it is projected onto the curved Earth.

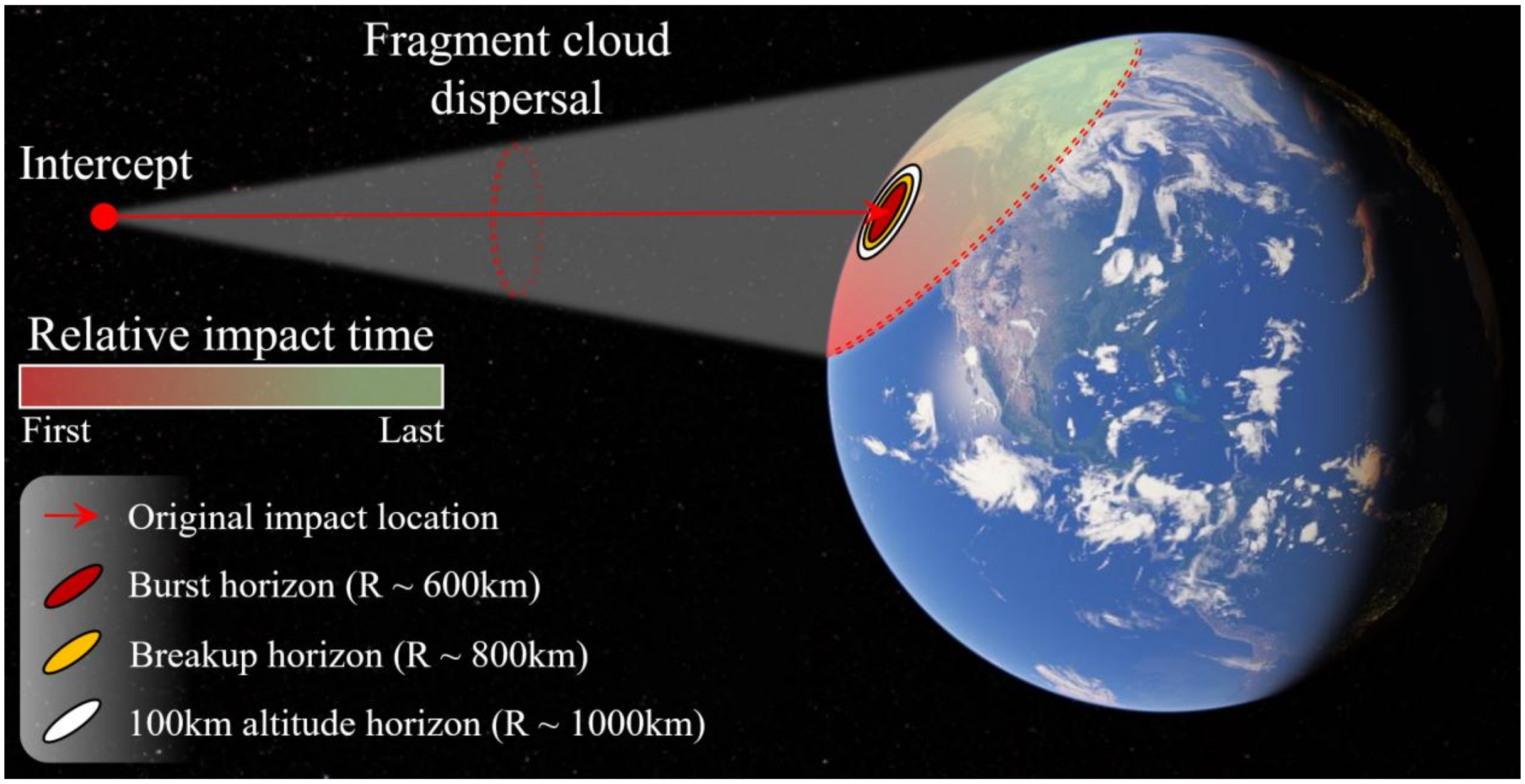

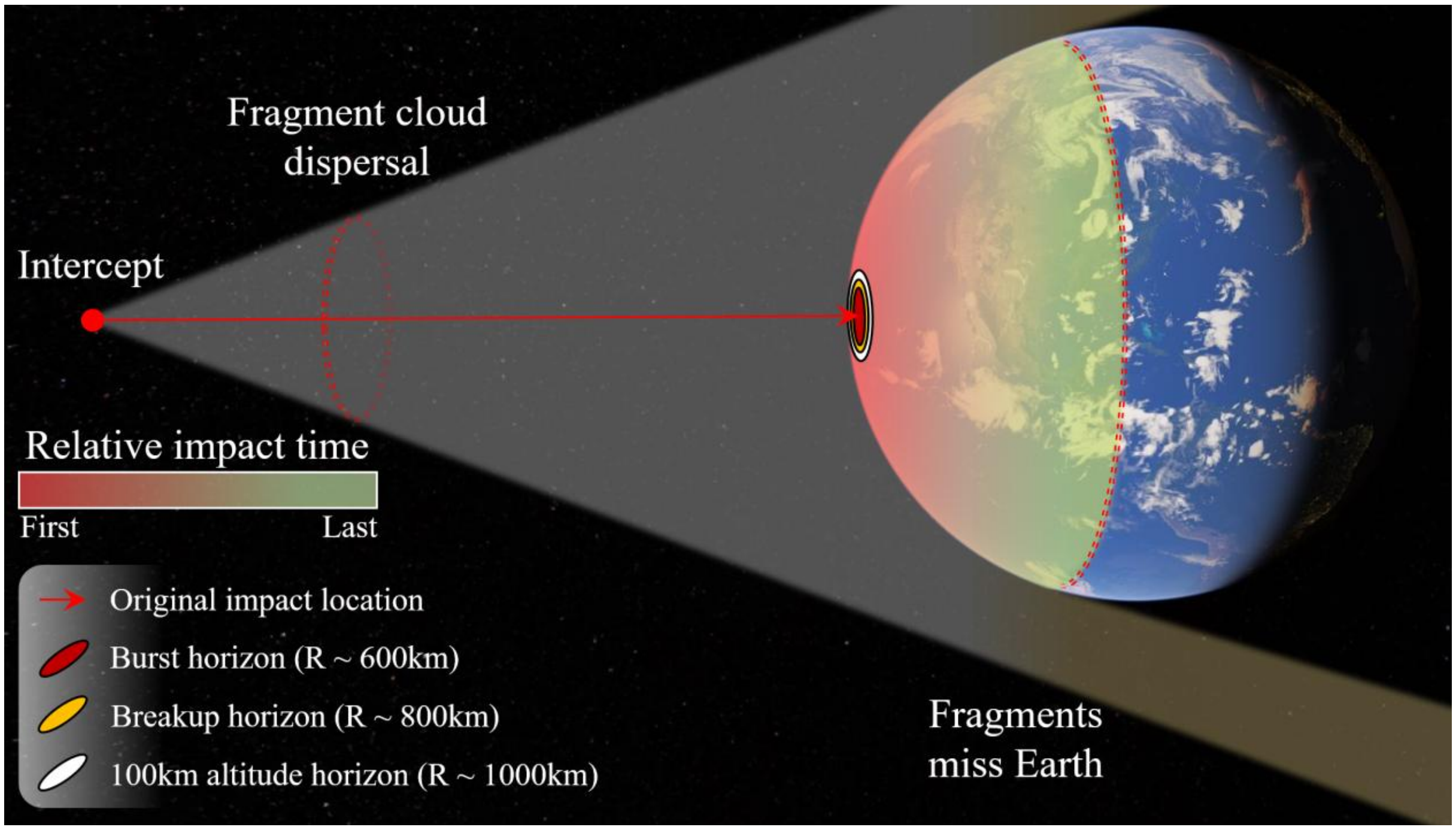

**Figure 110 –** Top: Projection of the fragment cloud onto the surface of the Earth. The superimposed color scale shows the distribution in impact time as a function of the impact location. (Bottom) Projection of the fragment cloud onto the surface of the Earth in the case of a much larger fragment cloud. In this case, a large fraction of the fragments miss the Earth entirely. This can be achieved either by earlier interdiction or by imparting a larger amount of kinetic energy to the fragments.

**Quantitative Analysis of the Fragment Cloud Impacting the Curved Earth' Atmosphere** – We make a model of the fragment cloud as a series of planes that are perpendicular to the parent bolide velocity vector with the velocity vector being at any attack angle relative to the unmitigated bolide. Note that for a vertical angle of attack of the parent bolide (90-degree attack angle relative to the horizon), the variation in the angle of attack for each fragment in the fragment cloud is much less than for a small angle of attack for the parent bolide. In the language of the diagram and calculations below, the case of a vertical angle of attacks is when the impact distance from the contact point of the tangent plane is zero ($r_o$=0).

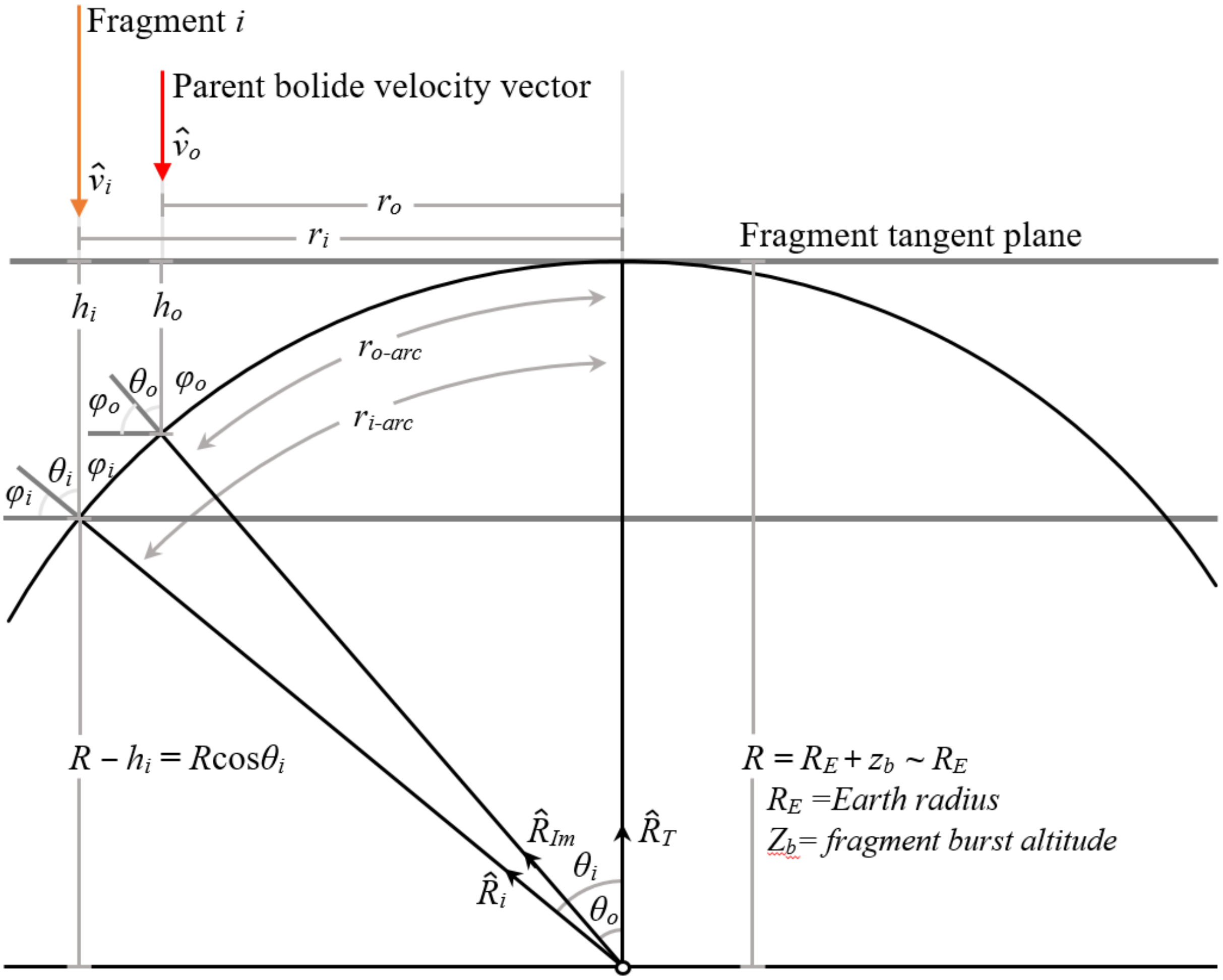

**Figure 111** – Diagram of projected fragment plane used in calculations below. Note that the angle of attack of the fragment relative to the horizon ($\phi_i$ ) varies as the position $r_i$ of the fragment varies. Note that the fragment "tangent plane" is defined such that the parent bolide velocity vector is parallel to the fragment plane normal which is the same as the local normal to the Earth at the "contact point" of the tangent plane and the Earth's surface. The angle of parent bolide velocity vector relative to the local horizon where the parent bolide would have hit had it not been mitigated is the angle of attack of the parent bolide and is the angle $\phi_0$ in this diagram. Once the parent bolide is intercepted and disassembled then each fragment will have its individual angle of attack $\phi_i$ as shown. The radius of each fragment relative to the parent bolide is the distance ($r_i - r_0$). A vertical impact of the parent bolide is when $r_0$=0.

Impact below refers to fragment impact with atmosphere - typically at burst

$R_E = Earth\,radius$

$z_b = burst\,altitude\,of\;\;fragment$

$R = R_E + z_b \sim R_E\;as\,z_b << R_E$

$\boldsymbol{R}_{\text{Im}}$ =unit radius vector from Earth center to parent bolide impact

$\boldsymbol{R}_i$ =unit radius vector from Earth center to fragment i

$\widehat{\boldsymbol{R}_T = unit}$ radius vector from Earth center to fragment plane tangent point

$\widehat{\boldsymbol{V}_o = unit}$ velocity vector from of parent bolide

$\theta_o = zenith\,impact\,angle = \pi - \cos^{-1}(\widehat{\boldsymbol{V}_o \bullet \boldsymbol{R}_{\text{Im}}})$

$\phi_o = horizon\,impact\,attack\,angle = \pi/2 - \theta_o = \cos^{-1}(\widehat{\boldsymbol{V}_o \bullet \boldsymbol{R}_{\text{Im}}}) - \pi/2$

$\theta_i = \;fragment\;\;i\,zenith\,impact\,angle = \pi - \cos^{-1}(\widehat{\boldsymbol{V}_o \bullet \boldsymbol{R}_i})$

$\phi_i = \;fragment\;\;i\,horizon\,impact\,attack\,angle = \pi/2 - \theta_i = \cos^{-1}(\widehat{\boldsymbol{V}_o \bullet \boldsymbol{R}_i}) - \pi/2$

$r_i$ = distance along tangent plane from tangent point intersection to fragment i impact

$r_0$ = distance along tangent plane from tangent point intersection to parent bolide impact

$r_{o-arc}$ = arc distance along Earth surface from tangent point intersection to parent bolide impact

$r_{i-arc}$ = arc distance along Earth surface from tangent point intersection to fragment i impact

$h_i$ = distance along parent bolide velocity from trangent plane to impact for fragment i

From $\widehat{\boldsymbol{V}_o},\;\widehat{\boldsymbol{R}_{\text{Im}}}$ we get $\theta_o, \phi_o\;or\;\;from\,\theta_o, \phi_o\;and\;\widehat{\boldsymbol{R}_{\text{Im}}}\;we\,get\,\widehat{\boldsymbol{V}_o\,(rel\,to\,\boldsymbol{R}_{\text{Im}})}$

From $\theta_o, \phi_o$ we get $r_0 = R\sin(\theta_0)\,and\,h_0 = R(1-\cos(\theta_0))$

From $r_0$ we get $r_i$ if we know the fragment spread speed $v_i$ and time before impact $\tau \to (r_i - r_0) = v_i\tau$

From $r_i$ we get $\theta_i$ from $r_i = R\sin(\theta_i)$

$r_{o-arc} = R\theta_o$

$r_{i-arc} = R\theta_i$

$$h_{i\text{-}0} = R - h_i = R\cos(\theta_i) = \sqrt{R^2 - r^2{}_i}$$

$$r_i = R\sin(\theta_i),\;\; h_i = R(1-\cos(\theta_i)) = R(1-\sqrt{1-(r_i/R)^2}) \to h_i/R \to \frac{1}{2}\left[\frac{r_i}{R}\right]^2\;\;when\,r_i << R$$

$$r_0 = R\sin(\theta_0),\; h_0 = R(1-\cos(\theta_0)) = R(1-\sqrt{1-(r_o/R)^2}) \to h_0/R \to \frac{1}{2}\left[\frac{r_0}{R}\right]^2\;\;when\,r_0 << R$$

Note the derivatives can become large particularly when $\theta_i, \theta_0$ are not small angles :

$dh_i/d\theta_i = R\sin(\theta_i) = r_i,\;\; dh_0/d\theta_0 = R\sin(\theta_0) = r_0,\; dr_i/d\theta_i = R\cos(\theta_i),\; dr_0/d\theta_0 = R\cos(\theta_0)$

$dh_i/dr_i = (dh_i/d\theta_i)/(dr_i/d\theta_i) = \tan(\theta_i) \to \infty\;\;as\;\;\theta_i \to \pi/2,\; dh_0/dr_0 = \tan(\theta_0) \to \infty\;\;as\;\;\theta_0 \to \pi/2$

$dh_i/dr_i = \tan(\theta_i) = \tan(\sin^{-1}(r_i/R)),\; dh_0/dr_0 = \tan(\sin^{-1}(r_0/R))$

$Another\;way\,to\,view\,this: (R-h_i)^2 + r^2{}_i = R^2 \to R^2 - 2h_iR + h_i^2 + r^2{}_i = R^2 \to h_i^2 - 2h_iR + r^2{}_i = 0$

$$h_i = \frac{2R \pm \sqrt{4R^2 - 4r^2{}_i}}{2} = R - \sqrt{R^2 - r^2{}_i} = R(1-\sqrt{1-(r_i/R)^2})$$

From the distance between the tangent plane and the impact of each fragment we can compute the relative time delay that each fragment experiences by $\Delta\tau_i = h_i/v_o$ where $v_o$ is the parent bolide speed. For extreme cases, where the fragment cloud extends beyond the radius of the Earth then $\theta_i = \pi/2$ and $h_i = R \sim R_E$ (radius of Earth). As an example, if the parent asteroid is coming in at $v_o = 20$ km/s then the fragment time delay can be as large as $\Delta\tau_i = h_i/v_o = R_E/v_o \sim 300$ sec (5 min). For a case where $\theta_i = \pi/2$ the fragment is spread at a distance equal to the Earth's radius with $r_i = h_i = R_E$ and hence the fragment detonation is far beyond the optical horizon for an observer near the first fragments to hit and thus the effects from these distant fragments are minimal. There may be some "over the horizon" weak acoustics as well as distant atmospheric scattering and even weaker lunar backscattering of the optical signature but it will be minimal.

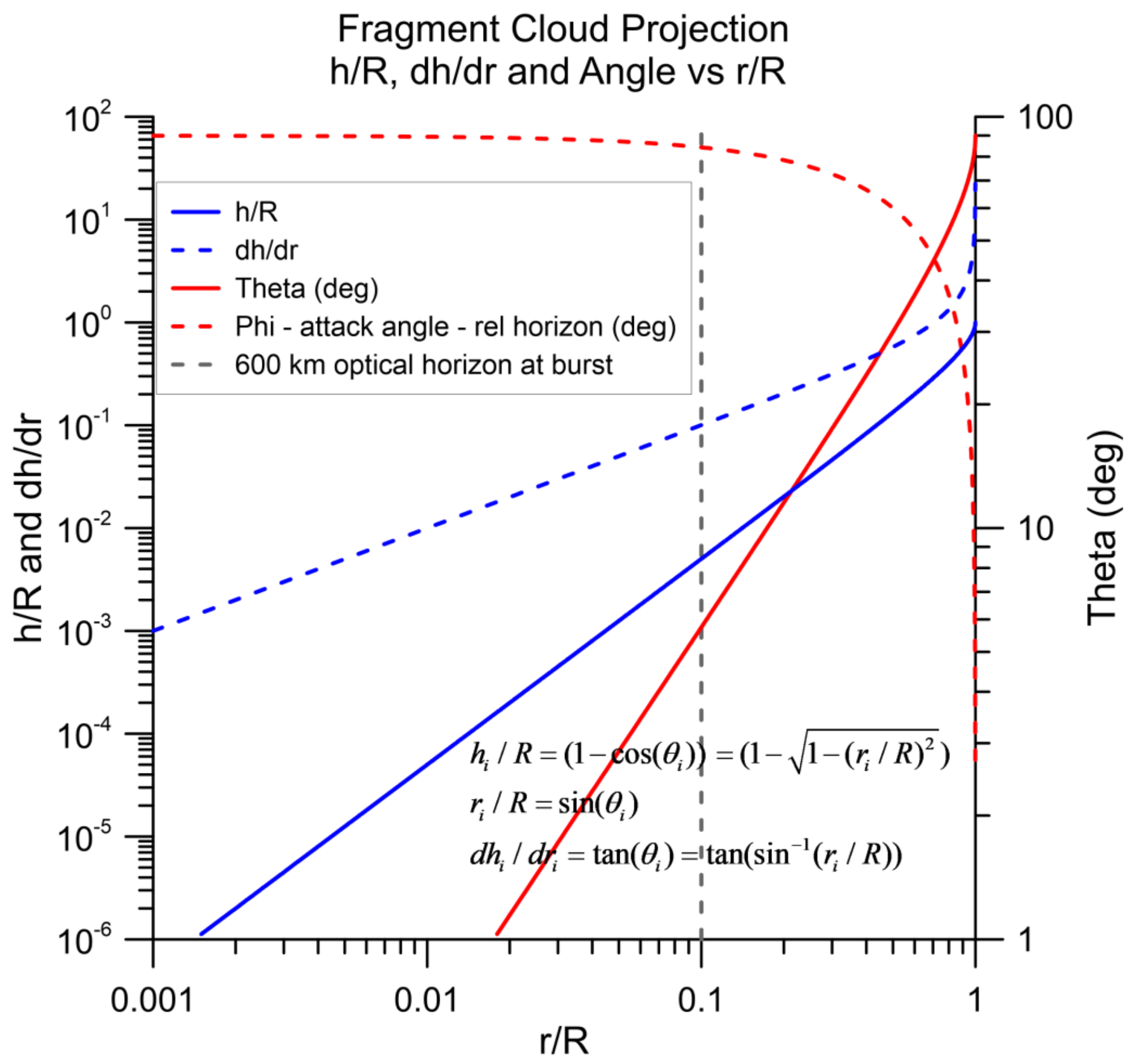

**Figure 112** – Plot of fragment cloud projected onto the Earth as outlined in the equations and accompanying figure in the section above. Shown here are the ratio of the fragment plane intersect distance ratio for a given fragment distance from the "center" relative to the Earth's radius including atmospheric height at typical burst which is completely dominated by the Earth's surface radius (ratio h/R) as well as the derivative dh/dr, angle (theta) relative to the unmitigated bolide and fragment angle of attack (phi) relative to the horizon. The angle of attack varies with the fragment position. The optical horizon at 600 km from the "center" of the tangent plane for a burst altitude of approx. 30 km is shown for reference.

As an example, we assume normal incidence to the observer ($\theta_0=0$) and set ri=640 km ($r_i/R \sim 0.1$) which is approximately the optical horizon for a burst at 30 km altitude. This yields an $h_i/R \sim 0.005$ or $h_i$ = 32 km. If the parent bolide has a speed of 20 km/s this would give a fragment transit time of about 1.5 seconds. This is very short compared to the acoustic blast wave travel across the 600 km distance of 1800 second but very long compared to the optical travel time across 600 km of 2 ms.  The angle $\theta_i$ for $r_i$=640 km is about 5.7 degrees (~0.1 radian).

### *Dynamic Penetrator Array Reconfiguration and Imaging at Intercept*

Due to the limited information that may be available at launch as to the target geometry and rotation state anticipated at intercept, the penetrator array can be configured once the interceptor gets close enough to the target to precisely measure the optimal array geometry. Non-circular geometries can be utilized in this approach. The same final approach targeting system, or an auxiliary imaging system, can also be used to make precision visual measurements after intercept to confirm target fragmentation.

### *Blast Wave Phase Speed Along the Earth's Surface*

When the blast wave intercepts the Earth's surface the speed of the blast phase front moves faster than the group speed of the blast wave. This is shown below. The phase speed diverges to infinity directly under the fragment (θ=0) and asymptotes to the speed of sound (or speed of the blast wave (group speed) at θ= π/2.

$$\tan(\theta) = r_g / z_b$$

$$v_p / v_s = 1/\sin(\theta)$$

$$r_s = (r_g{}^2 + z_b{}^2)^{1/2} = r_g(1+(r_g / z_b)^{-2})^{1/2}$$

$$dr_s / dr_g = r_g / r_s = (1+(r_g / z_b)^{-2})^{-1/2}$$

$$\to 0 \; for \, r_g \to 0$$

$$\to 1 \; for \, r_g \to \infty$$

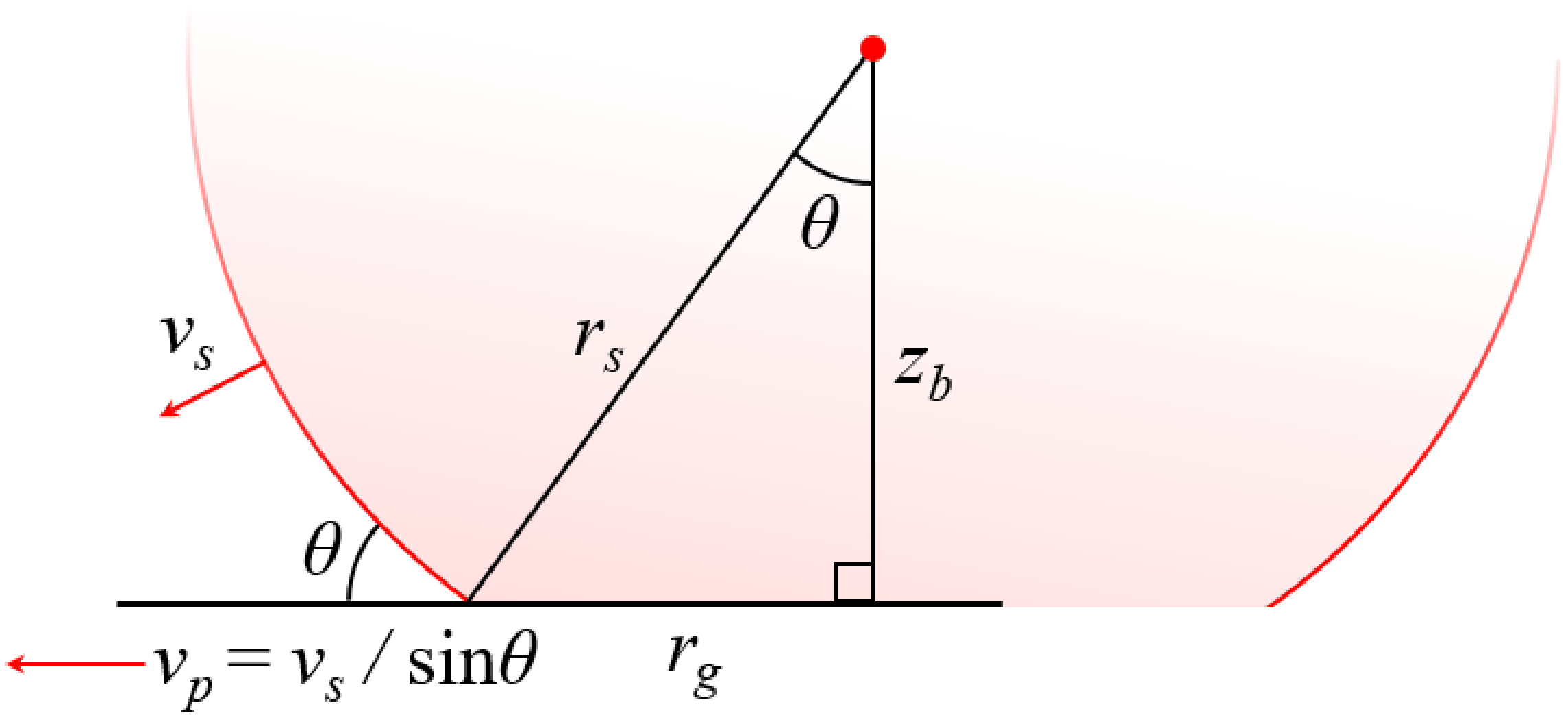

**Figure 113 -** Plot of blast wave phase speed $v_p$ projected onto the Earth surface. The fragment burst altitude is $z_b$ ground distance is $r_g$ and slant range is $r_s$ and the sound speed is $v_s$. The phase speed diverges at θ=0 and goes to the sound speed $v_s$ at θ=π/2.

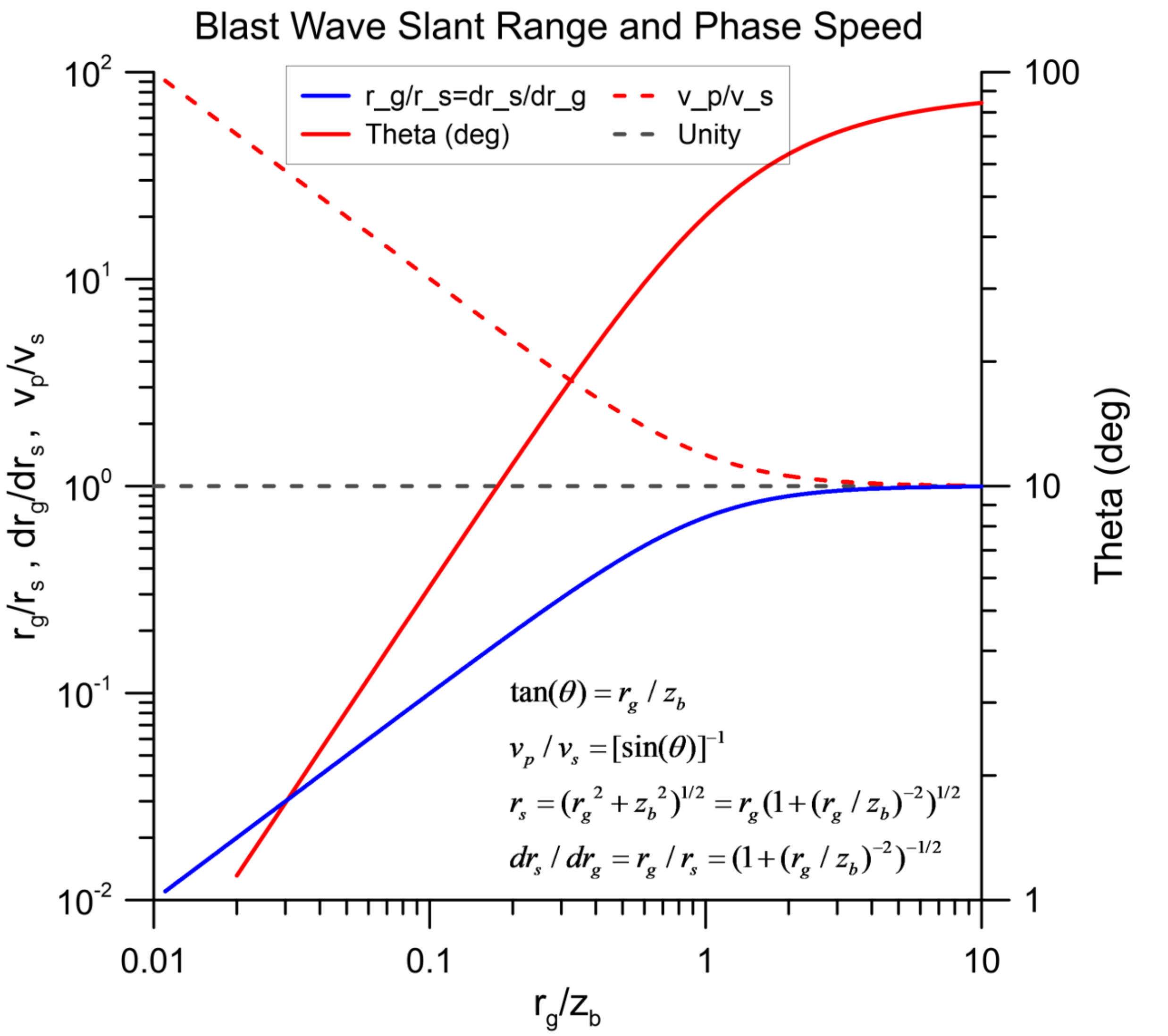

**Figure 114** Blast wave slant range, derivative, observer angle from burst and phase speed to sound (blast wave) speed vs ratio of ground distance to burst altitude.

### *Launch Time vs Intercept Time and Time to Impact after Intercept*

For serious threats, it will be generally desired to launch interceptors as early as possible, even if complete information is not available. In any realistic scenario, multiple launchers would be utilized and communications between interceptors, including imaging feedback, and the Earth would be part of any targeting system. We work out the various relationships between the relevant parameters. We ignore the speed increase of the bolide as it is pulled in by the Earth's gravity.

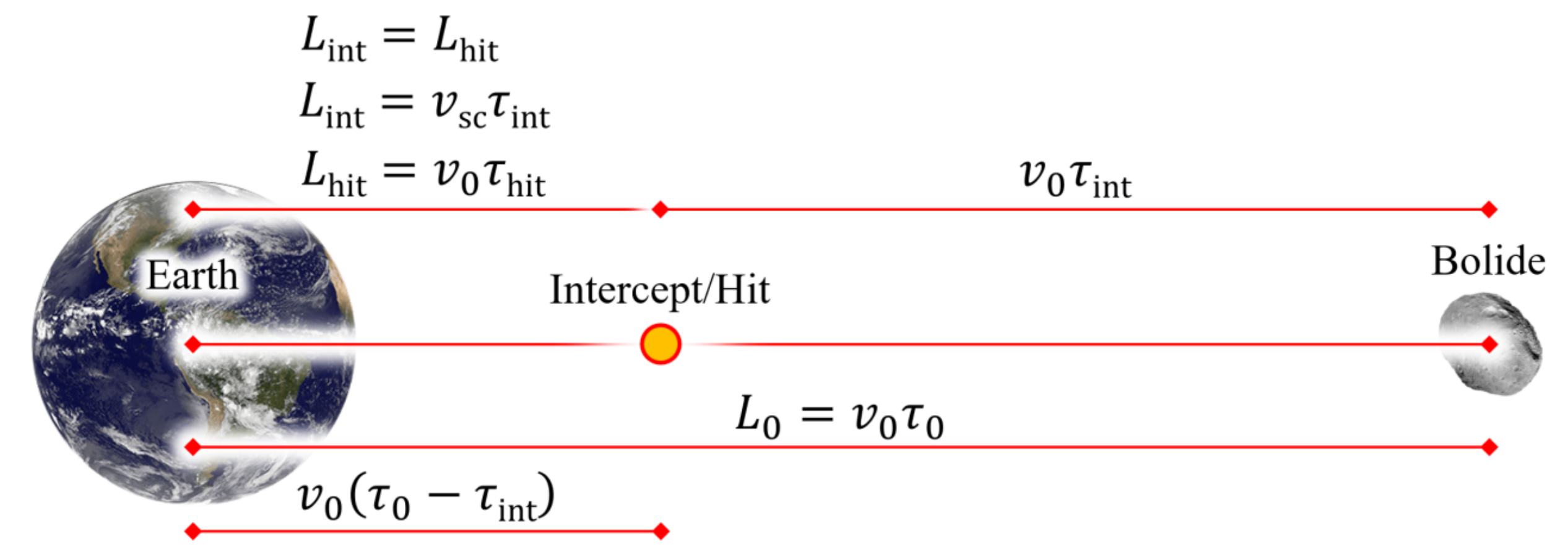

**Figure 115** - Relationship between intercept, spacecraft speed and initial bolide position. For simplicity this ignores the complexity of the actual orbital dynamics which is case specific.

$v_0 =$ target speed relative to Earth

$v_{sc} =$ spacecraft speed relative to Earth

$L_{\text{int}} = L_{hit} =$ Distance to target at intercept/hit, $L_0 =$ Distance to target at launch, $1\text{AU} \sim 1.5\text{x}10^{11} m \sim 395\,LD$

$\tau_{\text{int}} =$ Time for spacecraft to intercept starting from t=0 at launch

$\tau_0 =$ Time for target to collide with Earth from launch

$\tau_{hit} =$ Time for target to collide with Earth from intercept

$\tau_0 = L_0 / v_0,\ \ \tau_{\text{int}} = L_{\text{int}} / v_{sc},\ \ \tau_{hit} = L_{hit} / v_0 = L_{\text{int}} / v_0$

$L_{\text{int}} = v_{sc}\tau_{\text{int}} = L_0 - v_0\tau_{\text{int}} = v_0(\tau_0 - \tau_{\text{int}})$

$\tau_{\text{int}} = L_0 / (v_0 + v_{sc}) = \tau_0 v_0 / (v_0 + v_{sc}),\ \ \tau_{hit} = L_{hit} / v_0 = \tau_0 - \tau_{\text{int}}$

$$\frac{\tau_{hit}}{\tau_0} = \frac{v_{sc}}{v_0 + v_{sc}} \qquad \frac{\tau_{\text{int}}}{\tau_0} = \frac{v_0}{v_0 + v_{sc}} \qquad \tau_{hit} + \tau_{\text{int}} = \tau_0 \qquad L_{hit} = v_0\tau_{hit} = v_{sc}\tau_{\text{int}} = L_{\text{int}}$$

Examples for short warning asteroid and longer warning for very large asteroid or comet.

$Ex: v_0 = 20km/s, v_{sc} = 5km/s \rightarrow \tau_{hit}/\tau_0 = 0.2 \rightarrow IF\ \ \tau_0 = 25hr\ \ then\ \tau_{hit} = 5\,hr, \tau_{\text{int}} = 20hr,\ L_{\text{int}} \sim 0.0024AU \sim 0.95LD$

$Ex: v_0 = 20km/s, v_{sc} = 10km/s \rightarrow \tau_{hit}/\tau_0 = 0.33 \rightarrow IF\ \ \tau_0 = 5\,day\ \ then\ \tau_{hit} = 1.7\,day, \tau_{\text{int}} = 3.3day,\ L_{\text{int}} \sim 0.019AU \sim 7.5LD$

$Ex: v_0 = 20km/s, v_{sc} = 5km/s \rightarrow \tau_{hit}/\tau_0 = 0.2 \rightarrow IF\ \ \tau_0 = 5\,day\ \ then\ \tau_{hit} = 1day, \tau_{\text{int}} = 4day,\ L_{\text{int}} \sim 0.015AU \sim 4.6LD$

$Ex: v_0 = 40km/s, v_{sc} = 10km/s \rightarrow \tau_{hit}/\tau_0 = 0.2 \rightarrow IF\ \ \tau_0 = 5\,mo\ \ then\ \tau_{hit} = 1mo, \tau_{\text{int}} = 4mo,\ L_{\text{int}} \sim 0.69AU \sim 273LD$

$Ex: v_0 = 20km/s, v_{sc} = 5km/s \rightarrow \tau_{hit}/\tau_0 = \ \ 0.2 \rightarrow IF\ \ \tau_0 = 5\,mo\ \ then\ \tau_{hit} = 1mo, \tau_{\text{int}} = 4mo, L_{\text{int}} \sim 0.35AU \sim 137LD$

***50m diameter intercept with a one day prior to impact launch*** - Note that the example above for a 25 hour prior to impact launch with an intercept at 5 hours prior to impact allows the mitigation of a 50m diameter density 2.6 g/cc bolide travelling at 20km/s and disrupted at 1m/s. This means that a "system at the ready" could react rapidly (~1 day) prior to impact for 50m diameter or less class threats. This is comparable to a Tunguska class event. For this case the intercept would be at about 0.95 LD with a distance at the time of launch of about 4.8 LD.

***Threat Detection*** - To put this in the context of detectability for a 50m diameter threat we calculate how visible this object (4.8 LD at launch) would be at the relatively short response time. The flux of light received at the Earth's surface from an object of this diameter illuminated by the Sun with a 10% albedo is about 8 fW/m$^2$ or ~5x10$^4$ ph/s-m$^2$ in the visible band (0.4-0.7 microns) if fully solar illuminated, or roughly visible magnitude $m_v$=15. **This is easily detected in a 10cm diameter lens/ telescope with an integration time of 1 second using a modern low noise CMOS or CCD Si imaging array. This bodes well for a compact "all sky" high cadence ground or space/ lunar survey allowing a relatively short time response detection system. A space/lunar survey is much better as it avoids the case of a target being only visible during the night but starting on the ground is a first step. Note that even 10% solar illumination is easily detectable. A harder case in the visible is a threat "blinded" by coming from the direction of the Sun. Having both "all sky" imaging from Earth, lunar and Lagrange points or large orbit surveys would mitigate the solar blind cases given the large parallax due to the short distance targets we are searching for.** A visible multi location (Earth, lunar and band high cadence survey combined with a space based mid/ long wave (5-15 micron) thermal IR survey would be extremely effective for the detection of relatively small diameter threats that need fast mitigation response.

***General Threat Detection*** **-** Note that solar illuminated "ground flux" flux is proportional to threat diameter squared and inversely proportional to detection distance (and hence time) squared. A **key metric is the ratio of threat diameter to mitigation time. Larger threats and hence larger mitigation time allow longer integration time surveys which detect at larger distances.** ***Since PI can mitigate virtually any threat so rapidly we see that all the cases we cover (up to 1 km) are readily detectable in sufficient time with the same detection strategy. This is a significant advantage of PI.***

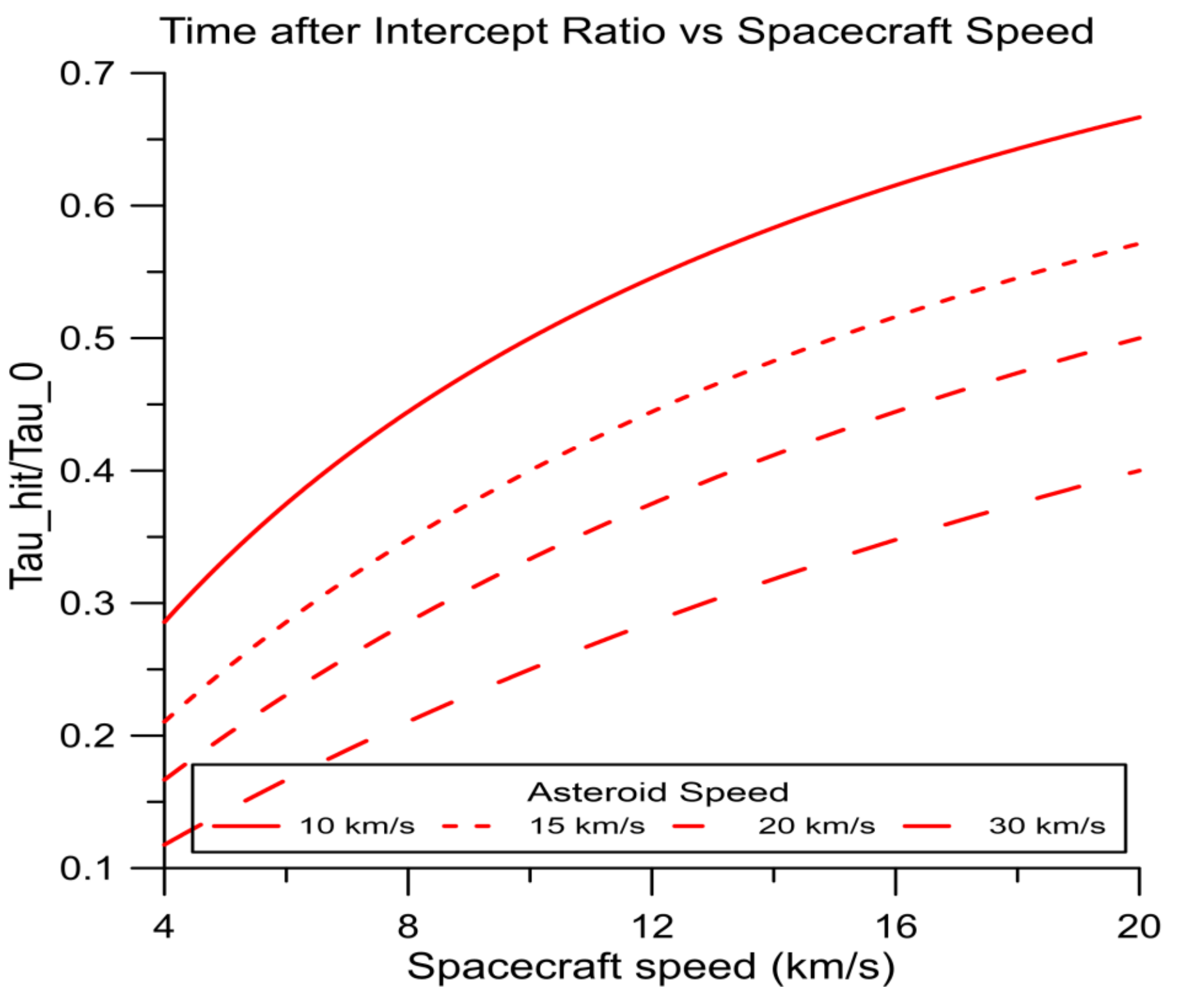

**Figure 116** – Ratio of $\tau_{int} / \tau_o$ versus spacecraft speed and asteroid speed where $\tau_{int}$ is the time to hit the Earth after intercept and $\tau_o$ is the time to hit the Earth after launch.

*Enhanced Deflection Mode*

Another mode of operation is to asymmetrically "eject" part of the target off to one side and thus push against the other larger portion. This can be used for an extremely effective, though long time scale, deflection option. This is not like conventional deflection techniques which use the relative momentum of the interceptor to "nudge" the bolide off an impact trajectory. In our deflection case we deliver energy and not momentum with the relative momentum of the spacecraft/ penetrators being very small compared to the momentum transfer to the remaining larger remaining fragment. The net change in momentum of the system (ejected portion + remaining portion) being essentially zero but the introduction of the energy in the internal detonation "pushed" the ejected fragment away with large momentum thus imparting the same but opposite momentum to the larger portion remaining.

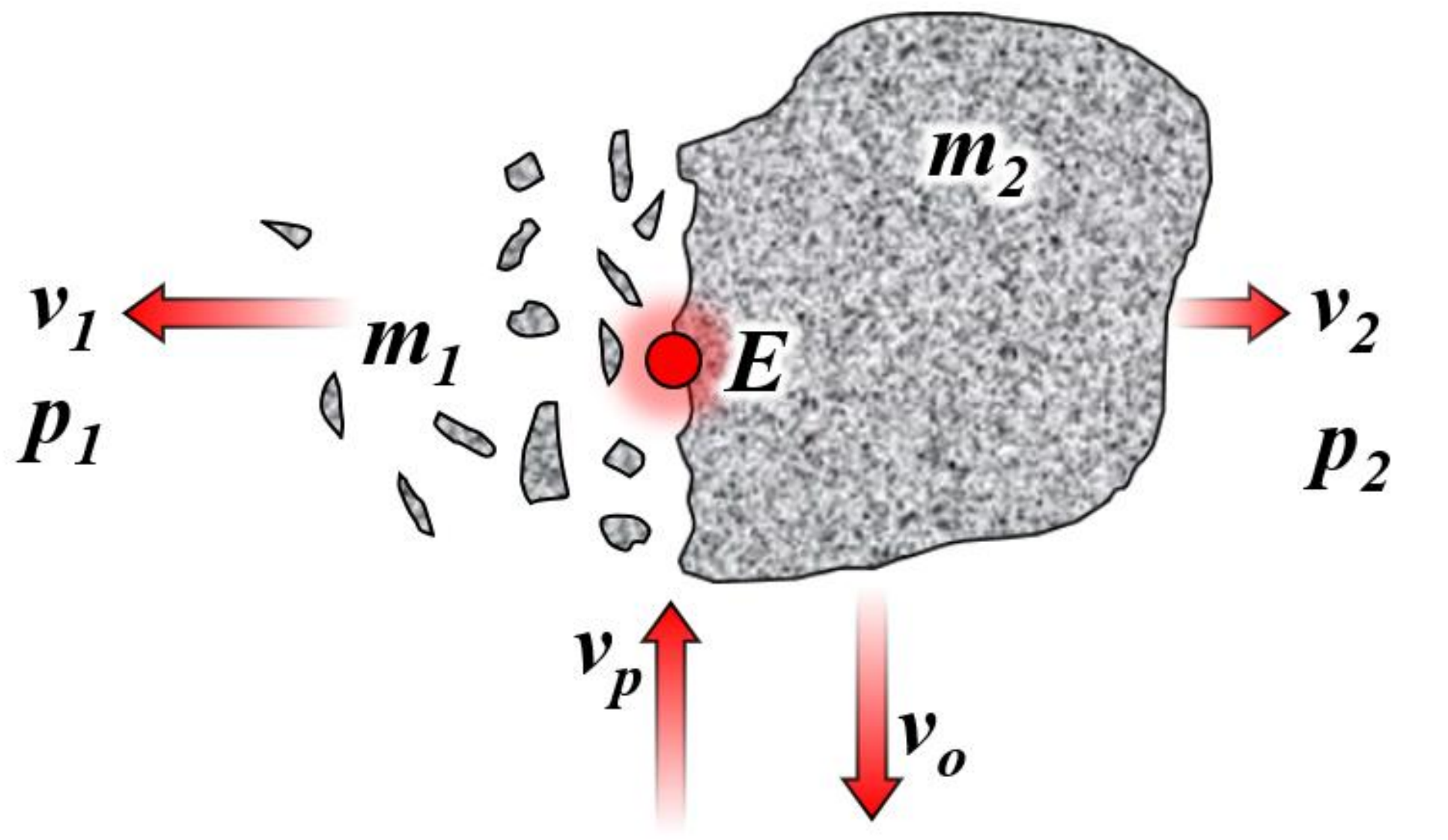

**Figure 117 –** Diagram of enhanced deflection via energy injection induced mass ejection. Note that while the diagram shows a transverse momentum mode, any other mode such as longitudinal (parallel/anti-parallel to parent bolide velocity vector) is viable.

***Transverse vs Longitudinal Momentum Modes*** – While the diagram shows a transverse momentum transfer, any other coupling mode is viable. In any realistic energy injection (detonation) there will be a wide variety of issues including geometry (view factor) that will be relevant. **As we discuss below the enhancement of momentum transfer can be vastly larger than that from a purely classical impactor momentum transfer for the same mass of the impacting system.** The basic reason is simple in that energy injection whether via the kinetic energy of the impactor or through a combination of the impactor kinetic energy and explosive charge. The same is true whether the explosive charge is chemical or an NED.

***Tradeoff between Deflection and Disruption (complete fragmentation) Modes*** – as shown below the use of the enhanced deflection mode generally only makes sense if the target is extremely large AND there is a very long warning time. We normally prefer to completely fragment a threat rather than deflect it but in some cases deflection may be desired. The tradeoff key metric is the energy injection to the system. As shown below only modest energy inputs are required for large deflection but as the energy is increased, we quickly come to a point where complete fragmentation is feasible. This is a quantitation issue as shown in the examples below.

$v_0 =$ target speed relative to Earth

$v_{sc} =$ spacecraft speed relative to Earth

$m_1, v_1, p_1 =$ mass, speed, momentum of ejected small portion of bolide (at $\infty$)

$m_2, v_2, p_2 =$ mass, speed, momentum of remaining large portion of bolide (at $\infty$)

$p_1 = m_1 v_1 = p_2 = m_2 v_2$

$R =$ radius of full bolide, $\rho(kg/m^3) =$ density of bolide

$$E_{esc} = gravity\ escape\ energy\ of\ ejected\ mass\ m_1 = \frac{Gm_1m_2}{R} = m_1 v_{esc}{}^2 \quad (v_{esc} = \sqrt{\frac{Gm_2}{R}} v_{esc} \simeq 2.36\times10^{-5}\sqrt{\rho}R)$$

$Ex: \rho = 2600\,(kg/m^3)(stony\ asteroid) \rightarrow v_{esc} = 0.12m/s\,(R = 100m), = 0.6m/s\,(R = 500m = 1km\ diam)$

$v_1 > v_{esc}$ to have gravitational escape of fragment $m_1$

$m_p =$ mass of penetrator/impactor

$E =$ energy deposited in bolide (kinetic, chemical or NED)

$\varepsilon =$ efficiency of energy deposition conversion to fragment KE (0-1) (some energy goes into heat)

$\alpha =$ momentum enhancement factor of impactor mass (1=completely inelastic, 2=completely elastic)

$\mu = (1/m_1 + 1/m_2)^{-1} \equiv reduced\ \ mass\ \ \mu \sim m_1\ \ when\ m_2 >> m_1$

$Momentum\ conservation\ \ m_1v_1 - m_2v_2 = 0 = p_1 - p_2 \rightarrow p_1 = p_2$

$Energy\ conservation\ \ (m_1v_1{}^2 + m_2v_2{}^2)/2 = p_1{}^2/2m_1 + p_2{}^2/2m_2 = p_2{}^2(1/m_1 + 1/m_2)/2 = p_2{}^2/2\mu = \varepsilon E - E_{esc}$

$\rightarrow p_2 = \sqrt{2\mu(\varepsilon E - E_{esc})}$

1) If all energy E comes from penetrator KE only $\rightarrow$ $E = m_p(v_0 + v_{sc})^2/2$ $(assumes\ direct\ head\ on\ impact)$

$\rightarrow p_{2-KE} = \sqrt{2\mu(\varepsilon E - E_{esc})} = \sqrt{2\mu(\varepsilon m_p(v_0 + v_{sc})^2/2 - E_{esc})}$

$p_{2-KE} = \sqrt{\mu\varepsilon m_p}(v_0 + v_{sc})$ $(grav\ small\ IF\ E_{esc} << \varepsilon E,\ v_{esc} << v_1)$

2) Compare to direct momentum transfer p' from impactor hit

$p' = \alpha m_p(v_0 + v_{sc})$ $(assumes\ direct\ head\ on\ impact)\ or\ \ p' = \alpha m_p v_{sc} (assumes\ side\ hit)$

$p' = (m_1 + m_2)v' \sim m_2 v'\ \ (m_1 << m_2)\ \ where\ v' = speed\ change\ of\ bolide\ from\ classical\ impact\ deflection$

3) $Ratio\ of\ enhanced\ deflection\ over\ normal\ deflection\ for\ general\ energy\ input\ E$

$p_2/p' = \sqrt{2\mu(\varepsilon E - E_{esc})}/[\alpha m_p(v_0 + v_{sc})] \sim v_2/v'$ $(assumes\ head\ on\ impact)$

4) $Ratio\ of\ enhanced\ deflection\ over\ normal\ deflection\ for\ E\ from\ penetrator\ KE\ from$ 1)

$p_{2-KE}/p' = \sqrt{2\mu(\varepsilon m_p(v_0 + v_{sc})^2/2 - E_{esc})}/[\alpha m_p(v_0 + v_{sc})] \sim v_2/v'$

$$p_{2-KE}/p' = \sqrt{\mu\varepsilon m_p}(v_0 + v_{sc})/[\alpha m_p(v_0 + v_{sc})] = \sqrt{\mu\varepsilon m_p}/(\alpha m_p) = \sqrt{\frac{\mu}{m_p}}\frac{\sqrt{\varepsilon}}{\alpha} \sim \sqrt{\frac{m_1}{m_p}}\frac{\sqrt{\varepsilon}}{\alpha}\ (m_1 << m_2)(grav\ small)$$

Note that for $m_1(ejected\ mass) >> m_p \rightarrow p_{2-KE}/p' \sim v_2/v' >> 1$

$\rightarrow$ *Bolide speed change much greater with energy injection vs classical deflection*

**Relationship between energy injection and mass blow off vs ejection speed** – For a given energy input E we model the energy fraction going into the blow off fragment KE by a factor of ε. The detailed calculations for ε depend on many factors that are not generally analytically modeled for complex geometries. For a well tamped explosive a typical value is ε~1/2 similar to the fraction of explosive energy going into a blast wave as discussed. The placement of the explosive (whether purely kinetic or from chemical explosives) and the energy injection timing is also critical. In general, the most momentum transfer will happen for the slowest speed mass ejection that exceeds the gravitational escape speed. If we assume an ejection of speed $v_1$ as modeled above then the amount of mass ejected $m_1$ for a given energy and ejection speed $v_1$ is given by:

$$\varepsilon E \sim m_1 v_1^2 / 2 \ \rightarrow m_1 = 2\varepsilon E / v_1^2$$

$$p_2 = \sqrt{2\mu(\varepsilon E - E_{esc})} \sim \sqrt{2m_1(\varepsilon E - E_{esc})} \ (m_1 << m_2)$$

$$\rightarrow v_2 = p_2 / m_2 = \sqrt{2m_1(\varepsilon E - E_{esc})} / m_2 \ \sim 2\varepsilon E \sqrt{(1 - \frac{E_{esc}}{\varepsilon E})} / (v_1 m_2) \ \ (m_1 << m_2)$$

$$v_2 \sim \frac{2\varepsilon E}{v_1 m_2} \ \ (m_1 << m_2, E_{esc} << \varepsilon E)$$

The mass $m_1$ blown off is:

$$m_1 = p_1 / v_1 = p_2 / v_1 = m_2 (v_2 / v_1) = \sqrt{2m_1(\varepsilon E - E_{esc})} / v_1 \ \sim 2\varepsilon E \sqrt{(1 - \frac{E_{esc}}{\varepsilon E})} / v_1^2 \ \ (m_1 << m_2)$$

$$m_1 \sim \frac{2\varepsilon E}{v_1^2} \ \ (m_1 << m_2, E_{esc} << \varepsilon E) \ (also\ from\ above\ \varepsilon E \sim m_1 v_1^2 / 2 \ \rightarrow m_1 = 2\varepsilon E / v_1^2)$$

As discussed above the momentum transfer from a conventional deflection technique is:

$$p' = \alpha m_p (v_0 + v_{sc}) \ (assumes\ direct\ head\ on\ impact) \ or \ \ p' = \alpha m_p v_{sc} (assumes\ side\ hit)$$

$$p_2 (= m_2 v_2) / p' = \frac{2\varepsilon E}{v_1 \alpha m_p (v_0 + v_{sc})} = v_2 / v', \quad m_1 \sim \frac{2\varepsilon E}{v_1^2} \ \ (m_1 << m_2, E_{esc} << \varepsilon E)$$

Ex: $v_1 = 1 m/s, v_0 = 20 km/s, v_{sc} = 3 km/s, E = 4.2x10^9 J (1 ton (1000 kg) TNT), m_p = 1000 kg$

$$\rightarrow p_2 / p' = v_2 / v' = 365 \frac{\varepsilon}{\alpha}, \quad m_1 = 8.4x10^9 kg >> m_p \quad (m_1 << m_2, E_{esc} << \varepsilon E)$$

*Note that*: $p_2 / p' = v_2 / v'$ *and* $m_1$ *both scale as E. Use smaller E for smaller bolide* $(Ex: 100m \sim 10^9\ kg)$

There is a tradeoff between deflection and complete fracturing.

If all energy E comes from penetrator KE only $\rightarrow E = m_p (v_0 + v_{sc})^2 / 2$ (*assumes direct head on impact*)

$$m_1 \sim \frac{2\varepsilon E}{v_1^2} = \varepsilon m_p \frac{(v_0 + v_{sc})^2}{v_1^2} = \varepsilon m_p \left[ \frac{v_0 + v_{sc}}{v_1} \right]^2 \ \ (m_1 << m_2)$$

$$p_2 / p' = \frac{2\varepsilon E}{v_1 \alpha m_p (v_0 + v_{sc})} = \frac{2\varepsilon m_p (v_0 + v_{sc})^2 / 2}{v_1 \alpha m_p (v_0 + v_{sc})} = \frac{\varepsilon}{\alpha} \frac{(v_0 + v_{sc})}{v_1} \sim v_2 / v'$$

This enhanced deflection factor can be extremely large depending on the energy coupling $\varepsilon$.

Note that this enhancement factor is largely independent of the impactor mass $m_p$.

Ex: $v_1 = 1 m/s, v_0 = 20 km/s, v_{sc} = 3 km/s \rightarrow p_2 / p' = 2.3x10^4 \frac{\varepsilon}{\alpha} \sim v_2 / v' \rightarrow$ *much greater speed change*

$$m_1 / m_p \sim \varepsilon \left[ \frac{v_0 + v_{sc}}{v_1} \right]^2 \sim 5.3x10^8 \varepsilon \ \ (m_1 << m_2) (\text{this is an extremely large mass enhancement ejection ratio})$$

Potentially an extremely large enhancement factor is possible IF there is good energy coupling ($\alpha$~unity)

**Launch Mass Comparison between Normal Deflection and PI** – A useful comparison is to compare the launch mass requirements for deflection vs PI. We plot intercept deflection mass required in metric tons (mt) vs intercept time prior to impact from 1 day to 300 years for 20 km/s bolides from 10m to 10km diameters for a 4x Earth radius deflection. Comparing the deflection mass needed to the mass required for the fragmentation (PI) method we propose, shows the extreme mass required for deflection, particularly for large objects compared to the much smaller launch mass required for PI. In addition to the vastly reduced mass needed for PI mitigation there is also the significant advantage that PI can operate in both a short term (terminal defense) mode as well as in a long term (long time scale) intercept mode. The deflection shown assumes a completely inelastic collision and no detailed orbital mechanics specific to each case. For existential threats (d~10km) using PI, as discussed previously in this paper, we cannot accept the full fragment cloud hit due to the resultant atmospheric temperature rise and thus would need to intercept early enough to fragment the threat such that the fragment cloud spread out to largely miss the Earth. Since PI depends on energy transfer and not momentum transfer for mitigation the two techniques are quite different in their mitigation time scale and mass needs. PI requires vastly less launch mass and can accommodate vastly shorter intercept time scales than deflection in general.

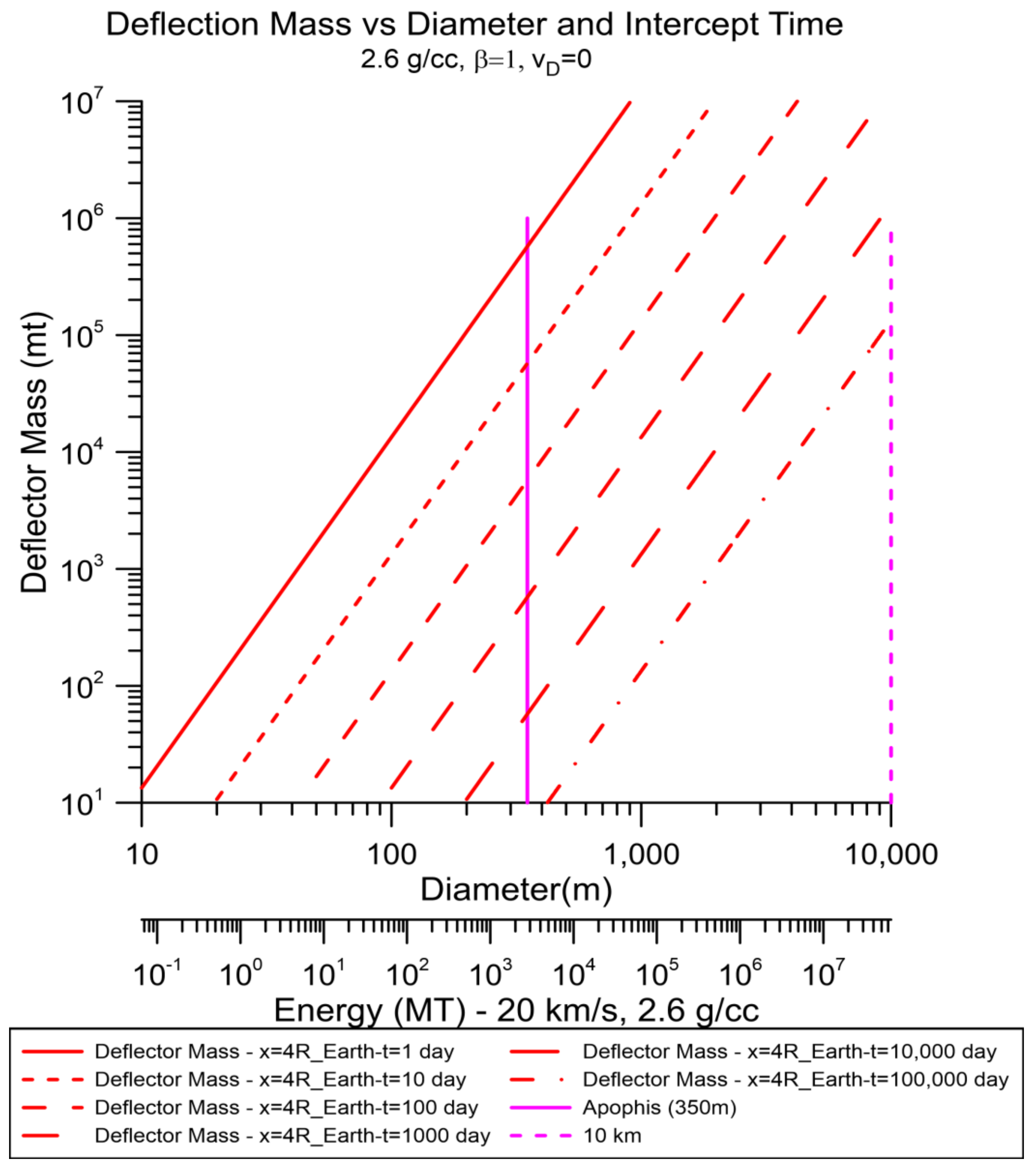

**Figure 118** – Deflection mass required for a completely inelastic momentum transfer for threats from 10m to 10km diameter assuming a 20 km/s and density 2.6 g/cc threat with a 4 $R_E$ deflection for deflection times from 1 day to 300 years prior to impact.

**Deflection**

$m_D$ = mass deflector, $m_B$ = mass bolide
$v_D$ = speed deflector relative to Earth just before impact
$v_B$ = speed bolide relative to Earth just before impact
$v_f$ = speed bolide relative to Earth after impact in head on collision
$v_E$ = orbital speed Earth - assume perpendicular to path of bolide (best *outcome* for deflection strategy)
$L$ = distance to bolide from Earth at impact
$\Delta t$ = time difference of bolide arrive at/ near Earth between no deflection and deflection
$\Delta v$ = speed difference of bolide arrive at/ near Earth between no deflection and deflection for head on collision
$\Delta v = v_B - v_f$
$t_B$ = time from deflection to impact (if no mitgation) =L/$v_B$
$v_E$ = speed of Earth in orbit (~30 km/s) - assume best case (Earth vel vector - $\perp$ threat vel vector)
$\beta$ = deflection enhancement factor (1=inelastic, 2=elastic, can be >2 if significant mass ejection)
$\rho$ = density bolide, $d$ = diameter bolide
$x$= "miss distance"
$m_B = \pi \rho d^3 / 6$
$P_i$ = initial momentum before deflection, $P_f$ = final momentum after deflection

## 1) Head on Impact

$$Complete\ inelastic\ collision\ \beta = 1$$

$$P_i = m_B v_B - m_D v_D = (m_D + m_B) v_f = P_f = P \rightarrow v_f = \frac{m_B v_B - m_D v_D}{m_D + m_B}$$

$$\rightarrow \Delta v = v_B - v_f = v_B - \frac{m_B v_B - m_D v_D}{m_D + m_B} = \frac{m_D (v_B + v_D)}{m_D + m_B} \sim (v_B + v_D) \frac{m_D}{m_B}$$

$$Not\ Complete\ inelastic\ collision\ \beta \neq 1$$

$$\Delta v = v_B - v_f = \beta (v_B + v_D) \frac{m_D}{m_B}$$

$$Both\ cases\ (arbitrary\ \beta)$$

$$\Delta t = \frac{L}{v_f} - \frac{L}{v_B} = L \frac{v_B - v_f}{v_B v_f} = L \frac{\Delta v}{v_B v_f} = \frac{L}{v_B} \frac{\Delta v}{v_f} = t_B \frac{\Delta v}{v_f} \sim t_B \frac{\Delta v}{v_B} = \beta t_B \frac{(v_B + v_D)}{v_B} \frac{m_D}{m_B}$$

$$x = v_E \Delta t = \beta v_E t_B \frac{(v_B + v_D)}{v_B} \frac{m_D}{m_B} = \beta v_E t_B (1 + v_D / v_B) \frac{m_D}{m_B}$$

$$m_D = \frac{x m_B}{\beta v_E t_B} \frac{v_B}{v_B + v_D} = \frac{x m_B}{\beta v_E t_B} \frac{1}{1 + v_D / v_B} = \frac{x \pi \rho d^3}{6 \beta v_E t_B} \frac{1}{1 + v_D / v_B}$$

$$m_D \rightarrow \frac{x m_B}{\beta v_E t_B} = \frac{x \pi \rho d^3}{6 \beta v_E t_B}\ for\ v_D << v_B$$

## 2) Perpendicular Side Hit of Bolide $(bolide\ acquires \perp speed)$

$$m_B v_\perp = \beta m_D v_D\ (momentum\ transfer)$$

$$v_\perp = \frac{\beta m_D v_D}{m_B}$$

$$x = v_\perp t_B = \beta v_D t_B \frac{m_D}{m_B}\ (Bolide\ moves \perp)$$

$$m_D = \frac{x m_B}{\beta v_D t_B} = \frac{x \pi \rho d^3}{6 \beta v_D t_B}$$

In general $v_E >> v_D$ and hence head on deflection is preferred as required deflector mass is less for head on.

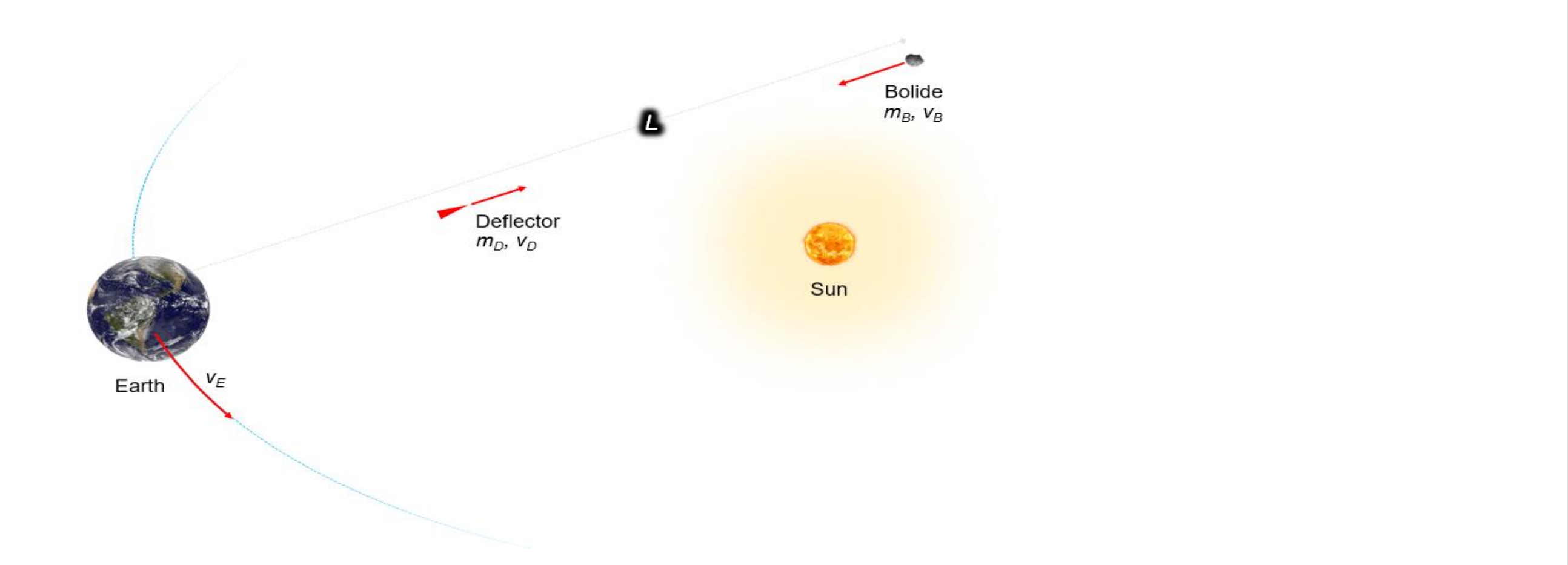

**Figure 119** – Setup for calculating bolide deflection intercept time and deflection distances.

**Enhanced Momentum Transfer from Ejecta – Induced Rocket Mode** – There is another mode of momentum transfer that should be noted. This is the momentum imparted to the bolide during the conversion of the KE of the penetrator into hot vaporized bolide material that is rapidly ejected through the entrance cavity caused by the penetrator impact. This will become clearer in our upcoming papers that show the details of the penetrator physics. During the penetration process the penetrator KE is rapidly transferred into both shock waves in the bolide causing rapid disintegration of the bolide and into rapid conversion of the penetrator KE into vaporization and plasma formation inside the bolide. This then leads to a high pressure largely adiabatic expanding gas/ plasma inside the bolide causing additional material failure, expansion of and eventual dispersal of the bolide fragments generated but there is a time period prior to complete disintegration where the small entrance hole from the penetrator widens into what is essentially a "rocket nozzle" out of which hot bolide gas and bolide fragments are ejected. The bolide mass itself is the "rocket fuel" that both destroys and propels itself. This is similar in some respects to our prior work on laser ablation planetary defense but on a vastly shorter time scale. The typical time scale over which this process happens is seconds. As an example, the penetrator KE per unit mass is E=5x10$^5$ J/(km/s)$^2$/kg. For mass ejection time scales of one second this leads to power levels (per penetrator mass) of P~0.5 MW/(km/s)$^2$/kg**.** For a 20 km/s impact this is P~ 200MW/kg. In a number of our penetrator simulations we use a 100 kg penetrator which leads to instantaneous power levels of roughly 20 GW assuming a 20km/s impact. **This is comparable to the chemical propulsion power of a heavy lift launch vehicle at liftoff.** This can lead to large momentum transfers in addition to the catastrophic destruction of the bolide. The full process is extremely complex as will be seen in upcoming papers that show full 4D simulations of the penetrator impact. A simple order of magnitude calculation assuming constant ejecta speed ($v_{rel}$) is as follows.

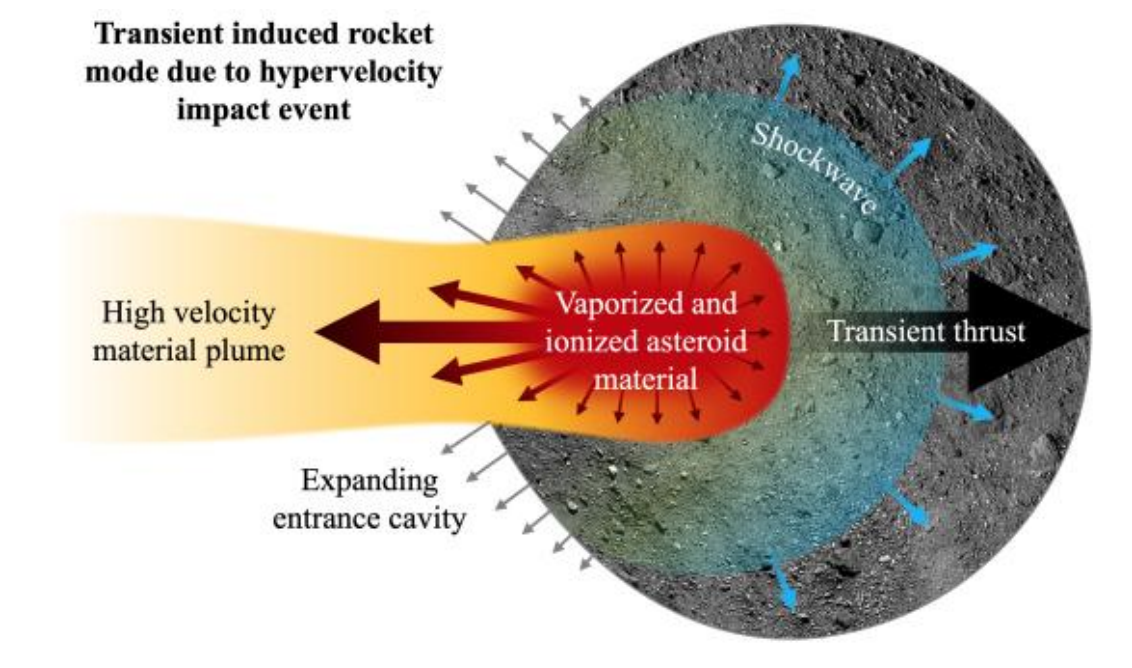

**Figure 120** – Induced rocket mode model.

$$v_{rel} = exhaust\ speed\ ,\ T(thrust) = \dot{m} v_{rel}\ ,\ p_{ejecta}(thrust\ power) = 0.5 \dot{m} v_{rel}{}^2\ ,\ T / p_{ejecta} = 2 / v_{rel}$$

$$P_{ejecta}(bolide\ momentum\ change\ due\ to\ ejecta) = 2E_{ejecta} / v_{rel}\ ,\ E_{ejecta} = total\ energy\ in\ ejecta$$

$$P_T = P_{penetrator} + P_{ejecta} = P_{penetrator} + 2E_{ejecta} / v_{rel}\ ,\ P_{penetrator} = m_{penetrator} v_{impact}\ ,\ E_{ejecta} = \varepsilon E_{penetrator} = \varepsilon P_{penetrator}{}^2 / 2m_{penetrator}$$

$$\varepsilon = eff\ of\ conversion\ of\ impactor\ KE\ to\ ejecta\ energy,\ \ v_{impact} >> v_{rel}\ ,\ \beta_{eff} = momentum\ enhancement\ factor$$

$$\beta_{eff} \equiv P_T / P_{penetrator} = \left[1 + 2E_{ejecta} / v_{rel} / P_{penetrator}\right] = \left[1 + 2\varepsilon P_{penetrator}{}^2 / (2m_{penetrator} v_{rel}) / P_{penetrator}\right] = \left[1 + \varepsilon v_{impact} / v_{rel}\right]$$

Depending on the efficiency of the conversion of penetrator KE to ejecta KE as the speed of the ejecta, this can potentially lead to very large momentum transfer enhancement. Yet another example of the system use. There is similarly here to the enhanced deflection mode we discussed above.

## 13. Lunar or orbital LIDAR for early detection and target designation

One of the primary problems facing planetary defense is the detection of and the orbital parameter determination of threatening objects. Detection to date has been done with visible wide field astronomical surveys via sunlight illumination of and subsequent detection of the object. The orbital determination is needed to determine if there is any significant chance of impact. While large objects (> km) are relatively well characterized, the situation for smaller objects (<100m) is poor. While we have addressed the mitigation portion in this paper, you cannot mitigate that which you cannot see. Long range radar is helpful, but the detection range is very poor for smaller objects due to the small radar cross section and the classic $1/r^4$ problem with active radar. Other methods of detection include upcoming wide field thermal IR imaging space missions. We proposed another option which may be better suited for small object detection, and that is long range LIDAR with an orbital or lunar deployment. The advantage of LIDAR is the wavelength is vastly shorter than RADAR and thus the flux on target ($\sim P/\lambda^2$) can be vastly higher with LIDAR than with RADAR, allowing for much smaller body detection and much larger range. Large aperture or multi-aperture with large baseline phased array laser systems operating at 1 micron combined with large aperture wide field photon counting narrow band IR imaging are all options for better detection of small solar system bodies. We discussed this in a previous paper and will address a more detailed design in a future paper. The same LIDAR system used for detection can also be subsequently used for target illumination during intercept. While detection and mitigation can be completely separate areas, we propose that a serious effort to take control over our vulnerability to extraterrestrial threats would combine the two to form an effective operational detection and mitigation system.

## 14. Earth and Lunar Launch Interceptor Options

Launch vehicles launched from the surface of the Earth and designed to take a payload and escape the Earth's gravity are often parameterized by their characteristic energy, or the parameter known as $C_3$. This is ultimately related to detailed designs of the propulsion system, the most important being the propulsion specific impulse, or $I_{sp}$, and the details of the staging, atmospheric drag, etc. This way of characterizing launch vehicles is specific to ground-based launches from Earth and thus if we desire to analyze the same or similar vehicles that "start" from orbit, such as a refueled Space X Starship, or launched from the lunar surface, the final speed obtained for a given payload mass can be much larger. In general, there are few strategies that would use the same launcher in orbit or on the Moon as is used for an Earth surface launch, but it is useful to understand the general relationships and possibilities this would enable. For planetary defense applications, this can be an important consideration. For example, solid fueled boosters often used for ICBM's are not suitable for Earth surface-to-LEO due to their lower $I_{sp}$ compared to liquid fueled boosters. However, this situation changes dramatically for a lunar or orbital launcher where solid fuel boosters become viable. This is extremely important for a long-term vision of planetary defense where basing on other than the Earth's surface may be considered. Having a solid fueled booster, similar to a silo-based ICBM, is vastly easier to envision than a LOX - LH, $LCH_4$, L-hydrocarbon booster. While these are "practical" considerations, they are important ones to understand. We outline some of the critical parameters for interceptors of a variety of types below.

***Earth launch payload mass capability*** – Due to the Earth's significant gravitational well, the launch of payloads to targets at lunar and greater distances is vastly more difficult than launching to LEO. Launchers suitable for carrying payloads to the moon and beyond are often characterized by their "characteristic energy" which is designated as $C_3$, given in units of $(km/s)^2$ and is a function of the payload mass being lifted. The square root of $C_3$ is the residual speed at infinite distance from the Earth and hence $C_3=0$ is the minimum requirement for escaping the Earth, which is 11.2km/s immediately above the atmosphere and decreases to zero at infinite distance as the payload converts its kinetic energy into potential energy from the Earth's gravity. $C_3 > 0$ means the booster can escape he Earth, while $C_3 < 0$ means it cannot escape. Interestingly, the Saturn V with its payload for the Apollo program had a $C_3 \sim -1$ meaning it could not escape the Earth, though it could be captured by the Moon's gravity. Below we show some of the current and near future (SLS) boosters and their $C_3$. Note that while the Space X Falcon 9 is outstanding at lifting payloads into LEO, it has a modest (but positive) $C_3$ as it was not designed for deep space missions. The Falcon 9 Heavy has significantly higher lift and larger $C_3$ capability than the normal Falcon 9 and both are candidates for interceptors. The Space X Starship with refueling capability (not shown) should have excellent deep space high load capability as it is being optimized for heavy lift missions to Mars. The NASA SLS program could be an outstanding choice for a heavy lift Earth launcher depending on its future development program. The SLS use of $LH_2$/LOX and the resulting high $I_{sp}$ along with the solid rocket boosters allows it to have a high $C_3$ potentially. Both the SLS and Space X Starship are candidates for larger threats. Use of NEDs equipped penetrators would greatly enhance the capability of any of these launchers for very large threats if required. Even a Falcon 9 with NED's penetrators would be a potent mitigation launch system.

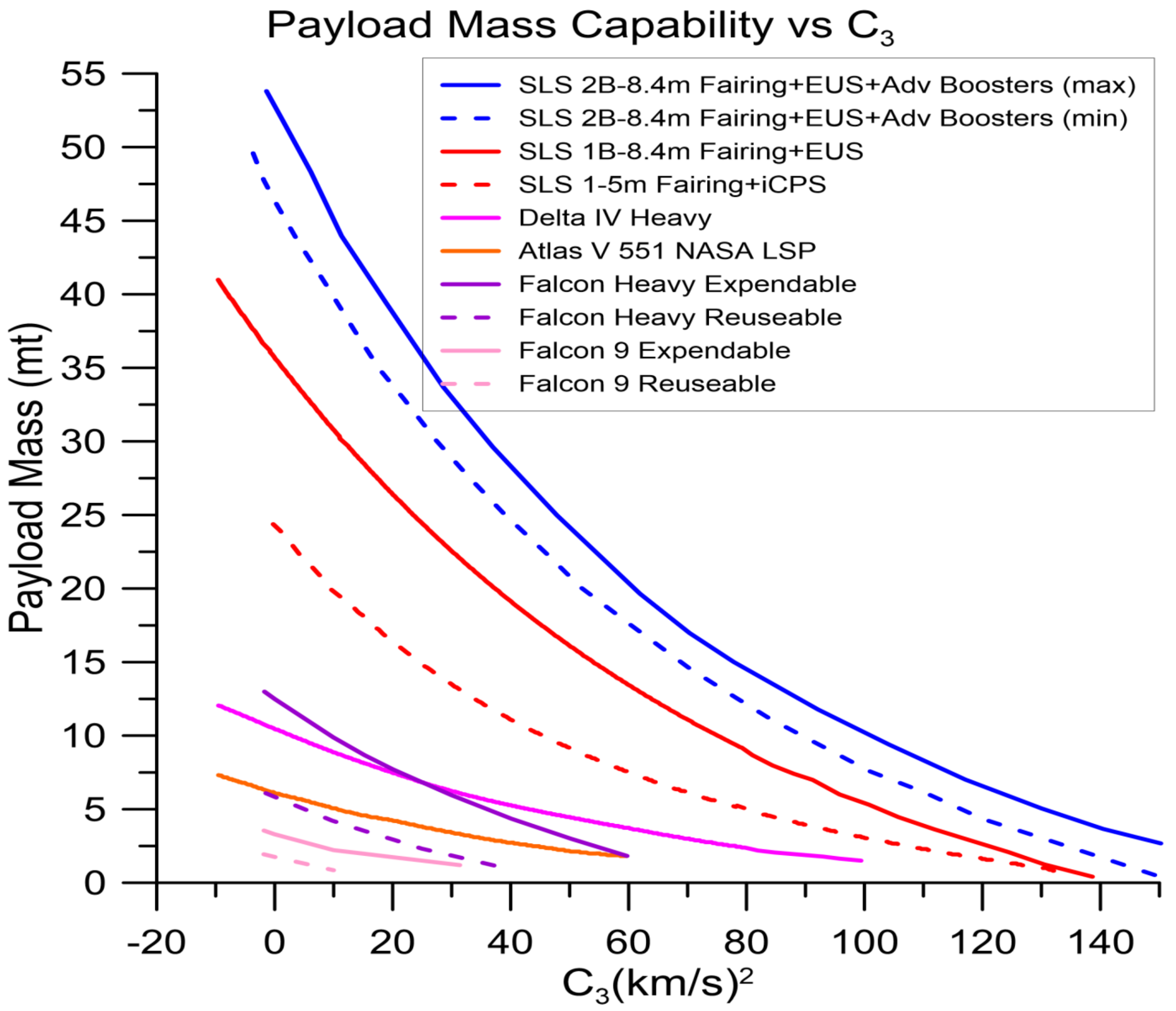

**Figure 121** – Payload mass capability for various positive $C_3$ boosters. The NASA SLS is still in development and the upcoming Space X Starship $C_3$ specifications depend on the use of in orbit refueling capability which is not currently published.

***Comparing Earth and Lunar Launch capabilities*** – The much smaller gravitational well and subsequent lower escape speed from the Moon (2.3km/s vs 11.2km/s for the Earth), makes the moon a potentially attractive place as an option for a long term layered planetary defense capability. As we will see, this is particularly true in that solid boosters become a viable option for lunar based interceptors while they are not suitable for Earth launch interceptors. A full system involving both Earth and lunar/orbital interceptors would be optimal depending on the threat. Lunar or orbital based solid fueled interceptors would be analogous to current solid fueled ICBM capabilities. We show the general relationship between Earth and lunar based capability below. Note that this is only notional as no reasonable operational capability exists for lunar based equivalents of high $C_3$ Earth-based launchers. The main takeaway will be that lunar-based interceptors using modest solid fueled $I_{sp}$ become attractive future options. This become clear when comparing the performance of the specific example of the solid rocket booster (SRB) with an $I_{sp}$ =242 sec used in the previous Shuttle program which is not suitable by itself for a high $C_3$ mission when launch from the Earth, but would be suitable for a high $C_3$ mission if an equivalent type of SRB were lunar-based.

***High Performance Solid Fuel Rockets – Feasibility for Lunar or Orbital Launch Option*** – While all boosters that are launched from the surface of the Earth that can achieve LEO, let alone high $C_3$, use one or more stages of liquid fuel/oxidizer, there are a number of very high-performance solid fuel options that have been developed for ICBM use that may be feasible options for lunar or orbital based interceptors. These include HMX, $C_4H_8N_4(NO_2)_4$, used in the propellant of the Peacekeeper ICBM and is the main ingredient in NEPE-75 propellant used in the Trident II D-5 Fleet Ballistic Missile [52]. Because of the explosive hazard that solid propellants containing HMX possess, these are generally not used for commercial launch vehicles except when it is an ICBM adapted for launching. Examples of this include the Minotaur IV and V. A more recently developed solid propellant with promise is $C_6H_6N_6(NO_2)_6$, commonly called CL-20, is less sensitive to detonation and safer to store. Compared to HMX, CL-20 has 14% more energy per mass, 20% more energy per volume, and a higher oxygen-to-fuel ratio [53]. Examples of specific impulse of high-performance solid fuels using HMX have achieved 309s demonstrated in the Peacekeeper's second stage. CL-20 propellant is expected to increase specific impulse to about 320s in ICBM upper stage applications [54]. As we will see below, though not suitable or Earth based interceptors for our applications, these solid fuels are sufficient to allow high speed lunar or orbital based interceptors due to the greatly reduced escape speed needed for launch outside the deep gravitational well of the surface of the Earth.

Assume rapid burn of propulsion - impulsive burn

$E_i = initial\ state\ energy$

$E_f = final\ state\ energy$

$v_i$ = initial speed after impulsive burn at distance $r_i$ from center of gravity well (planet, moon etc)

$v_f$ = final speed at distance $r_f$ from center of gravity well (planet, moon etc)

The calculations allow for speed $v_f$ vs distance ($r_f$).

$m = payload\ mass$

$$v_{esc}(M,R) = escape\ speed\ from\ mass\ M\ at\ R = \sqrt{\frac{2GM}{R}}$$

$$v_{orb}(M,R) = \sqrt{\frac{GM}{R}} = v_{esc}(M.R)/\sqrt{2} \rightarrow_{esc} (M.R) = \sqrt{2}v_{orb}(M,R)$$

$\rightarrow$ Escape speed from distance R is $\sqrt{2}$ higher than orbital speed at distance R

$$v_{esc} \equiv escape\ speed\ from\ Earth\ surface = v_{esc}(M_E, R_E Earth\ surface) = \sqrt{\frac{2GM_E}{R_E}}$$

$M_E = 5.97x10^{24} kg,\ R_E = 6378 km$

$$v_{esc} = \sqrt{\frac{2GM_E}{R_E}} = 11.2\, km/s\ ,\ v_{esc} = 125.4\, (km/s)^2$$

$$v_{orb}(LEO) = \sqrt{\frac{GM_E}{R_E}} = v_{esc}/\sqrt{2} = 7.92\, km/s$$

$$C_3(m) \equiv (v_i(m)^2 - v_{esc}{}^2(Earth\ surface)) \rightarrow v_i(m)^2 = C_3(m) + v_{esc}{}^2 = C_3(m) + \frac{2GM_E}{R_E}$$

$C_3 = positive = escapes\ Earth\ gravity$

$C_3 = negative = does\ not\ escapes\ Earth\ gravity$

For a given rocket $C_3$ depends on the mass ratio to obtain speed $v_i(m)$

General mass M for planet, moon, asteroid etc:

$$\frac{E_i}{m} = \frac{T_i + U_{grav-i}}{m} = \frac{v_i^2}{2} - \frac{GM}{r_i}$$

$$\frac{E_f}{m} = \frac{T_f + U_{grav-f}}{m} = \frac{v_f^2}{2} - \frac{GM}{r_f}$$

$$E_i = E_f$$

$$v_f^2 = v_i^2 + \frac{2GM}{r_f} - \frac{2GM}{r_i} = C_3(m) + v_{esc}^2 + \frac{2GM}{r_f} - \frac{2GM}{r_i}$$

$$v_f^2 = v_i^2 + v_{esc}^2(from\, r_f) - v_{esc}^2(from\, r_i)$$

Recall $v_i^2 = C_3(m) + v_{esc}^2 \quad where \;\; v_{esc} \equiv v_{esc}(Earth\, surface)$

1) *Earth surface launch*

$$M = M_E, r_i = R_E$$

$$v_f^2 = v_i^2 + \frac{2GM}{r_f} - \frac{2GM}{r_i} = v_i^2 + \frac{2GM_E}{r_f} - \frac{2GM_E}{R_E} = C_3(m) + v_{esc}^2 + \frac{2GM_E}{r_f} - \frac{2GM_E}{R_E}$$

$$= C_3(m) + \frac{2GM}{r_f} = C_3(m) + v_{esc}^2(from\, r_f)$$

$$\rightarrow v_f^2 = C_3(m) + \frac{2GM_E}{R_E} = C_3(m) + v_{esc}^2 = v_i^2 \; for\; r_f = R_E$$

$$\rightarrow v_f^2 = C_3(m) \; for\; r_f = \infty$$

Recall $v_i(m)^2 = C_3(m) + v_{esc}^2 = C_3(m) + \frac{2GM_E}{R_E}$

$$\rightarrow v_f^2 = C_3(m) + v_{esc}^2(from\, r_f) = C_3(m) + v_{esc}^2(Earth\;escape\,speed)\;(r_f = R_E)\;(speed\;at\;Earth\,surface)$$

$$\rightarrow v_f^2 = C_3(m) + v_{esc}^2(from\, r_f) = C_3(m) + 125.4(km/s)^2 \;(r_f = R_E)\;(speed\;at\;Earth\,surface)$$

$$\rightarrow v_f^2 = C_3(m)\; at\; r_f = \infty$$

2) *Lunar surface launch*

$$Assume\ M = M_{moon}, r_i = R_{moon}$$

$$v_f^{\,2} = v_i^{\,2} + \frac{2GM}{r_f} - \frac{2GM}{r_i} = C_3(m) + v_{esc}^{\,2} + \frac{2GM_{moon}}{r_f} - \frac{2GM_{moon}}{R_{mooni}} = v_i^{\,2} + v_{esc}^{\,2}(from\, r_f) - v_{esc}^{\,2}(from\, r_i)$$

$$v_f^{\,2} = v_i^{\,2} + \frac{2GM}{r_f} - \frac{2GM}{r_i} = v_i^{\,2} - \frac{2GM}{r_i} = v_i^{\,2} - v_{esc}^{\,2}(moon)\ \ for\ \ r_f = \infty\ \ (ignores\ Earth\ grav)$$

Recall $v_i^{\,2} = C_3(m) + v_{esc}^{\,2}$

$$v_{esc}^{\,2} \equiv v_{esc}^{\,2}(Earth\ surface)$$

$$v_{esc}^{\,2}(lunar\ surface) = \frac{2GM_{moon}}{R_{mooni}}$$

$$M_{moon} = 7.35x10^{22}\,kg,\ R_{moon} = 1737\,km$$

$$v_{esc}(moon) = \sqrt{\frac{2GM_{moon}}{R_{moon}}} = 2.38\,km/s\ ,\ v_{esc}^{\,2} = 5.66(km/s)^2$$

$$\rightarrow v_f^{\,2} = C_3(m) + v_{esc}^{\,2}(Earth\ surface) - v_{esc}^{\,2}(lunar\ surface) + \frac{2GM_{moon}}{r_f}$$

$$= C_3(m) + 119.7\,(km/s)^2 + \frac{2GM_{moon}}{r_f}$$

$$\rightarrow v_f^{\,2} = C_3(m) + v_{esc}^{\,2}(Earth\ surface) = C_3(m) + 125.4\,(km/s)^2\ (for\ r_f = R_{moon})\ (speed\ at\ Moon\ surface)$$

Initial Speed at Moon surface = Initial Speed at Earth surface (after impulsive burn)

$$\rightarrow v_f^{\,2} = C_3(m) + v_{esc}^{\,2}(Earth\ surface) - v_{esc}^{\,2}(lunar\ surface) = C_3(m) + 119.7\,(km/s)^2\ \ (r_f = \infty)$$

3) *Lunar surface launch with addition of Earth's gravity potential in worst case of going away from Earth and moon. At $\infty$ direction does not matter.*

$$Assume\ M = M_{moon}, M_E = 81.2M_{moon}, r_i = R_{moon}, r_f = \infty, L_D = 384Mm = 60.3R_E = 221R_{moon}$$

$$v_i(surface\ moon) = v_i(surface\ Earth) = \left[C_3(m) + v_{esc}^{\,2}\right]^{1/2}\ for\ same\ propulsion\ system$$

$$\frac{E_i}{m} = \frac{T_i + U_{grav-i}}{m} = \frac{v_i^{\,2}}{2} - \frac{GM_{moon}}{R_{moon}} - \frac{GM_{Earth}}{L_D}$$

$$= (C_3(m) + v_{esc}^{\,2})/2 - \frac{GM_{moon}}{R_{moon}} - \frac{81.2GM_{moon}}{221R_{moon}} = \left[C_3(m) + \frac{2GM_E}{R_E} - 1.37\frac{2GM_{moon}}{R_{moon}}\right]/2$$

$$= \left[C_3(m) + v_{esc}^{\,2}(Earth\ surface) - 1.37v_{esc}^{\,2}(lunar\ surface)\right]/2$$

$$\frac{E_f}{m} = \frac{v_f^{\,2}}{2} - \frac{GM}{r_f} = \frac{v_f^{\,2}}{2}\ \ (r_f = \infty)$$

$$v_f^{\,2} = v_i^{\,2} - \frac{2GM_{moon}}{R_{moon}} - \frac{2GM_{Earth}}{L_D} = v_i^{\,2} - v_{esc}^{\,2}(lunar\ surface) - v_{esc}^{\,2}(Earth\ at\ Moon)\ (r_f = \infty)$$

$$\rightarrow v_f^{\,2} = C_3(m) + v_{esc}^{\,2}(Earth\ surface) - 1.37v_{esc}^{\,2}(lunar\ surface) = C_3(m) + 117.6\,(km/s)^2\ \ (r_f = \infty)$$

$\rightarrow$ Correction due to earth's grav field at moon is a small corretion to $v_f^{\,2}$ at $\infty$

$\rightarrow$ Correction due to earth's grav field is direction dependent at finite distances but not at $\infty$

$$v_{esc}(lunar\ surface) = \sqrt{\frac{2GM_{moon}}{R_{moon}}} = 2.38\,km/s\ ,\ v_{esc}^{\,2} = 5.66(km/s)^2$$

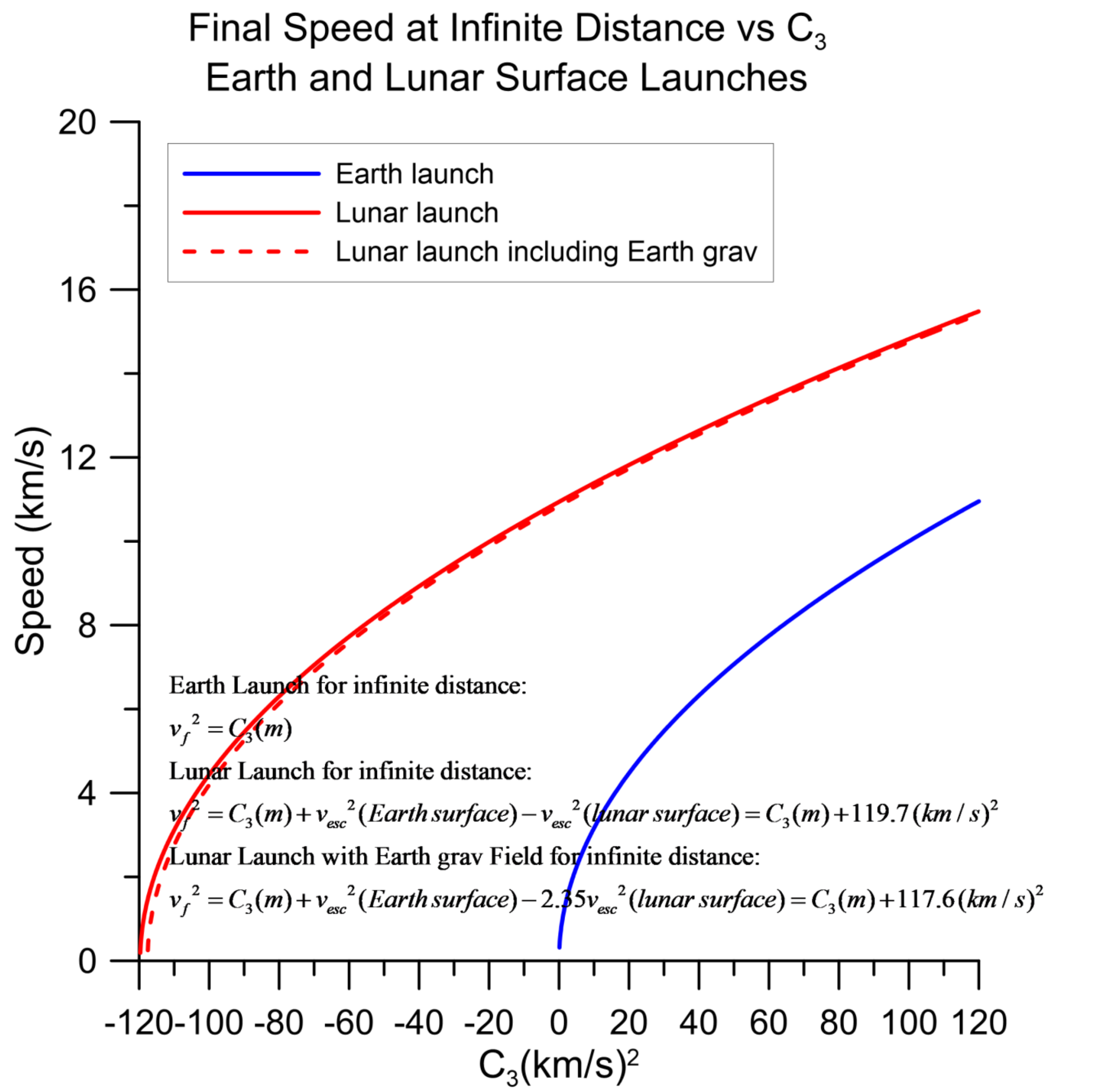

**Figure 122** – Payload speed for Earth and lunar launch boosters vs $C_3$.

Mass ratio and Specific Impulse:

$$v_f - v_i = \Delta v = -v_{exh}\ln(m_f / m_i)$$

$$m_f = m_i e^{-\Delta v/v_{exh}} = m_i e^{-(v_f - v_i)/v_{exh}}$$

$$v_{exh} = gI_{sp}$$

For surface launch on Earth or moon or other body $\rightarrow \mathrm{v}_i = 0$

$$\rightarrow m_f / m_i = e^{-\Delta v/v_{exh}} = e^{-(v_f - v_i)/v_{exh}} = e^{-v_f/v_{exh}}$$

$$v_f = speed\ for\ impulsive\ launch\ (at\ surface)\ NOT\ at\ \infty\ (ie\ not\ fighting\ grav\ potential)$$

For multiple stage boosters:

$$N = \#\, stages$$

$$\Delta v_{final} = final\ payload\ speed\ at\ end\ of\ all\ stages\ burnout$$

$$\Delta v_i = delta\ speed\ of\ stage\ i\ at\ burnout$$

$$v_{exh-i} = exhaust\ speed\ for\ stage\ i$$

$$v_{exh-i} = gI_{sp-i}$$

$$I_{sp-i} = specific\ impulse\ of\ stage\ i\ propellant$$

$$m_p = final\ payload\ mass$$

$$m_{begin-i} = mass\ stage\ i\ at\ beginning\ of\ its\ burn$$

$$m_{end-i} = mass\ stage\ i\ at\ the\ end\ of\ its\ burn$$

$$Define\ \alpha_i \equiv m_{end-i} / m_{begin-i} \rightarrow m_{end-i} = \alpha_i m_{begin-i}$$

$$\Delta v_i = -v_{exh-i}\ln\left[ (m_p + m_{end-i} + \sum_{j=i+1}^{N} m_{begin-j}) / (m_p + \sum_{j=i}^{N} m_{begin-j}) \right]$$

$$\Delta v_{final} = \sum_{i=1}^{N}\Delta v_i = -\sum_{i=1}^{N} v_{exh-i}\ln\left[ (m_p + m_{end-i} + \sum_{j=i+1}^{N} m_{begin-j}) / (m_p + \sum_{j=i}^{N} m_{begin-j}) \right]$$

$$\Delta v_{final} = -\sum_{i=1}^{N} v_{exh-i}\ln\left[ (m_p + \alpha_i m_{begin-i} + \sum_{j=i+1}^{N} m_{begin-j}) / (m_p + \sum_{j=i}^{N} m_{begin-j}) \right]$$

$$\Delta v_{final} = speed\ for\ impulsive\ launch\ (at\ surface)\ NOT\ at\ \infty\ (ie\ not\ fighting\ grav\ potential)$$

$$In\ grav\ field\ starting\ at\ r_i\ and\ ending\ at\ r_f$$

$$v_f^{\,2} = v_i^{\,2} + \frac{2GM}{r_f} - \frac{2GM}{r_i}\ where\ v_i^{\,2} = (\Delta v_{final})^2 \rightarrow v_f^{\,2} = (\Delta v_{final})^2 + \frac{2GM}{r_f} - \frac{2GM}{r_i}$$

$$v_f^{\,2} = (\Delta v_{final})^2 + 2v_{orb}^{\,2}(M, r_f) - 2v_{orb}^{\,2}(M, r_i) = (\Delta v_{final})^2 + v_{esc}^{\,2}(M, r_f) - 2v_{esc}^{\,2}(M, r_i)$$

$$\rightarrow (\Delta v_{final})^2 = v_f^{\,2} - \frac{2GM}{r_f} + \frac{2GM}{r_i} = v_f^{\,2} - 2v_{orb}^{\,2}(M, r_f) + 2v_{orb}^{\,2}(M, r_i) = v_f^{\,2} - v_{esc}^{\,2}(M, r_f) + v_{esc}^{\,2}(M, r_i)$$

$$For\ escape\ \ r_i = R, r_f = \infty$$

$$v_f^{\,2} = (\Delta v_{final})^2 - \frac{2GM}{R} = (\Delta v_{final})^2 - 2v_{orb}^{\,2}(M, R) = (\Delta v_{final})^2 - v_{esc}^{\,2}(M, R)$$

$$\rightarrow (\Delta v_{final})^2 = v_f^{\,2} + 2v_{orb}^{\,2}(M, R) = v_f^{\,2} + v_{esc}^{\,2}(M, R)$$

For the special case of all stages being equal in mass fraction (alpha) and $I_{sp}$. Generally, a multi-stage rocket as an optimized design would have lower mass stages "at the top" with optimized individual stages.

$$1)\ \textit{Case with all stages equal } I_{sp} \textit{ and } \alpha$$

$$v_{exh-i} = v_{exh},\ \alpha_i = \alpha$$

$$\Delta v_{final} = -\sum_{i=1}^{N} v_{exh-i} \ln\left[ (m_p + \alpha_i m_{begin-i} + \sum_{j=i+1}^{N} m_{begin-j}) / (m_p + \sum_{j=i}^{N} m_{begin-j}) \right]$$

$$\Delta v_{final} / v_{exh} = -\sum_{i=1}^{N} \ln\left[ (m_p + \alpha m_{begin-i} + \sum_{j=i+1}^{N} m_{begin-j}) / (m_p + \sum_{j=i}^{N} m_{begin-j}) \right]$$

$$= -\sum_{i=1}^{N} \ln\left[ ((m_p / m_{begin-i}) + \alpha + \sum_{j=i+1}^{N} (m_{begin-j} / m_{begin-i})) / ((m_p / m_{begin-i}) + \sum_{j=i}^{N} (m_{begin-j} / m_{begin-i})) \right]$$

$$2)\ \textit{Case with all stages equal } I_{sp},\ \alpha \textit{ and stage masses} - \textit{not optimal in practise}$$

$$v_{exh-i} = v_{exh},\ m_{begin-i} = m_{begin},\ \alpha_i = \alpha$$

$$\Delta v_{final} / v_{exh} = -\sum_{i=1}^{N} \ln\left[ ((m_p / m_{begin}) + \alpha + \sum_{j=i+1}^{N} (1)) / ((m_p / m_{begin}) + \sum_{j=i}^{N} (1)) \right]$$

$$= -\sum_{i=1}^{N} \ln\left[ ((m_p / m_{begin}) + (N - i + \alpha))) / ((m_p / m_{begin}) + (N - (i-1))) \right]$$

***Lunar Single Stage to Orbit (SSTO) and Single Stage to Infinity (SSTI)*** – SSTO is not possible from the Earth's surface with any currently existing technology and thus no SSTI is possible from Earth. This is primarily due to the deep gravity well of the Earth. However, the case for SSTO and SSTI from the lunar surface is completely different. The ability for even a single stage to achieve positive speed at infinity, even with solid boosters, opens up a vastly different set of options for fast response planetary defense if desired. In this sense it is much more like existing ICBM single stage boosters. A lunar Two Stage to Infinity TSTI is also an option where a high speed at infinity is desired.
We have modeled various existing multi-stage Earth launched liquid fueled rockets with the detailed specifications of each stage to confirm our analysis and we achieve nearly identical results to the measured performance. This is not particularly surprising given that the Earth's atmosphere is a relatively small correction in performance for large rockets. This allows us to consider much simpler designs for lunar mitigation approaches. While the desire to use the Moon as a base for planetary defense is more than a technical one and would involve both political policy issues and the practical costs of maintaining operational ability from the Moon, it is tempting to look at planetary defense in some ways as we look at terrestrial defense systems with solid fuel launchers "at the ready". A possible future would have interceptors in lunar silos for protection against meteorite impacts, low temperatures during the lunar night and high temperatures during the lunar day. Rapid response from the Moon would not be feasible for threats that are close to Earth due to the long travel time (~ a day) from Moon to Earth, Lunar basing could be part of a human presence on the Moon that would allow a synergistic base of operations for Earth or lunar defense. The low temperatures in lunar darkness and possible PSR would be another advantage for IR detection system whether passive or actively illuminated.

**Planetary Defense on Mars** – When not the primary target of this paper, we have also modeled using the same approach on Mars using solid fueled boosters. Though the Martian atmosphere is about 1% of the Earth at ground level, the same basic technic works well on Mars. We have modeled the largely $CO_2$ Martian atmosphere with much of the same computational analysis as presented here.

***Notional Lunar, LEO and GEO Solid Fuel Launcher Based Interceptor Design based on Current ICBM's***– As a concrete example of a solid fuel booster we have designed a nominal lunar, LEO, GEO **interceptor from the first two stages of the current generation Minuteman III ICBM**. We give the specification in the table below. Note that the requirements for the intercept of a threat is specific in each case. A full orbital dynamics analysis is needed for each case but we can draw general conclusions about the capability to achieve the required speeds to break away from the gravity well in each case. In each of the possible launch sites (Earth, LEO, GEO, lunar) there is also an orbital component of the initial velocity that needs to be considered to each intercept. We ignore this initial velocity as it can either be a positive or negative effect depending on the specific intercept scenario.

Note that for a lunar launch the lunar surface gravity is only 1.62 $m/s^2$ so even with a 5 mt payload (max shown in table below), the total launch mass is only 35,109 kg for a lunar surface "weight" of 56,880 N. The first stage thrust is 792 kN for an acceleration at launch of (792,000-56,880)/35,109=20.9 $m/s^2$ ~ 2.1 "gee" where "gee" is referenced relative to Earth "g loading" – ie the interceptor has a decently fast "off the launch pad" acceleration which increases in time until first stage burnout. Note that the first stage quoted thrust of 792 kN is the Earth "sea level thrust" with the "vacuum thrust" being about 890 kN as discussed below. This would give a slightly higher initial acceleration on the lunar surface of (890,000-56,880)/35,109=23.7 $m/s^2$ ~ 2.4 "gee" At second stage ignition the total mass of the second stage plus the maximum 5000 kg payload shown would have a mass of 12,032 kg for a lunar "weight" of 19,490 N. With a second stage thrust of 268 kN this gives an initial second stage acceleration of (268,000-19,490)/12,032=20.7 $m/s^2$ ~ 2.1 "gee" or essentially the same as the initial "off the pad" acceleration of the first stage. Like the first stage burn the second stage acceleration will increase during the second stage burn due to propellant mass ejection. In general, a small vectored thrust third stage or target intercept course correction propulsion will be needed to allow fine steering during the extremely high closing speed intercept. This will be one of the challenges of any interceptor scenario. While the closing speeds are not orders of magnitude above that of current KKV terrestrial interceptors (typ ~ 6km/s) the challenge of bolide closing speeds in excess 20 km/s will require significant thought and testing. Luckily there are no decoys nor active (non-cooperative) target maneuvers available to the bolide plus the cross section of relevant bolide targets is large (>20 m diam) compared to KKV targets. Still, this will be a part of the R&D phase of the program.

We will ignore the following points in the general analysis though they are important in specific orbital intercept calculations for a given launch site. For a LEO based interceptor the initial orbital velocity vector direction at launch relative to the desired trajectory vector is particularly critical and may require waiting until the launcher "comes around" in a LEO scenario to a more optimum initial velocity vector. In some very time critical rapid response scenarios this delay would be unacceptable for extremely short intercepts (< 1 hour). This type of short time response scenario would favor a GEO or ground based or many LEO based assets. This would need to be factored into any agile and robustly layered system.

- Earth rotation of about 0.46 km/s equatorial
- LEO speed of 7.7 km/s at 400 km altitude (ISS) – very important depending on target and timing
- GEO speed of 3.07 km/s - important depending on target and timing
- Lunar speed around Earth of 1 km/s

**Table 4. –** Example of using existing ICBM technology for lunar and geosync launch capability to intercept targets. Two stage solid fuel booster using Minuteman III ICBM engines**.** We model a modified Minuteman III using two of the nominal three stages (upper stage removed). Data from our model shows this interceptor would be suitable for lunar launch to infinity (far from Moon and Earth) with payloads up to 5 mt. Earth referenced $C_3$ is shown. For Earth surface launch booster does not achieve orbital speeds let alone escape even for small payloads while for lunar launch, the launcher achieves escape to intercept distant targets. **Note that using an ICBM from LEO can work for getting positive speed far from the Earth but this depends on the desired velocity vector for intercept. However, for short time threats where positive velocity is not required for intercept, having an ICBM interceptor in a LEO "planetary defense space station" could be viable. Note that the manufacturer thrust values (T) quoted for stages 1,2 in columns 4,5 are consistent with a first principles calculation of propellant mass flow from the quoted burn time (τ).**

$T = dm/dt * v_{rel} = dm/dt * gI_{sp} = [(m_{begin} - m_{end})/\tau]\ gI_{sp} = [m_{begin}(1-\alpha)/\tau]\ gI_{sp}$

($dm/dt = m_{begin}(1-\alpha)/\tau$ = 347 kg/s (stage 1), 95 kg/s (stage 2) assuming constant rate)

T= 805kN (stage 1 sea level), 890kN (stage 1 vac), 268kN (stage 2 vac)

| **Stage 1**<br><br>**Thiokol Tu-122 Thiokol**<br><br>**$I_{sp}$ (s)** | **Stage 2**<br><br>**Aerojet SR19**<br><br>**$I_{sp}$ (s)** | **Payload(kg)** | **Stage 1 Delta_v (km/s) vs payload**<br><br>**Thrust 792 kN (sea level)**<br><br>**Burn 60 sec** | **Stage 2 Delta_v (km/s) vs payload**<br><br>**Thrust 268 kN (vac)**<br><br>**Burn 66 sec** | **Total delta_v (km/s) No grav vs payload** | **$C_3$ $(km/s)^2$ vs payload** | **Lunar Surface Launch**<br><br>**Speed far from Moon w/Earth grav (km/s)**<br><br>**vs payload** | **Earth Geosync Launch**<br><br>**Speed far from Earth (km/s)**<br><br>**vs payload** |
|---|---|---|---|---|---|---|---|---|
| 262 (vac) | 288 (vac) | 500 | 2.92 | 5.02 | 7.94 | -62.5 | 7.43 | 6.64 |
| 237 (sea level) | | 1000 | 2.84 | 4.26 | 7.10 | -75.1 | 6.53 | 5.61 |
| **Stage 1 Alpha** | **Stage 2 Alpha** | 1500 | 2.75 | 3.73 | 6.49 | -83.4 | 5.86 | 4.81 |
| 0.099 | 0.11 | 2000 | 2.68 | 3.33 | 6.01 | -89.3 | 5.33 | 4.15 |
| | | 3000 | 2.54 | 2.76 | 5.30 | -97.4 | 4.51 | 3.03 |
| **Stage 1 m_begin (kg)** | **Stage 2 m_begin (kg)** | 4000 | 2.41 | 2.36 | 4.78 | -102.6 | 3.88 | 1.98 |
| 23077 | 7032 | 5000 | 2.30 | 2.07 | 4.38 | -106.3 | 3.37 | 0.49 |

***Earth Orbital Defense Layer*** – A possible layered system with Earth surface, Earth orbital and lunar could be a long-term strategy. The advantage of an Earth orbital layer is rapid response with a combined detection and mitigation "planetary defense space station" platform in LEO or possibly GEO. Having "at the ready" solid launcher in orbit with both LEO equatorial and polar platforms would start with an 8 km/s "start" and thus have much more flexibility, particularly for rapid response threat requirements. GEO would have the advantage of lower escape speeds required but the operational disadvantage of deployment and maintenance.

***Escape Speed - Whose Infinity*** - The concept of escape speed at "infinity" is always in the context of the particular gravitational well, that is being escaped. The problem is that there are various layers of gravity wells when we are looking at the solar system. For the lunar surface the escape speed from just the lunar gravity well is 2.38 km/s and for the Earth escape speed from just Earth gravity the escape speed is 11.2 km/s but we live in a much deeper gravity well if we are looking at interdiction at distances large compared to the Earth – Moon distance. We have to consider the Sun's gravity well which at the distance of the Earth from the Sun, the escape speed is about 42 km/s ($2^{1/2}$ times the Earth orbital speed around the Sun) not to mention galactic escape (~ 400 km/s) …. The latter is of course not relevant to planetary defense. If we look at interdiction scenarios that are at AU scales (for example at Mars) then we need to take into account, the solar gravity well. For most of the interdictions for the PI system the distances from the Earth are modest due to the short time interdictions enabled but for existential threats (10 km diam class) the solar gravity well is relevant, though not overwhelming in general. This is the reason that missions like Voyager took so long as gravity assist was needed given the purely chemical propulsion available at the time. In our plots when we use the term "speed at infinity" in this paper, it refers to escape from the Earth alone, Moon alone or Moon plus Earth alone.

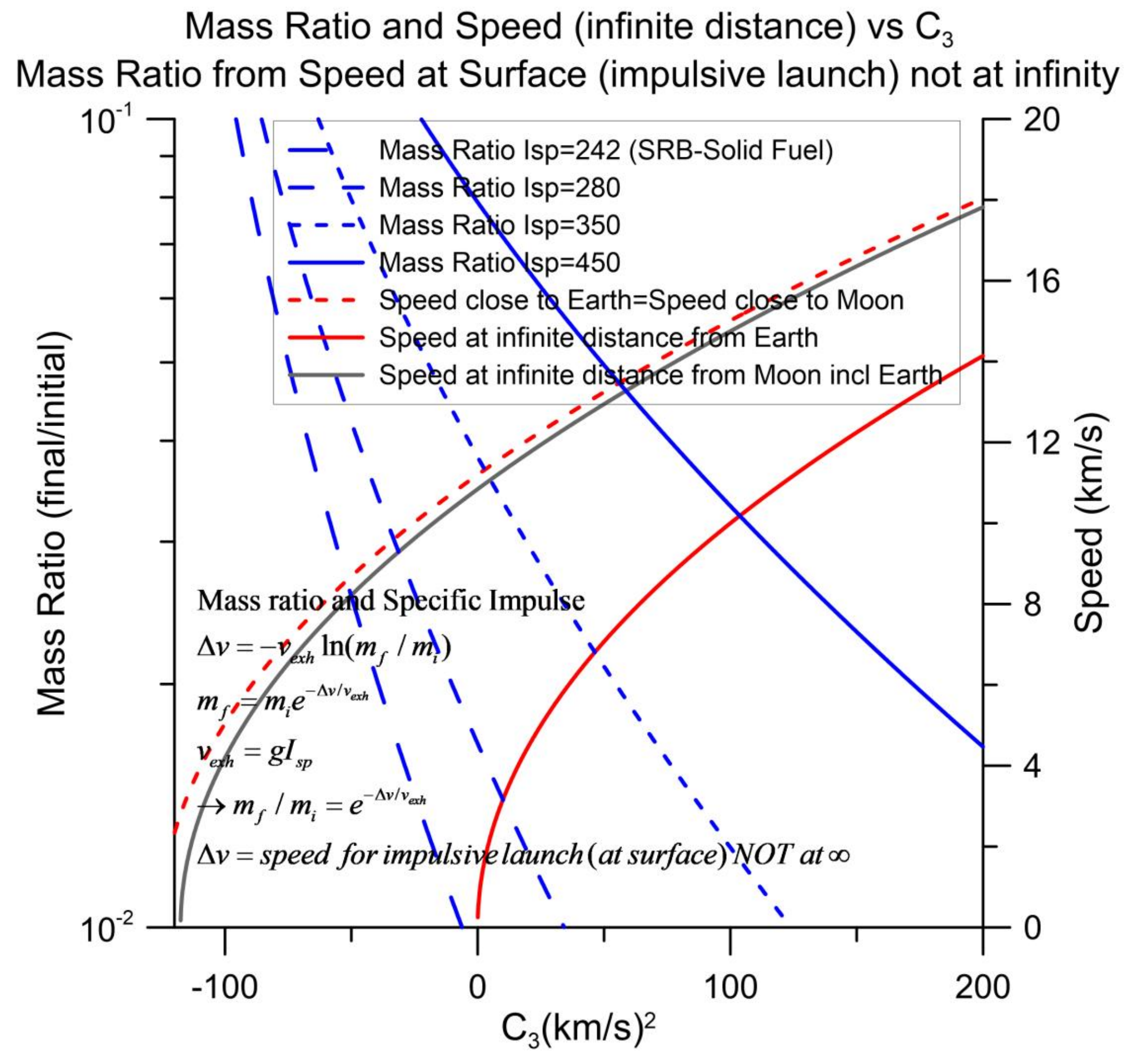

**Figure 123** – Mass ratio of final to initial vs $C_3$ for various specific impulse engines for both Earth and lunar launches.

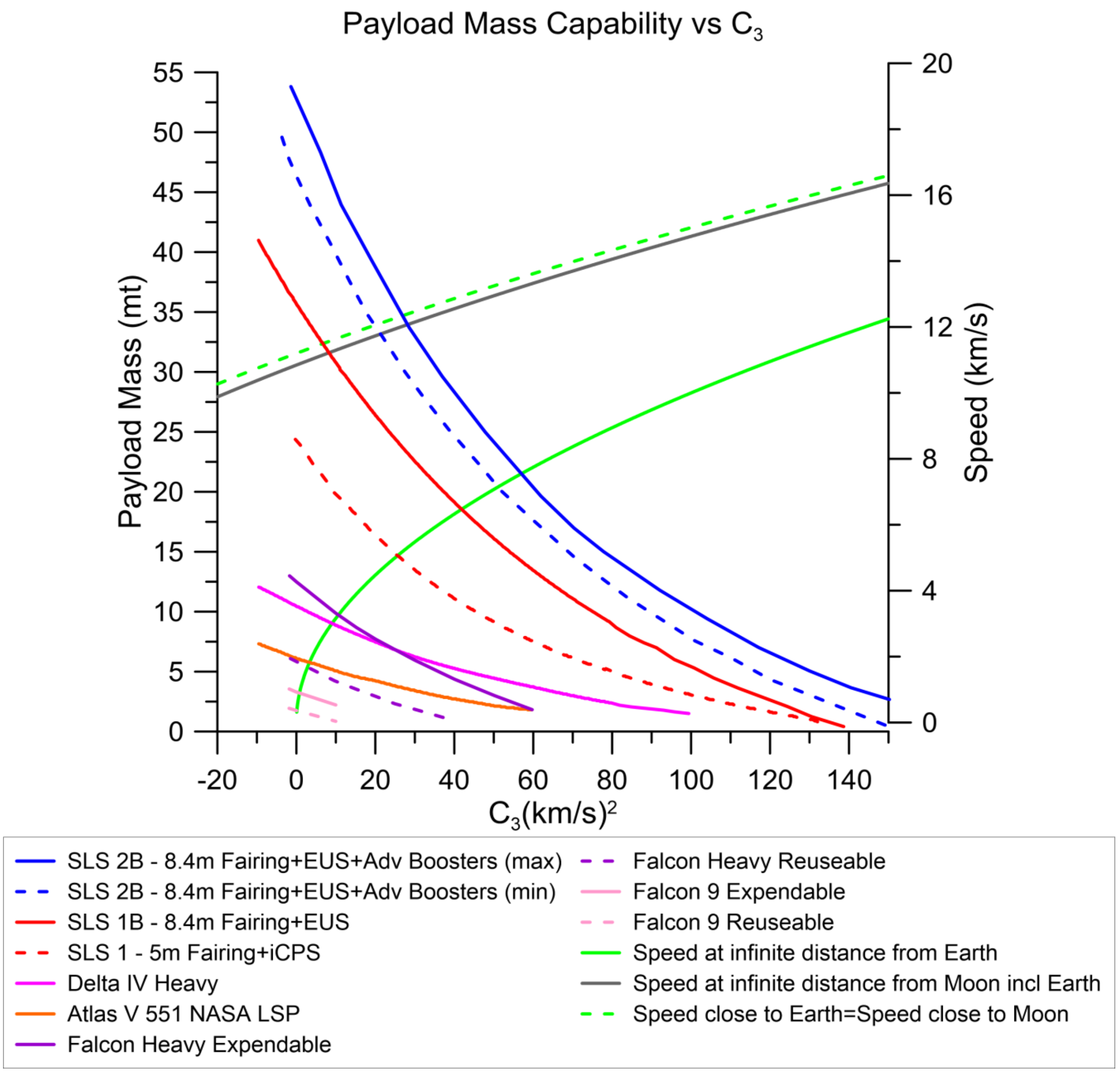

**Figure 124** – Payload mass and speed for both Earth and lunar launcher vs $C_3$ IF the same launcher were used in both an Earth and lunar launch scenario. Note that the payload speed for the same launcher from a lunar launch capability is vastly larger. While it does not make sense to have the same launcher on the Moon for the launchers shown, it does show that lower performance launchers, in particular the use of solid fuel launchers, would become viable on the Moon while they are not viable on the Earth.

***Disruption Energy Needed vs threat Size*** – As discussed earlier, one of the key optimizations to be explored in our technique is the efficiency with which the high closing speed (~20km/s for asteroids and 40-60km/s for comets) penetrator kinetic energy in the bolide frame is deposited into the desired low speed fragmentation speed. Our studies of a wide range of internal threat structures and hypervelocity penetrators using 3D supercomputer codes such a ALE3D and CTH form another ongoing part of our program which will enable us to compute the coupling efficiencies we can expect in different scenarios. To bound this problem, we adopt a pessimistic value for the energy coupling per unit mass of the parent bolide (**κ (J/kg)**) from studies of asteroids hitting other asteroids from both observed cratering and from SPH studies of low strength spherical asteroids hitting other low strength spherical asteroids. This is NOT how our system is designed as we use very high strength optimized penetrators but nonetheless this serves an upper limit of **κ =100 J/kg** corresponding to a 14.4m/s disruption vs the theoretical (too optimistic value) of κ =0.5 (J/kg) for 1m/s disruption with unity coupling. In the accompanying plot we show the penetrator mass required to deliver the required disruption energy assuming values of κ from 0.5 to 100 J/kg. The conclusion is that even for the pessimistic value of κ =100 J/kg, we can still mitigate even very large threats in the km size range using the upcoming heavy lift, positive $C_3$ launch vehicles such as SLS and Starship (refueled in orbit). For a threat such as Apophis, a single launcher is sufficient assuming the pessimistic value of κ =100 J/kg. For threats such as a 1km threat, multiple SLS (14xBlock 2) or Starship (7x) would be required for κ =100 J/kg. These would be large, though still within a near future launch capability. We have much more to explore on optimizing the energy coupling but even the pessimistic value of κ =100 J/kg is encouraging in allowing mitigation scenarios for even large threats.

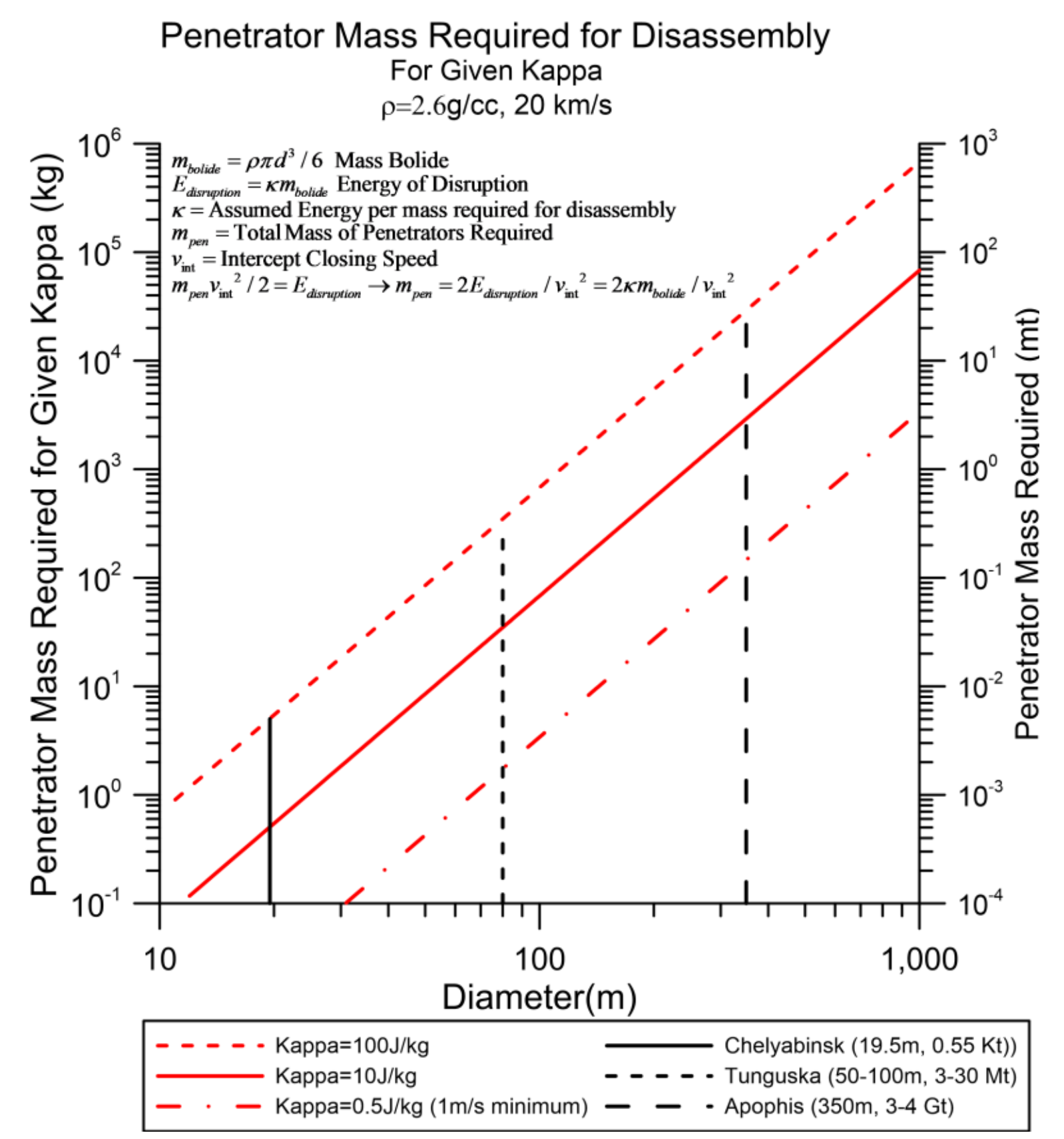

**Figure 125** – Penetrator mass required vs target diameter and a metric (κ) that describes the energy per unit of target mass for disruption. In theory, κ=0.5 J/kg for 1m/s disruption even for km scale objects (gravitational energy is sub-dominant) while observation and SPH studies of asteroids hitting asteroids (NOT purposefully designed high strength penetrators), indicated κ~100J/kg. We show the total penetrator mass required for κ=0.5, 10 and 100 J/kg to bound the problem. Even with the pessimistic case of κ=100J/kg we can still mitigate even very large threats with the new generation of heavy lift, positive $C_3$, launchers becoming available. This is extremely encouraging.

For Bennu, with an estimated mass of $8x10^{10}$ kg and assuming κ =100J/kg to disrupt, we get $8x10^{12}$ J = 1.9kt TNT ~ 40 tons of penetrators @ 20 km/s to disrupt. The trade space of using NED penetrators, explosive packed penetrators vs passive penetrators is part of our ongoing research. The value to society of trading the purely economic damage caused by a large threat such as Apophis, Bennu or a hypothetical 1km threat is another subject though in any such trade space the loss of life and disruption of societies would have to be weighed against the cost of the launch vehicles. It is difficult to imagine a scenario where a very large threat, such as a 1km (~100 Gt yield) would not be mitigated even if it meant launching multiple launch vehicles, whether passive or NED equipped penetrator. Another issue, often misunderstood in the comparison between mitigation via deflection, is that our technique allows for mitigation of even km class objects on the time scale of intercept of months prior to impact for the largest threats (1 km) while other techniques often require vastly longer time scales. This is an important distinction.

***Disruption Energy and Cohesive Strength*** – The term disruption, is used to define many different scenarios when it comes to planetary defense. In this paper we use the terms disassembly and disruption to mean taking apart a threat to a level where the largest fragments are no larger than approximately 10m in diameter. The energy to disassemble or disrupt a threat consists of five basic elements

- Residual kinetic energy of the fragments far from the original intact bolide
- Gravitational binding energy
- Cohesive internal binding energy
- Internal heating of the bolide due to the disruption method
- Energy used to vaporize material that escapes – ie escaping molecular level fragments
    - As opposed to internal vaporization that acts as a "gas expansion engine"

$$E_{disruption} = E_{KE-fragments} + E_{gravitational-BE} + E_{cohesion} + E_{bultk-thermal} + E_{vaporization-escaping}$$

***Cohesive Internal Binding Energy*** – The bolide can be viewed as a complex object with internal compressive, expansion and shear strengths. These manifest themselves as a heterogeneous elastic body that ultimately can break. These include more complex processes such as fracturing. Realistic bolides are not homogeneous bodies and thus the complexity of a real threat has to be approximated. The area of most interest to us in this paper is understanding mitigation methods that minimize the required energy to disassemble the threat (break apart) in such a way that no remaining fragment is much larger than about 10m (for rocky threats) and spread the fragments so they are far enough apart to produce de-correlated blast waves in the Earth's atmosphere or miss the Earth completely. We DO NOT need to reduce the threat to a microscopic level nor do we need to vaporize it to mitigate it. One question is "what is the least mass of interception" that breaks the threat into fragments of roughly 10m or less AND is applicable to a wide range of possible and largely unknown internal structure. Based on the observed rotation of asteroids, it is fairly well established that asteroids larger than about 100m are gravitationally bound rubble piles with very low internal cohesion at large scales. This does not mean that there is no cohesion (molecular binding) at small scales but only that at large scale there is very little binding (rubble pile). For example, a self-gravitating sand, gravel, rock and boulder pile is a good example of the possible internal structure of large asteroids (>100m). At smaller scales the rotation of asteroids indicates non-trivial internal cohesion. Examples of the latter are the estimated cohesive strength of the Chelyabinsk (20m) and Tunguska (~50-100m) bolides that are estimated to have had 2 MPa and 1 MPa effective internal (compressive) strength respectively. In any realistic internal bolide composition, the cohesion will be very scale dependent with much lower cohesion at larger scale.
The equivalent energy density corresponding to the strength is useful to understand the order of magnitude of internal cohesive binding. Knowing the internal macroscopic material strength, we can compute the equivalent internal energy density.

$Example\ (Chelyabinsk):$

$$\sigma = 2MPa,\ \rho = 3.3g/cc\ \ (3300kg/m^3)$$

$$\rho_{cohesive}(J/kg) = \sigma/\rho = 606J/kg$$

How should we interpret this internal cohesive energy density? One question to ponder is what are we really trying to do here? In our case we want to disrupt (fragment) the threat into pieces that are not larger than ~ 10m in size with a practical system. The extremes of disruption are to do nothing (no energy required) to the other extreme of fragmentation to the molecular level by vaporizing the threat (very larger amount of energy of roughly 4-6MJ/kg). Neither of these is desired. We can gain some simple insight by asking the following simple question. Suppose I have a tree with a cross grain yield strength of 2 MPa (typical for wood) that is 30m long and 1m in diameter and want to expend the minimum amount of energy to fragment the tree so that no single fragment is great than 1m. What is the lowest energy solution? A couple of possible solutions to fragmenting the tree are explosives, "sledgehammer" and sawing. If we ponder this simple problem, we get insight into what we are trying to go in fragmenting a bolide. In the case of the tree a very low energy solution is to use a saw whose blade width is a parameter as well as the size of the tree fragments after disassembly. Since we desire no tree fragment be greater than 1m, we will "saw" the tree into 1m lengths. If we ponder the issue of the blade width as a parameter, it is easy to understand that the least energy in breaking the cohesive bonds (tearing or breaking the wood fibers) is to choose the minimum blade width possible but understand the issue of "chip removal". Imagine a chain saw, vs a rotary hand saw vs a thin blade dicing/ band saw. The thin width dicing saw (band saw) gives the minimum energy. Fundamentally, the saw breaks the wood fibers allowing detachment of each fragment (or 1 m log in this case). In a realistic bolide intercept however, we do not have the ability to "saw" the bolide into pieces.
If we imagine a real bolide consisting of a heterogeneous distribution of boulders each of which is weakly bound to its nearest neighbors that are touching it. The energy required to break the cohesive bonds allows for disassembly. We ignore the gravitational binding and final kinetic energy in this discussion. Since we only desire to fragment to level compatible to acceptable blast wave damage we can set both an "upper limit" ($d_{frag-max}$) on the fragments (to limit the blast wave damage) and a "lower limit" below which we do not have to go as this would generally require extra energy. In realistic hypervelocity impacts we do not have precise control on the lower size limit as this may come for "free" in the realistic interception interaction. We do not have the realistic "luxury" of precision slicing of the bolide so we will generally only have control over the maximum size scale. The latter could be set by the spacing between the penetrators for example.
Imagine the "real bolide" as consisting of a range of "rock" sizes from sand grains to large boulders. The smaller elements are essentially of no concern to us. It is only the larger boulders we need to fracture so that the largest fragment remaining is no larger than $d_{frag-max}$. In this sense the minimum energy required is the energy required to fracture any portion that is "solid or molecularly bound" larger than $d_{frag-max}$ and as the same time we have to "plough through" the smaller sized material to get to the larger well bound material. Since we lack knowledge of the internal structure of a threat in general, we will need to be conservative in our assumptions as to what is required to reduce the largest remaining fragment to less than $d_{frag-max}$. Since $d_{frag-max}$ ~10-15m for rocky bolides, it is also possible that there is no such "solid or molecularly" object inside but we cannot assume that without probing the interior. The most conservative operational solution is to assume that the entire interior is solid though we know that this is not consistent with the rubble pile observed nature of large (>100m diam) bolides. In our current hypervelocity simulations, we are building models that are both homogeneous as well as complex heterogeneous structure. Much more will be forthcoming as we simulate the hypervelocity penetrator simulations with both ALE3D (LLNL code) and with CTH (Sandia/ LANL code) to study a wide range of scenarios to bound the problem. In the most conservative case of complete molecular binding of a homogeneous solid then we are reduced to fracturing this into 10m class remnants. To some extent this is similar to the issue of "fracking" used in the petrochemical industry.
The general issue of fracturing can be visualized as the breaking of molecular bonds that have a restoring force (spring constant or modulus of elasticity). Once a critical width of the crack is formed then the "pieces can come apart". There is no general mathematical solution to the problem of fracturing, except for the case of a flat circular geometry and some special elliptical case.

***Fracture Microphysics*** - The microphysics of the problem is much like a van der Waals force between molecules except in 3D with large number of bonds that are not all the same. Typically, the length scale required to separate the pieces so that the molecular forces become negligible or even slightly repulsive is relative small (sub-micron – typically less than 10nm) and we can make simple estimates to bound the energy but assuming the parent bolide needs to be "sliced up" into pieces of size $d_{frag-max}$ which can be translated into a "total slide area" $A_{slides-total}$ and a minimum separation distance $h_{slice-min}$ between the "sliced pieces". The product of the slice area and the slice thickness gives a total slice volume $V_{slice-total} = A_{slides-total} * h_{slice-min}$. The total energy of this process can be thought of as $E_{slide-total} = \rho_{cohesive} * V_{slice-total}$. **If we set $h_{slice-min} = 10^{-6}$ m to be conservative** and imagine slicing into cubes of size 10m as a specific example we get (per 10m cube fragment) $V_{slice-total} = A_{slides-total} * h_{slice-min} = 6x10^2 (m^2) x10^{-6} (m) = 6x10^{-4} m^3$. If we use the example above for Chelyabinsk compressive "binding" strength of $\sigma$=2 MPa =$\rho$(J/m$^3$) = 2x10$^6$ J/m$^3$ we get for 10m cube (with all 6 sides) a total "fracture energy" per 10m cube of $E_{slide-total} = \rho_{cohesive} * V_{slice-total}$ (per 10m cube) = $2x10^6$ J/m$^3$ x $6x10^{-4}$ m$^3$= 1200 J per 10m cube. We can normalize this by the mass of the cube to get energy required per mass of the fragment. The actual type of material is critical to be able to do a more detailed analysis. For example, if we had a homogeneous piece of metal that could "stretch" until it broke then we would approach the problem by using a bulk modulus of elasticity and apply this to the failure point "integrating the spring force" to get the energy of disruption. We "sweep this all under the rug" by assuming a bulk internal energy density and determine the volume of material we must remove to break the cohesive bonds. This is seen in the cutting of rock with a diamond saw or using a hammer and a chisel to fracture the rock vs "stretching the rock" until it breaks. This is easier to imagine with the energy required to cut a rubber band into pieces with a scissors vs stretching the rubber band until it breaks and then repeating this process until the rubber band pieces are small enough to meet our fragment criteria. The reality of a hypervelocity "chisel or scissors" is clearly a complex problem critically dependent on the actual internal structure and the penetrator and bolide solid to liquid to gas to plasma transitions and the subsequent "gas expansion engine" explosion inside the bolide that fractures it. In a sense this is like taking apart a mountain "one slice at a time". This is what our penetrator array simulations do. They follow the complex evolution of the hypervelocity penetrators as they "slice" the bolide apart.

***Chelyabinsk Example*** - For Chelyabinsk the density was about 3300 kg/m$^3$ and thus the 10m cube would be $3.3x10^6$ kg. This would yield a fracture energy per mass of the 10m cube of 1200J/ $3.3x10^6$ kg~ $4x10^{-4}$ J/kg. This is negligible compared to a 1 m/s disruption speed (for example) of 0.5J/kg.

***Energy per unit mass scale with fragment size*** – The scaling of the mass of a fragment scales as $d^3$ while the fracture surface area scale as the fragment surface area or as $d^2$. The fracture energy scales as the fragment surface area or as $E_{fracture-fragment} \sim d^2$ and hence the fracture energy per mass of the fragment scales as $d^2/d^3=1/d$. The larger the fragment the less fracture energy per unit fragment mass. The number of fragments for the parent bolide diameter of D scales as $N_{frag} = (D/d)^3 \sim 1/d^3$. Hence the total amount of fracture energy for fracturing the parent bolide into pieces of size d scales as $E_{fracture-total-bolide} \sim N_{frag} * E_{fracture-fragment} \sim (1/d^3) * d^2 = 1/d$ … less fracture energy needed for larger fragments and more fracture energy is needed for small scale fragmentation as expected.

***Energy that goes into Bulk Heating of the Threat*** – As discussed earlier the specific heat of rock is ~ 1kJ/kg-C and thus we do not want to bulk heat the overall threat very much (much less than 1 C desired).

***Hypervelocity Intercept Actualization*** - The reality of any realistic hypervelocity intercept and fragmentation however is very different than this idealized "fracturing" or "fracking" of the hypothetical bolide above. While in theory the energy per unit mass of the large (10m in this case) fragments is extremely small and indeed negligible compared to the dispersion KE of the fragments, the reality is that we cannot currently land on a threat and "frack it" so we are limited to hypervelocity intercepts and the energy to accomplish a hypervelocity "fracking" remains one of our major research areas for the future.
As we mentioned the SPH simulations of Jutzi et al 2009 for "asteroid on asteroid" disruption give an energy per unit mass of about 100 J/kg. We currently take this as an upper limit to what an optimized penetrator "fracking" can accomplish. Even with this extremely conservative number (100 J/kg) we can accomplish the mitigation needed but we will learn much more as we explore numerous penetrator scenarios.

***Cohesive Strength and Maximum Bolide Rotation Rate*** – Large asteroids (d>~300m) are known to have maximum rotation rates that appear to imply they are largely dominated by gravitational bound rubble pile structures rather than being dominated by strong molecular bonding. We can model the effect of a bulk cohesive binding and calculate the effects on the maximum spin rate. We assume a spherical parent bolide as well as a spherical distribution of fragments that the parent bolide is made of. The bolide is both self-gravitating as well as cohesively bound. In parametrizing the cohesive strength, we note this is the tensile (pulling) strength or pressure require to pull away a fragment off the surface. The tensile strength is vastly smaller than the compression strength. This is easily visualized as the difference between picking up (tensile) a small rock off a large boulder vs pushing the rock (compression) into the boulder by pushing "down". For example, above we examined the compressive yield strength for the Chelyabinsk asteroid due to the extreme atmospheric entry ram pressure. The compressive strength is in the MPa range while the tensile strength of pulling the various internal constituents apart is likely orders of magnitude smaller. Below we examine the tensile strength vs spin.

$$r_o = bolide\ radius$$

$$\rho_o = bolide\ density\ average$$

$$m_o = mass\ bolide = \rho_o 4\pi r_o^3 / 3$$

$$\omega_o = angular\ freq\ at\ equatorial\ breakup\ of\ bolide\ with\ no\ cohesion\ (grav\ only)$$

$$\tau_o = rotation\ period\ at\ equatorial\ breakup\ of\ bolide\ with\ no\ cohesion\ (grav\ only)$$

$$v_o = r_o \omega_o = surface\ speed\ at\ equator\ of\ bolide\ with\ no\ cohesion\ (grav\ only)$$

$$\vec{r}_{o-i} = radius\ vector\ from\ fragment\ to\ center\ of\ bolide$$

$$r_i = fragment\ radius$$

$$\rho_i = fragment\ density$$

$$m_i = fragment\ mass = \rho_i 4\pi r_i^3 / 3$$

$$\sigma_i = cohesive\ binding\ pressure\ (strength)\ at\ fragment\ i - tensile\ (pulling)\ not\ compression$$

$$f_i = fraction\ of\ cohesion\ at\ fragment\ i - includes\ fractional\ area\ and\ projection\ angle$$

$$\omega_i = \left|\vec{\omega}_i\right|\ angular\ freq\ at\ equatorial\ breakup\ of\ bolide\ with\ no\ gravity\ (cohesion\ only)$$

$$\omega_{i-T} = angular\ freq\ at\ equatorial\ breakup\ of\ bolide\ with\ gravity + cohesion\ only$$

$$F_i = cohesive\ bonding\ force\ on\ fragment\ i = \pi r_i^2 f_i \sigma_i$$

$$f_i = 1 \text{ assuming } \vec{F_i} \text{ vector is along } \vec{r}_{o-i}\ (equator\ and\ \vec{F_i} \perp \vec{\omega}_i)\ and\ full\ hemispherical\ cohesive\ contact$$

$$f_i = 0 \text{ assuming } \vec{F_i} \text{ vector is perpendicular to } \vec{r}_{o-i}\ (\vec{F_i} \parallel \vec{\omega}_i)$$

$$F_{grav-i} = gravitational\ force\ on\ fragment\ i\ at\ surface = G m_o m_i / r_o^2 = G \frac{4\pi r_o^3 \rho_o}{3 r_o^2} m_i = G \frac{4\pi r_o \rho_o}{3} m_i$$

$$F_i / m_i = \frac{\pi r_i^2 f_i \sigma_i}{\rho_i 4\pi r_i^3 / 3} = \frac{3 f_i \sigma_i}{4 r_i \rho_i}$$

$$F_{grav-i} / m_i = G \frac{4\pi r_o \rho_o}{3}$$

$$\frac{F_i}{F_{grav-i}} = \frac{9 f_i \sigma_i}{16\pi G r_o r_i \rho_o \rho_i}$$

$$\omega_i = \left[ \frac{F_i / m_i}{r_o} \right]^{1/2} = \left[ \frac{3 f_i \sigma_i}{4 r_i \rho_i r_o} \right]^{1/2}$$

$$\omega_o = \left[ \frac{F_{grav-i} / m_i}{r_o} \right]^{1/2} = \left[ G \frac{4\pi \rho_o}{3} \right]^{1/2} (\text{independent } of\ r_o) = 1.672 x 10^{-5} \sqrt{\rho_o} (MKS)$$

$$\omega_{i-T} = \left[ \frac{(F_i + F_{grav-i}) / m_i}{r_o} \right]^{1/2} = \left[ \omega_i^{\ 2} + \omega_o^{\ 2} \right]^{1/2},\ k \equiv \omega_{i-T} / \omega_o = \left[ 1 + \left( \frac{\omega_i}{\omega_o} \right)^2 \right]^{1/2}$$

$$\frac{\omega_i}{\omega_o} = \left[ \frac{F_i}{F_{grav-i}} \right]^{1/2} = \left[ \frac{9 f_i \sigma_i}{16\pi G r_o r_i \rho_o \rho_i} \right]^{1/2} = \left[ \frac{9 f_i \sigma_i}{16\pi G r_o^{\ 2} (r_i / r_o) \rho_o \rho_i} \right]^{1/2} = \left[ \frac{9 f_i \sigma_i}{16\pi G r_o^{\ 2} \varepsilon_{frag} \rho_o \rho_i} \right]^{1/2}$$

$$\varepsilon_{frag} = r_i / r_o = \textit{fragment fractional size}$$

$$\rightarrow f_i \sigma_i = \left( \frac{\omega_i}{\omega_o} \right)^2 \frac{16\pi G r_o r_i \rho_o \rho_i}{9} = \left( \frac{\omega_i}{\omega_o} \right)^2 \frac{16\pi G r_o^{\ 2} \varepsilon_{frag} \rho_o \rho_i}{9}$$

$$v_o = r_o \omega_o = r_o \left[ G \frac{4\pi \rho_o}{3} \right]^{1/2} = 1.672 x 10^{-5} r_o \sqrt{\rho_o}\ (m / s)\ (MKS)$$

$$\tau_o = 2\pi / \omega_o = 2\pi \left[ G \frac{4\pi \rho_o}{3} \right]^{-1/2} = \left[ G \frac{\rho_o}{3\pi} \right]^{-1/2} = 3.759 x 10^5 \rho_o^{\ -1/2} (s) = 104.4 \rho_o^{\ -1/2} (hrs)\ (MKS)$$

Note that this period is consistent with the observed shortest period ($\tau_o \sim 2.2\, hrs$) for large asteroids (diam > 1km) implying a rubble pile with density ( $\rho_o \sim 2250\, kg / m^3$) which is roughly that of rock.

See observed asteroid rotation curve plot below.

*Note that some smaller asteroids with short observed periods imply significant cohesive (molecular) binding (ie NOT a rubble pile).*

As an example in Fig. 126 below, max k= $\omega_{i-T}$ / $\omega_o$ ~100 for diameters from 10 to 100m

$$\rightarrow f_i \sigma_i = \left( \frac{\omega_i}{\omega_o} \right)^2 \frac{16\pi G r_o^{\ 2} \varepsilon_{frag} \rho_o \rho_i}{9} \sim 630 Pa\, \varepsilon_{frag} (r_o = 5m), 63kPa\, \varepsilon_{frag} (r_o = 50m)\ (\rho_o = \rho_i = 2.6 g / cc)$$

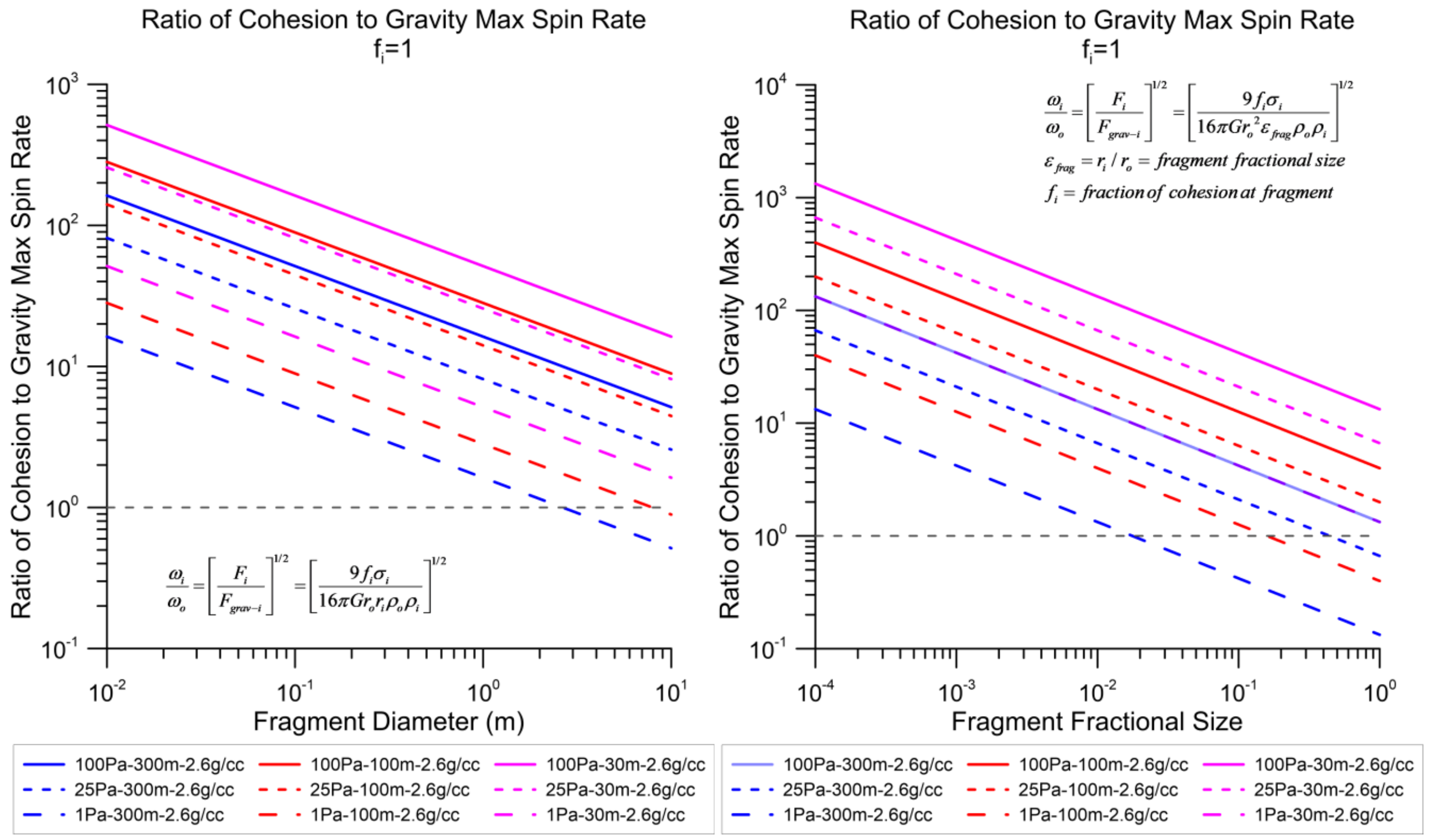

**Figure 126** – L: Ratio of Cohesive Binding to Gravitational Binding maximum rotation rate vs bolide diameter and surface fragment (rock) diameter. R: Ratio of Cohesive Binding to Gravitational Binding maximum rotation rate vs bolide diameter and surface fragment (rock) fractional diameter of parent bolide diameter.

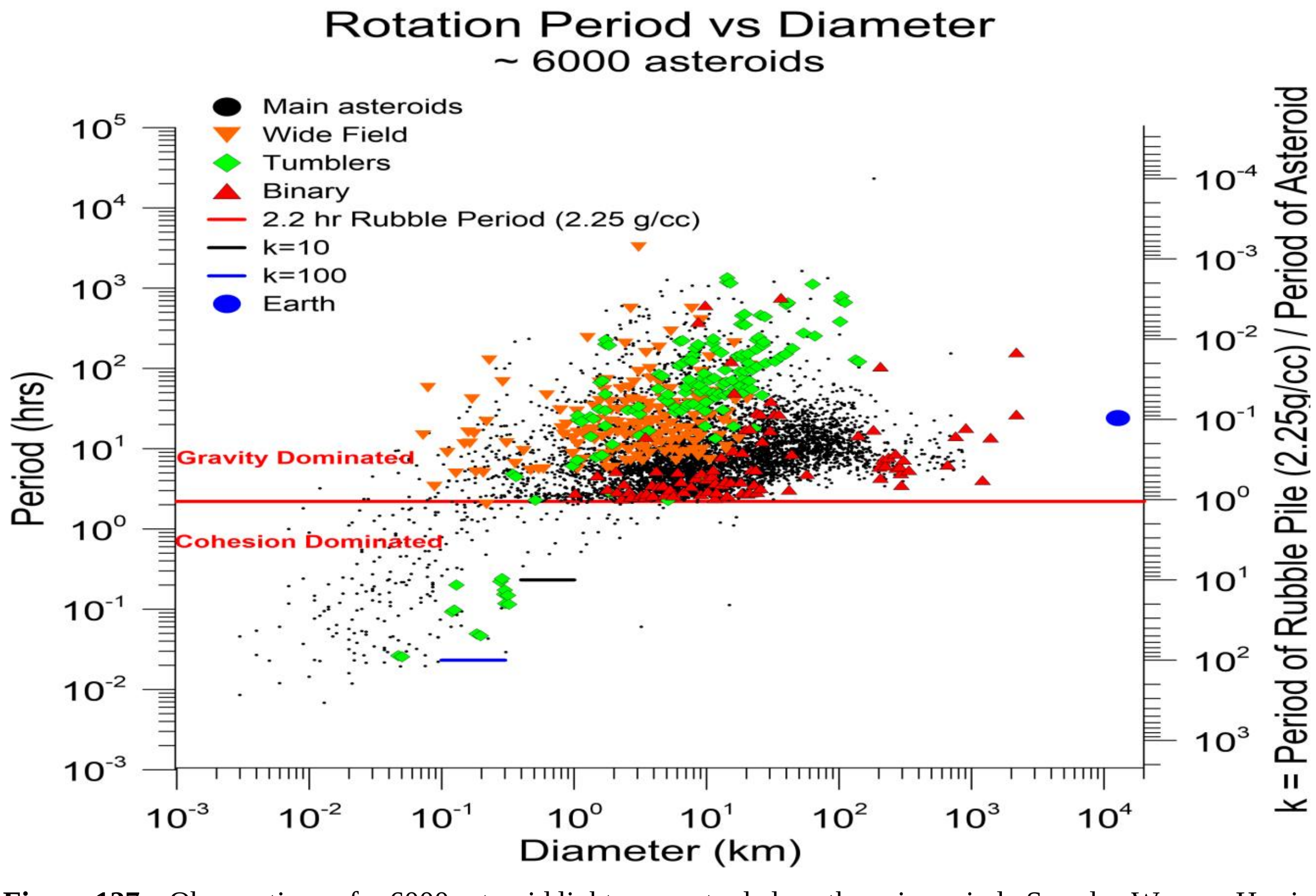

**Figure 127 –** Observations of ~ 6000 asteroid light curves to deduce the spin period. See also Warner, Harris and Pravec, Icarus 202, 134 (2009). Note sharp cutoff at about 2.2hrs (~10/day spin rate) for large (rubble pile) bolides. Conclusion is large asteroids are rubble piles & smaller asteroids have significant cohesive binding.

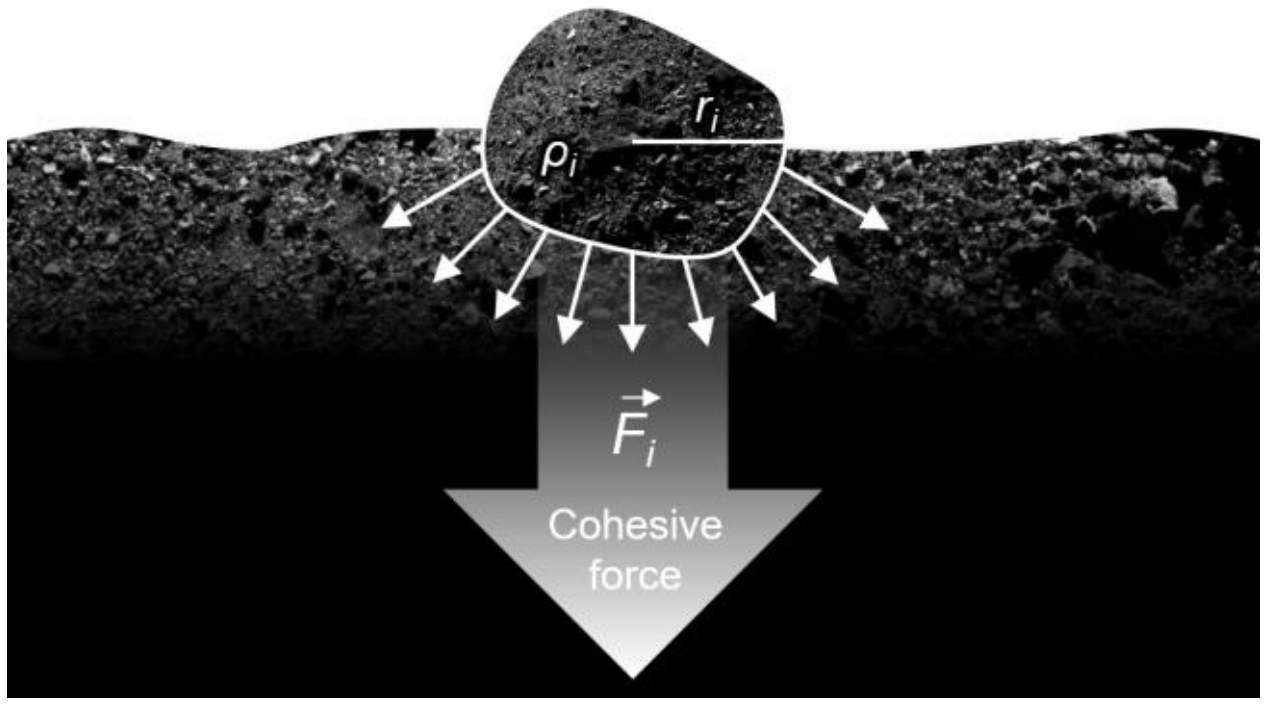

**Figure 129** – Modeling of cohesive tensile strength of a "rock" embedded in a fine binder material that holds the rock via molecular surface binding. Here we show the rock at the surface of the bolide with ½ of the rock attached via the fine binder and ½ above as an example.

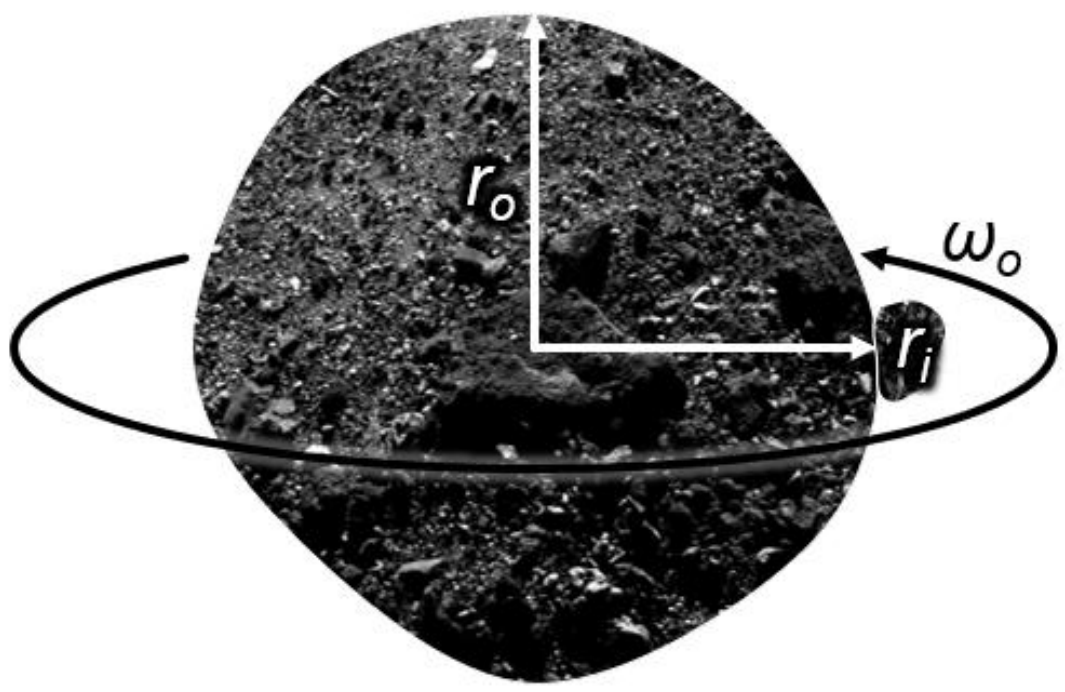

**Figure 128** - Example of rock at surface of a rotating bolide that is bound via a fine binder with surface molecular forces as in the previous figure.

### *Asteroid Rotation and Sequential Penetrators*

In some extreme cases, we propose the use of sequential penetrators for "hole drilling" to allow the use of a final explosive (conventional or nuclear). To most effectively "drill deeply" it is important that the sequential penetrators impact close enough on the surface and hence the rotation of the asteroid becomes can become an issue for a guidance and control system on each penetrator, though penetrators can also be semi-rigidly linked via rods. This is both an issue of the impact point as well as the penetrator velocity vector and body rotation vector. A simple analysis can be done by assuming a rotating spherical asteroid. Our ALE3D simulations show that the typical diameter of the entrance hole produced from a hypervelocity penetrator impact is roughly a few meters for a small diameter penetrator (typ ~ 0.1m diam). Thus, we want the distance between penetrator impacts due to asteroid rotation (d below) to be <1m. The measured value for k (incl cohesion) is shown in the figure above on observed asteroid rotation periods. The critical asteroid diameters of interest are roughly 0.1-1km with typical k<1 and a max k=100 for outliers at 0.1km and a max of k=10 for outliers at 1km diameter.

$\theta =$ angle between asteroid spin axis and first penetrator surface impact point

$v_o =$ surface speed at asteroid equator assuming rubble pile

$v_i =$ surface speed at asteroid equator assuming additional cohension

$v_{imp} =$ surface speed at impact point assuming rubble pile $= v_o \sin(\theta)$

$v_{imp-T} =$ surface speed at impact point assuming additional cohension $= kv_{imp}$

(*this assumes the cohesion is independent of* $\theta$)

$\tau =$ time between penetrator impacts

$d =$ surface distance between penetrator impacts on bolide

$$k \equiv v_{imp-T} / v_o = \omega_{i-T} / \omega_o = \left[1+\left(\frac{\omega_i}{\omega_o}\right)^2\right]^{1/2} \left[1+\left(\frac{9 f_i \sigma_i}{16\pi G r_o{}^2 \varepsilon_{frag} \rho_o \rho_i}\right)\right]^{1/2} \quad \text{(includes effects of both gravity and cohesion)}$$

$$v_{imp} = v_o \sin(\theta) = r_o \omega_o \sin(\theta) = r_o \left[G\frac{4\pi\rho_o}{3}\right]^{1/2} \sin(\theta) = 1.672x10^{-5} r_o \sqrt{\rho_o}\ \sin(\theta)\,(m/s)\ (MKS)$$

$$d = \tau v_{imp-T} = k\tau v_{imp} = k\tau v_o \sin(\theta) = 1.672x10^{-5} r_o \sqrt{\rho_o}\ \sin(\theta) k\tau \ \ (m)$$

*For data where we have the asteroid radius* $r_o$ *and rotation period* $\tau_o \rightarrow v_o = 2\pi r_o / \tau_o \rightarrow d = \tau v_o \sin(\theta) = 2\pi r_o \sin(\theta)\tau / \tau_o$

Assume worst case of impact at equator, $(\theta = \pi/2)$, $\rho_o$ =2250 kg/m$^3$ , →d(m) ~ 7.93x10$^{-4}$ $r_0 k \tau$

**Ex: τ=1ms (10m apart@10km/s),(0.1,1km diam,k=1)** $r_0$ **=50, 500m→d~0.04, 0.4mm – ie << size hole**

For the largest k~100 for 0.1km diam, k=10 for 1km diam, in a small number of cases → **d ~4mm << hole**

**Ex: τ=1s (10km apart@10km/s)** largest k~100 for 0.1km diam, k=10 for 1km diam, → **d ~4m ~ size hole**

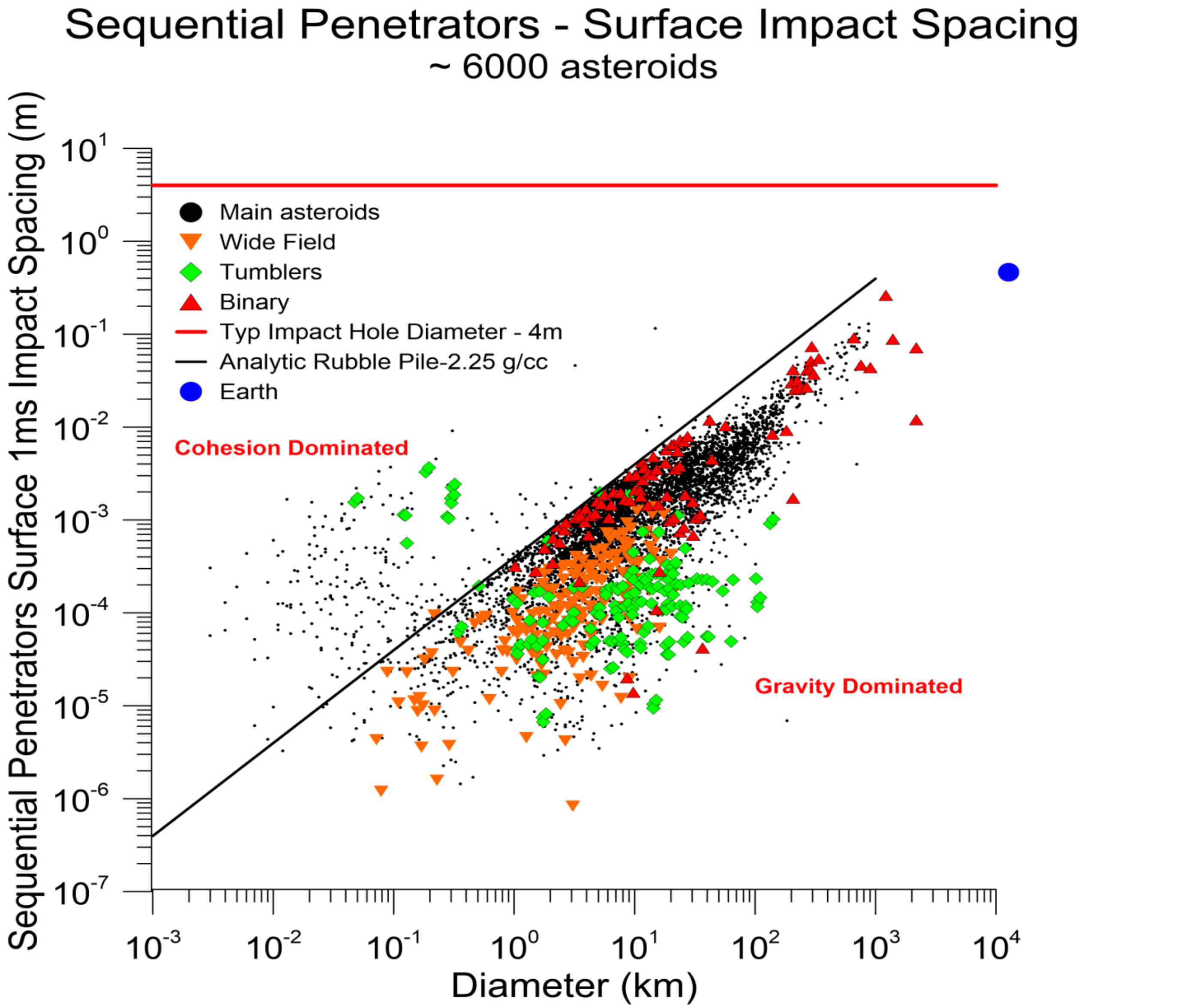

**Figure 130 –** Measured rotation curves from more than 6000 asteroids from asteroid light curves to deduce the spin period. Data from Warner, Harris and Pravec, Icarus 202, 134 (2009) which are then used to calculate the equatorial surface speed (max surface speed/ worst case) and hence the surface spacing between sequential penetrators for equatorial impacts. This becomes important for the use of sequential penetrators for both passive and active penetrators (including NED's) used for larger threats. The vertical axis is the surface spacing normalized to 1ms time difference between penetrator impacts. As an example, a 1ms time difference between impacts corresponds to 20m penetrator spacing for a closing speed of 20 km/s. This becomes important for fine guidance and control of sequential penetrators used for "hole drilling" for multiple purposes including deep penetration to maximize the use of active penetrators for fragmenting large threats. This can be extremely useful for active penetrators including using NED's where deep sub-surface detonations greatly increasing the explosive coupling. The surface spacing is compared to the typical size of the penetrator induced hole diameter of ~4m from our ALE3D simulations.

### *Asteroid Angular Rotation Rate and Sequential Penetrator Entry Angle*

In addition to guidance and control (GC) to allow sequential penetrator to "hit" close to the same bolide surface spot, there is an additional concern as to the entry angle of each sequential penetrator relative to the first and subsequent penetrators. This entry angle includes both the penetrator body symmetry vector (typically along the "long axis" of the penetrator as well as the center of mass velocity vector of each penetrator relative to all previous penetrators. Thus there are three issues to consider for the case of using multiple sequential penetrators for "hole drilling" for both passive and final active (including NED) penetrators. These are the surface impact position, penetrator body angle and center of mass velocity vector. All but the last can be controlled with relatively low GC thruster impulse. Since the "use case" for sequential penetrators is usually only needed for large threats (typ > 100m diameter) as smaller threats can normally be completely fragmented by single or multiple "non-sequential" penetrators. We show the results for known rotations of asteroids and consider the entry angle of subsequent sequential penetrators relative to the radial vector, from the center, for hypothetical, approximately spherical, asteroids. Assuming we can control the impact point of multiple sequential penetrator using the GC thrusters, this allows us to compute the angle of subsequent sequential penetrators, relative to the first, and thus compute the angle relative to the initial penetrator radial vector for the impact spot. This angle which advances due to the asteroid rotation, allow us to compute the final projected impact position in the "bottom of the bored hole". The figure of merit is that relative change of this final impact position deep inside the bored hole should be small compacted to the hole diameter. In the figure below, we computer the relative angle with the product of this angle and bored depth, gives the "delta" of the subsequent penetrator projected impact at the bottom of the bore hole. As can be seen from the figure below, the rotation rate for asteroids from 0.1-1km diameter is <0.01 radians/s. As an example, sequential penetrators with an entry "time delta" of 1ms this correspond to an angle of $<10^{-5}$ radians. For the example of a 10m deep bore hole this corresponds "hole bottom delta" d< $10m*10^{-5}$rad = 0.1mm. Increasing the time delta to an extreme 1s → d<0.1m. This is acceptable. Other cases for different "bore hole depths" can be easily computed.

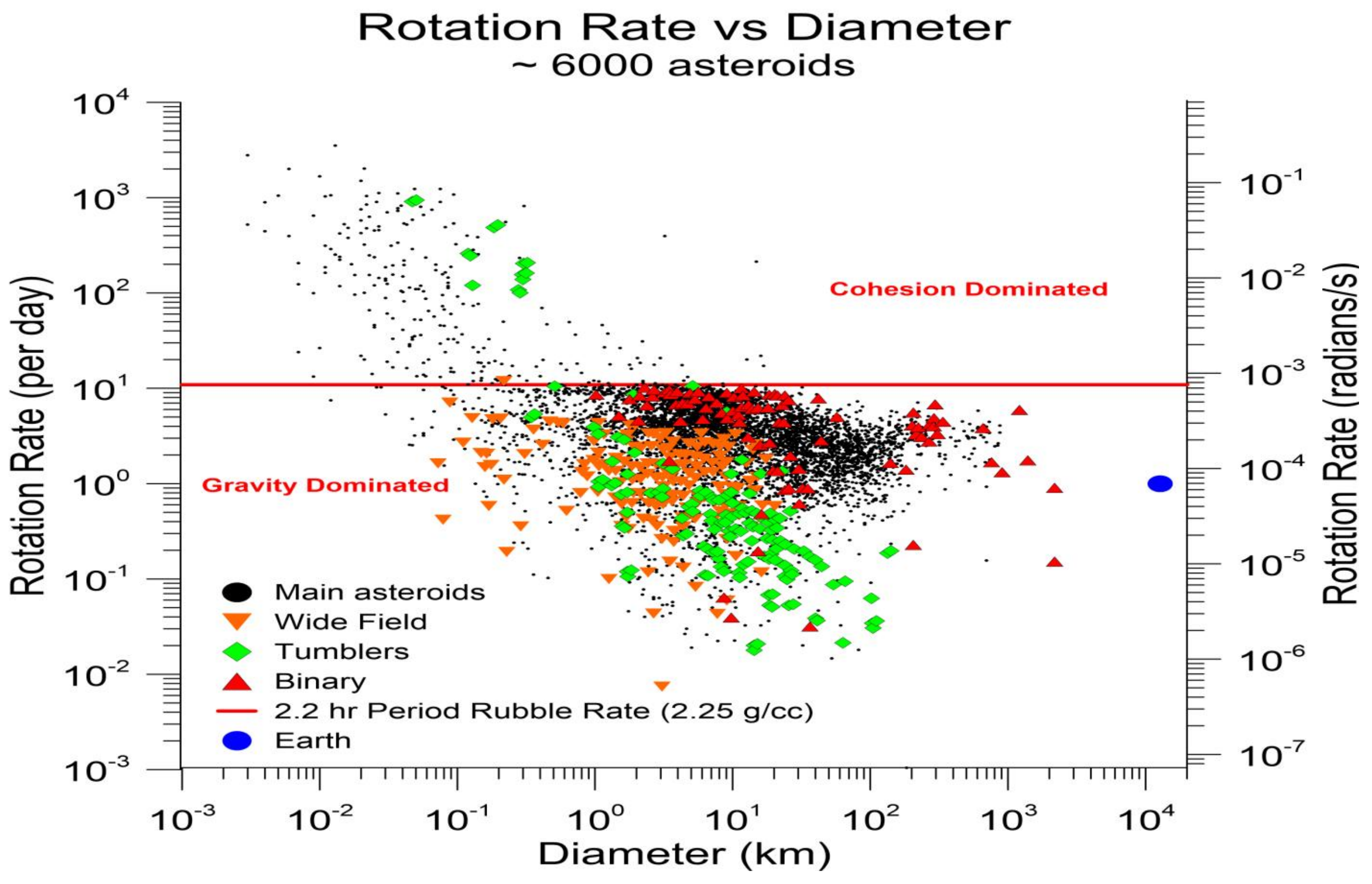

**Figure 131 –** Measured rotation rate of more than 6000 asteroids (+ Earth) used to compute sequential penetrator angles as discussed in the text.

## 15. A Simple Analogue

The reason this technique works is that the hazard is not just due to the total energy of the asteroid impact, but it is due to the high, near-instantaneous power of the un-fragmented impact. In our technique, the total energy released into the blast wave is similar whether the material is dispersed or not, BUT the instantaneous power received by a ground observer at any given time is vastly reduced due to the de-coherence of the many blast wave detonations of each fragment. A simple and convincing analogue as to why this technique works is to consider the following: imagine four scenarios, each of which contains the same amount of explosive energy.

1) One kg of TNT explosive in a compacted form detonated 10m away from you.
2) One thousand 1g "firecrackers" all compacted together and detonated simultaneously 10m away from you. **Effect: same as 1).**
3) One thousand 1g "firecrackers" dispersed randomly in a 10-meter diameter sphere whose center is 20 meters from you and detonated simultaneously. **Effect: vastly reduced from 1) due to pulse duration and shock travel time.**
4) **One thousand 1g "firecrackers" dispersed randomly in a ten-meter diameter sphere whose center is 20 meters from you and detonated randomly over a 10 second interval. This is the analogue to our technique. Effect: vastly reduced from 1) due to pulse duration, shock travel time, and randomization (de-coherence) of detonation times.**

**Another simple explanation** – Consider that the Earth is hit by about 100 tons per day of small-scale meteoritic debris that we do not notice. A 20m diameter asteroid density 3g/cc, like the Chelyabinsk event, is roughly equal to the average mass hitting the Earth in 4 months while a 50m, like the Tunguska 1908 event, is roughly equal 5 years of average debris mass hitting the Earth, a 100m is roughly equal to 40 years. Clearly, we do not worry about the "average debris rate", we worry about the tail of the temporal rate distribution. It is not the total mass (energy) hitting us but the time scale over which that total mass is deposited. When the temporal and spatial spread of the debris cloud is considered then it becomes "obvious" that this technique works.

### *Fragment size optimization*

We have chosen a very conservative fragmentation model with a fragment size of roughly 10m and a distribution function allowing larger fragments with decreasing probability. We have shown that this allows extremely effective planetary defense on very short time scales, allowing for a radical change in our ability to protect the Earth, and we can do even better if needed. We have run statistical fragmentation models from mean fragmentation sizes below 1m to greater than 15m with various size distribution functions. For loosely bound materials, the energy required to fragment to a given size is dependent on the coupling efficiency of the penetrator energy (KE plus possible explosive energy) in coupling to acoustical waves that fragment and eject the subsequent fragment of the bolide. For example, if we were to choose a smaller mean fragment of 5m (vs the current 10m in this paper) with a tighter upper size cutoff of, for example, 10m, then the conclusions of this paper remain largely the same, but the acoustic signature effects, and to a less extent the optical signature, become even more favorable (ie less). Smaller fragments are better if feasible. This is clearly seen in the figures showing the acoustic signature vs fragment size.

### *Spreading fragment cloud to largely miss hitting earth*

An alternative to using the Earth's atmosphere to absorb the hit is to disrupt as described above, but do so early enough so that the fragments largely miss the Earth altogether. If the mean radius of the fragment cloud is larger than the Earth's radius, then we largely mitigate by missing the Earth. This is shown in the plot below for several disruption speeds. For disruption speeds of 1m/s, as used for most of the preceding simulations, the intercept time required for the fragment cloud to miss the Earth is about 75 days assuming "ground zero" projected in a plane is "centered" on the Earth whereas the worst case is for "ground zero" to be projected at an "edge" in which case the fragment cloud will largely miss the Earth with a 150 day intercept for mean 1m/s disruption speed. A small number of fragments will still hit due to the "disruption noise" in the fragmentation process, but the vast majority can miss if there is sufficient time for intercept prior to impact. There is a clear trade space between the energy required for disruption at higher speed vs the time available to accomplish this scenario.

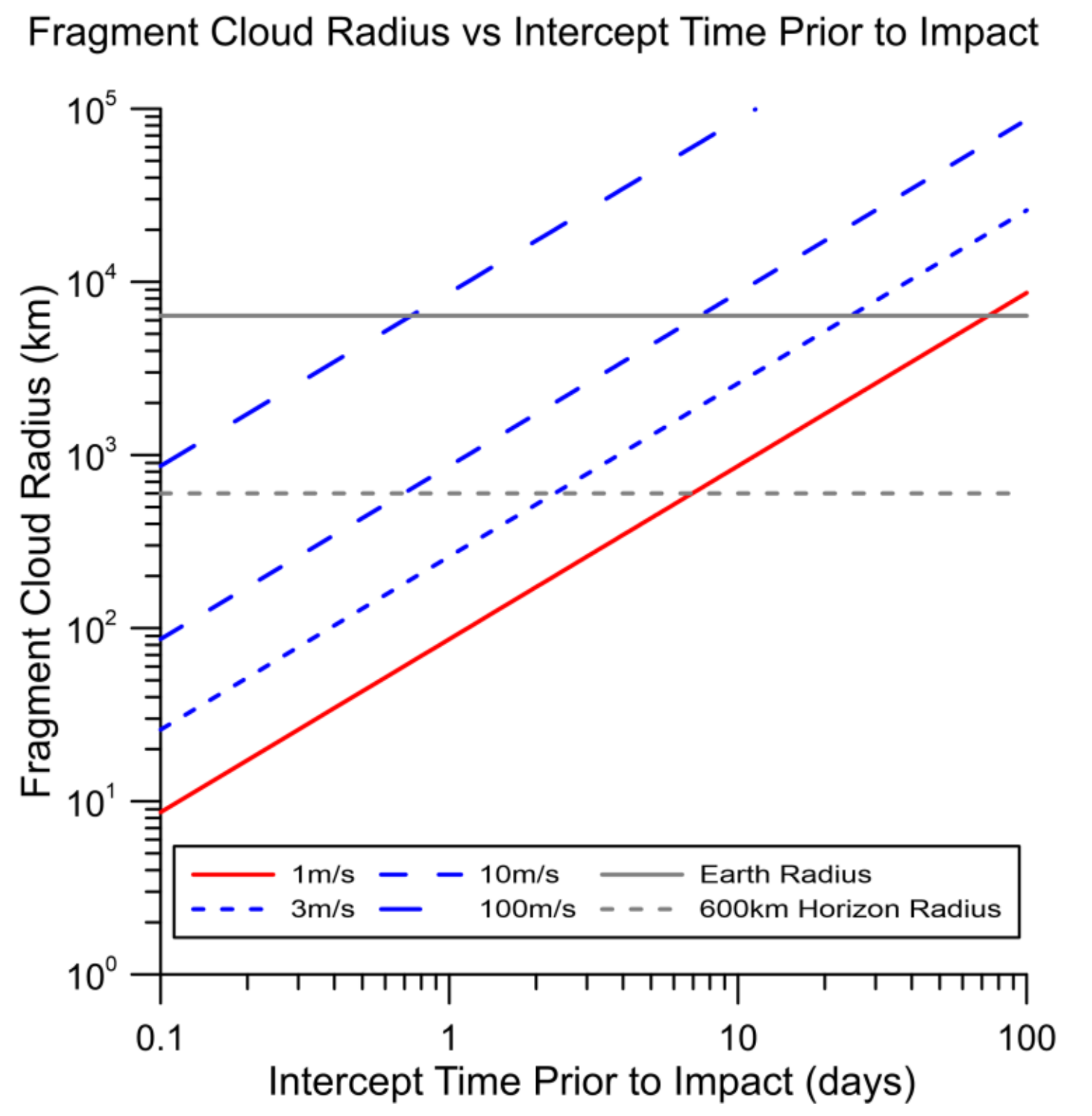

**Figure 132** – Fragment cloud radius vs intercept time. Note that at intercept times greater than about 75 days to 150 days depending on ground zero impact position and angle of attack with 1m/s disruption, the fragment cloud grows to be larger than the Earth and most fragments miss the Earth completely. Shorter intercepts are always possible with higher disruption speed.

### *Civil defense*

The issue of civil preparedness and large-scale notifications via electronic notification from cellular and internet sources is capable of rapid notification that an "event" is expected at a given time. Similar to the CD of the 1950 and 1960's to the present via sirens to notify of an imminent situation such as a tsunami, simple civil defense can alert the public to move away from windows and seek shelter from both the acoustical and optical signatures. While the public is sensitized to this for nuclear events, it would be relatively simple to begin an educational campaign to explain simple methods to minimize causalities. Very simple methods such as closing our eyes and ears, moving indoors, but away from windows, brick walls etc. could be easily enabled. The simple use of adhesive plastic film or even "packing or duct tape" on windows will largely mitigate damage from broken glass.

### *In-space testing – synthetic bolides in space*

Eventually, space testing of this technique, particularly for terminal guidance system lock and to test the effectiveness of the penetrator fragmentation of the parent bolide will be needed after extensive ground testing and design iterations. For space testing, the options could be small (1000kg-class) synthetic bolides that could be tested in space to compare to ground testing. This type of test would have to be done at either very low altitudes so the debris would de-orbit rapidly and not contribute to additional space debris or very far from the Earth.  Another possible test is to use an inflated "bolide" similar to the Echo satellite (1960,1964 - 70 kg, 30 m diameter gas inflated Aluminized thin plastic sphere) to create a low mass but "at scale" target. With orbital counter rotating target and inflated target, the closing speed would be approximately 16 km/s which would be an excellent test of realistic terminal guidance systems. With such a large surface area and low areal mass density the "fragments would de-orbit relatively rapidly thus minimizing additional space debris. Having multiple inflated targets could also be used to simulate a set of comet fragments as one complex example. In some ways this is analogous to ICBM MIRV decoys.

### *Apophis as a Test*

The options for actual asteroid testing will depend on deciding on "close in" vs long distance asteroid testing such as the DART mission. An option that may be controversial would be to use Apophis as the test target. Apophis comes close to the Earth frequently as the orbital period of Apophis is about 0.9 years. There will be a close encounter in April 2029 that could be a possible penetrator test opportunity using an instrumented penetrator that would measure the interior structure of Apophis via measurements of the deceleration vs time. This has never been done before and would require the development of extremely high "g loading" penetrators with electronics and a way to transmit back the data. This would be a challenging development, but given that the acceleration levels are not vastly different from artillery, the development could use similar technology. Getting the data out during the short time scale (~ 10ms) that the penetrator is inside the target will be challenging, though lower frequency RF could penetrate the asteroid and be received. Towing a wire behind the penetrator as an antenna is another option, though "wire" would likely be destroyed during entry. We would gain an enormous amount of data pertaining to the interior structure of asteroids independent of planetary defense, but we would also understand Apophis much better, which would allow optimizing the strategy to mitigate the threat Apophis poses. A related issue is that Apophis is particularly dangerous as it is a "repeat offender" and has an orbital keyhole that we would have to be careful not to nudge it into, unless we are willing to commit to actual full mitigation as discussed in this paper.

One option for a proposed set of missions exploring Apophis in particular is:

1) Friday, April 13, 2029 – 38Kkm (geo center ref) - Mission to launch a small array of penetrators and observe penetrator performance, ideally by in-situ deceleration measurements. The closest distance is extremely short, within the GEO belt (geo ref radius 42.16 Kkm), and thus this is an excellent time for a test. There is nearly 8 years to plan and execute this mission. While challenging, it is about the same as the entire Apollo program. This encounter also has the advantage that a very modest launcher could be used.
2) March 27, 2036 – 46Mkm – Large-scale testing of penetrators. The distance for this intercept is large so a suitable scale of mass of interceptors delivered would need to be carefully considered.
3) April 19, 2051 – 6.2 Mkm – 16 lunar distances - PI Apophis. Completely mitigate the risk so future generation do not have to worry about it. This would be a first for humanity.

**Table 5.** We show the next five passes of Apophis taken from the JPL Small Body Database (SBDB) for the orbital parameters [55]. The Friday April 13, 2029 close approach would be an ideal opportunity for penetrator probing and system testing though the time scale to design and launch a mission is short (~ 8 years from now).

| Date | Geocentric Distance (AU) | Current 3 sigma position error |
|---|---|---|
| April 13, 2029 | 0.0002541 (38.01 Mm) < GEO | 2.6 km |
| March 27, 2036 | 0.309797 (46.3450 Gm) | 99 Mm |
| April 19, 2051 | 0.041378 (6.1901 Gm) | 190 Mm |
| September 16, 2066 | 0.069847 (10.4490 Gm) | 700 Mm |
| April 12, 2116 | 0.023645 (3.5372 Gm) | 7 Gm |

***Planetary Defense Conference (PDC) Simulation Scenarios***

The Planetary Defense Conference (PDC) is a bi-annual international conference hosted by the International Academy of Astronautics (IAA) dedicated to studying the threat posed by asteroid impacts. For each meeting, a mock asteroid threat is fabricated and attending teams attempt to hypothetically mitigate the threat as more precise information about the object is released during the course of the 5-day exercise. For the 2021 PDC meeting, a hypothetical threat is discovered only 6 months prior to its potential Earth impact. Upon its initial discovery, the asteroid's size is uncertain and initial estimates range from as small as 35m to as large as 700m in diameter. During the course of the five days after the asteroid's discovery, increasing numbers of observations are performed and its impact probability increases from 0.04% to about 5%, and two weeks after its initial discovery the impact probability is 100%, while its size remains uncertain. Four months prior to impact, the asteroid size is limited to less than 500m, with an average estimated diameter of 140m. The impact zone is limited to a large region in central Europe. Finally, 6 days prior to impact, radar observations of the asteroid limit its size to approximately 105m and narrow the impact location to a region 23km across bordering the Czech Republic, Germany, and Austria. Its impact velocity is estimated to be 15.2km/s.

Given the brevity of the timeframe and uncertainty of the posed threat, no adequate mitigation strategy was produced for the 2021 PDC mock asteroid threat, and civil defense response and evacuation strategies were discussed instead.

As we will show, the PI terminal interdiction method is capable of mitigating threats such as the 2021 PDC mock threat in even shorter timeframes than those proposed during the exercise. Since the time between detection and impact in this hypothetical scenario was only 6 months, no feasible reconnaissance opportunities existed to further study the size and composition of the asteroid until it is within radar range. Given that the PI interdiction method can accommodate extremely short response times (as little as 1 day for a 100m asteroid), even waiting until the asteroid size is certain (~6 days prior to impact) would allow for adequate response time. However, since 100m asteroids pose a significant threat if inadequately mitigated (~100 Mt blast yield assuming 2.6g/cc density and 20km/s exo-atmospheric velocity), an intercept mission designed for the worst-case scenario could be launched before the size of the asteroid is completely determined. For example, if we assume the intercept mission is launched 4 months prior to impact when the maximum size estimate is 500m (note that even at this point, the average size estimate is much smaller), we can mitigate the threat with less than 80 tons of purely passive (or possibly chemical explosive equipped) penetrators, even in the very pessimistic energy coupling regime of 100 J/kg. Even with this extremely pessimistic assumption, mitigation is within the capabilities of upcoming heavy lift launch vehicles such as Starship and SLS Blocks 1B and 2, though multiple launchers may be needed in large threat cases and for "backup". Note that in this case, assuming 1m/s fragment dispersion velocity, we could achieve intercept times as short as 20 days for a 500m asteroid. If we opt for a lower dispersal speed, the energy required goes down rapidly (quadratic with dispersal speed) and thus the mass of the kinetic penetrators goes down correspondingly, though the fragment spread at the Earth is less. There is a trade space between the energy available for disruption, the launcher speed vs payload mass, and the intercept time. It is interesting to note that although the blast yield for a 500m asteroid is comparable to the world's nuclear arsenal if no interdiction is performed, nuclear devices are not absolutely necessary here and the kinetic energy of the penetrators alone is sufficient to fragment the asteroid into ~10m fragments which disperse at ~1m/s. The use of NED equipped penetrators, even for fission only (W82 class – 70 tons TNT/kg for the "bare" NED), would dramatically reduce the launch mass needed if high "g" capable NED penetrators were developed as discussed earlier. It must also be noted that no serious planetary threat mitigation mission would consist of only one mitigation attempt. Thus, while the initial intercept would be equipped for the worst-case scenario, it could also be equipped with instrumentation which would inform subsequent penetrator waves that would further ensure the fragmentation of the asteroid.

**PDC 2023 Threat -** The upcoming 2023 PDC threat will be approx 800m diameter, within the range of our program to mitigate. This will be discussed in detail in an upcoming paper.

**Single-launcher solution** – We have studied the possible use of a single launcher solution for passive, conventional explosives and NED modes allowing for a combined solution using one launch vehicle as a path forward to an operational planetary defense program. For extremely large threats NED penetrators would use the "sequential penetrator mode" for "hole drilling" to bring NEDs deep below the surface. This potentially allows for detonation deep within a large threat allowing for much better effective "tamping" and coupling efficiency. This will be studied as a natural extension of our ALE3D and CTH simulations, initially using an "energy pill" explosive approximation approach for both conventional and NED explosives. **To date our simulations, show that a single launch vehicle, such as a Falcon 9, Falcon Heavy, or SLS, can be used for a wide range of threats up to 1 km** and possibly beyond (largest require NED's) allowing for a vastly simplified approach to operational planetary defense. One example that we are exploring in detail is the use of a small Falcon 9 launcher that is capable of modest payload capability with positive $C_3$ (~2500kg @ 3 km/s beyond escape - $C_3$=10 (km/s)$^2$ ) that would allow mitigation of up to about 100-150m diameter rocky threats without NED's and much larger with NED's. A Falcon 9 is an example of a relatively low "buy-in threshold" with a proven track record. Having such an at-the-ready or reserved launcher of this class would be one possible approach towards an operational program.

## Conclusions

We have shown that very short time terminal defense against asteroids and comets is possible through fragmentation via hypervelocity penetrators. This technique dramatically reduces the threat by using the Earth's atmosphere to convert the parent object kinetic energy primarily into acoustical waves and secondarily into an optical signature, both of which have acceptable fluxes. We have shown that fragmentation into sizes of less than about 10m diameter is sufficient for most threatening objects. The energy required to gravitationally de-bind and scatter the fragments sufficiently with a speed of approximately 1m/s is small. With available and upcoming launch vehicles, the mitigation of targets up to 500m in diameter appears feasible with relatively short time scales of threat notice and with modest penetrators (non-nuclear). Depending on the type of asteroid/comet, the penetrators can be purely passive or enhanced with chemical or nuclear explosives for even larger targets. Given the extreme impact speeds, passive penetrators carry much more energy per mass than chemical explosives. Though not needed except against extremely large threats, current nuclear explosives would have to be redesigned given the extreme accelerations upon penetration or weakly coupled or sequential NED augmented penetrator designs ("inline drilling") would be needed. This is an area we are currently exploring. The availability to test the non NED penetrator system on the Earth, as well as using close approach asteroids for targets is a significant advantage of our mitigation method. The system could be Earth-, orbital-, and lunar-based, as could an enhanced threat detection system. If successful, such a system would allow humanity to take control of its fate for the first time.

## Acknowledgements

I gratefully acknowledge funding from NASA grants 80NSSC22K0764 and 80NSSC23K0966, NASA California Space Grant NNX10AT93H and funding from the Emmett and Gladys W Technology Fund. The assistance and technical comments of Alexander Cohen have been extremely helpful. I thank the many students involved in discussions on this project and who are authors on upcoming papers as well as being acknowledged on our website.

## History of this paper

The first version of this paper was April 7, 2015 which outlined the basic idea of hypervelocity penetrator disassembly and fragmentation and the resulting threat mitigation via using the Earth's atmosphere in the terminal mode. This was a follow on to our directed energy ablation planetary defense deflection studies we published from 2013-2016. We were searching for a much shorter time scale response which then formed the basis for this paper. We shelved the paper until reactivating it in late 2019 when we began to solve the general case of blast wave production and evolution and the optical pulse signatures for arbitrary fragment sizes. This led to large scale simulations of various threat scenarios. Much of the delay was also the "tyranny of the immediate" in our other programs that is so common in life.

## Website and additional materials

This paper and additional papers, simulations and visualizations can be found on our website:
www.deepspace.ucsb.edu under the Projects area. Data on our high flux laser tests of ablative materials relevant to supersonic atmospheric entry are also on our website. The PI website area will be updated as new materials and simulations are developed. The detailed simulations are on our group YouTube channel:
See our web resources:
https://www.researchgate.net/profile/Philip_Lubin
www.deepspace.ucsb.edu/projects/pi-terminal-planetary-defense
www.youtube.com/@UCSBDeepspace

# Appendices

## *Modeling of Fragment Cloud*

For simplicity we model the fragment cloud spatial and temporal distribution as a thick-walled cylinder. The mean radius of the cylinder depends on the mean radial speed of the fragments and the time from intercept to Earth impact. We allow for a stochastic component to both the radial speed distribution as well as a longitudinal (along the parent asteroid velocity vector) speed distribution. We also allow for a stochastic component of the fragment size distribution and density We model the fragments as spheres. For computational purposes we truncate the lower size limit of the fragments to be about 2 m in diameter.

$L_{orig}$ = parent asteroid diameter

$v_0$ = parent asteroid speed

$\rho_0$ = parent asteroid average density

$N$ = number of fragments

i= fragment number $= 1, 2, ...N$

$\rho_i$ = fragment average density

$\sigma_{rho}$ = density dispersion

$x_i, y_i, z_i =$ fragment Earth impact position in atmosphere relative to ground coordinates

$z_i = z_{b_i} =$ *fragment i burst altitude*

$z_{i-bolide} =$ fragment z (longitudinal) position relative to unfragmented bolide center of mass

$(v_{x_i}, v_{y_i}, v_{z_i}) = (x_i, y_i, z_{i-bolide}) / \tau =$ *fragment velocity vector relative bolide center of mass*

$v_i = (v_{x_i}{}^2 + v_{y_i}{}^2 + v_{z_i}{}^2)^{1/2} = (x_i^2 + y_i^2 + z_{i-bolide}{}^2)^{1/2} / \tau =$ *speed of fragment i relative bolide center of mass*

$x = y = z = 0 =$ Ground origin for ground zero of parent asteroid

$x_{obs}, y_{obs}, z_{obs} =$ Observer ground coordinates

$\theta_i =$ angle of fragment i at burst in cylindrical coordinates

$$\theta_i = \frac{2\pi}{N} i$$

$\tau =$ time from intercept to impact of parent asteroid if not fragmented

$\tau_i =$ time from intercept to impact of fragment i - allows for longitudinal speed dispersion

$t_i = z_{i-bolide} / v_0 = \tau - \tau_i =$ time difference of arrival of fragment i from mean arrival time

$L_{ave}$ = average fragment diam = $L_{orig} N^{-1/3}$

$\sigma_{diam}$ = fragment size dispersion

$L_{0_i}$ = diameter of fragment i

$v_{rad}$ = average fragment radial speed in asteroid cm frame

$v_{long}$ = average fragment longitudal (z axis) speed in asteroid cm frame

$R$ = ave cylinder radius = $v_{rad}\tau$

$\sigma_{rad-v}$ = radial fragment speed dispersion

$\sigma_{long-v}$ = longitudinal fragment speed dispersion

$NORMINV(RAND(),\mu,\sigma)$ = Gaussian random number generator with mean $\mu$ and dispersion $\sigma$
this distribution function is truncated depending on the variable so that non physical values
such as negative radii are prevented

$L_{0_i} = NORMINV(RAND(), L_{ave}, \sigma_{diam})$

$x_i = NORMINV(RAND(), R\cos(\theta_i), \tau\sigma_{rad-v})$

$y_i = NORMINV(RAND(), R\sin(\theta_i), \tau\sigma_{rad-v})$

$z_{i-bolide} = \tau * NORMINV(RAND(), 0, \sigma_{long-v})$

$t_i = z_{i-bolide} / v_0 = \tau * NORMINV(RAND(), 0, \sigma_{long-v}) / v_0$

$\rho_i = NORMINV(RAND(), \rho_0, \sigma_{rho})$

$m_i = \rho_i \pi L_{0_i}{}^3 / 6 = mass\ of\ fragment\ i,\quad v_i = (x_i{}^2 + y_i{}^2 + z_{i-bolide}{}^2)^{1/2} / \tau = speed\ of\ fragment\ i\ rel\ bolide\ cm$

$KE_{i-bolide} = KE\ of\ fragment\ i\ in\ bolide\ frame = m_i v_i^2 / 2 = m_i (x_i{}^2 + y_i{}^2 + z_{i-bolide}{}^2) / 2\tau^2$

**Computing Azimuth and Zenith Angle of each fragment relative to the Observer**

$x_i = y_i = 0$ − Origin for ground zero of parent asteroid

$x_i, y_i, z_i$ = fragment Earth impact position in atmosphere relative to ground coordinates

$x_{obs}, y_{obs}, z_{obs}$ = observer ground coordinates

$z_i = z_{b_i} = burst\ altitude\ for\ fragment\ i$

$z_i - z_{obs}$ = burst altitude of fragment relative to observer (normally observer is at $z_{obs} = 0$)

$r_o$ = horizontal ground range from observer to fragment = $\sqrt{(x_i - x_{obs})^2 + (y_i - y_{obs})^2}$

$r_s$ = slant range from observer to fragment = $\sqrt{(x_i - x_{obs})^2 + (y_i - y_{obs})^2 + (z_i - z_{obs})^2}$

$\theta_{zenith-i}$ = zenith angle of $i_{th}$ fragment relative to observer

$\phi_{az-i}$ = azimuth angle of $i_{th}$ fragment relative to observer

$x_i - x_{obs} = r_s \sin(\theta_{zenith-i})\cos(\phi_{az-i})$

$y_i - y_{obs} = r_s \sin(\theta_{zenith-i})\sin(\phi_{az-i})$

$z_i - z_{obs} = r_s \cos(\theta_{zenith-i})$

$\cos(\theta_{zenith-i}) = (z_i - z_{obs}) / r_s = (z_{b_i} - z_{obs}) / r_s$

For $z_{obs} = 0$ = obs on ground level $\rightarrow \cos(\theta_{zenith-i}) = z_{b_i} / r_s$

$\sin(\theta_{zenith-i}) = r_o / r_s$

$\tan(\phi_{az-i}) = (y_i - y_{obs}) / (x_i - x_{obs})$

***Fraction of energy in blast wave*** – As outlined in the section describing the details of the transition from exo-atmospheric entry to the burst phase, the fraction of energy in the blast wave is computed as the energy fraction dissipated between entry and burst. As discussed, our model conservatively neglects ablation and radiative terms which are more important for smaller asteroids/ fragments than for the larger ones of interest to us. The energy that appears in the shock wave is computed from the energy lost between the initial KE at exo-atmospheric entry and the KE at burst, For the fragments we are considers (typ <20m) and typical speeds, this fraction is a significant fraction of the exo-atm energy. The conclusions of this paper do not depend critically on the details of this.

$$E_{blast} = \frac{m_i v_0^2}{2} - \frac{m_i v(z_b)^2}{2} = E_0[1-(v(z_b)/v_0)^2]$$

$$E_0 = \frac{m_i v_0^2}{2} = \text{exo-atmopsheric KE}$$

$$v(z_b) = \textit{speed at burst}$$

We show the results for some relevant simulations below.

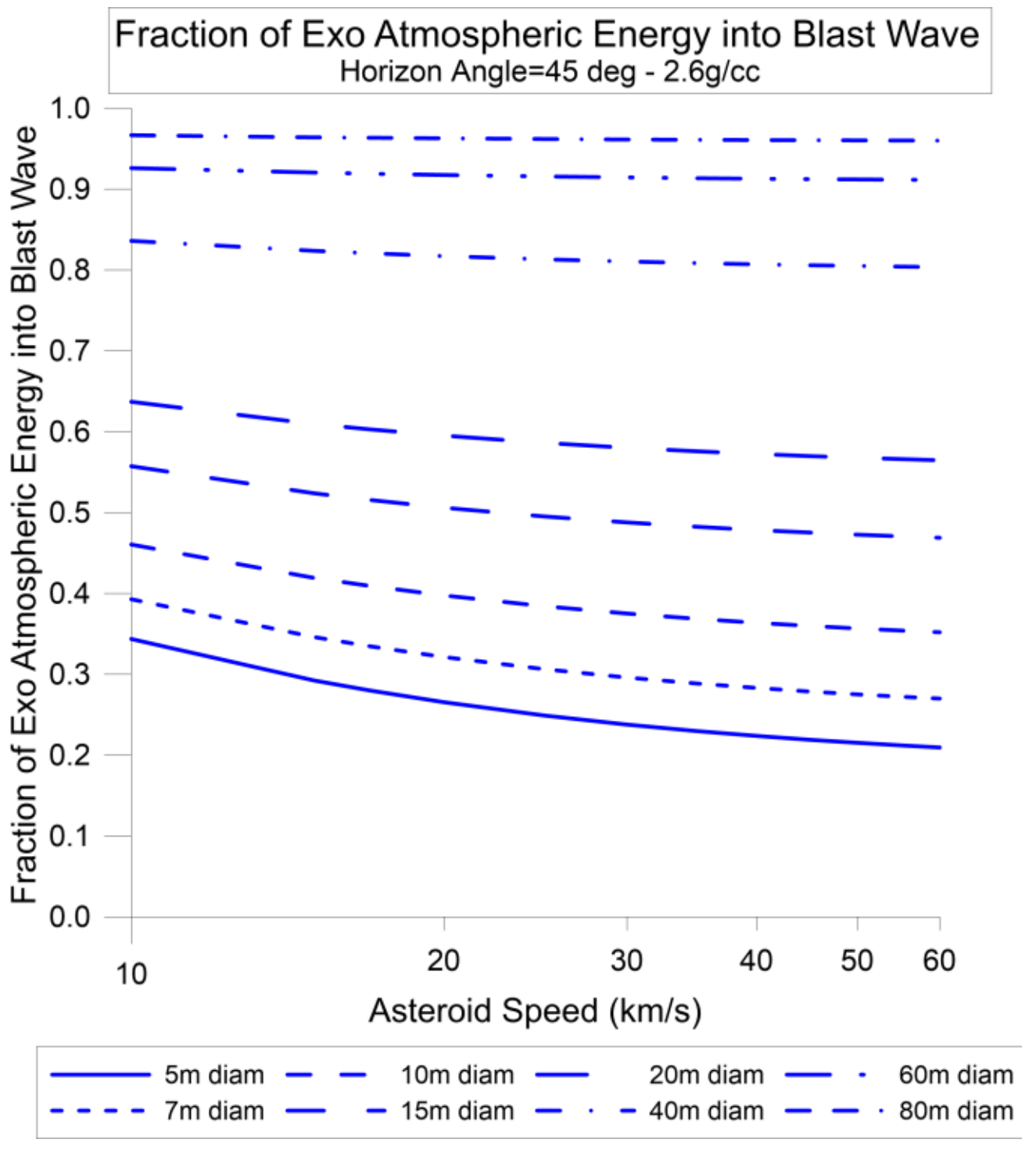

**Figure 133** – Fraction of exo-atmospheric kinetic energy that goes into the blast wave as a function of impact speed

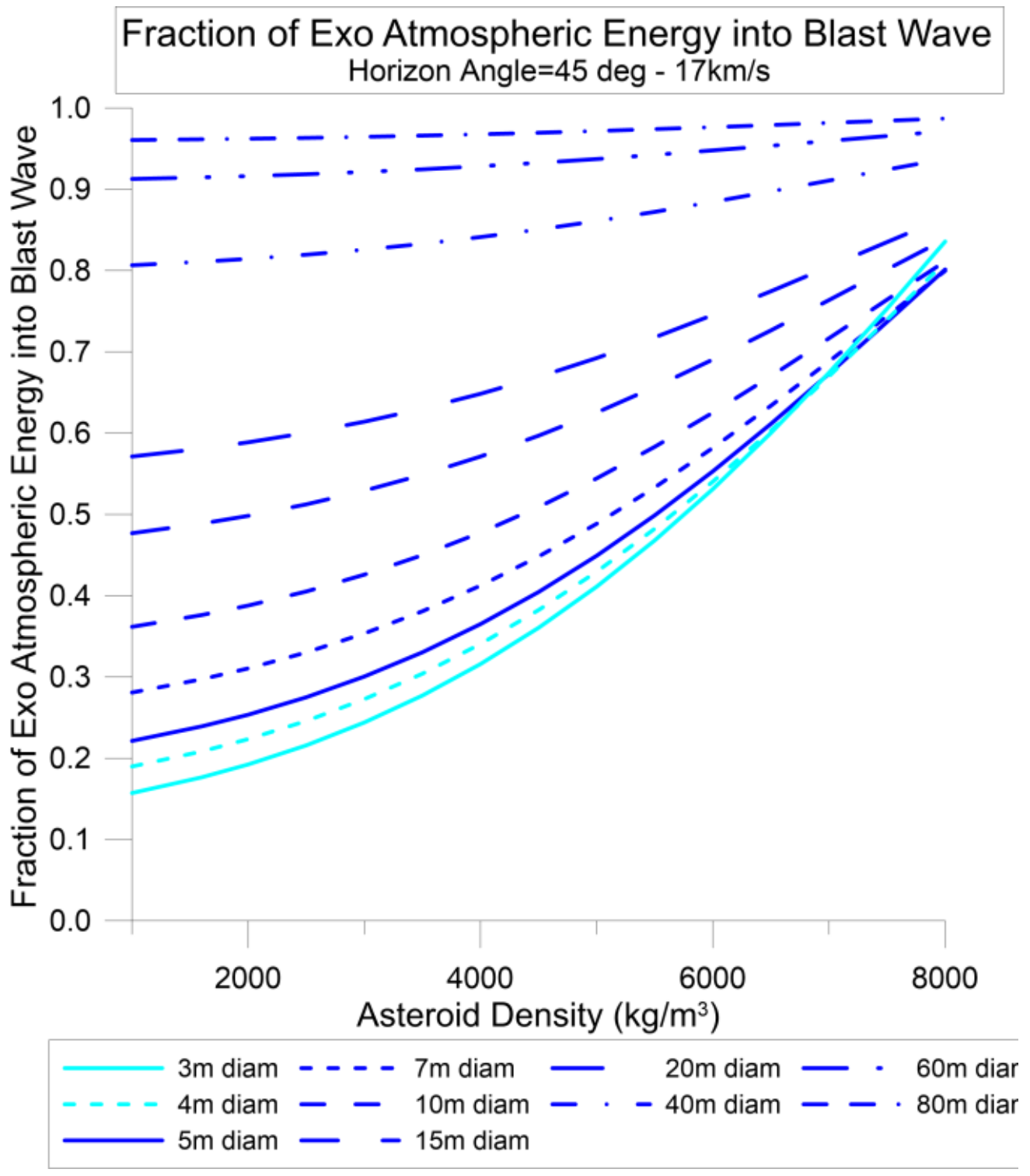

**Figure 134 -** Fraction of exo-atmospheric kinetic energy that goes into the blast wave as a function of impact speed and densitv.

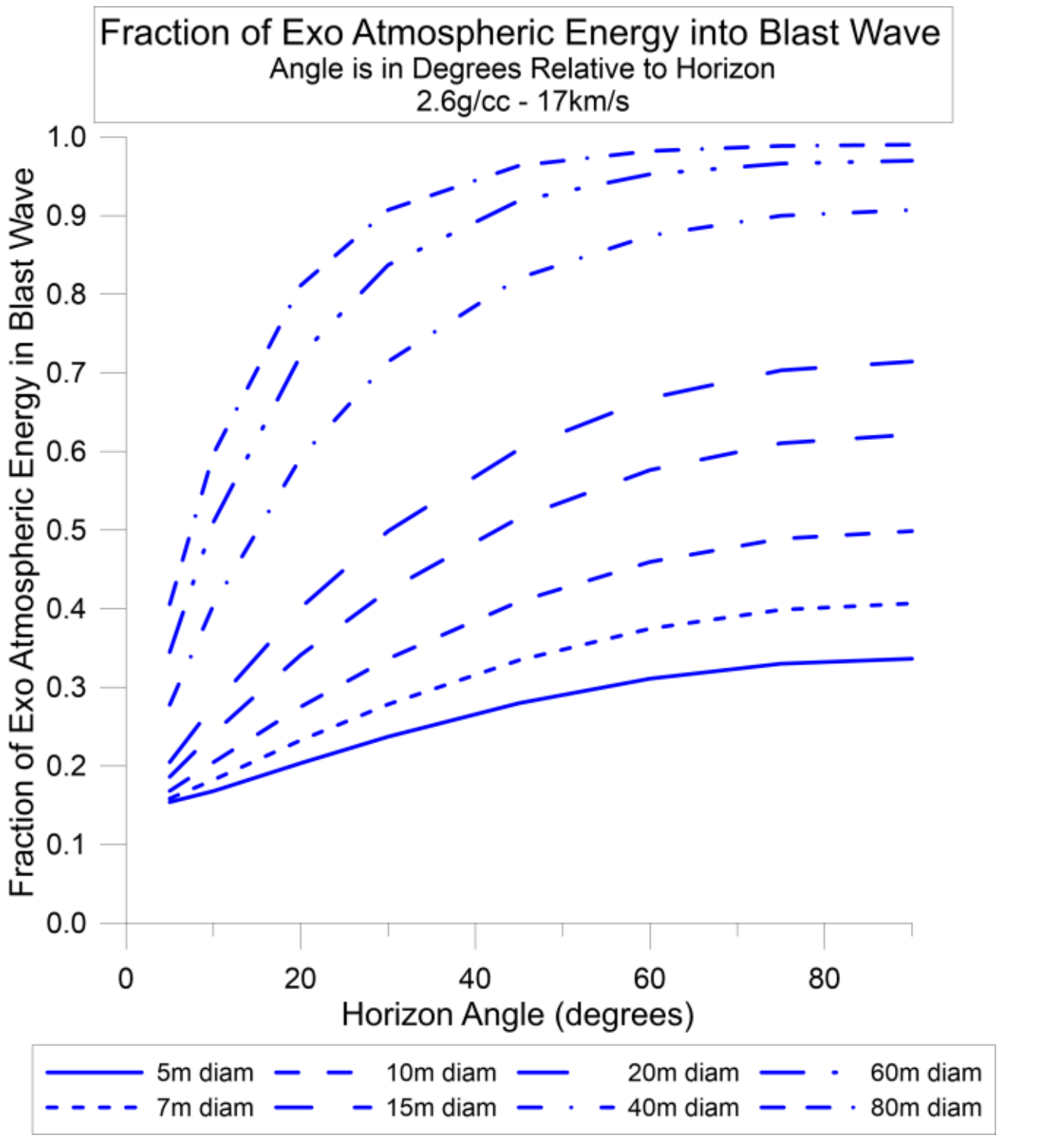

**Figure 135** -Fraction of exo-atmospheric kinetic energy that goes into the blast wave as a function of impact speed and attack angle.

***Blast wave scaling with asteroid speed, density and attack angle*** – We have analyzed a large parameter space exploring multiple scenarios. We show the results of some of these simulations, in particular showing the scaling of the blast overpressure with speed, density and angle of attack relative to the horizon for asteroids and fragments from 3 to 80 m diameter. For higher density bolides, we prefer to keep the fragments sizes smaller as seen below.

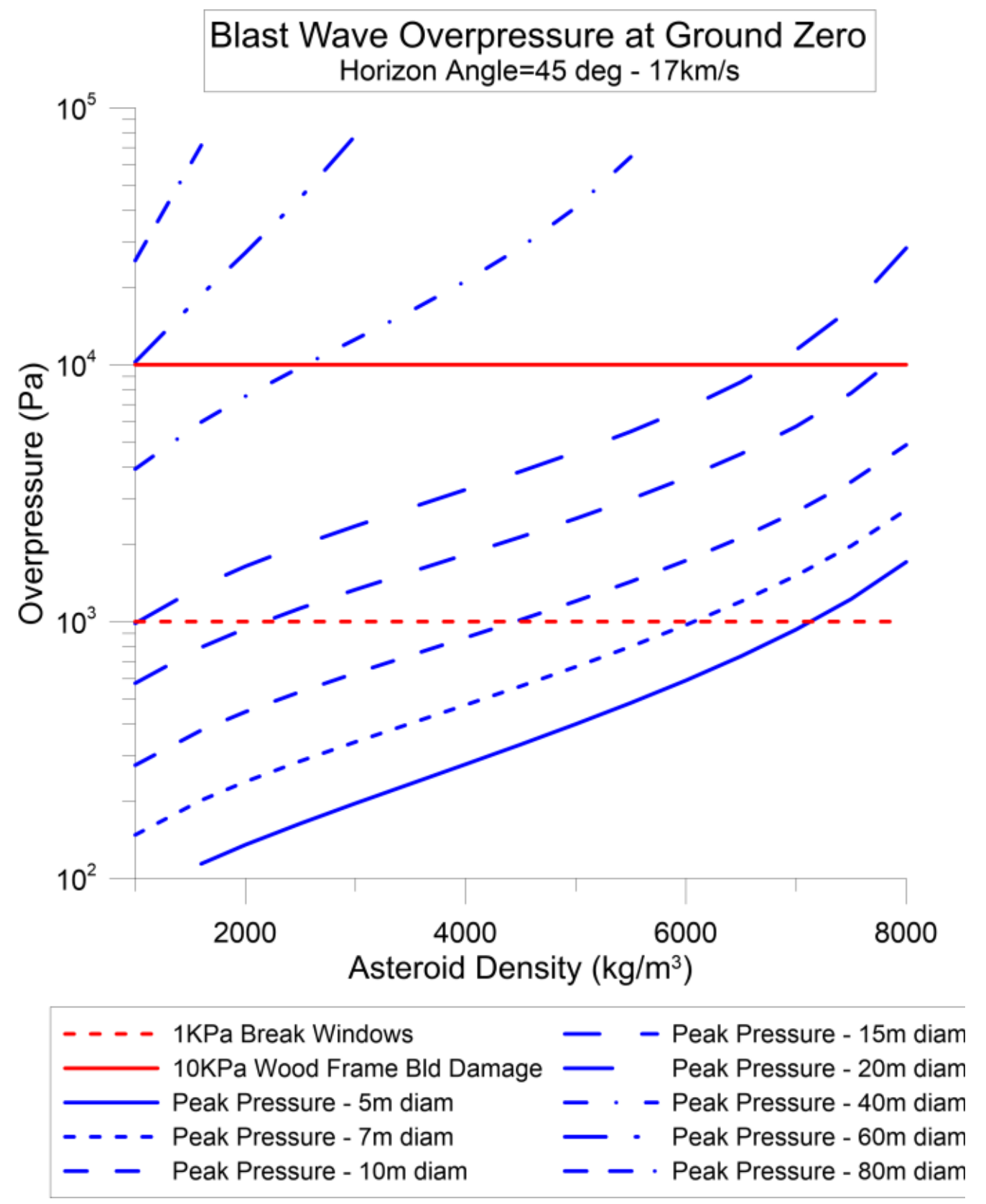

**Figure 136** - Blast wave peak pressure (overpressure) vs asteroid density at ground zero. Note the strong dependence on density. For higher density threats (rare large NiFe bolides for example – 8 g/cc), it is preferable to keep the fragment size below about 5m while for typ rocky asteroids ( <4g/cc) fragment sizes less than ~12m are acceptable. **Smaller fragments are always preferred if possible.**

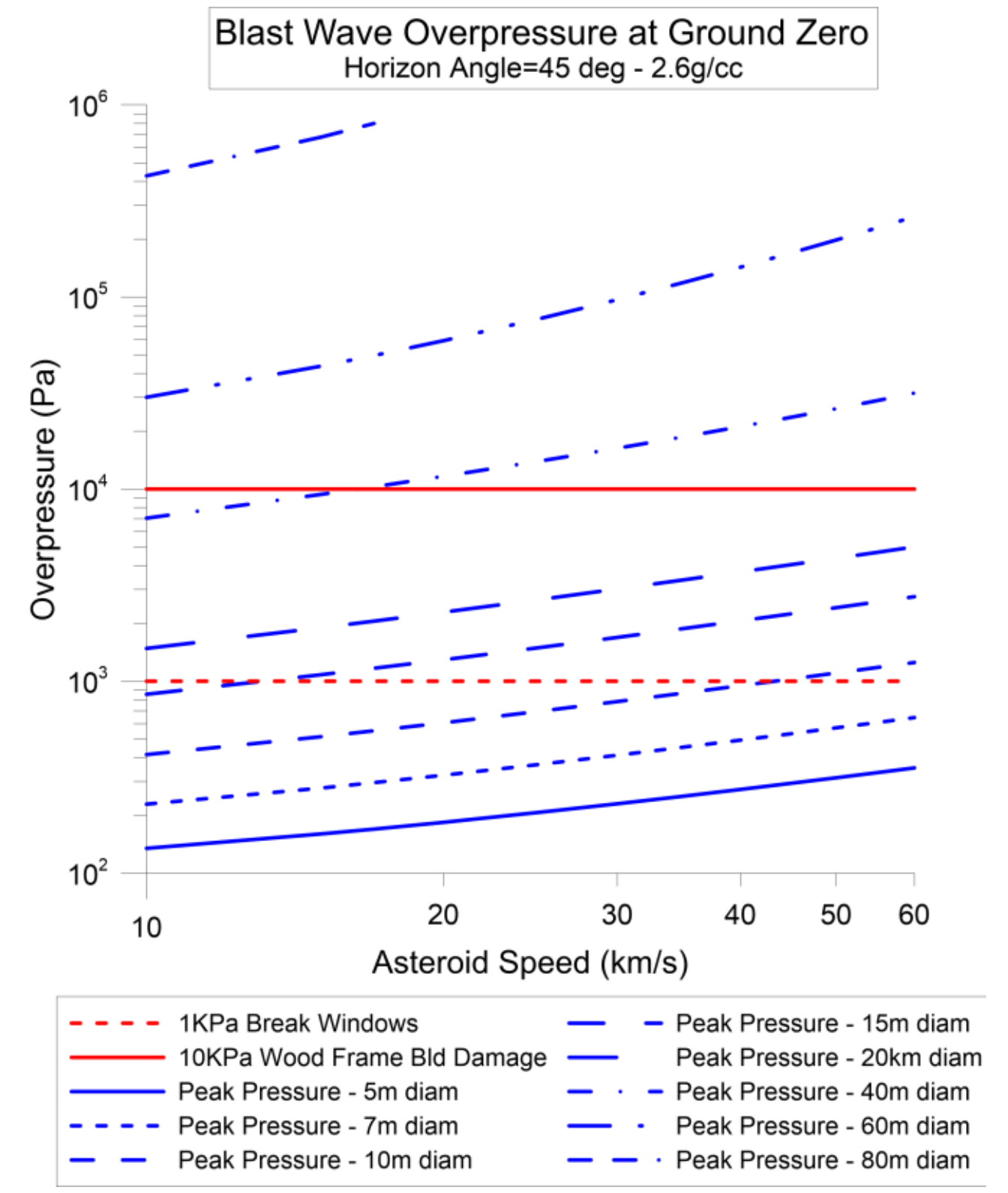

**Figure 138** – Blast wave peak pressure (overpressure) vs asteroid speed at ground zero. Note the very week dependence on speed. Higher speed leads to early bursts with a lower fraction of initial KE in blast.

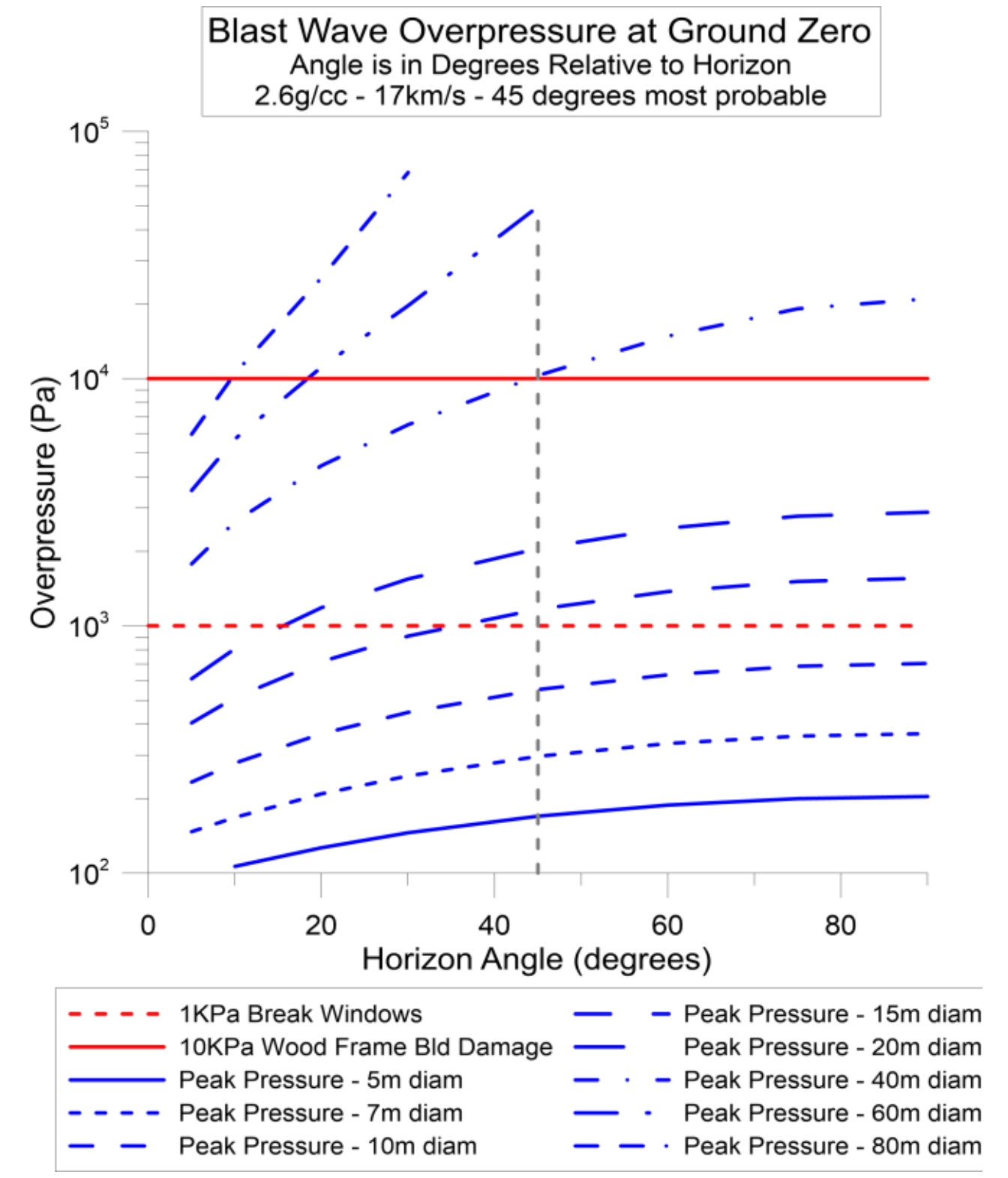

**Figure 137 -** Blast wave peak pressure (overpressure) vs horizon angle of attack at ground zero. Note the very week dependence on angle for small diameters.

*Additional atmospheric and thermal models*

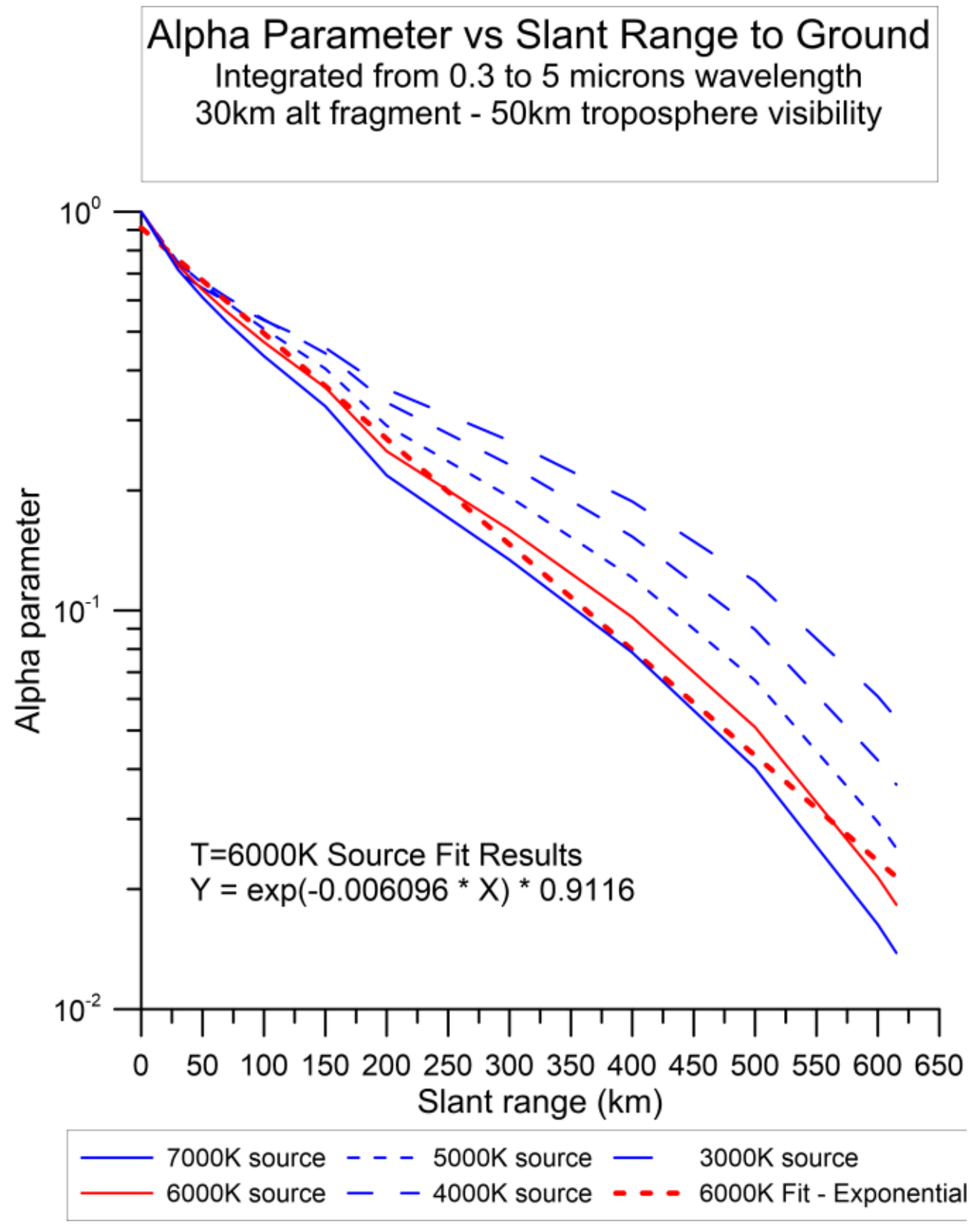

**Figure 139** – Alpha parameter (BB convolved with atmosphere) vs slant range for 30km alt bolide. for 50km troposphere visibility.

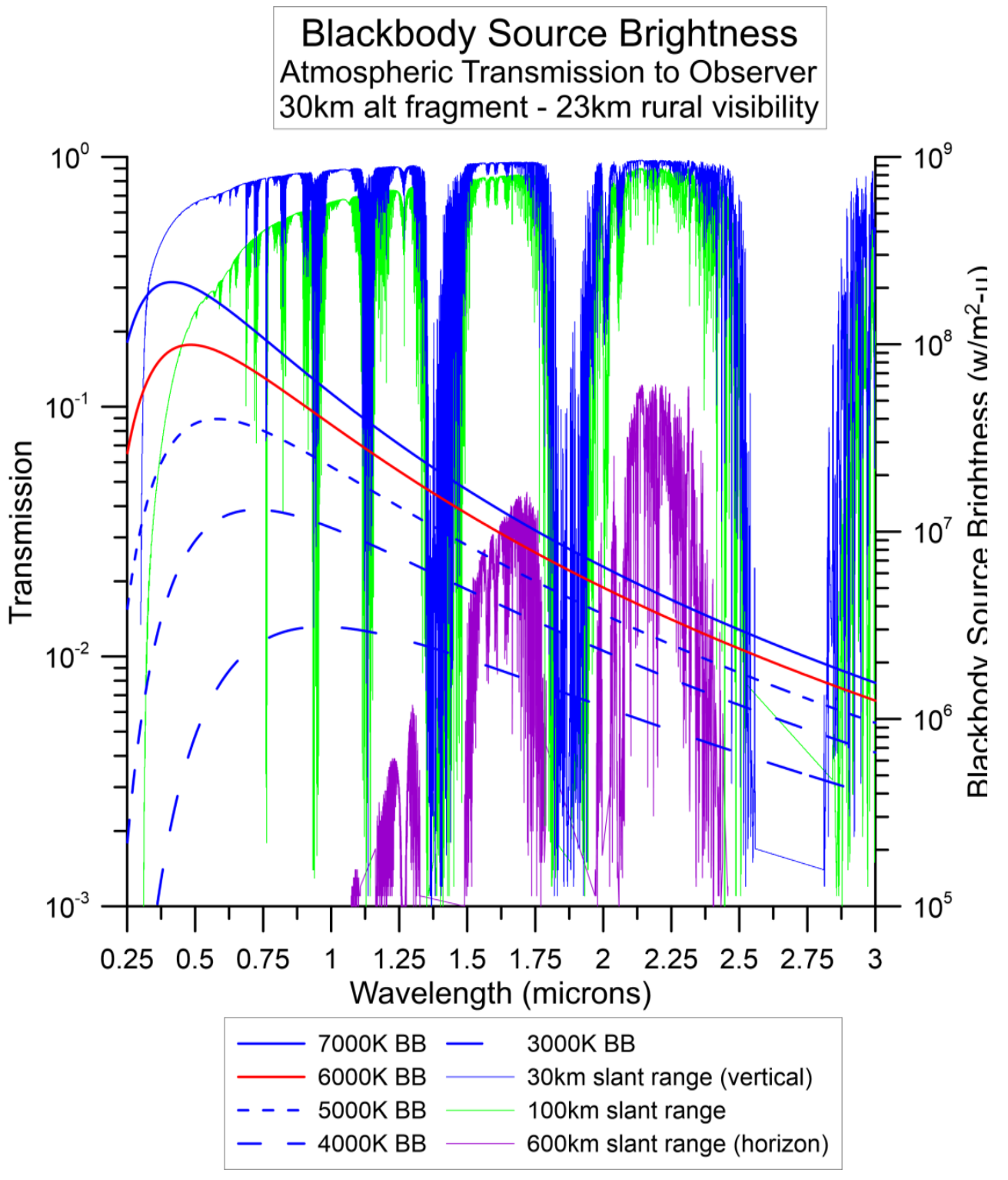

**Figure 140** – Atmospheric transmission vs wavelength from 0.3 to 2 microns with BB source brightness vs T.

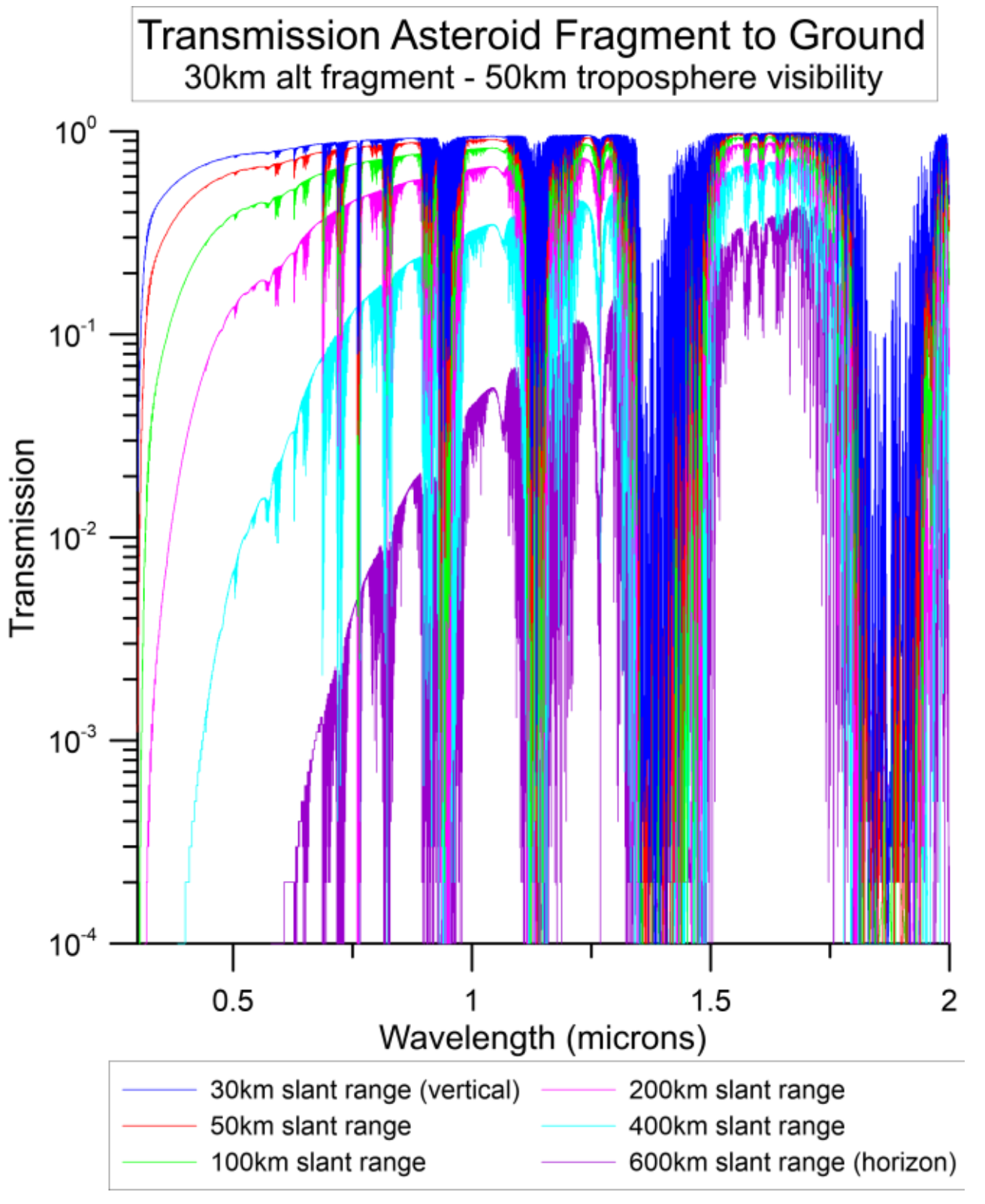

**Figure 141** – Atmospheric transmission vs wavelength from 0.3 to 2 microns vs slant range for 30km alt bolide with 50km troposphere visibility..

***Modeling the Earth's Upper Atmosphere***

The key to the success of the PI program in the terminal mode (fragments hit the Earth) is the interaction of the fragments in the Earth's atmosphere, primarily at high altitudes (>25 km). It is critical to understand the structure of the atmosphere to properly model the interaction between the hypervelocity fragments and the atmosphere. We can model the Earth's atmosphere to first order as an isotropic gravitationally bound ideal gas system assuming a spherical Earth with altitude above the surface of x where $x \ll R_E$ with density $\rho(x)$. The real atmosphere is not isothermal and varies with season, latitude and time, but this is a reasonable approximation to start with. In addition, the assumption of uniform mixing of species is a good approximation at lower altitudes (below the tropopause) where bulk motion effects such as thermal and pressure driven winds are dominant but at high altitude (above the tropopause) it is less valid. We assume uniform mixing for simplicity. Various atmospheric profiles are shown in the figures below. The range of altitudes critical for fragment interactions are 25-100km which includes the stratosphere, mesosphere and thermosphere (below).

The various layers of the atmosphere we model are commonly referred to as:

- Troposphere
  - 0 to 12-17 km depending on latitude/season. 80% of the atmosphere's mass in in this region
- Stratosphere
  - 12-55 km
- Mesosphere
  - 50-85km
- Thermosphere
  - 80-1000km with upper limit depending on solar activity, typ 700km in low solar activity
  - Densities very low
  - Temperatures (kinetic) ~ 1000K driven by and dependent on solar activity
- Exosphere
  - >1000km
  - Densities are extremely low. Concept of a "gas" is less appropriate than a particle collection
  - Mean free path extremely large (often comparable to the size of the Earth)
  - Temperatures are really kinetic temperature driven by solar activity. T~500-2000K

***Uniform mixing – Turbopause*** – The turbopause is the altitude (typ ~ 110km) below which the molecular species are relatively uniformly mixed (constant fractional density with altitude) and above which molecular separation begins. The region below the turbopause is referred to as the homosphere. As shown in the plot below, there are modest differences in the various seasonal atmospheric models of temperature (+- 5%) and pressure (+-20%) vs altitude, particularly in the critical 30-70km range where bolide fragment breakup and burst occurs. As we show below, these have only a modest effect in the blast wave pressures on the ground and thus do not significantly change the overall ground effects.

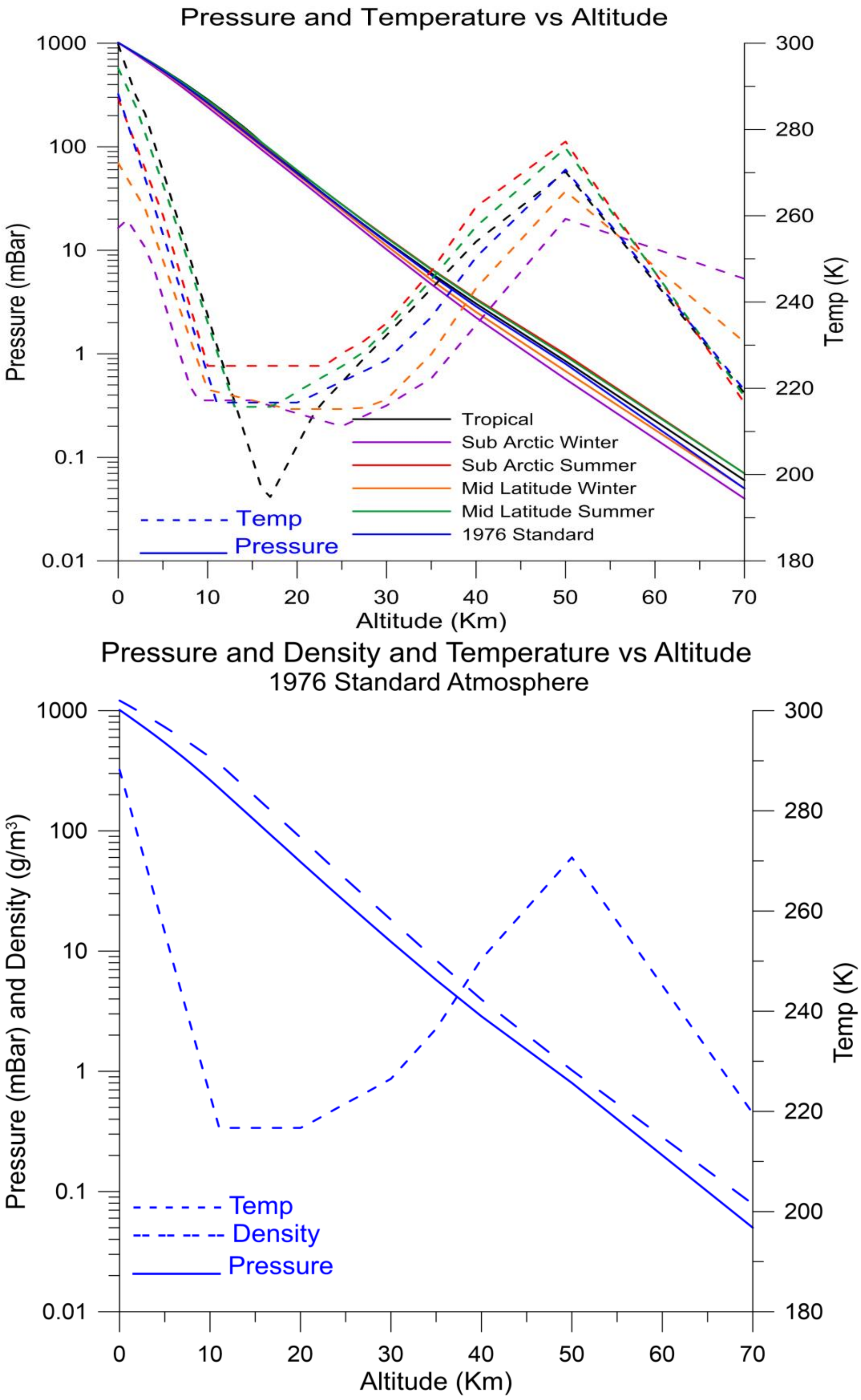

**Figure 142 -** Upper: Atmospheric Temperature and Pressure vs altitude vs atmospheric model. Note that in the critical regime of 30 to 70km altitude there is significant difference in the temperature profile between seasonal atmospheric models resulting in modest pressure differences. There are also latitude differences that are not shown. Lower: Pressure, Density and temperature for 1976 standard atmosphere which we will use as a reasonable representation in our modeling.

$$dP/dz = -\rho g$$
$$PV = nkT \rightarrow n/V = P/kT$$
$$\rho = \frac{n}{V}\sum_{i=1}^{N} f_i m_i = \frac{P}{kT}\sum_{i=1}^{N} f_i m_i \sim P/T$$
$$n/V(\#\, per\, vol) = \rho / \sum_{i=1}^{N} f_i m_i \sim P/T$$

$f_i$ = fraction in gas species i
$m_i$ = mass of gas species i

$$dP/dz = -\rho g = \frac{-Pg}{kT}\sum_{i=1}^{N} f_i m_i$$

If we assume an isothermal atmosphere with uniform mixing
[T, $f_i$ independent of x] then:

$$P(x) = P_0 e^{-\frac{gz}{kT}\sum_{i=1}^{N} f_i m_i} = P_0 e^{-z/h_o}$$

$$h_o = \frac{kT}{g\sum_{i=1}^{N} f_i m_i} = \text{scale height (e folding height)}$$

For the Earth's atmosphere uniform mixing
(species fraction independent of altitude) is a good assumption
except for water molecules. Except for $H_2O$ we have:

$f_{N_2} = 0.7809 \quad m_{N_2} = 28.01\ amu$
$f_{O_2} = 0.2095 \quad m_{O_2} = 32.00\ amu$
$f_{Ar} = 0.0093 \quad m_{Ar} = 39.95\ amu$
$f_{CO_2} = 0.0004 \quad m_{CO_2} = 44.01\ amu$
$f_{Ne} = 0.000018 \quad m_{Ne} = 20.18\ amu$
$f_{He} = 0.0000052 \quad m_{He} = 4.00\ amu$
$f_{Kr} = 0.0000011 \quad m_{Kr} = 83.80\ amu$
$1\ amu\,(atomic\ mass\ unit) \sim 1.661x10^{-27}\ kg$

$$\sum_{i=1}^{N} f_i m_i \sim 28.97 amu \sim 4.81x10^{-26}\ kg \rightarrow 2.08x10^{25}\ particles/kg$$

$$dh_o/dT = \frac{k}{g\sum_{i=1}^{N} f_i m_i} = 29.3 m/K$$

This give a nominal scale height (T~250K average) of

$$h_o = \frac{kT}{g\sum_{i=1}^{N} f_i m_i} \sim 7.3\text{km}$$

### *Space Weather Effects on the Upper Atmosphere*

For altitudes above about 100km the atmosphere can have significant temperature and density and thus scale height variations due to solar flares and other space weather effects. This only has a minor effect on the ground blast wave and optical signatures and thus is not a significant issue for our program.

### *Molecular and Atomic Species Collision Cross Sections*

We use a simple model and model each species as a sphere of radius $R_{species}$ yielding a cross section of:

$$\sigma = \pi(2R_{species})^2 = 4\pi R_{species}{}^2$$

$$Ex: N_2, O_2 \quad R \sim 0.15nm \rightarrow \sigma \sim 2.8x10^{-19} m^2$$

$$Mean\ free\ path\ \ \lambda = 1/(n\sigma)\ \ where\ n = \#\ particles/volume\ \ (\#/m\wedge 3)$$

$$Mean\ time\ between\ collisions = \lambda / v\ \ where\ v = speed\ of\ particle$$

***Ram Pressure - Ram Flux – Equivalent Temperature***

$v = bolide\ speed\ in\ atmosphere$

$Ram\ mass\ flux\ impinging : \Gamma_{mass-flux}(kg/m^2 - s) = \rho v$

$Ram\ momentum\ flux - impinging : \Gamma_{mom-flux-impinging}(kg - m/s/m^2 - s = kg/m - s^2) = \Gamma_{mass-flux} v = \rho v^2$

$Ram\ momentum\ flux - effective : \Gamma_{mom-flux-effective}(kg - m/s/m^2 - s = kg/m - s^2) = \beta \Gamma_{mom-flux-impinging} = \beta \rho v^2$

$\beta = momentum\ transfer\ coefficent\ (1 = inelastic, 2 = pure\ elastic, can\ be > 2\ with\ mass\ blowoff)$

$Ram\ pressure : P_{ram}(Pa) = \Gamma_{mom-flux-effective} = \beta \rho v^2$

$Ram\ Power\ Flux : F_{ram}(w/m^2) = Ram\ mass\ flux\ impinging * v^2/2 = \rho v * v^2/2 = \rho v^3/2$

Equivalent ram temperature (no ablation nor ionization) assuming "*forward edge*" *looking at* "*Earth*:

$$\sigma T_{ram}{}^4 = F_{ram} + \sigma T_{Earth}{}^4 \rightarrow T_{ram} = \left[\frac{F_{ram} + \sigma T_{Earth}{}^4}{\sigma}\right]^{1/4} \text{ (generally dominated by } F_{ram}\text{)}$$

**Temperatures at low densities (high altitudes)**

The interpretation of “temperature” in regimes where the mean free path is large compared to the structure size or where the mean free time between species collisions is large, leads to a definition of temperature based on the kinetic energy of the species. In this realm the term “kinetic temperature” is often used though often interchangeably with simply the term “temperature”.

**Molecular and Atomic Speeds vs Temperature**

We use a Maxwell-Boltzmann distribution function for each species to calculate the speed distributions. The most probable (maximum in the speed probability distribution, the average speed and the RMS speed are all reasonably close (~ +-10%). We use the average speed in computing the molecular collision rates vs altitude.

$Maxwell - Boltzmann\ speed\ distribution$

$M = species\ mass\ per\ mole\ (kg/mol)$

$R = N_A * k \sim 8.314 J/mol - K$

$Most\ probable\ speed\ (MB\ peak\ probability)$

$$v_{mp} = \left[\frac{2RT}{M}\right]^{1/2}$$

$Average\ molecular\ speed$

$$v_{ave} = \left[\frac{8RT}{\pi M}\right]^{1/2}$$

$RMS\ Speed$

$$v_{rms} = \left[\frac{3RT}{M}\right]^{1/2}$$

$Ex : N_2\ @\ 298K\ (25C):$

$v_{mp} = 421 m/s,\ v_{ave} = 475 m/s,\ v_{rms} = 515 m/s$

***Species Impact Energy, Shock Heating, Ionization and Plasma Cloud around Fragment***

At the typical bolide-Earth closing speeds relevant, the molecular and atomic energy per species can be above the ionization level for outer shell ionization. The results are that both the air species and the bolide surface become ionized. In addition, strong shock waves result in additional ionization with the net effect being a significant ionization cloud formed around the bolide fragment at high altitudes. This is similar to the ionization cloud formed around a re-entry vehicle during their high altitude and high speed trajectories.

For example, the energy of a nitrogen molecule (28 amu) is 0.145 eV/(km/s)$^2$ or 5.18 meV//(km/s)$^2$ per AMU. The energy of a oxygen molecule (32 amu) is 0.166 eV/(km/s)$^2$

For a nitrogen molecule at 10 km/s closing speed this is 14.5 eV and 58.0 eV for a 20km/s closing speed, 131eV for 30km/s and 232 eV for 40km/s closing speed. For a oxygen molecule at 10 km/s closing speed this is 16.6 eV and 66.4.0 eV for a 20km/s closing speed, 149eV for 30km/s and 266 eV for 40km/s closing speed. Note that for LEO re-entry speeds of 7.8km/s the $N_2$ impact energy is about 8.82 eV and 10.1 eV for $O_2$. (1 eV~1.602x10$^{-19}$J). The first ionization energy of O is 13.5 eV while that of $O_2$ is 12.2 eV while the first ionization of N is 14.5 eV while that of $N_2$ is 15.5 eV. The direct impact energy is very close to the first ionization energy for LEO re-entry speed and just at threshold for a speed of 10km/s to ionize the dominant species of air ($N_2$/$O_2$). Another effect that increases the ionization rate is the speed of the air molecules in the rest frame of the atmosphere. Typical speeds are about 500m/s at a nominal temperature of about 300K while in the upper atmosphere (above 120 km) the temperatures can be over 1000K with molecular and atomic speeds over 1 km/s. A part of the speed distribution will have greater than 2x average giving a closing speed for LEO re-entry (and 300K atmosphere) of 7.8+1=8.8 km/s giving an impact KE of about 11.2 eV for $N_2$ and 12.9eV for $O_2$ which is just above the first ionization energy. It is clear that even at LEO re-entry speed in the cooler parts of the atmosphere, ionization will begin from the "tail" of the molecular speed distribution. This all assumes completely inelastic scattering, yet some fraction of the molecules will undergo partial elastic scattering and then "hit" other molecules which will rapidly initiate ionization. Yet another effect is involved, namely that of shock wave heating which rapidly compresses the air molecules (a form of extreme adiabatic compression) and this effect can dissociate and ionize the air which results in more ionization. The combination of rapid compression (shock heating), direct impact energy via both inelastic and quasi elastic collisions as well as the additional impact speed from the speed of the air molecules in the ret frame of the air is sufficient to ionize air at re-entry speeds (7.8 km/s) and above producing an ionization (plasma) cloud around the "incoming" object even for LEO re-entry let alone bolide speeds which often have far greater closing speeds that LEO speeds. In our solutions for the speed of the fragments slowing down with air drag there is still an ionization cloud, often all the way until burst.

### *Light production from Bolide Impact with the Atmosphere*

In addition to the plasma cloud around the entering bolide fragments there is also light production from both direct molecular impacts and from shock heating causing electronic transitions and recombination radiation resulting is direct light production. The sum of all the processes including direct heating and ablation of the bolide fragments, direct impact and shock heating ionization and electronic transitions, rotational and vibrational (far IR) from molecular impacts and shock heating yields a large variety of mechanisms that produce light in the UV, visible and IR from the high speed fragment entry. This gives a complex spectrum of emitted radiation emission that we "interpret" with our eyes as a "white hot" glowing object but the actual spectrum is much more complex than a pure blackbody, though a blackbody of roughly 6000K is a reasonable approximation for our purpose of computing ground effects.

### *Ground Effects from Varying High Altitude Modeling*

In the figures below we summarize the various high altitude atmospheric modeling issue from ground to 1000km altitude. The critical bolide interaction regime is between 25 and 120km, for the fragment sizes we are consider (up to 20m diameter). We also show the ground effect shock wave pressures for fragment sizes from 2-20 meter for various models of the high altitude atmosphere and **show there is only a modest effect on the ground shock wave for reasonable variations in the atmosphere.**

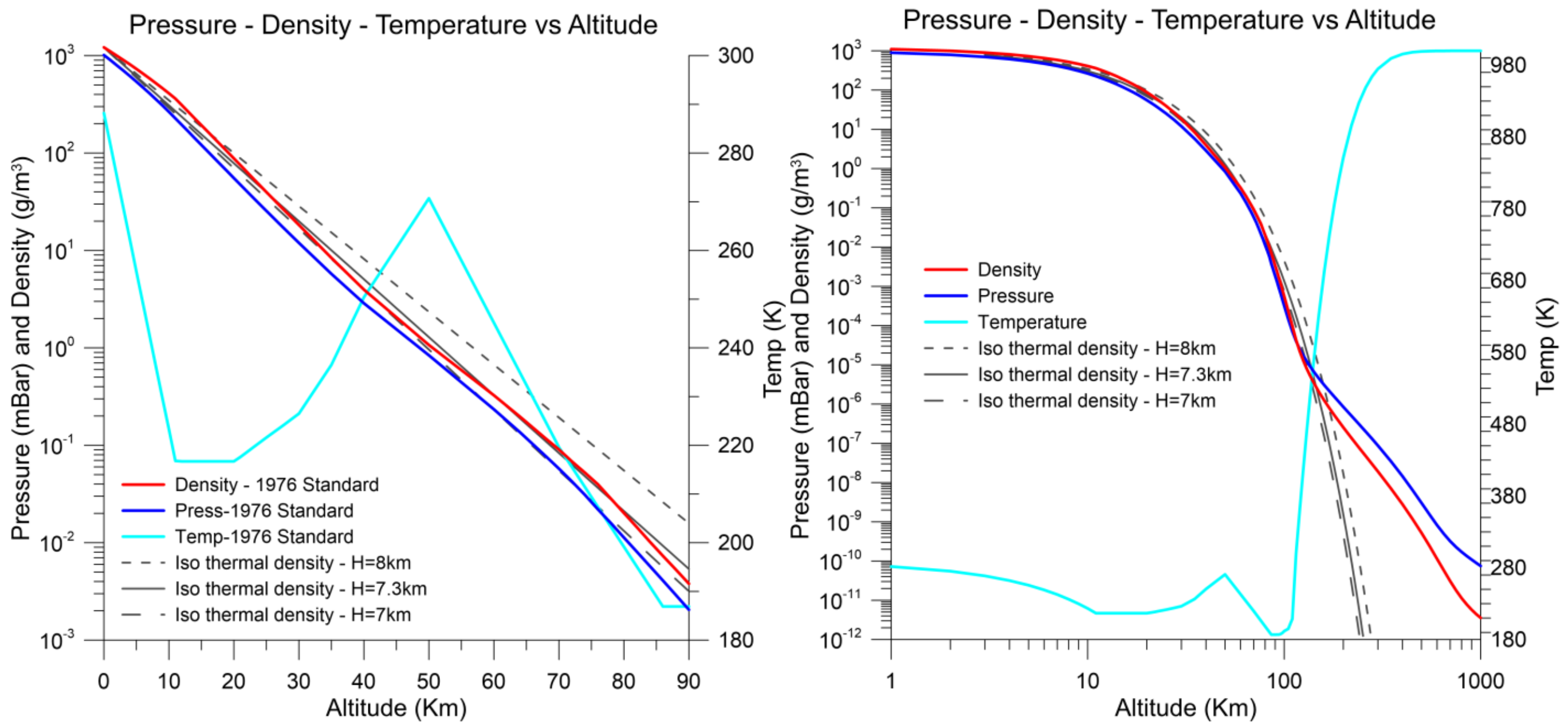

**Figure 144 –** Left: Atmospheric pressure, density and temperature up to 90 km. Note that pressure and density are relatively well fit by an iso-thermal self-gravitating model with a scale height of H=7.3km. As we show later, the ground effect for H from 7 to 8km is only modest. Right: Pressure, density and temperature up to 1000km. Above about 120km the atmosphere is strongly heated and affected by the sun (UV, solar wind, flares) with significant space weather effects.

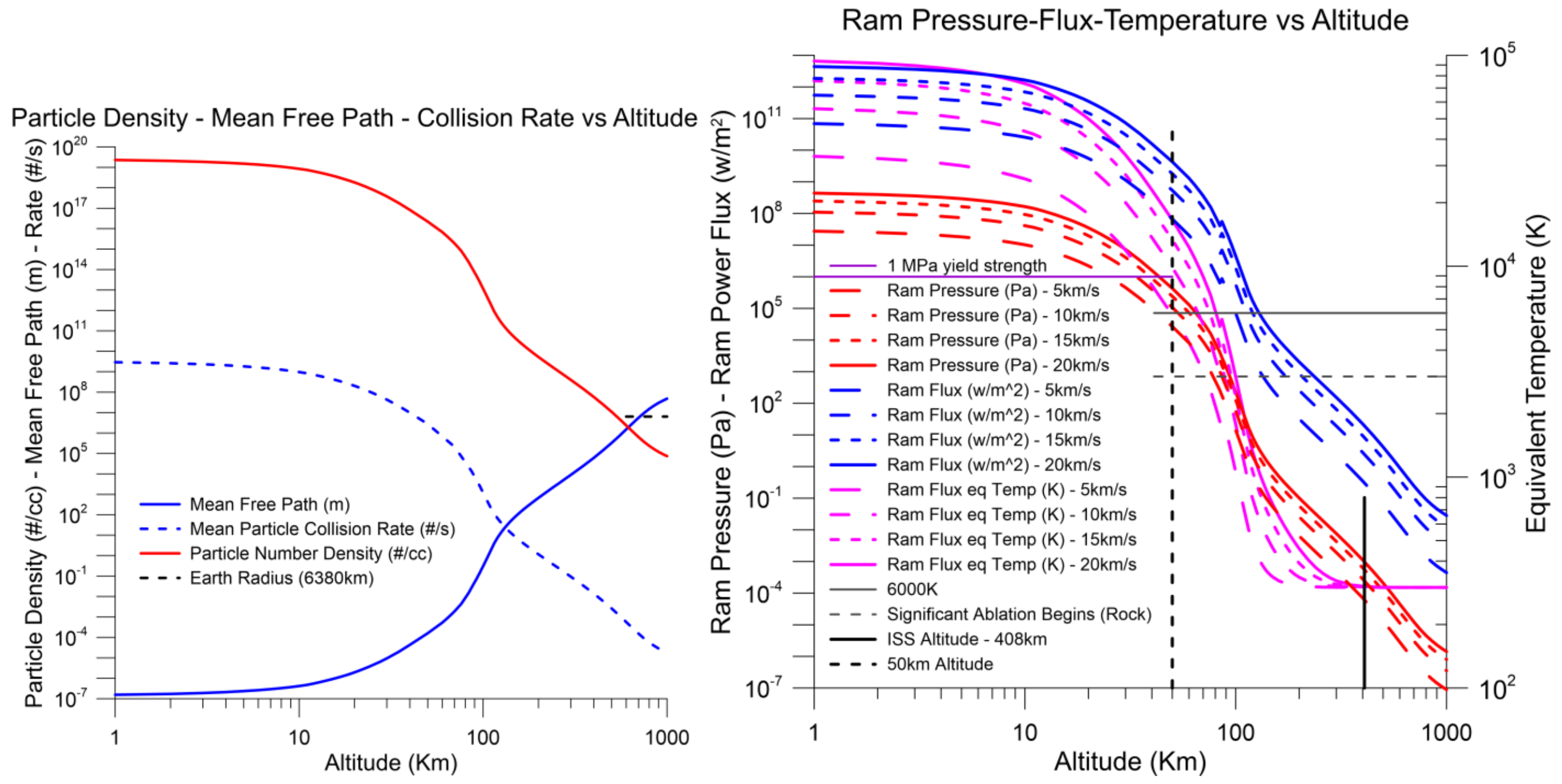

**Figure 143 –** Left: Atmospheric particle density, mean free path between collisions and collision rate up to 1000km. Right: Bolide atmospheric ram pressure, power flux and equivalent temperature for closing speeds of 5,10, 15 and 20km/s. The speeds are shown independent of altitude to show the various effects. In practice the bolide speed slows down with decreasing altitude. Note that the equivalent temperature assumes no ablation. Ablation (see next figures) will limit the actual temperature depending on the bolide thermophysical properties. For rocky type asteroids, ablation will typically limit the physical surface temperature to about 4000-7000K depending on the material composition and closing speed. An example of this is shown earlier in the paper. Vapor pressure vs flux for various refractory compounds as well as water are shown in the next figure from our work on laser ablation. Note the ram pressure at 50km altitude for a 20km/s bolide is about 1 MPa which is consistent with the Chelyabinsk and Tunguska event bolide yield strength estimates. Higher strength bolides will breakup and burst at lower altitudes. Note the very rapid rise with decreasing altitude (increasing air density) so that even high yield strength asteroids will break up and burst typically above 10 km for even very high strength asteroids (10-100 MPa yield) even as they slow down.

*Vapor Pressure, Temperature, and Mass Ejection Flux from Bolide in Upper Atmosphere*

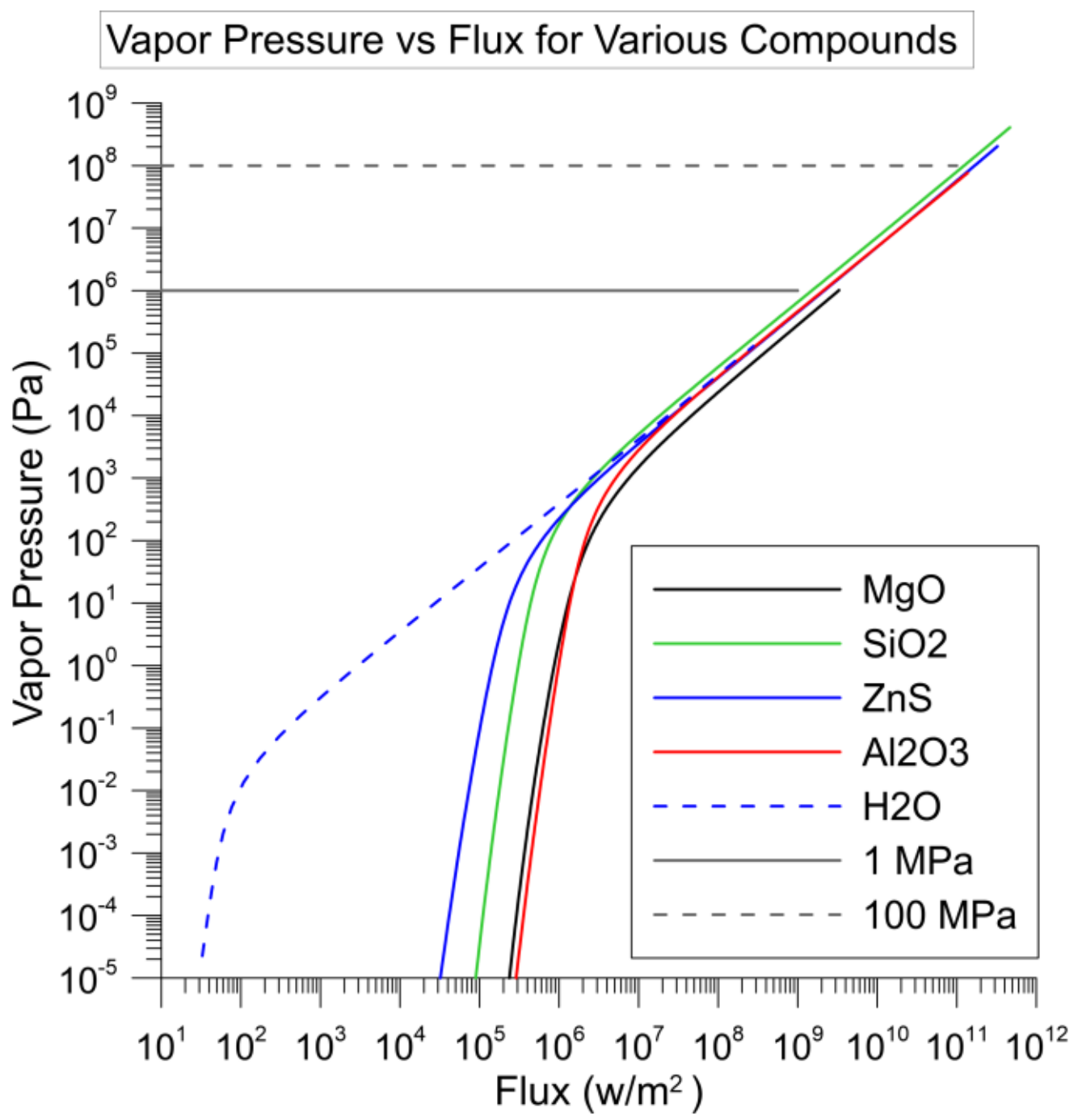

**Figure 145 –** Bolide ablation vapor pressure vs atmospheric ram flux for various compounds including high temperature refractory compounds and low temp ($H_2O$).

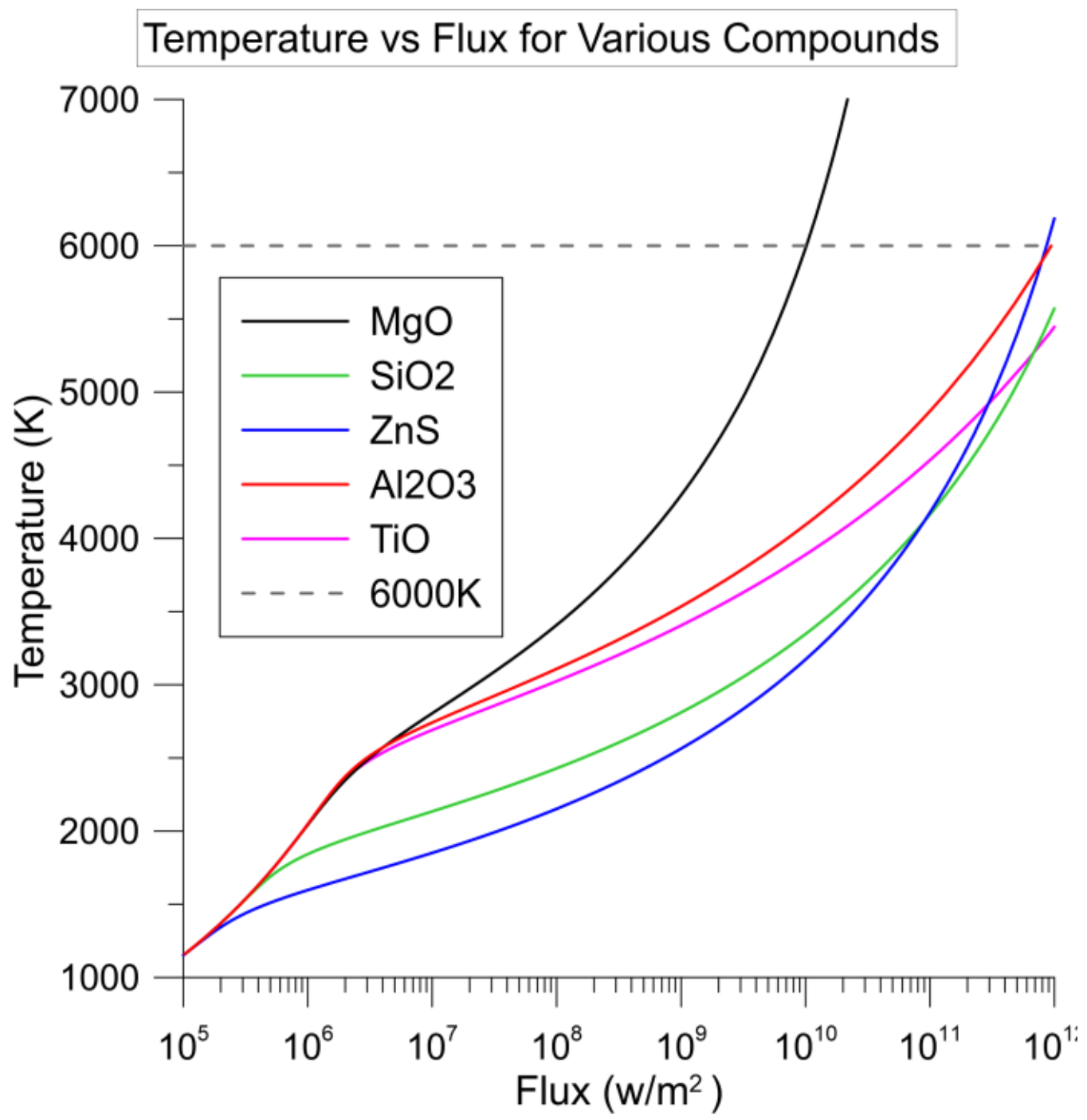

**Figure 146 –** Bolide ablation temperature vs atmospheric ram flux for various high temperature refractory compounds.

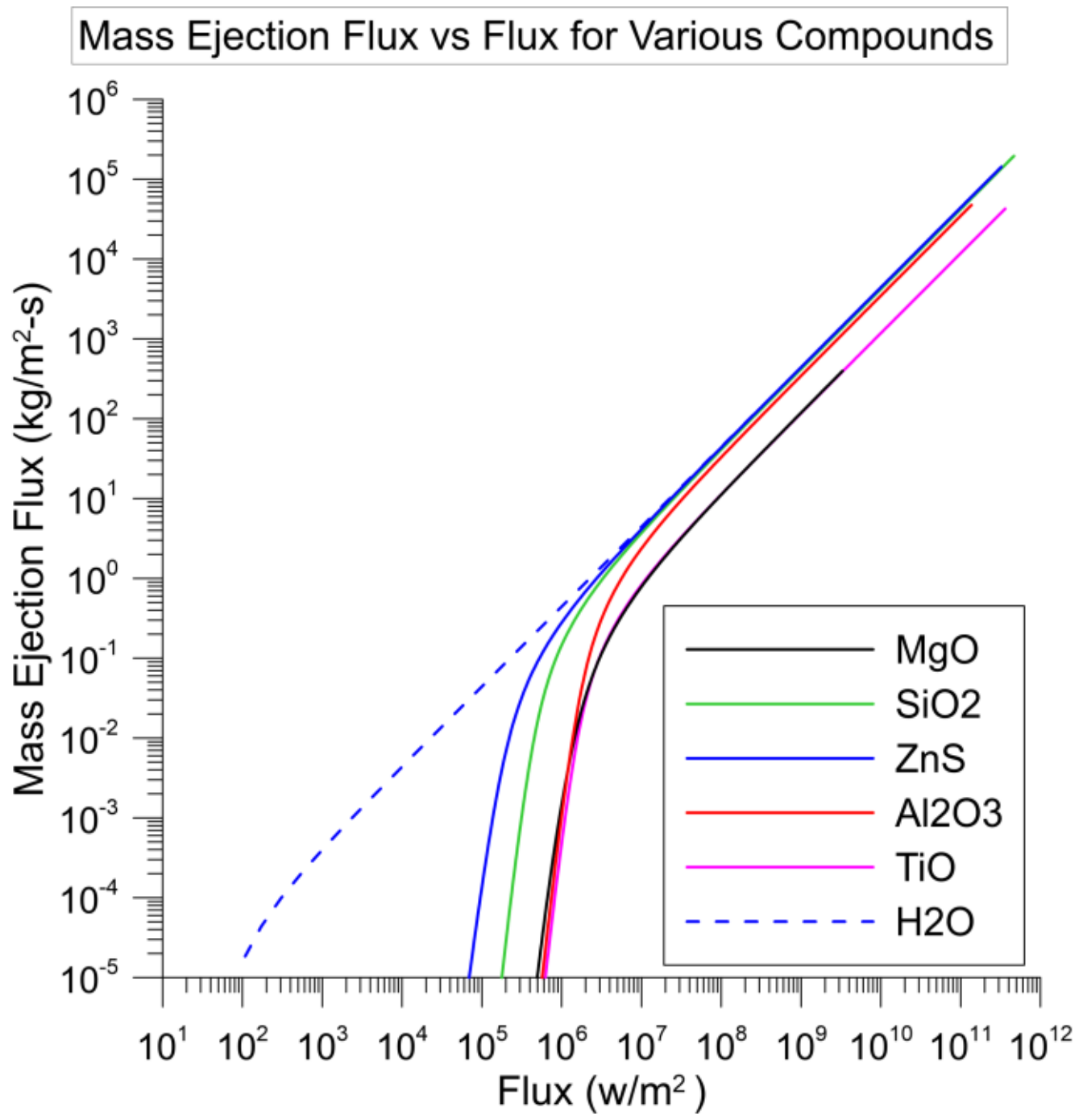

**Figure 148** – Ablation mass ejection flux from bolide vs atmospheric ram flux.

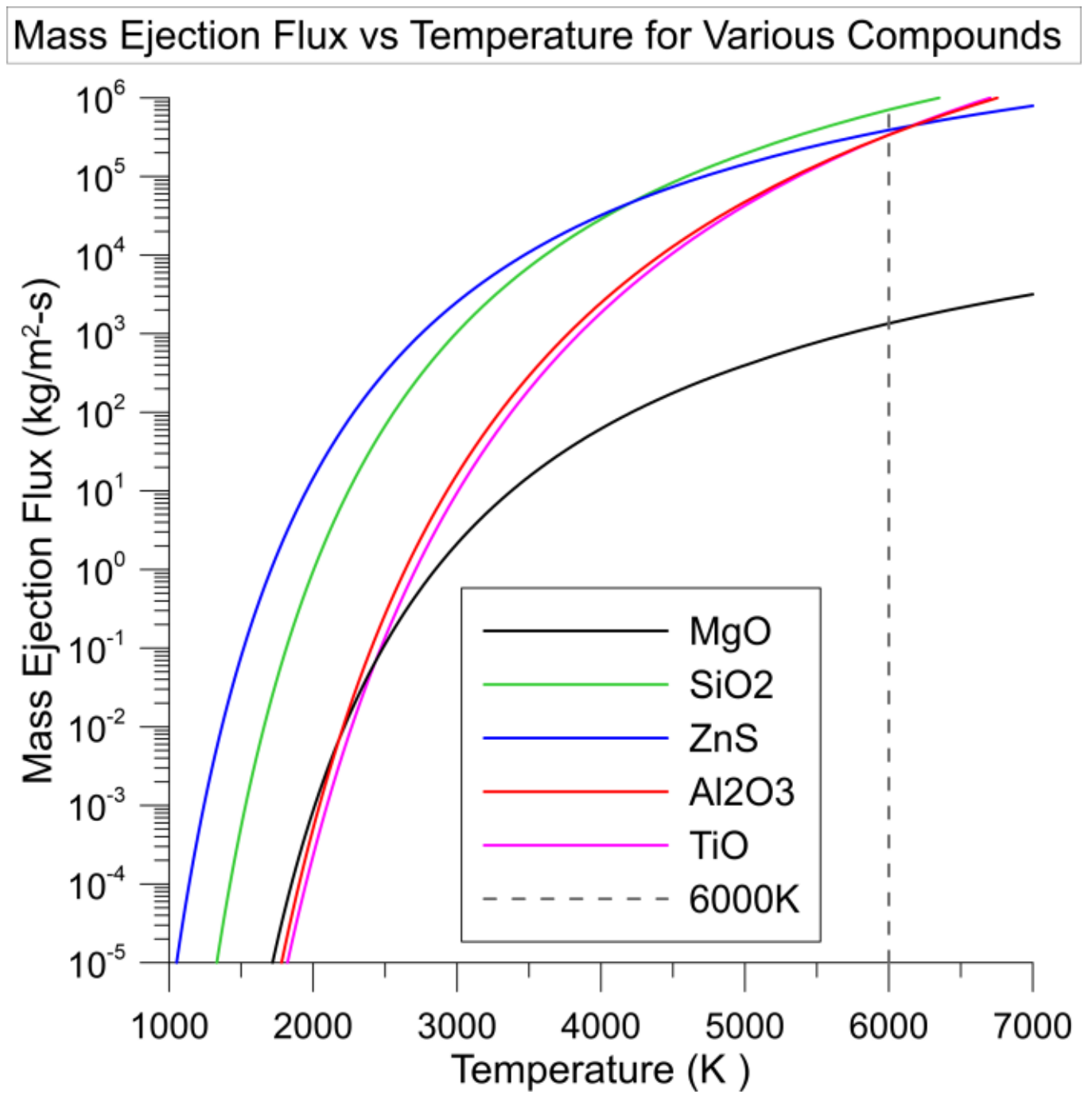

**Figure 147 –** Ablation mass ejection flux from bolide vs bolide temperature at forward edge due to atmospheric ram flux.

## *Atmos Scale Height and Temperature vs Altitude – Bolide Breakup, Burst Altitude, Blast Pressure and Blast Yield vs Diameter*

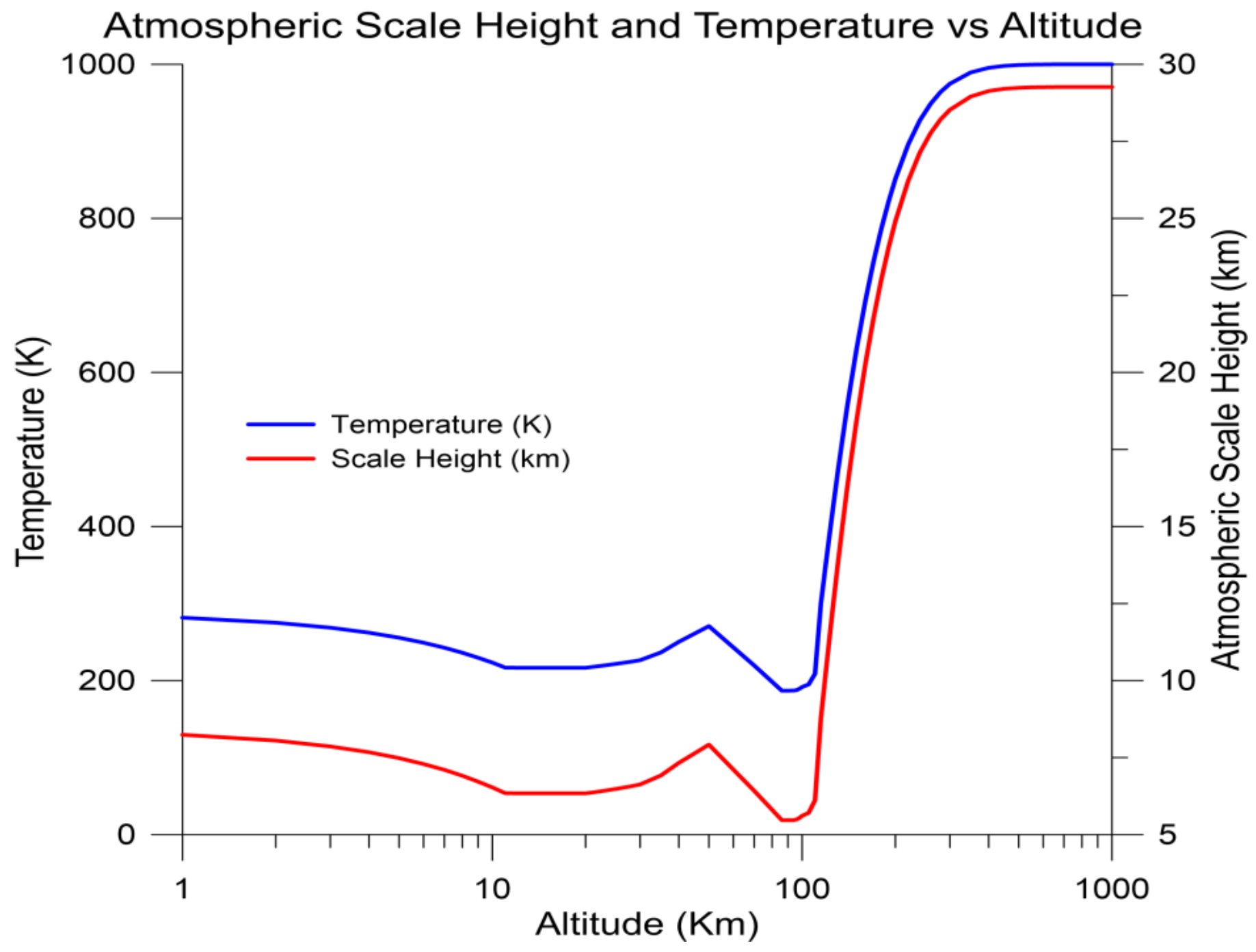

**Figure 149 -** Atmospheric temperature and equivalent scale height vs altitude. Note rapid increase in temperature and scale height above 120km altitude.

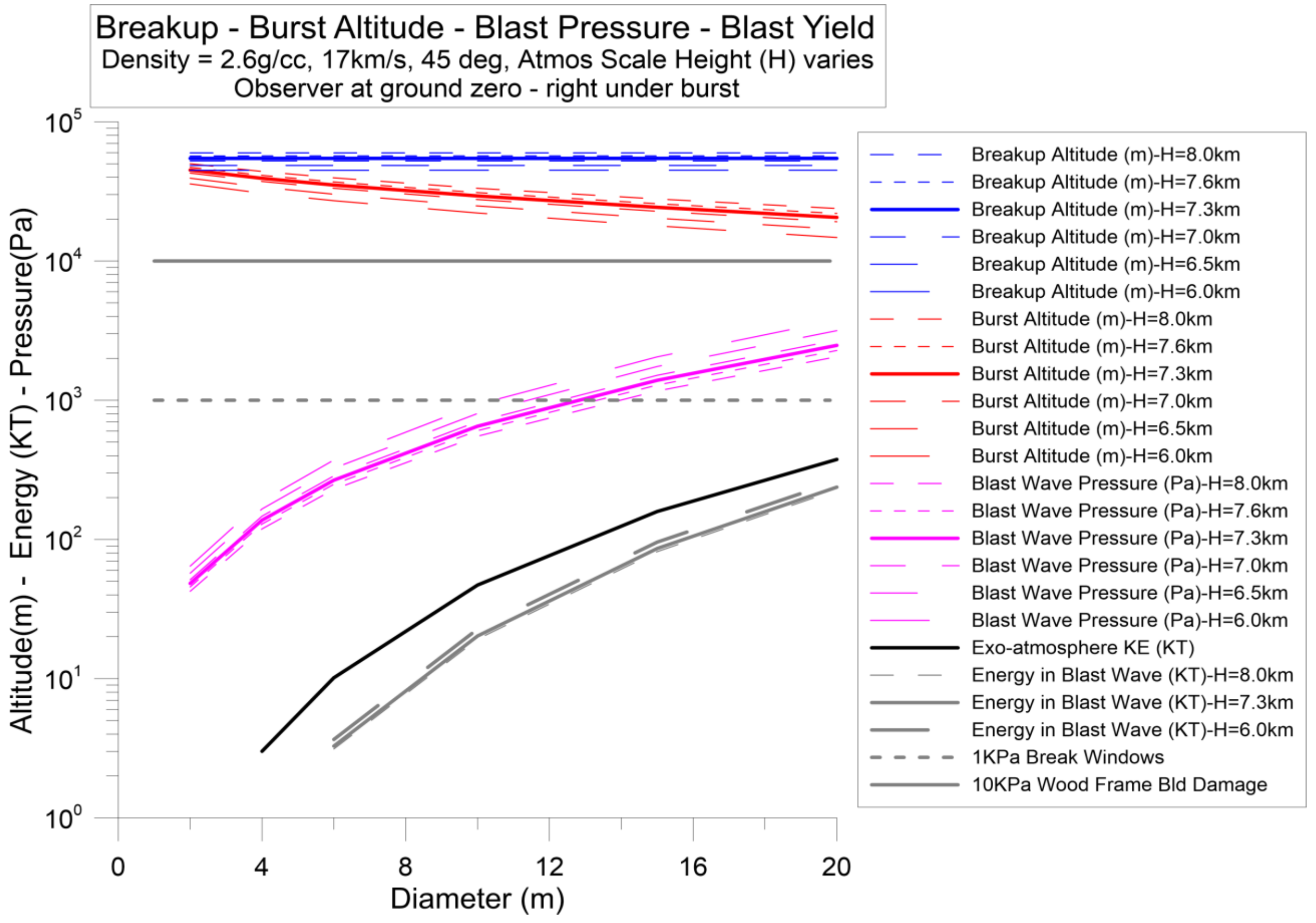

**Figure 150 –** Fragment breakup and burst altitude as well as ground zero blast pressure and total blast wave energy for varying scale heights from H=6 to H=8km. The difference in ground effects is modest for varying H. We assume H~7.3 to 8km as good fits to atmospheric profiles up to 100km altitude below which the critical atmospheric interactions occur.

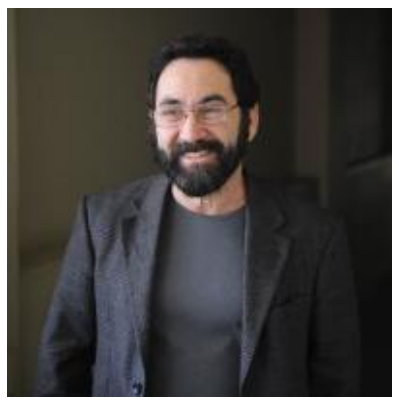

**Philip Lubin** is a professor of Physics at the University of California, Santa Barbara. He received his Ph.D. in Physics from UC Berkeley. His primary work is in studies of the early universe and he is co-recipient of the Gruber Prize for Cosmology in 2006 and 2018. He has more than 4 decades of experience in designing, building and deploying far IR and mm wave systems for ground, airborne and orbital applications.

# List of Figures